\documentclass[11pt,a4paper,twoside]{report}
\usepackage{psfrag,epsfig}
\usepackage{latexsym}
\usepackage{a4}
\usepackage{fancyheadings}
\usepackage{amsmath}
\usepackage{amssymb}
\usepackage{cite}
\usepackage{caption}
\usepackage{alltt}
\usepackage{verbatim}
\usepackage{lscape}

\newcommand{\ttt}{{\cal T}}
\newcommand{\ord}{{\cal O}}
\newcommand{\xmin}{x_{\rm{min}}}
\newcommand{\xmax}{x_{\rm{max}}}
\newcommand{\itot}{I_{\rm{tot}}}
\newcommand{\mtot}{M_{\rm{tot}}}
\newcommand{\xitot}{\xi_{\rm{tot}}}
\newcommand{\deff}{D_{\rm eff}}
\newcommand{\veff}{V_{\rm eff}}
\newcommand{\pin}{P_{\rm in}}
\newcommand{\entspr}{\,\,\hat = \,\,}
\newcommand{\bldx}{\mbox{\boldmath $\xi$}}
\newcommand{\perint}{\int_{-L/2}^{L/2}}

\newcommand{\sect}{section }
\newcommand{\sects}{sections }
\newcommand{\ch}{chapter }
\newcommand{\Ch}{Chapter }
\newcommand{\chs}{chapters }
\newcommand{\Chs}{Chapters }
\newcommand{\eq}{equation }

\newcommand{\eqs}{equations }
\newcommand{\fig}{figure }

\newcommand{\figs}{figures }

\def\figeins{
\begin{figure}[ht] 
\centerline{\epsfig{file=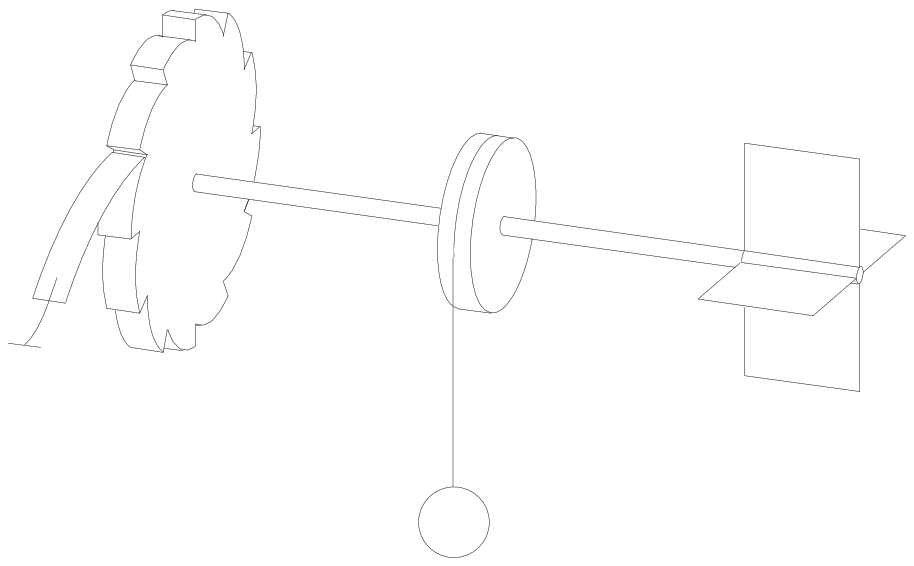, width = 10 cm} }
\caption{
Ratchet and pawl. The ratchet is connected by an axle 
with the paddles and with a spool, which may lift a load. 
In the absence of the pawl (leftmost object) and the load,  
the random  collisions of the surrounding gas molecules 
(not shown) with the paddles cause an unbiased 
rotatory Brownian motion.
The pawl is supposed to rectify this motion so as to lift 
the load.
} 
\label{fig1} 
\end{figure} 
}

\def\figzwei{
\begin{figure}[ht] 
\centerline{\epsfig{file=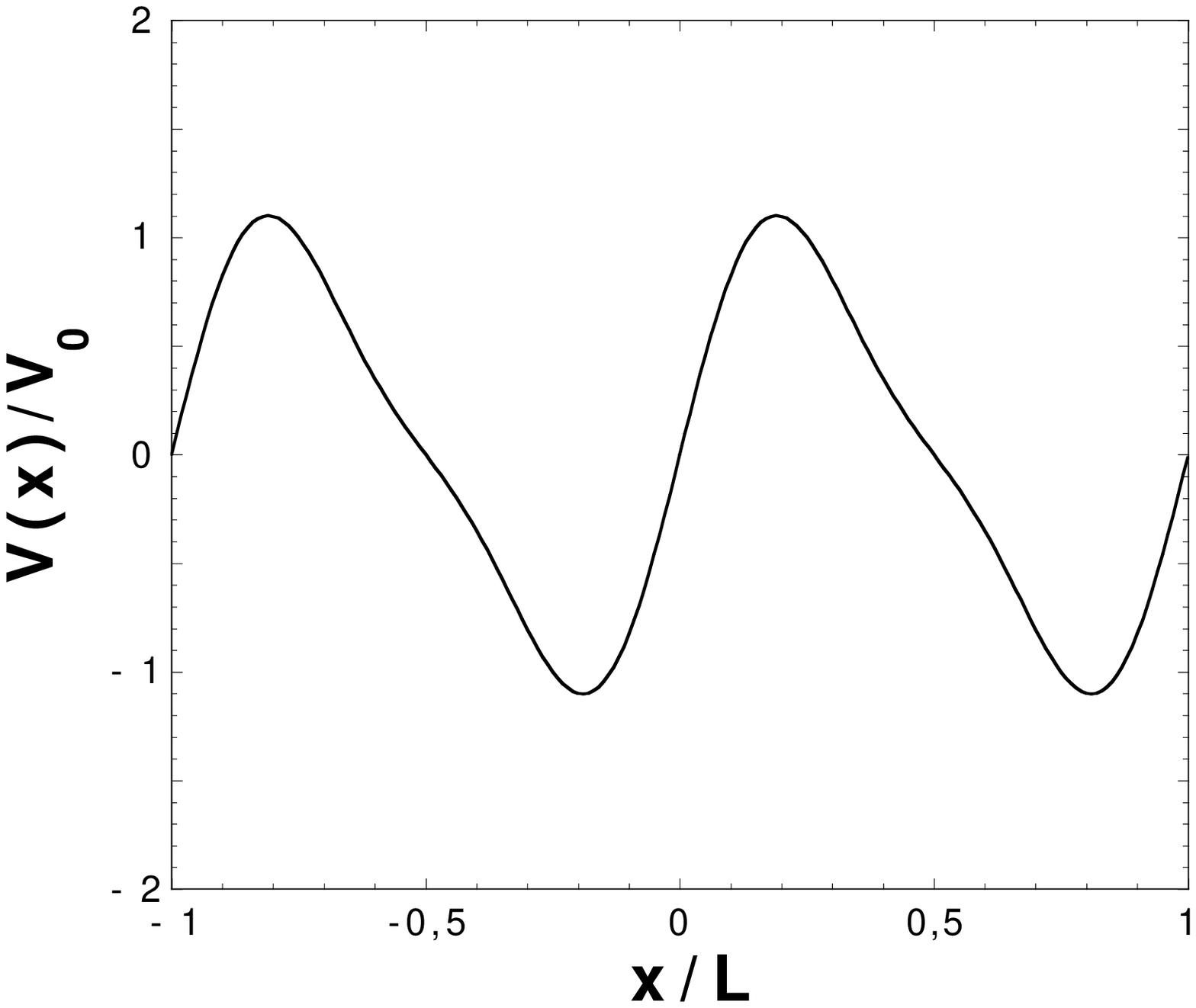, width = 10 cm} }
\caption{
Typical example of a ratchet-potential $V(x)$, periodic in space
with period $L$ and with broken spatial symmetry.
Plotted is the example from (\ref{2.1''}) in dimensionless units.
} 
\label{fig2} 
\end{figure} 
}

\def\figdrei{
\begin{figure}[ht] 
\centerline{\epsfig{file=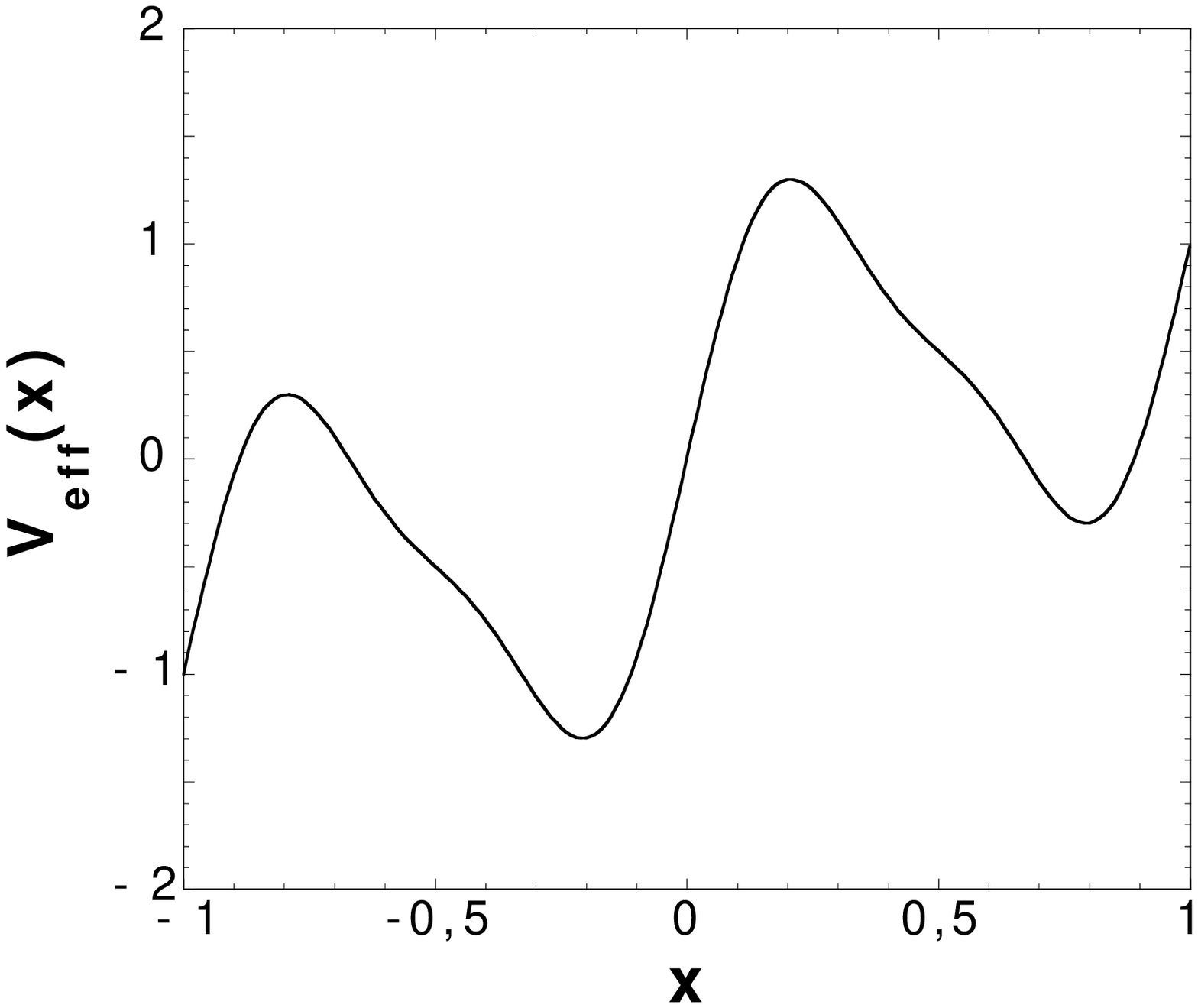, width = 10 cm} }
\caption{
Typical example of an effective potential from
(\ref{2.21'}) ``tilted to the left'', i.e. $F<0$.
Plotted is the example from (\ref{2.1''}) in dimensionless units
(see \sect \ref{sec2.1.2.4} in Appendix A)
with $L=V_0=1$ and $F=-1$, i.e. 
$V_{\rm{eff}} (x)=\sin(2\pi x)+0.25\,\sin(4\pi x) + x$.
} 
\label{fig3} 
\end{figure} 
}

\def\figvier{
\begin{figure}[ht] 
\centerline{\epsfig{file=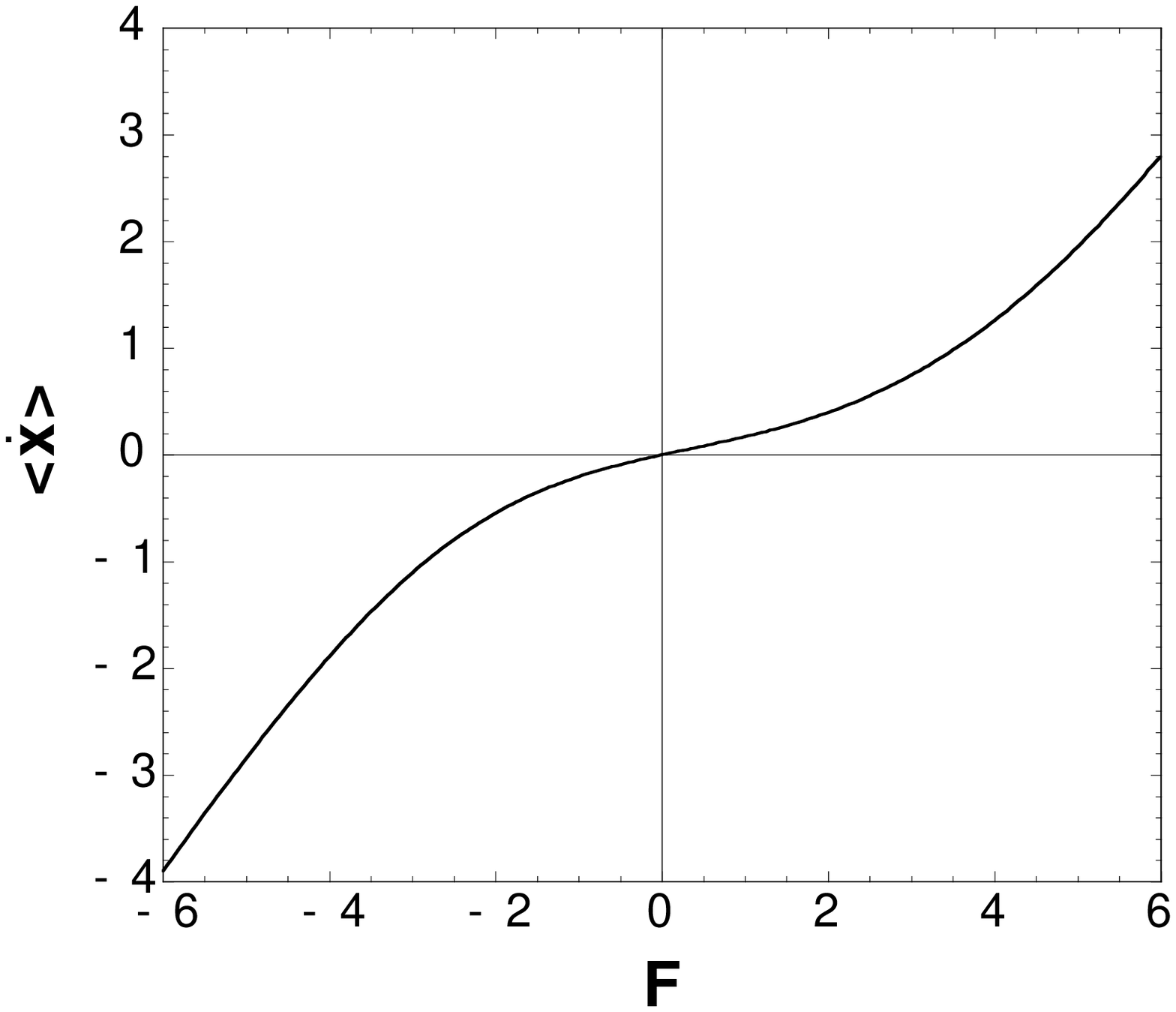, width = 10 cm} }
\caption{
Steady state current $\langle\dot x\rangle$ from (\ref{2.23}) 
versus force $F$
for the tilted Smoluchowski-Feynman ratchet dynamics 
(\ref{2.3}), (\ref{2.21}) with the potential
(\ref{2.1''}) in dimensionless units 
(see \sect \ref{sec2.1.2.4} in Appendix A) with 
$\eta=L=V_0=k_B=1$ and $T =0.5$.
Note the broken point-symmetry.
} 
\label{fig4} 
\end{figure} 
}

\def\figfunf{
\begin{figure}[ht] 
\centerline{\epsfig{file=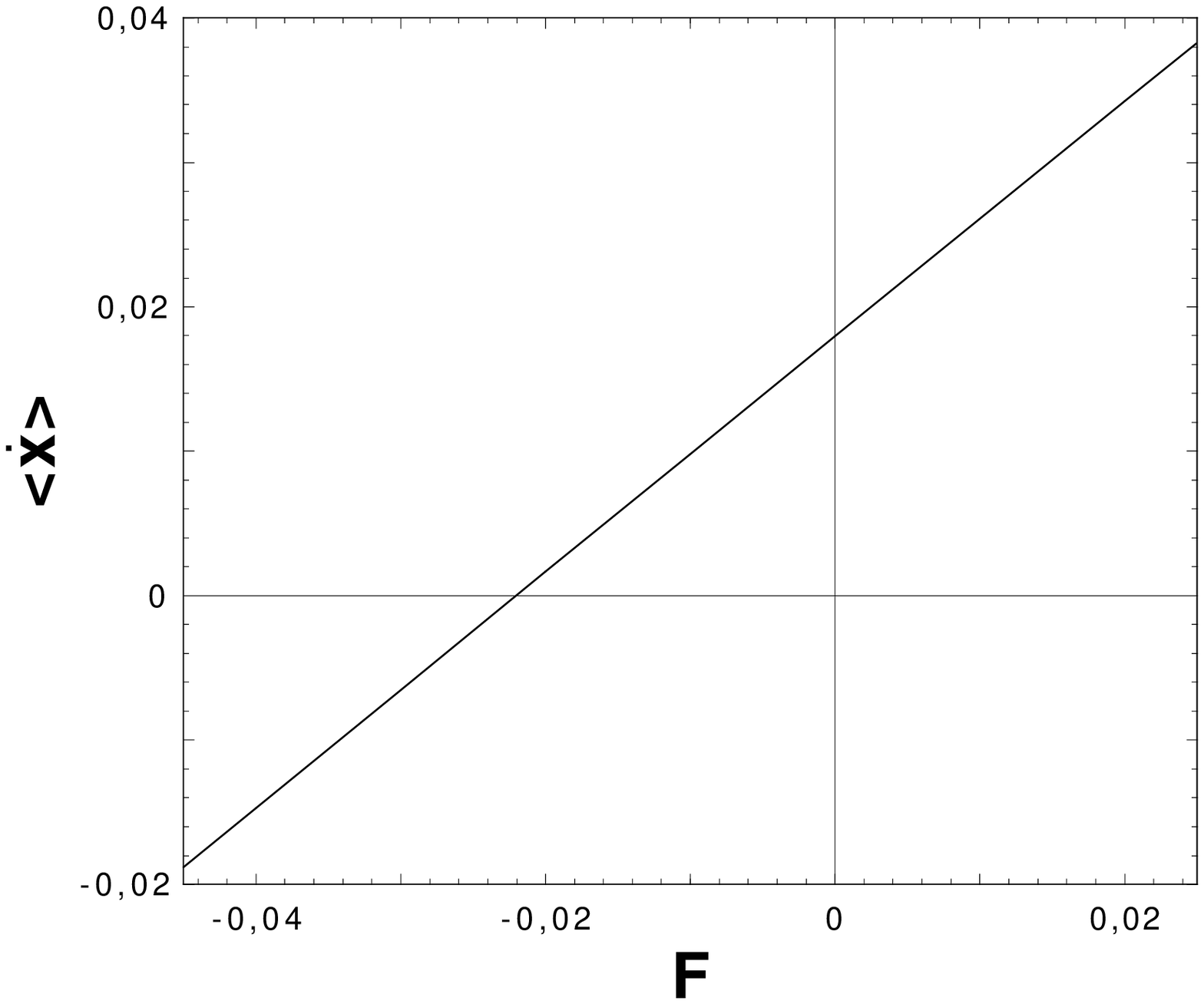, width = 10 cm} }
\caption{
Average particle current $\langle\dot x\rangle$
versus force $F$ for the temperature ratchet dynamics
(\ref{2.1''}), (\ref{2.21}), (\ref{2.23'}), (\ref{2.24a})
in dimensionless units (see \sect \ref{sec2.1.2.4} in Appendix A).
Parameter values are 
$\eta = L = \ttt = k_B = 1$, $V_0=1/2 \pi$, 
$\overline{T} = 0.5$, $A=0.8$.
The time- and ensemble averaged current (\ref{2.24''}) has been
obtained by numerically evolving the Fokker-Planck equation
(\ref{2.24'}) until transients have died out.
} 
\label{fig5} 
\end{figure} 
}

\def\figsechs{
\begin{figure}[ht] 
\centerline{\epsfig{file=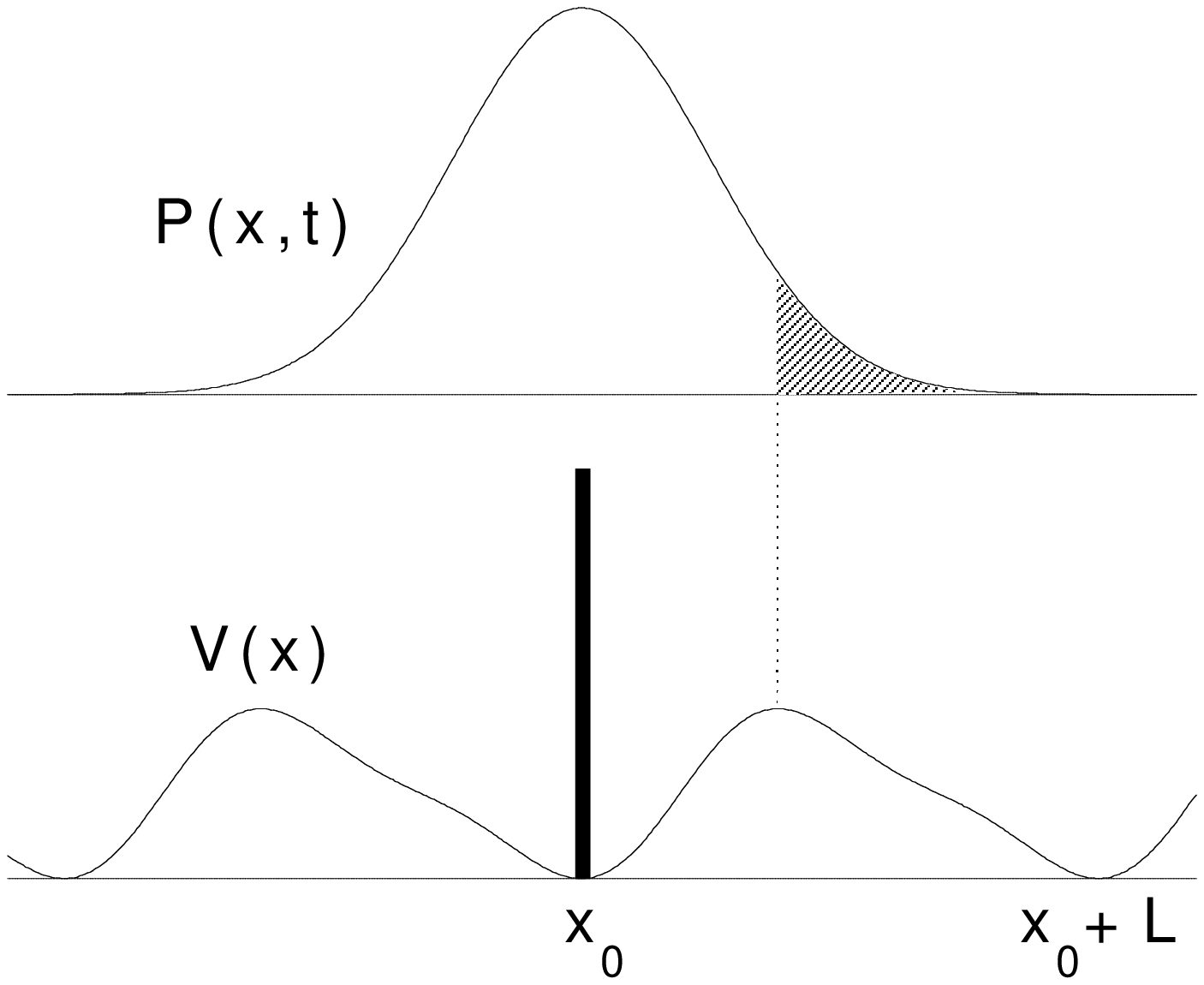, width = 10 cm} }
\caption{
The basic working mechanism of the the temperature ratchet
(\ref{2.21}), (\ref{2.23'}), (\ref{2.24a}).
The figure illustrates how Brownian particles, 
initially concentrated at $x_0$ (lower panel), spread out when the
the temperature is switched to a very high value (upper panel).
When the temperature jumps back to its initial low value,
most particles get captured again in the basin of attraction
of $x_0$, but also substantially in that of $x_0+L$ (hatched area).
A net current of particles to the right, 
i.e. $\langle\dot x\rangle >0$ results.
Note that practically the same mechanism is at
work when the temperature is kept fixed and instead
the potential is turned ``on'' and ``off'' (on-off ratchet,
see \sect \ref{sec4.1}).
} 
\label{fig6} 
\end{figure} 
}

\def\figsieben{
\begin{figure}[ht!] 
\centerline{\psfrag{ttt}{{\large $\ttt$}}\epsfig{file=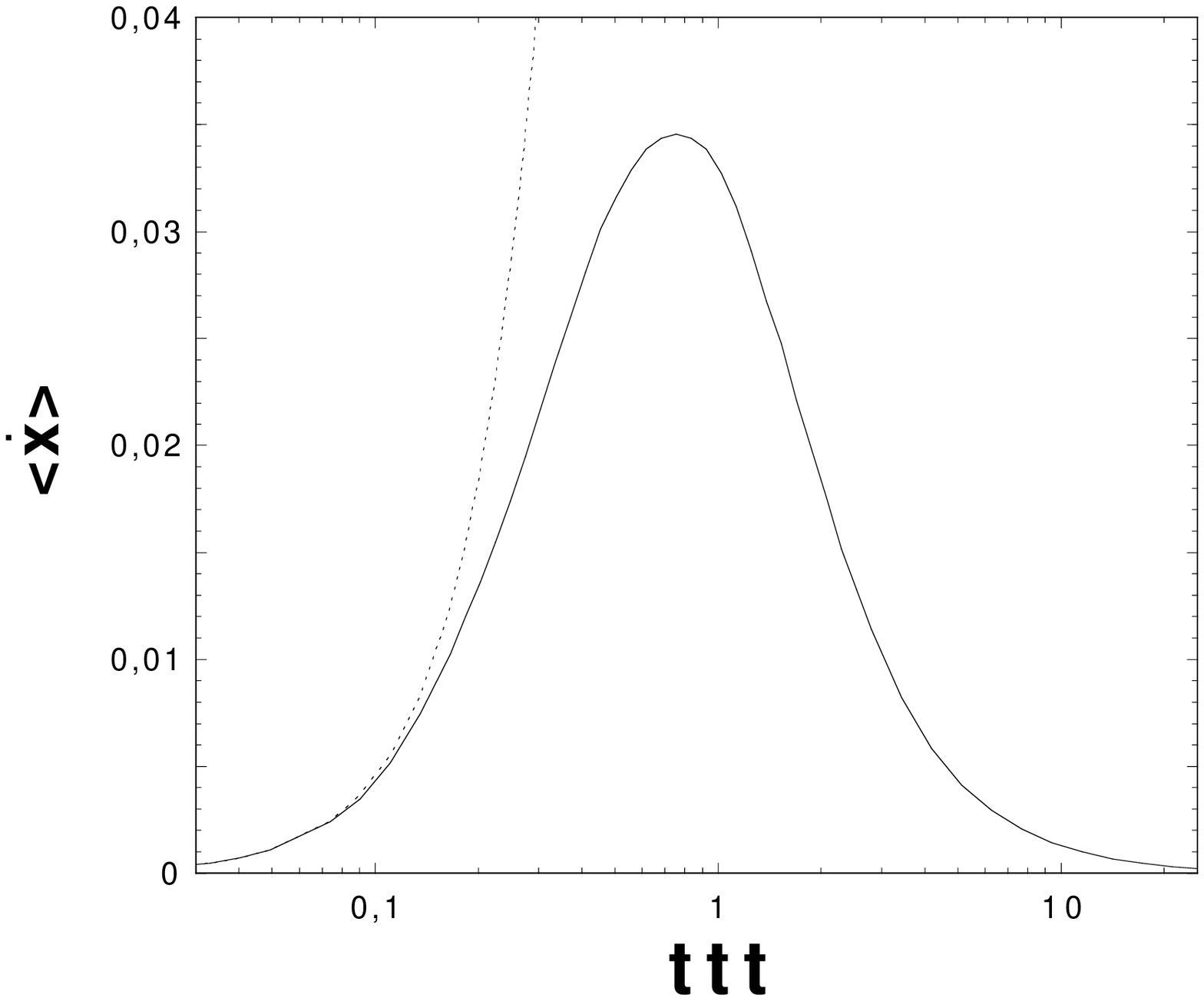, width = 10 cm} }
\caption{
Average particle current $\langle\dot x\rangle$
versus period $\ttt$ for the temperature ratchet dynamics
(\ref{2.1''}), (\ref{2.21}), (\ref{2.23'}), (\ref{2.24b})
in dimensionless units (see \sect \ref{sec2.1.2.4} in Appendix A).
Parameter values are $F=0$,
$\eta = L = k_B = 1$, $V_0=1/2 \pi$, 
$\overline{T} = 0.1$, $A=0.7$.
Solid: Time- and ensemble averaged current (\ref{2.24''})
by numerically evolving the Fokker-Planck equation
(\ref{2.24'}) until transients have died out.
Dotted: Analytical small-$\ttt$ 
asymptotics from (\ref{2.3.2}).
} 
\label{fig7} 
\end{figure} 
}

\def\figachtpot{
\begin{figure}[p] 
\centerline{\epsfig{file=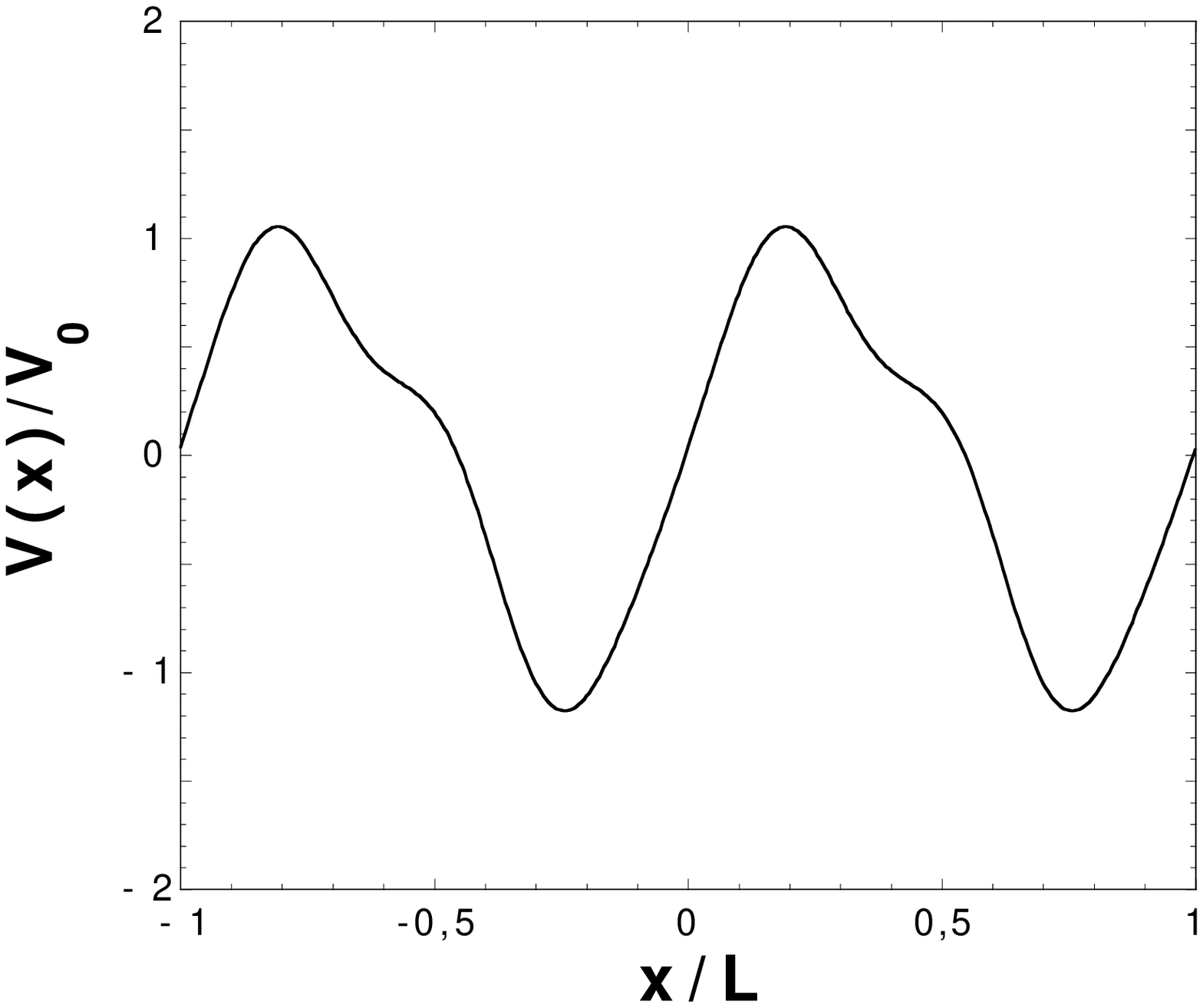, width = 10 cm} }
\caption{
The ratchet potential 
$V(x)=V_0\,[\sin(2\pi x/L)+0.2\,\sin(4\pi(x/L-0.45))
+0.1\,\sin(6\pi(x/L-0.45))]$.
} 
\label{fig8pot} 
\end{figure} 
}

\def\figacht{
\begin{figure}[p] 
\centerline{\psfrag{ttt}{\large{ $\ttt$}}\epsfig{file=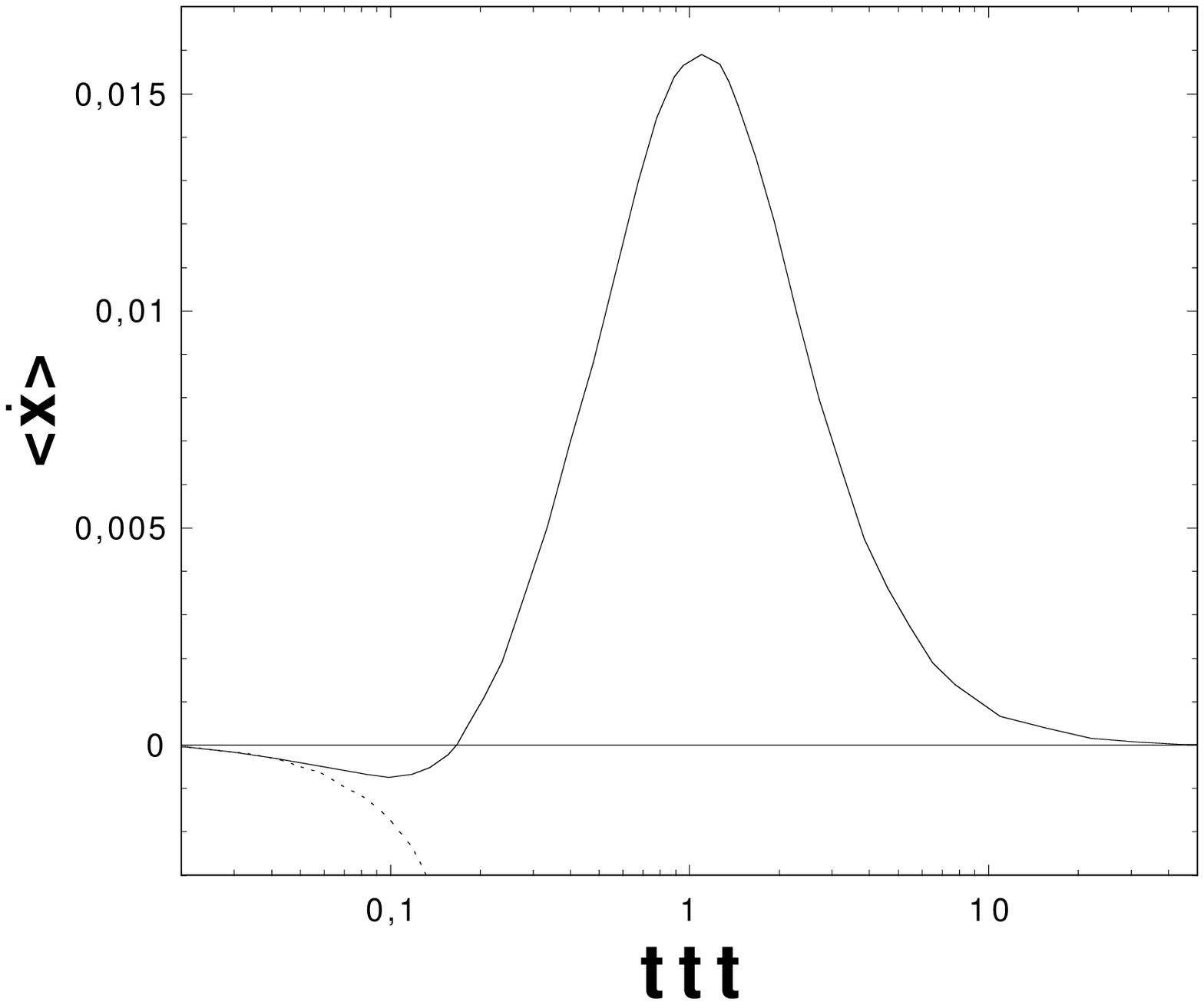, width = 10 cm} }
\caption{
Same as \fig \ref{fig7} but for the ratchet potential
from \fig \ref{fig8pot}.  
} 
\label{fig8} 
\end{figure} 
}
\def\figneun{
\begin{figure}[ht] 
\centerline{\psfrag{ttt}{{\large $\ttt$}}\epsfig{file=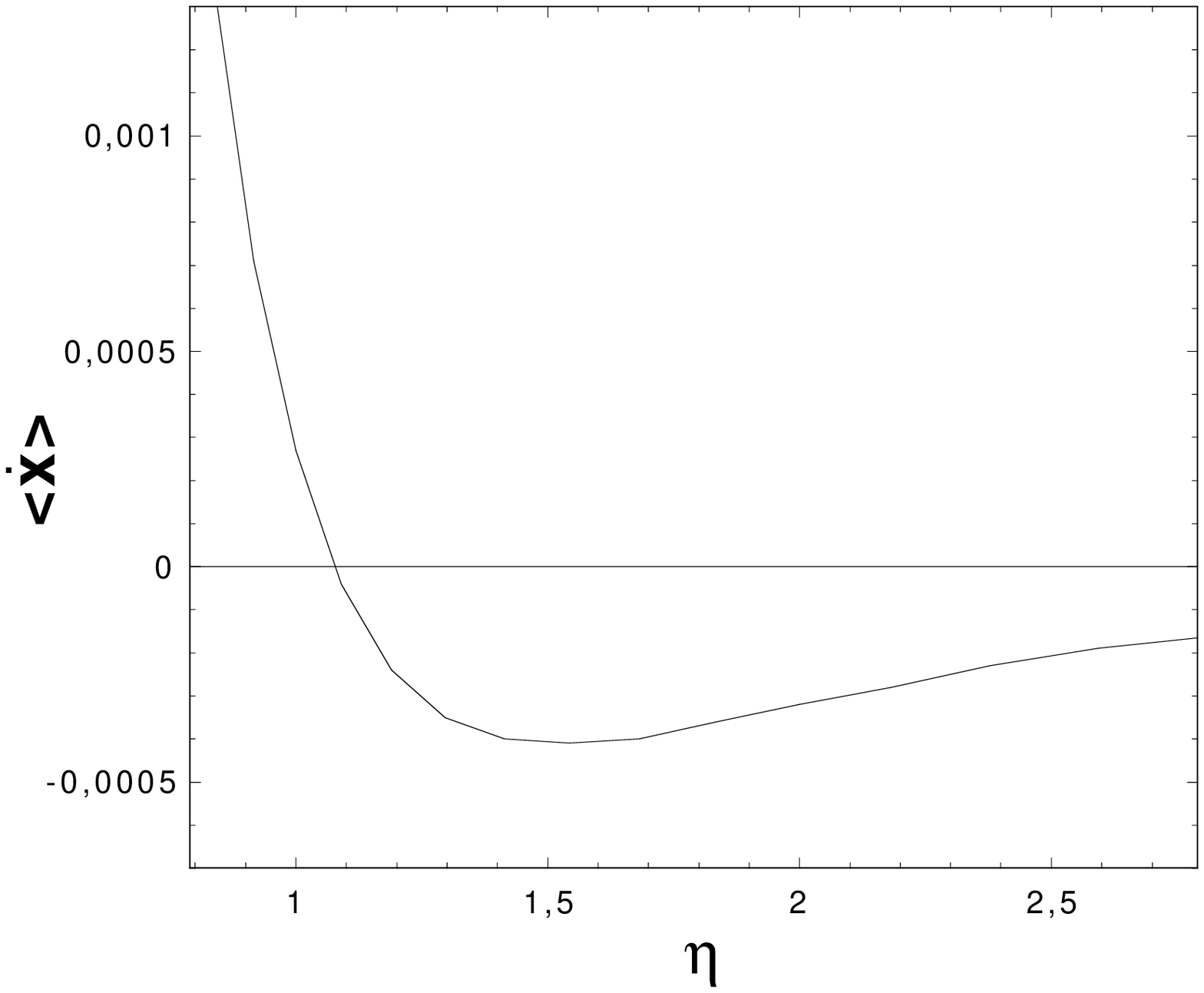, width = 10 cm} }
\caption{
Same as in \fig \ref{fig8} but with a fixed period $\ttt=0.17$ 
(i.e. close to the
inversion point in \fig \ref{fig8}) and instead with
a varying friction coefficient $\eta$.
} 
\label{fig9} 
\end{figure} 
}

\def\figviereins{
\begin{figure}[ht] 
\centerline{\epsfig{file=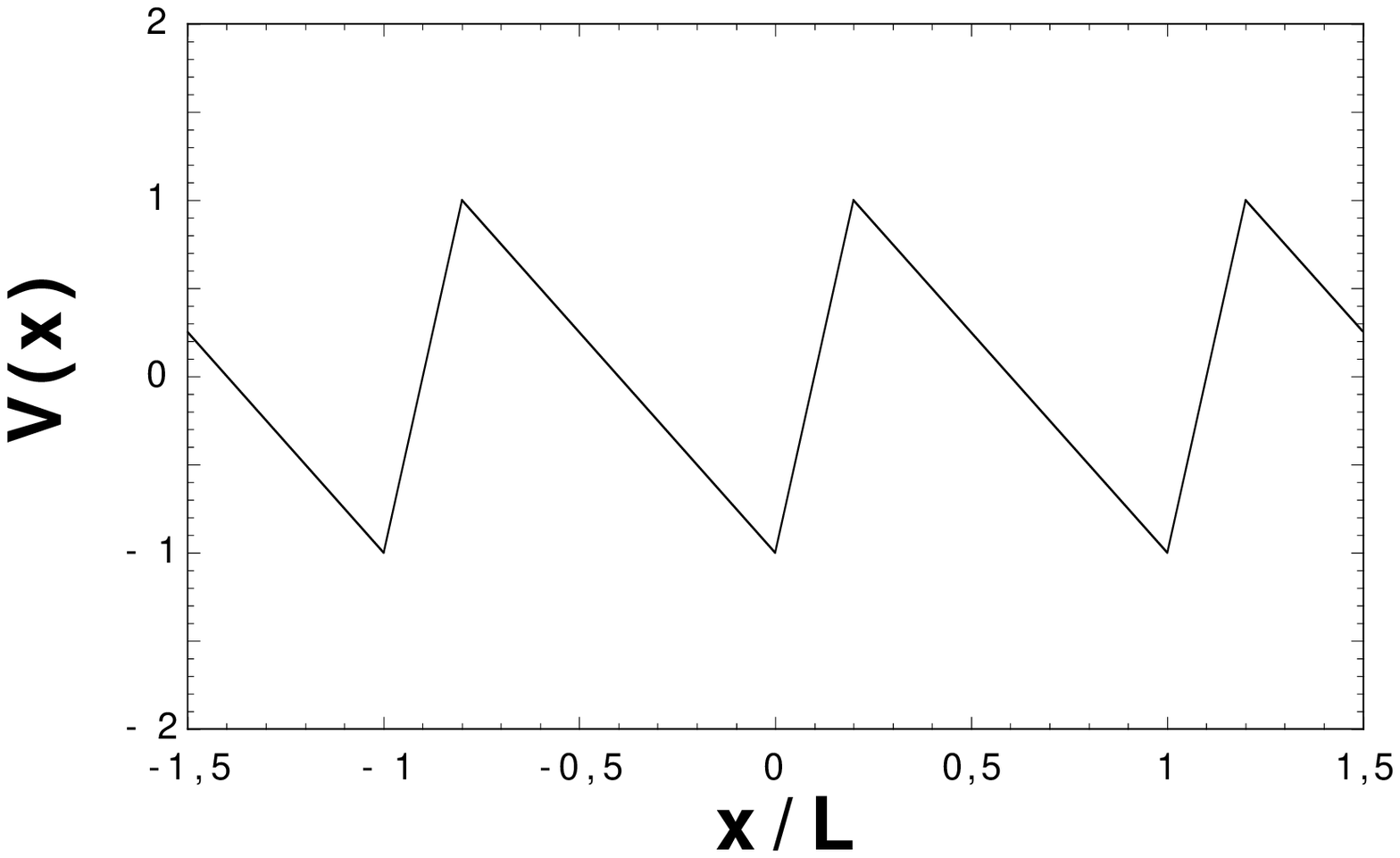, width = 12 cm} }
\caption{
Schematic illustration of a piecewise linear
``saw-tooth'' ratchet potential $V(x)$ (in arbitrary units), 
consisting of two continuously matched
linear pieces per period $L$, one with negative and one
with positive slope, but otherwise asymmetric.
} 
\label{fig4.1} 
\end{figure} 
}

\def\figfeyeins{
\begin{figure}[ht] 
\centerline{\epsfig{file=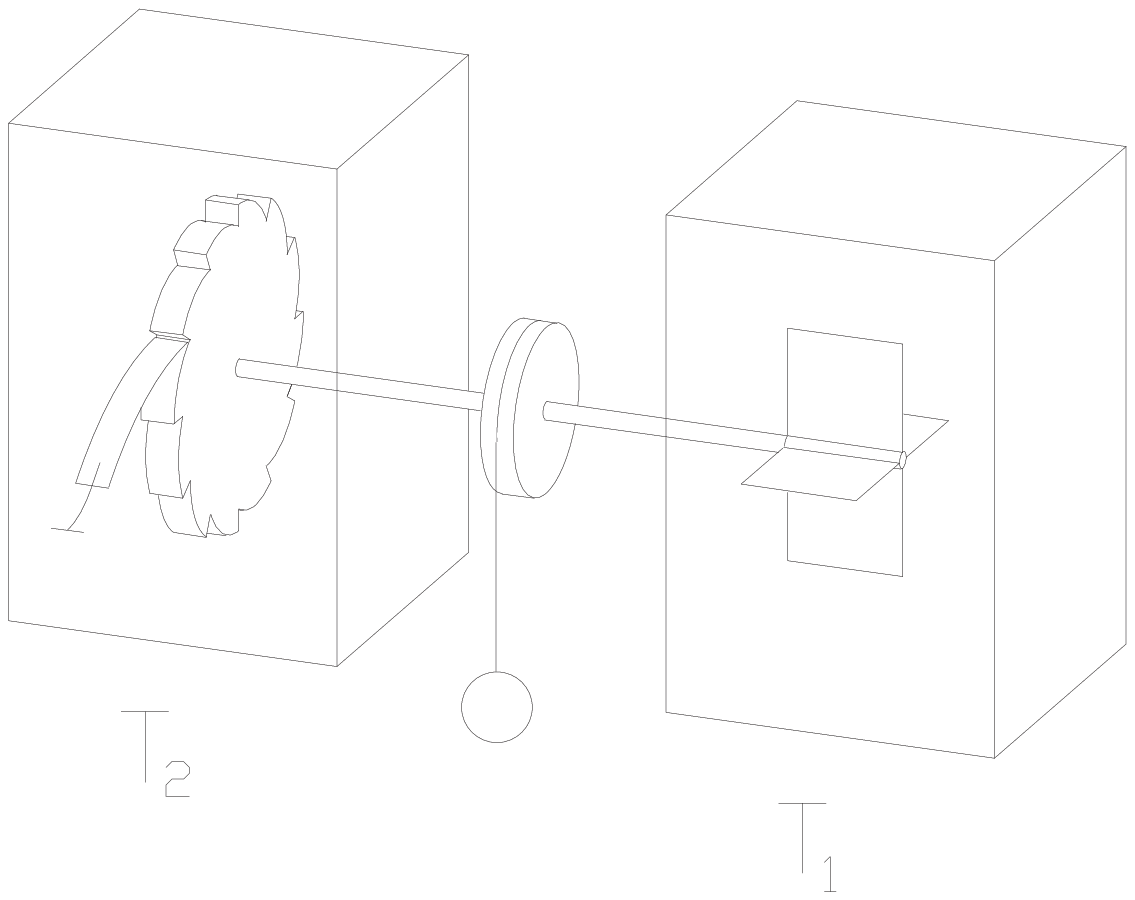, width = 10 cm} }
\caption{
Same as \fig \ref{fig1} but with the ratchet and pawl
kept at a different temperature than the paddles and 
its surrounding gas.} 
\label{figfey1} 
\end{figure} 
}

\def\figdiffeins{
\begin{figure}[ht] 
\centerline{
\psfrag{L2}{{\footnotesize $\!\! L/2$}}
\psfrag{LL}{{\footnotesize $L$}}
\psfrag{V0}{{\footnotesize $V_0$}}
\psfrag{yt}{{\footnotesize $\!\!\! y(t)$}}
\psfrag{y0}{{\footnotesize $\! y_0$}}
\psfrag{my0}{{\footnotesize $\! -y_0$}}
\epsfig{file=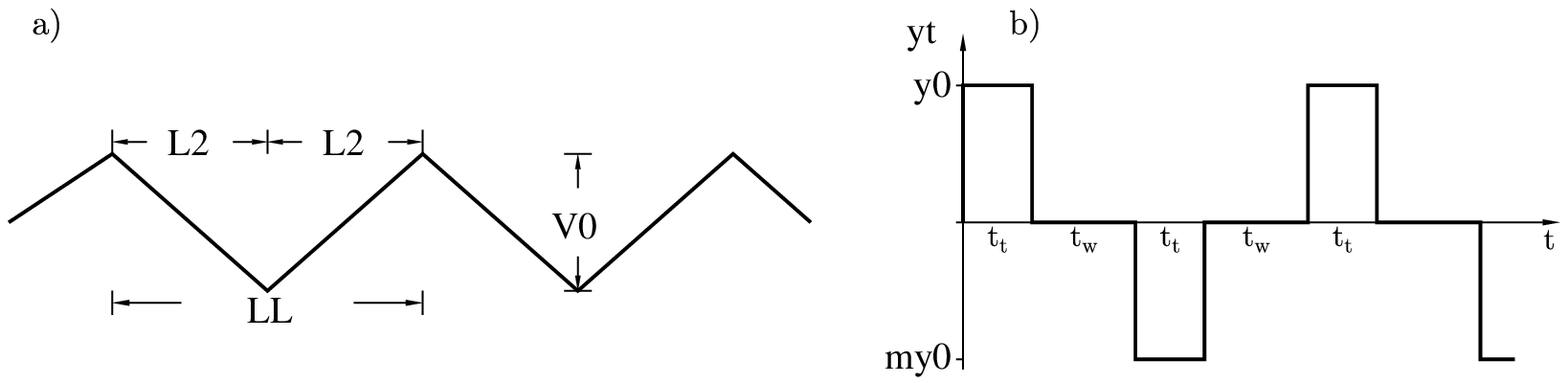, width = 14 cm} } 
\caption{a) Symmetric sawtooth potential $V(x)$ with period $L$ and barrier 
height $V_0$. b) Time-periodic, piecewise 
constant driving force $y(t)$ with model-parameters $y_0$ (``tilt''),
$t_{\rm t}$ (``tilting-time''), and $t_{\rm w}$ (``waiting-period'').
} 
\label{figdiff1} 
\end{figure} 
}

\def\figdiffzwei{
\begin{figure}[ht] 
\centerline{
\psfrag{DD}{{\small $\!\!\! \!\!\!D_{{\rm eff}}/D$}}
\epsfig{file=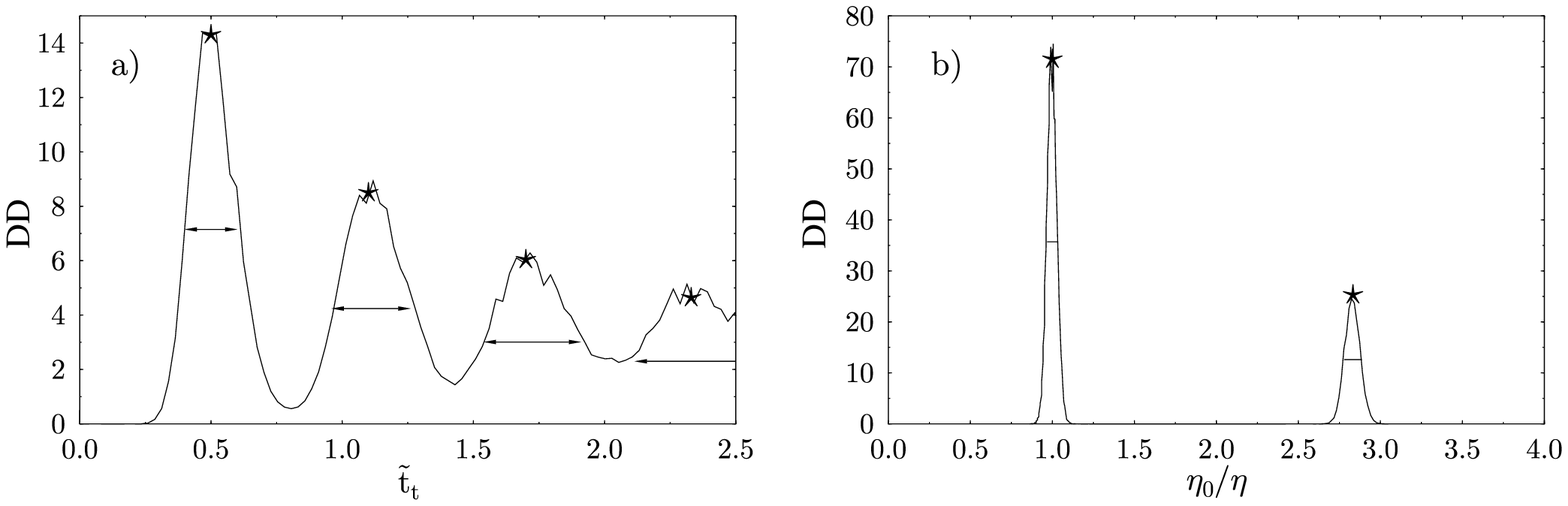, width = 14 cm} } 
\caption{a) Effective diffusion coefficient (\ref{4j}) in units of the 
``bare'' $D$ from (\ref{2.4}) versus scaled ``tilting-time'' 
$\tilde t_{\rm t} := t_{\rm t} V_0/\eta L^2$ from
numerical simulations of the stochastic dynamics
(\ref{6.1}).  
The wiggles reflect the statistical uncertainty due to the finite 
though extensive number of realizations.
The relevant dimensionless  parameters in \fig \ref{figdiff1} are  
$k_BT/V_0=0.01$, $t_{\rm w} V_0/\eta L^2=0.375$, and $y_0L/V_0 =3$. 
Theoretical predictions for the height of the peaks 
from (\ref{diff1}) are indicated by stars.
In addition, the theoretical estimate for the peak-widths 
at half height from \cite{sch98} are indicated by arrows.
b) Effective diffusion coefficient
versus friction coefficient $\eta$ from simulations
of (\ref{6.1}) with $k_BT/V_0=0.005$ and $y_0L/V_0 =22$. The 
times $t_{\rm t}$ and $t_{\rm w}= t_{\rm t}$ are 
both kept at fixed values and also define $\eta_0$
via $\eta_0=(2Ly_0-4V_0)/L^2 t_{\rm t}$.
Theoretical predictions are indicated analogous to a).
} 
\label{figdiff2} 
\end{figure} 
}

\def\figsseins{
\begin{figure}[ht] 
\centerline{\epsfig{file=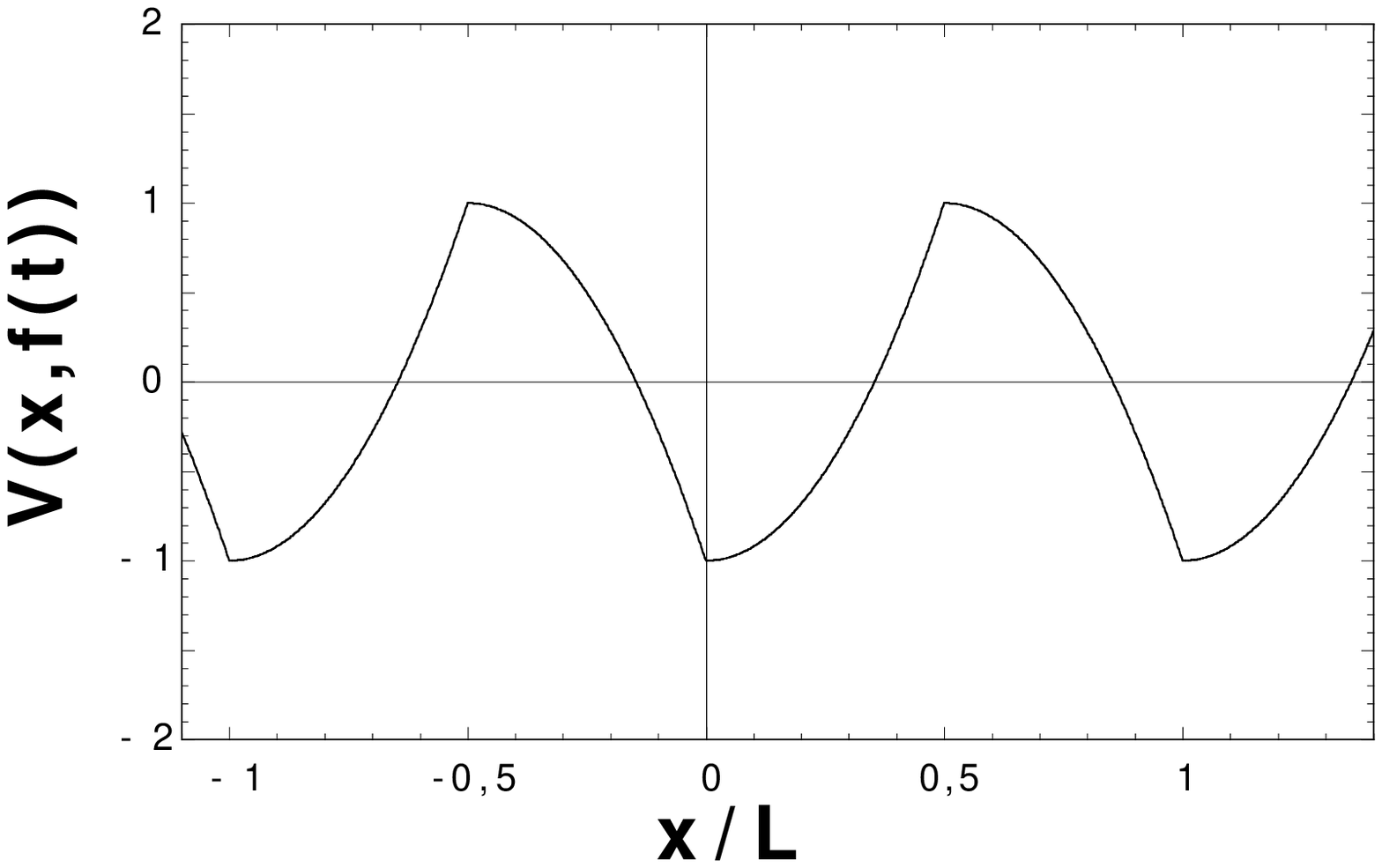, width = 12 cm} }
\caption{Example of a supersymmetric potential $V(x,f(t))$ (in arbitrary units)
of the type (\ref{ss7}) at an arbitrary but fixed $f(t)$-value.
} 
\label{figss1} 
\end{figure} 
}

\def\figsszwei{
\begin{figure}[ht] 
\centerline{\psfrag{ttt}{{\Large ${\bf {\rm t}/\ttt}$}}\epsfig{file=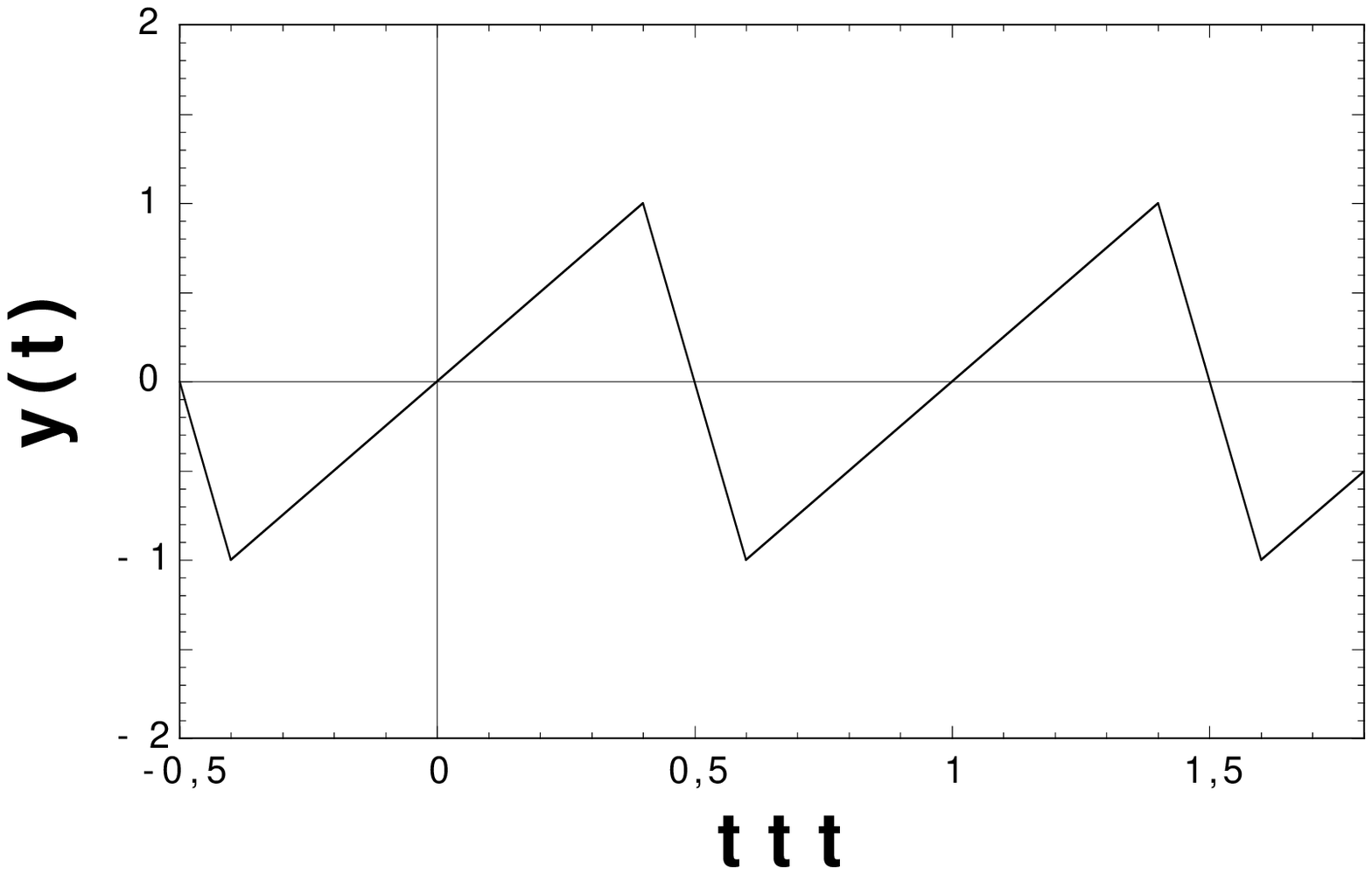, width = 12 cm} }
\caption{Example of a supersymmetric $\ttt$-periodic
driving $y(t)$ (in arbitrary units)
of the type (\ref{ss13}).
} 
\label{figss2} 
\end{figure} 
}

\def\figssdrei{
\begin{figure}[ht!] 
\centerline{\psfrag{ttt}{{\Large ${\bf {\rm t}/\ttt}$}}\epsfig{file=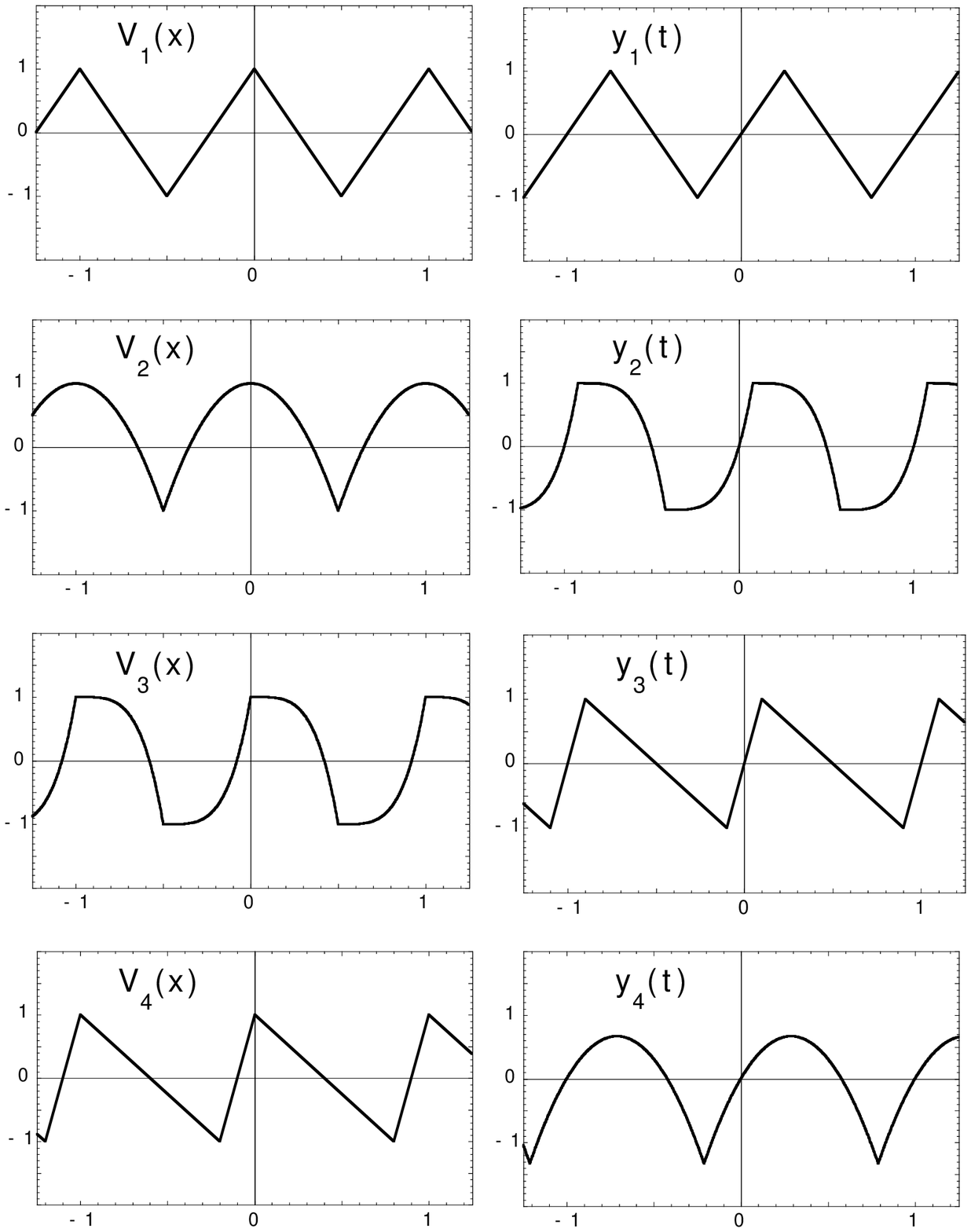, width = 12 cm} }
\caption{Summary of the symmetry considerations for tilting
ratchets with potentials $V_i(x)$ and periodic 
drivings $y_i(t)$ (in arbitrary units).
$i=1$: symmetric and supersymmetric.
$i=2$: symmetric but not supersymmetric.
$i=3$: supersymmetric but not symmetric.
$i=4$: neither symmetric nor supersymmetric (but still satisfying (\ref{4g})).
The particle current (\ref{4c1}) vanishes for arbitrary
combinations of potentials and drivings which are
either both symmetric or both supersymmetric.
For any other combination of potentials and drivings,
a finite current arises generically.
} 
\label{figss3} 
\end{figure} 
}

\def\figketzwei{
\begin{figure}[ht] 
\centerline{\epsfig{file=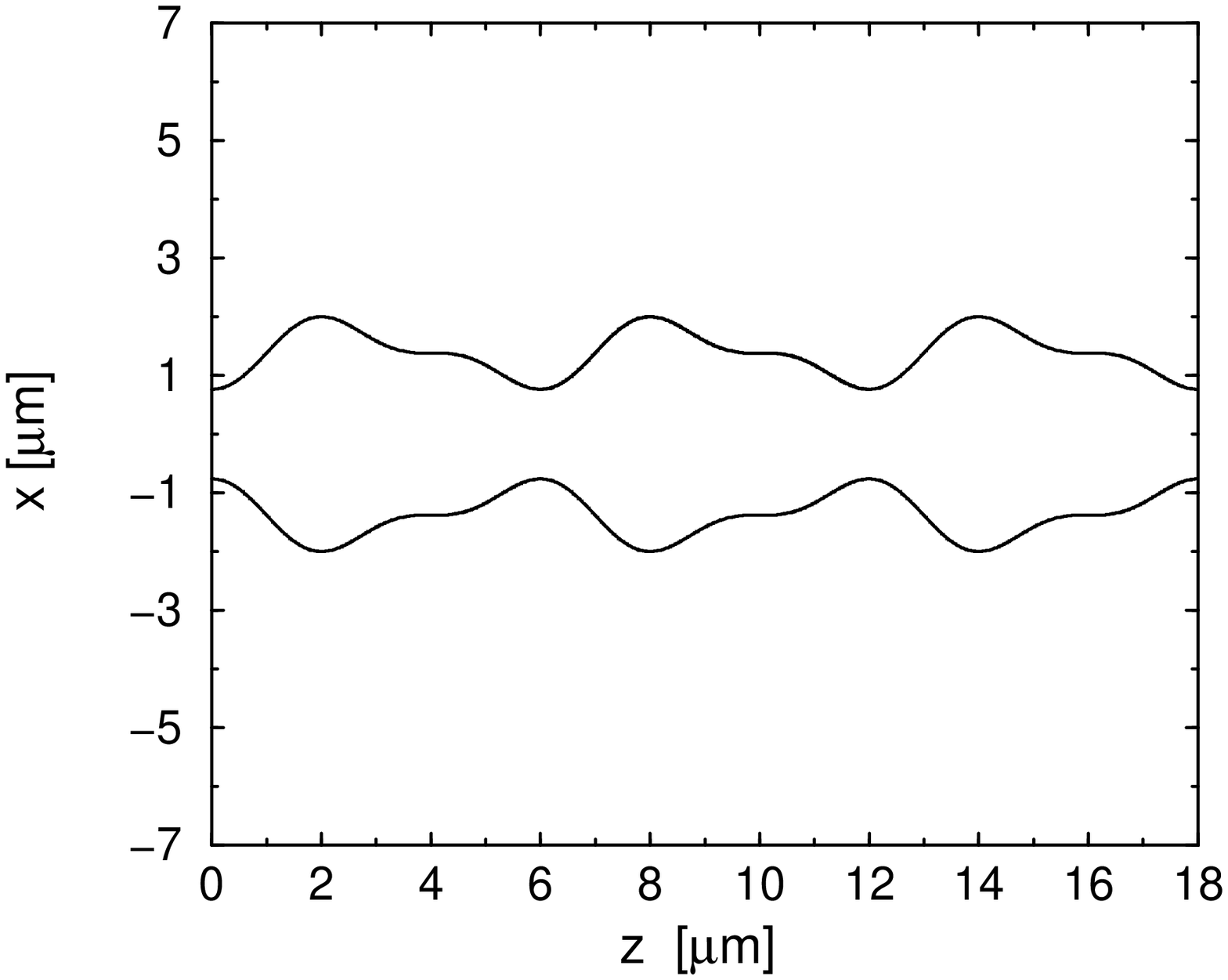, width = 10 cm} }
\caption{Schematic cross-section ($x$-$z$-plane) through a single pore with
an experimentally realistic, ratchet-shaped variation of the diameter
along the pore axis ($z$ axis).
} 
\label{figket2} 
\end{figure} 
}

\def\figketdrei{
\begin{figure}[ht] 
\centerline{\epsfig{file=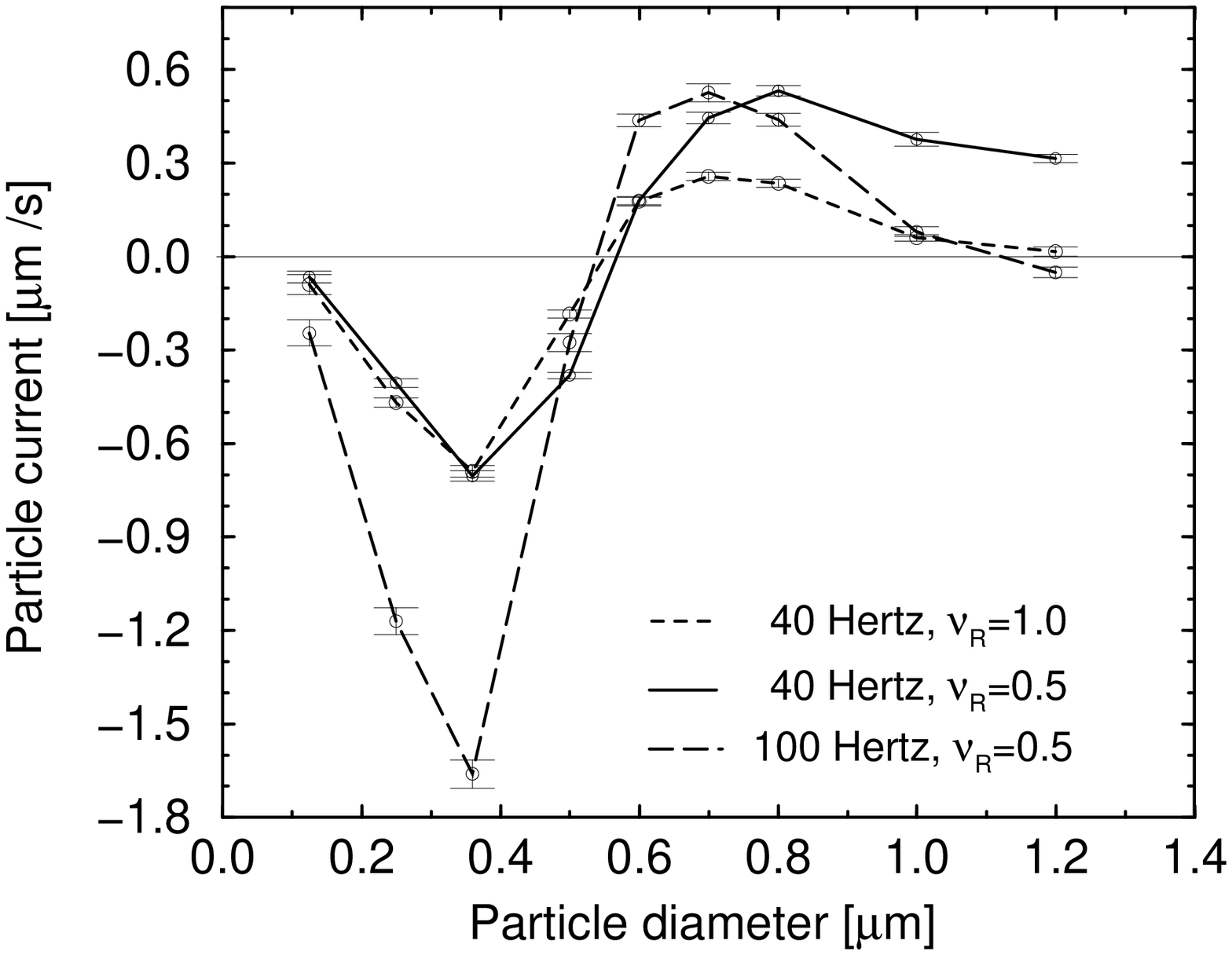, width = 10 cm} }
\caption{Numerical simulation of the stochastic dynamics (\ref{ket1}), 
(\ref{ket2}) for a pore shape as depicted in \fig \ref{figket2}, 
at room temperature ($T=293\, K$).
The friction coefficient $\eta$  in (\ref{ket2}) is
given by Stokes law
$6 \pi  R \, \nu$, where $R$ is the particle radius and
$\nu=\nu_{\rm R}\,\nu_{water}$ the viscosity of the liquid
in units of the viscosity $\nu_{water}$ of water. 
The velocity field in (\ref{ket1}) has been obtained numerically 
with a sinusoidal pumping of the liquid
at a frequency of 40 Hertz and 100 Hertz. 
The pumping amplitude $A$ 
is chosen as $A=2L$, where $L=6\mu m$ is the period of the 
ratchet-shaped pore in \fig \ref{figket2}.
Depicted is the time- and ensemble-averaged particle current 
$\langle \dot z\rangle$ along the pore axis ($z$ axis) versus the
particle diameter for various driving frequencies and
viscosities.
} 
\label{figket3} 
\end{figure} 
}

\def\figketvier{
\begin{figure}[ht] 
\centerline{\epsfig{file=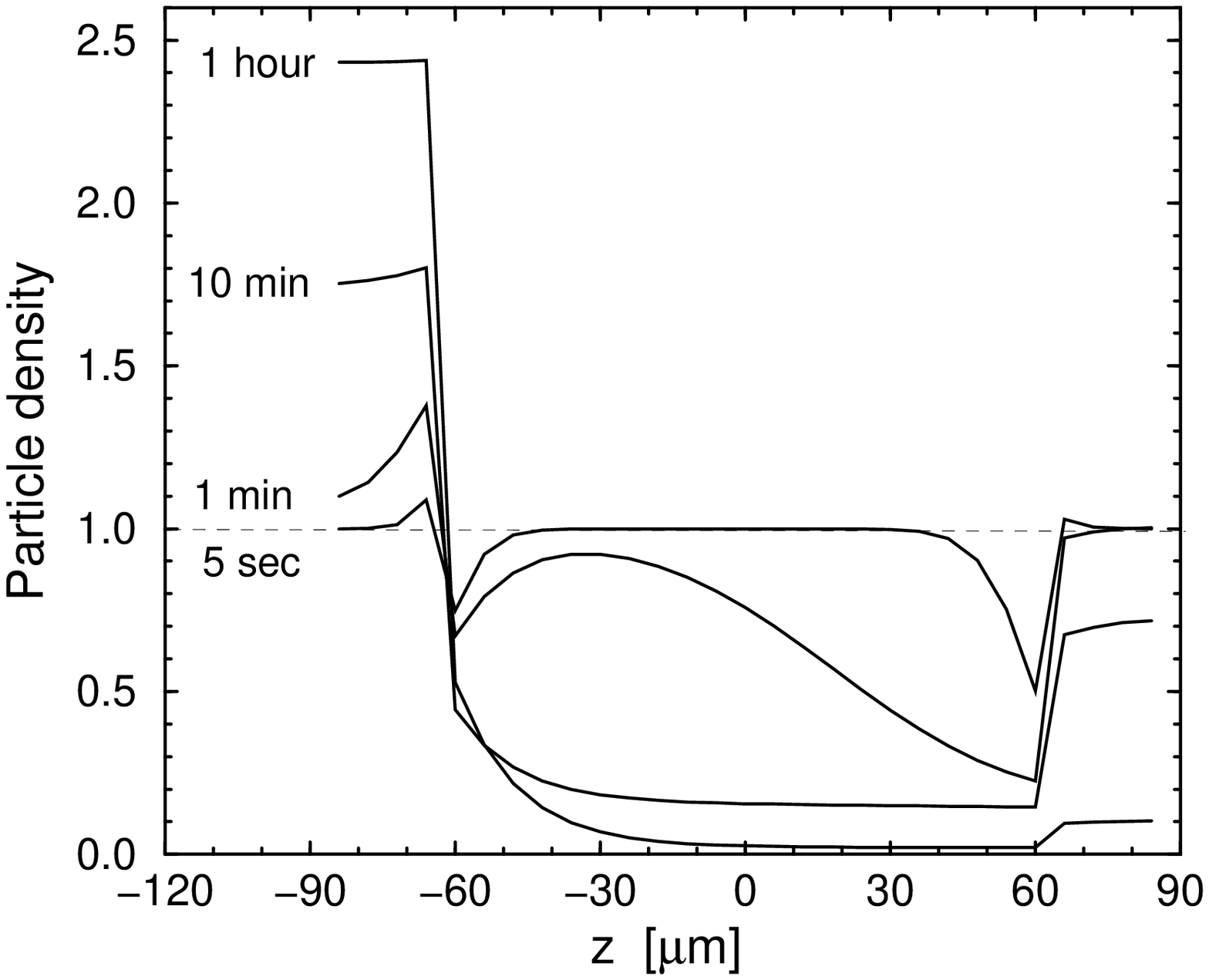, width = 10 cm} }
\caption{
Time evolution of the particle density 
(within the liquid-plus-particle suspension)
along the $z$-axis
starting with a homogeneous initial distribution (normalized to unity).
The pore length (along the $z$ axis) is 126$\mu m$ and the
extension $\Delta z$
of each of the two adjacent basins along the $z$ axis
is 24$\mu m$. 
Other details are like in \fig \ref{figket3} with
particle radius $R=$0.36$\mu m$,  
pumping frequency 100 Hertz,
pumping amplitude $ A = L$,
and relative viscosity $\nu_R = 0.5$.
} 
\label{figket4} 
\end{figure} 
}

\def\figketfunf{
\begin{figure}[ht] 
\centerline{\epsfig{file=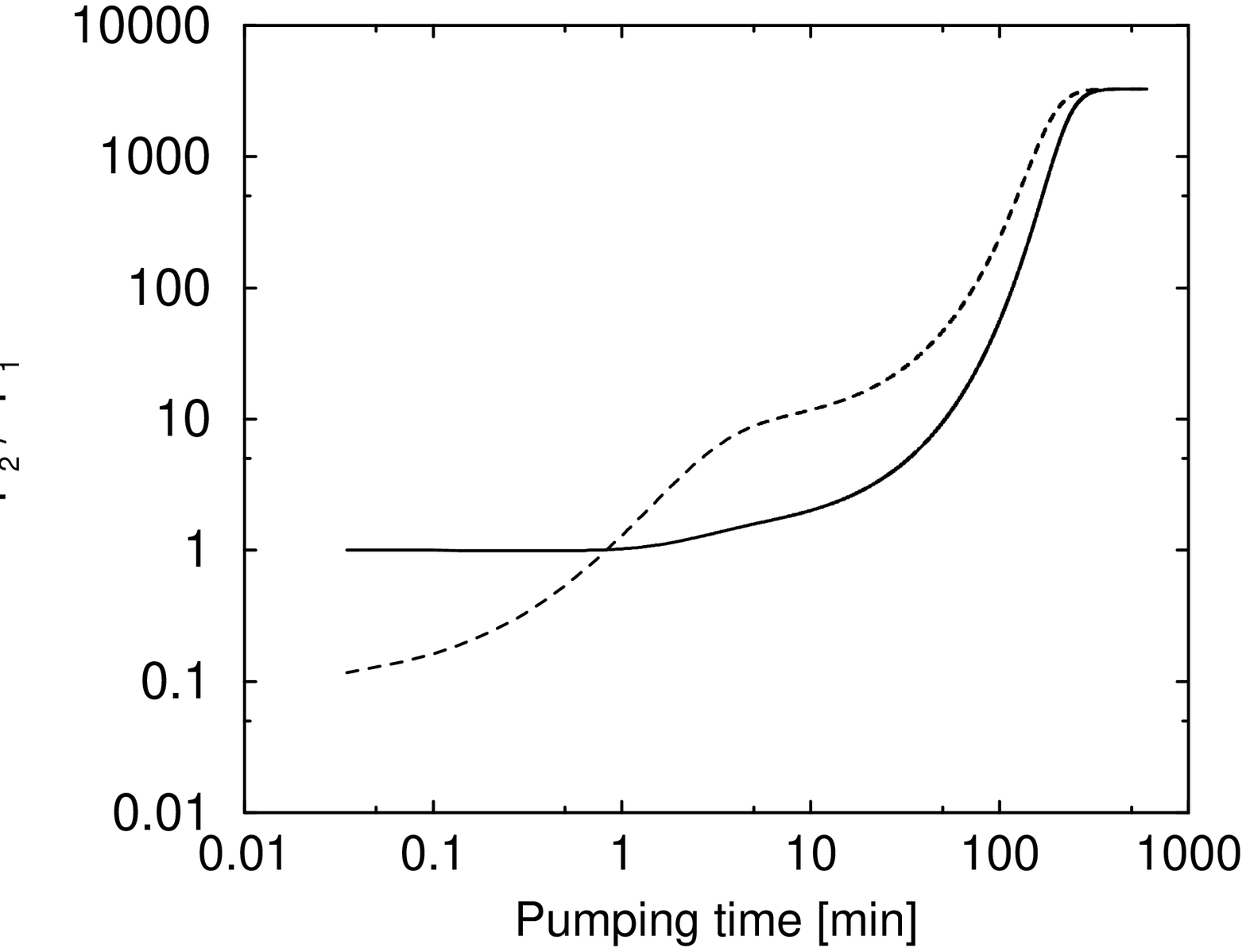, width = 10 cm} }
\caption{
Ratio $P_2/P_1$ of particle densities
for two types of particles versus time $t$.
The setup is the same as in \fig \ref{figket4}
but with a pumping amplitude $A=2L$
and with radii of the two types of particles
$R_1=$0.36$\mu m$ and $R_2=$0.7$\mu m$ (corresponding to
opposite current directions in \fig \ref{figket3}).
The ratio of the densities $P_2/P_1$ refers to the
border of the right basin at 
$z=$87$\mu m$ ($=$126/2$\mu m +$24$\mu m$).
Solid line: Overall homogeneous initial densities. 
Dashed line: Initially homogeneous densities 
in the pore region and vanishing densities 
in the two basin regions.
} 
\label{figket5} 
\end{figure} 
}

\def\figqmeins{
\begin{figure}[ht] 
\centerline{\epsfig{file=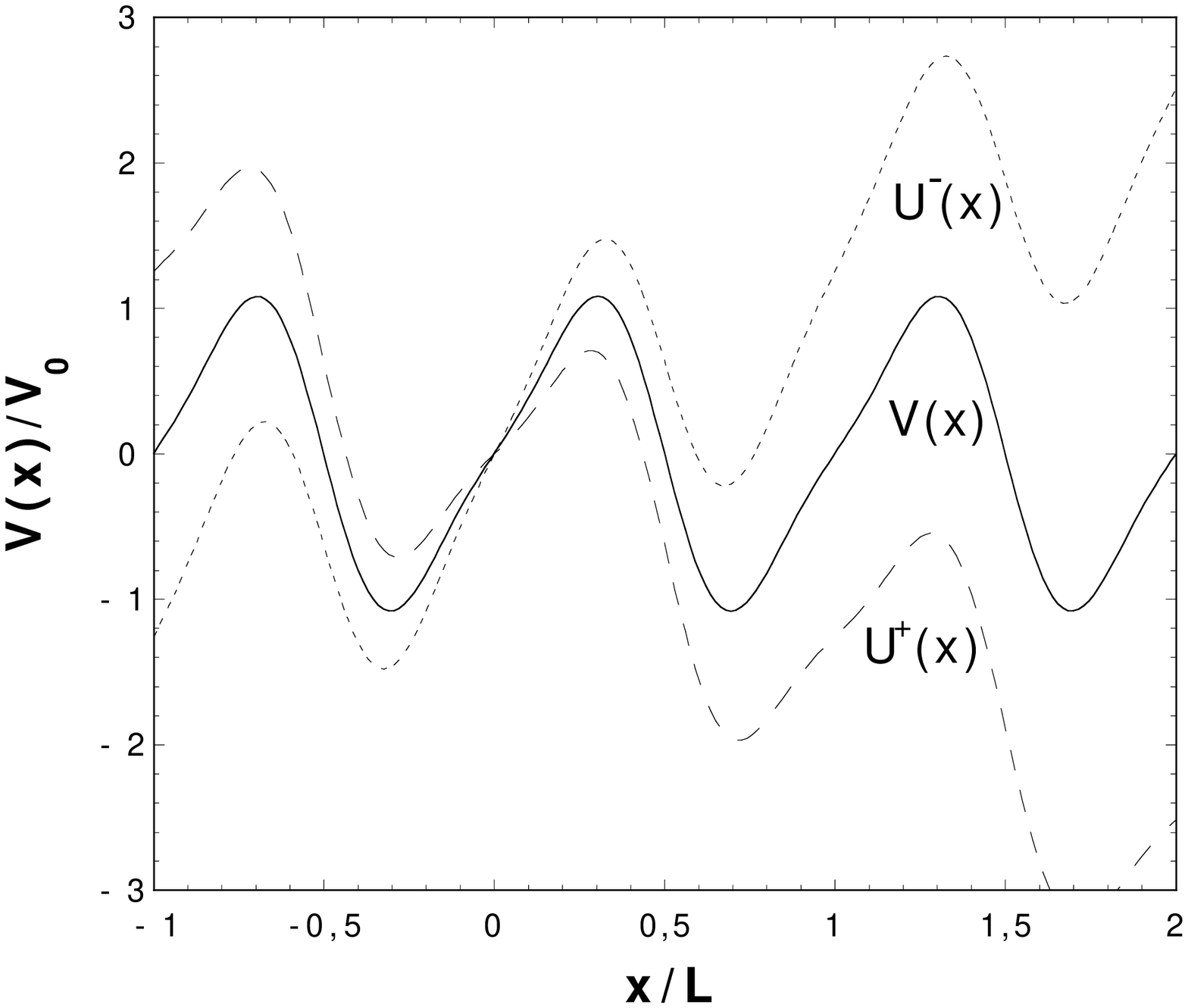, width = 14 cm} }
\caption{Solid: The ratchet potential 
$V(x)=V_0\,[\sin(2\pi x/L)-0.22\,\sin(4\pi x/L)]$.
Note that this potential is almost identical to
the {\em spatially inverted } potential from
(\ref{2.1''}), see also \fig \ref{fig2}.
Dashed and dotted: ``tilted
washboard potentials'' $U^\pm(x)$ in
(\ref{qm3}) with $F\, l=0.1\, V_0$, $l=L/2\pi$.
} 
\label{figqm1} 
\end{figure} 
}

\def\figqmzwei{
\begin{figure}[ht] 
\centerline{
\psfrag{J1}{{\large{\bf $\langle\dot x\rangle/\Omega_0 L$}}}
\psfrag{J2}{{\normalsize $+\langle\dot x\rangle_{\rm qm}$}}
\psfrag{J3}{{\normalsize $\!\!\!\!-\langle\dot x\rangle_{\rm qm}\ \ $}}
\psfrag{J4}{{\normalsize $\ \ \ \ \ \ \  +\langle\dot x\rangle_{\rm cl}$}}
\epsfig{file=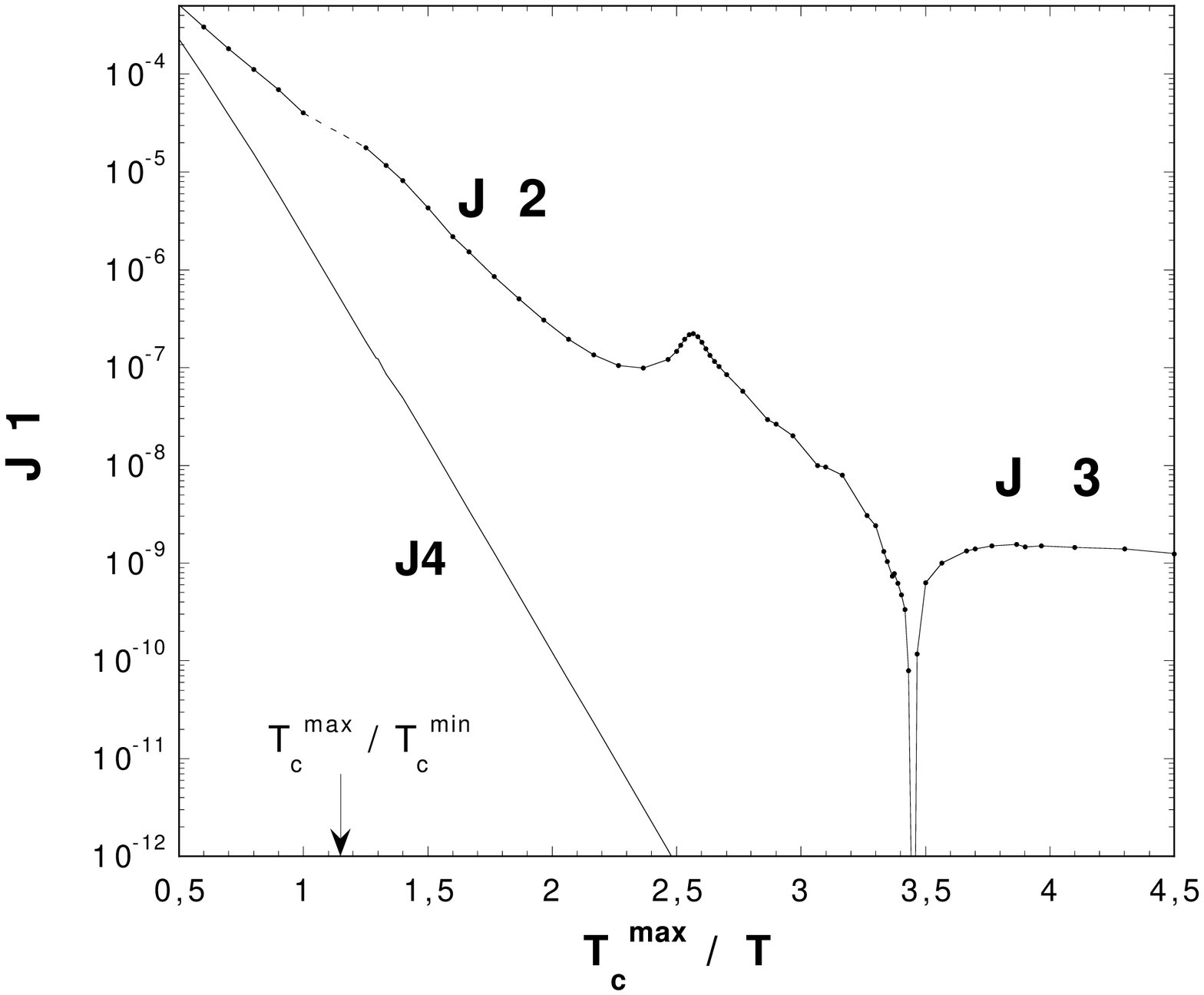, width = 14 cm} }
\caption{The classical steady state current $\langle\dot x\rangle_{{\rm cl}}$
and its quantum mechanical counterpart 
$\langle\dot x\rangle_{{\rm qm}}$ 
for the ratchet potential from \fig \ref{figqm1} 
in dimensionless units $\langle\dot x\rangle/ L\, \Omega_0$.
Note that in the present Arrhenius plot (logarithmic ordinate)
the observed behavior of the quantum current near $T_c^{{\rm max}}/T=3.5$
is not the signature of a singularity but rather of a current inversion.
Further worth mentioning features are the non-monotonicity of
$\langle\dot x\rangle_{{\rm qm}}$ 
and that
$\langle\dot x\rangle_{{\rm qm}}$ 
tends towards a finite limit when $T\to 0$.
} 
\label{figqm2} 
\end{figure} 
}

\def\figqmdrei{
\begin{figure}[p] 
\centerline{\epsfig{file=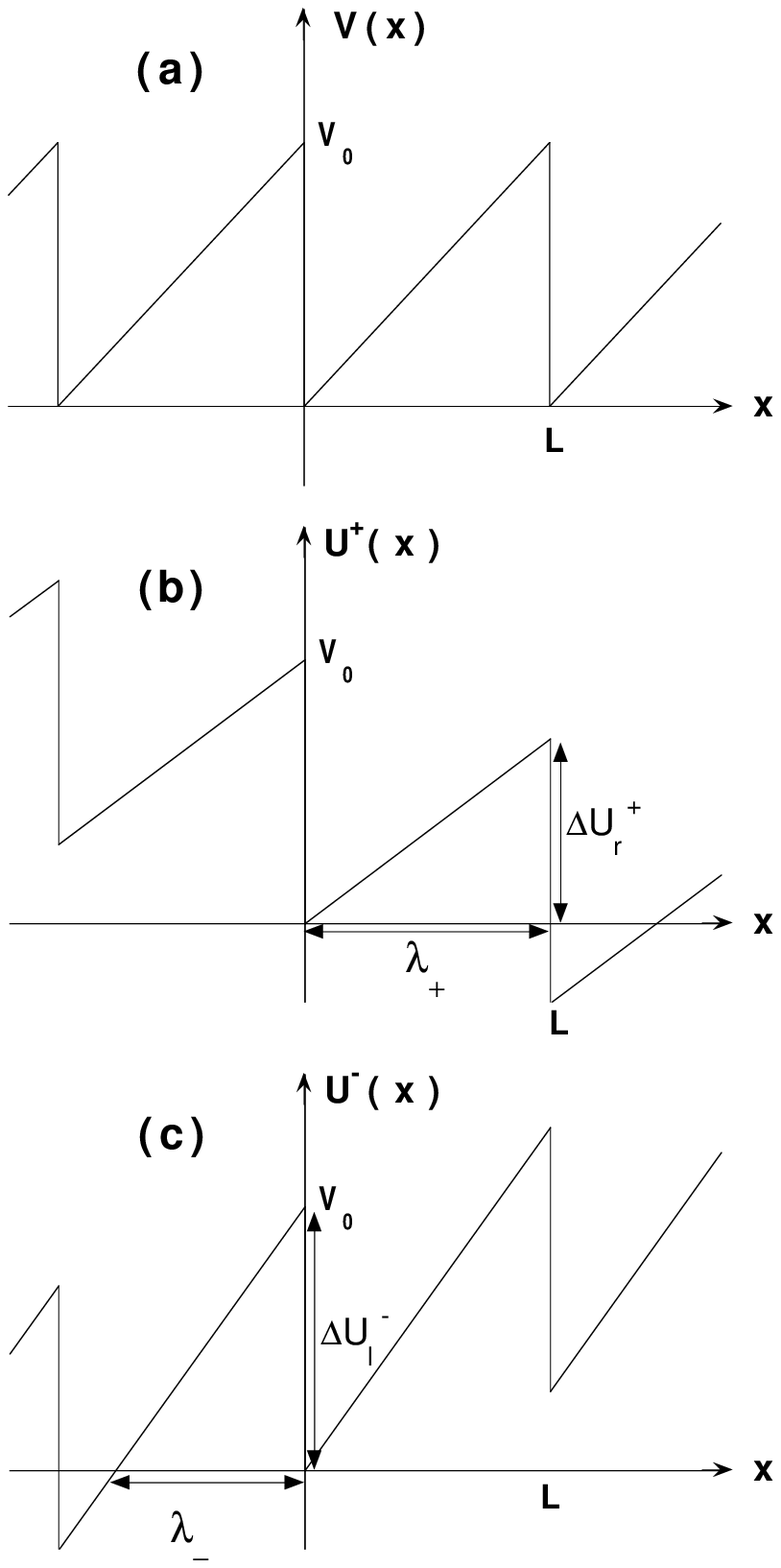, width = 23 cm} }
\caption{(a): Stylized sawtooth ratchet-potential $V(x)$
with spatial
period $L$ and barrier height $V_0$. (b): The tilted ratchet
potential $U^+(x)=V(x)-Fx$ from (\ref{qm3}) with the ``tunneling-length''
$\lambda_+$ and the potential barrier $\Delta U_r^+$, relevant for
the tunneling rate $k_r^+$ from the local minimum $x_0=0$ to its 
neighboring local minimum to the right. (c): Same for the 
potential $U^-(x)=V(x)+Fx$. [The depicted $F$-value is $V_0/3L$.]
} 
\label{figqm3} 
\end{figure} 
}

\def\figqmvier{
\begin{figure}[ht] 
\centerline{\epsfig{file=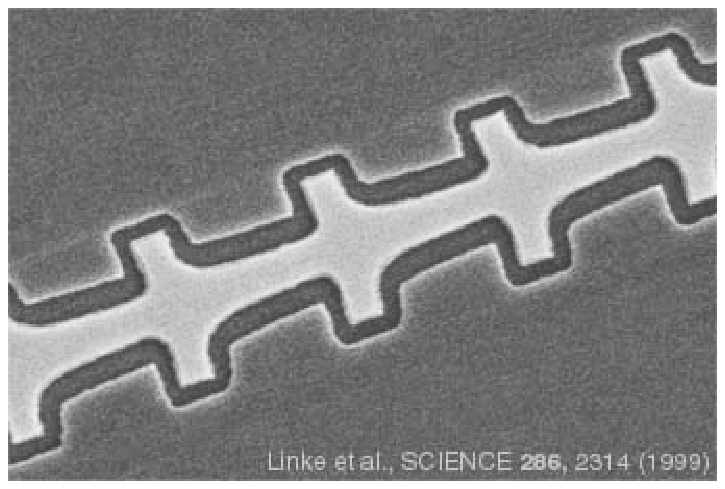, width = 12 cm} }
\caption{Scanning electron micrograph of an array of triangular
shaped quantum dots, etched from a GaAs/AlGaAs semiconductor
heterostructure.
The depicted top view defines the $x$-$y$-plane accessible to the 
two-dimensional conducting electron gas.
The etched areas (dark regions) are insulating domains for the 
electrons.
Shown are 4 out of the 10 triangles used in the actual experiment \cite{lin99a}.
The period $L$ of the triangles is about $1.2\mu m$.
} 
\label{figqm4} 
\end{figure} 
}

\def\figqmfunf{
\begin{figure}[ht] 
\centerline{\epsfig{file=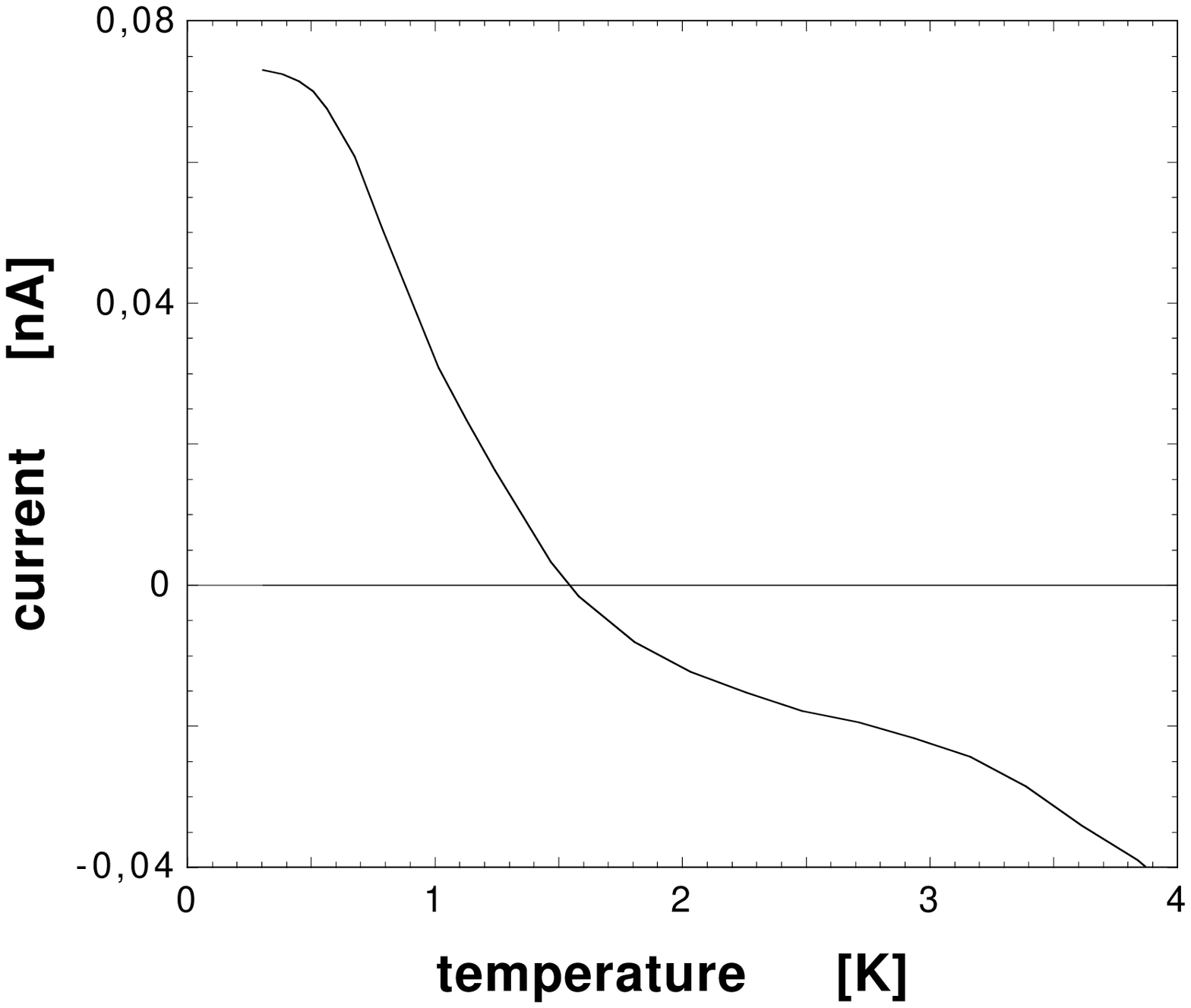, width = 10 cm} }
\caption{Electrical current $I=-e\,\langle\dot x\rangle$ 
along the quantum dot array from \fig \ref{figqm4}
versus temperature 
for an unbiased rocking voltage $y(t)$ which periodically
jumps between $\pm 1mV$ at a frequency of 191Hz \cite{lin99a}.
} 
\label{figqm5} 
\end{figure} 
}

\def\figsiebeneins{
\begin{figure}[ht] 
\centerline{\epsfig{file=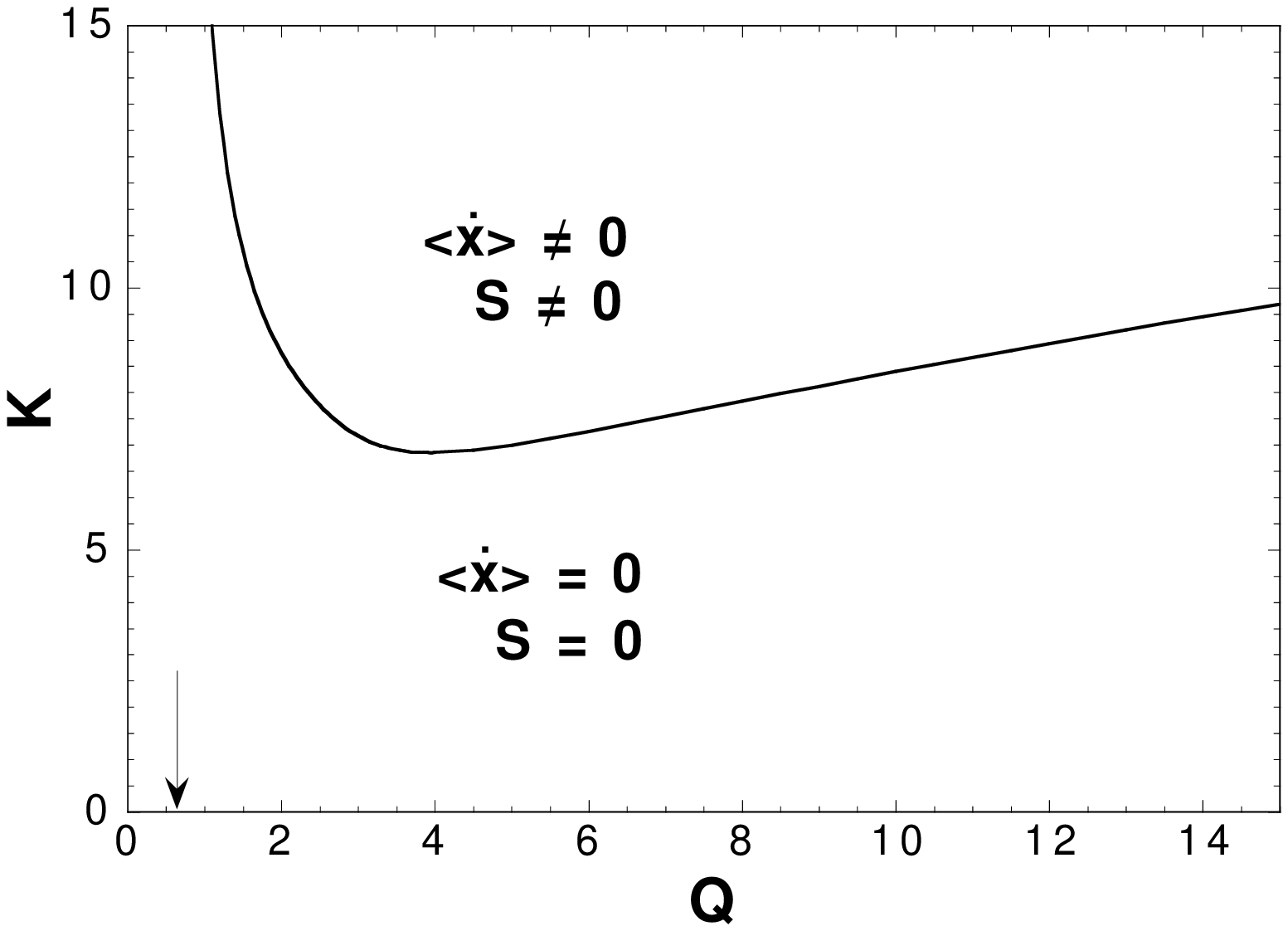, width = 10 cm} }
\caption{Phase diagram for the model
(\ref{7.10})-(\ref{7.10b}) with $V(x)=-\cos x - 0.15\, \cos(2x)$ (cf.
\eq (\ref{7.25}) below) and $T=2$ in the thermodynamic limit
$N\to\infty$ by numerically evolving the non-linear Fokker-Planck 
equation (\ref{7.13}) until a steady state was reached.
$\langle\dot x\rangle$ is the particle current,
$S$ the order parameter from (\ref{7.12}), and the
arrow indicates the asymptotic phase boundary (\ref{7.21})
for $K\to\infty$.
} 
\label{fig7.1} 
\end{figure} 
}

\def\figsiebenzwei{
\begin{figure}[ht] 
\centerline{\epsfig{file=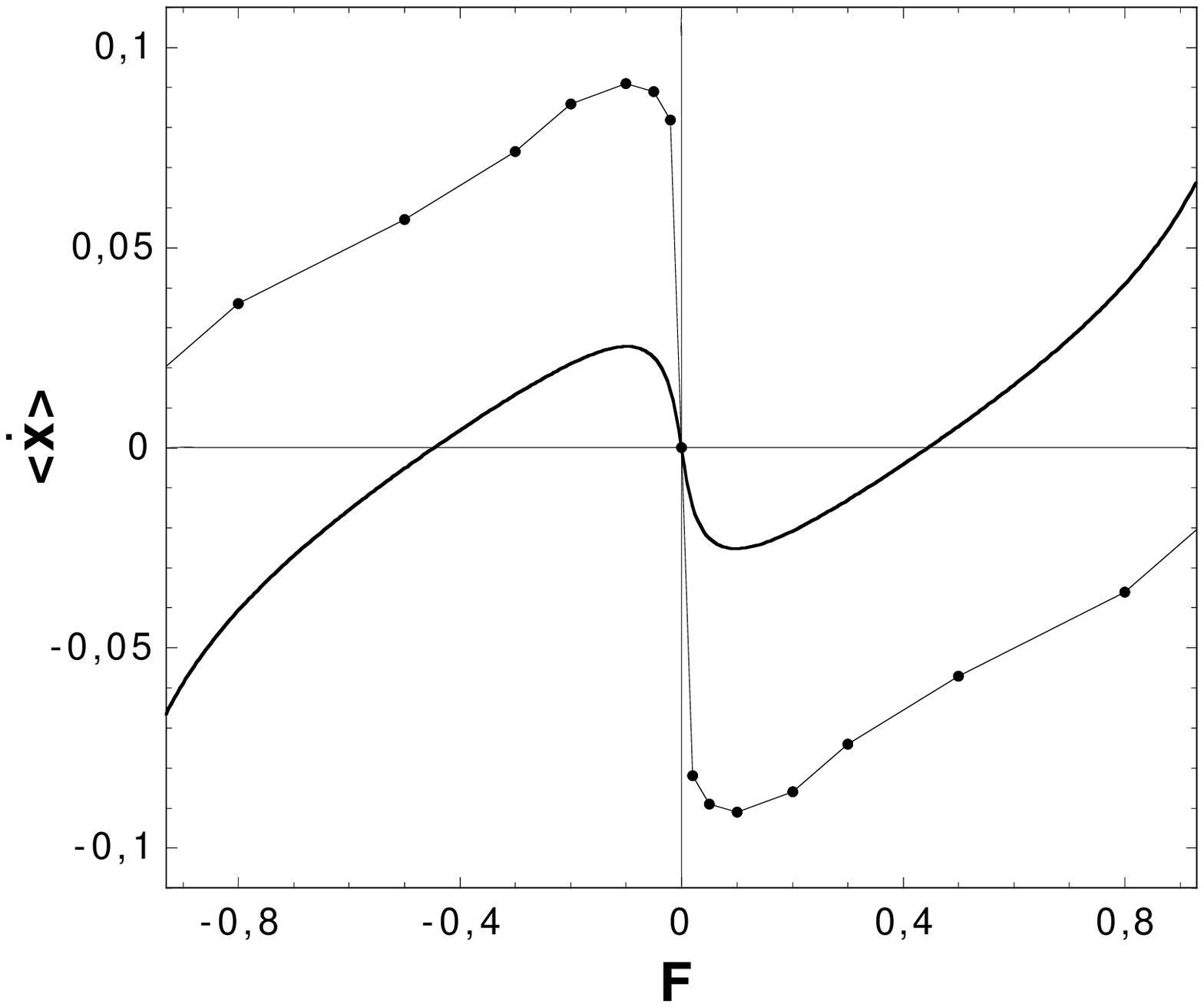, width = 10 cm} }
\caption{
Solid line: Steady state current $\langle\dot x\rangle$ versus force $F$ for the
model (\ref{7.10})-(\ref{7.10b}), (\ref{7.25}), (\ref{7.29a})
with $T=2$, $Q=2$, $K=8$, $A=0.15$ in the thermodynamic limit
$N\to\infty$ by solving the non-linear Fokker-Planck 
equation (\ref{7.13}).
Interconnected dots:
Simulations of (\ref{7.10}) with nearest
neighbor instead of global coupling
(\ref{7.30}) for a 64*64 square lattice with periodic
boundary conditions and modified parameters $Q=6$, $K=15$,
averaged over 10 realizations.
} 
\label{fig7.2} 
\end{figure} 
}

\def\figsiebendrei{
\begin{figure}[ht] 
\centerline{\epsfig{file=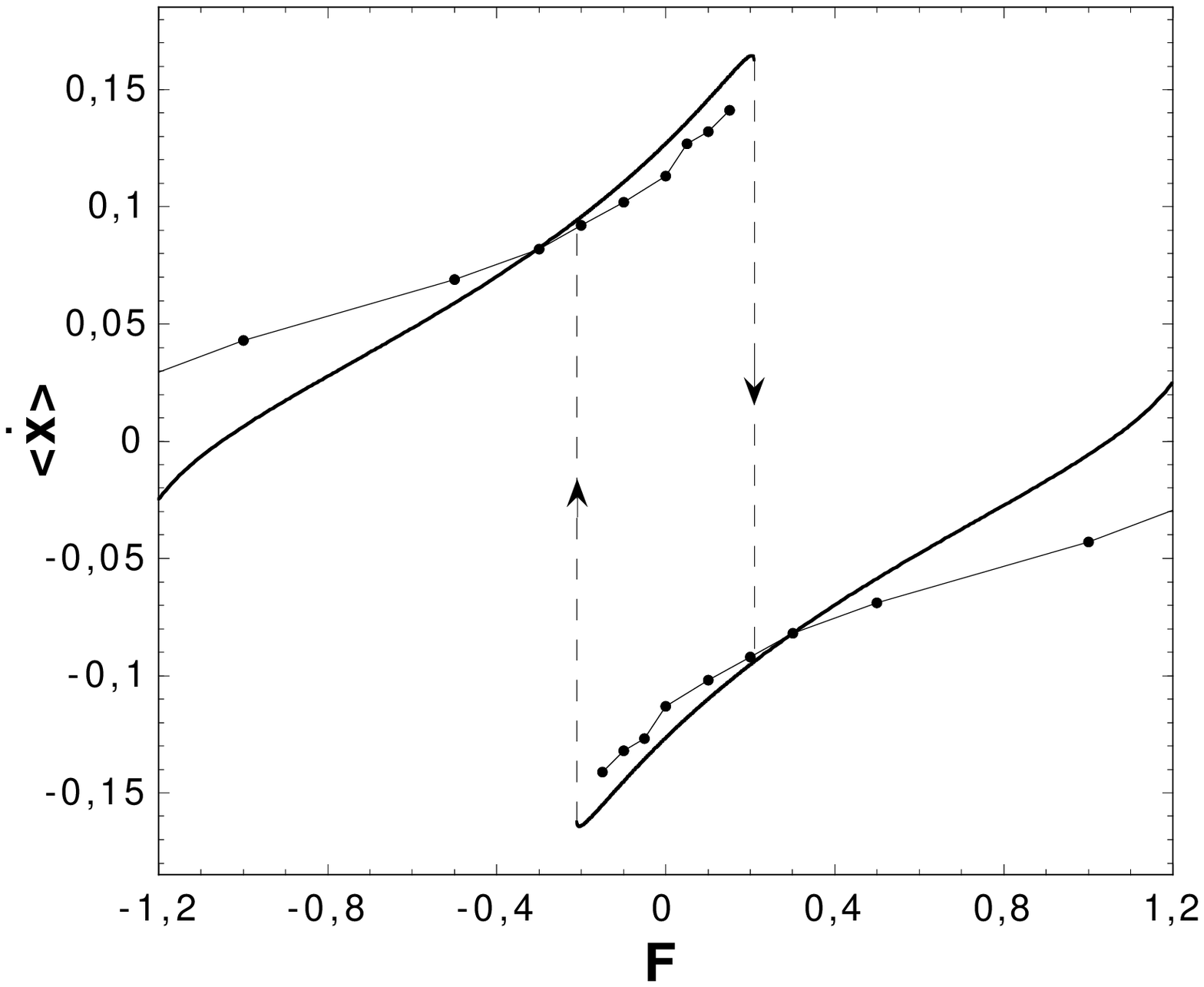, width = 10 cm} }
\caption{
Same as in \fig \ref{fig7.2} but for $Q=4$, $K =10$ (global coupling)
and $Q=10$, $K =20$ (nearest neighbor coupling).
} 
\label{fig7.3} 
\end{figure} 
}

\title{Brownian motors: noisy transport far from 
equilibrium\footnote{Revised version, accepted for publication 
in Physics Reports}}
\author{Peter Reimann\\
Universit\"at Augsburg, Institut f\"ur Physik, Universit\"atsstr. 1, 86135 Augsburg, Germany}
\date{August 2001}

\begin{document}
\thispagestyle{empty}
\maketitle

\setcounter{page}{1}
\pagenumbering{roman}
\tableofcontents
\newpage

\begin{abstract}
Transport phenomena in spatially periodic systems far from thermal
equilibrium are considered.
The main emphasize is put on directed transport in so-called 
Brownian motors (ratchets), i.e. a dissipative dynamics in the 
presence of thermal noise and some prototypical perturbation that drives
the system out of equilibrium without introducing a priori 
an obvious bias into one or the other direction of motion.
Symmetry conditions for the appearance (or not) of directed current,
its inversion upon variation of certain parameters,
and quantitative theoretical predictions for specific
models are reviewed as well as a wide variety of experimental realizations
and biological applications, especially the modeling of molecular motors.
Extensions include quantum mechanical and collective effects,
Hamiltonian ratchets, the influence of spatial disorder, 
and diffusive transport.
\\
\\
PACS:  
5.40.-a, 
5.60.-k, 
87.16.Nn 

\end{abstract}

\pagenumbering{arabic}
\setcounter{page}{1}
\pagestyle{headings}

\chapter{Introduction}\label{cha1}
\section{Outline and Scope}
The subject of the present review are transport phenomena in 
{\em spatially periodic} systems out of thermal equilibrium.
While the main emphasis is put on {\em directed transport}, also
some aspects of {\em diffusive transport} will be addressed.
We furthermore focus mostly on small-scale systems for which 
{\em thermal noise} plays a non-negligible or even dominating role.
Physically, the thermal noise has its origin in the thermal environment
of the actual system of interest. 
As an unavoidable consequence, the system dynamics is then always subjected
to dissipative effects as well.

Apart from transients, directed transport in a spatially periodic system
in contact with a single dissipation- and noise-generating thermal heat 
bath is ruled out by the second law of thermodynamics.
The system has therefore to be driven {\em away from thermal equilibrium}
by an additional deterministic or stochastic perturbation.
Out of the infinitely many options, we will mainly focus on 
either a {\em periodic driving} or a restricted selection of 
{\em stochastic processes} of prototypal simplicity.
In the most interesting case, these perturbations are furthermore
{\em unbiased}, i.e. the time-, space-, and ensemble-averaged
forces which they entail are required to vanish.
Physically, they may be either externally imposed
(e.g. by the experimentalist) or of system-intrinsic origin,
e.g. due to a second thermal heat reservoir at a different temperature
or a non-thermal bath.

Besides the breaking of thermal equilibrium,
a further indispensable requirement for directed transport 
in spatially periodic systems is clearly the {\em breaking of the 
spatial inversion symmetry}.
There are essentially three different ways to do this, and we will
speak of {\em a Brownian motor}, or equivalently, {\em a ratchet system} 
whenever a single one or a combination of them is realized:
First, the spatial inversion symmetry of the periodic system
itself may be broken intrinsically, that is,
already in the absence of 
the above mentioned non-equilibrium perturbations.
This is the most common situation and typically involves some 
kind of periodic and asymmetric, so-called ratchet-potential.
A second option is that the non-equilibrium perturbations,
notwithstanding the requirement that they must be unbiased,
bring about a spatial asymmetry of the dynamics.
A third possibility arises as a collective effect
in coupled, perfectly symmetric non-equilibrium systems, 
namely in the form of spontaneous symmetry breaking.

As it turns out, these two conditions (breaking of thermal
equilibrium and of spatial inversion 
symmetry) are generically sufficient
for the occurrence of the so-called {\em ratchet effect},
i.e. the emergence of directed transport in a spatially periodic system.
Elucidating this basic phenomenon in all its facets is the
central theme of our present review.

We will mainly focus on two basic classes of ratchet systems,
which may be roughly characterized as follows
(for a more detailed discussion see \sect \ref{sec3.1.3}):
The first class, called {\em pulsating ratchets}, are
those for which the above mentioned periodic or stochastic
non-equilibrium perturbation gives rise to a time-dependent
variation of the potential shape without affecting its spatial
periodicity.
The second class, called {\em tilting ratchets},
are those for which these non-equilibrium perturbations
act as an additive driving force, which is 
unbiased on the average.
In full generality, also combinations of
pulsating and tilting ratchet schemes are possible,
but they exhibit hardly any fundamentally new
basic features (see \sect \ref{sec3.2.2}).
Even within those two classes,
the possibilities of breaking thermal equilibrium and
symmetry in a ratchet system are still numerous and
in many cases, predicting the actual direction of the
transport is already far from obvious, not to 
speak of its quantitative value.
In particular, while the occurrence of a ratchet effect is 
the rule, exceptions with zero current are still possible.
For instance, such a non-generic situation may be created
by fine-tuning of some parameter.
Usually, the direction of transport then exhibits a change
of sign upon variation of this parameter, called 
{\em current inversions}.
Another type of exception can be traced back to {\em symmetry reasons}
with the characteristic signature of zero current without 
fine tuning of parameters.
The understanding and control of such exceptional cases is clearly 
another issue of considerable theoretical and practical interest that we
will discuss in detail (especially in \sects 
\ref{sec3.4} and \ref{sec3.5}).

\section{Historical landmarks}\label{sec3.3}
Progress in the field of Brownian motors
has evolved through
contributions from rather different directions and
re-discoveries of the same basic principles
in different contexts have been made repeatedly.
Moreover, the organization of the much more 
detailed subsequent chapters will not
always admit it to keep the proper historical order.
For these reasons, a brief historical {\em tour d'horizon}
seems worthwhile at this place.
At the same time, this gives a first flavor of the
wide variety of applications of Brownian motor concepts.

Though certain aspects of the ratchet effect are contained implicitly 
already in the works of Archimedes, Seebeck, Maxwell, Curie, and others,
Smoluchowski's {\em Gedankenexperiment} from 1912 \cite{smo12}
regarding the {\em prima facie} quite astonishing absence of 
directed transport in spatially asymmetric systems in contact with 
a single heat bath,
may be considered as the first seizable major contribution
(discussed in detail in \sect \ref{sec2.1}).
The next important step forward represents Feynman's 
famous recapitulation and
extension \cite{fey63} to the case of two thermal heat baths 
at different temperatures (see \sect \ref{sec4.8.2}).

Brillouins paradox \cite{bri50} from 1950 (see \sect \ref{sec2.2.3c}) 
may be viewed as a variation of Smoluchowski's 
counterintuitive observation.
In turn, Feynman's prediction that in the presence of a second heat bath
a ratchet effect will manifest itself, has its Brillouin-type 
correspondence in the Seebeck
effect (see \sect \ref{sec4.8.1}), 
discovered by Seebeck in 1822 of course without
any idea about the underlying microscopic ratchet effect.

Another root of Brownian motor theory leads us into the realm of
intracellular transport research, specifically the biochemistry of
molecular motors and molecular pumps.
In the case of molecular motors, the concepts which we have in mind here
have been unraveled in several steps, starting with A. Huxley's
ground-breaking 1957 work on muscle contraction \cite{hux57}, 
and continued in the late 80-s by Braxton and Yount 
\cite{bra88,bra89}
and in the 90-s by Vale and Oosawa \cite{val90}, Leibler and Huse
\cite{lei91,lei93}, Cordova, Ermentrout, and Oster \cite{cor92},
Magnasco \cite{mag93,mag94},  
Prost, Ajdari, and collaborators \cite{pro94,jul97a},
Astumian and Bier \cite{ast94,ast96a}, 
Peskin, Ermentrout, and Oster \cite{pes94,pes95}
and many others, see \ch \ref{cha5}.
In the case of molecular pumps, the breakthrough came with the theoretical
interpretation of previously known experimental findings \cite{ser83,ser84}
as a ratchet effect in 1986 by Tsong,
Astumian and coworkers \cite{tso86,wes86}, see \sect \ref{sec4.5}.
While the general importance of asymmetry induced rectification,
thermal fluctuations, and the coupling of non-equilibrium enzymatic 
reactions to mechanical currents according to Curie's principle
for numerous cellular transport processes is long known \cite{bio83,fri86},
the above works introduced for the first time a quantitative microscopic
modeling beyond the linear response regime close to thermal equilibrium.

On the physical side, 
a ratchet effect in the form of voltage rectification by a dc-SQUID
in the presence of a magnetic field and an unbiased ac-current 
(i.e. a tilting ratchet scheme) has been
experimentally observed and theoretically interpreted
as early as in 1967 by De Waele, Kraan, de Bruyn Ouboter, and Taconis
\cite{wae67,wae69}.
Further, directed transport induced by unbiased,
time-periodic driving forces in spatially periodic structures with 
broken symmetry has been the subject of several hundred 
experimental and theoretical papers since the mid 70-s.
In this context of the so-called photovoltaic and photorefractive effects in 
non-centrosymmetric
materials, a ground breaking experimental
contribution represents the 1974 paper by
Glass, von der Linde, and Negran \cite{gla74}. 
The general theoretical framework was elaborated a few
years later by Belinicher, Sturman and coworkers, 
as reviewed -- together with the above mentioned numerous experiments --
in their capital works \cite{bel80,stu92}.
They identified
as the two main ingredients for the occurrence of the ratchet
effect in periodic systems the breaking of thermal equilibrium
(detailed balance symmetry) and of the spatial symmetry,
and they pointed out the much more general validity
of such a tilting ratchet scheme beyond
the specific experimental systems at hand, see \sect \ref{sec6.2.1}.

The possibility of producing a dc-output by two superimposed 
sinusoidal ac-inputs at frequencies $\omega$ and $2\omega$ in a spatially periodic, 
symmetric system, exemplifying a so-called asymmetrically tilting ratchet mechanism, 
has been observed experimentally 1978 by Seeger and Maurer
\cite{see78} and analyzed theoretically 1979 by Wonneberger \cite{won79}, 
see \sect \ref{sec6.3.1}.
The occurrence of a ratchet effect has been theoretically predicted 1987
by Bug and Berne \cite{bug87} for the simplest variant of
a pulsating ratchet scheme, termed on-off ratchet (see \sect \ref{sec4.1}).
A ratchet model with a symmetric
periodic potential and a state-dependent temperature (multiplicative
noise) with the same periodicity but out of phase,
i.e. a simplified microscopic model for the Seebeck 
effect (see \sect \ref{sec4.8.1}),
has been analyzed 1987 by B\"uttiker \cite{but87}.

The independent re-inventions of the on-off ratchet scheme 1992
by Ajdari and Prost \cite{ajd92} and of the tilting
ratchet scheme 1993 by Magnasco \cite{mag93}
together with the seminal 1994 works (ordered by date of receipt)
\cite{pro94,ast94,doe94,hon94,mil94},\cite{mag94,rou94,ajd94a,cha94},\cite{pes94},\cite{ajd94,bar94}
provided the inspiration for a whole
new wave of great theoretical and experimental activity and 
progress within the statistical physics community
as detailed in the subsequent chapters and reviewed e.g. in
\cite{mad1,mad2,mad3,lei94,pop94,bar95a,doe95,faz95,han96,kos96,luc96,ast97,jul97a,bie97,bie97z,cha98,luc99,ast00,nzz01,ast01a}.
While initially the modeling of molecular motors has served as one of the main
motivations, the scope of Brownian motor studies has subsequently been 
extended to an ever increasing number of physical and technological
applications, 
along with the re-discovery of the numerous pertinent works from before 1992.
As a result, a much broader and unified conceptual basis has been
achieved, new theoretical tools have been developed which
lead to the discovery of many interesting and quite astonishing
effects, and a large variety of exciting new experimental realizations
have become available.

Within the realm of noise-induced or -assisted 
non-equilibrium phenomena,
an entire family of well-established major 
fields are known under the labels of 
stochastic resonance \cite{gam98},
noise induced transitions \cite{hor84} and 
phase transitions \cite{vdb97,gar99}, 
reaction rate theory \cite{han90,han95,rei97ra}, 
and driven diffusive systems \cite{sch95c,sch00c}, to name
but a few examples.
One objective of our present review is to show that
the important recent contributions of many
workers to the theory and application of Brownian 
motors has given rise to another 
full-fledged member of this family.

\section{Organization of the paper}
This review addresses two readerships:
It may serve as an introduction to the
field without requiring any specialized preknowledge.
On the other hand, it offers to the active researcher
a unifying view and guideline through the very rapidly
growing literature.
For this reason, not everything will be of equal interest
for everybody.
The following outline together with the table of contents 
may help to make one's selection.

Essentially, the subsequent 8
chapters can be divided into three units 
of rather different characters: 
The first unit (\ch \ref{cha2}) is predominantly
of {\em introductory} and pedagogical nature,
illustrating the
basic phenomena, concepts, and applications by way of examples.
Technically, the discussion is conducted on a rather elementary
level and the calculations are to a far extent self-contained.
``Standard'' lines of reasoning and
the derivation of basic working tools are discussed rather
detailed in
mathematically heuristic but physically suggestive terms.
While these parts of \ch \ref{cha2} are not meant
to replace a systematic introduction to the field of stochastic
processes, they may hopefully serve as a minimal basis for
the technically less detailed subsequent chapters.

\Ch \ref{cha3} is devoted to general and {\em systematic} 
considerations which are relevant for the entire subsequent
parts of the paper.
The main classes of ratchet models and their physical 
origin are discussed with particular emphasize on
symmetries, current inversions, and asymptotic regimes.
\Chs \ref{cha4}-\ref{cha06} represent the main body of the 
present work and are to a large extent of {\em review} character.
It was only during the completion of these chapters that the amount
of pertinent literature in this context became clear.
As a consequence, specific new aspects of the considered ratchet systems 
and of the obtained results could only be included for a 
selection of particularly significant such studies. 
Even then, the technical procedures and the detailed 
quantitative findings had to be mostly omitted.
Besides the conceptual theoretical considerations and the
systematic discussion of various specific model classes, 
a substantial part of chapters \ref{cha4}-\ref{cha06}
has also been reserved for the manifold
{\em experimental} applications of those ideas.

\Chs \ref{cha5}-\ref{cha7} represent the third main unit 
of our work, elaborating in somewhat more detail
three major instances of {\em applications and extensions}.
Of methodic rather than review character are the 
first 3 sections of \ch \ref{cha5}, 
illustrating a typical stochastic
modeling procedure for the particularly important example
of intracellular transport processes by molecular motors.
The remainder of \ch \ref{cha5} presents a survey of
the field with particular emphasize on cooperative
molecular motors and the character of the 
mechanochemical coupling.
\Ch \ref{sec6.5} is devoted to the discussion of 
theoretically predicted new characteristic
quantum mechanical signatures of Brownian motors
and their experimental verification on the basis of a quantum
dot array with broken spatial symmetry.
Finally, \ch \ref{cha7}
deals with collective effects of interacting ratchet systems.
On the one hand, we review modifications of the directed transport
properties of single ratchets caused by their 
interaction (\sect \ref{sec7.1}).
On the other hand (\sect \ref{sec7.2}) we exemplify genuine
collective transport phenomena by a somewhat more detailed discussion
of one specific model of paradigmatic simplicity --
meant as a kind of  ``normal form'' description
which still captures the essence of more realistic
models but omits all unnecessary details, 
in close analogy to the philosophy usually adopted in 
the theory of equilibrium phase transitions.

Concluding remarks and future perspectives are presented in
\ch \ref{cha9}. Some technical details from the introductory
\ch \ref{cha2} are contained in the appendices.

Previously unpublished {\em research} represent 
the considerations about supersymmetry in \sect \ref{sec3.4}, 
the method of tailoring current inversions in \sect \ref{sec3.5},
the general treatment of the linear response regime 
in \sect \ref{sec3.1.4},
the approximative molecular motor model with two 
highly cooperative ``heads'' in \sect \ref{sec5.5},
as well as a number of additional minor new results which are 
indicated as such throughout the text, e.g.,
various exact mappings between different classes 
of ratchet systems.
New, mainly by the way of presentation but to some degree also
by their content, are 
parts of \sects \ref{2.1}-\ref{sec2.1.4} and \ref{sec4.8.1}-\ref{sec4.8.4}, 
the systematic ratchet classification scheme 
and its physical basis in \sects \ref{sec3.1.3} and \ref{sec3.2},
the unified fast fluctuating force asymptotics in \sect \ref{sec6.1.1},
as well as the coherent historical review in the preceding 
\sect \ref{sec3.3}.

A kind of red thread through the entire 
review consists in the {\em asymptotic analysis} of
the so-called fast-driving limit.
By collecting and rewriting the various results spread out in the 
literature and completing the missing pieces,
a unified picture of this asymptotic regime
emerges for the first time.
The structural similarity of these analytical
results in view of the rather different underlying 
models is remarkable. 
For instance, within our standard working model --
the overdamped Brownian motion in a periodic non-equilibrium 
system involving some ratchet-potential $V(x)$ of period $L$ --
the direction of the average particle current is governed
under very general circumstances by a factor of the form
$\int_0^L V'(x)\, [d^n V(x)/dx^n]^2 dx$ with a
model dependent $n$-value between $1$ and $3$.
Especially, already within this asymptotic regime,
the intriguingly complicated dependence of the directed 
transport, e.g. on the detailed potential-shape $V(x)$, 
becomes apparent -- a typical feature of systems far from 
thermal equilibrium.

Basically, the review is organized in three levels (chapters,
sections, subsections). While from the logical viewpoint,
additional levels would have been desirable,
the present rather ``flat'' structure simplifies 
a quick orientation on the basis of the table of contents.
Throughout the main text,
cross-referencing to related subsections
is used rather extensively.
It may be ignored in case of a systematic reading,
but is hopefully of use otherwise.

\chapter{Basic concepts and phenomena}\label{cha2}

This chapter serves as a motivation and first exposition of the
main themes of our review,
such as the absence of directed transport in ratchet systems 
at thermal equilibrium, its generic occurrence away from equilibrium, 
and the possibility of current inversions upon variation of some parameter.
These fundamental phenomena are exemplified in their simplest form in
\sect \ref{sec2.1}, \sects \ref{sec2.2.2}-\ref{sec2.2.3c}, and 
\sect \ref{sec2.5},
respectively, and will then be elaborated in more generality and depth 
in the subsequent chapters.
At the same time, this chapter also introduces the basic 
stochastic modeling concepts as well as the mathematical 
methods and ``standard arguments'' in this context.
These issues are mainly contained in \sects \ref{sec2.1.3}-\ref{sec2.2.1} 
and \ref{sec2.4.1},
complemented by further details in the respective appendices.
Readers which are already familiar with these basic physical
phenomena and mathematical concepts may immediately proceed to 
\ch \ref{cha3}.

\section{Smoluchowski-Feynman ratchet}\label{sec2.1}
Is it possible, and how is it possible to gain useful work out
of unbiased random fluctuations?
In the case of macroscopic fluctuations, the task can indeed
be accomplished by various well-known types of mechanical and 
electrical rectifiers.
Obvious daily-life examples are the wind-mill or the self-winding
wristwatch.
More subtle is the case of microscopic fluctuations, as demonstrated
by the following {\em Gedankenexperiment} about converting Brownian motion
into useful work.
The basic idea can be traced back to a conference talk by Smoluchowski 
in M\"unster 1912 (published as 
proceedings-article
in Ref. \cite{smo12}) and was later popularized and extended in Feynman's
Lectures on Physics \cite{fey63}.

\subsection{Ratchet and pawl}\label{sec2.1.1}
The main ingredient of Smoluchowski and Feynman's {\em Gedankenexperiment} 
is an axle with at one
end paddles and at the other end a so-called {\em ratchet}, reminiscent 
of a circular saw with asymmetric saw-teeth (see \fig \ref{fig1}). The whole 
device is surrounded by a gas at thermal equilibrium.
So, if it could
freely turn around, it would perform a rotatory Brownian motion
due to random impacts of gas molecules on the paddles.
The idea is now to rectify this unbiased random motion with the help 
of a {\em pawl} (see \fig \ref{fig1}).
It is indeed quite suggestive that the pawl will admit the saw-teeth to
proceed without much effort into one direction (henceforth
called ``forward'') but practically exclude a rotation in the 
opposite (``backward'') direction. 
In other words, it seems quite convincing that
the whole gadget will perform on the average a systematic 
rotation in one direction, and this in fact even if a small 
load in the opposite direction is applied. 

\begin{center} 
\figeins
\end{center} 

Astonishingly enough, this naive expectation is wrong:
{\em In spite of the built in asymmetry,
no preferential direction of motion is possible}.
Otherwise, such a gadget would represent a
{\em perpetuum mobile} of the second kind, in contradiction to the
second law of thermodynamics.
The culprit must be our assumption about the working of the pawl, 
which is indeed closely resembling Maxwell's 
demon\footnote{Both Smoluchowski and Feynman have pointed out 
the similarity between the working principle of the
pawl and that of a valve.
A valve, acting between two boxes of gas, 
is in turn one of the simplest realizations of a Maxwell
demon \cite{max71}. 
For more details on Maxwell's demon, especially the history
of this apparent paradox and
its resolution, we refer to the commented collection of reprints in \cite{lef90}.}.
Since the impacts of the gas molecules take place on a microscopic scale,
the pawl needs to be extremely small and soft in order to admit a rotation
even in the forward direction. 
As Smoluchowski points out, the pawl itself is therefore also
subjected to
non-negligible random thermal fluctuations. So, every once in a while the 
pawl lifts itself up and the saw-teeth can freely travel underneath.
Such an event clearly favors on the average a rotation in the
``backward'' direction in \fig \ref{fig1}. At overall thermal
equilibrium (the gas surrounding the paddles and the pawl being at
the same temperature)
the detailed quantitative analysis \cite{fey63} indeed results in the 
subtle probabilistic balance which just rules out the functioning
of such a {\em perpetuum mobile}.

A physical system as described above will be called after
Smoluchowski and Feynman.
We will later go one step further and consider the case that 
the gas surrounding the paddles and the 
pawl are not at the same temperature (see \sect \ref{sec4.8.2}).
Such an extension of the original {\em Gedankenexperiment}
appears in Feynman's lectures, but has not been discussed
by Smoluchowski, and will therefore be named after Feynman only.

Smoluchowski and Feynman's ratchet and pawl has been 
{\em experimentally} realized on a molecular scale by 
Kelly, Tellitu, and Sestelo \cite{kel97,kel98,dav98,seb00}.
Their synthesis of triptycene$[4]$helicene
incorporates into a single molecule all essential components:
The triptycene ``paddlewheel'' functions simultaneously as circular
ratchet and as paddles, the helicene serves as pawl and provides
the necessary asymmetry of the system.
Both components are connected by a single chemical bond, giving rise to
one degree of internal rotational freedom.
By means of sophisticated NMR (nuclear magnetic resonance)
techniques, the predicted absence of a preferential direction of rotation
at thermal equilibrium has been confirmed experimentally. The behavior
of similar experimental systems beyond the realm of thermal 
equilibrium will be discussed at the end of \sect \ref{sec4.4.2}.

\subsection{Simplified stochastic model}\label{sec2.1.2}
In the sense that we are dealing merely with a specific instance
of the second law of thermodynamics, the situation 
with respect to Smoluchowski-Feynman's ratchet and pawl  
is satisfactorily
clarified. On the other hand, the obvious intention of Smoluchowski and
Feynman is to draw our attention to the amazing content and
implications of this very
law, calling for a more detailed explanation of what is going on.
A satisfactory modeling and analysis of the relatively complicated
ratchet and pawl gadget as
it stands is possible but rather involved,
see \sect \ref{sec4.8.2}.
Therefore, we focus on a considerably simplified model which,
however, still retains the basic qualitative features:
We consider a Brownian particle in one dimension with coordinate $x(t)$
and mass $m$, which is governed by Newton's equation of 
motion\footnote{Dot and prime indicate differentiations with respect 
to time and space, respectively.}:
\begin{equation}
m\ddot x(t)+V'(x(t)) = -\eta\,\dot x(t) + \xi (t)\ .
\label{2.1}
\end{equation}
Here $V(x)$ is a periodic potential with period $L$,
\begin{equation}
V(x+L)=V(x)\ ,
\label{2.1'}
\end{equation}
and broken spatial 
symmetry\footnote{Broken spatial symmetry means that there is no 
$\Delta x$ such that $V(-x) = V(x+\Delta x)$ for all $x$.}, 
thus playing the role of the ratchet in \fig \ref{fig1}.
A typical example is
\begin{equation}
V(x) = V_0 \, [\sin (2\pi x/L) + 0.25\, \sin(4 \pi x/L)] \ ,
\label{2.1''}
\end{equation}
see \fig \ref{fig2}.

The left hand side in (\ref{2.1}) represents the deterministic,
conservative part of the particle dynamics, while the right hand side
accounts for the effects of the thermal environment. These are energy 
dissipation, modeled in (\ref{2.1}) as
viscous friction with
friction coefficient $\eta$, and randomly fluctuating forces
in the form of the thermal noise $\xi(t)$.
These two effects are not independent of each other
since they have both the same origin, 
namely the interaction of the particle $x(t)$ with a huge number of 
microscopic degrees of freedom of the environment.
As discussed in detail in \sects \ref{sec2.1.2.1} and \ref{sec2.1.2.2} 
of Appendix A, our assumption that the environment is an equilibrium
heat bath with temperature $T$ and that its effect on the system
can be modeled by means of the phenomenological {\em ansatz} 
appearing on the right hand side of (\ref{2.1}) completely fixes
\cite{ein05,ein10,joh28,nyq28,cal51,ber55,leb57,mag59,rub60,leb63,res64,ull66,zwa73,hyn75,spo78,gra80,gra82a,cal83,for88,han90,wei99a,rei01a}
all statistical properties of the fluctuations $\xi(t)$ without referring
to any microscopic details of the environment 
(see also \sects \ref{sec2.2.3c}, \ref{sec3.2.1}, \ref{sec6.5.1}).
Namely, $\xi(t)$ is a {\em Gaussian white noise} of zero mean,
\begin{eqnarray}
\langle\xi(t)\rangle & = & 0 \ , 
\label{2.2}
\end{eqnarray}
satisfying the {\em fluctuation-dissipation relation}
\cite{nyq28,joh28,cal51}
\begin{eqnarray}
\langle\xi(t)\xi(s)\rangle & = & 
2\,\eta\, k_B T \, \delta(t-s) \ ,
\label{2.3}
\end{eqnarray}
where $k_B$ is Boltzmann's constant, 
$2\eta k_BT$ is the {\em noise intesity} or {\em noise strength},
and $\delta(t)$ is Dirac's delta-function.
Note that the only particle property which 
enters the characteristics of the noise is the friction 
coefficient $\eta$, which may thus be viewed as the coupling strength
to the environment.

\begin{center} 
\figzwei
\end{center} 

For the typically very small systems one has in mind, and for which
thermal fluctuations play any notable role at all, the dynamics (\ref{2.1})
is {\em overdamped}, that is, the inertia term $m\, \ddot x(t)$ is 
negligible (see also the more detailed discussion of this point 
in \sect \ref{sec2.1.2.4} of Appendix A).
We thus arrive at our ``minimal'' {\em Smoluchowski-Feynman ratchet} model
\begin{eqnarray}
& & \eta\, \dot x(t) = -V'(x(t))+\xi(t)\ .
\label{2.5}
\end{eqnarray}

According to (\ref{2.3}), the Gaussian white noise $\xi(t)$ is 
uncorrelated in time, i.e., it is given by
independently sampled Gaussian random numbers at any time
$t$.
This feature and the concomitant infinitely large second moment
$\langle\xi^2(t)\rangle$ are mathematical idealizations. In physical reality,
the correlation time is meant to be
finite, but negligibly small in comparison with
all other relevant time scales of the system. In this spirit, we may
introduce a small time step $\Delta t$ and consider a time-discretized 
version of the stochastic dynamics (\ref{2.5}) of the form
\begin{equation}
x(t_{n+1}) = x(t_{n}) -  \Delta t \, [V'(x(t_{n}))+\xi_n]/\eta \ ,
\label{2.7}
\end{equation}
where $t_n := n\,\Delta t$ and where the $\xi_n$ are 
independently sampled,
unbiased Gaussian random numbers with second moment
\begin{equation}
\langle\xi^2_n\rangle = 2\,\eta\, k_B T/\Delta t\ .
\label{2.7'}
\end{equation}
The continuous dynamics (\ref{2.5}) 
with uncorrelated noise is then to be
understood \cite{gar83,ris84,kam92}
as the mathematical limit of 
(\ref{2.7}) for $\Delta t\to 0$. Moreover, this discretized dynamics 
(\ref{2.7}) is a suitable starting point for a numerical simulation 
of the problem. 
Finally, a derivation of the so-called Fokker-Planck equation 
(see \eq (\ref{2.10}) below) based on (\ref{2.7}) is given in Appendix B.

\section{Fokker-Planck equation}\label{sec2.1.3}
The following 4 sections are mainly of methodological nature 
without much new physics.
After introducing the Fokker-Planck equation in the present section,
we turn in \sects \ref{sec2.1.3'} and \ref{sec2.1.4} to the
evaluation of the particle current $\langle\dot x\rangle$, with
the result that even when the spatial symmetry is broken
by the ratchet potential $V(x)$, there arises no 
systematic preferential 
motion of the random dynamics in one or the other direction.
Finally, in \sect \ref{sec2.2.1} the effect of an additional
static ``tilting''-force $F$ in the Smoluchowski-Feynman
ratchet dynamics (\ref{2.5}) is considered,
with the expected result of a finite particle current
$\langle\dot x\rangle$ with the same sign as the applied force $F$.
Readers which are already familiar with or not interested in 
these standard techniques are recommended to continue with 
\sect \ref{sec2.2.2}.

Returning to (\ref{2.5}), a quite natural next step is to
consider a statistical ensemble of these stochastic processes 
belonging to independent realizations of the random 
fluctuations $\xi(t)$. The corresponding probability density
$P(x,t)$ in space $x$ at time $t$ describes the distribution of the 
Brownian particles and follows as an ensemble average\footnote{To
be precise, an average over the initial conditions $x(t_0)$ according to
some prescribed statistical weight $P(x,t_0)$ together with an average over 
the noise is understood on the right hand side of (\ref{2.8a}).}
of the form
\begin{equation}
P(x,t) :=\langle\delta(x-x(t))\rangle \ .
\label{2.8a}
\end{equation}
An immediate consequence of this definition is the normalization
\begin{equation}
\int_{-\infty}^\infty dx\, P(x,t) = 1 \ .
\label{2.11}
\end{equation}
Another trivial consequence is that $P(x,t)\geq 0$ for all $x$ and $t$.

In order to determine the time-evolution of
$P(x,t)$, we first consider in (\ref{2.5}) the special 
case $V'(x)\equiv 0$.
As discussed in detail in \sect \ref{sec2.1.2.3} of Appendix A,
we are thus dealing with the force-free thermal diffusion
of a Brownian particle with a diffusion coefficient $D$
that satisfies Einstein's relation \cite{ein05}
\begin{equation}
D = k_B T / \eta \ .
\label{2.4}
\end{equation}
Consequently, $P(x,t)$ is governed by the diffusion equation
\begin{equation}
\frac{\partial}{\partial t} P(x,t) = \frac{k_B T}{\eta} 
\, \frac{\partial^2}{\partial x^2}P(x,t)
\qquad \mbox{if}\qquad  V'(x)\equiv 0 \ .
\label{2.8}
\end{equation}
Next we address the deterministic dynamics $\xi(t)\equiv 0$ in (\ref{2.5}).
In complete analogy to classical Hamiltonian mechanics, one then finds that
the probability density $P(x,t)$ evolves according to a 
Liouville-equation of the 
form\label{fot2.3}\footnote{Proof: Let $x(t)$ be a solution of 
$\dot x(t)=f(x(t))$ and define $P(x,t):=\delta(x-x(t))$.
Note that the variable $x$ and the function $x(t)$ are mathematically completely 
unrelated objects.
Then $\frac{\partial}{\partial t} P(x,t)$ 
$\!= - \dot x(t)\, \frac{\partial}{\partial x} \delta(x-x(t))$
$\!= - f(x(t))\, \frac{\partial}{\partial x} \delta(x-x(t))$
$\!= - \frac{\partial}{\partial x} \{ f(x(t)) \delta(x-x(t)) \}$
$\!= - \frac{\partial}{\partial x} \{ f(x) \delta(x-x(t)) \}$
(the last identity can be verified 
by operating with $\int dx\, h(x)$ on both sides, where
$h(x)$ is an arbitrary test function with $h(x\to\pm\infty )=0$, and then
performing a partial integration).
Thus (\ref{2.9}) is recovered for a $\delta$-distributed initial condition.
Since this \eq (\ref{2.9}) is linear in $P(x,t)$, the case of a general initial
distribution follows by linear superposition.}
\begin{equation}
\frac{\partial}{\partial t} P(x,t) = 
\frac{\partial}{\partial x}\left\{\frac{V'(x)}{\eta}\, P(x,t)\right\}
\qquad \mbox{if}\qquad  \xi(t) \equiv 0 \ .
\label{2.9}
\end{equation}
Since both (\ref{2.8}) and (\ref{2.9}) are linear in $P(x,t)$
it is quite obvious that the general case follows by combination of
both contributions, i.e., one obtains the so-called 
{\em Fokker-Planck equation} \cite{han82b,ris84}
\begin{equation}
\frac{\partial}{\partial t} P(x,t) = 
\frac{\partial}{\partial x}\left\{\frac{V'(x)}{\eta}\, P(x,t)\right\}
+ \frac{k_BT}{\eta} \, \frac{\partial^2}{\partial x^2}P(x,t) \ ,
\label{2.10}
\end{equation}
where the first term on the right hand side is called ``drift term'' and
the second ``diffusion term''.

While our above derivation of the Fokker-Planck equation is admittedly
of a rather heuristic nature, it is appealing due to its extreme
simplicity and the intuitive physical way of reasoning.
A more rigorous calculation, based on the discretized dynamics
(\ref{2.7}) in the limit $\Delta t\to 0$ is provided in Appendix B.
Numerous alternative derivations
can be found e.g. in \cite{kra40,han82b,gar83,ris84,han84,zwa90,kam92,rys97}
and further references therein.
A brief historical account of the Fokker-Planck equation has
been compiled in \cite{kam97}, see also \cite{cha43}.

\section{Particle current}\label{sec2.1.3'}
The quantity of foremost interest in the context of transport
in periodic systems is the {\em particle current} $\langle\dot x\rangle$,
defined as the time-dependent ensemble average over the velocities 
\begin{equation}
\langle\dot x\rangle := \langle \dot x(t)\rangle \ .
\label{2.10-1}
\end{equation} 
For later convenience, 
the argument $t$ in $\langle\dot x\rangle$ is omitted.
Obviously, the probability density $P(x,t)$ contains the entire
information about the system; in this section we treat
the question of how to extract 
the current $\langle\dot x\rangle$ out of it.

The simplest way to establish such a connection
between $\langle\dot x\rangle$ and $P(x,t)$ follows by
averaging in (\ref{2.5}) and taking into account (\ref{2.2}),
i.e., $\langle\dot x\rangle = - \langle V'(x(t)) \rangle /\eta$.
Since the ensemble average means by definition an average with respect to
the probability density $P(x,t)$ we arrive at our first basic observation, 
namely the connection between $\langle\dot x\rangle$ and $P(x,t)\ $:
\begin{equation}
\langle\dot x\rangle = 
- \int_{-\infty}^\infty dx\, \frac{V'(x)}{\eta}\, P(x,t) \ .
\label{2.10'}
\end{equation}

The above derivation of (\ref{2.10'}) has the disadvantage that
the specific form (\ref{2.5})
of the stochastic dynamics has been exploited.
For later use, we next sketch an alternative, more general
derivation:
From the definition (\ref{2.8a})
one obtains, independently of any details
of the dynamics governing $x(t)$, a so-called {\em master equation}
\cite{kam92,han82b,ris84}
\begin{eqnarray}
& & \frac{\partial}{\partial t}\, P(x,t) + 
\frac{\partial}{\partial x}\, J(x,t) = 0 
\label{2.10a}\\
& & J(x,t) := \langle \dot x(t)\,\delta (x-x(t))\rangle\ .
\label{2.10b}
\end{eqnarray}
Note that the symbols $x$ and $x(t)$ denote here completely unrelated
mathematical objects.
The master equation (\ref{2.10a}) has the form of a continuity 
equation for the probability density 
associated with the conservation
of particles, hence $J(x,t)$ is called the 
{\em probability current}.
Upon integrating (\ref{2.10b}), the following
completely general connection between the probability current and
the particle current is obtained:
\begin{equation}
\langle\dot x\rangle = \int_{-\infty}^\infty dx\, J(x,t)\ .
\label{2.10c}
\end{equation}

By means of a partial integration, the current in (\ref{2.10c})
can be rewritten as $-\int dx\, x\, \partial J(x,t)/\partial x$ and by
exploiting (\ref{2.10a}) one recovers the relation
\begin{equation}
\langle\dot x \rangle = \frac{d}{dt}\, \int_{-\infty}^\infty dx\, x\, P(x,t)\ ,
\label{2.10c'}
\end{equation}
which may thus be considered as an alternative definition
of the particle current $\langle\dot x\rangle$.

For the specific stochastic dynamics (\ref{2.5}), we find
by comparison of the Fokker-Planck equation (\ref{2.10})
with the general master equation (\ref{2.10a}) the explicit 
expression for the probability current
\begin{equation}
J(x,t) = -\left\{\frac{V'(x)}{\eta} 
+ \frac{k_B T}{\eta}\,\frac{\partial}{\partial x}\right\}\, P(x,t)
\label{2.10d}
\end{equation}
up to an additive, $x$-independent function.
Since both, $J(x,t)$ and $P(x,t)$ approach zero
for $x\to\pm \infty$, it follows that this function must be identically
zero. By introducing (\ref{2.10d}) into (\ref{2.10c}) we
finally recover (\ref{2.10'}).

\section{Solution and discussion}\label{sec2.1.4}
Having established the evolution equation (\ref{2.10})
governing the probability density $P(x,t)$ our next goal is to 
actually solve it and determine the current
$\langle\dot x\rangle$ according to (\ref{2.10c}).
Such a calculation is illustrated in detail in this section.

We start with
introducing the so-called {\em reduced probability density}
and {\em reduced probability current}
\begin{eqnarray}
\hat P (x,t) & := & \sum_{n=-\infty}^\infty P(x+nL,t)
\label{2.12}\\
\hat J (x,t) & := & \sum_{n=-\infty}^\infty J(x+nL,t) \ .
\label{2.12a}
\end{eqnarray}
Taking into account (\ref{2.11}), (\ref{2.10c}) it follows that
\begin{eqnarray}
& & \hat P(x+L,t) = \hat P(x,t)\label{2.13}\\
& & \int_0^L dx\, \hat P(x,t) = 1\label{2.14}\\
& & \langle\dot x\rangle = \int_0^L dx\, \hat J(x,t)\ .\label{2.18}
\end{eqnarray}
With $P(x,t)$ being a solution of the Fokker-Planck equation (\ref{2.10}) it follows with 
(\ref{2.1'}) that also $P(x+nL,t)$ is a solution for any integer $n$.
Since the Fokker-Planck equation is linear, it is also satisfied by the 
reduced density (\ref{2.12}).
With (\ref{2.10d}) it can furthermore be recast into the form
of a continuity equation:
\begin{equation}
\frac{\partial}{\partial t}\hat P(x,t) + 
\frac{\partial}{\partial x}\hat J(x,t) = 0 \ ,
\label{2.16}
\end{equation}
with the explicit form of the reduced probability current
\begin{equation}
\hat J(x,t) = -\left\{\frac{V'(x)}{\eta}
+\frac{k_BT}{\eta}\,\frac{\partial}{\partial x}\right\}\,
\hat P(x,t) \ .
\label{2.17}
\end{equation}
In other words, {\em as far as the particle current $\langle\dot x\rangle$ is 
concerned, it suffices to solve the
Fokker-Planck equation with periodic boundary (and initial) conditions}.

An interesting counterpart of (\ref{2.10c'}) arises by operating with
$\int_{x_0}^{x_0+L} dx\, x \dots$ on both sides 
of (\ref{2.16}), namely
\begin{equation}
\langle\dot x\rangle = 
\frac{d}{dt}\,\left[\int_{x_0}^{x_0+L} dx\, x\, \hat P(x,t)\right]
+L\,\hat J(x_0,t) \ ,
\label{2.17'}
\end{equation}
where $x_0$ is an arbitrary reference position.
In other words, the total {\em particle current} 
$\langle\dot x\rangle$ is composed of the motion of the ``center of mass''
$\int_{x_0}^{x_0+L} dx\, x\, \hat P(x,t)$ plus $L$ times the reduced 
{\em probability current} $\hat J(x_0,t)$ at the reference
point $x_0$.
Especially, if the reduced dynamics assumes a 
{\em steady state}, characterized by $d \hat P(x,t)/dt =0$, then the reduced
probability current $\hat J(x_0,t)=\hat J^{st}$ becomes independent 
of $x_0$ and $t$ according to (\ref{2.16}), (\ref{2.17}),
and the particle current takes the suggestive form
\begin{equation}
\langle\dot x\rangle = L\, \hat J^{st} \ .
\label{2.17''}
\end{equation}

We recall that in general the current $\langle\dot x\rangle$ is time dependent
but the argument $t$ is omitted (cf. (\ref{2.10-1})).
However, the most interesting case is usually its behavior in the long-time limit,
corresponding to a steady state in the reduced description (unless an external
driving prohibits its existence, see e.g. \sect \ref{sec2.2.2.1} below).
In this case, no implicit $t$-dependent of $\langle\dot x \rangle$
is present any more, see (\ref{2.17''}).

We have tacitly assumed here that the original problem
(\ref{2.5}) extends over the entire real $x$-axis.
In some cases, a periodicity condition after one or several periods $L$
of the potential $V(x)$ may represent a more natural modeling, for
instance in the original Smoluchowski-Feynman ratchet of circular shape
(\fig \ref{fig1}).
One readily sees, that in such a case (\ref{2.13})-(\ref{2.17''})
remain valid without any change.
We furthermore remark that the specific form of the stochastic dynamics
(\ref{2.5}) or of the equivalent master equation (\ref{2.10a}), (\ref{2.10d})
has only been used in (\ref{2.17}), while 
\eqs (\ref{2.12})-(\ref{2.16}), (\ref{2.17'}), (\ref{2.17''})
remain valid for more general stochastic dynamics.

For physical reasons we expect that the reduced probability density $\hat P(x,t)$
indeed
approaches a {\em steady state} $\hat P^{st}(x)$ in the long-time limit $t\to\infty$
and hence $\hat J(x_0,t)\to \hat J^{st}$.
From the remaining ordinary first order differential equation (\ref{2.17})
for $P^{st}(x)$ in combination with (\ref{2.13}) it follows that 
$\hat J^{st}$ must be zero and therefore the solution is
\begin{eqnarray}
& & \hat P^{st}(x) = Z^{-1}\, e^{-V(x)/k_B T}\label{2.18'}\\
& & Z := \int_0^L dx\,  e^{-V(x)/k_B T}\ , \label{2.19}
\end{eqnarray}
while (\ref{2.18}) implies for the steady state particle current the result
\begin{equation}
\langle\dot x\rangle = 0 \ .
\label{2.20}
\end{equation}
It can be shown that the long-time asymptotics of a Fokker-Planck equation like 
in (\ref{2.16}), (\ref{2.17}) is 
{\em unique} \cite{lan54,ber55,leb57,sch80,kam92}.
Consequently, this unique solution must be (\ref{2.18'}),
independent  of the initial conditions.
Furthermore, our assumption that a steady state is approached for $t\to\infty$ is 
self-consistently confirmed.

The above results justify a posteriori our proposition that
(\ref{2.5}) models an overdamped Brownian motion under the influence of a thermal
equilibrium heat bath at temperature $T$:
indeed, in the long time limit (steady state), \eq (\ref{2.18'}) 
correctly reproduces 
the expected Boltzmann distribution
and the average particle current vanishes (\ref{2.20}), as
required by the second law of thermodynamics.
The importance of such consistency checks when modeling thermal
noise is further discussed in \sect \ref{sec2.2.3c}.

Much like in the original ratchet and pawl gadget (\fig \ref{fig1}), 
the absence
of an average current (\ref{2.20}) is on the one hand a simple
consequence of the second law of thermodynamics.
On the other hand, when looking at the  stochastic motion
in a ratchet-shaped potential like in \fig \ref{fig2},
it is nevertheless quite astonishing that 
{\em in spite of the broken spatial
symmetry there arises no systematic preferential motion of the
random dynamics in one or the other direction.}

Note that if the original problem (\ref{2.5}) extends over
the entire real axis (bringing along natural boundary
conditions), then the probability density $P(x,t)$ will
{\em never approach a meaningful
\footnote{The trivial long time behavior $P(x,t)\to 0$
does not admit any further conclusions and is therefore not
considered as a meaningful steady state.} steady state}.
It is only the reduced density $\hat P(x,t)$,
associated with periodic boundary conditions, which tends
toward a meaningful $t$-independent long-time limit.
In particular, 
only after this mapping, which leaves the particle 
current unchanged, are the concepts of equilibrium
statistical mechanics applicable.

Conceptionally, the simplified Smoluchowski-Feynman ratchet model 
(\ref{2.5}) has one crucial advantage in comparison with the
original full-blown ratchet and pawl gadget from \fig \ref{fig1}:
The second law of thermodynamics
has not to be invoked as a kind of {\em deus ex machina},
rather the absence of a current (\ref{2.20})
now follows directly from the basic model (\ref{2.5}),
without any additional assumptions.
As a consequence, modifications of the original situation, for which
the second law of thermodynamics
no longer applies, are
relatively straightforward to treat within a correspondingly
modified Smoluchowski-Feynman ratchet model (\ref{2.5}), but
become very cumbersome \cite{par96,mag98} for the more
complicated original ratchet and pawl gadget from \fig \ref{fig1}.
A first, very simple such modification of the 
Smoluchowski-Feynman ratchet model will be addressed 
next.

\section{Tilted Smoluchowski-Feynman ratchet}\label{sec2.2.1}
In this section we consider the
generalization of the overdamped Smoluchowski-Feynman 
ratchet model (\ref{2.5}) in the presence of an additional
homogeneous, static force $F$:
\begin{equation}
\eta\, \dot x(t) = - V'(x(t)) + F + \xi (t)\ .
\label{2.21}
\end{equation}
This scenario represents a kind of ``hydrogen atom''
in that it is one of the few exactly solvable cases
and will furthermore turn out to be equivalent to the archetypal 
ratchet models considered later in \sects 
\ref{sec4.2.2}, 
\ref{sec4.3.1},
\ref{sec6.0.2},
\ref{sec4.8.1}, and
\ref{sec7.2}.
For instance, in the original ratchet and pawl gadget (\fig \ref{fig1})
such a force $F$ in (\ref{2.21}) 
models the effect of a constant external torque due to a load.

\begin{center} 
\figdrei
\end{center} 

We may incorporate the ratchet potential $V(x)$ and the force $F$
into a single effective potential 
\begin{equation}
V_{\rm{eff}} (x) := V(x)-x\,F
\label{2.21'}
\end{equation}
which the Brownian particle (\ref{2.21}) experiences.
E.g. for a negative force $F<0$, pulling the particles to the left,
the effective potential will be tilted to the left as well,
see \fig \ref{fig3}. In view of $\langle \dot x\rangle =0$ for $F=0$ 
(see previous section) it is 
plausible that in such a potential the particles will move on the 
average ``downhill'', i.e., $\langle \dot x \rangle < 0$ for
$F<0$ and similarly $\langle \dot x \rangle > 0$ for
$F>0$. This surmise is confirmed by a detailed calculation along
the very same lines as for $F=0$ (see \sect \ref{sec2.1.4}), 
with the result (see \cite{str58,iva69,amb69}
and also Vol.2, Chapter 9 in \cite{str69})
that in the steady state (long time limit)
\begin{eqnarray}
& & \hat P^{st}(x) = {\cal N}\, 
\frac{\eta}{k_BT}\, e^{-V_{\rm{eff}}(x)/k_B T}\, 
\int_x^{x+L} dy \, e^{V_{\rm{eff}}(y)/k_B T}
\label{2.22}\\
& & \langle\dot x\rangle = L\, {\cal N}\, 
[ 1 - e^{[V_{\rm{eff}}(L)-V_{\rm{eff}}(0)]/k_B T}]
\label{2.23}\\
& & {\cal N} := \frac{k_BT}{\eta}\left[ 
\int_0^L dx\, \int_x^{x+L} dy \, e^{[V_{\rm{eff}}(y)-V_{\rm{eff}}(x)]/k_B T}\right]^{-1} 
\ .
\label{2.23z}
\end{eqnarray}

Note that for the specific form (\ref{2.21'}) of the effective potential
we can further simplify (\ref{2.23}) by exploiting 
that $V_{\rm{eff}}(L)-V_{\rm{eff}}(0) = -LF$. However, the result 
(\ref{2.22})-(\ref{2.23z}) is valid for completely general effective potentials
$V_{\rm{eff}}'(x)$ provided $V_{\rm{eff}}'(x+L)=V_{\rm{eff}}'(x)$.

Our first observation is that 
{\em a time independent probability density $\hat P^{st}(x)$ does not exclude
a non-vanishing particle current $\langle \dot x\rangle$}.
Exploiting (\ref{2.21'}), one readily sees that -- as expected --
the sign of this current (\ref{2.23}) agrees with the sign of $F$.
Furthermore one can prove that the current is a monotonically increasing
function of $F$ \cite{cec96} and that for any fixed $F$-value, the
current is maximal (in modulus) when $V(x)=const.$ (see \sect \ref{sec4.3.1}).
The typical quantitative behavior of the steady state current
(\ref{2.23}) as a function of the applied force $F$
(called ``response curve'', ``load curve'', or (current-force-)
``characteristics'') is exemplified in \fig \ref{fig4}.
Note that the leading order (``linear response'') behavior is
symmetric about the origin, but not the higher order contributions.

\begin{center} 
\figvier
\end{center}

The occurrence of a non-vanishing particle current in (\ref{2.23}) signals
that (\ref{2.22}) describes a steady state which is not in thermal
equilibrium, and actually far from equilibrium unless $F$ is very 
small\footnote{In particular, the effective diffusion coefficient is 
{\em no} longer related to the mobility via a generalized Einstein relation 
(\ref{2.4}), i.e. $\deff = k_B T\,\partial \langle\dot x\rangle/\partial F$
only holds for $F=0$ \cite{rei01}.}.
As mentioned already at the end of the previous section,
while at (and near) equilibrium one may question the need
of a microscopic model like in (\ref{2.21})
in view of the powerful principles of equilibrium statistical mechanics,
such an approach has the advantage of remaining valid far from 
equilibrium\label{fot2.5}\footnote{Note that there is no inconsistency of a thermal
(white) noise $\xi(t)$ appearing in a system far from thermal equilibrium:
any system (equilibrium or not) can be in contact with a thermal heat bath.}, 
where no such general statistical mechanical principles are available.

As pointed out at the end of the preceding section,
only the reduced probability density $\hat P(x,t)$
approaches a meaningful steady state, but not the original
dynamics (\ref{2.21}), extending over the entire $x$-axis.
Thus, {\em stability criteria} for steady states, 
both mechanical and thermodynamical, can only be discussed
in the former, reduced setup.
As compared to the usual reflecting boundary conditions
in this context,
the present periodic boundary conditions entail some
quite unusual consequences:
With the definition $\mu(F,x_0):= \int_{x_0}^{x_0+L} dx\, x\, \hat P^{st}(x)$
for the ``center of mass'' in the steady state (cf. (\ref{2.17'})), 
one can infer from the periodicity $\hat P^{st}(x+L)=\hat P^{st}(x)$ and 
the normalization $\int_0^L dx\, \partial \hat P^{st}(x)/\partial F=0$
that $\partial \mu(F,x_0+L)/\partial F = \partial \mu(F,x_0)/\partial F$,
where $x_0$ is an arbitrary reference position.
Furthermore, one finds that
\begin{equation}
\int_0^L dx_0 \, \frac{\partial \mu(F,x_0)}{\partial F} =
\int_0^L dx_0 \int_0^L dx \, (x+x_0)\, 
\frac{\partial \hat P^{st}(x+x_0)}{\partial F} = 0
\label{2.23p}
\end{equation}
Excluding the non-generic case that $\partial \mu(F,x_0)/\partial F$
is identically zero, it 
follows\footnote{Note that we did not exploit any specific property
of the underlying stochastic dynamics.}
that $\partial \mu(F,x_0)/\partial F$ 
may be negative or positive, depending on the choice of $x_0$. 
In other words, {\em the ``center of mass'' may move either in the 
same or in the opposite direction of the applied force $F$},
and this even if the unperturbed system is at thermal equilibrium.
Similarly, also with respect to the dependence of the steady state 
current $\langle \dot x\rangle$ upon the applied force $F$, no 
general {\em a priori} restrictions due to certain 
``stability criteria'' for steady states exist.

\subsection{Weak noise limit}\label{sec2.2.1.1}
In this section we work out the simplification of the 
current-formula (\ref{2.23}) for small thermal energies 
$k_BT$ -- see \eq (\ref{2.23u}) below -- and its quite 
interesting physical interpretation, repeatedly 
re-appearing later on.

Focusing on not too large $F$-values, such that $V_{\rm{eff}}(x)$ in
(\ref{2.21'}) still exhibits at least one local minimum and maximum
within each period $L$, one readily sees that the function
$V_{\rm{eff}}(y)-V_{\rm{eff}}(x)$ has generically a unique global 
maximum within the two-dimensional integration domain
in (\ref{2.23z}), say at the point $(x,y)=(\xmin,\xmax)$,
where $\xmin$ is a local minimum of $V_{\rm{eff}}(x)$ and
$\xmax$ a local maximum, sometimes called metastable and activated 
states, respectively. Within $(\xmin,\xmin +L)$ the point $\xmax$ 
is moreover a global maximum of $V_{\rm{eff}}(x)$ and similarly $\xmin$ a global
minimum within $(\xmax-L,\xmax)$, i.e.
\begin{equation}
\Delta V_{\rm{eff}} :=V_{\rm{eff}}(\xmax)-V_{\rm{eff}}(\xmin)
\label{2.23y}
\end{equation}
is the effective potential barrier that the particle has to
surmount in order to proceed from the metastable state $\xmin$ to 
$\xmin+L$. Likewise,
\begin{equation}
V_{\rm{eff}}(\xmax-L) - V_{\rm{eff}}(\xmin) = \Delta V_{\rm{eff}} - [ V_{\rm{eff}}(L)- V_{\rm{eff}}(0)]
\label{2.23x}
\end{equation}
is the barrier between $\xmin$ and $\xmin-L$.
For small thermal energies
\begin{equation}
k_B T \ll \{ \Delta V_{\rm{eff}},\, \Delta V_{\rm{eff}} -[ V_{\rm{eff}}(L)- V_{\rm{eff}}(0)]\,\}
\label{2.23w}
\end{equation}
the main contribution in (\ref{2.23z}) stems from a small vicinity
of the absolute maximum $(\xmin,\xmax)$ and we thus can employ the 
so-called {\em saddle point approximation}
\begin{equation}
 V_{\rm{eff}}(y)- V_{\rm{eff}}(x)\simeq \Delta V_{\rm{eff}} 
- \frac{| V_{\rm{eff}}''(\xmax)|}{2} (y-\xmax)^2 
- \frac{| V_{\rm{eff}}''(\xmin)|}{2} (x-\xmin)^2  \ ,
\label{2.23v}
\end{equation}
where we have used that $ V_{\rm{eff}}'(\xmax)= V_{\rm{eff}}'(\xmin)=0$ and
$ V_{\rm{eff}}''(\xmax)<0$, $ V_{\rm{eff}}''(\xmin)>0$.
Within the same approximation, the
two integrals in (\ref{2.23z}) can now be
extended over the entire real $x$- and $y$-axis.
Performing the two remaining Gaussian integrals in (\ref{2.23z})
yields for the current (\ref{2.23}) the result
\begin{eqnarray}
\langle\dot x\rangle & = & L\, [k_+ - k_- ]\label{2.23u}\\
k_+ & := & \frac{| V_{\rm{eff}}''(\xmax) V_{\rm{eff}}''(\xmin)|^{1/2}}{2\,\pi\,\eta}\,
e^{-\Delta V_{\rm{eff}}/k_BT}\label{2.23t}\\
k_- & := & 
k_+ \, e^{[ V_{\rm{eff}}(L)- V_{\rm{eff}}(0)]/k_B T}
\nonumber\\
& = & 
\frac{| V_{\rm{eff}}''(\xmax-L) V_{\rm{eff}}''(\xmin)|^{1/2}}{2\,\pi\,\eta}\,
e^{-[ V_{\rm{eff}}(\xmax-L)- V_{\rm{eff}}(\xmin)]/k_BT}
\ , \label{2.23s}
\end{eqnarray}
where we have exploited (\ref{2.23x}) and the periodicity of $ V_{\rm{eff}}''(x)$
in the last relation in (\ref{2.23s}).

One readily sees that $k_+$ is identical to the so-called
{\em Kramers-Smoluchowski rate} \cite{han90} for transitions from
$\xmin$ to $\xmin+L$, and similarly $k_-$ is the escape rate
from $\xmin$ to $\xmin-L$.
For weak thermal noise (\ref{2.23w}) these rates are small
and the current (\ref{2.23u}) takes the suggestive form of a net 
transition frequency (rate to the right minus rate to the left)
between adjacent local minima of $ V_{\rm{eff}}(x)$ times the step
size $L$ of one such transition.

\section{Temperature ratchet and ratchet effect}\label{sec2.2.2}
We now come to the central issue of the present chapter,
namely the phenomenon of directed transport in a 
spatially periodic, asymmetric system away from equilibrium.
This so-called ratchet-effect is very often illustrated by
invoking as an example the on-off ratchet model, as introduced 
by Bug and Berne \cite{bug87} and by Ajdari and Prost \cite{ajd92},
see \sect \ref{sec4.1}.
Here, we will employ a different example, the so-called temperature ratchet,
which in the end will however turn out to be actually very closely
related to the on-off ratchet model (see \sect \ref{sec4.8.3}).
We emphasize that the choice of this example is not primarily 
based on its objective or historical significance but rather on
the author's personal taste and research activities.
Moreover, this example appears to be particularly suitable 
for the purpose of illustrating besides the ratchet effect 
{\em per se} also many other important concepts 
(see \sects \ref{sec2.2.2.3}-\ref{sec2.5})
that we will encounter again in much more generality 
in later chapters.

\subsection{Model}\label{sec2.2.2.1}
As an obvious generalization of the tilted
Smoluchowski-Feynman ratchet model (\ref{2.21})
we consider the case that the temperature of the Gaussian white noise
$\xi(t)$ in (\ref{2.3}) is subjected to periodic temporal 
variations with period $\ttt$ \cite{rei96}, i.e.
\begin{eqnarray}
& & \langle\xi (t)\xi (s)\rangle  = 2\,\eta\, k_B T(t)\,  \delta(t-s)\label{2.23'}\\
& & T(t) = T(t+\ttt )\ , \label{2.23''}
\end{eqnarray}
where $T(t)\geq 0 $ for all $t$ is taken for granted.
Note that due to the time-dependent temperature in (\ref{2.23'})
the noise $\xi(t)$ is strictly speaking no longer stationary.
A stationary noise is, however, readily recovered by rewriting
(\ref{2.21}), (\ref{2.23'}) as
\begin{equation}
\eta \dot x(t) = - V'(x(t)) + F + g(t)\,\hat\xi(t) \ ,
\label{2.23'''}
\end{equation}
where $\hat \xi (t)$ is a  Gaussian white noise with
$\langle\hat\xi(t)\hat\xi(s)\rangle=2 \delta(t-s)$ and 
$g(t):=[\eta k_BT(t)]^{1/2}$.
Two typical examples which we will adopt for our numerical 
explorations below are
\begin{eqnarray}
T(t) & = & \overline{T} \, [1+A\,{\rm sign}\{\sin (2\pi t/\ttt)\}]
\label{2.24a}\\
T(t) & = & \overline{T} \, [1+A\,\sin (2\pi t/\ttt)]^2\ ,
\label{2.24b}
\end{eqnarray}
where ${\rm sign}(x)$ denotes the signum function and $|A|<1$.
The first example (\ref{2.24a}) thus jumps between 
$T(t) = \overline{T} \, [1+A]$ and $T(t) = \overline{T} \, [1-A]$
at every half period $\ttt/2$.
The motivation for the square on the right hand side of
(\ref{2.24b}) becomes apparent when rewriting
the dynamics in the form (\ref{2.23'''}).

Similarly as in \sect \ref{sec2.1.3}, one finds that the reduced particle density
(\ref{2.12}) for this so-called {\em temperature ratchet model} 
(\ref{2.21}), (\ref{2.23'}), (\ref{2.23''}) is governed 
by the Fokker-Planck equation
\begin{equation}
\frac{\partial}{\partial t} \hat P(x,t) = 
\frac{\partial}{\partial x}\left\{\frac{V'(x)-F}{\eta} \, \hat P(x,t)\right\}
+ \frac{k_B\, T(t)}{\eta} \, \frac{\partial^2}{\partial x^2} \hat P(x,t) \ .
\label{2.24'}
\end{equation}
Due to the permanent oscillations of $T(t)$, this equation does
not admit a time-independent solution.
Hence, the reduced density $\hat P(x,t)$ will not approach a steady state but
rather a unique periodic behavior in the long time 
limit\label{fot2.4}\footnote{Proof:
Since $T(t+\ttt )= T(t)$ we see that with $\hat P(x,t)$
also $\hat P(x,t+\ttt )$ solves (\ref{2.24'}). Moreover, for the
long time asymptotics of (\ref{2.24'}) the general proof of uniqueness from
\cite{leb57,sch80} applies. Consequently, $\hat P(x,t+\ttt )$ must converge towards
$\hat P(x,t)$, i.e. $\hat P(x,t)$ is periodic and unique for $t\to\infty$.}.
It is therefore natural to include a time average into the definition 
(\ref{2.10-1}) of the
particle current. Keeping for convenience the same symbol 
$\langle\dot x\rangle$, the generalized expression (\ref{2.18}),
(\ref{2.17}) for this current becomes
\begin{equation}
\langle\dot x\rangle = \frac{1}{\ttt }\int_t^{t+\ttt }dt 
\int_0^L dx\, \frac{F - V'(x)}{\eta} \,\hat P(x,t)\ . \label{2.24''}
\end{equation}

Note that in general, the current $\langle\dot x\rangle$ in 
(\ref{2.24''}) is still $t$-dependent. Only in the long time
limit, corresponding in the reduced description to a $\ttt$-periodic
$\hat P(x,t)$, this $t$-dependence disappears. Usually, this latter
long-time limit is of foremost interest.

\subsection{Ratchet effect}\label{sec2.2.2.2}
After these technical preliminaries, we return to the physics of our model 
(\ref{2.21}), (\ref{2.23'}), (\ref{2.23''}): In the case of the tilted 
Smoluchowski-Feynman ratchet (time-independent temperature $T$),
\eq (\ref{2.23}) tells us that for a given force, say $F<0$, the 
particle will move ``downhill'' on the average, i.e. 
$\langle\dot x\rangle < 0$, and this for any fixed (positive)  
value of the temperature $T$. Turning to the temperature ratchet with $T$ 
being now subjected to periodic temporal variations,
one therefore should expect that the particles still move
``downhill'' on the average. The numerically calculated ``load curve''
in \fig \ref{fig5} demonstrates 
that the opposite is true within an entire
interval of negative $F$-values.
Surprisingly indeed, {\em the particles are climbing ``uphill''
on the average, thereby performing work against the load force $F$},
which apparently can have no other origin than the white thermal 
noise $\xi(t)$. 

\begin{center} 
\figfunf
\end{center}

A conversion (rectification)
of random fluctuations into useful work as exemplified
above is called {\em ``ratchet effect''}.
For a setup of this type, the names
{\em thermal ratchet} \cite{val90,cor92,mag93},
{\em Brownian motor} \cite{bar95a,rei96}, 
{\em Brownian rectifier} \cite{han96} (mechanical diode \cite{mag93}), 
{\em stochastic ratchet} \cite{iba97,doe98a}, 
or simply 
{\em ratchet} are in use\footnote{The notion ``molecular motor'' 
should be reserved for models focusing specifically
on intracellular transport processes, see 
\ch \ref{cha5}.
Similarly, the notion ``Brownian ratchet'' has been introduced
in a rather differen context, namely as a possible operating 
principle for the translocation of proteins accross 
membranes \cite{sim92,pes93,kuo00,els00c,lie01}.}.
Since the average particle current $\langle\dot x\rangle$ usually
depends continuously on the load force $F$, 
it is for a qualitative analysis sufficient to consider the 
case $F=0$: {\em the occurrence of the 
ratchet effect is then tantamount to a finite
current}
\begin{equation}
\langle\dot x\rangle\not = 0 \ \ \ \mbox{for}\ \ \  F=0 \ ,
\label{2.24'''}
\end{equation}
i.e. the unbiased Brownian motor
implements a {\em ``particle pump''}.
The necessary force $F$ which leads to an exact cancellation of
the ratchet effects, i.e $\langle\dot x\rangle =0$, 
is called the {\em ``stopping force''}.
The property (\ref{2.24'''}) is the distinguishing feature
between the ratchet effect and the somewhat related
so-called {\em negative mobility} effect, encountered later in 
\sect \ref{sec7.2.4}.

\subsection{Discussion}\label{sec2.2.2.3}
In order to understand the basic physical mechanism behind the ratchet
effect at $F=0$,
we focus on the dichotomous periodic
temperature modulations from (\ref{2.24a}).
During a first time interval, say $t\in[\ttt/2,\ttt]$,
the thermal energy $k_B T(t)$ is kept at a constant value 
$\overline{T} \, [1-A]$ much
smaller than the potential barrier $\Delta V$ between two 
neighboring local minima of $V(x)$. 
Thus, all particles will have accumulated
in a close vicinity of the potential minima at the end of this
time interval, as sketched in the lower panel of
\fig \ref{fig6}. 
Then the thermal energy jumps
to a value $\overline{T} \, [1+A]$
much larger than $\Delta V$ and remains there
during another half period, say $t\in[\ttt,3\ttt/2]$. 
Since the particles then hardly feel 
the potential any more in comparison to the violent thermal noise,
they spread out practically like in the case of free thermal diffusion
(upper panel in \fig \ref{fig6}).
Finally, $T(t)$ jumps back to its original low value 
$\overline{T} \, [1-A]$
and the particles 
slide downhill towards the respective closest local minima of $V(x)$.
Due to the asymmetry of the potential $V(x)$, the original
population of one given minimum is re-distributed asymmetrically
and a net average displacement results after one time-period 
$\ttt$.

\begin{center} 
\figsechs
\end{center}

In the case that the potential $V(x)$ has exactly one minimum and maximum
per period $L$ (as it is the case in \fig \ref{fig6}) it is quite obvious that
if the local minimum is closer to its adjacent maximum to the right
(as in \fig \ref{fig6}), a positive particle current $\langle\dot x\rangle >0$
will arise, otherwise a 
negative current. For potentials with additional extrema, the determination 
of the current direction may be less obvious.

As expected, a 
qualitatively similar behavior is observed for more general temperature 
modulations $T(t)$ than in \fig \ref{fig6}
provided they are sufficiently slow. The effect 
is furthermore robust with respect to the potential shape \cite{rei96}
and persists even for (slow) random instead of deterministic changes of
$T(t)$ \cite{luc97,li97}, e.g. (rare) random flips between the two possible
values in \fig \ref{fig6},
as well as for a modified dynamics with a discretized state space
\cite{sok97,sok98}.
The case of finite inertia and of
various correlated (colored) Gaussian noises
instead of the white noise in (\ref{2.21}) or (\ref{2.23'''})
has been addressed in \cite{bao00} and \cite{bao99a}, respectively.
A somewhat more detailed quantitative analysis will
be given in \sects \ref{sec2.4.1} and \ref{sec2.5} below.

In practice, the required magnitudes and time scales of the temperature 
variations may be difficult to realize {\em experimentally} by directly adding
and extracting heat, but may well be feasible indirectly, e.g. by
pressure variations.
An exception are point contact devices with a defect which
tunnels incoherently between two states and thereby changes the coupling
strength of the device to its thermal environment
\cite{ral84,mul92,gol92,ral92,kei96,kog96,smi96}.
In other words, when incorporated into an electrical circuit,
such a device exhibits random dichotomous jumps both of its 
electrical resistance and of the intensity of the thermal fluctuations which 
it produces \cite{bri97}.
The latter may thus be exploited to drive a temperature ratchet system 
\cite{luc97}.

Further, it has been suggested \cite{mul83,mul95} that microorganisms
living in convective hot springs may be able to extract energy out of the 
permanent temperature variations they experience; the temperature ratchet
is a particularly simple mechanism which could do the job.
Moreover, a temperature ratchet-type modification of the experiment by
Kelly, Tellitu, and Sestelo \cite{kel97,kel98,dav98}
(cf. \sect \ref{sec2.1.1})
has been proposed in \cite{seb00}.

Finally, it is known that certain enzymes 
({\em molecular motors}) in living cells
are able to travel along polymer filaments by hydrolyzing
ATP (adenosine triphosphate).
The interaction (chemical ``affinity'') between molecular motor 
and filament is
spatially periodic and asymmetric, and thermal fluctuations play a significant
role on these small scales. On the crudest level,
hydrolyzing an ATP molecule may be viewed as converting a certain amount
of chemical energy into heat, thus we recover all the essential
ingredients of a temperature ratchet.
Such a temperature ratchet-type model for intracellular transport has 
been proposed in \cite{val90}.
Admittedly, modeling the molecular motor 
as a Brownian particle without any relevant internal
degree of freedom\footnote{A molecular motor is a very complex enzyme 
with a huge number of degrees of freedom (see \ch \ref{cha5}).
Within the present temperature ratchet model,
the ATP hydrolyzation energy is thought to be quickly converted
into a very irregular vibrational motion of these degrees of freedom,
i.e. a locally increased apparent temperature.
As this excess heat spreads out, the temperature decreases
again.
Thus, the internal degrees of freedom play a crucial role but are
irrelevant in so far as they do not give rise to any additional
slow, collective state variable.}
and the ATP hydrolysis as a mere production of heat is a gross
oversimplification from the biochemical point of view, see \ch \ref{cha5},
but may still be acceptable as a primitive sketch of the basic physics.
Especially, quantitative estimates indicate \cite{lei93,hun94,how96}
that the temperature variations (either their amplitude or their
duration) within such a temperature ratchet model may not be
sufficient to reproduce {\em quantitatively} the observed traveling 
speed of the molecular motor.

\section{Mechanochemical coupling}\label{sec2.2.3a}
We begin with pointing out
that the ratchet effect as exemplified by the temperature ratchet model 
is not in contradiction with the second law
of thermodynamics\footnote{We also note that a current 
$\langle \dot x\rangle$ opposite to the force $F$ is not in contradiction 
with any kind of ``stability criteria'', cf. the discussion 
below (\ref{2.23p}).}
since we may consider the changing temperature
$T(t)$ as caused by several heat baths at different 
temperatures\footnote{In passing we notice that
the case $F=0$ in conjunction with
a time-dependent temperature $T(t)$ is conceptually quite interesting:
It describes a system which is at any 
given instant of time an equilibrium system in a non-equilibrium 
(typically far from equilibrium) state.}.
From this viewpoint, our system is nothing else than an extremely 
primitive and small heat engine \cite{mag94}.
Specifically, the example from (\ref{2.24a}) and \fig \ref{fig5}
represents the most common case with just two equilibrium
heat baths at two different temperatures.
The fact that such a device can produce work
is therefore not a miracle but still amazing.

At this point it is crucial to recognize that there is also 
one fundamental difference between the usual types of
heat engines and a Brownian motor as
exemplified by the temperature ratchet: 
To this end we first note that the two
``relevant state variables'' of our present system are 
$x(t)$ and $T(t)$.
In the case of an ordinary heat engine, these state variables
would always cycle through one and the same periodic sequence
of events (``working strokes'').
In other words, the evolutions of the state variables $x(t)$ and 
$T(t)$ would be tightly coupled together (interlocked, synchronized).
As a consequence, a single suitably defined effective state variable
would actually be sufficient to describe the 
system\footnote{Note that a fixed sequence of events does not 
necessarily imply a deterministic evolution in time. In particular,
small (``microscopic'') fluctuations which can be described by
some environmental (equilibrium or not) noise are still admissible.}.
In contrast to this standard scenario, {\em the relevant
state variables of a genuine Brownian motor are loosely coupled}:
Of course, some degree of interaction is indispensable
for the functioning of the Brownian motor, but while $T(t)$
completes one temperature cycle, $x(t)$ may evolve in several
essentially different ways (it is not ``slaved'' by $T(t)$).

The loose coupling between state variables is a salient point
which makes the Brownian motor concept more than just a
cute new look at certain very small and primitive, but otherwise
quite conventional thermo-mechanical or even purely mechanical
engines.
In most cases of practical relevance, 
the presence of some amount of (not necessarily thermal)
random fluctuations
is therefore an indispensable ingredient of a genuine 
Brownian motor;
exceptionally, deterministic chaos may be a substitue
(cf. \sects \ref{sec6.2.3.2} and \ref{sec6.3.2}).

We remark that most of the specific ratchet models
that we will consider later on do have a second
relevant state variable besides\footnote{While this second
state variable obviously influences $x(t)$ in some or the
other way, a corresponding back-reaction may or may not
exists. The latter case is exemplified by the 
temperature ratchet model.}
$x(t)$.
One prominent exception are the so-called Seebeck ratches,
treated in \sect \ref{sec4.8.1}.
In such a case the above condition of a loose coupling 
between state variables is clearly meaningless.
This does, however, not imply that those are no genuine 
Brownian motors.

The important issue of whether the coupling between relevant 
state variables is loose or tight has been mostly discussed
in the context of molecular motors 
\cite{mit88,mag94,ast96a} and has
been given the suggestive name {\em mechanochemical coupling},
see also \sect \ref{sec5.4.3} and \ref{sec5.8}.
The general fact that such couplings of non-equilibrium enzymatic 
reactions to mechanical currents play a crucial role
for numerous cellular transport processes is long 
known \cite{bio83,fri86}.

\section{Curie's principle}\label{sec2.2.3b}
The main, and {\em a priori} quite counterintuitive observation from \sect
\ref{2.1} is the fact that no preferential direction
of the random dynamics (\ref{2.3}), (\ref{2.5})
arises in spite of the broken spatial symmetry of the system.
The next surprising observation from \sect \ref{sec2.2.2} is the
appearance of the ratchet effect, i.e.
of a finite current $\langle\dot x \rangle$, for the slightly modified
temperature ratchet model (\ref{2.5}), (\ref{2.23''})
in spite of the {\em absence of any macroscopic static forces, 
gradients (of temperature, concentration, chemical potentials etc.), 
or biased time-dependent perturbations}. 
Here the word ``macroscopic'' refers to ``coarse grained'' effects
that manifest themselves over many spatial periods $L$. Of course, on the
``microscopic'' scale, a static gradient-force  $-V'(x)$ is 
acting in (\ref{2.5}), but that averages
out to zero for displacements by multiples of $L$.
Similarly, at most time-instants $t$, a non-vanishing thermal
force $\xi(t)$ is acting in (\ref{2.5}), but again that averages out to zero
over long times or when an entire statistical ensemble is considered.

The first observation, i.e. the absence of a current at thermal equilibrium, 
is a consequence of the second law of thermodynamics. In the second 
above mentioned situation, giving rise to a ratchet effect,
this law is no longer applicable, since the system is not in a thermal
equilibrium state.
So, in the absence of this and any other prohibitive
{\em a priori} reason, and in view of the
fact that, after all, the spatial symmetry of the system {\em is}
broken, the manifestation of a preferential direction for the
particle motion appears to be an almost unavoidable educated guess.

This  common sense hypothesis, namely
that {\em if a certain phenomenon is not ruled out by symmetries then it 
will occur}, is called Curie's 
principle\footnote{In the biophysical literature \cite{bio83,fri86} the
notion of Curie's principle (or Curie-Prigogine's principle) is mostly 
used for its implications in the special case of linear response
theory (transport close to equilibrium) in isotropic systems, 
stating that a force can couple only to currents of the same tensorial order,
see also \cite{bie00a,bie00}.}
\cite{curie}.
More precisely, the principle postulates the absence of accidental
symmetries in the generic case.
That is, an accidental symmetry may still occur as an exceptional coincidence 
or by fine-tuning of parameters, but typically it will not.
Accidental symmetries are structurally unstable, an arbitrarily 
small perturbation destroys them \cite{mag94},
while a broken symmetry is a structurally stable situation.

In this context it may be worth noting that the absence of
a ratchet effect at thermal equilibrium in spite of the spatial
asymmetry is no contradiction to Curie's principle:
The very condition for a system to be at thermal equilibrium can also
be expressed in the form of a symmetry condition,
namely the so-called {\em detailed balance symmetry}\footnote{To be precise,
detailed balance is necessary but not sufficient for thermal equilibrium
\cite{gra71a,han82b}.
Conversely, detailed balance is sufficient but not necessary for
a vanishing particle current $\langle\dot x\rangle$.}
\cite{ons31,gre52,kam57,gra71a,gra71b,han82b,gar83,ris84,kam92}.

\section{Brillouin's paradox}\label{sec2.2.3c}
As mentioned in \sect \ref{sec2.1.1}, both Smoluchowski and Feynman
have already pointed out the close similarity of the
ratchet and pawl gadget from \fig \ref{fig1} with a Maxwell demon and
also with the behavior of a mechanical valve.
But also the analogy of such a ratchet device with an electrical rectifier,
especially the asymmetric response to an external static force field
(cf. \fig \ref{fig4}), has been pointed out in Feynman's 
Lectures \cite{fey63}, see also Vol. III, \sect 14-4 therein.
In this modified context of an electrical rectifier, the astonishing
fact that random thermal fluctuations cannot be rectified into useful work is
called Brillouin's paradox \cite{bri50}
and has been extensively discussed e.g. in 
\cite{kam65,mcf71,kam92,str92,lan98,sok98a}.

The main point of this discussion can be most easily understood by comparison
with the corresponding tilted Smoluchowski-Feynman ratchet model (\ref{2.21}).
Furthermore, we focus on the case of an electrical circuit with a 
{\em semiconductor diode}\footnote{A 
tube diode requires permanent heating and it is not obvious how
to reconcile this with the condition of thermal equilibrium.}.
With the entire circuit being kept at thermal equilibrium,
at any finite temperature and conductance,
a random electrical noise arises and it is {\em prima facie } indeed quite
surprising that its rectification by the diode is impossible.
The stepping stone becomes apparent in the corresponding
Smoluchowski-Feynman ratchet model (\ref{2.21}).
While its response to an external force $F$ in \eq (\ref{2.23}) and \fig
\ref{fig4} shares the typical asymmetric shape with a diode,
it is now clearly wrong to {\em phenomenologically} describe
the effect of the thermal noise in such a system by simply averaging
the current $\langle\dot x\rangle$ from (\ref{2.23}) with respect
to $F$ according to the probability with which
the thermal noise takes these values $F$.
Rather, the correct modeling, which in particular consistently incorporates the
common {\em microscopic} 
origin of friction and thermal noise, is represented by (\ref{2.21}) (with $F=0$).
In contrast, the response characteristics (\ref{2.23}) is already the result
of an averaging over the thermal noise under the additional
assumption that $F$ is practically constant on the typical
transient time scales of the emerging current $\langle\dot x\rangle$.
It is clear, that we do not recover the full-fledged noisy dynamics
(\ref{2.21}) by replacing {\em phenomenologically } $F$ by $\xi(t)$
in (\ref{2.23}), notwithstanding the fact that in (\ref{2.21})
these two quantities indeed appear in the same way.
The close analogy of this situation with that in a semiconductor
diode becomes apparent by considering that also in the latter case
the asymmetric response characteristics is the result of a thermal
diffusion process of the electrons near the interface of the 
n-p junction under quasi-static conditions and after averaging out the
thermal fluctuations.

This example (see also \cite{ast87} for another such example)
demonstrates that {\em the correct modeling of the thermal environment
is not always obvious}.
Especially, taking the averaged macroscopic behavior of the system
as a starting point for a phenomenological modeling of the noisy
dynamics may be dangerous outside the linear response regime,
as van Kampen and others are emphasizing since many years \cite{kam92}.
Much safer is a {\em microscopic staring point} in order to
consistently capture the common origin of the dissipation and
the fluctuations in the actual system of interest, as exemplified
in \sects \ref{sec2.1.2}, \ref{sec3.2.1}, and \ref{sec6.5.1}.

Away from thermal equilibrium, the realization of the ratchet effect
by diodes and other semiconductor heterostructures is further
discussed in \sects \ref{sec4.8.1} and \ref{sec6.5.5}.

\section{Asymptotic analysis}\label{sec2.4.1}
In the remainder of this chapter, 
we continue our exploration of the temperature ratchet
model (\ref{2.21}), (\ref{2.23'}), (\ref{2.23''}) with the objective to
understand in somewhat more detail the behavior of the particle current
$\langle\dot x\rangle$ at zero load $F=0$
as a function of various parameters of the model.
Since a closed analytical solution of the Fokker-Planck equation
(\ref{2.24'}) is not possible in general, we have to recourse to
asymptotic expansions and qualitative physical arguments,
complemented by accurate numerical results for a few typical cases.
In the present, somewhat more techical
section we analyse the behavior of the particle current
for asymptotically slow and fast temperature
oscillations.

For asymptotically slow temporal oscillations
in (\ref{2.23''}) the time- and ensemble-averaged particle current
$\langle\dot x\rangle$ approaches 
zero\footnote{In the following we tacitly restrict ourselves to smooth $T(t)$,
like e.g. in (\ref{2.24b}).
For discontinuous $T(t)$, for instance (\ref{2.24a}),
the conclusion $\langle\dot x\rangle\to 0$ for $\ttt\to \infty$ remains valid,
but the reasoning has to be modified.}.
Considering that $\ttt\to \infty$ means a constant $T(t)$
during any given, finite time interval, this
conclusion $\langle\dot x\rangle\to 0$ is physically
quite obvious. It can also be be formally confirmed by the
observation that 
\begin{equation}
\hat P^{{\rm ad}}(x,t) := Z(t)^{-1}\, e^{-V(x)/k_B T(t)}
\label{2.3.0}
\end{equation}
with $Z(t):=\int_0^L dx\, e^{-V(x)/k_B T(t)}$
solves the Fokker-Planck equation (\ref{2.24'}) in arbitrarily good 
approximation for sufficiently large $\ttt$ and $F=0$.
Comparison with (\ref{2.18'}) shows that this so-called 
{\em adiabatic approximation} (\ref{2.3.0}) represents an
accompanying or instantaneous equilibrium solution in which the
time $t$ merely plays the role of a parameter.
Introducing (\ref{2.3.0}) into (\ref{2.24''}) with $F=0$ indeed
confirms the expected result $\langle\dot x\rangle = 0$.

Turning to finite but still large $\ttt$, one expects that 
$\langle\dot x\rangle$ decreases
proportional to $\ttt^{-1}$ in the general case.
In the special case that $T(t)$ is symmetric under time 
inversion\footnote{Time inversion symmetry
means that there is a $\Delta t$ such that $T(-t)=T(t+\Delta t)$ for all $t$.}, 
as for instance
in (\ref{2.24a}), (\ref{2.24b}),
the current $\langle\dot x\rangle$ must be an even
function of $\ttt$ and thus typically decreases proportional to 
$\ttt^{-2}$ for large $\ttt$.
Furthermore, our considerations along the lines of \fig \ref{fig6} 
suggest that,
at least for potentials with only one maximum and minimum per period $L$,
the current $\langle\dot x\rangle$ approaches zero from above if the
minimum is closer to the adjacent maximum to the right, and from 
below otherwise. Here and in the following, we tacitly assume that apart form
the variation of the time-period $\ttt$ itself,
the shape of $T(t)$ does not change, i.e. 
\begin{equation}
\hat T(h):=T(\ttt\, h)
\label{2.3.1'}
\end{equation}
is a $\ttt$-independent function of its dimensionless argument $h$
with period $1$.

Addressing small $\ttt$, i.e. fast temperature oscillations,
it is physically plausible that the system cannot follow any more
these oscillations and thus behaves for $\ttt\to 0$ like in the
presence of a constant averaged temperature
\begin{equation}
\overline{T} := \frac{1}{\ttt}\int_0^\ttt dt\, T(t) = \int_0^1 dh\, \hat T( h) .
\label{2.3.1}
\end{equation}
Within this so-called {\em sudden approximation} we thus recover an effective
Smoluchowski-Feynman ratchet dynamics (\ref{2.5}).
In other words, we expect that $\langle\dot x\rangle \to 0$  for
$\ttt\to 0$. This behavior is confirmed by the analytical perturbation 
calculation in Appendix C, which yields moreover the leading order 
small-$\ttt$ result \cite{rei96}
\begin{eqnarray}
& & \langle\dot x\rangle = \ttt^2\, B\, \int_0^L dx\, V'(x)\, [V''(x)]^2
+ \ord(\ttt^3 )\label{2.3.2}\\
& & B := \frac{4\,L\int_0^1 dh\, 
 \left[\int_0^h d\hat h\, ( 1- \hat T( \hat h)/\overline{T})\right]^2}
 {\eta^3\, \int_0^L dx\, e^{V(x)/k_B\overline{T}}\, \int_0^L dx\, e^{-V(x)/k_B\overline{T}}}\ .
 \label{2.3.3} 
\end{eqnarray}
Note that $B$ is a strictly positive functional of $T(t)$ and $V(x)$ and 
is independent of $\ttt$. 

The most remarkable feature of (\ref{2.3.2})
is that there is no contribution proportional to $\ttt$ independently of
whether $T(t)$ is symmetric under time inversion or not.
More according to our expectation is the fact that the current vanishes 
for very weak thermal noise, as a closer inspection of 
(\ref{2.3.2}) implies: Similarly as for the weak noise analysis in 
\sect \ref{sec2.2.1.1}, for $\overline{T}\to 0$ the particles can never 
leave the local minima of the potential $V(x)$. In the opposite limit, i.e.
for $\overline{T}\to\infty$, the potential should play no role any more
and one expects again that  $\langle\dot x\rangle\to 0$, 
cf. \sect \ref{sec3.1.4}.
A more careful perturbative analysis of the high-temperature limit
confirms this expectation.
On the other hand, \eq (\ref{2.3.2}) predicts a finite
limit for $\overline{T}\to\infty$, implying that the 
limits $\overline{T}\to\infty$ and $\ttt\to 0$ cannot be interchanged
in this perturbative result.
In other words, the correction of order 
$\ord(\ttt^3 )$ in (\ref{2.3.2}) approaches zero for any finite $\overline{T}$ 
as $\ttt\to 0$, but is no longer negligible if we keep $\ttt$ fixed 
(however small) and let $\overline{T}\to \infty$.

The above predictions are
compared with accurate numerical solutions in \fig  \ref{fig7} for
a representative case, showing very good agreement.

\begin{center} 
\figsieben
\end{center}

\section{Current inversions}\label{sec2.5}
The most basic qualitative prediction, namely that generically
$\langle\dot x\rangle \not =0$, is a consequence of Curie's principle.
In this section we show that under more general conditions than
in \sect \ref{sec2.2.2.3}, even the sign of the current
$\langle \dot x\rangle$ may be already very difficult to 
understand on simple intuitive grounds, not to speak of its 
quantitative value.
This leads us to another basic phenomenon in Brownian motor
systems, namely the inversion of the current direction upon
variation of a system parameter.
Early observations of this effect
have been reported in \cite{koc75,asn79,doe94,mil94,ajd94a,bar94}; 
here we illutrate it once more for our stantard example of the
the temperature ratchet.

Since the quantity $B$ from (\ref{2.3.3}) is positive, it is
the sign of the integral in (\ref{2.3.2}) which
determines the direction of the current. For the specific ratchet
potential (see \eq (\ref{2.1''}) and \fig \ref{fig2})
used in \fig \ref{fig7} this sign is positive, but one can
easily find other potentials $V(x)$ for which this sign is negative.
By continuously deforming one potential into the other one can infer
that there must exist an intermediate $V(x)$ with the property that
the particle current $\langle\dot x\rangle$ is zero at
some finite $\ttt$-value. In the generic case, the
$\langle\dot x\rangle$-curve passes with a finite slope through
this zero-point, implying \cite{rei96} the existence of a so-called
{\em ``current inversion''} of $\langle\dot x\rangle$
as a function of $\ttt$. 
An example of a potential $V(x)$ exhibiting such a current inversion
is plotted in \fig \ref{fig8pot} and the resulting current in \fig 
\ref{fig8}. As compared to the example from \fig \ref{fig2},
the modification of the ratchet potential in \fig \ref{fig8pot}
looks rather harmless.
Especially, the explanation of a positive current $\langle\dot x\rangle >0$ 
for large $\ttt$ according to \fig \ref{fig5} still applies.
However, for small-to-moderate $\ttt$ this modification of 
the potential has dramatic consequences for the current in \fig
\ref{fig8} as compared to \fig \ref{fig7}.

\begin{center} 
\figachtpot
\figacht
\end{center} 

Once a current inversion upon variation of one parameter of the
model ($\ttt$ in our case) has been established, 
an {\em inversion upon variation of any other parameter } (for instance
the friction coefficient $\eta$) can be inferred along the
following line of reasoning \cite{rei98}:
Consider a current inversion upon variation of $\ttt$, say at $\ttt_0$, 
as given, while $\eta$  is kept fixed, say at $\eta_0$. 
Let us next consider $\ttt$ as fixed to $\ttt_0$ and
instead vary $\eta$ about $\eta_0$. In the generic case the current
$\langle\dot x\rangle$ as a function of $\eta$ will then
go through its zero-point at $\eta_0$ with a finite slope, meaning that
we have obtained the proposed current inversion upon
variation of $\eta$, see \fig \ref{fig9}.

\begin{center} 
\figneun
\end{center} 

In other words,
Brownian particles with different sizes will have different friction 
coefficients $\eta$ and will thus move in opposite directions 
when exposed to the same
thermal environment and the same ratchet potential. Had we not neglected 
the inertia effects $m\ddot x(t)$ in (\ref{2.1}), such a particle 
separation mechanism also with respect to the mass $m$ could
be inferred along the above line of reasoning,
and similarly for other dynamically relevant particle properties!

Promising applications of such current inversion effects for
{\em particle separation} methods, based on the ratchet effect,
are obvious.
Another interesting aspect of current inversions arises from the
observation that structurally very similar {\em molecular motors} 
may travel in opposite
directions on the same intracellular filament (see \ch \ref{cha5}).
If we accept the temperature ratchet as a crude qualitative model in this context
(cf. \sect \ref{sec2.2.2.3}), it is amusing to note that also
this feature can be qualitatively reproduced:
If two types of molecular motors are known to 
differ in their ATP consumption rate $1/\ttt$,
or in their friction coefficient $\eta$, or in any other parameter 
appearing in our temperature ratchet model,
then it is 
possible\footnote{See also \sect \ref{sec3.5} 
for a detailed proof.} to figure out a ratchet potential $V(x)$ 
such that they move indeed in opposite directions.

For a more general discussion of current inversion effects we
refer to \sect \ref{sec3.5} below. Additional material on the
temperature ratchet model is contained in \sect \ref{sec4.8.3}.


\chapter{General Framework}\label{cha3}
In \chs \ref{cha3}-\ref{cha7} we will review theoretical 
extensions and their experimental realizations of the
concepts which were introduced by means of 
particularly simple examples in \ch  \ref{cha2}.
In the present chapter,
we provide a first overview and general framework for
the more detailed discussion in the subsequent chapters:
The main classes of ratchet models and their physical 
origin are introduced.
Symmetry considerations regarding the occurence or not
of a finite particle current (ratchet effect) are
a second important issue, complemeted by
a general method of tailoring current inversions.
Finally, a general treatment is provided for the asymptotic
regimes of weak and strong noise-strength and of
weak non-equilibrium perturbations.
Specific examples and applications of these general
concepts are mostely postponed to later chapters


\section{Working model}\label{sec3.1.1}
In hindsight, the essential ingredient of the ratchet effect 
from \sect \ref{sec2.2.2.2} was a modification of the 
Smoluchowski-Feynman ratchet model (\ref{2.5})
so as to permanently keep the system away from thermal equilibrium. 
We have exemplified this procedure
by a periodic variation of the temperature (\ref{2.23''})
but there clearly exists a great variety of other options.
In view of this example, the following
{\em guiding principles} should be observed also in more general cases:
(i) We require spatial periodicity and 
either invariance or periodicity under translations in time.
(ii) All forces and  gradients have to vanish after averaging over 
space (``coarse graining'' over many spatial periods),
over time (in the case of temporal periodicity), and over 
statistical ensembles (in the case of random fluctuations).
(iii) The system has to be driven permanently
out of thermal equilibrium and there should be no symmetries
which prohibit a ratchet effect {\em a priori}.
According to Curie's principle we can therefore expect the generic 
appearance of a finite particle current $\langle\dot x\rangle$.
(iv) In view of the title of our present study, we will mostly 
(not exclusively) focus on models with a finite amount of thermal 
noise\footnote{Note that (iv) is not a consequence of (iii),
as demonstrated by any dissipative driven system at zero temperature.}.

According to these preliminary considerations,
we adopt as our {\em basic working model}
the overdamped one-dimensional stochastic dynamics
\begin{eqnarray}
& & \eta\,\dot x(t) = -V'(x(t),f(t)) + y(t) + F + \xi(t) \label{4a}\\
& & \langle\xi (t)\,\xi(s)\rangle = 
2\, \eta\, k_B T\, \delta (t-s)\ , \label{4b}
\end{eqnarray}
where $\eta$ is the viscous friction coefficient and 
$V'(x,f):=\partial V(x,f)/\partial x$.
With respect to its spatial argument $x$, the potential
is periodic for all possible arguments $f(t)$, i.e.
\begin{equation}
V(x+L,f(t)) = V(x,f(t))
\label{4e}
\end{equation}
for all $t$ and $x$.
Along the same line of reasoning 
as in \sect \ref{sec2.1.2.4} of Appendix A, 
inertia effects are neglected and 
thermal fluctuations are modeled by uncorrelated (white) 
Gaussian noise $\xi (t)$ 
of zero average and intensity $2\,\eta\, k_B T$ (see also \sect
\ref{sec3.2.1} below).
Finally, $F$ is a constant ``load'' force. Since such a bias violates
the above requirement (ii), it should not be considered
as part of the system but rather as an externally imposed 
perturbation in order to study its response behavior.

We furthermore assume that  $f(t)$ and $y(t)$ are
either {\em periodic} or {\em stochastic} functions of time $t$.
In the case that one or both of them are a stochastic process, 
we make the simplifying assumption
that this process is {\em stationary},
and in particular statistically independent of the 
thermal noise $\xi(t)$ and of the state of system $x(t)$.
With the symbol $\langle\cdot\rangle$ we
henceforth indicate an ensemble average
over realizations of the stochastic dynamics (\ref{4a}), 
i.e. a statistical average with
respect to the thermal noise $\xi (t)$ and in addition with respect to
$f(t)$ and/or $y(t)$ if either of them is a stochastic process.

The quantity of central interest is the average {\em particle current}
(cf. (\ref{2.10-1}))
\begin{equation}
\langle\dot x\rangle := \langle\dot x(t)\rangle \ . 
\label{4c'}
\end{equation}
In most cases\footnote{There are only very few investigations on
transient features of ratchet systems \cite{ari96,han97a,kol98,yev00,goy00,goy01}.}
we will furthermore
focus on the behavior of the particle current in the
{\em long-time limit} $t\to\infty$ (cf. \sect \ref{sec2.1.4}).
If both $f(t)$ and $y(t)$ are random processes in time, then the existence
of a {\em stationary } long-time limit and its uniqueness
are taken for granted. If $f(t)$ and/or $y(t)$ 
is a periodic function of $t$,
then the existence of a unique {\em periodic} long-time behavior is
assumed and a time average is tacitly incorporated
into $\langle\dot x\rangle $ (cf. \eq (\ref{2.24''})).
Both, for random and periodic processes, this long time limit of the current 
can usually be identified, due to ergodicity reasons, with the time averaged
velocity of a single realization $x(t)$ of the stochastic dynamics (\ref{4a}),
i.e. with probability $1$ we have that
\begin{equation}
\langle\dot x\rangle =\lim_{t\to\infty}\,\frac{x(t)}{t} \ ,
\label{4c1}
\end{equation}
independent of the initial 
condition\footnote{Proof: The time averaged current from (\ref{4c'}) 
can be rewritten as 
$\langle\dot x\rangle 
= \langle \lim_{t\to\infty}t^{-1}\int_0^t\dot x(t')dt'\rangle
= \lim_{t\to\infty}\langle x(t)-x(0)\rangle/t$.
The random process $x(t)-x(0)$ exhibits on top of the systematic
drift $\langle x(t)-x(0)\rangle/t$ a certain random
dispersion (or diffusion) of the order
$\sqrt{2\deff t}$ for large $t$, cf. \eq (\ref{4.3}).
Due to the division by $t$ it follows that this dispersion is negligible, i.e.
$\langle\dot x\rangle 
= \lim_{t\to\infty} [x(t)-x(0)]/t
= \lim_{t\to\infty} x(t)/t$ with probability 1 for any realization $x(t)$.}
$x(0)$.

A further quantity of interest is the {\em effective diffusion coefficient}
\begin{equation}
\deff  := \lim_{t\to\infty}\,  \frac{1}{2t} 
\, \langle \, [x (t) -\langle x(t)\rangle ]^2\, \rangle\ .
\label{4c2}
\end{equation}
For $-V'(x(t),f(t)) + y(t) \equiv 0$, the effective diffusion
coefficient (\ref{4c2}) agrees with the bare coefficient (\ref{2.4}),
independent of $F$.
In general, its determination is a difficult time-dependent
problems\footnote{While for the current it is sufficient to consider an 
auxiliary dynamics with periodic boundary conditions, which approaches
a stationary (if $f(t)$ and $y(t)$ are random processes) or periodic
long time limit (cf. \sect \ref{sec2.1.4}), no such simplification is 
possible with respect to the effective diffusion coefficient.
In particular, the effective diffusion coefficient is in general 
{\em no} longer related to the mobility via a generalized Einstein relation 
(\ref{2.4}), i.e. $\deff = k_B T\,\partial \langle\dot x\rangle/\partial F$
only holds when $f(t)\equiv 0$, $y(t)\equiv 0$, and $F=0$.}.
and we will restrict ourselves to a few special cases.

On a sufficiently coarse grained scale in space,
the motion of the particle $x(t)$ 
takes the form of single ``hopping''
events which are independent of 
each other and equally distributed.
According to the central limit theorem \cite{kam92}, a statistical
ensemble of particles $x(t)$ with initial condition $x(0)=x_0$ thus 
approaches a Gaussian distribution \cite{cox67,vdb89,jun96,har97,fre99,con99,lin01,rei01}
\begin{equation}
P(x,t)\simeq \frac{1}{\sqrt{4\pi\deff t}}\, 
\exp\left\{-\frac{[\langle\dot x\rangle t-x_0]^2}{4\deff t}\right\}
\label{4.3}
\end{equation}
for large times $t$.
As far as the objective of particle separation is concerned, 
we see that not only a large difference or even opposite sign 
of the velocities $\langle\dot x\rangle$ is important (cf. \sect \ref{sec3.5}),
but also the effective diffusion coefficients and the time $t$
(or, equivalently, the length $\langle\dot x\rangle t$ of the
experimental device) play a crucial role 
\cite{ajd92,jun96,fre99,ket00,lin01}, see also \sect \ref{sec6.4}.
A purely diffusive ($\langle\dot x\rangle =0$) particle separation
scheme will be discussed in \sect \ref{sec6.2.5}.

Once in a while, certain extensions
of the above framework will appear, e.g. an additional finite
inertia term $m\,\ddot x(t)$ on the left hand side of (\ref{4a})
or two instead of one spatial dimensions, see e.g. in \sects
\ref{sec6.2.3.2} and \ref{sec6.2.4}, respectively.
Furthermore, 
{\em models with a time or space dependent temperature} in (\ref{4b})
will be discussed in \sects \ref{sec4.8.1}-\ref{sec4.8.3},
and similarly in \sect \ref{sec4.8.4}) 
{\em models with a time or space dependent friction}.
Deviations of the spatial periodicity (\ref{4e}) may arise in the
form of some amount of quenched spatial disorder (\sect \ref{sec3.7})
or as a superposition of several periodic contributions
with incommensurate periods (\sect \ref{sec4.4.1}).
The case of a spatially discretized state variable is
reviewed in \sect \ref{sec3.9}.
A class of models with  a non-trivial dependence of the 
process $f(t)$ upon the state $x(t)$ of the system appears
in \sect \ref{sec4.8.2} and in \ch  \ref{cha5}.
Generalizations of a more drastic nature are addressed in 
\chs \ref{sec6.5} and \ref{cha7}.

If $y(t)$ is a periodic function of time, say
\begin{equation}
y(t+\ttt)=y(t)
\label{4f}
\end{equation}
then we can assume without loss of generality that
\begin{equation}
\int_0^\ttt dt\,  y(t) = 0
\label{4g}
\end{equation}
thanks to the free constant $F$ in (\ref{4a}).
Similarly, if $y(t)$ is a stationary stochastic process then we can assume that
\begin{equation}
\langle y(t) \rangle = 0 \ .
\label{4h}
\end{equation}
Without loss of generality, it is also sufficient to concentrate 
on $f(t)$ which are unbiased in the same sense as in 
(\ref{4g}), (\ref{4h}).

As far as unbiased stationary random processes are concerned,
two examples are of particular importance due to their archetypal simplicity.
To be specific, we will use the symbol $f(t)$, while completely analogous
considerations can of course be immediately transcribed to $y(t)$ as well.
The first example is a so-called 
{\em symmetric dichotomous noise} or telegraphic noise
\cite{kly77,han83,hor84,vdb84}, 
i.e. a stochastic process
which switches back and forth between two possible ``states''
$+\sigma$ and $-\sigma$ with a constant probability $\gamma$ per time unit.
In the stationary state the distribution of the noise
\begin{equation}
\rho (f) :=\langle\delta(f-f(t))\rangle
\label{6k1}
\end{equation}
is thus given by
\begin{equation}
\rho (f) =\frac{1}{2}\, [\delta(f-\sigma) + \delta(f+\sigma) ]
\label{6k2}
\end{equation}
independent of the time $t$ in (\ref{6k1}).
One furthermore finds that the correlation is given by
\begin{equation}
\langle f(t)\, f(s)\rangle = \sigma^2\, e^{-|t-s|/\tau}\ ,
\label{6l}
\end{equation}
where $\tau:=1/2\gamma$ is the correlation time 
and\footnote{Note that $\sigma$ in (\ref{6l'}) is consistent with 
(\ref{6k2}) and (\ref{6m}).}
\begin{equation}
\sigma^2 := \langle f^2(t)\rangle = \int_{-\infty}^\infty df\, f^2\rho(f)
\label{6l'}
\end{equation}
is the variance (independent of $t$).

Being abundant in natural systems as well as in 
technological applications, a stationary Gaussian distributed noise $f(t)$
is clearly a second type of random fluctuations that warrants to be
analyzed in more detail. In the simplest case, these stationary Gaussian
fluctuations are furthermore unbiased, and
Markovian\footnote{The future 
of $f(t)$ only depends on its present state, not
on its past \cite{han82b}.}. According to Doob's theorem \cite{kam92}, $f(t)$ is
thus a so-called {\em Ornstein-Uhlenbeck process} \cite{han82b,ris84},
characterized by a stationary probability distribution
\begin{equation}
\rho(f) = (2\pi\sigma^2)^{-1/2}\, e^{-f^2/2\sigma^2}
\label{6m}
\end{equation}
and the same correlation as in (\ref{6l}).
So, the variance $\sigma^2$ and the correlation time $\tau$ are the 
model parameters for both, dichotomous noise and Ornstein-Uhlenbeck noise.

\section{Symmetry}\label{sec3.1.2}
\subsection{Definitions}
The potential $V(x,f(t))$ is called  spatially symmetric or simply
{\em symmetric} if there exists a 
$\Delta x$ such that
\begin{equation}
V(-x,f(t))=V(x+\Delta x,f(t))
\label{s1}
\end{equation} 
for all
$x$ and $t$.
In other words, we will use the notions ``symmetry'' and ``asymmetry'' as
synonyms for ``spatial isotropy'' and ``anisotropy'', respectively.

A further important symmetry regards the unbiased tilting process $y(t)$:
If $y(t)$ is {\em periodic} in time and there exists a $\Delta t$ such that
$-y(t) = y(t+\Delta t)$ for all $t$ then we call $y(t)$ ``inversion symmetric''
or simply {\em symmetric}.
By performing the symmetry transformation twice,
it 
follows that $y(t) = y(t+2\Delta t)$ and under the assumption that $\ttt$
is the fundamental time-period, i.e. the smallest $\tau >0$ such that 
$y(t+\tau)=y(t)$, the symmetry condition takes the form \cite{ajd94a}
\begin{equation}
-y(t) = y(t+\ttt /2) \ .
\label{s2}
\end{equation}
If $y(t)$ is a stationary {\em stochastic} process, then 
we call it symmetric if
all statistical properties of the process $-y(t)$ are the same as 
those of $y(t)$, symbolically indicated as
\begin{equation}
-y(t)\entspr y(t) \ .
\label{s3}
\end{equation}
Examples are the symmetric dichotomous noise and the Ornstein-Uhlenbeck 
process as introduced at the end of the preceding subsection,
or the symmetric Poissonian shot noise from \sect \ref{sec6.1}.
Note that the assumption of an unbiased $y(t)$, see
(\ref{4g}), (\ref{4h}), does not
yet imply that $y(t)$ is symmetric.

Regarding nomenclature, 
an asymmetric potential is also called a {\em ratchet potential}.
On the other hand, the
dynamics (\ref{4a}) will be termed 
{\em Brownian motor}, {\em ratchet dynamics}, or simply {\em ratchet} 
not only if the potential  $V(-x,f(t))$ is asymmetric  but also if
the driving $y(t)$ is asymmetric, while the potential may 
then be symmetric.

\subsection{Conclusions}
From the definition (\ref{s1}) it follows that a $L$-periodic
potential $V(x,f(t))$ is symmetric if and only if it is of the 
general form
\begin{equation}
V(x,f(t)) = \sum_{n=1}^\infty a_n(f(t))\, \cos (2\pi n x /L)
\ .
\label{ss1}
\end{equation}
Here and in the following, trivial freedoms in the choice of the $x$- and $V$-origins
are neglected. In the specific case (\ref{s1}) this means
that we have silently
set $\Delta x =0$ and $a_0(f(t))=0$ in (\ref{ss1}).
Similarly, one sees that the symmetry condition (\ref{s2})
for a {\em periodic}, deterministic driving $y(t)$ is equivalent to
a Fourier representation of the general form
\begin{equation}
y(t) = \sum_{n=1,3,5,...}
b_n \cos \left( \frac{2\pi nt}{\ttt}+\phi_n \right)
\ .
\label{ss2}
\end{equation}
In the case that $y(t)$ is a {\em stochastic} process, the symmetry
condition (\ref{s3}) is equivalent to the requirement that all its odd
moments vanish \cite{han96a,chi97}, i.e.
\begin{equation}
\langle y(t_1)\, y(t_2)\cdots  y(t_{2n+1})\rangle = 0 
\label{ss3}
\end{equation}
for all integers $n\geq 0$ and all times\footnote{Here and in what follows
we tacitly assume that all multiple-time moments of the precess $y(t)$
exists.}
$t_1,\, t_2,...,t_{2n+1}$.
Basically, the reason for this is that the stochastic process $y(t)$ is
completely specified by the set of all its multiple-time
joint probability distributions (Kolmogorov-theorem) and those are in turn
completely fixed by all its moments \cite{kam92}.
On this basis, the equivalence of (\ref{s3}) and (\ref{ss3}) follows.

If the potential $V(x,f(t))$
respects the symmetry condition (\ref{ss1}) {\em and} the driving $y(t)$
either (\ref{ss2}) or (\ref{ss3}) then we can conclude that the long time
averaged particle current (\ref{4c1}) vanishes in the absence of a static 
tilt $F$ in (\ref{4a}), i.e.
\begin{equation}
\langle\dot x\rangle = \lim_{t\to\infty} \frac{x(t)}{t} = 0 \ . 
\label{ss4}
\end{equation}
For a proof, we recall that the current in (\ref{4c1}) is independent of the
initial condition $x(0)$ and we may thus choose $x(0)=0$.
If both, $V(x,f(t))$ and $y(t)$ are symmetric according to 
(\ref{s1})-(\ref{s3}), or equivalently
(\ref{ss1})-(\ref{ss3}),
then it follows that a realization $x(t)$ of the random process (\ref{4a}), (\ref{4b})
with $F=0$ in (\ref{4a}) occurs with the same probability as its mirror image 
$-x(t)$. Hence, we can infer from (\ref{4c1}) that 
$\langle\dot x\rangle= - \langle\dot x\rangle$, implying (\ref{ss4}).
In other words, the main conclusion of this subsection is that
{\em if both, the potential $V(x,f(t))$ and the driving $y(t)$ 
are symmetric according to (\ref{s1})-(\ref{ss3}) then the 
average particle current (\ref{4c1}) is zero.}

If the potential and the driving $y(t)$ do not both satisfy their respective
symmetry criteria, then, according to Curie's principle, a finite average
current is expected in the generic case.
The exceptional (non-generic) cases with zero current 
(\ref{ss4}) in spite of a broken symmetry
are either in some sense ``accidental'' \cite{mag94}
(analogous to the current inversion in 
\fig \ref{fig8}) or can be traced back to certain ``hidden''
symmetry reasons of a more
fundamental and systematic nature. Examples of the latter type 
will be the subject of \sects \ref{sec3.4} and \ref{sec4.8.4a},
see also the concluding remarks in \ch \ref{cha9}.

The generalization of these symmetry considerations to the
case of a quasiperiodic driving $y(t)$ is due to \cite{neu01},
while an extension to two-dimensional systems (cf. \sect \ref{sec6.2.4})
and models with an internal degree of freedom (cf. \sect \ref{sec5.7})
is contained in \cite{wei99,wei00} and \cite{cil01}, respectively.

\section{Main ratchet types}\label{sec3.1.3}
In this section we introduce the classification scheme underlying the 
organization of \chs \ref{cha4}-\ref{cha06}.
Some general physical considerations complementing this abstract
classification are summarized in \ref{sec3.2}.

As already discussed in \sect \ref{sec2.2.2}, of foremost
interest is usually the current $\langle\dot x\rangle$ 
in the long time limit in the absence of a static tilt $F$ in 
(\ref{4a}).
If both, the potential $V(x,f(t))$ and the tilting force 
$y(t)$ are symmetric, then a vanishing current will be the 
result (see preceding subsection).
The following classification of the different types of 
ratchet models is on the one hand based on the systematic
breaking of this symmetry, on the other hand
it follows to some extent the historically grown, non-systematic
nomenclature.

There are two fundamental classes of ratchet models
arising from (\ref{4a}). The first one are models
with $y(t)\equiv 0$, which we denote as {\em pulsating ratchets}.
The second are models with $f(t)\equiv 0$, called 
{\em tilting ratchets} \cite{doe95}.

Within the realm of pulsating ratchets ($y(t)\equiv 0$),
the first main subclass is obtained when $f(t)$ in (\ref{4a})
is additive, i.e. 
\begin{equation}
V(x,f(t)) = V(x)\, [1+f(t)] \ .
\label{s4}
\end{equation}
Such models carry the name {\em fluctuating potential ratchets}.
The summand $1$ is a matter of convention, reflecting
a kind of ``unperturbed'' contribution to the total potential.
The class of fluctuating potential ratchets contains as 
special case the {\em on-off ratchets} when $f(t)$ can take only two possible 
values, one of them being $-1$ (potential ``off''). 
Without loss of generality
the other value can then be assumed to be $+1$.

One readily sees that the potential $V(x,f(t))$ on the left hand side
of (\ref{s4}) satisfies the symmetry condition
(\ref{s1}) if and only if $V(x)$ on the right hand side of (\ref{s4})
is symmetric as well. Furthermore, it is
obvious that a symmetric $V(x)$ in (\ref{s4}) 
always results in
a vanishing current $\langle\dot x\rangle$, 
whatever the properties of $f(t)$ are.
We will therefore focus on the simplest non-trivial scenario, namely
{\em asymmetric potentials $V(x)$ in combination 
with symmetric $f(t)$}.
As the word ``fluctuating potential'' already suggests, we
will mainly focus on {\em random} $f(t)$, though periodic $f(t)$ are 
in principle meant to be equally covered by this name.

A second subclass of pulsating ratchets, called 
{\em traveling potential ratchets}, 
have potentials of the form 
\begin{equation}
V(x,f(t))=V(x-f(t)) \ .
\label{s5}
\end{equation}
The most natural choice, already suggested by the name ``traveling potential'',
are $f(t)$ with a systematic long time drift $u:=\lim_{t\to\infty}f(t)/t$.
As a consequence, $f(t)$ can only be a veritable periodic function
or stationary stochastic process after subtraction of this systematic drift.
We will call such a model a {\em genuine traveling potential ratchet} scheme.
This slight extension of our general framework will be justified by our
demonstration that such a model is exactly equivalent either
to a tilting ratchet or to a so-called {\em improper traveling potential
ratchet}, for which already the ``original'' $f(t)$ is a periodic 
function or a stationary stochastic process.
Within a traveling potential ratchet scheme, the potential $V(x,f(t))$
on the left hand side of (\ref{s5}) never satisfies the
symmetry criterion (\ref{s1}), independently of whether the potential
$V(x)$ on the right hand side is symmetric or not.
Both, the genuine and improper schemes are therefore interesting to 
study since {\em a symmetric potential $V(x)$ 
is sufficient for current generation}.
Especially, the word ``ratchet'' does not necessarily 
refer to an asymmetric potential $V(x)$ in this context.

Next we turn to the tilting ratchet scheme, characterized by
$f(t)\equiv 0$ and thus 
\begin{equation}
V(x,f(t))=V(x)
\label{s6}
\end{equation}
in (\ref{4a}).
When $V(x)$ is a ratchet potential, then we will restrict ourselves mostly
to symmetric $y(t)$.
If $y(t)$ is a {\em stochastic} process, 
we speak of a {\em fluctuating force ratchet}.
The case of a tilting ratchet with a {\em periodic} driving $y(t)$ is of
particular experimental relevance and carries the obvious
name {\em rocking ratchet} \cite{bar94}.

Coming to symmetric potentials $V(x)$ in (\ref{s6}),
a broken symmetry of $y(t)$ turns out to be
necessary and 
generically also sufficient for a finite current $\langle\dot x\rangle $.
We will use the name {\em asymmetrically tilting ratchet} if
$y(t)$ is not symmetric, independently of whether it is a 
periodic function or a stochastic process, and independently of whether
$V(x)$ is symmetric or not.

A further important class of ratchets 
is given by models of the form (\ref{4a}), (\ref{4e})
with both $f(t)\equiv 0$ and $y(t)\equiv 0$ but instead with a 
{\em space or time dependent temperature} $T$ in (\ref{4b}). 
They carry the names {\em Seebeck ratchets} and  
{\em temperature ratchets}, respectively. 
In the case of a space dependent temperature, $T(x)$ 
is assumed to have the same periodicity $L$ as
the potential $V(x)$.
In the case of a time-dependent temperature $T(t)$, 
again a periodic or stochastic, stationary behavior is assumed.
We anticipate that
models of this type are obviously not pulsating ratchets in the original sense,
but -- as will be demonstrated in \sects \ref{sec4.8.1}  and \ref{sec4.8.3} --
{\em they can be mapped onto genuine pulsating ratchets}.
Also discussed in this context (\sect \ref{sec4.8.2}) will be so-called
{\em Feynman ratchets}, i.e. the extension of the isothermal
Smoluchowski-Feynman ratchet and pawl from \fig \ref{fig1}
to the non-equilibrium
case involving simultaneously two thermal baths at different temperatures.
Starting with a faithful two-dimensional model, which is in fact equivalent
to a generalized fluctuating potential scheme, additional simplifications
give rise to a one-dimensional,
{\em Seebeck ratchet-like approximative description}.
Finally, the case of a varying friction coefficient in (\ref{2.4})
(temporal and/or spatial) is denoted as {\em friction ratchet}.
In \sect \ref{sec4.8.4a},
we show that such a modification of the Smoluchowski-Feynman ratchet
model (\ref{2.3}), (\ref{2.5})
does {\em not} break the detailed balance symmetry and thus does {\em not}
admit a ratchet effect\footnote{Especially, such a modification 
requires a correct handling of the non-trivial overdamped 
limit $m\to 0$ in (\ref{2.1}), see \sect \ref{sec4.8.4a}.},
in contrast to a modified, so-called memory-friction 
modelling as discussed in 
\sect \ref{sec4.8.4c}.

We remark that the main idea of the above classification scheme is the
identification of different basic {\em minimal models}. Clearly,
there are many possible combinations and generalizations, e.g. a 
simultaneously pulsating and tilting ratchet or the simultaneous breaking
of more than one symmetry. Especially, there exist numerous pulsating
ratchet schemes involving potentials
$V(x,f(t))$ which go beyond the special cases of 
fluctuating potential and traveling potential ratchets.
Such generalizations will not be {\em systematically}
analyzed since no fundamentally new phenomena are expected.
They are, however, realized in some interesting experimental
systems and will be discussed in such {\em specific} 
contexts.

\section{Physical basis}\label{sec3.2}
The physical situations in which a model of the form (\ref{4a})-(\ref{4e}) may arise
are extremely diverse. Therefore, a systematic discussion makes little sense 
and we restrict ourselves in this section
to a few general remarks before turning to the
various concrete systems in the subsequent chapters.

The stochastic process $x(t)$ in (\ref{4a}) has as state space
the entire real axis and for simplicity is often called a ``Brownian particle''.
While in some cases, $x(t)$ indeed represents the position of a true
physical particle, in others it may also
refer to some quite different type of
collective degree of freedom or relevant (slow) state variable.
Examples which we will encounter later on 
are the chemical reaction coordinate of an enzyme,
the geometrical configuration or some other internal degree of freedom
of a molecule, the position of the circular ratchet
in \fig \ref{fig1} with respect to the pawl, the Josephson phase in a SQUID
(superconducting quantum interference device),
and the collective angular variable in phenomenological models for pinned
charge density waves.
In many cases the state variable $x(t)$ is thus originally of a phase-like nature
with a circle as state space.
The expansion to the real axis is immediate and has the additional
advantage of counting the number of revolutions.
Accordingly, the periodicity (\ref{4e}) -- a central property of our model -- 
may have its root either in a true spatial periodicity of the physical 
system or in the phase-like nature of the original state variable.

\subsection{Thermal environment}\label{sec3.2.1}
Another central feature in our working model (\ref{4a}), (\ref{4b}) 
is the presence of a thermal environment.
In this section we continue and extend our discussion from
\sect \ref{sec2.1.2} (see also \sect \ref{sec2.1.2.1} 
in Appendix A) regarding the physical origin of
the particularly simple form of the system-bath interaction
in (\ref{4a}), (\ref{4b}), namely
an additive white Gaussian noise and an additive viscous 
dissipation proportional to the instantaneous system velocity.

Adopting a phenomenological approach, in many cases \cite{han82b,ris84,han90}
such an {\em ansatz} has proven to provide a rather faithful modeling,
justified by its agreement with experimental measurements and the 
intuitive physical picture that has emerged on the basis of
those observations.

A different approach starts with a microscopic modeling of the 
system of actual interest and its thermal environment.
In the following we briefly sketch the main steps of such an
approach. For a somewhat more detailed illustration of these 
general concepts for specific physical examples we also refer to
\sects \ref{sec5.2}, \ref{sec5.3}, and \ref{sec6.5.1}.
On the one hand, such a microscopic foundation provides a
physical picture of why the phenomenological modeling 
(\ref{4a}), (\ref{4b}) is successful in such a
wide variety of different systems. On the other hand,
a feeling for the conditions under which such a modeling is 
valid is acquired as well as an idea of how to modify the 
model when they break down.

Our starting point is a Hamiltonian of the general form
\begin{equation}
H= \frac{p^2}{2\, m} + V_s(x) + \sum_{j=1}^N \frac{p^2_j}{2\, m_j} + V_b(x,x_1,...,x_N) \ ,
\label{q1}
\end{equation}
where $x$ and $p$ are the coordinate and momentum of the actual system of 
interest, while $x_j$ and $p_j$ are those of the numerous
($N\gg 1$) microscopic degrees of freedom of the environment.
The last term in (\ref{q1}) is a general interaction potential, including 
the coupling between system and environment.
To keep things simple, we restrict ourselves to a single 
relevant (i.e. ``slow'') state variable $x(t)$, e.g. the 
cartesian coordinate of a particle in the absence of magnetic 
fields or the Josephson phase in a SQUID.
We remark that in other cases, e.g. the chemical reaction coordinate of 
an enzyme, the geometrical configuration, or some other internal degree 
of freedom of a molecule, the respective ``slow'' relevant state variable
$x(t)$ is usually a generalized coordinate (a non-trivial function of the
cartesian coordinates of the nuclei, cf. \sect \ref{sec5.2}), and
similarly for the ``fast'' bath degrees of freedom $x_j(t)$.
As a consequence, the kinetic energy terms are of a more complicated
form than in (\ref{q1}) and with respect to the potential terms there
exists no longer a meaningful distinction between the ``actual system
of interest'' and the ``environment plus the system-bath-coupling''.
In those cases, our general line of reasoning remains still
valid, but the detailed calculations become more 
involved \cite{gre52,gra80,gra82a,mil83}.

\subsubsection{Elimination of the bath degrees of freedom}
Having set the stage (\ref{q1}), our next goal is to
get rid of the environmental degrees of freedom $x_j(t)$.
To this end, we start by formally solving the respective equations 
of motions for any prescribed function $x(t)$ and initial conditions 
$\phi_0:=(x_1(0),p_1(0),...,x_N(0),p_N(0))$ at time $t_0=0$.
In other words, we can write down (formal) solutions $x_j(t,[x(t')],\phi_0)$
which are at the same time functions of $t$ and $\phi_0$ and
functionals of the (explicitly still unknown) system dynamics $x(t')$
for $t' \in [0,t]$.
Introducing these solutions into the equation of motion for the system 
$x(t)$ is equivalent to a Newtonian dynamics of the general 
structure\footnote{The explicit but formal expression of
$f(x(t),t,[x(t')],\phi_0)$ in terms of the potentials in (\ref{q1}) and 
the formal solutions $x_j(t,[x(t')],\phi_0)$ is straightforward but
of no further use, see below. 
Especially, $f(x(t),t,[x(t')],\phi_0)$ in
(\ref{q2}) has nothing to do with $f(t)$ from (\ref{4a}).}
\begin{equation}
m\, \ddot x(t) = f(x(t),t,[x(t')],\phi_0) \ .
\label{q2}
\end{equation}

In most cases, an explicit analytical expression for 
$f(x(t),t,[x(t')],\phi_0)$ is not 
available\footnote{The only solvable exception -- 
the so-called harmonic oscillator bath --
arises when $V_b(x,x_1,...,x_N)$ in (\ref{q1})
is a quadratic function of its arguments and thus the 
bath-dynamics is not chaotic, see \sect \ref{sec6.5.1}.}
since this would require
analytical solutions $x_j(t,[x(t')],\phi_0)$ of a high dimensional 
chaotic dynamics and would in fact comprise the derivation
of the basic principles of equilibrium statistical mechanics 
as special case.
Rather, one proceeds the other way round, exploiting the 
fact that the environment is a thermal equilibrium heat bath 
and thus statistical mechanical principles can be invoked.
Namely, one assumes that the systems initial conditions $x(0)$ and $p(0)$
are arbitrary but fixed, while the initial state of the bath $\phi_0$
is randomly sampled from a canonical probability 
distribution\footnote{The physical origin of this canonical description
is a ``superbath'' to which the bath of actual interest is {\em weakly} coupled.}
$P(\phi_0)\propto \exp\{-H(x(0),p(0),\phi_0)/k_BT\}$.
It is via this randomness of the environmental initial conditions
$\phi_0$ that the system dynamics (\ref{q2}) 
acquires itself a stochastic nature.
Denoting the average over those initial conditions by
\begin{equation}
\tilde f(x(t),t,[x(t')]) := \langle f(x(t),t,[x(t')],\phi_0)\rangle
\label{q3}
\end{equation}
we can decompose the right hand side of (\ref{q2}) into a sum of three terms,
\begin{equation}
m\, \ddot x(t) = -V'(x(t))- h(x(t),t,[\dot x(t')])+ \xi (x(t),t,[x(t')],\phi_0) \ ,
\label{q4}
\end{equation}
where the first term is determined by the instantaneous state of the system,
the second by its past history, and the third term is of microscopic origin,
giving rise to the stochastic nature of the dynamics.
Their explicit definitions are:
\begin{eqnarray}
& & V'(x(t)):= - \tilde f(x(t),t,[x(t')\equiv x(t)])
\label{q5a} \\
& & h(x(t),t,[\dot x(t')]):= - \tilde f(x(t),t,[x(t')]) + \tilde f(x(t),t,[x(t')\equiv x(t)])
\label{q5b} \\
& & \xi (x(t),t,[x(t')],\phi_0):= 
f(x(t),t,[x(t')],\phi_0) - \tilde f(x(t),t,[x(t')]) \ .
\label{q6}
\end{eqnarray}
Here, $[x(t')\equiv x(t)]$ means that the function $x(t')$ keeps the same
value $x(t)$ for all times $t'\in [0,t]$ and is understood as a formal 
functional argument rather than an actual solution of the real 
system dynamics (\ref{q4}).
Further, the modified functional argument $[\dot x(t')]$ on the left hand 
side of (\ref{q5b}) is justified by the fact that
any function $x(t')$ with $t' \in [0,t]$
can be reconstructed from the knowledge of $x(t)$ and $\dot x(t')$.
Finally, we remark that the souce of randomness $\phi_0$
enters via the ``noise'' (\ref{q6}), which has a vanishing mean value by construction.

Observing that for $x(t')\equiv x(t)$ the bath keeps its initial
canonical probability distribution and expressing the force on the
right hand side of (\ref{q2}) in terms of the potentials in (\ref{q1})
one can infer from (\ref{q3}) and (\ref{q5a}) that
\begin{equation}
V(x) = V_s(x) - k_B T \ln 
\int \prod_{j=1}^N dx_j\, \exp\{-V_b(x,x_1,...,x_N)/k_B T\} \ .
\label{q7}
\end{equation}
In general, the bare system potential is thus renormalized (dressed) 
by the eliminated degrees of freedom of the environment 
and plays a role similarly to a free energy rather than
a (bare) energy \cite{gre52,gra80,gra82a,kel00}.
However, if the potential $V_b(x,x_1,...,x_N)$ is translation invariant
(i.e. equal to $V_b(x+\Delta,x_1+\Delta,...,x_N+\Delta)$ for all $\Delta$)
then the renormalization in (\ref{q7}) boils down to an irrelevant 
additive constant.

\subsubsection{Linearized friction and thermal fluctuations}
While all so far formal manipulations are still exact, we finally make two
approximations with respect to the ``friction'' term (\ref{q5b}).
First, we functionally expand $h(x(t),t,[\dot x(t')])$ with respect to
$\dot x(t')$. Considering that $h(x(t),t,[\dot x(t')\equiv 0])=0$ (cf. (\ref{q5b}))
and that $t'\in [0,t]$, the leading order approximation is
\begin{equation}
h(x(t),t,[\dot x(t')])\simeq \int_0^t \, ds\
\frac{\delta h(x(t),t,[\dot x(t')\equiv 0])}{\delta \dot x(s)} \ \dot x(s) \ .
\label{q8}
\end{equation}
Second, we exploit the assumed property that the relevant state 
variable $x(t)$ changes ``slowly'' in comparison with the environment, 
hence $\dot x(s)\simeq \dot x(t)$ for all $s$-values which notably
contribute in (\ref{q8}) (Markov approximation).
By closer inspection one sees that within the same approximation
the remaining intergal does no longer explicitly depend on $t$.
As a result, we approximately find a friction term of the 
following general form
\begin{equation}
h(x(t),t,[\dot x(t')])\simeq \eta(x(t))\, \dot x(t) \ .
\label{q9}
\end{equation}

As far as the omitted corrections on the right hand side of (\ref{q9})
are not incidentally identically zero, 
by neglecting them we are tampering with the original equilibrium 
environment with the consequence of
a (possibly very small but generically non-vanishing) breaking of thermal 
equilibrium and thus a violation of the second law of thermodynamics,
see also \sects \ref{sec2.2.3c} and \ref{sec6.5.1}.
This shortcoming can only be remedied by a corresponding adjustment
of the fluctuations in (\ref{q6}) in the following way:
Along a similar line of reasoning as in \cite{rei01a}
(see also \sect \ref{sec2.1.2.1} in Appendix A)
one can show that the specific structure (\ref{q4}), (\ref{q9})
of the dynamics together with the requirement that the environment is
at thermal equilibrium (respects the second law of thermodynamics)
uniquely determine all statistical properties of
those properly adjusted fluctuations appearing in (\ref{q4}).
Namely, they are necessarily an unbiased Gaussian white noise whose
correlations satisfy a fluctuation-dissipation relation of the form
\begin{equation}
\langle \xi (x(t),t,[x(t')],\phi_0)\, \xi (x(s),s,[x(s')],\phi_0)\rangle
= 2\, \eta(x(t)) \, k_B T\,\delta(t-s) \ .
\label{q10}
\end{equation}
As already noticed below (\ref{2.3}), the function
$\eta (x)$ may thus be viewed as the coupling strength 
to the thermal environment.

If the potential $V_b(x,x_1,...,x_N)$ is known
to be translation invariant then not only the renormalization of the 
potential in (\ref{q7}) reduces to an irrelevant additive 
constant but also the spatial dependence of the 
friction coefficient $\eta(x)$ disappears.
In the overdamped limit $m\to 0$ we thus exactly recover our 
``unperturbed'' working model\footnote{For the sake of notational 
simplicity only we have not included $f(t)$, $y(t)$, 
and $F$ into the definition of $V_s(x)$ form (\ref{q1}).}
(\ref{4a}), (\ref{4b}).
This omission of the inertia term in (\ref{4a}) is usually 
a quite satisfactory approximation for the typically
very small systems under consideration, 
cf. \sect \ref{sec2.1.2.4} in Appendix A.
A noteworthy exception is the case of a SQUID 
system\footnote{The reason is that the ``effective 
inertia'' in a SQUID has a ``macroscopic'' origin,
namely the capacitance of the considered circuit, cf.
\sect \ref{sec6.2.4a}.}, for which 
both the overdamped limit (\ref{4a})
as well as the case with finite inertia describe 
realistic experimental situations of 
interest, see \sect \ref{sec6.2.4a}.

The translation invariance of $V_b(x,x_1,...,x_N)$ 
and thus the $x$-independence of the system-bath coupling $\eta$ 
arises naturally if the periodic potential in (\ref{4a}) and the 
thermal environment have different physical origins.
Since this is the case in most concrete examples
which we will consider or at least it can be assumed
without missing basic new effects, we will mostly focus 
on an $x$-independent friction coefficient $\eta$ henceforth.
Prominent examples with $x$-dependent friction coefficients
$\eta (x)$ are discussed in 
\sects \ref{sec4.8.4b} and \ref{sec5.3}.

One basic assumption in our so far discussion has been
the existence of a clear-cut separation between the
characteristic time scales governing the ``slow'' system
variable and those of the environment, with the
consequence of a memoryless friction mechanism 
and uncorrelated thermal fluctuations.
However, there exist physical systems for which this 
assumption is not fulfilled. One reason may be
that one has overlooked additional relevant ``slow''
state variables and thus one simply has to go over 
to a higher dimensional vector $x(t)$ in the above
calculations. However, in some cases the necessary 
dimensionality of $x(t)$ may become very high, while
those additional dimensions are actually of no further
interest, so that keeping a memory-friction and correlated
noise may be more convenient.
Restricting ourselves to the simples case with a
translation invariant potential
$V_b(x,x_1,...,x_N)$, the approximation (\ref{q8})
takes the general form
\begin{equation}
h(x(t),t,[\dot x(t')])\simeq \int_{-\infty}^t \, ds\
\hat \eta (t-s)\, \dot x(s) \ ,
\label{q11}
\end{equation}
where we have assumed that $\dot x(t)=0$ for 
all\footnote{This can be physically realized by means of a time
dependent potential $V_s(x,t)$ in (\ref{q1}) which keeps
$x(t)$ at a fixed position for $t\leq 0$ and switches to the
actual potential of interest for $t>0$.}
$t\leq 0$ in order to uniquely define the evolution of the
integro-differential
equation (\ref{q4}), and hence the lower integration limit could been 
extended to $-\infty$.
Similarly as in (\ref{q10}), the assumption of thermal equilibrium
then implies that the 
properly adjusted fluctuations appearing in (\ref{q4})
are necessarily an unbiased Gaussian 
noise whose correlation satisfies a fluctuation-dissipation relation 
of the form
\begin{equation}
\langle \xi (x(t),t,[x(t')],\phi_0)\, \xi (x(s),s,[x(s')],\phi_0)\rangle
= \hat \eta(t-s) \, k_B T \ .
\label{q12}
\end{equation}
Examples of this type will be discussed in 
\sects \ref{sec4.8.4c} and \ref{sec6.5.1}.

It should be emphasized that the dynamics (\ref{q4})
reproduces the correct equilibrium distribution
$P(x,p)\propto \exp\{-[p^2/2m + V(x)]/k_BT\}$ in the long time limit,
independently of the choice of $\eta(x)$ or $\hat \eta (t)$ in
(\ref{q9})-(\ref{q12}).
Especially, this distribution is exactly identical to the
steady state result for the original system (\ref{q5a})-(\ref{q6})
before making any approximations \cite{gra82a}.
Moreover, the second law of thermodynamics is strictly satisfied in
all cases.
It is only away from equilibrium that the specific choice
of $\eta(x)$ or $\hat \eta (t)$ becomes
important\footnote{An obvious example is the mobility in the absence
of the system potential $V_s(x)$,
independently of whether the system is close to or far from equilibrium. 
Other observables which significantly
depend on the choice of $\eta(x)$ or $\hat \eta (t)$ are
escape rates (even so-called equilibrium rates) \cite{han90},
as well as the particle current and the effective diffusion
from (\ref{4c'}) and (\ref{4c2}).} and that the approximations
made in (\ref{q9})-(\ref{q11}) may have a noticeable effect.

In general, the above program of identifying ``slow'' and ``fast''
variables, establishing the microscopic model (\ref{q1}), and
determining\footnote{The explicit detemination of $V(x)$ according to
(\ref{q7}) may still be feasible.}
$\eta(x)$ or $\hat \eta (t)$ according
to (\ref{q8}) cannot be practically carried out \cite{gra82a}.
The same applies for a well-controlled justification of the 
approximation (\ref{q8}), although this linearization 
turns out to provide remakably
good approximations in a large variety of different systems.
One reason may be the fact that in most cases only terms of odd
order in the system velocity will contribute to the omitted corrections
on the right hand side of (\ref{q8}) due to symmetry reasons.
In view of those practical difficulties we are thus
in some sense back at a phenomenological modeling which
draws its legitimation from the comparison with experimental 
findings.
However, as already mentioned, the microscopic modeling provides
a general framework (functional form) for a large class of
approximate models and a feeling for their wide range
of applicability as well as for possible reasons in case they 
fail.

\subsection{Nonequilibrium perturbations}\label{sec3.2.2}
There are two main types of possible ``perturbations'' of the 
``unperturbed'' equilibrium system
(\ref{4a}) with $f(t)\equiv 0$, $y(t)\equiv 0$, and $F=0$.
The first acts essentially like the force $F$ in (\ref{4a}),
i.e. the system $x$ gains (or looses) energy if it is displaced by one
spatial period $L$.
For instance, this may be a homogeneous force acting on a true Brownian particle 
or an angular momentum-type perturbation if $x$ was originally 
of a phase-like nature.
In any case, such a perturbation interacts directly with the state 
variable $x$.
The unbiased, time-dependent part of such a perturbation gives rise to the
``tilting force'' $y(t)$ and the systematic part to the 
``static force'' $F$ in (\ref{4a}).
The second possible type of perturbations interacts directly with the system
variable $x$ but does not lead to an energy change if $x$ is
displaced by one period $L$.
A simple example is an electrical dipole with a single rotational degree of
freedom in a homogeneous electrical field.

Another option is a perturbation which does not directly interact
with the state variable $x$, but rather affects the physical
mechanism responsible for the periodic potential in (\ref{4a}).
Either some ``internal degree of freedom''
of the system $x$ is excited, which modifies the interaction with the periodic 
potential \cite{pro94}, or the periodic potential itself may be affected 
by the perturbation 
\cite{ast96b}. 
For instance, an electrical field may change the internal charge distribution 
(electrical polarization) 
of a neutral Brownian particle or of the periodic substrate with which 
it interacts.
This type of perturbation gives rise to a ``pulsating potential''
$V(x,f(t))$ in (\ref{4a}).
Depending on the details of the system, either one
of the three basic types (fluctuating, improper traveling, 
or genuine traveling  
potential) arises in its pure form, or a combination thereof,
possibly even with a tilting ratchet admixture, is encountered.

One possible origin of those different types of ``perturbations'' may be
an {\em experimentally applied external field}.
While periodic signals then clearly represent the standard case, 
random perturbations have been realized as well \cite{liu90}.
Another possibility is a system-intrinsic source of ``perturbations'',
usually of stochastic nature.
The origin
of such an intrinsic noise source may be either
a {\em non-equilibrium heat bath} or a 
{\em second thermal heat bath}\footnote{Microscopic models for
two (or more) coexisting thermal heat baths at different temperatures have been
discussed in \cite{fey63b,mil95,jay96,han97}. In the case
of a tilting ratchet scheme it turns out that
a ratchet effect (cf. \sect \ref{sec2.2.2.2}) is only possible
for a correlated (non-white) thermal noise 
$y(t)$ and a concomitant memory friction term, see \sect \ref{sec4.8.4c}.
A generalization of these microscopic models to the case that
one bath is out of equilibrium is also possible.}
at a different temperature than the $\xi(t)$-bath.
As far as the tilting ratchet scheme is concerned, 
the coexistence of such an extra heat bath and the thermal $\xi(t)$-bath, 
which both interact directly with the state variable $x$ but practically 
not with each other,
may be experimentally tailored, but is not very common in natural systems.
Exceptions are electrical circuits, where non-equilibrium fluctuations,
e.g. dichotomous noise \cite{zap98} or shot noise (see Vol.1 of \cite{str69}) 
may naturally arise, and experimental analog electronic circuits
for dichotomous \cite{pos96,nik98} or Gaussian \cite{arr00} colored noise.
More common are sources of noise which manifest themselves via an internal degree 
of freedom and thus lead to a pulsating ratchet scheme.
Examples are catalytic chemical reactions with reactant and product
concentrations far from their equilibrium ratio,
or excitations induced by electromagnetic irradiation.
Another example is a modified Feynman ratchet as discussed in \sect \ref{sec4.8.2}.
In such cases, the coexistence of two practically independent sources of the
noises $\xi(t)$ and $f(t)$ in (\ref{4a})  is indeed realistic.

At first glance, the property (\ref{4e}) that the potential is changing
its shape in perfect synchrony over arbitrary distances $x$ might appear
somewhat strange.
However, this is in fact very natural
if either $x$ is of a phase-like character or if the pulsating potential
mechanism is caused by an internal degree of freedom of the system $x$.
Also experimentally imposed external perturbations usually do not
cause an asynchronous pulsating potential scheme.
{\em Asynchronously pulsating potentials} \cite{schi97,par98a,par98b,hoh99,par00,hoh01}
can only be expected if $x$ is a space-like variable
and if the potential is subjected to independent ``local'' nonequilibrium
noise sources, or in a specifically tailored experimental setup.

In the case of stochastic ``perturbations'' $f(t)$ or $y(t)$ in (\ref{4a}),
we have assumed stationarity and especially $x$-independence of their statistical properties.
Similarly as for the thermal noise $\xi(t)$, this reflects the assumption that their origin
is a ``huge'' heat bath which is practically not influenced by the behavior of the ``small''
system $x(t)$.
A more drastic assumption in (\ref{4a}) is the implicit 
{\em omission of a back-coupling} mechanism 
(``active decoupling'') \cite{mag93,mag94,ast94}
to the $f(t)$- or $y(t)$-heat bath, analogous
to the dissipation mechanism in the case of the equilibrium 
$\xi(t)$-bath.
This means that the coupling to this former bath is very weak and that 
this bath is very far away (``highly excited'') from equilibrium 
with respect to the $\xi(t)$-bath at temperature $T$.
Only then, the effect of the fluctuations $f(t)$ or $y(t)$ are still appreciable
while the corresponding back-coupling effects are
negligible\footnote{For example, the origin of $f(t)$ or $y(t)$ may be 
a second thermal equilibrium bath at a temperature much higher than $T$.
Though such a model may not be very realistic it is of great
conceptual appeal as one of the simplest models for a system
far from equilibrium \cite{fey63b}: Two thermal equilibrium
baths are connected through a single degree of freedom $x(t)$ and can be
exploited to do work.
A concrete example is the Feynman ratchet in \sect \ref{sec4.8.2}.}.
In \ch \ref{cha5} we will encounter a specific model where such a back-coupling
mechanism is fully taken into account (see \sect \ref{sec5.3.1}).
Furthermore, it will be demonstrated explicitly how this back-coupling may again become
negligible as the corresponding source of noise is driven far away from 
equilibrium (see \sect \ref{sec5.4.2}).
Another 
example with a nontrivial 
back-coupling appears in \sect \ref{sec4.8.2}.

Note that if $f(t)$ or $y(t)$ are such that a periodic
perturbation of the system arises, then those
back-coupling effects are also omitted in our model (\ref{4a}) but
in this case such an omission is very common.
The actual justification for doing so, however, follows in
fact along the same line of reasoning as for random perturbations!

In all those various cases, it is clear that the system can never reach a thermal 
equilibrium state even in the long time limit:
either this is prohibited by a permanent periodic perturbation or a second heat bath
out of equilibrium or at equilibrium but with a temperature different from $T$.
In either case, the second law of thermodynamics cannot be applied,
i.e. the symmetry of detailed balance is violated.
In the absence of any other prohibitive symmetries, 
which we have systematically broken 
by our classification scheme (cf.  \sect \ref{sec3.1.3}), 
we thus expect the generic 
occurrence of the ratchet effect $\langle\dot x\rangle\not =0$ according to Curie's
principle. The corresponding intuitive microscopic picture is a permanent
energy flow from the source of the perturbations $f(t)$ or $y(t)$ -- be it
a periodic external driving or a second heat bath --
into the thermal bath at temperature $T$ via the single common
degree of freedom $x(t)$.

\section{Supersymmetry}\label{sec3.4}
In this section we continue our symmetry considerations 
from \sect \ref{sec3.1.2}, where we have seen that breaking
thermal equilibrium, or equivalently, breaking the symmetry of detailed
balance in whatever way, in a periodic, asymmetric system, is 
generically sufficient for the ratchet effect to manifest itself:
In general, the occurrence of a finite current in such systems
is the rule rather than the exception, in accord with Curie's principle.
We thus more and more return to Smoluchowski and Feynman's
point of view that {\em  away from thermal equilibrium,
the absence rather than the presence
of directed transport in spite of a broken symmetry is the 
truly astonishing situation}.
In this section, an entire class of such intriguing 
exceptional cases is identified which do not exhibit a
ratchet effect in spite of broken thermal equilibrium and
broken symmetry.
Especially, such systems (cf. \eqs (\ref{4a}), (\ref{4b}) with $F=0$)
exhibit zero current $\langle\dot x\rangle$ for
{\em any} choice of the friction $\eta$, the temperature $T$,
the amplitude and characteristic time scale of the drivings
$f(t)$ and $y(t)$ etc.,
much like the symmetric systems from \sect \ref{sec3.1.2}.
In contrast to usual current inversions 
(cf. \sects \ref{sec2.5} and \ref{sec3.5}),
no fine-tuning of those parameters is thus 
required in order that $\langle\dot x\rangle = 0$.

\subsection{Definitions}\label{sec3.4.1}
We begin with the following definitions:
We call a {\em potential} $V(x,f(t))$ with a {\em periodic} function $f(t)$
{\em supersymmetric} if there exist 
$\Delta x$, $\Delta t$, $\Delta V$ such that 
$-V(x,f(t))= V(x+\Delta x, f(-t+\Delta t))+\Delta V$ for all $x$ and $t$.
If $f(t)$ is a {\em stochastic} process then we call the potential
$V(x,f(t))$ supersymmetric if for any $x$ all statistical
properties of $-V(x,f(t))$ and  
$V(x+\Delta x, f(-t))+\Delta V$ are the same (no $\Delta t$
is needed since $f(t)$ is stationary).
Especially, a static potential is supersymmetric if
$-V(x)= V(x+\Delta x)+\Delta V$ for all $x$.
Note that while we can and will choose the $t$- and $V$-origins such that
$\Delta t=0$ and $\Delta V=0$, the same is not possible for $\Delta x$.
In fact, by applying the above defined supersymmetry transformation 
twice, we can conclude
that $V(x+2\Delta x,f(t))=V(x,f(t))$ for all $x$ and $t$.
Under the assumption that $L$ is the fundamental period of $V(x,f(t))$, i.e.
the smallest $z>0$ with $V(x+z,f(t))=V(x,f(t))$, we can henceforth focus on
$\Delta x = L/2$. In summary, the {\em supersymmetry criterion} can thus be
symbolically indicated (cf. (\ref{s3})) for both, 
periodic and stochastic $f(t)$ as
\begin{equation}
-V(x,f(t)) \entspr V(x+L/2,f(-t)) \ .
\label{ss5}
\end{equation}

Turning to the {\em driving} $y(t)$, we will call it {\em supersymmetric}
if for a {\em periodic} $y(t)$ we have that $-y(t)=y(-t+\Delta t)$ for all
$t$ and an appropriate $\Delta t$, which can be transformed to zero as usual.
For a {\em stochastic} $y(t)$ we speak of  supersymmetry if $-y(t)$ 
and $y(-t)$ are statistically equivalent. In other words, {\em supersymmetry} 
means for both, periodic and stochastic $y(t)$,
a parity-time-invariance of the form
\begin{equation}
-y(t)\entspr y(-t) \ .
\label{ss6}
\end{equation}

Regarding our above introduced 
notion of supersymmetry we remark that for undriven
($f(t)\equiv 0$, $y(t)\equiv 0$, $F=0$) systems (\ref{4a}), a connection with 
supersymmetric quantum mechanics
\cite{wit81,dut88} has been first pointed out  in \cite{ben83} and
has been further developed in \cite{ber84,mar88}, see
also the \cite{jun96a} for a review.
The basic idea is to transform the
Fokker-Planck equation (cf. \sect \ref{sec2.1.3})
associated with the undriven stochastic dynamics (\ref{4a})
into a Schr\"odinger-type equation \cite{fav67,tom76,han82b,ris84,kam92}. 
By replacing in this equation
the potential by its supersymmetric partner 
potential (in the quantum mechanical sense) a new
Schr\"odinger equation emerges which can be
transformed back into a new Fokker-Planck equation.
The potentials of the original and the new Fokker-Planck 
equations then coincide (up to irrelevant shifts $\Delta x$ 
and $\Delta V$) if and only if 
the supersymmetry condition (\ref{ss5}) is satisfied. 
In the presence of a periodic driving $y(t)$ (but still
$f(t)\equiv 0$, $F=0$) in the stochastic dynamics (\ref{4a}), 
a similar line of reasoning has been developed in 
\cite{jun91a}, yielding the supersymmetry condition (\ref{ss6}).
The case of various stochastic drivings $y(t)$ has been 
addressed in \cite{lei87,lei88}.
Here, {\em we will borrow
the previously established notion of ``supersymmetry''
for the conditions (\ref{ss5}), (\ref{ss6}),
but we will neither exploit nor further discuss their 
connection with quantum mechanical concepts}.

\subsection{Main conclusion}\label{sec3.4.2}
We now come to the central point of this section:
We consider the general stochastic dynamics (\ref{4a}), (\ref{4b})
with $F=0$ together with the usual assumptions on $f(t)$
and $y(t)$ from \sect \ref{sec3.1.1}. 
By introducing $z(t):=x(-t)+L/2$, we can infer that
$\dot z(t):=-\dot x(-t)$, i.e. the averaged currents
satisfy\footnote{Note that it is not possible to derive this conclusion
$\langle\dot z\rangle = - \langle\dot x\rangle$ 
from (\ref{4c1}). The reason is that the initial
and final times exchange their roles when going over from $x(t)$ to $z(t)$
and thus the implicit assumtion in (\ref{4c1}) that the initial time is kept
fixed while $t\to \infty$ is no longer fulfilled for $z(t)$.
The properly generalized version of (\ref{4c1}) reads
$\langle\dot x\rangle = \lim_{t-t_0 \to\infty}[x(t)-x(t_0)]/[t-t_0]$, from which
one readily recovers $\langle\dot z\rangle = - \langle\dot x\rangle$.}
$\langle\dot z\rangle = - \langle\dot x\rangle$.
In doing so, we have exploited that only deterministic and/or 
{\em stationary} stochastic processes appear in (\ref{4a}), (\ref{4b}),
hence the evolution of the dynamics backward in time does not give rise 
to any problem. 
Especially, $-\xi(-t)$ is statistically equivalent to the
forward Gaussian white noise $\xi(t)$.
On the other hand, if both $V(x,f(t))$ and $y(t)$ are supersymmetric
according to (\ref{ss5}), (\ref{ss6}) then one can readily see that
$z(t)$ satisfies the same
dynamics (\ref{4a}) as $x(t)$.
Due to the self-averaging property of the current
in (\ref{4c1}) it follows that
$\langle\dot z\rangle = \langle\dot x\rangle$.
In view of our previous finding 
$\langle\dot z\rangle = - \langle\dot x\rangle$ 
we arrive at our main conclusion:
{\em if both $V(x,f(t))$ and $y(t)$ are 
supersymmetric according to (\ref{ss5}), (\ref{ss6}) then the
average particle current $\langle\dot x\rangle$ is zero},
see also \cite{kan99,yev01,yan01,rei01c}.

We emphasize again that the conclusion $\langle\dot x\rangle =0$ only 
holds true
if either {\em both}, the potential and the driving are symmetric or if 
{\em both} of them are supersymmetric.
In any other case, $\langle \dot x\rangle \not = 0$ is expected
generically.
Especially, a symmetric but not
supersymmetric potential in combination with a supersymmetric but not
symmetric driving generically implies
$\langle\dot x\rangle\not =0$ 
(see \sects \ref{sec3.4.3} and \ref{sec6.3} for 
more details and examples).

\subsection{Examples}\label{sec3.4.3}
Next we turn to the discussion of examples.
Our first observation is the following
completely general implication of the
supersymmetry condition (\ref{ss5}):
For any minimum of $V(x,f(t))$, say at $x=\xmin$, there exists
a corresponding maximum at $x=\xmin+L/2$ and vice 
versa\footnote{Since this property holds separately
for any given $f(t)$-value, $\xmin$ and $\xmax$ may in
general still depend on $t$.}.
For the rest, the condition (\ref{ss5}) 
is still satisfied by a very large class of potentials and their 
exhaustive characterization on an intuitive level seems
rather difficult.
Here, we restrict ourselves to two sufficient (but not necessary)
simple criteria, which are still very general, namely:

1. The potential $V(x,f(t))$ is of the general form
\begin{eqnarray}
& & V(x,f(t))  = \sum_{n=1,3,5,...}
\alpha_n(f(t))\,  \cos \left(\frac{2\pi n x}{L} + \psi_n(f(t))\right)
\nonumber\\
& & \mbox{and $f(t)$ time-inversion invariant,}
\label{ss7}
\end{eqnarray}
where time-inversion invariance of $f(t)$ means, 
in the same sense as in (\ref{ss6}),
that $f(-t)\entspr f(t)$.
A typical example of this type (\ref{ss7}) of supersymmetric potential
$V(x,f(t))$ is depicted in \fig \ref{figss1}.
Note that in general not only
the shape of $V(x,f(t))$ but also the location of the 
extrema may still be different for any $f(t)$-value.
One readily sees that (\ref{ss7}) indeed implies (\ref{ss5}).
For fluctuating potential ratchets, i.e. $V(x,f(t))=V(x)[1+f(t)]$,
and especially for static potentials $V(x)$, also the inverse can be
shown, that is, (\ref{ss7}) is an exhaustive characterization of 
supersymmetric potentials in these special cases, but not in general.

\figsseins

2. A second class of supersymmetric potentials $V(x,f(t))$ 
is obtained by means of the representation
\begin{eqnarray}
V(x,f(t)) & = & V_+(x,f(t)) + V_-(x,f(t)) \label{ss8}\\
V_\pm(x,f(t)) & := & \frac{V(x,f(t)) \pm V(x,-f(t))}{2} \ , \label{ss9}
\end{eqnarray}
i.e., the potential is decomposed into symmetric and antisymmetric
contributions with respect to $f(t)$, 
\begin{equation}
V_\pm(x,-f(t))=\pm V_\pm(x,f(t)) \ . 
\label{ss9'}
\end{equation}
Then, the following conditions are sufficient for the potential
$V(x,f(t))$ to be supersymmetric:
\begin{eqnarray}
\!\!\!\!\!\!\!\!
& & V_+(x,f(t))  = \sum_{n=1,3,5,...}
\alpha_n(f(t))\, \cos \left(\frac{2\pi n x}{L} + \psi_n(f(t))\right) 
\nonumber
\\
\!\!\!\!\!\!\!\!
& & \mbox{and  }
V_-(x+L/2,f(t)) =  V_-(x,f(t))
\nonumber
\\
\!\!\!\!\!\!\!\!
& & \mbox{and $f(t)$ supersymmetric} \ .
\label{ss11'}
\end{eqnarray}
One readily verifies that these conditions (\ref{ss11'})
in combination with (\ref{ss8}), (\ref{ss9'}) indeed imply (\ref{ss5}),
i.e. $V(x,f(t))$ is supersymmetric.
A simple example is a supersymmetric $f(t)$ and
\begin{equation}
V(x,f(t)) = V_1(x)+V_2(x)\, f(t) \ , 
\label{ss12}
\end{equation}
where $V_1(x)$ is a static supersymmetric potential (cf. \eq (\ref{ss7})
and \fig \ref{figss1}) and where $V_2(x)$ is an {\em arbitrary}
$L/2$-periodic function.
In other words, in (\ref{ss8}) the potential 
$V_+(x,f(t))$ is independent of $f(t)$ and $V_-(x,f(t))$ is 
linear in $f(t)$.

\figsszwei

Next we come to the supersymmetry conditions\footnote{They can of
course be immediately transcribed into corresponding supersymmetry conditions
for $f(t)$ as well.} 
for the driving $y(t)$.
If $y(t)$ is {\em periodic} then the condition (\ref{ss6}) of
supersymmetry is equivalent to a Fourier representation 
of the general form
\begin{equation}
y(t) = \sum_{n=1}^\infty \gamma_n \sin(2\pi nt/\ttt) 
\ .
\label{ss13}
\end{equation}
A typical example of such a supersymmetric $y(t)$
is depicted in \fig \ref{figss2}.
For a {\em stochastic} $y(t)$ we can rewrite (\ref{ss6}) as
\begin{equation}
\langle y(t_1)\, y(t_2)\cdots  y(t_{n})\rangle = 
(-1)^n\, \langle y(-t_1)\, y(-t_2)\cdots  y(-t_{n})\rangle
\label{ss14}
\end{equation}
for all integers $n\geq 1$ and all times $t_1,\, t_2, ... ,\, t_n$
(see also the discussion below \eq (\ref{ss3})).
Note that out of the three possible symmetry properties of $y(t)$,
namely
(ordinary) symmetry, supersymmetry, and time-inversion invariance,
two always imply the third.
All three invariance properties are indeed satisfied for many particularly simple
examples $y(t)$ which we will treat in more detail below, for instance symmetric 
dichotomous noise and Ornstein-Uhlenbeck noise (see \sect \ref{sec3.1.1}),
as well as symmetric Poissonian shot noise (see \sect \ref{sec6.1}).
Note also that arbitrary linear combinations of supersymmetric
drivings are again supersymmetric.

A few specific examples which, {\em prima facie} quite astonishingly, produce
zero current due to supersymmetry reasons are worth mentioning:
The first set of examples are tilting ratchets ($f(t)\equiv 0$) with a
supersymmetric potential like in \fig \ref{figss1} and a periodic driving
$y(t)$ like in (\ref{ss13}), see also \fig \ref{figss2},
or with a symmetric dichotomous noise $y(t)$, an Ornstein-Uhlenbeck noise $y(t)$,
or a symmetric Poissonian shot noise $y(t)$.
On the other hand, a symmetric, but not supersymmetric potential $V(x)$
(e.g. (\ref{ss1}) with $a_1\not = 0$ and $a_2\not = 0$) in combination
with a supersymmetric but not symmetric driving $y(t)$
(e.g. (\ref{ss13}) with $\gamma_1\not = 0$ and $\gamma_2\not = 0$) {\em does}
generically produce a finite current $\langle\dot x\rangle$, 
see \sect \ref{sec6.3}.

A summary of the symmetry considerations
for tilting ratchets with periodic drivings
(i.e. rocking ratchets and asymmetrically tilting ratchets)
is depicted in \fig \ref{figss3}.
In order to bring out the essential features as clearly as
possible, we have chosen in this figure
stylized, non-smooth potentials $V(x)$
and drivings $y(t)$ and we have restricted ourselves to
time-periodic $y(t)$.

\figssdrei

In the case of a pulsating ratchet ($y(t)\equiv 0$), 
a symmetric dichotomous noise $f(t)$, an Ornstein-Uhlenbeck noise $f(t)$,
a symmetric Poissonian shot noise $f(t)$, or a periodic $f(t)$ of the
form (\ref{ss13}) yields $\langle\dot x\rangle = 0$ if one of
the following two conditions is met: (i)
The potential $V(x,f(t))$ is for any given $f(t)$-value of the form (\ref{ss5}),
see also \fig \ref{figss1}. 
We recall that not only
the shape of $V(x,f(t))$ but also the location of the extrema may be different
for any $f(t)$-value, i.e. both fluctuating potential ratchet and
(improper) traveling potential ratchets are covered. (ii) The potential
$V(x,f(t))$ respects supersymmetry when $f(t)=0$ and is augmented 
for $f(t)\not\equiv 0$ by
a fluctuating potential term $V_2(x)\, f(t)$ with an arbitrary
$L/2$-periodic function $V_2(x)$, see (\ref{ss12}).

\subsection{Discussion}\label{sec3.4.4}
As long as $V(x,f(t))$ and $y(t)$ are supersymmetric, the
property $\langle\dot x\rangle =0$ is robust with respect to
any change of the friction $\eta$, temperature $T$, amplitude and 
characteristic time scale of the drivings $f(t)$ and $y(t)$ etc. 
Much in contrast to ordinary current inversions,
we thus find $\langle\dot x\rangle =0$ 
{\em without fine-tuning any of these model parameters}.
The same is of course true for symmetric
instead of supersymmetric $V(x,f(t))$ and $y(t)$.
Note that this conclusion 
is no contradiction to Curie's principle since a {\em generic}
variation within the entire class of admitted ratchet models
also involves a change of $V(x,f(t))$ and $y(t)$ such that these 
symmetries are broken, see also the concluding discussion in \ch \ref{cha9}.

In the above respect, but also upon comparison of (\ref{ss1})-(\ref{ss3})
with (\ref{ss7}), (\ref{ss13}), (\ref{ss14}), the formal
structure and the consequences of symmetry and supersymmetry are
remarkably similar.
There is, however, also one fundamental difference which appears
if an additional inertia term $m\ddot x(t)$ is included
on the left hand side of the general ratchet dynamics (\ref{4a}):
{\em While symmetry implies $\langle\dot x\rangle =0$ even in the
presence of inertia effects, the same
conclusion no longer applies in the case of supersymmetry}.
For instance, a rocking ratchet with a cosine potential
$V(x)$ and a driving $y(t)$ like in \fig \ref{figss2}
implies $\langle\dot x \rangle = 0 $ in the overdamped limit
\cite{won79,bre82,won84}
but generically $\langle\dot x \rangle \not= 0 $ if inertia is
included \cite{bre84}.
In the opposite limit of a deterministic Hamiltonian rocking
ratchet dynamics (finite inertia, vanishing dissipation and 
thermal noise) a condition \cite{fla00} reminiscent of supersymmetry
will be discussed in \sect \ref{sec6.2.3.2}.
In the intermediate regime of finite inertia and dissipation,
no comparable symmetry concept is known. 
Since the current changes always continuously upon variation
of any model parameter, it follows that
for any sufficiently small deviations from a perfectly 
supersymmetric situation, e.g.
in the presence of a very small ineria term, the current 
$\langle\dot x\rangle$ will still be arbitrarily small \cite{yev01}.
In the following we focus again on the overdamped limit.

\subsection{Generalizations}\label{sec3.4.5}
We close with a brief look at the ratchet classes with both
$f(t)\equiv 0$ and $y(t)\equiv 0$ in (\ref{4a}) but instead with
a varying temperature $T$ in (\ref{4b}):
In the case of {\em Seebeck ratchets}, characterized by a space dependent,
$L$-periodic temperature $T(x)$, we speak of a (spatially) symmetric system
if both, $V(x)$ and $T(x)$ satisfy the symmetry condition (\ref{s1}) with the
{\em same} $\Delta x$, which may be transformed to zero as usual,
i.e.
\begin{equation}
V(-x)= V(x) \ \ \mbox{and}\ \ T(-x)=T(x)
\ \ \ \ \mbox{[symmetry].}
\label{ss15a}
\end{equation}
Similarly, supersymmetry is defined by the following condition
for the potential together with a modified such condition for the
temperature:
\begin{equation}
-V(x)= V(x+L/2) \ \ \mbox{and}\ \ T(-x)=T(x+L/2)
\ \ \ \ \mbox{[supersymmetry].}\label{ss15b}
\end{equation}
Along the same line of reasoning as in \sect \ref{sec3.1.2}, i.e.
by considering the mirror image $-x(t)$ of $x(t)$,
one readily finds that the average current $\langle\dot x\rangle$ indeed
vanishes if the symmetry conditions (\ref{ss15a})
are satisfied.
On the other hand, by considering $z(t):=x(-t)+L/2$, one verifies that 
$\langle\dot x\rangle = 0$ if supersymmetry (\ref{ss15b}) is respected.

Finally, in the case of {\em temperature ratchets}, 
characterized by a time-dependent temperature $T(t)$, a zero current
$\langle\dot x\rangle = 0$ is recovered provided that either
\begin{equation}
V(-x) = V(x)
\ \ \ \ \mbox{[symmetry]} \ ,
\label{ss15c}
\end{equation}
independently of the properties of $T(t)$, or that
\begin{eqnarray}
& & -V(x) = V(x+L/2)
\nonumber\\
& & \mbox{and $T(t)$ time-inversion invariant}
\ \ \ \ \mbox{[supersymmetry]} \ .
\label{ss15d}
\end{eqnarray}
Comparison with (\ref{s1}) and (\ref{ss7}) confirms
once more the similarity between
pulsating ratchets and temperature ratchets 
(see also \sect \ref{sec4.8.3}).

Further generalization to higher dimensional systems are
also possible but not further pursued here, see also
\sect \ref{sec7.2.3}.

\section{Tailoring current inversions}\label{sec3.5}
The argument which we have invoked at the end of \sect
\ref{sec2.5} can be considerably generalized as follows:
We consider {\em any} ratchet model of the general form 
(\ref{4a})-(\ref{4e}),
usually (not necessarily\footnote{Note that current inversions
upon variation of any model parameter can obviously be enforced by
applying an appropriately chosen external force $F$ 
\cite{lin97,ast97,nik98,lin99}.}) with $F=0$ in (\ref{4a}), and possibly
also with an $x$- and/or $t$-dependent temperature $T$ in (\ref{4b}).
Next we focus on an arbitrary parameter $\mu$ of the model
and we prescribe an arbitrary reference value $\mu_0$. 
Under the only assumption that
two potentials $V_i(x,f(t))$, $i=0,1$, with opposite currents 
$\langle\dot x\rangle$ at $\mu=\mu_0$ exist, we can then construct
a third {\em potential}, say $V_{\lambda_0}(x,f(t))$, with the property 
that {\em the current
$\langle \dot x\rangle$ as a function of the parameter
$\mu$ exhibits a current inversion at the prescribed reference
value $\mu_0$}.

The proof of this proposition is almost trivial.
Namely, we define a set of potentials
\begin{equation}
V_\lambda(x,f(t)) := \lambda\, V_1(x,f(t)) + (1-\lambda)\, V_0(x,f(t))\ ,
\label{2.3.101}
\end{equation}
parametrically dependent on $\lambda\in [0,1]$.
In other words, the potentials $V_\lambda(x,f(t))$ continuously
interpolate between the above defined 
two potentials $V_i(x,f(t))$, $i=0,1$, with opposite current 
directions at $\mu=\mu_0$.
Under the tacit assumption that the current 
$\langle\dot x\rangle$ changes continuously upon variation 
of $\lambda$, it follows that it vanishes at a certain 
intermediate potential $V_{\lambda_0}(x,f(t))$.
We remark that this assumption is very
weak: For instance, one can show that a non-vanishing
thermal noise $\xi(t)$ in (\ref{4a}) is sufficient, 
but by no means necessary.
Since the sign of $\langle\dot x\rangle$ is robust against
small changes of $V_i(x,f(t))$, it can furthermore be taken 
for granted that $V_{\lambda_0}(x,f(t))$ is a generic potential in the
sense that the dynamics (\ref{4a}) is neither symmetric 
nor supersymmetric, nor exhibits any other ``accidental'' symmetry.
In other words, we are dealing with the generic case that, upon
variation of the parameter $\mu$, the current 
$\langle\dot x\rangle$ exhibits an isolated zero,
i.e. a genuine current inversion, at $\mu=\mu_0$.

If the condition that two potentials $V_i(x,f(t))$ with opposite current 
directions at $\mu=\mu_0$ exist is not fulfilled, then also a current 
inversion at $\mu_0$ is obviously not possible, i.e. 
this condition is both necessary and sufficient.
For instance, if the driving $y(t)$ is symmetric and the 
temperature $T$ independent of $x$ then we know that $V(-x,f(t))$ 
yields a current opposite to that associated with $V(x,f(t))$. 
Hence we can choose 
as $V_0(x,f(t))$ any potential with $\langle\dot x\rangle \not =0$ and
as $V_1(x,f(t))$ a slightly deformed 
modification of $V_0(-x,f(t))$ to conclude that a current inversion 
exists always.

We may also consider some characteristic property of
the {\em driving} $y(t)$ as variable and instead 
leave all the other ingredients (especially the potential) of the 
ratchet dynamics
(\ref{4a}) fixed.
If the existence of two special drivings $y_i(t)$, $i=0,1$, with opposite 
currents at $\mu=\mu_0$
is known, then we can prove along the same line of reasoning as
above that there exists at least one $\lambda\in(0,1)$,
say $\lambda_0$, such that
\begin{equation}
y_\lambda (t):=\lambda\, y_1(t)+(1-\lambda)\, y_0(t)
\label{ta1}
\end{equation}
produces a current 
inversion at the arbitrarily prescribed reference parameter
value $\mu_0$. For instance, if $V(x,f(t))$ is symmetric
(and the temperature $T$ constant) then we
know that an asymmetric $y(t)$ generically produces a 
current $\langle\dot x\rangle\not=0$ and
$-y(t)$ a current in the opposite direction. Hence,
an appropriate asymmetric driving
$y(t)$ can be tailored which separates particles on opposite
sides of an arbitrarily prescribed parameter value $\mu_0$.
Since $y(t)$ is typically generated by an externally applied field,
such a separation scheme may be of considerable experimental interest.

At this point it is worth recalling that 
{\em once a current inversion upon variation
of one model parameter has been established, the existence
of an inversion upon variation of any other parameter follows}
along the same line of reasoning as in \sect \ref{sec2.5}.

Current inversions upon changing certain parameters 
of the system have been studied for the first time
in the context of photovoltaic effects in non-centrosymmetric 
materials \cite{koc75,asn79}.
Early observations in simple theoretical models as we study
them here are due to \cite{doe94,mil94,ajd94a,bar94}. 
Since then the search
and control of current inversions
has been attracting much attention with
respect to the possibility of new particle separation technologies
based on the ratchet effect.
Moreover,
multiple current inversions have been exemplified e.g. in 
\cite{jun96,bie96c,der96z,ber97,bar97,sch98,aba98,lin99,mat00,mat01,kos01}.
In the latter case, 
particles with parameter values within a characteristic ``window''
may be separated from all the others.
The first systematic investigation of such multiple inversions 
from \cite{kos01} suggest that it may always be possible to
tailor an arbitrary number of current inversions at prescribed
parameter values. However, a corresponding generalization of our
rigorous proof has not yet been established.

Our method of tailoring current inversions implies that, in
general, the direction of the particle current,
and even more so its quantitative magnitude,
depends in a very complicated way on many details of the ratchet 
potential $V(x,f(t))$ and/or on the driving $y(t)$.
In this respect,
the leading order small-$\ttt$ behavior of the 
temperature ratchet in (\ref{2.3.2}) is still a rather simple example.
Therefore, any heuristic ``explanation'' or
simple ``rule'' regarding current directions should make
us suspicious unless it is accompanied by a convincing (and usually
rather severe) restriction on the admitted potentials $V(x,f(t))$
and/or the driving $y(t)$.
Otherwise, one can typically construct
even quite innocent looking counter-examples of such a ``rule''.

The above described procedure is a very simple and universal tool for the
construction of current inversions {\em per se}.
However, little control over the more detailed dependence of the current 
as a function of the considered parameter $\mu$
is possible in this way.
For instance, the maximal magnitudes of the currents may be very different in
the positive and negative directions.
Likewise, we can hardly avoid ending up with a quite complicated 
looking potential $V_{\lambda_0}(x)$ and/or driving $y_{\lambda_0}(t)$.
For both purposes, ``symmetrically'' shaped current inversions as well as
``simple'' potentials and/or drivings which do the job, 
more detailed analytical predictions are invaluable.
For instance, the results depicted in \fig \ref{fig8} have not been 
obtained directly by the above construction scheme. 
Rather, the approximation (\ref{2.3.2})  has been exploited
in order to obtain such an ``innocent'' looking ratchet 
potential with a current inversion.

\section{Linear response and high temperature limit}\label{sec3.1.4}
For vanishing $f(t)$, $y(t)$, and $F$ we recover a Smoluchowski-Feynman
ratchet in (\ref{4a}), yielding zero current (\ref{4c1})
in the long time limit (steady state), see \sects \ref{sec2.1}-\ref{sec2.1.4}.
In the case of a tilting ratchet scheme, an interesting question
regards the linear response behavior in the presence of a weak
but finite driving $y(t)$ (while $f(t)$ and $F$ are still zero), i.e. the
behavior of the averaged long-time current (\ref{4c1}) in linear 
order\footnote{Formally, this amounts to replacing $y(t)$ by $\epsilon\, y(t)$
and then performing a series expansion in $\epsilon$ while keeping $y(t)$ 
fixed.}$y(t)$. In the case of a symmetric driving $y(t)$ 
(cf. (\ref{s2}) and (\ref{s3}))
no such linear contribution
can arise, since the averaged long-time current is clearly
invariant under inversion of the driving amplitude, $y(t)\mapsto -y(t)$.
In the general (asymmetric) case, we can expand $y(t)$ into a Fourier series,
which does not exhibit a constant term since $y(t)$ is unbiased, 
see (\ref{4g}) or (\ref{4h}).
In linear order $y(t)$, the net current follows simply by summing up all
the contributions of the single Fourier modes.
Since each Fourier mode corresponds to a symmetric driving, the resulting
net current is zero.
A similar line of reasoning applies for pulsating ratchets
with a weak driving $f(t)$ (and $y(t)=F=0$).
An exception is a genuine traveling potential scheme 
with a systematic long time drift $u:=\lim_{t\to\infty}f(t)/t$.
In this case, the above Fourier expansion of $f(t)$ 
cannot be applied any more
and indeed a finite linear order $f(t)$ contribution
to the current is observed generically,
see \sect \ref{sec4.3.1}.
In other words, {\em for tilting ratchets, fluctuating force ratchets,
and improper traveling potential ratchets no directed current occurs 
within the linear response regime} (linear order $y(t)$ and $f(t)$, 
respectively).
The same conclusions obviously extends to systems with
simultaneously small
but finite $y(t)$ and $f(t)$ (but still $F=0$).
Due to their equivalence with fluctuating potential ratchets,
{\em the conclusion also carries over to temperature ratchets}
(cf. \sect \ref{sec4.8.3}) with a small perturbation of the temperature
$T(t)$ about the (finite) average value $\overline{T}$.

In the above line of reasoning we have tacitly assumed analyticity
of the current with respect to the amplitude of the perturbations
$y(t)$ and $f(t)$ and that a Fourier 
expansion\footnote{Or any other series expansion in 
terms of symmetric basis functions.} of these perturbations
is possible. (Especially, for a stochastic process,
the word ``weak perturbation'' refers 
to its intensity, but not necessarily to its instantaneous value 
at any given time $t$.)
Though this may be difficult to rigorously justify in general,
a more careful analysis of each specific case 
(known to the present author)
shows that the conclusion of vanishing linear response 
remains indeed correct.

Another limit which admits a completely general conclusion is that
of asymptotically large temperatures $T$ in (\ref{4a}), (\ref{4b}): 
Again one finds that
{\em the current $\langle \dot x \rangle$ always
approaches zero in the limit $T\to\infty$}.
While this result is physically rather suggestive
(the effect of the potential $V(x,f(t))$ is completely 
overruled by the noise $\xi(t)$)
the technical details of the mathematical proof go 
beyond the scope of this review.
In those numerous cases for which the current also vanishes
for $T\to 0$, a bell-shaped $\langle\dot x\rangle$-versus-$T$
curve is thus recovered.

We finally remark that in the limit of asymptotically strong 
drivings $y(t)$ and/or $f(t)$, 
no generally valid predictions are possible.

\section{Activated barrier crossing limit}\label{sec3.6}
For many of the above defined classes of ratchets (\ref{4a})
it may turn
out that in the absence of the thermal Gaussian noise ($T=0$),
the particle $x(t)$ is confined to a restricted part of one spatial period
for all times.
In the presence of a small amount of thermal noise, the particle will be
able to cross the previously forbidden regions by thermal activation.
Yet, such events will be rare and after each thermally activated transition 
from one spatial period into an adjacent one, the particle will again remain there
for a long time. Since the duration of the actual transition events is negligible 
in comparison with the time the particle spends in a {\em quasi steady state}
(metastable state)
between the transitions, it follows that the probability for a transition
per time unit can be described in very good approximation by a constant 
rate\footnote{If $f(t)$ or $y(t)$ is periodic in time, then we will have a
quasi periodic behavior between transition events and the transition probability
is only given by a constant after ``coarse graining'' over one time-period.}.

Denoting the rate for a transition to the right by $k_+$ and to the left by $k_-$,
the average particle current readily follows\label{fot51}\footnote{We tacitly
focus here on the simplest and most common case with just one metastable state per period $L$.
For more general cases, the current and the effective diffusion coefficient can still 
be expressed in closed analytical form in terms of all the involved 
transition rates, but the formulas become more complicated,
see \cite{der82,der83,koz99} and further references therein.
For special cases, such formulas have been repeatedly re-derived, 
e.g. in \cite{keh97,der98z}.} as
\begin{equation}
\langle\dot x\rangle = L\, [k_+ - k_- ]\ .
\label{4i}
\end{equation}
For a special case, this result has been derived explicitly already
in \eq (\ref{2.23u}). Exploiting the rate description for the transitions
once more, also the effective diffusion coefficient (\ref{4c2}) 
can be readily evaluated \cite{kam92} with the result
\begin{equation}
D_{\rm{eff}} = \frac{L^2}{2}\, [k_+ + k_- ]\ .
\label{4j}
\end{equation}
The very same conclusions (\ref{4i}), (\ref{4j}) 
hold true if transitions are not excluded
but still very rare at $T=0$.
This may be the case for instance in a tilting ratchet ($f(t)\equiv 0$)
when $y(t)$ is a Gaussian random process with a small intensity
$\int dt\, \langle y(t)\, y(0)\rangle$.
On the other hand, genuine traveling potential ratchets will turn out to
support an appreciable particle current typically even for $T=0$ and are henceforth 
excluded.

The evaluation of the current  (\ref{4i}) and the diffusion coefficient
(\ref{4j}) has thus
been reduced to the determination of certain rates $k$ across rarely 
visited regions between some type of effective local potential wells or 
periodic attractors (if $f(t)$ or $y(t)$ is periodic in $t$).
In the case that both $f(t)$ and $y(t)$ are stochastic processes (possibly one
of them identically zero), this problem of thermally activated 
surmounting of a potential barrier with randomly fluctuating shape
has attracted considerable attention since the discovery of the so-called
{\em resonant activation} effect \cite{doe92}, see \cite{rei97ra} for a review.
On condition that a rate description of the barrier crossing problem is possible,
i.e. the transitions are rare events especially in comparison with the time 
scale of the barrier fluctuations \cite{rei97ra},
the fluctuating barrier crossing problem is thus
equivalent to determining the current and diffusion in a ratchet model.
A large body of analytical results on the former problem
\cite{han80ra,ste90,zur93,bie93,pec94,han94ra,rei94ra,rei95ra1,rei95ra2,mad95,bar95,rei96ra,iwa96,rei98ra,rei99ra,ank99}
are thus readily applicable for our present purposes.
Particularly closely related to the resonant activation
effect are the theoretical works
\cite{ast87,che87} on externally driven molecular pumps
(cf. \sect \ref{sec4.5.1}) and their experimental 
counterparts in \cite{xie94,xie97}.

If $f(t)$ and/or $y(t)$ are periodic in time with a {\em large} 
period $\ttt$ then a close connection to the phenomenon of 
{\em stochastic resonance}  \cite{gam98} 
can be established. If the time period $\ttt$ is {\em not very large}, 
then this problem
of thermally activated escape over an oscillating potential barrier
represents a formidable technical challenge 
\cite{jun93}.
The few so far available results on
weak \cite{dyk97,sme99}, 
slow (but beyond the adiabatic approximation) \cite{tal99},
fast \cite{gra84}, and general oscillations
\cite{leh00,leh00a}
can again be readily
adapted for our present purposes via (\ref{4i}), (\ref{4j}).

One basic conclusion of all the above mentioned analytical studies of the
thermally activated escape over fluctuating or oscillating barriers is that 
those rates become exponentially small with decreasing thermal
noise strength,
much like in the simple explicit example (\ref{2.23t}), (\ref{2.23s}).
On top of that, also the ratio $k_+/k_-$ generically becomes either exponentially
small or large.
In other words, practically only transitions in one direction occur
in the course of time and $x(t)$
basically simplifies to a unidirectional 
Poissonian hopping process \cite{kam92}.
This type of {\em unidirectionality is a distinct weak-noise feature}.
As soon as the thermal noise-strength increases, the
stochastic trajectory $x(t)$ always displays appreciable
displacement in {\em both} directions.

Though a substantial part of the present author's contributions to the
field under review is concerned with the evaluation of such 
thermally activated rates over fluctuating or oscillating 
barriers,
we desist from going into any further details of
these technically rather involved theories.

\chapter{Pulsating ratchets}\label{cha4}

In this chapter we focus on the
pulsating ratchet scheme, i.e., we
consider a stochastic dynamics of the form
\begin{equation}
\eta\, \dot x(t) = - V'(x(t),f(t)) + \xi (t) \ ,
\label{6j}
\end{equation}
complemented by the fluctuation-dissipation relation (\ref{4b})
for the thermal noise $\xi(t)$ and the periodicity condition 
(\ref{4e}) for the potential.
Further, $f(t)$ is assumed to be
an unbiased time-periodic function or stationary stochastic process. 
As compared to the general working model (\ref{4a}) we have
set the load force $F$ equal to zero on the right hand side
of (\ref{6j}) since this case is usually of foremost interest.

\section{Fast and slow pulsating limits}\label{sec4.1a}
We consider an arbitrary 
(unbiased, i.e. $\langle f(t)\rangle =\langle f(0)\rangle = 0$)
{\em stochastic} process $f(t)$ and assume that its stationary distribution
$\rho (f)$ (see (\ref{6k1}))
and thus especially its variance 
$\langle f^2(t)\rangle = \langle f^2(0)\rangle$ 
(see (\ref{6l'})) is always the same, while its correlation 
time\footnote{By means of a calculation similar to that in 
\sect \ref{sec2.1.2.2} of Appendix A one can show 
that the intensity $\int dt\,\langle f(t)f(s)\rangle$ and 
hence the correlation time (\ref{6n1}) is always non-negative.}
\begin{equation}
\tau := \frac{\int_{-\infty}^\infty dt\, \langle f(t) f(s)\rangle}
{2\langle f^2(t)\rangle} \ ,
\label{6n1}
\end{equation}
characterizing the decay 
of the correlation 
$\langle f(t)\, f(s)\rangle = \langle f(t-s)\, f(0)\rangle$, 
can be varied over the entire positive real axis.
More precisely, we assume that the process $f(t)$ can be written in the
form
\begin{equation}
f(t) = \hat f (t/\tau) \ ,
\label{6n2}
\end{equation}
where $\hat f( h )$ is a suitably defined, fixed reference process with
dimensionless time-argument $h$, cf. (\ref{2.3.1'}). 
The statistical properties of the process
$f(t)$ then depend solely on the parameter $\tau$,
while $\rho(f)$ is $\tau$-independent.
One readily sees that the examples of dichotomous and 
Ornstein-Uhlenbeck noise from (\ref{6k2})-(\ref{6m}) are of this
type.
This so-called {\em constant variance scaling} assumption \cite{bar96}
is ``natural'' in the
same sense as it is natural to keep in a time-periodic perturbation $f(t)$
the amplitude fixed while the frequency is varied, cf. \sect \ref{sec2.4.1}.

For very small $\tau$, the noise $f(t)$ changes very quickly, while its strength
$\int dt\, \langle f(t)\, f(s)\rangle$ tends to zero (see also previous footnote 1).
Thus, for $\tau\to 0$  (fast pulsating limit) there will be no effect of
$f(t)$ in (\ref{6j}), i.e. we recover a Smoluchowski-Feynman ratchet
with $\langle\dot x\rangle =0$.
Similarly, for $\tau\to\infty$ 
the noise becomes very slow, $f(t)\simeq f = const.$
(adiabatic approximation, see also \sect \ref{sec2.4.1}).
Since for
any fixed value of $f$ we have again a Smoluchowski-Feynman ratchet
in (\ref{6j}), 
the result $\langle\dot x\rangle =0$ subsists after 
averaging over all those fixed
$f$-values according to $\rho(f)$. The very same conclusion $\langle\dot x\rangle =0$
holds for periodic functions $f(t)$ in the limits of asymptotically
long and short periods $\ttt$
and can be furthermore extended also to generic traveling potential ratchets
(cf. \sect \ref{sec4.3.1}).
In other words, we arrive at the completely general result that 
{\em for any type of pulsating ratchet (\ref{6j}), 
the current $\langle\dot x\rangle$ 
disappears both in the fast and slow pulsating limit}.

We exemplify the leading order correction to this asymptotic result
$\langle\dot x\rangle =0$ for the case that $f(t)$ is an arbitrary 
stationary stochastic process with very small correlation time $\tau$.
Generalizing an argument form
\cite{rei95ra1,rei97ra} we can, within a leading order approximation,
substitute in (\ref{6j})
the random process $V(x,f(t))$ for any fixed $x$ by an uncorrelated Gaussian 
process with the same mean value 
\begin{eqnarray}
V_0(x) & := & \langle V(x,f(t))\rangle = 
\int_{-\infty}^\infty df\, \rho(f)\, V(x,f)
\label{6p}
\end{eqnarray}
and the same intensity $\int dt\, C(x,t)$, where the correlation $C(x,t)$ 
is defined as
\begin{eqnarray}
C(x,t) & := & \langle V'(x,f(t))V'(x,f(0))\rangle - [V'_0(x)]^2\ . 
\label{6q}
\end{eqnarray}
An alternative representation of $C(x,t)$ analogous to the 
second equality in (\ref{6p}) is given by
\begin{eqnarray}
C(x,t) & = &\int_{-\infty}^\infty df_1\, df_2\, 
\rho(f_1,f_2,t)\, V'(x,f_1)\,V'(x,f_2) \ ,
\label{6q1a}
\end{eqnarray}
where $\rho(f_1,f_2,t)$ is the joint two-time distribution of the stationary process
$f(t)$, i.e.
\begin{equation}
\rho(f_1,f_2,t) := \langle \delta(f(t)-f_1)\, \delta(f(0)-f_2)\rangle \ . 
\label{6q2}
\end{equation}
Referring to \cite{rei95ra1,rei97ra} for the calculational
details, one obtains in this way the result
\begin{equation}
\langle\dot x\rangle = \frac{L}{\eta^2k_B T}\, 
\frac{\int_0^L dx\, V_0'(x)\int_{-\infty}^\infty
 dt\, C(x,t)}{\int_0^L dx\, e^{V_0(x)/k_B T} 
\int_0^L dx\, e^{-V_0(x)/k_B T} } \ .
\label{6o}
\end{equation}
Note that the effective potential $V_0(x)$ is again $L$-periodic.
Under the conditions
that $T>0$ and that (\ref{6q}) decays exponentially in time,
the result (\ref{6o}) gives the leading order contribution for small correlation
times $\tau$ of $f(t)$. 
Using (\ref{6n2}) we can infer that
\begin{equation}
\int_{-\infty}^\infty dt\, C(x,t) = \tau\int_{-\infty}^\infty
\, dh\,  \hat C(x,h) \ ,
\label{60'}
\end{equation}
where $\hat C(x,t)$ is defined like in (\ref{6q}) but with
$\hat f(h)$ instead of $f(t)$ and thus the integral on the
right hand side of (\ref{60'}) is $\tau$-independent.
In other words, the asymptotic current (\ref{6o}) grows linearly with
the correlation time $\tau$.

In the special case that $V(x,f(t))$ is of the 
fluctuating potential ratchet
form $V(x)\, [1+f(t)]$, the small-$\tau$ result from \cite{mie95b} 
is readily recovered from (\ref{6o}), namely
\begin{equation}
\langle\dot x\rangle = \frac{ 2\, L\, \tau\, \langle f^2(t)\rangle\,
\int_0^L dx\, [V'(x)]^3}{\eta^2 k_B T\, \int_0^L dx\, e^{V(x)/k_B T} 
\int_0^L dx\, e^{-V(x)/k_B T} } \ ,
\label{6q1}
\end{equation}
see also \cite{bao00a}.

A similar leading order correction of the adiabatic approximation
$\langle\dot x\rangle=0$ is possible for large $\tau$ \cite{mie95b}
but leads to a complicated formal
expression which still depends on much more details
of the noise than in (\ref{6o}).

For {\em periodic} 
driving $f(t)$, expansions along the lines of Appendix C can be
undertaken, see \sect \ref{sec4.8.3}.
Technically similar calculations can also be found in
\cite{pla98,mil99} for fast and \cite{tal99} for slow periodic driving.
We will not pursue this task here any 
further\footnote{In the special case of slow on-off and slow,
traveling potential ratchets, the qualitative behavior will 
become intuitively clear later on.}$^,$\footnote{If $V(x,f(t))$ is 
of the fluctuating potential ratchet form $V(x)\, [1+f(t)]$ it is possible
to show that $\langle\dot x\rangle$ vanishes faster than proportional to
$\tau^{-1}$ in the slow driving limit $\tau\to\infty$ for both,
periodic and stochastic $f(t)$. In the latter case, this conclusion
is also contained implicitly in the calculations of \cite{mie95b}.}
since the main conclusion follows from the already complicated enough
result (\ref{6o}), namely that a finite current 
$\langle\dot x\rangle$ is generically
expected for driving signals $f(t)$ with a finite characteristic time scale
and that its sign and magnitude depends very sensitively on the details of
the potential $V(x)$ and the driving $f(t)$.

\section{On-off ratchets}\label{sec4.1}
The on-off ratchet scheme has been introduced 
in a specific theoretical context in 1987 by Bug and Berne \cite{bug87}
and has been independently re-invented as a general theoretical concept
in 1992 by Ajdari and Prost \cite{ajd92}.
In its latter form it is of archetypal simplicity and the predicted occurrence of the 
ratchet effect has been verified by several experimental implementations.
The model is a special case of the overdamped one-dimensional
Brownian motion dynamics (\ref{6j}), namely
\begin{equation}
\eta\,\dot x(t) = -V'(x(t))\,[1+f(t)] + \xi(t)\ ,
\label{4.1}
\end{equation}
where $V(x)$ is spatially periodic,
asymmetric ``ratchet''-potential.
The function $f(t)$ is restricted to the two values $\pm 1$, so the 
potential in (\ref{4.1}) is either ``on'' or ``off''. 

In the simplest case the potential $V(x)$ has one maximum and
minimum per spatial period $L$ (for examples see \figs 
\ref{fig2} and \ref{fig4.1}), the potential difference between 
maxima and minima is much larger than the thermal energy $k_B T$,
and $f(t)$ is a time-periodic function with {\em long} sojourn-times
in the $+1$-state (potential ``on'').
Under these premises 
the analysis of the model proceeds in complete analogy to \fig \ref{fig6}.
Qualitatively, a net pumping of particles into the positive direction will
occur if the minima of $V(x)$ are closer to their neighboring maxima
to the right than to the left (``forward on-off ratchet''),
otherwise into the negative direction\footnote{A computer 
animation (Java applet) which graphically
visualizes the effect is available on the internet under
\cite{elmvideo}.}.
Quantitatively, the average current $\langle\dot x\rangle$
can be readily expressed in closed analytical form \cite{ajd92}
apart from an error function, which has to be evaluated numerically.
Similarly, one can readily
evaluate \cite{ajd92} the effective diffusion coefficient (\ref{4c2}).
Depending on their friction coefficient $\eta$, different species of particles
which are initially mixed (say $x(0)=0$ for all of them) will 
exhibit after a time $t$ different displacements
$\langle\dot x\rangle\, t$ and dispersions $\sqrt{2\deff t}$ and thus
can be separated for sufficiently large $t$, see also \eq (\ref{4.3}).

\begin{center} 
\figviereins
\end{center} 

For more general potentials $V(x)$ and drivings $f(t)$, things become more
complicated. As seen in \sect \ref{sec4.1a}, the current $\langle\dot x\rangle$
approaches zero both for very fast and slow switching of $f(t)$ between $\pm 1$. 
As long as the potential $V(x)$ is sufficiently similar to the
simple examples from \figs \ref{fig2} and \ref{fig4.1}, a single
maximum of $\langle\dot x\rangle$ develops at some intermediate
switching time. For more complicated, 
suitably chosen potentials $V(x)$, the existence of current inversions 
follows from \sect \ref{sec3.5} and has been exemplified in \cite{cha95},
see also \cite{che99,yan01}.
For a few additional quantitative analytical results
we also refer to the subsequent \sect \ref{sec4.2} on fluctuating
potential ratchets, which includes the on-off scenario as
a special case.

\subsection{Experimental realizations}\label{sec4.1.1}
The theoretically predicted pumping effect $\langle\dot x\rangle \not = 0$
has been demonstrated  experimentally by
Rousselet, Salome, Ajdari, and Prost \cite{rou94}
by means of colloidal polysterene latex spheres, suspended
in solution and exposed to a dielectric ``ratchet''-potential,
created by a series of interdigitated ``Christmas-tree'' 
electrodes which were deposited on a glass slide by photolithography
and which were turned on and off periodically.
With one adjustable parameter, accounting for the uncertainty
about the shape of the one-dimensional ``effective'' potential
$V(x)$, the agreement of the measured currents
$\langle\dot x\rangle$ with the theory from \cite{ajd92}
turned out to be quite good for all considered particle
diameters between $0.2\, \mu m$ and $1\, \mu m$.

A very similar experimental setup was used by Faucheux and 
Libchaber \cite{fau95a} 
(see also \cite{sch00d})
but they went one step further in that they studied 
solutions containing two different species of particles at a time,
and demonstrated
experimentally that they can be separated.
Again, the quantitative findings depend on the particle size not only
via the corresponding viscous friction coefficient $\eta$ but also via 
slightly different ``effective'' potentials $V(x)$.

A further experimental verification of the on-off ratchet scheme by 
Faucheux, Bourdieu, Kaplan, and Libchaber \cite{fau95b}
works with single $1.5\,\mu m$ diameter polysterene spheres, 
confined to an effective one-dimensional ``ratchet''-potential
by laser-optical trapping methods (optical tweezers).
The characteristics of such an optical ratchet potential
can be determined with satisfactory accuracy, leading to a quite
good quantitative agreement of the the observed ratchet effect 
with the simple theory from \cite{ajd92} without any adjustable parameters.

A first step towards a practical usable pumping and separation device
was achieved by
Gorre-Talini, Jeanjean, and Silberzan \cite{gor97} 
(see also \cite{gor98}).
Due to their well defined geometry, also in
this experiment latex spheres diluted in water
with diameters mostly between $0.5\, \mu m$ and $2.5\, \mu m$
were used, but in principle nothing speaks against
replacing them e.g. by micrometer sized biological objects
like DNA, viruses, or chromosomes.
The main ingredient for creating the (electrostatic)
``ratchet''-potential is an optical diffraction grating,
commercially available in a variety of periods and asymmetries.
This setup overcomes many of the practical drawbacks of its
above mentioned
predecessors, agrees well with the simple theory from \cite{ajd92}
without any fitting, and the predicted separating power may well
lead to the development of a serious alternative to existing
standard particle separation methods.

As regards transport and separation of DNA, all the 
so far discussed setups
are expected to cease working satisfactorily for DNA 
fragments below one kilobase. By means of
an interdigitated electrode 
array\label{fot2.8}\footnote{About $100$ metallic strips perpendicular to the
transport axis with alternating smaller and larger spacings are 
either alternatingly charged (``on'') or all uncharged (``off'').}
Bader and coworkers \cite{bad99,ham00}
successfully micromachined an on-off ratchet on a silicon-chip, capable 
to transport DNA molecules of $25-100$ bases in aqueous solution.
With some improvements, such a ``lab-on-a-chip'' device could
indeed provide a feasible alternative to the usual electrophoresis
methods for nucleic acid separation, one of the central preprocessing
tasks e.g. in the Human Genome Project \cite{bad99,ham00,row97}.

As pointed out in \cite{gor98}, not only the separation of large DNA 
molecules by present standard methods seems to have become one of the major 
bottlenecks in sequencing programs \cite{3inchaos}, but also
chromosomes, viruses, or cells exhibiting major biological
differences may only differ very little from the physicochemical viewpoint, thus
making their separation highly challenging \cite{4inchaos,ert98}.

\section{Fluctuating potential ratchets}\label{sec4.2}
The fluctuating potential ratchet model has been introduced practically
simultaneously\footnote{The closely related work by Peskin, Ermentrout,
and Oster \cite{pes94} came somewhat later and mainly focuses on
the special case of an on-off ratchet.}
in 1994 by 
Astumian and Bier \cite{ast94}
and independently by
Prost, Chauwin, Peliti, and Ajdari \cite{pro94}.
The model is given by the same type of overdamped dynamics as
the on-off ratchet in (\ref{4.1}) except that  the amplitude modulations
$f(t)$ are no longer restricted to the two values
$\pm 1$. In other words, the time-dependent
potential $V(x)\,[1+f(t)]$, to
which the Brownian particle $x(t)$ is exposed, has always the same shape but its
amplitude changes  in the course of time between two or more values.
Thus the on-off scheme is included as a special case throughout the
considerations of this section.

One central and completely general feature of the fluctuating potential
ratchet scheme readily follows from (\ref{4.1}).
Namely, within each spatial period there are $x$-values where $V'(x)$ is zero.
These points cannot be crossed by $x(t)$ in the overdamped
dynamics (\ref{4.1}) without the help of the fluctuations $\xi(t)$.
In other words,
{\em in an overdamped fluctuating potential ratchet, thermal noise is 
indispensable for a finite current $\langle\dot x\rangle$.}
A second, completely general conclusion has been pointed out already
in \sect \ref{sec3.1.3}, namely that {\em only asymmetric
potentials $V(x)$ admit a ratchet effect}.

For the latter reason, we will mainly concentrate 
on the simplest non-trivial case
of an asymmetric potential $V(x)$ in combination with a symmetric driving 
$f(t)$.
As usual, one option in (\ref{4.1}) is a {\em periodic} driving $f(t)$, 
see e.g. \cite{cha95,tar98,ajd00}.
For asymptotic results for fast periodic driving $f(t)$
we refer to \sect \ref{sec4.8.3}.
In the following we focus, in accordance with most of the existing literature,
on the two particularly simple examples of a stationary 
{\em random} processes $f(t)$
as introduced in \sect \ref{sec3.1.1},
see also the asymptotic result (\ref{6q1}) for general fast stochastically 
fluctuating potentials.

A well-known {\em experimental} phenomenon which 
can be interpreted by means of a fluctuating potential 
ratchet scheme is the photoalignement (absorbtion-induced optical
reorientation) of nematic liquid crystals \cite{jan94,pal98,kos00a,kre00},
exemplifying the general considerations in \cite{pro94}.
Another realization by means of a Josephson junction decive
similarly to that proposed in \cite{gol01} 
(cf. \sect \ref{sec6.2.3.1}) 
is presently planned by the same authors.
The large variety of potential applications of the fluctuating 
potential ratchet mechanism for small-scale pumping devices is 
the subject of \cite{ajd00}.

\subsection{Dichotomous potential fluctuations}\label{sec4.2.1}
The case that the amplitude modulations $f(t)$ in (\ref{4.1}) are given by
a symmetric dichotomous noise
(see (\ref{6k2}), (\ref{6l}))
has been considered in \cite{ast94,pro94,pes94}.
A numerical simulation of the time discretized stochastic dynamics
(\ref{4.1}) similarly as in \eq (\ref{2.7}) is straightforward.
Equivalent to this stochastic dynamics is the following master
equation for the joint probability densities $P_\pm(x,t)$ that
at time $t$ the particle resides at the position $x$
and the dichotomous process $f(t)$ is in the states $\pm \sigma$, respectively:
\begin{eqnarray}
\frac{\partial}{\partial t} P_+(x,t) & = & 
\frac{\partial}{\partial x}\left\{\frac{(1+\sigma )\, V'(x)}{\eta}\, P_+(x,t)\right\}
+ \frac{k_BT}{\eta} \frac{\partial^2}{\partial x^2} P_+(x,t)\nonumber \\
& & - \gamma P_+(x,t) + \gamma P_-(x,t) \label{4.4}\\
\frac{\partial}{\partial t} P_-(x,t) & = & 
\frac{\partial}{\partial x}\left\{\frac{(1-\sigma )\, V'(x)}{\eta}\, P_-(x,t)\right\}
+ \frac{k_BT}{\eta} \frac{\partial^2}{\partial x^2} P_-(x,t)\nonumber\\
& & - \gamma P_-(x,t) + \gamma P_+(x,t) \label{4.5}\ .
\end{eqnarray}
The derivation of the first two terms on the right hand side of
(\ref{4.4}) and (\ref{4.5}) is completely analogous to 
the derivation of the Fokker-Planck equation (\ref{2.10}):
The first terms (drift terms) account for the Liouville-type
evolution of the probability densities $P_\pm(x,t)$ under the action
of the respective force fields $-V'(x)[1\pm \sigma ]$.
The second terms (diffusion terms) describe the diffusive broadening of the
probability densities due to the thermal white noise $\xi(t)$ 
of strength $2\,\eta\, k_B T$ in (\ref{4.1}).
The last two terms in (\ref{4.4}) and (\ref{4.5}) are loss and gain
terms due to the transitions of $f(t)$ between the two
``states'' $\pm \sigma $ with transition probability (flip rate) $\gamma$.

For the marginal density $P(x,t):= P_+(x,t)+P_-(x,t)$ one recovers from
(\ref{4.4}), (\ref{4.5}) a master equation of the general form
(\ref{2.10a}), whence the particle current (\ref{2.10c}) follows.
Like in \sect \ref{sec2.1.4} one sees that it suffices to 
focus on spatial periodic ``reduced'' distributions 
$\hat P_\pm(x+L,t)=\hat P_\pm(x,t)$ in
order to calculate the average current $\langle\dot x\rangle$ and that in 
the long time limit $t\to\infty$ a unique steady state $\hat P_\pm^{st}(x)$ is
approached.
However, {\em explicit analytical expressions for the probability densities
and the particle current in the steady state can only be obtained
in special cases}.
A prominent such case is that of a ``saw-tooth'' ratchet-potential
as depicted in \fig \ref{fig4.1}.
Within each interval of constant slope $V'(x)$, the steady state solutions
of (\ref{4.4}), (\ref{4.5}) can be readily determined which then have
to be matched together.
For the straightforward but rather tedious technical details of such
a calculation we refer to \cite{ast94,kul98a}.
The resulting expression for the current $\langle\dot x\rangle$ is
awkward and not very illuminating.
Qualitatively, the
results are like for the on-off scheme (see also \fig \ref{fig6}):
If the local minima of the saw-tooth potential are closer to
their neighboring maxima to the right than to the left,
then the current is positive for all finite temperatures $T$ 
and flipping rates $\gamma$ of the dichotomous noise,
and (cf. \sect \ref{sec3.1.1}) the current approaches zero if 
$T$ or $\gamma$ tends either to zero or to infinity \cite{ast94,pro94}.

As long as the potential is sufficiently similar to a saw-tooth
potential such that one can identify essentially
one steep and one flat slope per period $L$, 
qualitatively unchanged results are obtained, and similarly for more 
general potential fluctuations $f(t)$. In this sense,
the {\em ``natural''
current direction in a pulsating ratchet is given by the sign of
the steep potential slope}.

For more general potentials $V(x)$, the same qualitative
asymptotic features are expected for large and small $T$ and $\gamma$. 
However, as shown in \sect \ref{sec3.5},
the current direction may change  as a function 
of any model parameter. So, general hand weaving predictions about 
the sign are impossible, not to mention quantitative estimates 
for $\langle\dot x\rangle$, as already the small-$\tau$ result (\ref{6q1})
demonstrates.

A case of particular conceptual interest is 
the singular perturbation expansion about the unperturbed situation 
with vanishing thermal noise, $T=0$, and vanishing particle current
$\langle\dot x\rangle=0$.
On condition that the local potential extrema subsist for all values of
$f(t)$, i.e. $|\sigma |<1$, the transitions between neighboring
minima of the potential will be very rare for small $T$. As a 
consequence, the current $\langle\dot x\rangle$ can be described 
along the activated barrier crossing approach from \sect \ref{sec3.6}.
Specifically, for our present situation of thermally
assisted transitions across barriers with dichotomous amplitude fluctuations, 
the results of the singular perturbation theory from \cite{ank99}
are immediately applicable.

While for the case of a sawtooth ratchet potential $V(x)$, which is
subjected to  symmetric dichotomous amplitude fluctuations $f(t)$ in 
(\ref{4.1}), current 
inversions can be ruled out \cite{ast94}, for more complicated fluctuations 
$f(t)$ this is no longer the case:
The occurrence of a current inversion 
in a sawtooth potential $V(x)$ has been
demonstrated for asymmetric (but still unbiased)
dichotomous noise \cite{cha95,che99,yan01}, for sums of dichotomous
processes \cite{mie95b}, and for a three-state noise 
\cite{bie96a}. Originally, the inversion effect has been
reported in those works \cite{cha95,mie95b,bie96a,che99,yan01}
for certain parameters characterizing the noise $f(t)$,
but according to \sect \ref{sec3.5} the effect immediately carries over
to {\em any} parameter of the model.

\subsection{Gaussian potential fluctuations}\label{sec4.2.2}
Next we turn to the second archetypal driving $f(t)$, namely correlated
Gaussian noise of the Ornstein-Uhlenbeck type (\ref{6l}), (\ref{6m}).
If both, the thermal noise and the Ornstein-Uhlenbeck noise driven
barrier fluctuations are sufficiently weak, transitions between
neighboring potential wells of $V(x)$ are rare, and a considerable
collection of analytical results 
\cite{han94ra,pec94,rei95ra1,rei95ra2,mad95,rei97ra} 
(see also further references in \cite{rei97ra}) --
obtained in the context of
the ``resonant activation'' effect  --
can be immediately applied via (\ref{4i}).

\subsubsection{Fast fluctuations (multiplicative white noise)}
In \sect \ref{sec4.1a} we have discussed for a general flashing
ratchet scheme the asymptotics for a stochastic $f(t)$ with small correlation
time $\tau$ under the assumption that the distribution (\ref{6k1})
of the noise is kept fixed upon variation of $\tau$ (constant 
variance scaling). In the special case that the general flashing
potential $V(x,f(t))$ in (\ref{4a}) depends linearly on its argument
$f(t)$, as it is the case in our present fluctuating potential ratchet 
model (\ref{4.1}),
a second interesting scaling for small $\tau$ is possible:
Namely, one keeps the intensity $\int dt\,\langle f(t) f(0)\rangle$ of the noise
constant upon variation of $\tau$. The distinguishing feature of this so-called
{\em constant intensity scaling} \cite{bar96} is the
emergence of a sensible white noise limit $\tau\to 0$
in the sense that the effect of the noise $f(t)$ approaches a 
non-trivial limiting behavior (for constant variance scaling the
noise has no effect for $\tau\to 0$). Since the
limit depends on the detailed properties of the noise $f(t)$, we focus on the
archetypal example of Ornstein-Uhlenbeck noise (\ref{6l}), (\ref{6m}) with a constant
($\tau$-independent) intensity
\begin{equation}
Q:=\sigma^2/\tau \ ,
\label{4.7a}
\end{equation}
i.e. the variance $\sigma^2$ acquires an implicit dependence on $\tau$.
In other words, the correlation (\ref{6l}) takes the form
\begin{equation}
\langle f(t)\, f(s)\rangle = 2\, Q\,\delta_\tau (t-s) \ ,
\label{4.7b}
\end{equation}
where
\begin{equation}
\delta_\tau (t) := \frac{1}{2\tau}\, e^{-|t|/\tau}
\label{4.7c}
\end{equation}
approaches a Dirac-delta function for $\tau\to 0$.

While this choice of $f(t)$ is clearly of little practical relevance, it
gives rise to a model that shares many interesting features with more 
realistic setups, but, in contrast to them, can be solved in closed 
analytical form \cite{rei97c}.
Furthermore, this exactly solvable model will serve a the basis
for our discussion of collective phenomena in \sect \ref{sec7.2}.
A mathematically similar setup, however, with as starting point a
rather different physical system, will also be encountered in 
\sect \ref{sec4.8.1}.

The fact that the coupling strength $V'(x(t))$ of the noise $f(t)$ in (\ref{4.1})
depends on the state $x(t)$ of the system (so-called {\em multiplicative noise})
makes the white noise limit $\tau\to 0$ for constant intensity scaling subtle
(so-called Ito-versus-Stratonovich problem \cite{han82b,ris84,hor84,kam92}).
The basic reason is \cite{gra82} that the $\tau\to 0$ limit does not
commute with the $\hat m\to 0$ limit in 
(\ref{2.1}) and the $\Delta t\to 0$ limit
in (\ref{2.7}).
We will always assume in the following that the $\hat m\to 0$ and $\Delta t\to 0$
limits are performed first and that $\tau$ becomes small only afterwards.
For the sake of completeness we mention that this sequence of limits
amounts \cite{gra82}
to treating the multiplicative noise in (\ref{4.1})
in the so called {\em sense of Stratonovich} \cite{han82b,ris84,hor84,kam92},
though we will not make use of this fact in the following but 
rather implicitly derive it.

No such problem arises as far as the (additive) thermal
noise $\xi(t)$ in (\ref{4.1}) is concerned. Hence, we can replace for the moment
the Dirac delta in (\ref{4b}) by the pre-Dirac-function $\delta_\tau(t)$ from (\ref{4.7c})
and postpone the limit $\tau\to 0$ according to our purposes.
It follows that in (\ref{4.1}) 
the sum of the two independent, unbiased, $\delta_\tau$-correlated Gaussian
noises $-V'(x(t))\, f(t)$ and $\xi(t)$ are statistically 
equivalent\footnote{Proof:
The two noises $\xi_1(t):= - V'(x(t))\, f(t)+\xi(t)$ and 
$\xi_2(t):=\sqrt{2\, g(x(t))}\xi_\tau(t)$ with $g(x)$ from (\ref{4.7e})
and $\xi_\tau(t)$ from (\ref{4.7f}) are
both Gaussian, have zero mean, and the same correlation, thus all their
statistical properties are the same.}
to a single unbiased Gaussian multiplicative noise $\xi_\tau(t)$ of 
the form 
\begin{eqnarray}
& & \dot x(t) = - \frac{V'(x(t))}{\eta} + g(x(t))\,\xi_\tau(t)\label{4.7d}\\
& & g(x) := 
\left[ \frac{k_BT}{\eta} + Q\left(\frac{V'(x)}{\eta}\right)^2\right]^{1/2}
\label{4.7e}
\end{eqnarray}
with correlation
\begin{equation}
\langle\xi_\tau(t)\, \xi_\tau(s)\rangle = 2\,\delta_\tau(t-s) \ .
\label{4.7f}
\end{equation} 
By means of the auxiliary variable
\begin{equation}
y(x):=\int_0^x\frac{d\bar x}{g(\bar x)}
\label{a1}
\end{equation}
it follows from (\ref{4.7d}) that $y(t):=y(x(t))$ satisfies the stochastic dynamics
\begin{equation}
\dot y(t) = - \frac{d}{dy}\,\phi (x(y(t))) + \xi_\tau(t) \ ,
\label{a2}
\end{equation}
where $x(y)$ is the inverse of (\ref{a1}) (which obviously exists) and where
\begin{equation}
\phi(x):=\int_0^x d\bar x\,\frac{V'(\bar x)}{\eta\, g^2(\bar x)} \ .
\label{4.10}
\end{equation}
Next we perform the white noise limit $\tau\to 0$ in (\ref{a2}).
The corresponding Fokker-Planck equation for $P(y,t)$ follows by comparison
with (\ref{2.5}) and (\ref{2.10}), reading
\begin{equation}
\frac{\partial}{\partial t}\, P(y,t) = \frac{\partial}{\partial y}
\left\{\left[\frac{d}{dy}\phi(x(y))\right]\, P(y,t)\right\} + 
\frac{\partial^2}{\partial y^2} \, P(y,t) \ .
\label{4.7g}
\end{equation}
The probability densities $P(x,t)$ and $P(y,t)$ are connected by 
the obvious relation \cite{han82b}
\begin{equation}
P(x,t)=\int_{-\infty}^\infty dy\, \delta(x-x(y))\, P(y,t) = P(y(x),t)/g(x) \ . 
\label{4.7h}
\end{equation}
Upon introducing (\ref{4.7h}) into
(\ref{4.7g}) we finally obtain a Fokker-Planck equation (\ref{2.10a}) for
$P(x,t)$ with probability current
\begin{equation}
J(x,t)= - \left\{\frac{V'(x)}{\eta} + g(x)\frac{\partial}{\partial x} g(x)\right\} P(x,t) \ .
\label{4.8}
\end{equation}
Along the same calculations as 
in \sect \ref{sec2.1.4} one now can derive for the reduced density (\ref{2.12})
in the steady state the result \cite{str58,iva69,amb69,str69}
(see also (\ref{2.22}))
\begin{eqnarray}
& & \hat P^{st}(x) = {\cal N}
\frac{e^{-\phi(x)}}{g(x)}
\int_x^{x+L} dy\,
\frac{e^{\phi(y)}}{g(y)} \ ,
\label{4.9}
\end{eqnarray}
where the normalization ${\cal N}$ is fixed through (\ref{2.14}),
and for the particle current (\ref{2.18}) one finds
\begin{equation}
\langle\dot x\rangle = L\, {\cal N}\, [ 1 - e^{\phi(L)-\phi(0)}]\ . 
\label{4.11}
\end{equation}

\subsubsection{Discussion}
Our first observation is \cite{bao96a,rei97c}
that the sign of the current is completely determined by the (reversed) sign of 
$\phi(L)-\phi(0)=\phi(L)$ (note that $\phi(0)=0$).
Especially, the current vanishes if and only if $\phi(L)=\phi(0)=0$. 
As expected, this is the case in the absence of the potential
fluctuations ($Q=0$, cf. \sect \ref{sec2.1.4}) or if $V(x)$ is symmetric. 
In any other case, we can infer that the current will be typically non-zero,
notwithstanding the fact that only white noises are acting.

These basic qualitative conclusions 
become immediately obvious upon inspection
of the transformed dynamics (\ref{a2}): The effective potential
$\phi(x(y))$ from (\ref{4.10})
is periodic in $y$ if and only if $\phi(L)=0$.
In this case an effective Smoluchowski-Feynman ratchet (\ref{2.5}) arises with
the result $\langle\dot y\rangle =0$. If $\phi(L)\not =0$, we are dealing
in (\ref{a2}) with a tilted Smoluchowski-Feynman ratchet (cf. \sect \ref{sec2.2.1}),
yielding a current $\langle\dot y\rangle$ with a sign opposite to that of $\phi(L)$.
Considering that the ensemble average $\langle\dot x\rangle$ is equivalent to
the time average of a single realization (\ref{4c1})
and similarly for $y(t)$,
it follows from (\ref{a1}) that 
\begin{equation}
\langle\dot y\rangle = \frac{\langle\dot x\rangle}{L}\,
\int_0^L \frac{dx}{g(x)} \ ,
\label{4.11a}
\end{equation}
especially the sign of $\langle\dot x\rangle$ must be
equal to that of $\langle\dot y\rangle$.
Remarkably, this exact relation (\ref{4.11a})
between the currents in the
original (\ref{4.7d}) and the transformed (\ref{a2}) dynamics
remains valid for arbitrary (not necessarily Gaussian white)
noises $\xi_\tau(t)$ \cite{bao96b}.

As an example we consider a piecewise linear potential with
three continuously matched linear pieces per period $L$ 
with the following slopes:
$V'(x)=-1$ for $-2<x<0$, $V'(x)=\lambda$ for $0<x<1$, and  
$V'(x)=2-\lambda$ for $1<x<2$, where $\lambda\in(0,2)$ is a parameter
that can be chosen arbitrarily. Hence, the potential $V(x)$ has a 
minimum at $x=0$ and barriers of equal height $2$ at $x=\pm 2$.
Outside this `fundamental cell' of length $L=4$ the potential
is periodically continued\footnote{In doing so we are working 
in dimensionless units
(cf. \sect \ref{sec2.1.2.4} in Appendix A) with $\Delta V=2$, $L=4$, 
and $\eta=k_B=1$.}. 
One then finds from (\ref{4.10}) that 
\begin{equation}
\phi(L)=\frac{2\, Q\, (1-\lambda^2)\, [Q\,\lambda(2-\lambda)-3T]}
{(T+Q)\,(T+Q\,\lambda^2)\,(T+Q\,(2-\lambda)^2)}\ .
\label{4.12}
\end{equation}
As expected, $\phi(L)$ and thus the current (\ref{4.11}) vanishes for $\lambda =1$,
while for any $\lambda\not =1$ the quantity $\phi(L)$
changes sign at $Q=3T/\lambda(2-\lambda)$. It follows that
the current $\langle\dot x\rangle$, considered as a function of either 
$T$ or $Q$ undergoes a current reversal. Similar reversals as a function
of any model parameter follow according to \sect \ref{sec3.5}.

A somewhat similar observation of a current inversion arising upon
considering a slight modification of a symmetric
sawtooth-potential $V(x)$ has been
reported for dichotomous (and other) potential fluctuations in \cite{aba98},
see also \cite{cha95,lee99a,che99,yan01}.

A modified ratchet model dynamics, without any thermal noise 
and no particular relation between $g(x)$ and $V(x)$ comparable 
to (\ref{4.7e}), but instead 
with a Gaussian multiplicative colored noise of finite correlation 
time $\tau$ in (\ref{4.7d}) has been studies numerically and 
by means of various approximations in \cite{bao96a,bao96b}.
The physical viewpoint in these works is thus closely related
to that of the Seebeck ratchet scheme from \sect \ref{sec4.8.1}
rather than the fluctuating potential 
ratchet model of the present section.

\section{Traveling potential ratchets}\label{sec4.3}
In this section we consider a special case of the stochastic 
dynamics (\ref{6j}) of the form
\begin{equation}
\eta\,\dot x(t) = - V'(x(t) - f(t)) + \xi (t)\ .
\label{4.101}
\end{equation}
As usual, we are concentrating on the overdamped limit and the thermal
fluctuations are modeled by unbiased Gaussian white noise $\xi (t)$ of
strength $2\, \eta\, k_B T$.
Further, $V(x)$ is a periodic, but {\em not necessarily asymmetric} 
potential
with period $L$.
Thus, the effective potential experienced by the particle in (\ref{4.101}) is
traveling along the $x$-axis according to the function $f(t)$, which may be either of
deterministic or stochastic nature.

Upon introducing the auxiliary variable
\begin{equation}
y(t) := x(t) - f(t)
\label{4.103}
\end{equation}
in (\ref{4.101}), we obtain
\begin{equation}
\eta\,\dot y(t) = - V'(y(t)) -\eta\, \dot f(t) + \xi (t)\ ,
\label{4.104}
\end{equation}
whence the average velocity $\langle\dot x\rangle$ of the original
problem follows as
\begin{equation}
\langle\dot x\rangle = \langle\dot f\rangle + \langle\dot y\rangle \ .
\label{4.104'}
\end{equation}

\subsection{Genuine traveling potentials}\label{sec4.3.1}
As a first example of a so-called 
{\em genuine traveling potential ratchet} 
we consider a deterministic function $f(t)$ of the form
\begin{equation}
f(t) = u\, t \ .
\label{4.105}
\end{equation}
In other words, the potential $V(x - f(t))$ in (\ref{4.101}) is given by
a periodic array of traps (local minima of the potential),
traveling at a constant velocity $u$ along the $x$-axis.
Hence, (\ref{4.101}) models basically the working principle
of a screw or a screw-like pumping device -- both invented by Archimedes \cite{arc250} --
in the presence of random perturbations.
Qualitatively, we expect that the Brownian particle $x(t)$ will be dragged into the
direction of the traveling potential traps. 
Next, we note that with (\ref{4.105}), the auxiliary $y$-dynamics
(\ref{4.104}) describes the well-known overdamped motion in a 
``tilted washboard'' potential\footnote{Note that
a positive velocity $u$ corresponds to a washboard 
``tilted to the left''.} \cite{str58,iva69,amb69,str69,ris84}.
Quantitatively, upon comparison of (\ref{4.103})-(\ref{4.105}) with 
(\ref{2.21})-(\ref{2.23}) the average velocity in the steady state 
takes the form
\begin{equation}
\langle\dot x\rangle = u - 
\frac{L\,k_B T\,\left[ e^{\eta u L /k_B T} - 1 \right]}
{\eta\int_0^L dx\int_x^{x+L} dy\, e^{[V(y)-V(x)+(y-x)\eta u]/k_B T}} \ .
\label{4.106}
\end{equation}
Though this formula looks somewhat complicated, one sees that typically
$\langle\dot x\rangle \not =0$ with the expected 
exceptions that either $u=0$ or $V'(x)\equiv 0$.
Especially, {\em a broken spatial symmetry of the potential $V(x)$ is not
necessary for a finite current} $\langle\dot x\rangle$.
Furthermore, {\em thermal noise is not necessary either}:
For $T\to 0$ one obtains directly from (\ref{4.103}), (\ref{4.104}) that
\begin{eqnarray}
& & \langle\dot x\rangle = 
\left \{ \begin{array}{ll} 
u - \frac{L}{\int_0^L\frac{dx}{u+V'(x)/\eta}} & \mbox{if}\ \ 
u+V'(x)/\eta \not = 0 \ \ \mbox{for all}\ \ x \\
u & \mbox{otherwise.}
\end{array} \right .
\label{4.107}
\end{eqnarray}
This deterministic behavior already captures the essential features of the 
more involved finite-$T$ expression (\ref{4.106}):
Namely, {\em $\langle\dot x\rangle$ has always the same sign 
as $u$ but is never larger in modulus}, 
in agreement with what one would have naively 
expected\label{fot4.3.2}\footnote{Particles $x(t)$, 
moving opposite to the traveling potential in (\ref{4.101}), 
or, equivalently,
particles $y(t)$ sliding down the tilted washboard faster than 
in the absence of a
potential $V(y)$ (faster than $-\dot f(t) = -u$) would 
indeed be quite unexpected.
Since we did not find a rigorous proof in the literature,
we give one herewith:
If a function $f(x)$ is concave on an interval $I$, i.e. 
$f(\lambda x +(1-\lambda)y)\leq \lambda f(x) + (1-\lambda) f(y)$
for all $\lambda\in [0,1]$ and $x,y\in I$, then it follows by
induction that $f(N^{-1}\sum_{i=1}^N X_i)\leq N^{-1}\sum_{i=1}^N f(X_i)$
for all $X_{1,...,N} \in I$. 
Choosing 
$u>V'(x)/\eta$ for all $x$ (especially $u>0$),
$f(x)=1/x$, $I=[0,\infty]$,
and $X_i=u+V'(i/N)$ it follows for $N\to\infty$ that
$u^{-1} = [\int_0^1 dx\, (u+V'(x))]^{-1}\leq \int_0^1 dx\, (u+V'(x))^{-1}$.
Working in dimensionless units with $L=\eta=1$ (cf. \sect \ref{sec2.1.2.4} in 
Appendix A) 
we can infer that $\langle\dot x\rangle\geq 0 $ in (\ref{4.107})
if $u\geq 0$, and similarly
$\langle\dot x\rangle\leq 0 $ if $u\leq 0$.
Choosing
$f(x)=\exp \{-x\}$, $I={\bf R}$, $X_i=V(i/N)-V(i/N+z)$ it follows
that $1=\exp\{\int_0^1 dx\, [V(x+z)-V(x)]\}\leq
\int_0^1 dx\, \exp\{V(x+z)-V(x)\}$ for any $z$. 
Hence,
$\int_0^1 dx \int_x^{x+1} dy\, \exp\{u(y-x)+V(y)-V(x)\}
=\int_0^1 dz \, \exp\{uz\} \int_0^{1} dx\, \exp\{V(x+z)-V(x)\}
\geq \int_0^1 dz \, \exp\{uz\} = [\exp\{u\} -1 ]/u$.
It follows that $\langle\dot x\rangle\geq 0 $ in (\ref{4.106})
if $u\geq 0$, and similarly
$\langle\dot x\rangle\leq 0 $ if $u\leq 0$.}.
Furthermore, there are two basic ``modes'' of motion 
in (\ref{4.107}). In the first case in (\ref{4.107}), i.e. for large speeds $u$, 
the Brownian particle is only ``loosely  coupled'' to the traveling
potential (cf. \sect \ref{sec2.2.3a}), 
it behaves like a swimmer afloat on the surface of the ocean and may thus
be called a {\em Brownian swimmer} \cite{bor98}. In the second case in (\ref{4.107}), 
i.e. for small speeds $u$, 
we have a {\em Brownian surfer} \cite{bor98}, ``riding'' in a {\em tightly
coupled} way on the traveling
potential\footnote{In the corresponding noiseless 
tilted washboard dynamics (\ref{4.104}),
the particle $y(t)$ permanently travels downhill in the first case,
while it quickly comes to a halt in the second.}
in (\ref{4.101}), (\ref{4.105}).
{\em The current $\langle\dot x\rangle$ tends to zero for both very 
small and very large speeds $u$}, and has a maximum at the largest 
$u$ which still supports the surfing mode.

More general types of genuine traveling potential ratchets are
obtained by supplementing the uniformly traveling contribution
in (\ref{4.105}) by an additional unbiased
periodic function of $t$ or by a stationary random process.
In the transformed dynamics (\ref{4.104}) this yields
a tilting ratchet mechanism with an additional constant external force $-\eta u$
as treated in \ch \ref{cha6} in more details.

\subsubsection{Applications}
Out of the innumerable theoretical and experimental applications of the above
described simple pumping scheme (\ref{4.105})
we can only mention here a small selection.
The most obvious one is a particle, either suspended in a fluid or floating on its
surface,
in the presence of a traveling wave \cite{jan98,bor98,vdb99,bor99,li00,ben00}.
The resulting drift (\ref{4.107})
has been predicted in the deterministic case ($T=0$) already by Stokes \cite{sto47}.
In the presence of thermal fluctuations ($T>0$), 
the closed analytical solution (\ref{4.106})
has been pointed out in \cite{vdb99,li00}.
The effect of {\em finite inertia} has been discussed in \cite{bor98}, based
on the known approximative solutions for the corresponding
tilted washboard dynamics \cite{ris84}.
A model with an asymmetric potential $V(x)$ and
a dichotomous instead of a Gaussian white noise $\xi(t)$ in
(\ref{4.101}) has been studied in \cite{ben00} with the
possibility of {\em current inversions} as its most remarkable feature,
i.e. particles can now even move in the direction opposite 
to that of the velocity $u$.
This result can be readily understood as a consequence of the ratchet
effect in the equivalent fluctuating force ratchet 
(\ref{4.104}), (\ref{4.105}), cf. \sect \ref{sec6.1.2}.
Further generalizations for superpositions of traveling waves 
in arbitrary dimensions
are due to \cite{mes92,jan98}, which will be
discussed in somewhat more detail in \sect \ref{sec4.4.1}.

The mesoscopic analog of Archimedes' water pump is the adiabatic 
quantum electron pump by Thouless \cite{tho83}.
This theoretical concept has been realized in a quantum dot experiment in 
\cite{swi99}, triggering in turn 
further theoretical studies \cite{wag99,sol00,ast01}.
Similar single electron pumps, however, operating in the classical regime
\cite{swi99}, have been realized in
\cite{kou91a,kou91b,pot92,kel96}.
For additional closely related single electron pumping experiments, see also
\cite{wei95,xia99}.

A theoretical analysis of diffusion (unpredictability) in clocked reversible
computers in terms of a Brownian motion in a traveling potential is given in \cite{lan85}.

Experimental studies of Brownian particles 
($2\, \mu m$ diameter polysterene spheres in water),
moving on a circle in the presence of a traveling optical trap, have been reported 
in \cite{fau95} and are in good agreement with the simple theoretical
model (\ref{4.101}), (\ref{4.105}).

Single-electron transport by high-frequency surface acoustic 
waves in a semiconductor heterostructure
device has been demonstrated for example in \cite{tal97}.
A more sophisticated variant with excitons (electron-hole pairs)
instead of electrons, which is thus able to transport ``light'',
has been realized in \cite{roc97}.

Though an asymmetry in the periodic potential $V(x)$ is not necessary
to produce a current in (\ref{4.101}),(\ref{4.105}),
one may of course consider traveling ratchet-shaped potentials nevertheless.
This situation has been addressed for instance in \cite{pos98,mal98}, leading
to interesting effects for traveling wave
trains of finite spatial extension (i.e. $V'(x)\to 0$ for $x\to\pm\infty$)
which are reflected at a wall and then
pass by the same particle again in the opposite direction \cite{bor98}.

One basic effect of pumping particles by a traveling potential is 
a concentration gradient. The inversions, namely making a potential
travel by exploiting a particle flux (e.g. due to a concentration gradient)
is also possible \cite{sas98}, as exemplified by the chiral dynamics
of a ``molecular wind-mill'' \cite{fuk98}.
Note that useful (mechanical) work can be gained in either way.

A number of further applications in plasma physics and quantum optics have 
been compiled in \cite{bor98}.
In fact, a great variety of engines are operating in a cyclic manner
with a broken symmetry between looping forward and backward and may thus be
classified as ratchet systems, typically of the traveling
potential type.
The examples of screws, water pumps, propellers, or equally spaced traveling cars
(representing the traveling potential minima for the passengers)
demonstrate {\em a certain danger of invoking an overwhelming practical
relevance while the underlying basic principle may become trivial from the
conceptual viewpoint of contemporary theoretical physics.}
Furthermore, these largely mechanical
examples together with our above findings 
that neither a broken symmetry nor thermal noise are necessary, 
current inversions are impossible,
and the tight $x(t)$-to-$f(t)$-coupling (see \sect \ref{sec2.2.3a})
in the most important case of small speeds $u$ show that 
{\em many characteristic aspects of the Brownian motor
concept are actually absent in genuine traveling potential ratchets}.

While the general framework in \sect \ref{sec3.1.1} has been purposefully
set such as to include traveling potential ratchets, they are at the 
boarder line in so far as they often involve a quite obvious
{\em a priori} preferential direction of transport.
The boarder between the realm of Brownian motors and that of 
``pumping between reservoirs'' is definitely crossed when in 
addition the spatial periodicity and noise effects play no 
longer a central role.

\subsection{Improper traveling potentials}\label{sec4.3.2}
Next we turn to the simples type of a so-called 
{\em improper traveling potential ratchet}, arising through
a modification of the driving function $f(t)$ from 
(\ref{4.105}) of the  form
\begin{equation}
f(t) = u\, t - \int_0^t dt'\, 
\sum_{i=-\infty}^\infty n_i\, L\, \delta(t'-\tau_i )\ ,
\label{4.120}
\end{equation}
where, {\em the coefficients $n_i$ are either deterministically
fixed or randomly sampled integers} (not necessarily positive).
Thus, most of the time the function $f(t)$ changes at a constant rate $u$, but
at the special instants $\tau_i$ it jumps by an integer multiple of the spatial
period $L$.
These times are assumed to be ordered, $\tau_{i+1}>\tau_i$, and may be 
either regularly or randomly spaced. The main idea is to choose them such 
that $f(t)$ in (\ref{4.120}) becomes an {\em unbiased}
periodic function of time
or a stochastic process with zero average 
(hence the name ``improper traveling potential'').
In other words, extending the meaning of the symbol $\langle\dot f\rangle$
analogously as in (\ref{4c1}) we require that
\begin{equation}
\langle\dot f\rangle = \lim_{t\to \infty}\frac{f(t)}{t} = 0
\label{4.122}
\end{equation}
so that $\langle\dot x\rangle=\langle\dot y\rangle$ in (\ref{4.104'}),
i.e. on the long term the discontinuous jumps in (\ref{4.120}) have to
counterbalance the continuous change $u\, t$. 
This condition is satisfied if and only if
\begin{equation}
\ttt\,u\,\overline{n} = L \ ,
\label{4.120''}
\end{equation}
where we have introduced
\begin{eqnarray}
& & \overline{n} := \lim_{k\to\infty}\frac{1}{2k+1}\sum_{i=-k}^k n_i \ ,
\label{4.120'}\\
& & \ttt := \lim_{k\to\infty}\frac{1}{2k+1}\sum_{i=-k}^k (\tau_{i+1}-\tau_i)\ .
\label{4.120'''}
\end{eqnarray}
Especially, the so defined
limits (\ref{4.120'}), (\ref{4.120'''}) are
assumed to exist and to be independent of the considered
realization in the case that the summands are randomly sampled.

Regarding examples, the simples choice of the coefficients 
$n_i$ in (\ref{4.120}) is $n_i = 1$ for all $i$.
On the other hand, the simplest choice of the times $\tau_i$
arises if they are regularly spaced. Then
$\ttt$ in (\ref{4.120'''}) is obviously equal to this spacing, i.e.,
\begin{equation}
\tau_i=i\,\ttt +const.
\label{4.122'}
\end{equation}
Another simple option are random $\tau_i$ 
with a Poisson statistics, i.e.
the probability to have $m$ time points in a time 
interval of duration $t\geq 0$ is
\begin{equation}
P_m(t) = \frac{(t/\ttt )^m}{m!} \, e^{-t/\ttt} \ .
\label{4.123}
\end{equation}
Then, in accordance with (\ref{4.120'''}), $\ttt$ is again
the mean value of $\tau_{i+1}-\tau_i$. 

Returning to general processes (\ref{4.120}) with integers $n_i$
respecting (\ref{4.122}) or equivalently (\ref{4.120''}),
we now come to the central point of this section,
consisting in the following very simple observation:
Since the discontinuous jumps of the 
driving $f(t)$ in (\ref{4.120}) are always integer multiples of the period
$L$ and the potential $V(x)$ is $L$-periodic, {\em the jumps of 
$f(t)$ in (\ref{4.120}) do not have any effect whatsoever on the 
stochastic dynamics (\ref{4.101}) !}
In other words, the genuine traveling potential ratchet (\ref{4.101}), 
(\ref{4.105}) is equivalent to the improper traveling potential ratchet
(\ref{4.101}), (\ref{4.120}), (\ref{4.122}).
Especially, the results (\ref{4.106}), (\ref{4.107})
for the current $\langle\dot x\rangle$
can be taken over unchanged.
Due to (\ref{4.104'}), (\ref{4.122}) the same results are moreover valid for
$\langle\dot y\rangle$. With (\ref{4.120}) the term $\dot f(t)$ in the 
$y$-dynamics (\ref{4.104}) takes the form
\begin{equation}
\dot f(t) = u\ - \sum_{i=-\infty}^\infty n_i\, L\, \delta(t-\tau_i )\ .
\label{4.124}
\end{equation}
In other words, we have found that the dynamics 
(\ref{4.104}),(\ref{4.124})  with (\ref{4.122}) or equivalently (\ref{4.120''}), 
admits a closed analytic solution.
In the special
case of a Poisson statistics (\ref{4.123}) the random process (\ref{4.124})
is called a {\em Poissonian shot noise} \cite{ric54,han78,han80,vdb83,vdb84}.
Its mean value is zero owing to (\ref{4.122}), and its correlation is found to be
\begin{eqnarray}
& & \langle\dot f(t)\, \dot f(s)\rangle 
= \frac{\overline{n^2}\, L^2}{\ttt}\, \delta (t-s)\label{4.124'}\\
& & \overline{n^2} := \lim_{k\to\infty}\frac{1}{2k+1}\,\sum_{i=-k}^k n_i^2 \ ,
\label{4.124''}
\end{eqnarray}
i.e. the shot noise is uncorrelated in time (white noise).
The conclusion that a stochastic dynamics (\ref{4.104})
in the presence of a white shot noise \cite{luc95,cze97,cze00,cze01}
or an unbiased periodic driving 
of the form (\ref{4.124}) (with $n_i$ being integers)
and simultaneously a white Gaussian noise $\xi (t)$
is equivalent to the Brownian motion in a traveling potential or in
a tilted washboard potential and thus exactly solvable is to our knowledge new.

\subsubsection{Generalizations, equivalences, applications}
Generalizations of the above arguments are obvious and we only mention here a 
few of them.
First, an arbitrary {\em periodic} $f(t)$ in (\ref{4.101}) is equivalent to a 
dynamics (\ref{4.104}) with
a homogeneous periodic driving force $\dot f(t)$.
Due to the periodicity of $f(t)$, this force is unbiased 
in the sense of (\ref{4.122})
and due to (\ref{4.104'}) the currents of the original (\ref{4.101}) and the 
transformed dynamics (\ref{4.104}) are thus strictly equal.
Such a dynamics (\ref{4.104}) will be  considered 
under the labels ``rocking ratchets'' and ``asymmetrically tilting ratchets'' 
in \ch  \ref{cha6}.
Both these classes of ratchets are thus equivalent to a Brownian
motion in a back-and-forth traveling periodic potential.
As we will see, a finite current 
$\langle\dot x\rangle = \langle\dot y\rangle \not = 0$
generically occurs if $V(x)$ and/or
$\dot f(t)$ is asymmetric (see \sect \ref{sec3.1.2})
and unless both of them are supersymmetric according to \sect
\ref{sec3.4}.
An inversion of the current upon variation of an arbitrary parameter
of the model can be designed along the same line of reasoning as in 
\sect \ref{sec3.5}.

The above exemplified procedure 
of transforming a genuine into an improper traveling potential
ratchet is obviously very general.
Due to (\ref{4.122}) this transformed model can
then be mapped once more onto an unbiased tilting ratchet scheme
(\ref{4.104}). In short, {\em genuine and improper traveling potential models
are equivalent to each other and moreover equivalent to a tilting
ratchet}.

Finally, we turn to a modification of the (genuine) 
uniformly traveling potential 
(\ref{4.101}), (\ref{4.105}), namely the case that $f(t)$ and thus the 
{\em periodic potential advance in discrete steps}:
\begin{equation}
f(t) = \int_0^t dt'\, \sum_{i=-\infty}^\infty \Delta_i\, \delta(t'-\tau_i)\ .
\label{4.125}
\end{equation}
As seen before, steps $\Delta_i = n_i L$ do not have any effect on the
dynamics (\ref{4.101}).
Thus, the simplest nontrivial case arises when two subsequent steps
add up to one period $L$:
\begin{equation}
\Delta_{2i} = i\, L\ ,\ \ \Delta_{2i+1} = i\, L + \lambda\ ,\ \ \lambda\in(0,L)\ .
\label{4.126}
\end{equation}
We furthermore assume that the jumping times $\tau_i$ are regularly spaced
\begin{equation}
\tau_{2i} = i\, \ttt\ ,\ \ \tau_{2i+1} = i\, \ttt + \tau\ ,\ \ \tau\in(0,\ttt)\ .
\label{4.126'}
\end{equation}
For $\lambda=L/2$ and $\tau=\ttt/2$ the signal $f(t)$ in (\ref{4.125}) is thus
a discretized version of (\ref{4.105}) advancing at equidistant steps 
in time and in space with the same average speed $u=L/\ttt$.
For other values of $\lambda$ and $\tau$, every second step is modified.
More steps per period,
random instead of deterministic times $\tau_i$ and many
other generalizations are possible but do not lead to essential
new effects.

Recalling that jumps of $f(t)$ by multiples of $L$ do not affect the 
dynamics (\ref{4.101}), we can infer that (\ref{4.125})-(\ref{4.126'})
is equivalent to
\begin{eqnarray}
& & f(t) =
\left \{ \begin{array}{lll} 
0  & \mbox{if} & t\in [0,\tau) \\ 
 \lambda  & \mbox{if} & t\in [\tau, \ttt) \\ 
\end{array} \right . \nonumber\\
& & f(t+\ttt) = f(t) \ ,
\label{4.127} 
\end{eqnarray}
i.e., $f(t)$ periodically jumps between the two values $0$ and $\lambda$.
{\em Such a genuine
traveling potential advancing in discrete steps is thus
equivalent to a periodic switching between two shifted 
potentials}.
The periodicity of $f(t)$ in (\ref{4.127}) implies (\ref{4.122}),
thus the current in (\ref{4.104'}) agrees with that of 
the transformed dynamics (\ref{4.104}), featuring an unbiased additive force $\dot f(t)$ 
with $\delta$-peaks of weight $\lambda$ at $t=i\,\ttt$ and weight 
$-\lambda$ at $t=i\,\ttt+\tau$. As before, this equivalent dynamics establishes the 
connection of a stepwise traveling potential with the rocking and asymmetrically tilting ratchet
schemes from \ch  \ref{cha6}.

Without going into the details of the proofs we remark that:
(i) A symmetric potential $V(x)$ in combination with $\tau = \ttt/2$
in (\ref{4.126'}) implies $\langle\dot x\rangle =0$ for any $\lambda$ 
in (\ref{4.126}).
(ii) A symmetric potential $V(x)$ at temperature $T=0$ implies 
$\langle\dot x\rangle =0$ for any $\tau$ and $\lambda$.
(iii) For $T>0$, $\lambda\not = L/2$, $\tau\not = \ttt/2$
a current $\langle\dot x\rangle\not =0$ is generically expected
\cite{kan99}.
(iv) If $V(x)$ is asymmetric \cite{cha94,li00b} or $f(t)$ supports more than two
effectively different 
discrete states (cf. (\ref{4.127})), i.e. when
transitions between more than two shifted potentials are 
possible \cite{che97,li97a}, then a ratchet effect 
$\langle\dot x\rangle\not =0$ is expected generically.

Apart from those peculiarities, one expects 
{\em basically the same qualitative behavior as for the 
uniformly traveling potential case (\ref{4.105})}.
Quantitative results, confirming this expectation, 
are exemplified in \cite{cha94,li97a}.
Like in the continuously traveling potential case, there are again two
basic modes of motion (cf. the discussion below (\ref{4.107})):
One which is ``loosely'' coupled to the traveling potential (Brownian swimmer),
and one in which the particle ``rides'' on the traveling wave (Brownian surfer).
This clear cut distinction is washed out by the thermal noise. The
detailed dependence of $\langle\dot x\rangle$ on model parameters
like $\lambda$, $\tau$, or $\eta$ shows furthermore certain traces \cite{cha94}
of the
discrete jumps in (\ref{4.125}) which are not present in the continuously traveling
counterpart (\ref{4.105}).

An {\em experimental realization} of directed motion
by switching between two shifted ratchet shaped potentials has
been reported in \cite{gor97a}.
The moving particle is a mercury droplet of about $1mm$ in diameter and the
two shifted ratchet potentials are created by suitably positioned electrodes.
Both, for periodic and stochastic switching
between the two ratchet potentials, the measured displacements agree very well
with the simple $T=0$ theory from \cite{cha94}.

The same ratchet scheme has been implemented experimentally for $\mu m$-sized latex 
spheres in \cite{gor98}.
The setup is similar to the one by Rousselet, Salome, Ajdari, and Prost \cite{rou94}
and by Faucheux and Libchaber \cite{fau95a} described in \sect \ref{sec4.1.1}
and thus the same uncertainties when comparing measurements with
theory arise.
The main difference in \cite{gor98} is that now {\em two} superimposed 
``Christmas-tree electrodes'' are used, shifted against each other so as to generate
the two shifted ratchet potentials by switching the applied voltage.
A further difference with the on-off experiments \cite{rou94,fau95a} is that
in such a traveling potential-type setup \cite{gor98}
thermal fluctuations are negligible in very good approximation
(cf. \sect \ref{sec4.3.1}).
For two different species of highly diluted particles
(latex spheres with $0.2\,\mu m$ and $0.5\,\mu m$ diameters) the theoretically predicted
effect that for a suitable choice of parameters, only one of them appreciably moves, was
qualitatively verified in the experiment \cite{gor98}.

\section{Hybrids and further generalizations}\label{sec4.4}
\subsection{Superpositions of traveling potentials}\label{sec4.4.1}
In this section we consider combinations and other extensions 
of the fluctuating potential
and traveling potential ratchet schemes from (\ref{4.1}) and (\ref{4.101}).
As a first example, we consider a pulsating potential ratchet dynamics 
(\ref{6j}) involving superpositions of several traveling potentials
\cite{mes92,jan98}
\begin{eqnarray}
V(x,f(t)) & = & \sum_i V_i(x-u_it)\label{4.110}\\
V_i(x+L_i) & = & V_i(x) \label{4.111}\ . \
\end{eqnarray}
At variance with all other cases considered in this chapter, the potential
(\ref{4.110}) is thus not periodic in the
spatial variable $x$ unless the periods $L_i$ are all commensurable 
with each other.

The starting point for an approximate treatment is an expansion
of the single-potential result (\ref{4.106}) up to the first nontrivial
order in $V(x)/k_B T$, which turns out to be the second 
order \cite{jan98,vdb99}.
The next salient point is that for several potentials one simply can,
within the same approximation, add up the
contributions from all the single potentials
provided that their traveling
speeds $u_i$ and periods $L_i$ are, in modulus, different from
each other.
In other words, up to second order, no mixed contributions
in the amplitudes of the traveling
potentials appear \cite{mes92,jan98}.
Basically, the reason is that 
the mismatch of the different temporal and spatial periods
only leads to oscillating mixed terms which average out to zero in the long run.
Proceeding along this line of reasoning, the final
result for the net current $\langle\dot x\rangle$ is
\begin{eqnarray}
\langle\dot x\rangle \!\!\!\! & = & \!\!\!\! \sum_i u_i
\left[ \int\limits_0^{L_i} \frac{dx}{L_i}\frac{V_i^2(x)}{(k_B T)^2} -
\frac{\alpha_i}{e^{\alpha_i} -1 } 
\int\limits_0^{L_i} \frac{dx}{L_i} \int\limits_{x}^{x+L_i} \frac{dy}{L_i}
\frac{V_i(x)\, V_i(y)}{(k_B T)^2}\, e^{\alpha_i(y-x)/L_i}\right]
\nonumber\\ 
\alpha_i \!\!\!\! & := & \!\!\!\! \eta\, u_i L_i /k_B T\ ,
\label{4.112'}
\end{eqnarray}
where it is assumed that both terms in the square brackets are small
in comparison to unity, i.e., $V_i(x)/k_B T$ needs to be small, but
also $\alpha_i$ should be not too large in modulus.
Specifically, for sinusoidal traveling potentials of the general form
\begin{equation}
V(x,f(t)) = \sum_i A_i\sin\left(\frac{2\pi}{L_i}(x-u_i t) + \phi_i \right)
\label{4.113}
\end{equation}
one obtains
\begin{equation}
\langle\dot x\rangle =  \frac{1}{2(k_B T)^2}
\sum_i u_i\, A_i^2
\left[ 1+\left(\frac{L_iu_i\eta}{2\pi\, k_B T}\right)^2 \right]^{-1}\ . 
\label{4.113'}
\end{equation}
Thus, already with two superimposed potentials with opposite speeds $u_1$ 
and $u_2$ and $|u_1| \not = |u_2|$, $|L_1| \not = |L_2|$ one can
tailor the two amplitudes $A_1$ and $A_2$ such that the current
(\ref{4.113'}) will change its direction either upon variation of the
temperature $T$ or, at a fixed but finite $T$, upon variation of
the friction coefficient $\eta$.
While for transport based on a single traveling potential, 
thermal fluctuations are not important (cf. \sect \ref{sec4.3.1}), 
they are thus indispensable for this
type of particle separation scheme \cite{jan98}.

There is no reason to expect that the above effect is restricted
to potentials of small amplitudes, but beyond this regime quantitative
analytical progress becomes cumbersome.
Qualitatively, the following very simple prediction is worth mentioning:
We consider in (\ref{6j}) a potential that is given
as a linear combination of two potentials, moving uniformly in opposite
directions, i.e., $f(t)=u\, t$ and
\begin{equation}
V_\lambda (x,f(t)) :=  \lambda\, V_1(x+u t) + (1-\lambda)\, V_0(x-u t) \ ,
\label{4.114}
\end{equation}
where $\lambda$ is a control parameter and the spatial periods of 
$V_0(x)$ and $V_1(x)$ may or may not agree.
Similarly as in \sect \ref{sec3.5} 
one sees that for a ``generic'' choice of $V_0(x)$ and $V_1(x)$ 
(no ``accidental symmetries'' of $V_\lambda(x,f(t))$ 
for any $\lambda\in (0,1)$)
a $\lambda$-value must exists at which the 
current $\langle\dot x\rangle$ exhibits an inversion upon variation of
an arbitrarily chosen parameter of the model (\ref{6j}).
Note that in contrast to the prediction from (\ref{4.112'})
the present conclusion holds even if the thermal noise $\xi (t)$
in (\ref{6j}) is zero, see \eq  (\ref{4.107}).

Another variation with one static and one traveling potential, i.e.
\begin{equation}
V (x,f(t)) :=  V_0(x) + V_1(x-u t) \ ,
\label{4.114'}
\end{equation}
has been analyzed in \cite{por00} in the zero temperature limit
$\xi(t)\equiv 0$ in (\ref{6j}).
If at least one of the two potentials
$V_0(x)$, $V_1(x)$ is asymmetric and their relative amplitudes
are properly chosen then
the traveling potential is able to drag the particle $x(t)$ 
in (\ref{6j}) into one direction.
However, if the traveling direction is reversed ($u\mapsto -u$), 
the particle cannot be dragged in that direction anymore due to the 
asymmetry of the potential. The possibility of such a behavior becomes
particularly obvious in the case of small speeds $u$ such that the
particle tends to follow one of the instantaneous
local minima of the total potential $V(x,f(t))$ in (\ref{4.114'}): 
For a very small amplitudes of $V_1(x)$, the particle clearly cannot
be dragged into either direction, while for very large amplitudes
it can be dragged into both directions.
Thus there must be an intermediate amplitude of transition
from localized to commoving behavior.
Due to the spatial asymmetry, this transition amplitude is 
typically not the same
for positive and negative speeds and commoving behavior in only
one direction is recovered.

\subsection{Generalized pulasting ratchets and experimental realizations}\label{sec4.4.2}
In the remainder of this section we focus again on 
{\em periodic} potentials (\ref{4e}) 
in the genuine pulsating potential ratchet setting
(\ref{6j}).
Still, the various possibilities of how to choose $V(x,f(t))$ obviously 
rule out an exhaustive discussion.
We will restrict ourselves to a few representative examples
which cover most of the existing
experimental and theoretical literature and which already exhibit all main
features that one may possibly expect in more general cases.

The simplest example is a hybrid of a uniformly traveling
and simultaneously fluctuating potential ratchet of the form
\begin{equation}
V(x,f(t)) = V(x- u t) [1+\tilde f(t)]\ 
\label{4.203}
\end{equation}
where $f(t):= u\, t$ and the auxiliary function 
$\tilde f(t) := \tilde f(f(t)/u)$ is assumed to be a periodic 
function of its argument.
By means of the same transformation of variables as in
(\ref{4.103})-(\ref{4.104'}) one can map this model onto a purely
fluctuating potential ratchet with a superimposed tilt.
Thus a finite current $\langle\dot x\rangle$ is generic and the possibility of current
inversions is also immediately obvious.

A similar hybrid of a traveling and simultaneously fluctuating potential ratchet arises
if $f(t)$ is not given by $u\, t$ in (\ref{4.203}) but instead increases in 
discrete steps like in (\ref{4.125}).
In the simplest case, a model which switches either regularly or
randomly between two different potentials $V_m(x)$ ($m=1,2$)
arises (cf. \eq  (\ref{4.127})), which both have the same shape but
are shifted against each other and moreover
differ in their amplitudes.
Observing that the on-off ratchet is a special case and exhibits
current inversions for suitably tailored potentials \cite{cha94}
(cf. \sect \ref{sec4.1}), the same property follows for the present
more general situation.

Going just one step further, one may consider in (\ref{6j}) the case
of a periodic or random switching between two potentials
$V_m(x)$ ($m=1,2$) which have the same spatial period $L$ but are
otherwise completely independent of each other.
The generic occurrence of a non-vanishing
current $\langle\dot x\rangle$ and the existence of current inversions
for suitably chosen potentials is obvious.

An {\em experimental} study of such a system has been performed by
Mennerat-Robilliard, Lucas, Guibal, Tabosa, 
Jurczak,  Courtois,  and Grynberg \cite{men99}.
Laser cooled Rubidium atoms in the presence of two suitably
chosen counterpropagating electromagnetic waves can switch between
two effective optical potentials $V_m(x)$
with the above described properties.
The switching is caused by absorption-spontaneous emission cycles of the Rubidium atoms
and results in an average velocity $\langle\dot x\rangle$ of the atoms
of about $0.1\, m/s$. While
the simple stochastic dynamics (\ref{6j}) with two alternating
potentials $V_m(x)$ is sufficient for a qualitative explanation of the
observed results, a  quantitative comparison would require a 
semiclassical
or even full quantum mechanical treatment (see also \sect \ref{sec6.5.5}).

Another generalization are so-called asynchronously pulsating
ratchets \cite{schi97} (see also \sects \ref{sec3.2.2} and \ref{sec3.9})
and especially the so-called {\em sluice-ratchet} scheme 
\cite{par98a,par98b,par00}.
In this latter case, the amplitudes of every second potential 
barrier are periodically oscillating in perfect synchrony,
similarly as for a fluctuating potential ratchet.
The rest of the potential barriers are also synchronously
oscillating in the same way, but with a time-delay of $\ttt/4$ (where $\ttt$ is
the time-period).
Thus, the Brownian particle $x(t)$ moves forward somewhat similar to
a ship in an array of sluices and may achieve 100\% efficiency in the
adiabatic limit $\ttt\to\infty$ (see \sect \ref{sec3.8}).
An experimental realization by semiconductor superlattice 
heterostructures is due to \cite{hoh99,hoh01}.

We close with three promising experimental implementations
of a pulsating ratchet scheme on a molecular scale which have
so far been partially realized.
The first one 
is based on the single triptycene$[4]$helicene molecules which we
already encountered in \sect \ref{sec2.1}.
By means of certain chemical processes which basically play the role
of the non-thermal potential fluctuations in the
pulsating ratchet scheme, Kelly, De Silva, and Silva \cite{kel99,dav99,kel01}
achieved a unidirectional intramolecular rotary motion.
The system is so far only a ``partial'' Brownian motor in that
only rotations by $120^o$ have been realized.

Monodirectional rotation in a helical alcene molecule with a definite
chirality (broken symmetry) has been investigated by
Koumura, Zijistra, van Delden, Harada, and Feringa \cite{kou99,dav99}.
Basically, ultraviolet radiation induces transitions between two
ratchet-shaped potentials which are identical
except that they are 
shifted by half a period. In other words,
a photochemical two-state pulsating (or traveling) potential ratchet scheme
as anticipated theoretically in \cite{bel80,pro94} is realized.
Chemically, the light-induced switching between the two potentials 
corresponds to a cis-trans isomerization, and each such transition is followed
by a thermally activated relaxation process.
While experimentally, each of the partial steps of a full
cycle has so far been only demonstrated separately,
there seems no reason why the system should not be able
to also rotate continuously.

Finally, Gimzewski, Joachim and coworkers \cite{gim98,gim99}
have visualized single propeller-shaped molecular rotors
(hexa-tert-butyl decacyclene) deposited on a Cu-surface
by means of scanning tunneling microscopy (STM).
Under appropriate conditions, the molecule is observed
to perform a thermally driven rotary Brownian motion
within an environment which gives rise to a highly asymmetric,
ratchet-shaped effective potential of interaction with the rotor.
In principle, we are thus dealing with another molecular realization
of a Smoluchowski-Feynman ratchet and pawl gadget (cf. \sect \ref{sec2.1}), 
but in the present case the time resolution of the STM was too low to confirm
the absence of a preferential direction of rotation.
As the authors in \cite{gim98,gim99} propose, with the help of a
second non-thermal source of noise, for example a tunnel
current, it should be possible to realize a ratchet effect
in terms of a preferential direction of rotation. Considering that
such a tunnel current would not directly interact
with the angular state variable of the system
but rather with some internal degree of freedom (of the environment),
a pulsating ratchet scheme is expected according to our general 
analysis from \sect \ref{sec3.2.2}.

\section{Biological applications: molecular pumps and motors}\label{sec4.5}
Consider an isothermal chemical reaction in the presence of a 
catalyst protein, 
i.e. an enzyme. In the simplest case, the reaction can be described by
a single reaction coordinate, cycling through a number of chemical states.
A suitable working model is then an overdamped Brownian particle
(reaction coordinate) in the presence of thermal fluctuations in a
periodic potential.
The local minima within one period represent the chemical states and
looping once through the chemical cycle in one or the other direction is 
monitored by a forward or backward displacement of the reaction coordinate
by one spatial period.
Completing a cycle in one direction means that one entity of reactant 
molecules have been catalyzed into product molecules, while a cycle in the 
other direction corresponds to the reverse chemical reaction.

Since our so simplified model is nothing else than a Smoluchowski-Feynman 
ratchet (\ref{2.5}), the absence of a net current signals that we are
dealing with a chemical process at equilibrium, i.e. the concentrations
of reactants and products are at their equilibrium
values and are not changing on the average under the action of the enzyme.

If the concentrations of the reactants and products are away from their
equilibrium (detailed balance) ratio, then the catalyst molecule
will loop through the chemical reaction cycle preferably in one way, namely
such that the reaction proceeds towards equilibrium.
In the corresponding Smoluchowski-Feynman ratchet model, the
periodic potential has to be supplemented by a constant 
tilt\footnote{For a proper description of an out of equilibrium catalytic
cycle on a more sophisticated level, see \sect \ref{sec5.3.2}.
This description is in terms of a discrete chemical state variable $m$.
On the same level, a consistent description in terms of a continuous state variable $x$
does not seem to exist (see also \cite{kel00})
unless it is basically equivalent to the discrete description
(activated barrier crossing limit, see \sect \ref{sec3.6}).},
resulting in a stochastic dynamics of the form (\ref{2.21}).
Note that while the environment of the catalyst is out of equilibrium as far as the
concentrations of reactants and products are concerned,
the properties of the random environmental noise and of the dissipation
mechanism in (\ref{2.21}) are still the same as for the 
equilibrium system (\ref{2.5}).

Usually, one or several transitions between chemical states may also be
(in a probabilistic or deterministic sense) accompanied by a change of the
geometrical shape (mechanical configuration)
of the catalyst molecule (``mechanochemical coupling'').
Transitions between such configurations may then be exploited
for doing mechanical work.
Due to the preferential direction in which these transitions are
repeated as time goes on, one can systematically accumulate 
useful mechanical energy out of the chemical energy by keeping 
up the nonequilibrium concentrations of reactants and products.
This conversion of chemical into mechanical energy reminds one
of the working of a macroscopic combustion engine, except that 
everything is taking place on a molecular scale and thus thermal 
fluctuations must be added to the picture.

Similarly as for the chemical reaction coordinate,
in the simplest case the changes of the geometrical configuration 
can be described by a single mechanical coordinate, originally
living on a circle but easily convertible to a periodic description 
on the real axis.
In the absence of chemical reactions, another Smoluchowski-Feynman 
ratchet dynamics (\ref{2.5}) for the mechanical state variable arises.
One suggestive way to include the effect of the
chemical reaction is the traveling potential scheme (\ref{4.101}),
where $x(t)$ and $f(t)$ are the mechanical and chemical state variables, 
respectively. Thus, the traveling potential proceeds in a preferential
direction in accord with the chemical reaction and thereby is dragging the 
mechanical coordinate along the same preferential 
direction\footnote{Properly speaking, there is also a back-reaction of the 
mechanical on the chemical state variable. A more detailed discussion
of the present modeling framework is given in \ch  \ref{cha5}.}.
Another possibility is that, instead of producing a traveling potential,
the chemical process gives rise to a fluctuating potential to
which the mechanical coordinate is exposed, or an even more general
type of pulsating periodic potential (\ref{6j}).

This general scheme seems to be indeed exploited by nature for numerous
intracellular transport processes \cite{bio83,fri86}. An example
are ``molecular pumps'' (enzymes) in biological membranes, which transfer
ions or small molecules from one side of the membrane to the other
by catalyzing ATP (adenosine triphosphate) into ADP (adenosine diphosphate)
and P$_i$ (inorganic phosphate) \cite{alb94,ast96b}. 
Another example, also fueled by ATP,
are ``molecular motors'' which are able to
travel along intracellular polymer filaments.
A detailed discussion of the latter example will be presented in 
\ch  \ref{cha5}, especially in the final \sect \ref{sec5.8}.
Finally, we remark that in principle nothing speaks against the possibility that
the general scheme could be realized not only for enzymes (proteins)
but also for much simpler kinds of catalysts.

\subsection{Externally driven molecular pumps}\label{sec4.5.1}
Molecular pumps are enzymes that use energy from ATP
hydrolysis to create and maintain concentration gradients of ions
or other small molecules like sugars (glucose)
or amino acids across membranes.
As discussed before, such a chemical process requires that the concentrations
of reactants and products are kept away from their equilibrium ratio.
In living cells, this task is accomplished by intracellular
``energy factories'', maintaining the concentration of ATP
about 6 decades above its thermal equilibrium value.

Experimentally, there exists another interesting option 
\cite{ast96a,ast96b}, namely to suppress ATP hydrolysis either 
by low temperature or by bringing the ATP concentration close to 
its equilibrium value and instead apply an external time-dependent
electric field.
Without the field and in the absence of ATP hydrolysis, we thus
recover the Smoluchowski-Feynman model (\ref{2.5}) for the
mechanical state variable $x(t)$ of the molecular 
pump\footnote{We recall that the mechanical coordinate
represents the geometrical shape of the enzyme. Since ions or molecules
are mechanically transferred through the membrane, a displacement
of the mechanical coordinate monitors at the same time the pumping 
of ions or molecules.}.

Since ATP hydrolysis is suppressed, the chemical state variable, 
previously denoted by $f(t)$, can be omitted in the following 
discussion and the same symbol $f(t)$ is now used for the
external field.
We first consider the case that the field $f(t)$
only couples to the mechanical coordinate $x(t)$ 
of the enzyme but not to the pumped molecule
(e.g. ions are excluded if $f(t)$ is an electrical field).
As a consequence, the effective potential $V(x,f(t))$
experienced by the mechanical
state variable $x(t)$ changes its shape as a function of $f(t)$
but will maintain always the same spatial periodicity.
The corresponding model dynamics is thus of the general form 
(\ref{6j}). Though the detailed shape of the
fluctuating periodic potential $V(x,f(t))$ is usually not known,
the occurrence of a ratchet effect is generically expected for
a very broad class of periodic or
randomly fluctuating external driving signals $f(t)$.
In other words, the molecular pump starts to loop in one or the
other preferential direction and so pumps molecules across the
membrane. An external driving can thus substitute for the
chemical energy from ATP hydrolysis to power the molecular pump,
i.e., $f(t)$ in (\ref{6j}) may represent either the chemical reaction coordinate
or the external driving field, the main consequence 
$\langle\dot x\rangle\not = 0 $ is the same.

The more general case that the external field not only induces a pumping of molecules
across the membrane but also leads to a production of ATP out of ADP and P$_i$
is discussed in \cite{tso86,wes86}. Finally, ATP-driven pumping may also induce
electrical fields in the vicinity of the enzyme \cite{tso86,wes86}.

If the substance transported by the molecular pump
itself couples to the external field $f(t)$, 
e.g. a ion when $f(t)$ is an electrical field,
then the total potential experienced by the mechanical state 
variable $x(t)$ acquires a tilt in addition to the periodic 
contribution.
If the mechanical coordinate $x(t)$ does not couple to the field, 
then the periodic
contribution to the total potential is always the same
and we recover the tilting ratchet scheme from \ch  \ref{cha6}.
If the field affects both the transported substance and the enzyme,
a combined pulsating and tilting ratchet mechanism will result.

For {\em periodic} fields $f(t)$, this effect has been discovered 
{\em experimentally} in \cite{ser83,ser84,liu90}
and explained theoretically in \cite{tso86,wes86,ast89a,ast89b}
by means of a model with a
discretized mechanical coordinate, hopping between
four states at certain rates which change under the 
influence of the field $f(t)$.
Employing the same type of models, the possibility
that also a {\em randomly} fluctuating field $f(t)$
of zero average
may put molecular pumps to work was first predicted in \cite{ast87,che87}
and subsequently verified experimentally
in \cite{xie94,xie97}.
Though these and later discussions \cite{ful94,ful97,ful98}
are conducted mainly in the
language of discrete state kinetic models, the underlying
physical picture is equivalent to the spatially continuous
ratchet paradigm \cite{ast96a,ast96b,ast98,lee99a,tso00}.
Indeed, a plot reminiscent of a fluctuating potential ratchet
(restricted to a fraction of the full spatial period)
appears already in \cite{che87} and a full-fledged
traveling ratchet scheme is depicted in \cite{rob90}.
Note also the close connection to the resonant activation
effect from \sect \ref{sec3.6}.

\chapter{Tilting ratchets}\label{cha6}
\subsection{Model}\label{sec6.0.1}
At the focus of this chapter is the one-dimensional overdamped stochastic dynamics
\begin{equation}
\eta\,\dot x(t) = -V'(x(t)) + y(t) + \xi (t)\ .
\label{6.1}
\end{equation}
Here, as discussed in detail in \sect \ref{sec3.1.1},
$V(x)$ is a $L$-periodic potential, $\xi (t)$ is a white Gaussian noise 
of strength $2\,\eta\, k_BT$, and $y(t)$ is either an unbiased 
$\ttt$-periodic function or an
unbiased stationary random process (especially independent
of $\xi (t)$ and $x(t)$).
With respect to the load force $F$ from (\ref{4a}), we immediately
focus on the case of main interest $F=0$.

According to Curie's principle (\sect \ref{sec2.2.3a}), noise induced 
transport is expected when the system is permanently kept
away from thermal equilibrium and does not exhibit a
spatial inversion symmetry. Within the model (\ref{6.1}),
these requirements can be met in two basic ways:
The first option is an asymmetric ``ratchet-potential'' $V(x)$
in combination with a perturbation $y(t)$ which is
symmetric under inversion $y(t)\mapsto -y(t)$
(see \sect \ref{sec3.1.2} for a precise definition), amounting to a
``fluctuating force ratchet'' if $y(t)$ is a random process,
and to a ``rocking ratchet'' if $y(t)$ is periodic in $t$.
The second option is a spatially symmetric $V(x)$ in combination with a
broken symmetry of $y(t)$, 
called an ``asymmetrically tilting ratchet''.

A few models which go beyond the basic form (\ref{6.1}) are
also included in the present chapter:
Before all, this concerns the discussion of 
photovolatic effects in \sect \ref{sec6.2.1}.
Moreover, the influence of finite inertia is
discussed in \sect \ref{sec6.2.3.2}, while two-dimensional
generalizations are the subject of \sect \ref{sec6.2.4}.

\subsection{Adiabatic approximation}\label{sec6.0.2}
The simplest situation in (\ref{6.1}) arises if
the changes of $y(t)$ in the course of time
are extremely slow \cite{mag93}.
Then, at any given instant $t$, the particle current 
has practically the same value as the steady state current
(\ref{2.23}) for the tilted Smoluchowski-Feynman ratchet (\ref{2.21})
with a static tilt $F=y(t)$. Like in \sect \ref{sec2.4.1},
this so-called {\em adiabatic approximation} thus corresponds to
an accompanying steady state description in which
the time $t$ plays the role of a parameter.

For a {\em periodic} driving $y(t+\ttt )=y(t)$, the time averaged 
current (\ref{4c1}) in the adiabatic approximation thus
follows as \cite{mag93,bar94,kos96}
\begin{eqnarray}
& & \langle\dot x\rangle = \frac{1}{\ttt}\, \int_0^\ttt dt\, v(y(t)) 
= \int_0^1 dh\,  v(\hat y(h )) 
\label{6.2}\\
& & v(y) :=\frac{L\, k_BT\, [1-e^{-L\, y/k_BT}]}
{\eta\,\int_0^L dx \int_x^{x+L} dz\, e^{[V(z)-V(x)-(z-x)\,y]/k_BT}}
\label{6.3}\\
& & \hat y(h) := y(h\ttt)\label{6.3'} \ .
\end{eqnarray}
Similarly as in \sect \ref{sec2.4.1}, it is assumed that apart from the
variation of the time-period $\ttt$ itself, the shape of $y(t)$ does not 
change, i.e. $\hat y(h )$ is a $\ttt$-independent function of $h$.
As a consequence, the right hand side of (\ref{6.2}) is independent of
$\ttt$, in close analogy to \eq (\ref{2.3.1}).
In the zero-temperature limit $T\to 0$, one finds similarly as in 
(\ref{4.107}) that
\begin{eqnarray}
& & v(y,T=0) = 
\left \{ \begin{array}{ll} 
\frac{\eta\, L}{\int_0^L\frac{dx}{y-V'(x)}} & \mbox{if}\ \ 
y\not = V'(x) \ \ \mbox{for all}\ \ x \\
0 & \mbox{otherwise.}
\end{array} \right .
\label{6.4}
\end{eqnarray}
For finite but very small temperatures $T$ this result is only slightly
modified if $y\not = V'(x)$ for all $x$.
In the opposite case, there are ($y$-dependent) solutions $x=\xmax$ and
$x=\xmin$ of $y=V'(x)$  with the property that $\xmax$ maximizes
$V(x)-x y$ within the
interval $[\xmin,\xmin+L]$ and $\xmin$ minimizes $V(x)-x y$ within
$[\xmax-L,\xmax]$, cf. \sect \ref{sec2.2.1.1}. 
From (\ref{2.23y})-(\ref{2.23s})
we can read off that
\begin{eqnarray}
& & v(y) = L\, [k_+ - k_-] \nonumber \\
& & \qquad \ = \frac{L\, |V''(\xmax)\, V''(\xmin)|^{1/2}}{2\,\pi\,\eta}\, 
e^{-\Delta V(y)/k_BT}\, [1-e^{-yL/k_BT}]\label{6.5}\\
& & \Delta V := V(\xmax) - V(\xmin) - (\xmax -\xmin)\, y \label{6.6}
\end{eqnarray}
for sufficiently small temperatures such that
$k_BT\ll\Delta V(y),\Delta V(y)-yL$.

If $y(t)$ in (\ref{6.1}) is an unbiased {\em stochastic} process
with an extremely large correlation time (cf. (\ref{6n1}))
\begin{equation}
\tau := \frac{\int_{-\infty}^\infty dt\,\langle y(t)\, y(s)\rangle}
{2\,\langle y^2(t)\rangle}
\label{6.7}
\end{equation}
then one obtains along the same line of reasoning as in (\ref{6.2}) the
adiabatic approximation \cite{mie95a}
\begin{equation}
\langle\dot x\rangle = \int_{-\infty}^\infty dy\, \rho (y) v(y)\ . 
\label{6.8}
\end{equation}
Here, $\rho(y)$ is the distribution of the noise (cf. (\ref{6k1}))
\begin{equation}
\rho(y) := \langle\delta(y-y(t))\rangle
\label{6.8'}
\end{equation}
and it is required that $\rho(y)$ does not change upon
variation of the correlation time $\tau$. We have encountered this so-called
{\em constant variance scaling} assumption already in \sect \ref{sec4.1a}
and it is obviously the counterpart of the $\ttt$-independence of 
$\hat y (h)$ from (\ref{6.3'}) in the case of a periodic driving $y(t)$.

For general analytic conclusions, the adiabatic expressions (\ref{6.2})
or (\ref{6.8}) are still too complicated, one has to plot concrete
examples with the help of (\ref{6.3}) numerically.
Only in particularly simple special cases one may also be able to directly
predict the direction of the current.
Such an example arises if $y(t)$ can take only two possible values 
$\pm y_0$ with very rare deterministic or random flips, and $V(x)$
exhibits a very simple ratchet profile, 
consisting essentially of one steep and one flat slope
(see e.g. \fig \ref{fig2} or \ref{fig4.1}). Upon increasing
$y_0$, the condition $y\not=0$ for all $x$ in (\ref{6.4}) will be first
satisfied either for $y=y_0$
or $y=-y_0$ with a resulting $T=0$ current in (\ref{6.2})
or (\ref{6.8}) with a sign equal to that of the flat slope.
The intuitive picture is simple:
Out of the two tilted asymmetric potentials $V(x)\mp y_0x$, one does not
exhibit any extrema and thus supports a permanent downhill motion, while the
other still exhibits extrema which act as motion-blocking barriers.
One may go one step further and again decrease $y_0$ until both
$V(x)\mp y_0x$ exhibit potential barriers and thus prohibit 
deterministic motion.
One readily sees that the barrier induced by the steeper slope of $V(x)$
is higher than that induced by the flatter slope.
With (\ref{6.5}) it follows again that for weak thermal noise the current 
goes into the direction of the flat slope of $V(x)$.
Similarly, for an asymmetrically tilting ratchet with only two
possible values $y_\pm$ for $y(t)$ and a symmetric potential $V(x)$,
the sign of the bigger slope $y_\pm$ in modulus dictates the
sign of the current $\langle\dot x\rangle$.

Numerical evaluations 
\cite{mag93,ast94,bar94,mie95a,chi95,zap96,bar96,cec96,bie96c,mil96,keh97,zap98,sch98,wei99,wei00,sar99,lan99}
of the adiabatic expressions (\ref{6.2}) or (\ref{6.8}) for more complicated
drivings $y(t)$ but still relatively ``simple'' potentials like in
\figs \ref{fig2}, \ref{fig4.1} lead to analogous conclusions. 

Another noteworthy feature arises if only small $y$-values are known to play
a significant role in the expression for the adiabatic current 
(\ref{6.2}) or (\ref{6.8}).
For $T=0$ it immediately follows from (\ref{6.4}) that $\langle\dot x\rangle =0$
For $T>0$ and sufficiently small $y$, one may linearize (\ref{6.3})
to obtain
\begin{equation}
v(y) = y \, 
L^2 \left[\eta\int_0^L dx\, e^{-V(x)/k_B T} \int_0^L dx\, e^{V(x)/k_BT} 
\right]^{-1} \ .
\label{6.8''}
\end{equation}
Since $y(t)$ is unbiased, see (\ref{4g}) or (\ref{4h}), one recovers
again $\langle\dot x\rangle =0$ from (\ref{6.2}) or (\ref{6.8}), in agreement
with the general prediction from \sect \ref{sec3.1.4}.

\subsection{Fast tilting}\label{fasttilting}
In the case of a {\em stochastic} process $y(t)$ with a very small
correlation time (\ref{6.7}) one may proceed under the assumption
of constant variance scaling along the same line of reasoning
like for the fast pulsating ratchet scheme in \sect \ref{sec4.1a}.
Thus, we can replace in leading order $\tau$ the random
precess $y(t)$ by a white Gaussian noise with the same
intensity
$\int dt\,\langle y(t)y(0)\rangle = 2\,\tau\,\langle y^2(0)\rangle$.
Like in \sect \ref{sec4.2.2}, the resulting two independent Gaussian
white noises in (\ref{6.1}) can be lumped into a single Gaussian white
noise.
We thus recover an effective Smoluchowski-Feynman ratchet, implying
that in leading order $\tau$ no current $\langle\dot x\rangle$ is
obtained\footnote{Strictly speaking, our argument is only valid for $T>0$.
The conclusion, however, also remains true for $T=0$, see at the end
of \sect \ref{sec6.1}.}.
Since it is not possible to extend the above simple type of argument
to higher orders in $\tau$, such higher order results have to
be derived separately for each
specific type of noise $y(t)$.
Similarly, for {\em periodic} perturbations $y(t)$ one finds zero current in 
leading order of the period $\ttt$ and one has to proceed to higher orders.
The main conclusion of those various expansions, reviewed in more
detail in the next section, is that 
{\em the current $\langle\dot x\rangle$ for fast tilting ratchets vanishes in
leading order and depends on the detailed properties of $y(t)$ in 
higher orders}, i.e. no simple universal results as for the pulsating 
ratchets in \sect \ref{sec4.1a} are accessible.

\subsection{Comparison with pulsating ratchets}\label{sec6.0.3}
From \sect \ref{sec6.0.2} we can infer as a first major difference 
in comparison with the pulsating ratchet scheme that
{\em for tilting ratchets, a finite current $\langle\dot x\rangle$ is
generically observed in the limit of adiabatically slow tilting}.
Since in experiments it is often difficult to go beyond the adiabatic
regime, this feature is an invaluable advantage of the tilting
ratchet paradigm. An interesting exceptional class of asymmetrically
tilting ratchets will be discussed in \sect \ref{sec6.3}.

Our second conclusion is that {\em the ``natural'' current direction
in fluctuating force and rocking ratchets is given by the sign of the
flat potential slope}. Comparison with \sect \ref{sec4.2.1} shows that this
``natural'' direction is just {\em opposite to the ``natural'' direction
in a fluctuating potential ratchet}.
A similar ``natural'' direction can be identified for asymmetrically
tilting ratchets. However, precise criteria of ``simplicity'' such that
this natural current direction is actually realized are not 
available\footnote{Such precise criteria would probably be very 
complicated (in the worst case a huge lookup table)
and thus of little practical use and moreover different for any type 
of ratchet. On the other hand, there will be also many ``complicated'' 
examples which nevertheless lead to a ``natural'' current direction.}.
Opposite current directions can definitely been observed in more 
complicated potentials $V(x)$ and also for ``simple'' potentials 
outside the adiabatic regime.
Examples will be given later and can also been constructed
along the lines of \sect \ref{sec3.5}.

A third major difference in comparison with the fluctuating potential
ratchet model is that {\em thermal noise is not indispensable for the
occurrence of the ratchets effect} provided sufficiently
large tilting forces $y(t)$ appear in (\ref{6.2})
or (\ref{6.8}) such that a finite velocity in (\ref{6.4}) is possible.
This feature is of particular conceptual appeal in the case
of a stationary stochastic process $y(t)$ with unrestricted support
of $\rho (y)$, e.g. a Gaussian distributed noise.
In the absence of the thermal noise $\xi (t)$ in (\ref{6.1}) we
obtain a ratchet effect for a system in a non-equilibrium environment
of archetypal simplicity\footnote{We may always consider
$y(t)+\xi(t)$ in (\ref{6.1}) as a single noise, stemming from one and the
same non-equilibrium heat bath, but for $T\not =0$ this viewpoint is not
very ``natural''.}, see also \ch \ref{cha3}.
Such models have been extensively studied under the
label {\em colored noise problem}, see \cite{han95} for a review.

In \sect \ref{fasttilting} we have found that (within a constant variance 
scaling scheme) the current vanishes in leading order of the
characteristic time scale in the fast tilting limit.
Along a completely analogous line of reasoning one sees that
for a Gaussian noise driven fluctuating potential ratchet
within a constant intensity scaling scheme
the current still vanishes in the white noise limit, while it
remains finite for a fluctuating potential ratchet, see \sect \ref{sec4.2.2}
(for a traveling potential ratchet this limit is not well defined).
In other words, {\em both in the fast and slow driving limits, pulsating and
tilting ratchets behave fundamentally different}.

\section{Fluctuating force ratchets}\label{sec6.1}
In this section we consider the tilting ratchet scheme (\ref{6.1})
with a spatially asymmetric, $L$-periodic potential $V(x)$ and a fluctuating
force $y(t)$ which is given by a stationary stochastic process,
symmetric under inversion $y(t)\mapsto -y(t)$ (in the statistical sense, 
see \sect \ref{sec3.1.2}), hence in particular unbiased (\ref{4h}).

Physically, this gives rise to a model of paradigmatic simplicity
for a system under the influence of a non-thermal heat bath.
Similarly as for the so-called ``colored noise problem'' \cite{han95},
the setup is mainly of conceptual interest, while its direct
applicability to real systems is limited, see also 
\sects \ref{sec3.2.2} and \ref{sec6.1.2}.

As argued in the preceding section, if $y(t)$ is another
Gaussian white noise then we are dealing with an effective
Smoluchowski-Feynman ratchet. Hence, to obtain directed transport one either
has to invoke a correlated (non-white), Gaussian or non-Gaussian 
noise (``colored noise''),
giving rise to a so-called 
{\em correlation ratchet}\footnote{The same name has 
been introduced in \cite{pes94} for a fluctuating potential 
ratchet in our present nomenclature.}
\cite{chi95,bar96},
or a white, non-Gaussian noise.

As far a Gaussian colored noise is concerned, its properties are completely
fixed by the first and second moments 
$\langle y(t)\rangle = \langle y(0)\rangle$ and
$\langle y(t) y(s) \rangle = \langle y(t-s) y(0) \rangle$ 
\cite{kam92,han82b}. Focusing on unbiased stationary examples,
the distribution is thus always given by (\ref{6m}), while the
correlation $\langle y(t) y(s) \rangle$ can be chosen largely 
arbitrarily\footnote{One obvious restriction is that the intensity
$\int dt \, \langle y(t)y(s)\rangle$ and hence the correlation
time in (\ref{6.7}) must not be negative nor infinite.}.
The simplest example is Ornstein-Uhlenbeck noise with an
exponentially decaying correlation (\ref{6k1}).

A standard example of a non-Gaussian colored noise is the symmetric 
dichotomous noise from
(\ref{6k2})-(\ref{6l}). A further example of interest is its generalization
with three instead of two states \cite{els96}, i.e. the noise $y(t)$
can take three possible values, $y(t)\in\{-B,0,B\}$.
The transition rates from $\pm B$ to $0$ are defined as $1/\tau$
and the backward rates from $0$ to $\pm B$ as $\lambda/\tau$,
\begin{equation}
k_{\pm B\to 0}=1/\tau\ ,\ \ k_{0\to\pm B}=\lambda/\tau\ .
\label{6.9a}
\end{equation}
This so-called {\em three state noise} is thus characterized by the three 
parameters  $B ,\tau ,\lambda >0$.
Note that the so defined $\tau$ is proportional but in general not
identical to the correlation time defined in (\ref{6.7}).
The rather lengthy expression for the proportionality factor
follows from a straightforward calculation but is of no further interest for us.
In the context of the above three state noise, $\tau$ will always be
understood as given by (\ref{6.9a}) rather than (\ref{6.7}).
The special case of a dichotomous noise is recovered in the limit
$\lambda\to\infty$.

Finally, so-called {\em symmetric Poissonian shot noise} is defined as
\cite{ric54,han78,han80,vdb83,vdb84}
\begin{equation}
y(t) = \sum_{i=-\infty}^\infty y_i\,\delta (t-\tau_i)\ ,
\label{6.9b}
\end{equation}
where the ``spiking times'' $\tau_i$ are independently sampled
(thus Poissonian) random numbers with average interspike distance
\begin{equation}
\ttt :=\lim_{k\to\infty}\frac{1}{2k+1}\sum_{i=-k}^k (\tau_{i+1}-\tau_i)\ . 
\label{6.9c}
\end{equation}
Furthermore, the spiking amplitudes $y_i$ in (\ref{6.9b})
are random numbers, independent of each other and of the $\tau_i$, 
distributed according to some symmetric distribution $P(y_i)$.
Specifically, we will consider the example
\begin{equation}
P(y_i) = \frac{1}{2A}\, e^{-|y_i|/A} \ ,
\label{6.9d}
\end{equation}
yielding a correlation of the form
\begin{equation}
\langle y(t) y(s) \rangle = 2\, A^2\ttt^{-1}\,\delta(t-s)\ ,
\label{6.9e}
\end{equation}
i.e. this type of shot noise is uncorrelated (white noise) with
two model parameters $\ttt$ and $A$.
Yet, in close analogy to correlated noise
(cf. \sect \ref{sec4.1a} and \ref{sec6.0.2}),
the $\ttt$-dependence of $y(t)$ 
is of the form $\hat y(t/\ttt)$ with a suitably defined,
$\ttt$-independent Poissonian white shot noise $\hat y(t)$.

Note that a similar (but asymmetric) type of Poissonian shot noise has
already been encountered in (\ref{4.123})-(\ref{4.124'})
and will later appear again in the asymmetrically tilting ratchet
scheme in \sect \ref{sec6.3.2}. Throughout the present review,
Poissonian shot noise (symmetric or not) will be employed
as an interesting abstract example process of archetypal simplicity.
For concrete applications in various contexts of electronic devices and
solid state physics see for instance \cite{kog96}.
For models of chemical reactions and other transport processes 
in gases we refer to \cite{for73}.
We furthermore remark that the above Poissonian symmetric
shot noise can be recovered \cite{vdb83}
as a limiting case of the
three state noise (\ref{6.9a}) if
\begin{eqnarray}
& & \tau\to 0\ ,\ \ \lambda\to 0\ ,\ \ B\to\infty\nonumber\\
& & \ttt :=\tau/2\lambda \ , \ \ A:= \tau\, B \ \ \mbox{fixed} \ .
\label{6.9f}
\end{eqnarray}

\subsection{Fast fluctuating forces}\label{sec6.1.1}
We first address the case of a {\em correlation ratchet} (colored noise $y(t)$)
in the regime of small correlation times $\tau$ in (\ref{6.7}).
Examples are a dichotomous noise or an Ornstein-Uhlenbeck noise, cf. 
(\ref{6k2})-(\ref{6m}).
As mentioned before (see \sect \ref{fasttilting}), 
a simple leading-order $\tau$ argument as for the 
pulsating ratchet scheme in \sect \ref{sec4.1a} yields the
trivial result $\langle\dot x\rangle =0$, i.e. the correlation ratchet is in some
sense reluctant to obey Curie's principle in the fast noise regime.
Higher order $\tau$ contributions require a separate perturbation calculation
for each type of noise $y(t)$, similar in spirit as the example in
Appendix C.

The result of such perturbation calculations for 
various types of noises $y(t)$, among others symmetric dichotomous noise,
three-state noise, and Ornstein-Uhlenbeck noise, can be written in the general form
\cite{doe94,doe95,mie95a,els96,doe98}
\begin{equation}
\langle \dot x \rangle = - \frac{\tau^3\, L \,\langle y^2(0)\rangle}{\eta\, k_B T}
\,
\frac{ Y_1\frac{\langle y^2(0)\rangle}{(k_B T)^2}\,\int\limits_0^L dx\, [V'(x)]^3  +
Y_2\,\int\limits_0^L dx\, V'(x)\, [V''(x)]^2 }
{\int_0^L dx\, e^{V(x)/k_BT}\,\int_0^L dx\, e^{-V(x)/k_BT}} \ ,
\label{6.9}
\end{equation}
where $Y_{1,2}$ are dimensionless and $\tau$-independent characteristic
numbers of the specific noise $y(t)$ under consideration. For instance,
for a dichotomous process one has \cite{doe95,mie95a,doe98}
\begin{equation}
Y_1^{(DN)} = 1 \ \ , \ \ \ Y_2^{(DN)} = 1  \ .
\label{6.10a}
\end{equation}
For Ornstein-Uhlenbeck noise one can infer from
\cite{doe94,mil94,mie95a,bar96,bao96a,bao96b} that
\begin{equation}
Y_1^{(OU)} = 0 \ \ , \ \ \ Y_2^{(OU)} = 1  \ .
\label{6.10b}
\end{equation}
For the three-state noise from (\ref{6.9a}) one has \cite{els96}
\begin{equation}
Y_1^{(3)} = [3\phi -\phi^2 -1]/\phi^3 \ \ , \ \ \ Y_2^{(3)} =  1/\phi \ ,
\label{6.10c}
\end{equation}
where the so-called ``flatness'' is defined as 
\begin{equation}
\phi:= \langle y^4(0)\rangle/\langle y^2(0)\rangle^2  \ .
\label{6.10d}
\end{equation}
For the specific case of the three-state noise in (\ref{6.9a})
one obtains the result $\phi = 1+1/2\lambda$, which has
to be substituted in (\ref{6.10c}).

The following assumptions are crucial for the validity of (\ref{6.9}):
(i) constant variance scaling of the colored noise $y(t)$;
(ii) finite thermal noise strength $T>0$;
(iii) smooth potential $V(x)$.
It is not proven but may be expected as an educated guess
that the general form (\ref{6.9}) of the small-$\tau$ asymptotics
remains valid even beyond the various examples of colored noises
$y(t)$ so far covered in \cite{doe94,doe95,mie95a,bar96,els96,doe98}.

Turning to the case of the symmetric Poissonian white shot noise
(\ref{6.9b}), (\ref{6.9d}) one readily recovers the asymptotic 
behavior \cite{luc97}
for small characteristic times in (\ref{6.9c})
from the behavior of the three-state noise (\ref{6.9}), (\ref{6.10c})
in the limit (\ref{6.9f}).
Remarkably, the result is then again given by the same formula as in (\ref{6.9})
if one makes the ``natural'' replacement
\begin{equation}
\tau^3\,\langle y^2(0)\rangle^2 \, \mapsto \, \ttt^3 A^4
\label{6.10d'}
\end{equation}
and with 
\begin{equation}
Y_1^{(shot)} = -1 \ \ , \ \ \ Y_2^{(shot)} = 0 \ .
\label{6.10e}
\end{equation}

Our first observation in (\ref{6.9}) is that
$\langle\dot x\rangle$ vanished not only in leading order $\tau$, as already
mentioned above, 
but also in second order, i.e. {\em the fast fluctuating force ratchet 
is very reluctant to produce a current}.
Second, the functional dependence on the potential $V(x)$ in (\ref{6.9}) becomes
identical to the fluctuating potential asymptotics in (\ref{6q1}) when
$Y_2\to 0$ (e.g. for shot noise) and identical to
the asymptotics for the temperature ratchet in (\ref{2.3.2}) when $Y_1\to 0$ (e.g. for
Ornstein-Uhlenbeck noise).
This comparison gives also a quantitative flavor about the necessary caution
to be observed when comparing
``natural'' directions in fluctuation force and fluctuating potential ratchets.

Regarding the quantity $Y_1$ in (\ref{6.9}), it has been conjectured in
\cite{doe94,doe95} that, 
for a rather general class of colored noises $y(t)$, it is given by 
a simple function of the flatness (\ref{6.10d}),
e.g. $Y_1=2 - \phi$ for dichotomous and Ornstein-Uhlenbeck
noise, and by (\ref{6.10c}) for the three-state noise (\ref{6.9a}).
So far neither a proof nor a counterexample seems to be known.
The coefficient $Y_2$ depends in general on additional details of the noise $y(t)$.
E.g. for Gaussian noise (\ref{6m}) but with a correlation which is {\em not}
given by the pure exponential decay (\ref{6l}), the flatness 
in (\ref{6.10d}) is obviously always the same,
while the expression for $Y_2$ is 
in general different from the one in (\ref{6.10b}),
as can be concluded from \cite{mil94,bar97} (see also footnote
5 on page \pageref{fot115} below).

The direction of the current in (\ref{6.9}) is determined by the characteristics
$Y_{1,2}$ of the noise $y(t)$ and the integrals over $[V'(x)]^3$ 
and $ V'(x)\, [V''(x)]^2$.
The latter fact makes once more explicit the warning from \sect \ref{sec3.5}
that beyond the most primitive potential shapes, there exists
no simple rules and natural directions any more, the sign of
$\langle\dot x\rangle$ depends on all the details of $V(x)$ \cite{bar96}. 
Another surprising observation \cite{doe94} is that
{\em a current inversion solely upon changing the statistical properties
of the noise $y(t)$ is possible}.
An example \cite{els96} is the
three-state noise (\ref{6.9a}) which in the shot noise limit (\ref{6.9f})
gives, according to (\ref{6.10e}), rise to a current direction in (\ref{6.9})
opposite to that for the dichotomous noise limit $\lambda\to\infty$
(see (\ref{6.10a})), at least
for $\langle y^2(0)\rangle\gg k_BT$. In fact, when
$\langle y^2(0)\rangle\gg k_BT$ such an inversion upon 
changing the noise statistics will
occur for {\em any} potential $V(x)$ due to the factorization of the noise- 
and potential-properties in the numerator of
(\ref{6.9}) and is thus of a very different nature than the inversion-tailoring
procedure from \sect \ref{sec3.5}.

For the case of Ornstein-Uhlenbeck noise $y(t)$, the existence
of rather simple looking potentials $V(x)$ has been first pointed out 
in \cite{bar96} which give rise in the adiabatic limit $\tau\to\infty$
to a current $\langle\dot x\rangle$ in the
corresponding ``natural'' direction (see section \ref{sec6.0.3}), 
but in the opposite direction in the small-$\tau$ limit 
according to (\ref{6.9}), (\ref{6.10b}).
As a consequence, a current inversion upon variation of 
the correlation time $\tau$ has been predicted theoretically
and verified by precise numerical results in \cite{bar96}.
An analogous theoretical prediction and numerical verification
in the case of dichotomous noise $y(t)$ is due to \cite{rei96ra}.
Considering that for simple (saw-tooth-like, but smooth) potentials $V(x)$, 
the ``natural'' current direction will definitely be recovered in the
adiabatic limit $\tau\to\infty$ (cf. section \ref{sec6.0.2}),
a current inversion as a function of the correlation time $\tau$ follows
also for a three-state noise with suitably chosen parameters in 
(\ref{6.9a}) \cite{bie96c}, see also \cite{mie95a,els96,koh98,man00}.
A similar conclusion was reached in \cite{ber97} for a 
modified three-state noise $y(t)$ with broken
symmetry by cycling through the
three states in a preferential sequence (see also \cite{li97a}).
We remark that the three-state noise $y(t)$ from \cite{ber97}
is supersymmetric according to (\ref{ss6}),
hence $V(x)$ must not be supersymmetric (but may still be symmetric)
in order that $\langle\dot x\rangle \not = 0$.

We recall that mere the existence
of current inversions as exemplified above 
are just special cases of our general
current inversion tailoring procedure from section \ref{sec3.5}.
For a more detailed quantitative control of the effect, 
analytical approximations as exploited above are however invaluable.

We conclude our discussion of the fast potential fluctuation asymptotics
with some remarks regarding the validity conditions (i)-(iii) 
mentioned below (\ref{6.10d}).
First, if the potential $V(x)$ is not smooth,
then the second integral in (\ref{6.9}) is ill-behaving.
The adequate small-$\tau$ analysis becomes much more involved
and yields an ``anomalous'' $\tau^{5/2}$ leading order behavior 
\cite{doe95,doe98,klo99}.
Paradoxically, a piecewise linear saw-tooth 
potential (\fig \ref{fig4.1}), originally
introduced as a stylized approximation of more realistic, smooth potentials
in order to simplify the mathematics, actually makes the calculations
more difficult for $\tau\to 0$.
Second, we remark that while we are exclusively using here a constant variance
scaling for the noise $y(t)$,
in the literature on the small-$\tau$ asymptotics a constant intensity
scaling is often (but not always) employed.
Third, in the case $T=0$, which we excluded so far,
one finds within the realm of constant variance scaling
that for small $\tau$ the current $\langle\dot x\rangle$ 
approaches zero faster than any power of $\tau$
(for constant intensity scaling see \cite{koh98}).

\subsection{Specific types of fluctuating forces}\label{sec6.1.2}
Beyond the fast and adiabatically slow fluctuating force limits,
there has been a great variety of analytical and numerical
studies. We restrict ourselves to a brief overview
of the main analytical results and numerically observed effects
together with the few so far suggested or actually
realized experimental systems. 
For a more detailed discussion of the interesting special features
in each particular case we refer to the cited works.

\subsubsection{Dichotomous noise}
For a dichotomous process $y(t)$ (see (\ref{6k2}), (\ref{6l})) closed, though
not very transparent analytical solutions are possible for $T=0$ and
arbitrary $V(x)$ \cite{doe94,mie95a,kul96,zap98} and
for arbitrary $T$ and
piecewise linear $V(x)$ \cite{ast94,mie95a,kul98a,kul98b}
along the same lines as for the fluctuating potential scheme 
described in \sect \ref{sec4.2.1}.

For $T=0$ one sees from (\ref{6.4}), (\ref{6.8}), (\ref{6k2}) that
in the adiabatic limit the current vanishes for small amplitudes $\sigma$ of $y(t)$.
Upon increasing $\sigma$, the current $\langle\dot x\rangle$ as a function
of $\sigma$, sets in continuously but with a jump in its derivative
when one of the two effective potentials $V(x)\mp \sigma\, x$ in (\ref{6.1})
ceases to exhibit barriers against overdamped deterministic motion.
A similar discontinuous derivative appears when the extrema of the
other effective potential disappear. 
Upon adding in (\ref{6.1}) a load force $F$ (and keeping $\sigma$ fixed),
two analogous jumps in the ``differential resistance''
$\partial \, \langle\dot x\rangle /\partial F$ arise, while $\langle\dot x\rangle$
itself is always continuous.
The same features are recovered \cite{zap98,nik98} not only in the adiabatic limit but
for any finite correlation time $\tau$, basically because the noise $y(t)$ may remain
with small but finite probabilities
in the same state $+\sigma$ or $-\sigma$ for arbitrary long times.

If $\langle\dot x\rangle\not =0$ for $T=0$, then a straightforward perturbation 
expansion for small but finite $T$ is possible with the expected result of small
corrections to the unperturbed result $\langle\dot x\rangle$. More
challenging is the case that $\langle\dot x\rangle = 0$ for $T=0$, calling
for a so-named singular perturbation theory for small $T$,
see \sect \ref{sec3.6}.
This task has been solved in \cite{rei96ra} by a rate calculation based on
WKB-type methods which become asymptotically exact for small $T$ for both,
{\em arbitrary} correlation times $\tau$ and {\em arbitrary} (smooth)
potentials $V(x)$. The connection between the
rates obtained in this way and the
current then follows as usual from (\ref{4i}), yielding
a very good agreement with accurate numerical results \cite{rei96ra}.

An {\em experimental} ratchet system with additive dichotomous
fluctuations has been proposed by way of 
combining in an electric circuit two components 
that will both be discussed separately in more detail below:
On the one hand, an asymmetric dc-SQUID (superconducting quantum interference device)
threatened by a magnetic flux gives rise to an effective ratchet-shaped potential 
for the phase, see \cite{zap96} and \sect \ref{sec6.2.4a}.
On the other hand, a point contact with a defect, tunneling
incoherently between two states, can act as a source of dichotomous noise,
see \cite{zap98} and \sect \ref{sec6.3.2}.
Studies based on an experimental analog electronic circuit
have been performed for negligibly small thermal noise $T\to 0$ 
both in the overdamped limit as well as in the presence of a 
finite {\em inertia} term $m \ddot x(t)$  
on the right hand side of (\ref{6.1}) in \cite{pos96,nik98}.
Inertia-like effects have also been theoretically addressed, both for
dichotomously fluctuating potential and fluctuating force ratchets,
in \cite{ari98}.

\subsubsection{Gaussian noise}
The simplest type of Gaussian noise $y(t)$ is Ornstein-Uhlenbeck noise,
characterized by an exponentially decaying correlation (\ref{6l}), (\ref{6m}).
A first, numerical study of the corresponding correlation ratchet dynamics (\ref{6.1})
has been reported in \cite{mag93}, recognizing as main difference in comparison, e.g. 
with dichotomous noise, the fact that even in the absence of the thermal noise
($T=0$), a ratchet effect $\langle\dot x\rangle\not =0$ arises generically for any
finite intensity\footnote{The reason is the infinite support of the distribution
$\rho(f)$ in (\ref{6m}) as compared to
the bounded support e.g. for dichotomous noise in (\ref{6k2}),
thus the potential barrier cannot block completely any transport.}
of $y(t)$.

The case $T=0$ has been further studied analytically for small 
$\tau$ in \cite{doe94,bar96,bao96a,bao96b}
and especially in \cite{mil94}, indicating that even within the
restricted class of {\em Gaussian}
colored noises $y(t)$, the direction of the current
may change solely by modifying the statistical properties
of this Gaussian noise\label{fot115}\footnote{For an unbiased, 
stationary Gaussian process,
the statistical properties are completely determined by its correlation 
$\langle y(t) y(s)\rangle = \langle y(t-s) y(0)\rangle$. 
While $Y_1 =2-\phi$ in (\ref{6.9})
is always zero, $Y_2$ may change its sign upon modification of the correlation. 
The prediction from \cite{mil94} is that the
sign of $Y_2$ is given by that of 
$\int dt\, t^2\, \langle y(t) y(0)\rangle$.}.
This prediction has been numerically corroborated and extended to the finite-$T$ 
regime in \cite{bar97,bar98}, revealing moreover multiple current 
inversions beyond the realm of small $\tau$.
Additional details of the $T=0$ case have been studied
theoretically in \cite{dia97,bao99a,bao99b,arr00,cor00}
and by means of an experimental analog electronic circuit in \cite{arr00}.

Very accurate numerical results over extended parameter regimes
as well as two different
analytical approximations for {\em arbitrary} (smooth) potentials $V(x)$
and Ornstein-Uhlenbeck noise of {\em arbitrary} correlation time $\tau$
in the activated barrier crossing regime (i.e. weak noises $\xi (t)$ and $y(t)$)
are contained in
in \cite{bar96}.
These approximations exploit the connection (\ref{4i}) between the
particle current $\langle \dot x\rangle$ and the rate expressions from a path-integral
\cite{rei95ra2} and a so-called generalized unified colored noise approximation
\cite{mad95,bar95}, originally derived in the context of the so-called ``resonant
activation'' effect.
While the path integral method yields qualitatively the correct behavior
over the whole $\tau$ regime, including the occurrence and location
of current inversions, the generalized unified colored noise approximation
is limited to small $\tau$ values, where it is superior to the path integral
approach.
Supplementary studies along the same lines with particular emphasize
on the above mentioned accurate 
numerical methods and results are contained in \cite{bar97b,bar98}

For tilting ratchets driven by Ornstein-Uhlenbeck noise $y(t)$,
several groups have studied in detail the {\em effect of finite inertia}, i.e. if 
on the right hand side of (\ref{6.1}) a term $m\ddot x(t)$ is 
included\footnote{The case without a white Gaussian noise $\xi(t)$
but instead with an Ornstein-Uhlenbeck noise $y(t)$, an
additional periodic (rocking) force, and possibly also a memory 
friction (cf. \sect \ref{sec4.8.4c}) has been considered numerically in 
\cite{iba97}.}
\cite{lin97,iba97,mar98,lan98,lin99,lan01}.
Analytically, this problem represents a considerable technical challenge
and the results of various approximative approaches are not always
compatible.
The upshot of those analytical as well as numerical explorations is the 
convincing demonstration that also the particle mass is a parameter, upon
variation of which the current may change sign,
i.e. a mass-sensitive particle separation is feasible.
Similar conclusions have been reached in \cite{nik98,ari98}
for dichotomous instead of Ornstein-Uhlenbeck noise $y(t)$.

\subsubsection{Shot noise}
The symmetric Poissonian shot noise (\ref{6.9b}) is of interest for several reasons.
First, it demonstrates that the appearance of a net current in the fluctuating 
potential
scheme (\ref{6.1}) it is {\em not} necessary that the noise $y(t)$ is correlated 
in time \cite{luc97}.
Second, its ``natural direction'' is typically opposite to that of
correlated noise $y(t)$ in the adiabatic limit.
E.g. in a saw-tooth potential $V(x)$, the current direction
turns out to have the same qualitative features as for the 
on-off saw-tooth potential \cite{luc97} treated in 
\sect \ref{sec4.1}
if one identifies the characteristic time $\ttt$ from (\ref{6.9c}) with the
correlation time in the on-off scheme.
An intuitive explanation of this {\em prima facie} astonishing similarity
follows from the discussion of the three-state noise in \cite{doe94} in
combination with its shot noise limiting behavior according to (\ref{6.9f}).
Since for shot noise there is no correlation time, and the noise distribution
$\rho(y)$ is not well defined, an adiabatic limit in the sense
of (\ref{6.8}) does not exist. The regime of a slow time scale $\ttt$ in 
(\ref{6.9c}) is therefore of a fundamentally different nature.
Again analogous to the on-off scheme,
one finds \cite{luc97} that the current approaches zero as $\ttt$ becomes
very large (both for constant variance and constant intensity scaling).

\section{Photovoltaic effects}\label{sec6.2.1}
In this section we discuss {\em experimental} ratchet systems
which cannot be realistically captured by 
the simple model (\ref{6.1}) but are is physically
closely related to it.

In {\em non-centrosymmetric materials}, photocurrents are induced by short-wavelength 
irradiation (optical or x-ray illumination) in the absence of
any externally applied fields \cite{bal81}.
Experimental observations of this so-called
{\em photovoltaic effect}\footnote{Practically synonyms are 
``photorefractive effect'' and ``photogalvanic effect''.} 
in ferroelectrics, piezoelectrics, and pyrroelectrics such as
BaTiO$_3$ or LiNbO$_3$ can be traced back at least to the mid 60-s.
The basis of its correct theoretical explanation was laid 1974 by 
Glass, von der Linde, and Negran \cite{gla74},
recognizing that it is not a surface or interface effect -- in contrast
e.g. to photovoltaic effects occurring in n-p junctions 
(see \sects \ref{sec4.8.1} and \ref{sec6.5.5}) --
but rather a bulk phenomenon with the {\em asymmetry of the crystal 
lattice\footnote{Most of these systems exhibit a spatial periodicity,
but this is {\em de facto} not an indispensable prerequisite in this
context.} playing a central role}.
Furthermore, they already touch upon the points that
the absence of thermal equilibrium is another crucial precondition,
that the effect should be a general property of a large class of 
materials\footnote{Examples are monocrystalline piezoelectric materials,
such as ferroelectric ceramics, or liquids and gases showing natural optical
activity due to a chirality of their constituent molecules.
More recent systems are provided by asymmetric
semiconductor superlattices and heterostructures \cite{mag01}.},
and that the effect may be an attractive new
method of energy conversion in large-area pyroelectric
polymers or ceramics, acting e.g. as ``solar cells''.

These basic ideas have been subsequently developed into a full-fledged
theory by Belinicher, Sturman, and others. Several
hundred experimental and theoretical papers on the subject are 
reviewed in \cite{bel80,stu92} and various general conclusions
therein are remarkably similar to those
of our present paper. For instance, the counterpart of the
Smoluchowski-Feynman {\em Gedankenexperiment} in this context corresponds
to the question why a steady state photovoltaic effect cannot exist
under isotropic thermal blackbody irradiation. To answer such questions it is
pointed out\footnote{In the following we are quoting
from \cite{stu92}, but most of these statements can be found 
 already  in \cite{bel80}.} that in the absence of
\begin{quote}
``gradients in concentration, temperature, or light intensity ...
the current direction is controlled ... by the internal symmetry.
It constitutes the generation of a directed current in a
uniform medium on homogeneous illumination ... in any
medium (without exception) that lacks a center of symmetry ...
The absence of a center of symmetry ... results in a
current in virtually any nonequilibrium stationary state.
There is no current in thermodynamic equilibrium, in accordance with the
second law of thermodynamics ...
Under the nonequilibrium conditions provided by illumination,
that detailed balancing mechanism is violated and the asymmetry
in the elementary processes gives rise to a current ...
The photovoltaic effect is a kinetic effect and thus
has various extensions. Uniform illumination in the absence
of a center of symmetry may produce not only an electrical current 
but also fluxes of other quantities:
heat (photothermal effect), neutral particles, spin, etc. On the other hand,
light beams do not exhaust the nonequilibrium sources.
... The disequilibrium may be not only due to light but to sound or 
to colliding or isotropic particle fluxes etc.''
\end{quote}

Another point already recognized in various studies during
the 70-s and reviewed in \cite{bel80,stu92} is the
fact that the photovoltaic effect is a {\em nonlinear effect in the
irradiation field amplitude}, no current arises within the
realm of linear response (cf. \sect \ref{sec6.1.1}).
Furthermore, {\em current reversals} upon changing
the frequency or polarization of the irradiation \cite{koc75,asn79}
and upon changing the temperature \cite{asn79} have
been observed in this context.

The microscopic theoretical analysis is conducted in terms of electron
scattering processes in solids \cite{bel80,stu92} and goes beyond
our present scope. Though such an approach has little in common with our
present working model (\ref{6.1}), it is remarkable
that veritable {\em one-dimensional effective ratchet potentials}
exactly like in \fig \ref{fig1} are appearing in the discussion
along these lines.
We mention that it is not immediately obvious whether 
the effects of the irradiation,
treated on an adequate quantum mechanical level,
should be associated with a fluctuating force or rather with a 
rocking ratchet scheme:
On the one hand, besides the direct interaction with the electrons,
there may also be non-negligible effects of the
irradiation on the host material,
giving typically rise to a fluctuating potential 
ratchet mechanism \cite{pro94}.
On the other hand, the naive viewpoint that a signal, which is typically a
monochromatic electrical wave, induces an electrical current
suggests that the classification as a rocking ratchet -- as adopted
in the following -- may be justified.

The photovoltaic effect is practically exploited in holography, beam
amplification and correction, wavefront reversal etc. \cite{stu92}.
Basic research activity has somewhat decreased in recent
years, focusing e.g. on the so-called {\em mesoscopic photovoltaic effect}, 
where random impurities in conductors or microjunctions imitate
local symmetry breaking \cite{fal89,liu92},
on x-ray induced giant photovoltaic effects \cite{dal95},
and on photovoltaic effects in asymmetric semiconductor
heterostructures and superlattices \cite{mag01}.

Another variation of the photovoltaic effect has been theoretically
studied in \cite{kra93,aro93}.
Namely, in a mesoscopic, disordered normal-metal ring,
a breaking of the inversion symmetry can be achieved
by a static magnetic flux threatening
the ring, which survives even after averaging over
the quenched disorder of the individual samples.
As theoretically predicted in \cite{kra93,aro93}, in such a setup the
non-linear response to an additional high-frequency electromagnetic field 
is a {\em directed ring-current}.
While somewhat similar ``persistent currents'' may also exist
at thermal equilibrium, i.e. in the absence of the high-frequency
field, only away from equilibrium these currents can be exploited to do work,
i.e. we are dealing with a veritable ratchet effect.
Note that the basic ingredients are
remarkably similar to the SQUID ratchet systems from \sects 
\ref{sec6.2.3.1} and \ref{sec6.2.4a},
but the detailed physical mechanisms are completely different.

Finally, worth mentioning in this context is also
the generation of directed photocurrents in undoped, bulk semiconductors 
with an {\em intact centrosymmetry} by adjusting the relative phases of
{\em two} optical beams at frequencies $\omega$ and $2\omega$,
see \cite{shm85,ent89,ata96,hac97,ale99} and further
references therein and also the discussion at the end of
\sect \ref{sec6.5.4} below.
Such a modified photovoltaic effect leads us beyond the realm 
of the rocking ratchet scheme and will be treated in more detail under 
the label {\em asymmetrically tilting ratchets} in 
\sects \ref{sec6.3.1} and \ref{sec6.5.4}.

\section{Rocking ratchets}\label{sec6.2}
In this section we address the tilting ratchets dynamics 
(\ref{6.1}) with an $L$-periodic, asymmetric potential $V(x)$ and
a $\ttt$-periodic, symmetric external driving force $y(t)$.

\subsection{Fast rocking limit}\label{sec6.2.2}
In contrast to the slow rocking limit (adiabatic approximation),
the regime of very high frequencies has turned out to be rather
obstinate against analytical approximations or intuitive explanations.
Attempts have been made \cite{pla98,sch98,mil99} but cannot be
considered as fully satisfactory.
Numerical results, on the other hand, show \cite{bar94}
as a quite remarkable feature that in the fast rocking regime,
the ``natural'' current direction (i.e. the one realized for
``simple'' potentials $V(x)$ sufficiently similar to the asymmetric
saw-tooth potential from \fig \ref{fig4.1}) 
is just opposite to the one for slow rocking.
In order to finally conclude this issue, we sketch in the following
the main steps of an analytical solution of the fast rocking
asymptotics (details of this calculations will be 
presented in \cite{rei00a}).

Under the assumption that the $\ttt$-periodic function $y(t)$ in (\ref{6.1})
is of the form (\ref{6.3'}),
the asymptotics of the current $\langle\dot x\rangle$ in
(\ref{6.1}) for small $\ttt$ can be in principle determined along the same lines
as in Appendix C. In practice, the calculations become extremely
tedious since, as we will see, to obtain the first non-trivial 
contribution to the current,
one has to go up to the 4th order in $\ttt$.
Things can be simplified a lot by mapping (\ref{6.1}) onto
an equivalent improper traveling potential ratchet dynamics 
(cf. (\ref{4.104})) as follows: With the definition
\begin{eqnarray}
X(t) & := & x(t) - \frac{\ttt}{\eta}\, \hat y_1 (t/\ttt)\ ,
\label{6.21}
\end{eqnarray}
where (cf. (\ref{6.3'}))
\begin{eqnarray}
\hat y_0(h) & := & \hat y(h) = y(h \ttt )
\label{6.21'}\\
\hat y_i (h) & := & \int_0^h ds\, \hat y_{i-1} (s) + \int_0^1 ds\, s\, \hat y_{i-1} (s)
\ \ , \ \ \  i=1,2,...
\label{6.22}
\end{eqnarray}
one readily finds from (\ref{6.1}) that
\begin{equation}
\eta\, \dot X(t) = - V'\left(X(t)+\frac{\ttt}{\eta}\hat y_1 (t/\ttt)\right) 
+ \xi (t) \ .
\label{6.23}
\end{equation}
Since the relations $\hat y_i (h+1)=\hat y_i (h)$ and $\int_0^1 dh\, \hat y_i (h) = 0 $ 
are fulfilled for $i=0$, it follows with (\ref{6.22}) by induction that
the same relations are respected for $i=1,2,...$.
Using the self-averaging property
(\ref{4c1}) of the particle current, 
we can thus infer from (\ref{6.21}) that
\begin{equation}
\langle \dot x\rangle = \langle \dot X\rangle \ .
\label{6.24'}
\end{equation}

After expanding on the right hand side of (\ref{6.23})
\begin{equation}
V'\left( X(t)+\frac{\ttt}{\eta}\hat y_1 (t/\ttt)\right) = \sum_{k=0}^\infty
\frac{V^{(k+1)} (X(t))}{ k!} 
\left[\frac{\ttt \hat y_1 (t/\ttt)}{\eta}\right]^k \ ,
\label{6.24}
\end{equation}
one sees that in comparison with (\ref{6.1}) we have
``gained'' one order of $\ttt$, the ``perturbation'' in (\ref{6.24}) is
of leading order $\ttt$ only.
Due to this simplification, the approach from Appendix C is now applicable
with a reasonable effort and yields the leading-$\ttt$ result \cite{rei00a}
\begin{eqnarray}
& & \langle \dot x \rangle =
\frac{2\,\ttt^4 L\, Y\,\int_0^L dx\, V'(x)\, [V'''(x)]^2}
{\eta^5\,\int_0^L dx \, e^{V(x)/k_BT}\,\int_0^L dx \, e^{-V(x)/k_BT}}
\label{6.25}\\
& & Y:= \int_0^1 dh\,  [\hat y_2 (h)]^2 \ .
\label{6.26}
\end{eqnarray}
Here, we have exploited that $y(t)$ in (\ref{6.1}) is symmetric
(cf. (\ref{s2})), otherwise additional terms of order $\ttt^4$ would 
appear in (\ref{6.25}), see \sect \ref{sec6.3.1} below.

Our first conclusion from (\ref{6.25}) is that the {\em fast rocking 
ratchet is exceedingly reluctant to produce a current},
all contribution up to the order $\ttt^3$ are zero.
This fact suggest that also a simple intuitive explanation of the
current direction may be very difficult to figure out.
Second, for sufficiently simple (saw-tooth-like but smooth)
potentials $V(x)$, the sign of the current in (\ref{6.25})
is dictated by that of the steeper slope of $V(x)$,
and this independently of any further details of the driving $y(t)$.
Our result (\ref{6.25}) thus correctly reproduces the numerical 
observation \cite{bar94}
that {\em the ``natural'' current direction of the fast
and slow rocking ratchets are opposite}.
In other words, a current inversion upon variation
of $\ttt$ is typical (``natural'') in rocking ratchet
systems at finite temperatures $T>0$.

We finally remark that -- much like in the approximation 
(\ref{2.3.2}) for the temperature ratchet -- the limits
$\ttt\to 0$ and $T\to\infty$ do not commute, i.e.
(\ref{6.25}) is not valid for a fixed (however small) $\ttt$ 
if one lets $T\to\infty$, cf. \sect \ref{sec3.1.4}. 
In the special case of a fast sinusoidal driving $y(t)$
with asymptotically small amplitude
our result (\ref{6.25}) reproduces the one from 
\cite{mil99}.
Also worth noting is that (\ref{6.25}) is strictly 
quadratic in the driving amplitude (see (\ref{6.26})).
Deviations from this strictly quadratic behavior are
expected only in the next-to-leading order $\ttt$ contributions
that have been neglected in (\ref{6.25}).
For this reason, the limit of asymptotically large driving amplitudes
can once again not be interchanged with the fast 
driving limit $\ttt\to 0$.

\subsection{General qualitative features}\label{sec6.2.3}
A first remarkable feature of a periodically rocked ratchet
dynamics (\ref{6.1}) occurs if in the {\em deterministic limit}
($T\to 0$). Namely, the current $\langle\dot x\rangle$ as a 
function of the rocking amplitude $y(t)$, but also
as a function of other parameters,
displays a complex structure of constant ``plateaux'' which are
separated by discontinuous jumps 
\cite{mag93,bar94,ajd94a,han96,zap96,dia97,sch98}.
For a qualitative explanation we first note that the current
$\langle\dot x\rangle$, understood as a long time average (\ref{4c1}),
is independent of the initial condition\footnote{This property
readily follows from the fact that $x(t_0)$ and $x(t_0)+L$ obviously
lead to the same $\langle\dot x\rangle$ and that different trajectories
$x(t)$ cannot cross each other \cite{ajd94a}.}
$x(t_0)$ \cite{ajd94a}.
The emergence of the current-plateaux can be analytically
understood in detail for a saw-tooth potential $V(x)$ and a
driving which periodically jumps between a few discrete values 
\cite{ajd94a,sch98}, while in more complicated cases numerical 
solutions must be invoked \cite{mag93,bar94,zap96}.
Very loosely speaking, the deterministic dynamics (\ref{6.1}) with
periodic $y(t)$ and $\xi(t)\equiv 0$ is equivalent to a 
two-dimensional autonomous dynamics and thus admits as attractors 
generalized fixed points and periodic orbits, where the word ``generalized'' refers
to the fact that we identify $x$ and $x+L$ as far as the
attracting set is concerned.
Thus, in the long time limit, that is, after transient effects have died
out, the particle is displaced by some multiple $m$ of the spatial 
period $L$ after a certain multiple $n$ of the time period $\ttt$,
i.e. 
\begin{equation}
\langle\dot x\rangle = (L/\ttt)\ (m/n) \ .
\label{6.26a}
\end{equation}
Remarkably, though $\langle\dot x\rangle$ is independent of the
initial condition, several generalized periodic attractors (with the
same $m/n$) may still coexist \cite{bar94}.
The structural stability of these attractors implies 
that as a function of various model parameters, the ratio $m/n$ and thus
$\langle\dot x\rangle$ jumps only at discrete points and is constant in 
between. In other words, a kind of locking mechanism is at work,
closely related to the one responsible for the
Shapiro steps in symmetric
potentials with an extra tilt $F$ on the right hand side
of (\ref{6.1}) \cite{sha63,jun91}.
Further intriguing features, like the appearance of Devil's staircases
of current-plateaux or current reversals 
of $\langle\dot x\rangle$ as a function
of the driving amplitude $y(t)$, are discussed in detail in
\cite{bar94,ajd94a,han96,sch98}.

Upon including the {\em thermal noise} in (\ref{6.1}), the details
of the complex behavior of $\langle\dot x\rangle$ as a function of 
various model parameters is washed out. While for simple, 
saw-tooth-type
potentials $V(x)$ like in \figs \ref{fig2}, \ref{fig4.1} 
and not too large rocking amplitudes, the
deterministic ($T=0$) current $\langle\dot x\rangle$ is known
\cite{mag93,bar94,ajd94a,sch98} to always exhibit the
same direction, a current inversion for sufficiently fast
driving sets in as soon as a finite amount of thermal noise ($T>0$) is
added, as confirmed by our perturbative result (\ref{6.25}).

If the deterministic current $(T=0$) vanishes,
then for weak thermal noise (small $T$)
an activated barrier crossing problem arises
which can be reduced to an escape rate problem via (\ref{4i}).
In general, analytical progress requires technically sophisticated 
path-intergal and WKB-type singular perturbation methods which
are beyond our present scope, see also \sect \ref{sec3.6}.

Both, in the limits of small and large driving amplitudes one
can readily see that the current approaches zero.
Hence, there must be an ``optimal'' amplitude in between
for which the current is maximized. Typically, the dependence
of $\langle\dot x\rangle$ upon the amplitude is roughly speaking 
of a single-humped shape \cite{bar94,ajd94a}, onto which, however,
the previously described (non-monotonic) fine-structure
for small or zero thermal noise intensity is superimposed.

\subsection{Applications}\label{sec6.2.3.1}
An experimental realization of a rocking ratchet system
has been reported in \cite{gor96}:
A {\em mercury drop in a capillary} with a periodically but asymmetrically
varying diameter is subjected to an oscillating external electrical force
of electrocapillary nature.
While thermal fluctuations are negligible and the
experimental situation is at most qualitatively captured by the one-dimensional
model dynamics (\ref{6.1}), besides the directed transport
itself also the ``resonance-like'' dependence of the current
$\langle\dot x\rangle$ upon the rocking amplitude, as predicted theoretically,
has indeed been observed in the experiment.

Several further experimental realizations of the rocking ratchet scheme
have been proposed: In \cite{zap96} it has been demonstrated
that the phase across an {\em asymmetric SQUID} 
threatened by a magnetic flux
may be modeled by a rocking ratchet dynamics.
For more details we refer to \sect \ref{sec6.2.4a} below.

A second realization of the rocking ratchet scheme has been suggested in 
\cite{fal99}: The proposed system
consists of a one-dimensional\footnote{Practically,
a closed loop topology can replace the straight periodic setup
of infinite length.} 
{\em parallel array of Josephson-junctions} with
alternating critical currents and junction areas in the overdamped 
limit, see also \sect \ref{sec7.1}.
In such a system, it can be shown that the relevant
soliton-type solutions (also referred to as kinks, vortices, or fluxons) 
are approximately governed by a one-dimensional overdamped 
dynamics in an effective pinning potential which can be chosen ratchet-shaped.
In other words,
such a Josephson kink can be considered as quasi-particle (endowed with
effective mass, velocity, interaction with other kinks,
and other particle-like properties)
moving in an effective one-dimensional ratchet-potential along the array
and can be observed by measuring the time- and space-resolved
dc voltage along the array.
Taking into account an external periodic driving and thermal fluctuations,
a rocking ratchet setup is thus recovered.
The technical details of the problem are rather involved and finally
require a numerical evaluation, see \cite{fal99} for more details.
Significant experimental progress towards a realization of the
ratchet effect in such sorts of Josephson-junction arrays has been
accomplished in \cite{tri99}.
A modification, based on a continuous, one-dimensional 
{\em long Josephson junction}
(of annular shape), has been put forward in \cite{gol01}. An effective ratchet
potential for the kink dynamics emerges either by applying an external magnetic
field and choosing a properly deformed shape of the annular Josephson junction or
by modulating its width. 
A further option is to deposite a suitably shaped ``control line'' on top
of the junction in order to modulate the magnetic flux through it
\cite{car01}.
In either way, not only rocking ratchets 
-- as in \cite{fal99,tri99} -- but also 
{\em fluctuating potential ratchets} 
(not necessarily overdamped) can be realized \cite{gol01,car01}.
Further theoretical as well as experimental 
studies along closely related lines by several groups are currently 
in progress, see also \sects \ref{sec6.2.1} and \ref{sec6.5.5}.

As a third realization of the rocking ratchet scheme,
it has been proposed in \cite{lee99} that the application
of an alternating current to a
superconductor, patterned with an asymmetric
pinning potential, can induce a systematic directed {\em vortex motion}.
Thus, by an appropriate choice of the ratchet-shaped pinning
potential, the rocking ratchet scheme can be exploited to
continuously remove unwanted trapped magnetic flux lines
out of the bulk of superconducting materials.
Quantitative estimates \cite{lee99}
show that thermal fluctuations
are practically negligible in this application
of the rocking ratchet model (\ref{6.1}).
For a two-dimensional version \cite{wam99}
of the same idea see \sect \ref{sec6.2.4}.

Finally, it has been predicted \cite{der98z}
within a simplified hopping-model (activated
barrier crossing limit) for a crystalline surface, 
consisting of atomically flat terraces and 
monoatomic steps, that by application of an ac-field
a {\em surface-smoothening} can be achieved due to an 
underlying rocking ratchet mechanism.
First experimental findings which can be attributed
this theoretically predictied effect are due to \cite{pab00}.

For additional experimental realizations see also \sect \ref{sec6.5.5}.

\section{Influence of inertia and Hamiltonian ratchets}\label{sec6.2.3.2}
The rocking ratchet dynamics (\ref{6.1}) supplemented by a finite
inertia term $m\ddot x(t)$ on the right hand side is not only of
experimental interest (cf. the asymmetric SQUID model in \sect \ref{sec6.2.4a}
below) but exhibits also interesting new theoretical 
aspects\footnote{Regarding the issue of finite inertia in
traveling potential ratchets, Seebeck ratchets, 
fluctuating force ratchets, and quantum ratchets see \sects 
\ref{sec4.3.1}, \ref{sec4.8.1}, \ref{sec6.1.2}, and \ref{sec6.5.1}, 
respectively}.
Without the noise $\xi(t)$, the periodically driven deterministic
dynamics is equivalent to a three-dimensional autonomous dynamics and thus
in general admits chaotic attractors in the ``generalized'' sense
specified at the beginning of this section.
Numerical simulations \cite{jun96,mat00,bar00,mat01,bar01} 
show that a {\em chaotic} behavior
is indeed realized in certain parameter regions of the model.
As another crucial difference in comparison with the overdamped
case, the current in the long time average (\ref{4i}) in
general still depends on the initial conditions \cite{jun96,fla00}.

As a function of various model parameters, the current shows a still
much more complex behavior than in the overdamped case,
including multiple inversions even for a ``simple''
potential-profile like in \fig \ref{fig2}. For very weak
damping, the sign of the current is in fact predominantly
opposite to that in the overdamped limit \cite{jun96}.
These general features of $\langle\dot x\rangle$ are
expected to be robust also against a certain amount of noise.
The same is not expected for further interesting details of the 
deterministic dynamics reported in \cite{jun96,mat00,bar00,mat01,bar01,ari01},
some of them strikingly reminiscent of previous findings in the context
of deterministic diffusion in symmetric one-dimensional maps
\cite{fuj82,gei82,sch82,gei84,rei94a,rei94b,kla95}.

Of significant conceptual interest is the noiseless case in the
limit of vanishing dissipation, i.e. a 
{\em conservative (Hamiltonian) deterministic 
rocking\footnote{A Hamiltonian generalized traveling potential ratchet 
model has been considered in \cite{far98}.} ratchet dynamics}
\begin{equation}
m\,\ddot x(t)= -V'(x(t)) + y(t)\ .
\label{ham1}
\end{equation}
The salient difference in comparison with a dissipative system is the
time-inversion invariance provided the time-periodic driving 
$y(t)=y(t+\ttt)$ satisfies 
(after an irrelevant shift of the time origin) the symmetry 
condition \cite{fla00,yev01}
\begin{equation}
y(-t)=y(t) \ ,
\label{ham2}
\end{equation}
see also below\footnote{Note that in the present context 
of Hamiltonian ratchets the word ``rocking ratchet''-- 
unlike in the rest of this review -- is not necessarily reserved 
for symmetric drivings $y(t)$, i.e. $y(t)$ is $\ttt$-periodic but
need not satisfy (\ref{s2}).}
\eq (\ref{ss7}).
Another basic feature is the generic appearance of (Hamiltonian) chaos
with its complicated hierarchical fine structure of disjoint
stochastic (chaotic) layers, islands, KAM-tori etc. \cite{mei92,zas99,kov00}.
As a consequence, the behavior of the system depends in general
on the initial 
conditions\footnote{The dependence of the current
(\ref{4c1}) on the initial conditions $x(t_0)$, 
$\dot x(t_0)$, and especially on the ``initial phase'' $t_0$ in
$y(t_0)$ is obvious in the special case that 
$V'(x)\equiv 0$ in (\ref{ham1}). Though this special case is
untypical in that it does not exhibit chaos it still captures
some of the essential physics of the general case.}
unless one is in the limiting
case of strong (hyperbolic) Hamiltonian chaos \cite{far98,yev00,dit00,sch01}.
Strictly speaking, this case is not generic but it is often
adopted as an approximation for sufficiently strong 
perturbations of an integrable system with initial conditions
in that stochastic layer which contains $\dot x$-values of
either sign.
While {\em diffusive} transport with its intriguing
anomalous features (e.g. so-called L\'evy flights)
has been analyzed in great detail \cite{mei92,zas99,kov00},
our understanding of {\em directed} transport in such a system
with broken symmetry is 
considerably less well developed.

Under the assumptions that the {\em symmetry} (\ref{ham2}) is respected 
it has been predicted in \cite{fla00} that $\langle\dot x\rangle =0$
provided the initial condition $x(0),\dot x(0)$
is part of a stochastic layer which also contains an 
initial condition with $\dot x(0) = 0$.
Especially, this prediction is independent of whether the potential
$V(x)$ is asymmetric or not.
The basic reason is that such a trajectory $x(t)$, due to ergodicity
reasons, gives on the one hand rise to the same average current 
(\ref{4c1}) as its time inverted counterpart $z(t):=x(-t)$, i.e.
$\langle\dot x\rangle = \langle\dot z\rangle$. On the other hand, 
one also concludes that $\dot z(t)=-\dot x(-t)$ and thus
$\langle\dot x\rangle = - \langle\dot z\rangle$. As a consequence, 
it follows \cite{fla00} that $\langle\dot x\rangle = 0$.
A similar conclusion holds \cite{fla00} if the symmetry conditions
from \sect \ref{sec3.1.2} are respected by the potential $V(x)$ 
{\em and} the periodic driving $y(t)$ (cf. \eqs (\ref{s1}) and 
(\ref{s2})). Accordingly, 
the symmetry condition (\ref{ham2}) may be considered in some 
sense as the Hamiltonian counterpart of the supersymmetry
concept for overdamped systems (see \sect \ref{sec3.4.4}).
These different symmetries have been explored in quantitative detail
in \cite{yev01} by means of a kinetic {\em Boltzmann-equation} 
approach with special consideration of the weak and strong damping 
regimes. 
Returning to the limit of a Hamiltonian rocking ratchet,
we can conclude that
if neither of the above mentined symmetry conditions is 
satisfied then the occurrence of a finite current 
$\langle\dot x\rangle$ (ratchet effects) is generically 
expected (and numerically observed) \cite{fla00,goy00,yev01,den01}.

On the other hand, if the initial condition $x(0),\dot x(0)$
is not part of a stochastic layer which also contains an 
initial condition with $\dot x(0) = 0$
then generically $\langle\dot x\rangle\not=0$ even
if the symmetry conditions (\ref{ham2}) 
or (\ref{s1}) {\em and} (\ref{s2}) are respected.
Examples with a finite current in spite of the symmetry
property (\ref{ham2}) are discussed in \cite{yev00,dit00,sch01}
(see also the previous footnote).

Though it may be difficult in practice, in principle the entire
phase space of the Hamiltonian dynamics (\ref{ham1}) can be
decomposed into its different ergodic 
components\footnote{In the typical case, some of 
them are regular and some of them are chaotic.
Furthermore, the borderlines between them are the intact
KAM tori. Their number is infinite and they are arranged in a
very complicated hierarchical pattern \cite{mei92,zas99,kov00}.},
each of them characterized by its own particle current
$\langle\dot x\rangle$.
Next we observe \cite{dit00,sch01}
that the ``fully averaged particle'' current 
according to the uniform (microcanonical)
phase space density can be written as 
\begin{equation}
\int_0^\ttt dt \int_0^L dx \int_{-\infty}^\infty dp\, \dot x
= \lim_{p_0\to\infty} 
\int_0^\ttt dt \int_0^L dx \int_{-p_0}^{p_0} dp\, \frac{\partial H}{\partial p}
\ ,
\label{ham3}
\end{equation}
cf. \sects \ref{sec2.1.4} and \ref{sec3.1.1}.
Since the Hamiltonian of the dynamics (\ref{ham1}) is 
$H=p^2/2m +V(x)-xy(t)$
it follows that the microcanonically weighted average velocity
over all ergodic components in (\ref{ham3}) is equal to zero 
\cite{dit00,sch01}.
An immediate implication of this ``sum rule'' is that
a necessary requirement 
for directed transport is a mixed phase space
since the microcanonical distribution is the unique
invariant (reduced) density in this case and is
always approached in the long time limit.
In other words, even in the absence of the above mentioned symmetries,
systems with strong (hyperbolic) chaos
do not admit a ratchet effect \cite{dit00,sch01}.

While in \cite{fla00,den01} the above mentioned L\'evy flights
are proposed as the main reason for directed
transport in Hamiltonian ratchets, the emphasize in
\cite{dit00,sch01} is put on the picture that transport 
in the chaotic layers has its origin in the ``unbalanced''
currents within the regular islands.
The situation in systems with a 
more than two-dimensional phase space 
(bringing along Arnold diffusion) has so far 
not been considered at all.


\section{Two-dimensional systems and entropic ratchets}\label{sec6.2.4}
By explicitly keeping the dynamics that governs the
driving $y(t)$ or $f(t)$ in the basic ratchet model 
dynamics (\ref{4a}) -- independently of whether a back coupling
is absent (see \sect \ref{sec3.2.2}) or present (see \sect \ref{sec5.3.1}) --
one trivially ends up with a two-dimensional system.
In this section, however, genuine vectorial generalizations of
the basic model (\ref{4a}) are considered.
The simplest case of such a two-dimensional ratchet system consists 
of two completely independent equations of the form (\ref{6.1}), one 
for each spatial dimension $x_1$ and $x_2$. {\em Pro forma}, one may 
then define a common total potential $V(x_1,x_2)$
as the sum of the two individual 
potentials. Such a system offers the possibility to separate particles
with different ratios $\langle \dot x_1\rangle/\langle \dot x_2\rangle$
according to their traveling direction in the $x_1$-$x_2$-plane.

A more complicated situation arises if the dynamics involves a
non-trivial common potential $V(x_1,x_2)$, periodic and/or asymmetric 
in only one or both arguments.
An example of this kind (periodic in one component only)
has been treated already in the context
of Feynman's ratchet in \eqs (\ref{f01})-(\ref{f04}),
see also \cite{bao98c}.
Another example (periodic in both components) which, instead of the
usual linear directed transport, leads to a permanent circular
motion of particles, has been worked out in \cite{gho00}, see
also \cite{qia98,kos00}.
In fact, by giving up the requirement of a simple periodicity 
of the system along any straight spatial
direction, it should be in principle possible to 
{\em direct particles along arbitrarily pre-assigned pathways} 
in properly designed
two-dimensional systems \cite{han99,bro00}, possibly even along
different routes for different species of particles with
identical seeds.

In \cite{cec96} a two-dimensional potential landscape 
$V(x_1,x_2)$ was considered 
which consists of one straight ``valley'' along the $x_1$-axis and periodically
repeated ``side valleys'' of finite length (dead ends).
If the angle between those side-valleys and the $x_1$-axis is
different from
$\pm \pi/2$ then the spatial symmetry along the $x_1$-direction
is broken and a time-periodic rocking force generically induces
a finite current $\langle\dot x_1\rangle$.
Since this ratchet effect will occur even if there are no
potential barriers along the $x_1$-axis, i.e.  $V(x_1,x_2=0)=const.$, 
the name {\em entropic ratchet} has been coined for this system \cite{cec96}.
Moreover, if an additional bias $F$ is applied along the $x_1$-axis,
a non-monotonic behavior of $\langle \dot x_1\rangle$ as a function of $F$ 
may result \cite{cec96}.
This so-called {\em negative differential resistance} has also been
previously observed in the closely related context of networks 
with dead-ends, see \cite{bal95} and further references therein.
Very similar two-dimensional entropic ratchet schemes 
have been proposed in \cite{sla97} for the 
purpose of separating DNA molecules 
(see also \cite{tur90,des98,sla00,gri01} and \sect \ref{sec6.3.1}), 
in \cite{wam99} for the purpose of pumping, dispersing, and concentrating 
fluxons in superconductors by electrical ac-currents (cf. \sect \ref{sec6.2.3.1}),
and in \cite{sto01} for the purpose of
rectifying electronic currents with the help of
the Coulomb blockade effect, see also \cite{ven96}.

Another two-dimensional rocking ratchet scheme is obtained by choosing a
potential $V(x_1,x_2)$ which has basically the effect of a
two-dimensional, {\em periodic array of obstacles} (``scatterers'').
The spatial symmetry is broken by the shape of the single
obstacles, in the simplest case a triangle.
In its simplest form, such a setup can be imagined as a 
Galton-board-type device with a broken spatial (``left-right-'') symmetry.
This basic idea has been put forward already in the context of the
photovoltaic effect in non-centrosymmetric materials, see \sect \ref{sec6.1}.
For the purpose of separating macromolecules such as DNA, two-dimensional
arrays of obstacles (``sieves'') have been proposed and quantitatively
analyzed in \cite{ert98,duk98,duk98a,der98b,bie00,kos00}.
The technological feasibility of such sieves -- however with
symmetric obstacles -- has been demonstrated
already before these works in \cite{vol92}.
An {\em experimental} implementation of the same basic concept has been
realized in \cite{oud99} for the purpose of transporting 
and separating phospholipid molecules in a two-dimensional fluid 
bilayer.
In contrast to other standard separation methods,
such a rocking ratchet system is re-usable and enables continuous 
operation.

Experimentally, transport of electrons in two-dimensional periodic arrays
of triangular antidot scatterers under far-infrared irradiation has been 
demonstrated in \cite{lor98}.
With an approximative classical description of the system being
justified in the considered parameter regime, essentially  a 
two-dimensional rocking ratchet scheme is thus recovered.

A further two-dimesional SQUID ratchet system will be treated in
\sect \ref{sec6.2.4a} below.
Also the experimental ratchet devices described in
\sect \ref{sec4.1.1} and at the end 
of \sects \ref{sec4.3.2} and \ref{sec6.5.5} 
-- though admitting suggestive and rather faithful effective
one-dimensional descriptions -- are
strictly speaking of two-dimensional character.
A three-dimensional ratchet dynamics is discussed in \sect \ref{sec6.4}.
Further models with two degrees of freedom are treated in \sects
\ref{sec5.7} and \ref{sec5.6}.

\section{Rocking ratchets in SQUIDs}\label{sec6.2.4a}
In the {\em theoretical} work \cite{zap96} it has been demonstrated
that the phase across an asymmetric SQUID (superconducting
quantum interference device) threatened by a magnetic flux
may be modeled by a {\em one-dimensional rocking ratchet} dynamics
(cf. \sect \ref{sec6.2.3.1}).
The starting point
is the standard RSJ model (resistively shunted junction model, also
called Steward-McCumber model) for
the phase difference $\varphi$ of the macroscopic quantum mechanical 
wave function across a conventional Josephson junction
\begin{equation}
\frac{\Phi_0 C}{2\pi}\,\ddot\varphi(t) + 
\frac{\Phi_0}{2\pi R}\,\dot\varphi(t) + 
I_c\,\sin\varphi(t) = I(t)+\xi(t) \ ,
\label{j1'}
\end{equation}
where $C$, $R$, and $I_c$ are the capacitance, resistance, and
critical current of the junctions, $I(t)$ is the electrical current
flowing through the junction, and $\Phi_0:= h/2e$ is the flux quantum.
Thermal fluctuations are modeled by unbiased Gaussian white noise
$\xi(t)$ of strength $2 k_BT/R$. For the total phase difference
across a series of two identical such Josephson junctions
one recovers \cite{zap96} the same equation (\ref{j1'}) except that 
$\varphi(t)$ is replaced by $\varphi(t)/2$ and the noise strength
$2k_B T/R$ by $k_BT/R$.
Next, one considers a SQUID with the usual ``loop''-geometry,
formed by two conducting ``arms'' in parallel, but 
with {\em two} identical Josephson junctions in series in one ``arm'', 
and a third junction with characteristics $C'$, $R'$, and $I_c'$ in the
other ``arm''. The total current $\itot$ through the conducting loop 
follows by adding the currents through both arms.
Under the assumption that the loop inductance is much smaller
than $\Phi_0/(I_c+I_c'+\itot (t))$ (see also the discussion below (\ref{j10}) below),
the equation for the total phase difference $\varphi$
across the loop is then governed by the equation \cite{zap96}
\begin{eqnarray}
\!\!\!\!\!\!\!\!\!\!\!\!\!\!
& & \frac{\Phi_0}{2\pi}\!\left(\frac{C}{2}+C'\right)\! \ddot\varphi(t) + 
\frac{\Phi_0}{2\pi}\left(\frac{1}{2R}+\frac{1}{R'}\right)\!
\dot\varphi(t)  =  -V'(\varphi(t)) + \itot(t) + \xitot(t)
\label{j2'}\\
\!\!\!\!\!\!\!\!\!\!\!\!\!\!
& &  V(\varphi) := -\frac{I_c}{2}\,\cos (\varphi/2) 
- I_c'\cos (\varphi + 2\pi\Phi/\Phi_0) \ , 
\label{j3}
\end{eqnarray}
where $\Phi$ is the total magnetic flux threatening the loop
and where $\xitot (t)$ is a Gaussian white noise with correlation
\begin{equation}
\langle\xitot (t)\, \xitot(s)\rangle = 
2\, k_B T \left(\frac{1}{2R}+\frac{1}{R'}\right)\,\delta(t-s) \ .
\label{j4}
\end{equation}
The noise- and time-averaged ``phase current'' 
$\langle\dot\varphi\rangle$
is connected to the averaged voltage $U$ across the loop according 
to the second Josephson equation \cite{zap96}
\begin{equation}
U = \frac{\Phi_0}{2\pi}\, \langle\dot\varphi\rangle
\label{j5}
\end{equation}
and thus directly accessible to an experimental measurement.
In other words, for appropriately chosen external currents
$\itot (t)$ and (static) magnetic fields, a rocking ratchet dynamics
is recovered from (\ref{j2'}), which
is in particular of the overdamped form (\ref{6.1})
if $I_cR^2C,\, I'_c{R'}^{2}C'\ll\Phi_0$, cf. 
\sect \ref{sec2.1.2.4} in Appendix A.
Potentials (\ref{j3}) with additional Fourier modes may be obtained by
more complicated SQUIDs with additional ``arms'' in parallel.

Next we turn to {\em one of the first systems for which a ratchet 
effect has been theoretically described and experimentally measured}
\cite{wae67,wae69}.
While these early works focus on the realm of adiabatically slow rocking,
the extension beyond this regime has been realized {\em experimentally}
very recently in \cite{wei99,wei00}. The setup consists of the
following {\em two-dimensional} modification of the above
described rocking ratchet SQUID system 
(\ref{j1'})-(\ref{j5}):
The starting point is a SQUID with
the usual ``loop''-geometry, consisting of one Josephson junction
in each of the two parallel ``arms'' of the loop.
The phase across the junctions in the left (index ``$l$'')
and right (index ``$r$'') arm are thus both governed by an equation
of the form  (\ref{j1'}).
The difference between the two phases due to the vector potential
of the enclosed magnetic flux is governed by the ``flux quantization''
relation
\begin{equation}
\varphi_l-\varphi_r = 2\,\pi\,\Phi_{{\rm tot}}/\Phi_0 \ .
\label{j5'}
\end{equation}
The enclosed flux $\Phi_{{\rm tot}}$ is divided between an
{\em externally applied magnetic flux} $\Phi$ and the flux from the 
circulating current in the loop, yielding \cite{ear95}
\begin{equation}
\Phi_{{\rm tot}} = \Phi - [ L_l I_l - L_r I_r ] \ , 
\label{j5''}
\end{equation}
where $L_{l,r}$ are the inductances of the two junctions.
Under the simplifying assumptions that
\begin{equation}
C_l=C_r=:C\ \ , \ \ \ R_l=R_r =:R
\label{j6}
\end{equation}
and with the definitions
\begin{eqnarray}
& & I_c:=\frac{I_{c,l}+I_{c,r}}{2} \ \ , \ \ \ 
\alpha_I:=\frac{I_{c,l}-I_{c,r}}{I_{c,l}+I_{c,r}}\label{j6.1}\\
& & \bar L :=\frac{L_{l}+L_{r}}{2} \ \ , \ \ \ 
\alpha_L:=\frac{L_{l}-L_{r}}{L_{l}+L_{r}}\label{j6.2}\\
& & \varphi:=\frac{\varphi_l+\varphi_r}{2} \ \ , \ \ \ 
\psi:=\frac{\varphi_l-\varphi_r}{2} \label{6j.3}
\end{eqnarray}
it follows by adding and subtracting the two equations of the form 
(\ref{j1'}) with indices ``$l$'' and ``$r$'' that \cite{wei99,wei00}
\begin{eqnarray}
& & \frac{\Phi_0 C}{2\pi}\ddot\varphi(t) +
\frac{\Phi_0}{2\pi R}\dot\varphi(t) =
-\frac{\partial\, V(\varphi(t),\psi(t),t)}{\partial \varphi} + 
\frac{\itot(t)}{2}
+ \xi_1(t)\label{j7}\\
& & \frac{\Phi_0 C}{2\pi}\ddot\psi(t) +
\frac{\Phi_0}{2\pi R}\dot\psi(t) =
-\frac{\partial\, V(\varphi(t),\psi(t),t)}{\partial \psi} + \xi_2(t)
\label{j8}\\
& & V(\varphi,\psi,t):=-I_c[\cos\varphi\cos\psi-\alpha_I\sin\varphi\sin\psi]
\nonumber\\
& & \qquad\qquad\qquad + \frac{\pi}{4\Phi_0 \bar L}
\left[\psi\frac{\Phi_0}{\pi} - \Phi + \bar L\alpha_L\itot(t)\right]^2 \ .
\label{j9}
\end{eqnarray}
Here, $\itot(t):=I_l(t)+I_r(t)$ is the total electrical current flowing through
the SQUID and $\xi(t)$ ($i=1,2$) are two unbiased Gaussian white noises with
correlation
\begin{equation}
\langle\xi_i(t)\,\xi_j(s)\rangle = \frac{k_BT}{R}\,\delta_{ij}\,\delta(t-s) \ .
\label{j10}
\end{equation}
Finally, the time averaged voltage $U$ across the loop
is again given by (\ref{j5}).

Our first observation is that for $\bar L\ll\Phi_0/(I_c+\itot(t))$
it follows from (\ref{j9}) that $\psi\simeq\pi \Phi/\Phi_0$ and we
are left with an effective one-dimensional problem (\ref{j7}).
The same type of approximation has been made in the derivation of
(\ref{j2'})-(\ref{j4}).
In any case,
the potential (\ref{j9}) is periodic in the variable $\varphi$, while
the $\psi$-dependence is confined by the quadratic term on the right
hand side.
On condition that $\Phi$ is not a multiple of $\Phi_0/2$
and that either $\alpha_I\not =0$ or $\alpha_L\not =0$, the potential
(\ref{j9}) is neither inversion symmetric under 
$(\varphi,\psi)\mapsto(-\varphi,\psi)$ nor
$(\varphi,\psi)\mapsto(-\varphi,-\psi)$, thus a ratchet effect is 
theoretically predicted
and has been experimentally observed \cite{wei99,wei00}.
Especially, a non-vanishing externally applied magnetic field is necessary,
since otherwise $\Phi = 0$.

Given that the above conditions ($2\Phi/\Phi_0$ not an integer and
$\alpha_I\not = 0$ or $\alpha_L\not = 0$) are fulfilled, it is
instructive to rewrite (\ref{j7})-(\ref{j9}) in the form
\begin{eqnarray}
& & \frac{\Phi_0 C}{2\pi}\ddot\varphi(t) +
\frac{\Phi_0}{2\pi R}\dot\varphi(t) =
-\frac{\partial\, \tilde V(\varphi(t),\psi(t))}{\partial \varphi} 
+ \frac{\itot(t)}{2}
+ \xi_1(t)\label{j7'}\\
& & \frac{\Phi_0 C}{2\pi}\ddot\psi(t) +
\frac{\Phi_0}{2\pi R}\dot\psi(t) =
-\frac{\partial\, \tilde V(\varphi(t),\psi(t))}{\partial \psi} 
- \frac{\alpha_L \itot(t)}{2}
+ \xi_2(t)
\label{j8'}\\
& & \tilde V(\varphi,\psi):=
-I_c[\cos\varphi\cos\psi-\alpha_I\sin\varphi\sin\psi]
+ \frac{\Phi_0}{4\pi \bar L}
\left[\psi - \frac{\pi \Phi}{\Phi_0}\right]^2 \ .
\label{j9'}
\end{eqnarray}
In other words, a two-dimensional rocking ratchet scheme
is recovered, with a ``rocking force'' which acts along the
$\varphi$-direction if $\alpha_L=0$ and points into a more general
direction in the $\varphi$-$\psi$-plane if $\alpha_L\not = 0$.

Further studies on related Josephson ratchet systems are addressed in
\sects \ref{sec6.2.3.1} and \ref{sec7.1}, see also \sect \ref{sec6.2.1}.

\section{Giant enhancement of diffusion}\label{sec6.2.5}
In this section we return to the overdamped, one-dimensional tilting ratchet
scheme (\ref{6.1}), however, with the effective {\em diffusion coefficient}
({\ref{4c2}) rather than the particle current being the quantity of our
interest.
To this end, it turns out that the asymmetry of the potential
$V(x)$ in (\ref{6.1}) is not essential, and we will therefore 
focus on the simplest case of a {\em symmetric, periodic potential} $V(x)$.

In contrast to the investigation of directed transport in terms of 
$\langle\dot x\rangle$, studies of diffusive transport in periodic driven
systems are still rather scarce.
While the determination of the effective diffusion coefficient
is, in general, technically more demanding (cf. \sect \ref{sec3.1.1}) its
relevance e.g. for particle separation purposes may well be
comparable to the schemes based on directed transport.

The effective diffusion coefficient $\deff$ from (\ref{4c2}) in systems like
(\ref{6.1}) but in the absence of an external driving $y(t)$ has 
been considered in \cite{lif62} with the main result that $\deff$ is
for non-trivial potentials $V(x)$ 
always smaller than the bare diffusion coefficient (\ref{2.4}).
Diffusive separation of particles in the same system (\ref{6.1}) 
but with a 
static tilt $y(t)\equiv F$ (cf. (\ref{2.21})) has been addressed in
\cite{ajd91}, demonstrating an improvement of one to two orders of
magnitude in selectivity as compared with conventional continuous field 
free-flow electrophoresis methods, see also \cite{con99,lin01,rei01}.
Asymptotic results for fast pulsating and fluctuating force
ratchet schemes have been derived in \cite{gho94} and \cite{mal98b}, respectively.
Here, we will focus on the case of a deterministic, time-periodic perturbation
$y(t)$ in (\ref{6.1}), naturally arising in typical experimental settings.
Related studies are \cite{cla91,jun91,cla93,jun96,kim98} and especially the work
of  Gang, Daffertshofer, and  Haken \cite{gan96}. Our present
system is 
a conceptually simpler and more effective variation of the setup from 
\cite{jun91,gan96} which enables a controlled selective enhancement 
of diffusion that in principle can be made arbitrarily strong.

We focus on the simplest case of a {\em symmetric} sawtooth potential $V(x)$
with period $L$ and barrier height $V_0$ (\fig \ref{figdiff1}a) and a 
{\em time-periodic driving force $y(t)$} with three states $y_0$, $0$, and $-y_0$. As illustrated 
in \fig \ref{figdiff1}b, time-segments of length $t_{{\rm t}}$ with a constant tilt 
$y(t)=\pm y_0$ are separated by ``waiting-periods'' $t_{{\rm w}}$ with vanishing
$y(t)$. Further, we henceforth restrict ourselves to weak thermal noise $\xi(t)$,
i.e. $k_BT\ll V_0$.

\figdiffeins

We assume that $y_0>2V_0/L$ and that the initial particle distribution at time 
$t=0$ consists of a very narrow peak at a minimum of the potential $V(x)$, say at $x=0$.
As long as $t\leq t_{{\rm t}}$ we have $y(t)\equiv y_0$, so the peak
moves to the right under the action of the deterministic forces
and also broadens slightly due to the weak thermal noise in  (\ref{6.1}). The
deterministic time $t_n$ at which the peak crosses the $n$-th maximum of $V(x)$
at $x=(n-1/2)L$ while $y(t)=y_0$ is acting, can be readily figured out
explicitly \cite{sch98}.
If now $t_{{\rm t}}$ just matches one of those times $t_n$, then the original
single peak is split into two equal parts and if the subsequent
``waiting-interval'' $t_{{\rm w}}$ with $y(t)\equiv 0$ is sufficiently long  
the two parts will proceed towards the respective
nearest minimum of $V(x)$ at $x=(n-1)L$ and $x=nL$. The result consists in {\em two}
very sharp peaks after half a period $t=t_{{\rm t}}+t_{{\rm w}}$ of the driving
force $y(t)$. Similarly, after a full period $\tau:=2(t_{{\rm t}}+t_{{\rm w}})$
one obtains {\em three} narrow peaks at  $x=-L,\, 0,\, L$ with weights $1/4$,
$1/2$, $1/4$, respectively. For the variance 
$\langle x^2(t)\rangle -\langle x(t)\rangle^2$ one thus obtains the result
$L^2/2$. In the same way  one sees that after $n$ periods the variance amounts to
$n L^2/2$, yielding for the effective diffusion coefficient (\ref{4j}) the 
expression
\begin{equation}
\deff=L^2/8(t_{{\rm t}}+t_{{\rm w}})\ .
\label{diff1}
\end{equation}
In the case that $t_{{\rm t}}$ does {\em not} match any of the times $t_n$,
the initial single peak
is split after half a period $t=t_{{\rm t}}+t_{{\rm w}}$ into two peaks
with unequal weights. If $t_{{\rm t}}$ is sufficiently different from any 
$t_{n}$ and the thermal fluctuations are sufficiently weak, one
of those two peaks has negligible weight. Consequently, after a full 
period almost all 
particles will return to $x=0$. The effective diffusion coefficient $\deff$
is therefore very small, in particular much smaller than for free 
thermal diffusion (\ref{2.4}).

\figdiffzwei

An example of the effective diffusion coefficient $\deff$ as a function
of $t_{{\rm t}}$ is depicted in \fig \ref{figdiff2}a.
As usual (cf. \sect \ref{sec3.5}) such a multi-peak-structure of $\deff$ 
is not only expected upon variation of $t_{{\rm t}}$ but also by keeping 
$t_{{\rm t}}$ fixed and varying for instance the friction coefficient $\eta$,
corresponding to the
situation that different types of particles are moving in the same rocked 
periodic potential.
As \fig \ref{figdiff2}b demonstrates, the dynamics (\ref{6.1}) 
can indeed act as an extremely
selective device for separating different types of particles by 
{\em controlled, giant enhancement of diffusion}. 
Closer inspection shows \cite{sch98}
that the peaks in the effective diffusion coefficient $\deff$ can
in fact be made arbitrarily narrow and high by
decreasing the temperature or increasing $V_0$ at 
fixed $T$ while at the same time keeping $y_0 L/V_0$ large. 
Similarly as for the
friction coefficient $\eta$, particles can also be separated, {\em e.g.}, 
according to their electrical charge since this implies different values of
the ``coupling-parameters'' $V_0$ and $y_0$.

All these findings 
are obviously robust against various modifications of 
the model as long as one maintains periodicity in
space  and time and sufficiently long ``waiting-periods'' $t_{\rm t}$ with
$y(t)\equiv 0$  between subsequent ``tilting-times'' with non-vanishing $y(t)$.
A practical realization of such a particle separation device 
should be rather straightforward.

\section{Asymmetrically tilting ratchets}\label{sec6.3}
In this section we consider the ratchet model dynamics (\ref{6.1})
with a
symmetric, $L$-periodic potential $V(x)$ in combination with
a driving $y(t)$ of broken
symmetry, either periodic or stochastic.

If the characteristic time scale of the driving $y(t)$ is very large, the
adiabatic approximation (\ref{6.2}) for the periodic and (\ref{6.8}) for the
stochastic case can be applied.
Exploiting the symmetry of $V(x)$, a straightforward
calculation confirms the expected property that $v(y)$ in (\ref{6.3})
is an odd function of its argument.
In general, the contributions of $y$ and $-y$ in (\ref{6.2}) 
or (\ref{6.8}) will not cancel each other
and hence $\langle\dot x \rangle$ will generically be different from zero.
However, even though $y(t)$ is asymmetric, prominent examples exists
for which the contributions of $y$ and $-y$ do cancel each other, 
namely those respecting supersymmetry (\ref{ss6}).
Examples are a periodic driving $y(t)$ of the form (\ref{ss13}) with
$\gamma_1\not=0$ and $\gamma_2\not=0$ or the example depicted in \fig 
\ref{figss2}.
In this case, $\langle\dot x \rangle \to 0$ as the characteristic time
scale of $y(t)$ tends to infinity, which is a quite exceptional feature
within the class of tilting ratchets.
Since for fast driving the current approaches zero as well,
a qualitative behavior which in fact is reminiscent of a pulsating ratchet 
arises.
It is worth emphasizing, since it may appear counterintuitive at first glance,
that a symmetric, but not supersymmetric potential $V(x)$ (e.g. in (\ref{ss1})
with $a_1\not=0$ and $a_2\not=0$) combined with a supersymmetric but not
symmetric $y(t)$ (e.g. in \fig \ref{figss2}) generically {\em does}
lead to a ratchet effect\footnote{To dissolve any remaining doubts, we
have verified this fact by numerical simulations. A similar
prediction has been put forward previously in \cite{mah95,mah97}
without, however, recognizing the subtleties of supersymmetry in this context.}
$\langle\dot x\rangle\not=0$, see also \fig \ref{figss3}.
If moreover finite inertia effects $m\ddot x(t)$ are included on the
right hand side of (\ref{6.1}) then supersymmetry does no longer prohibit 
a current and thus even a pure sinusoidal potential $V(x)$ may be chosen.

\subsection{Periodic driving}\label{sec6.3.1}
The case of slow periodic driving is covered by the adiabatic approximation
(\ref{6.2}).
In the opposite case of a very small period $\ttt$, one finds along the same 
line of reasoning as in \sect \ref{sec6.2.2} the leading order asymptotics
\cite{rei00a}
\begin{eqnarray}
& & \langle \dot x \rangle =
\frac{\ttt^4 L\, [Y_{-}\,\int_0^L dx\, [V'''(x)]^2
+ Y_{+}\,\int_0^L dx\, [V''(x)]^3/2k_B T]}
{4\, \eta^5\,\int_0^L dx \, e^{V(x)/k_BT}\,\int_0^L dx \, e^{-V(x)/k_BT}}
\label{asy1}\\
& & Y_\pm := \int_0^1 dh\, 
[\hat y_0(h) \pm 2 \hat y_2(h)]\, [\hat y_2(h)]^2 \ ,
\label{asy2}
\end{eqnarray}
where $\hat y_0(h)$ and $\hat y_2(h)$ are defined in (\ref{6.21'}) and (\ref{6.22}).
Here we have exploited the symmetry of $V(x)$.
In the completely 
general case, the asymptotic current $\langle\dot x\rangle$ is given 
by the sum of the contributions in (\ref{6.25}) and (\ref{asy1}).

We notice that if the potential $V(x)$ is not only symmetric but 
also supersymmetric then $\int_0^L dx\, [V''(x)]^3=0$ and thus
the sign of the current in (\ref{asy1}) is dictated solely
by that of $Y_{-}$.
On the other hand, for a supersymmetric driving $y(t)$, both coefficients
$Y_{\pm}$ in (\ref{asy2}) vanish, that is, 
$\langle\dot x\rangle$ approaches zero even faster than
$\ttt^4$ as $\ttt\to 0$.

The possibility that a directed current, or, equivalently, a finite voltage 
under open circuit conditions, may emerge in a symmetric periodic
structure when driven by unbiased, asymmetric microwave signals
of the form
\begin{equation}
y(t) = \gamma_1\cos (2\pi t/\ttt) + \gamma_2\cos (4\pi t/\ttt + \Phi) 
\label{asy3}
\end{equation}
has been reported for the first time in the {\em experimental} work by Seeger 
and Maurer \cite{see78}.
From the traditional viewpoint of response theory in this context,
the basic mechanism responsible
for producing a dc-output by an unbiased ac-input 
(\ref{asy3})
then amounts to the so-called {\em harmonic mixing} of the two
microwaves of frequencies $2\pi/\ttt$ and $4\pi/\ttt$ in the
{\em nonlinear response} regime.
The electrical transport in such quasi-one-dimensional conductors is
usually described in terms of pinned charge density waves, which in turn 
are modeled phenomenologically as an overdamped Brownian particle in
a symmetric, periodic ``pinning'' potential \cite{won79,bre82,bre84,won84}.
The particle couples to the externally applied field
(\ref{asy3}) via an effective charge, i.e. we recover exactly the 
asymmetrically tilting ratchet model (\ref{6.1}).
We remark that both in the experimental work \cite{see78} and
in the subsequent theoretical studies \cite{won79,bre82,bre84,won84}
no emphasize is put on the fact of generating a dc-output by means
of an unbiased ac-input {\em per se}, and in this sense
the ratchet effect has been observed only implicitly.
Also worth mentioning is that the ``pinning''-potential $V(x)$ is usually
assumed to be of sinusoidal shape and thus respects supersymmetry.
Since the driving (\ref{asy3}) becomes supersymmetric for $\Phi = \pi/2$,
the current $\langle\dot x\rangle$ will exactly vanish at this 
point \cite{won79,bre82,won84}.
This feature does no longer arise for symmetric but not supersymmetric
potentials $V(x)$ or if finite inertia effects \cite{bre84}
become relevant, see also \cite{bar01,yev01}.

In the context of current generation by
photovoltaic effects (cf. at the end of \sect \ref{sec6.2.1})
very closely related theoretical and experimental
investigations are due to
\cite{shm85,ent89,ata96,hac97,ale99}.
The same basic idea to produce a directed current by means of the asymmetric
tilting ratchet scheme has also been exploited experimentally in a process
called zero-integrated field gel
electrophoresis\footnote{The gel network in which the DNA moves
does not exhibit the usual spatial periodicity but rather acts as
a random potential (due to basically static obstacles) 
in three dimensions.} 
which uses unbiased pulsed electric fields to separate chromosomal DNA 
\cite{vid85,tur90,sla97,des98,ser98,ser99,sla00,gri01}.

The ratchet effect in a periodically driven, asymmetrically
tilting ratchet has been independently re-discovered in \cite{ajd94a}.
Moreover, the complex structure of the current $\langle\dot x\rangle$ at
$T=0$, featuring plateaux and Devil's staircases, similarly as for
the rocking ratchet system in \sect \ref{sec6.2.3}, has been demonstrated for
especially simple examples of asymmetrically tilting ratchet models in
\cite{ajd94a}.
Further variations and extensions of such theoretical models,
the details of which go beyond our present scope, can be found in 
\cite{vid85,sla97,cha97,dyk97,des98,mog98,sar99,zol00,luc00a,sla00,gri01,yev01}.
For Hamiltonian (finite inertia, vanishing dissipation and thermal noise)
and quantum mechanical asymmetrically tilting ratchet systems we refer to
\sect \ref{sec6.2.3.2} and to \sects \ref{sec6.5.4} and \ref{sec6.5.5}, 
respectively.

\subsection{Stochastic, chaotic, and quasiperiodic driving}\label{sec6.3.2}
The generation of directed transport in symmetric, periodic potentials
$V(x)$ by an asymmetric stochastic driving $y(t)$ of zero average
in (\ref{6.1})
has been for the first time exemplified in \cite{luc95,cze97,cze00,cze01} for the
case of Poissonian white shot noise\label{fot6.3.1}\footnote{The specific
shot noise considered in \cite{luc95,cze97,cze00,cze01} is of the form 
(\ref{4.123})-(\ref{4.124}) but with the weights $n_i$ in (\ref{4.124}) 
not being integers but rather exponentially distributed, positive 
random numbers, see also (\ref{6.9b}), (\ref{6.9d}).}, 
see also \cite{hon96,han96a}.
At zero thermal noise $(T=0)$, a closed analytical solution is available 
\cite{luc95,cze00},
while for $T>0$ one has to recourse to asymptotic expansions, piecewise linear 
potentials, or numerical evaluations \cite{cze97,cze01}. Besides the fact of
a white-noise induced directed transport in symmetric potentials {\em per se},
the most remarkable finding is that the current always
points into the same direction as the $\delta$-spikes of the
asymmetric shot noise for any periodic potential (symmetric or not, but
different from the trivial case $V'(x)\equiv 0$).
We are thus facing one of the rare cases for which our procedure of
tailoring current inversions (see \sect \ref{sec3.5})
cannot be applied unless an additional 
systematic bias $F$ is included in (\ref{6.1}).
Leaving aside minor differences in the $\delta$-spikes statistics (cf. 
footnote 12) the basic reason for this 
unidirectionality can be readily understood by the mapping onto an improper 
traveling potential ratchet scheme according to (\ref{4.103}), (\ref{4.104})
and our discussion of the corresponding current 
(\ref{4.104'}), (\ref{4.106}), (\ref{4.122}).

The generic occurrence of a ratchet effect whenever $y(t)$ breaks the 
symmetry (\ref{s3})
has been pointed out in \cite{chi95} and 
exemplified by means of an asymmetric two-state
noise $y(t)$ in the adiabatic limit, cf. (\ref{6.8}).
Similar conclusions have been reached in \cite{han96a,chi97}.
In the case of an asymmetric dichotomous noise $y(t)$ and without thermal
fluctuations ($T=0$) in (\ref{6.1}), the exact analytical solution for 
arbitrary noise characteristics and potentials has been figured out
and discussed from different viewpoints in
\cite{wei93,kul96,mil96,zap98,ber97a}. 
Similarly as in (\ref{6.9f}), the above mentioned
shot noise model \cite{luc95,cze00} is
recovered as a special limit \cite{vdb83} from this analytical solution for 
dichotomous noise.
Approximations and analytically soluble particular cases in the presence of
a finite amount of thermal noise ($T>0$) have
been elaborated in \cite{kul98b,kul98a}. Regarding the asymptotics
of fast asymmetric tilting we remark that the expression (\ref{6.9}) vanishes
for symmetric potentials $V(x)$, hence a significantly different
structure of the leading order behavior is expected (compare also the
corresponding results (\ref{6.25}) and (\ref{asy1}) for periodic $y(t)$).
For asymmetric dichotomous noise
such an asymptotics has been derived in \cite{kul98b} within a constant
intensity scaling scheme, while for constant variance
scaling, as we mainly consider it in our present review, 
such an asymptotics has not yet been worked out.

Turning to {\em applications}, it has been argued in \cite{chi95} that
the absence of {\em a priori } symmetry reasons and thus
the appearance of an asymmetric noise $y(t)$ should be a rather common
situation in many systems far from equilibrium, especially
in biochemical contexts involving catalytic cycling 
(cf. \sect \ref{sec4.5} and \ch \ref{cha5}).
Specifically, if $y(t)$ represents a source of unbiased
nonequilibrium current fluctuations then an asymmetrically tilting ratchet
scheme can be readily realized by means of a Josephson junction
\cite{wei93,mil96,ber97a}, see (\ref{j1'}).
A concrete such source of current fluctuations has been
pointed out in \cite{zap98}. Namely, an asymmetric dichotomous noise
may arise intrinsically in point contact devices with a 
defect which tunnels incoherently between two states
\cite{ral84,mul92,gol92,ral92,kei96,kog96,smi96}.
A modified Josephson junction system with an asymetric 
total noise composed of two correlated symmetric noise
sources has been proposed in \cite{li98b,li98a}, see also
\cite{li98c,cao00,jia00}.

It is well-known 
\cite{fuj82,gei82,sch82,arg87,col89,bec90}
that in many situations, a low dimensional
dynamical system exhibiting deterministic {\em chaos} can induce
similar effects as a veritable random noise\footnote{In fact, we may
consider a noise (stochastic process) as generated by a chaotic deterministic
dynamics in the limit of infinitely many dimensions.
The close similarity between deterministic chaos and noise
is also exploited in any numerical pseudo-random number generator.}.
In the present case of the asymmetrically tilting ratchet scheme, the
emergence of directed transport (ratchet effect) when the
driving $y(t)$ is generated by a low dimensional chaotic dynamics has
been demonstrated
in \cite{hon94}, see also \cite{hon95,han96a,chi97}.

Another interesting intermediate between 
a stochastic and a periodic driving is
represented by the case of a 
{\em quasiperiodic driving} $y(t)$,
bringing along the possibly of a strange 
nonchaotic attractor \cite{gre84}.
Asymmetrically tilting ratchets of this type
have been studied in \cite{neu01}.

\chapter{Sundry Extensions}\label{cha06}
In this chapter we address various significant modifications
and extensions of the pulsating and tilting ratchet schemes
from \chs \ref{cha4} and \ref{cha6}
as well as an additional important observable in the
context of Brownian motors, namely their efficiency.
Remarkably, while most of those generalizations are conceptually
very different from a pulsating or tilting ratchet in the
original sense, an approximate or even exact mathematical
equivalence can be established in several cases.
In other cases, both the physics and the mathematics are 
fundamentally different.

\section{Seebeck ratchets}\label{sec4.8.1}
In this section we consider periodic systems under the
influence of thermal fluctuations, the intensity of which exhibits
a spatial variation with the same periodicity as the relevant
potential, while no other non-equilibrium perturbations are acting.

In a closed circuit composed of two dissimilar conductors
(or two dissimilarly doped semiconductors) a permanent electric current
arises when their junctions are kept at different temperatures \cite{pol92}.
This constitutes a thermoelectric circuit that converts thermal
energy into electrical energy. 
The effect has been discovered in 1822 by Seebeck and has been exploited,
e.g., to provide electrical power for satellites. In essence, the Seebeck effect
has the following microscopic origin:
Due to the different Fermi-levels prevailing in each of the conductors,
a kind of effective potential ramp for the electrons arises at
the junction\footnote{Within this very elementary picture we neglect
electron-electron interaction effects in the form of screening by
inhomogeneous charge densities around these potential ramps \cite{ash76}.}. 
Moving along the circuit in a definite direction, the electrons
will encounter at one junction an increasing potential ramp
and at the other junction a decreasing counterpart.
When looping in the opposite direction, the roles of the ramps is exchanged.
While sliding down a decreasing ramp is ``for free'', climbing
up an increasing ramp requires thermal activation. Therefore,
if one junction is kept at a higher temperature than the other, the
looping of electrons in one direction is more likely than in the other.

Expanding the circular motion through the closed  circuit
to the real axis yields a periodic effective potential $V(x)$
and a periodic temperature profile $T(x)$. Both have the same spatial period and 
each of them is typically {\em symmetric} under spatial inversion.
The spatial symmetry of the system is broken in that the two periodic
functions $V(x)$ and $T(x)$ are 
{\em out of phase}\footnote{A ratchet effect also arises for {\em asymmetric} 
$V(x)$ and/or $T(x)$ {\em in phase}, however, typically
in a quite different physical context, see \sect \ref{sec4.2.2}.}.
The simplest model for the electron motion consists
in an overdamped dynamics like in (\ref{4.7d})
with 
\begin{equation}
g(x)=[k_BT(x)/\eta]^{1/2}\ .
\label{see0}
\end{equation}
This model has been studied
by B\"uttiker \cite{but87} and independently by van Kampen \cite{kam88},
and has been further discussed by Landauer \cite{lan88}.
Later, similar models, either derived from a microscopic
description of the environment in terms of harmonic oscillators 
(cf. \sect \ref{sec6.5.1}), or based on a phenomenological approach
have been considered in \cite{mil95,jay96,jay96a,mah96} and 
\cite{bao96a,bao96b,bao99c,luc00}, respectively,
see also \sect \ref{sec4.8.4}.

Though the physical systems behind this {\em Seebeck ratchet} model and the
one in (\ref{4.7d}) are quite different, the mathematics is practically the 
same and in this sense {\em the Seebeck ratchet is closely related to a 
fluctuating potential ratchet} \cite{han96}.
One difference is that in one case the potential $V(x)$ is asymmetric
and the fluctuations of this potential of course ``in phase'' with the
``unperturbed'' (average) potential, while in the other case the symmetry
is broken due to a phase shift between $V(x)$ and $T(x)$.
A second possible difference is that after the white noise
limit $\tau\to 0$ in (\ref{4.7f}) the adequate treatment of the multiplicative
noise in (\ref{4.7d}) may not always be in the sense of Stratonovich.
For instance, if the dynamics (\ref{4.7d}) arises as limiting case with
negligible inertia effects (white noise limit $\tau\to 0$ in 
(\ref{4.7f}) {\em before} the limit of vanishing mass)
then \cite{gra82,jay96a,mah96,sek99,mat00a} a white noise $\xi(t)$
in the sense of Ito \cite{ris84,hor84} arises in (\ref{4.7d}). As a
consequence, the second summand in (\ref{4.8}) takes the modified
form $\partial g^2(x)/\partial x$ and the integrand in (\ref{4.9}) acquires an
extra factor $g(y)/g(x)$.
A still different treatment of the thermal noise $\xi(t)$ in (\ref{4.7d})
may be necessary in physical contexts without an inertia term right from the
beginning, see \cite{kam87,kam88,lan88} and further references therein.
Independent of these details, {\em the main conclusion is that 
$\langle\dot x\rangle\not = 0$ if and only if}
\begin{equation}
\int_0^L\frac{V'(x)}{T(x)}\, dx \not = 0
\label{see1}
\end{equation}
provided that both, $V(x)$ and $T(x)$ are 
$L$-periodic\footnote{Similarly as in \eq (\ref{4.11}), the sign
of the current $\langle\dot x\rangle$ is found to be
opposite to the sign of the intergal on the left hand side of 
(\ref{see1}).
Therefore, a current inversions upon variation,
e.g., of $\eta$ is not possible in this model.}.
One readily verifies that the two
``systematic'' conditions implying $\langle\dot x\rangle =0$ are 
indeed the
symmetry and supersymmetry criteria from (\ref{ss15a}) and (\ref{ss15b}),
respectively.

Though the Seebeck ratchet thus exhibits striking similarities
with a fluctuating potential ratchet, the equivalence
is not exact.
However, such {\em an exact equivalence can be readily established
with respect to the more general class of pulsating ratchet models} 
by choosing $\xi(t)\equiv 0$ 
and\footnote{Strictly speaking, 
(\ref{s5'}) does still not respect the $L$-periodicity
(\ref{4e}). To remedy this flaw, one has to multiply the square-root in
(\ref{s5'}) by a factor $\chi(x)$, defined as 
$\chi (x) := 1$ for $x\in[0,x_0)$,
$\chi (x) := -1$ for $x\in[x_0,L)$, and
$\chi (x+L) := \chi (x)$.
The reference position $x_0$ is then chosen such that 
$\int_0^L \chi(x)\, [T(x)]^{1/2} dx = 0$
with the result that (\ref{4e}) is indeed satisfied.
Note that this extra factor $\chi(x)$ in (\ref{s5'})
does not affect the stochastic dynamics 
(\ref{4a}) in any noticeable way.}
\begin{equation}
V'(x,f(t)) = V'(x) + \sqrt{2\eta k_B T(x)}\, f(t) \ ,
\label{s5'}
\end{equation}
with $f(t)$ being a $\delta$-correlated Gaussian noise.
The basic physical picture underlying this mathematical
equivalence is rather simple:
Thermal fluctuations with a spatially periodic variation
of their strength (temperature)
may equivalently be viewed as (very fast) potential fluctuations
(cf. \sect \ref{sec4.2.2}).
We furthermore remark that by first applying the transformation (\ref{s5'})
to a pulsating ratchet, and then considering the symmetry and supersymmetry
criteria (\ref{s1}) and (\ref{ss5}) for such a pulsating ratchet
model,
one indeed recovers the corresponding original criteria for
Seebeck ratchets in (\ref{ss15a}) and (\ref{ss15b}),
respectively.

Besides the Seebeck effect itself, another application of the
model may be the electron motion in a superlattice irradiated
by light through a mask of the same period but shifted with respect to the
superlattice \cite{but87}. In such a case, it
may no longer be justified to neglect inertia effects in the
stochastic dynamics (\ref{4.7d}).
The so-called underdamped regime of such a dynamics, i.e. 
friction effects are weak in comparison to the {\em inertia effects}, 
has been analytically treated
in \cite{bla98} by generalizing the methods developed in \cite{ris79}.

There are several well-known phenomena which may in fact be
considered a close relatives of the Seebeck effect and 
thus as further instances of the corresponding ratchet scheme:
First, we may augment our closed circuit, composed of
two differently doped semiconductors, by a piece of a
metal wire. In other words, we are dealing with an electrical
circuit that contains a semiconductor diode (n-p junction).
Again, an electrical current results if the diode is kept at a
temperature different from the rest of the circuit (thermogenerator),
see also \sects \ref{sec2.2.3c} and \ref{sec6.5.5}.
Second, the same device can also act as a photodiode or photoelement
by exposing the n-p junction to a source of light.
Especially, in the case of black-body irradiation, one basically
recovers the previous situation with two simultaneous heat baths 
at different temperatures.
Third, one may replace the semiconductor diode by a tube diode.
In this context, the two above mentioned ways of generating
an electrical current are then closely related to the so-called
Richardson-effect and photoeffect, respectively.

\section{Feynman ratchets}\label{sec4.8.2}
Throughout the discussion of Smoluchowski and Feynman's {\em Gedankenexperiment} in
\sect \ref{sec2.1.1} we have assumed that the entire gadget in \fig
\ref{fig1} is surrounded by a gas at thermal equilibrium. In his
lectures \cite{fey63}, Feynman also goes one step further in considering
the case that the gas around the paddles is in a
box at temperature $T_1$, while the ratchet and pawl are in contact with
a different bath (e.g. another gas in a box) at temperature $T_2\not = T_1$,
see \fig \ref{figfey1}.

\figfeyeins

While Feynman's discussion \cite{fey63} focuses on a thermodynamic analysis 
of this nonequilibrium system and apparently contains a misconception 
\cite{par96,sek97a,mag98,hon98},
here we concentrate on its microscopic modeling in terms of a stochastic process.
Our first observation is that there are essentially
two relevant (slow) collective coordinates: One
is an angle, which characterizes the relative position of the 
pawl and an arbitrary reference point on the circumference of 
the ratchet 
in \fig \ref{figfey1} and which we will henceforth
consider as expanded to the entire real axis and denoted as $x(t)$.
As we have seen in \sect \ref{sec2.1.1}, the possibility that
the pawl spontaneously (due to thermal fluctuations) lifts itself up so that the
ratchet can freely rotate underneath, is a crucial feature of the system.
Therefore, another relevant collective coordinate is the ``height'' 
$h(t)$ of the pawl,
i.e. its position in the direction perpendicular to $x$ 
(the ``radial'' direction in \fig \ref{figfey1}).

The next modeling step consists in taking into account the thermal
environment of the paddles, governing the state variable $x(t)$,
and the second heat bath, governing the dynamics $h(t)$ of the pawl.
A realistic description both of
the impacts of the gas molecules on the paddles
(e.g. by means of a Boltzmann-equation \cite{yev01})
and of the thermal fluctuations of the pawl
on a microscoping footing is very involved.
Along the general spirit of \sect \ref{sec2.1}, a phenomenological
modeling is the only realistically practicable modeling approach.
In a first approximation \cite{sek97a,mag98}, 
these environmental effects may be modeled
by an overdamped dynamics for both $x(t)$ and $h(t)$, i.e.
\begin{eqnarray}
\eta_1 \dot x(t) 
& = &  -\frac{\partial V(x(t),h(t))}{\partial x} + \xi_1(t)\label{f01}\\
\eta_2 \dot h(t) 
& = &  -\frac{\partial V(x(t),h(t))}{\partial h} + \xi_2(t)\label{f02}
\end{eqnarray}
with two independent white (thermal) Gaussian noises
\begin{equation}
\langle \xi_i(t)\, \xi_j(s)\rangle = 2\,\eta_i k_B T_i\, \delta_{ij}\,\delta(t-s)
\label{f03}
\end{equation}
at temperatures $T_1$ and $T_2$, respectively \cite{par96}.
The interaction between $x(t)$ and $h(t)$ arises through the common
potential $V(x,h)$ which incorporates the fact that the pawl is (weakly)
pressed against the ratchet (e.g. by a spring or due to its own elasticity)
and the constraint that the pawl cannot penetrate the ratchet. 
The latter, non-holonomous constraint can be included by appropriate
``potential walls'' into $V(x,h)$. An explicit example \cite{mag98}
is
\begin{equation}
V(x,h) = \kappa\, h + \frac{\mu}{h-H(x)}
\label{f04}
\end{equation}
where $\kappa$ is the ``spring constant'' of the pawl, $H(x)$ is the
geometrical profile of the ratchet, and $\mu$ is a parameter characterizing the 
``steepness'' of the potential walls which account for the constraint $h>H(x)$.

Note that (\ref{f01}) seems in fact to perfectly 
{\em fit into the general 
framework of a fluctuating potential ratchet scheme} (\ref{6j}).
However, it actually goes somewhat beyond this scheme in that our 
usual assumption of the ``potential fluctuations'' $h(t)$
being independent of the system $x(t)$ is no loner respected, there is
a ``back-coupling'' in (\ref{f02}).

In spite of the various so far made approximations, the model is still
only tractable by means of numerical simulations.
Detailed quantitative results of such simulations can be
found in \cite{mag98,sek97a}. 
Here, we proceed with the additional approximation that the
pawl remains permanently in contact with the ratchet, i.e. the constraint
$h>H(x)$ is replaced by $h=H(x)$.
Physical realizations of such a modified system with a fixed, one-dimensional
``track'' $(x,H(x))$ of the pawl can be readily figured out.
Moreover, it is clear that in those regions of the track
with a small slope $H'(x)$,
the noise acting on $x(t)$ dominates, while the noise acting
on $h(t)$ dominates for large slopes $H'(x)$.
In other words, an effective one-dimensional ratchet dynamics with a state 
dependent
effective temperature $T(x)$ is recovered \cite{mag98,sak98,hon98,sak00}:
{\em The Feynman ratchet can be approximately reduced to a Seebeck ratchet model.}

The main results of such a simplified one-dimensional description
are qualitatively the same as for the
more complicated two-dimensional original model
(\ref{f01})-(\ref{f04}) \cite{mag98,jar99}:
If the paddles experience a higher temperature than the pawl
($T_1>T_2$) then the rotation is in the direction naively
expected already in \fig \ref{fig1}. Remarkably, for $T_1<T_2$ the
direction is inverted, i.e. the pawl preferably
climbs up the steep slope of the ratchet profile $H(x)$.

{\em Experimental} realizations of the above Feynman ratchet and pawl
gadget are not known.
In order that thermal fluctuations will play any significant role, 
such an experiment
has to be carried out on a very small scale.
Quantitative estimates in \cite{mag98} indicate
that the necessary temperature differences in
order to achieve an appreciable ratchet effect are probably
not experimentally feasible.
However, modified two-dimensional settings of the general form 
(\ref{f01})-(\ref{f04}), e.g. with $\xi_2(t)$ consiting of a thermal noise
at the same temperature as $\xi_1(t)$ and a superimposed external driving, may
well be experimentally realizable, see \sect \ref{sec6.2.4}.
Finally, a Feynman ratchet-type model for a molecular motor 
(cf. \ch \ref{cha5}) has been proposed in \cite{bra88,bra89},
though this model was later proven unrealistic
by more detailed quantitative considerations \cite{hun94,how96}.

\section{Temperature ratchets}\label{sec4.8.3}
The properties and possible applications of the temperature ratchet 
\cite{rei96} with time-periodic 
temperature variations (\ref{2.5}), (\ref{2.23'})
have been discussed in detail in \sects \ref{sec2.2.2}, \ref{sec2.4.1}, and \ref{sec2.5}.
A modified model in which the temperature changes $T(t)$ are governed 
by a dichotomous random process (cf. \eqs (\ref{6k1})-(\ref{6l}))
has been studied in \cite{luc97,li97,bao99d,bao00b}. The resulting, so-called composite
noise $\xi(t)$ gives rise to a ``minimal'' ratchet model in (\ref{2.5})
in the sense that $\xi(t)$ is a stationary, unbiased, white noise
with correlation
\begin{equation}
\langle\xi(t)\,\xi(s)\rangle = 2\, \eta\, k_B T \, (1+\sigma^2)\,\delta(t-s) \ .
\label{d1}
\end{equation}
The noise is, however, not a thermal noise (e.g. it is not Gaussian distributed),
thus the generic appearance of the ratchet effect is not in 
contradiction to the second law of thermodynamics \cite{luc97,li97,bao99d,bao00b}.

Next, we consider again the general case that
$T(t)$ may be either
a periodic function or a random process, satisfying $T(t)\geq 0$ for all $t$.
Introducing the auxiliary time \cite{han81,rei96}
\begin{eqnarray}
& & \hat t(t) :=\int_0^t dt\, T(t)/\overline{T}\label{d2}\\
& & \overline{T} := \lim_{t\to\infty}\frac{1}{t}\int_0^t dt\, T(t)\label{d3}
\end{eqnarray}
it follows that the temperature ratchet dynamics (\ref{2.23''})
can be rewritten in terms of $y(\hat t) := x(t(\hat t))$ in the form
\begin{eqnarray}
& & \eta\, \dot y(\hat t) 
= - V'(y(\hat t))\, [1+f(\hat t)] + \bar\xi (\hat t)\label{d4}\\
& & f(\hat t) := \frac{ d\, t(\hat t)}{d\hat t} - 1  \ , \label{d5}
\end{eqnarray}
where $t(\hat t)$ is the inverse of (\ref{d2}) (which obviously exists)
and where $\bar \xi(\hat t)$ is a Gaussian white noise with correlation
\begin{equation}
\langle\bar \xi(\hat t)\, \bar \xi(\hat s)\rangle 
= 2\, \eta\, k_B \overline{T} \,\delta(\hat t-\hat s)
\label{d6}
\end{equation}
which is moreover statistically independent of $f(\hat t)$.
Exploiting (\ref{d2}), (\ref{d3}) one can furthermore 
show that $f(\hat t)$ is unbiased.

In general, if $T(t)$ is a {\em stochastic} process then the relation 
between properties of $y(\hat t)$ and $x(t)$ is not obvious, since the
time-transformation (\ref{d2}) is different for each realization of $T(t)$.
However, with respect to the steady state current we can infer
from the self-averaging property (\ref{4c1}) in combination with (\ref{d2})
that
\begin{equation}
\langle \dot y\rangle =\langle \dot x \rangle \ .
\label{d7}
\end{equation}
If $T(t)$ is a {\em periodic} function of $t$ then the very same conclusion follows
immediately.
In other words, from (\ref{4.1}), (\ref{d4}), (\ref{d7}) we can conclude that,
at least with respect to the particle current,
{\em the temperature ratchet
(\ref{2.5}), (\ref{2.23'}) is exactly equivalent to a fluctuating potential
ratchet (\ref{4.1})}, independently of whether the time variations of
$T(t)$ are given by a periodic function or a stochastic process.
On the other hand, a fluctuating potential ratchet can be mapped onto
a temperature ratchet, provided $f(t)>-1$ for all $t$ in (\ref{4.1}).
Especially, from the asymptotics (\ref{6q1}) for fast
stochastic potential fluctuations the corresponding result \cite{luc97,bao99d,bao00b} for
a temperature ratchet is recovered. Likewise, from the prediction
(\ref{2.3.2}) for a periodically modulated
temperature ratchet we can read off the asymptotics for ratchets with
fast, periodically fluctuating potentials. For similar reasons, the 
qualitative analysis of the temperature ratchet for slow dichotomous temperature variations
in \fig \ref{fig6} is practically the same as for the on-off ratchet scheme
\cite{ajd92}.

The basic physical picture behind this equivalence of a temperature ratchet
and a fluctuating potential ratchet is as follows: 
Very loosely speaking one may
mimic temperature modulations by potential modulations since,
under many circumstances, it is mainly the ratio of potential 
and temperature which plays the dominant role in transport 
phenomena (cf. \fig \ref{fig6}).

We finally recall that, apart from ``accidental'' cases,
the ``systematic'' conditions implying $\langle\dot x\rangle =0$
are the symmetry and supersymmetry criteria from
(\ref{ss15c}) and (\ref{ss15d}), respectively.
Not surprisingly, these are practically the same as the corresponding 
criteria of symmetry (\ref{s1}) and supersymmetry (\ref{ss5})
for a fluctuating potential ratchet $V(x,f(t))=V(x)\, [1+f(t)]$.

\section{Inhomogeneous, pulsating, and memory friction}\label{sec4.8.4}
\subsection{A no-go theorem}\label{sec4.8.4a}
In the preceding sections we have discussed modifications of
the Smoluchowski-Feynman ratchet model (\ref{2.5}) with either a
spatial or a temporal variation of the temperature $T$ in (\ref{2.3}).
In the generic case, a finite particle current $\langle\dot x\rangle$
results in such a model, as expected from Curie's principle.
In the following, we discuss an apparently rather similar modification
of the Smoluchowski-Feynman ratchet model (\ref{2.3}), (\ref{2.5}),
namely spatial and/or temporal variations of the friction coefficient
$\eta$, with the rather unexpected result that the average particle current
in the steady state is always zero.

In the case of a {\em non-constant friction coefficient} $\eta(x,t)$,
the overdamped limit $m\to 0$ is a subtle issue
\cite{ryt81,san82,kam88,jay96a,mah96} and one better keeps a finite mass $m$ in
(\ref{2.1}) right away.
The corresponding Fokker-Planck equation for the probability
density $P=P(x,v,t)$ follows along the same line of reasoning
as in \sect \ref{sec2.1.3} and Appendix B with the result \cite{ris84}
\begin{equation}
\frac{\partial}{\partial t}\, P = 
-v\, \frac{\partial}{\partial x}\, P
+\frac{1}{m}\,  \frac{\partial}{\partial v}
\left\{ V'(x)+\eta(x,t)\, v 
+\frac{\eta(x,t)\, k_B T}{m}\, \frac{\partial}{\partial v} \right\}
\, P \ ,
\label{fri1}
\end{equation}
where $v:=\dot x$. Going over to the reduced density $\hat P(x,v,t)$
(cf. (\ref{2.12})), which is periodic in $x$ but still satisfies (\ref{fri1}),
one readily verifies that the Boltzmann distribution
\begin{equation}
\hat P^{\rm st}(x,v) = Z^{-1}\, \exp\{ -[mv^2/2 + V(x)]/k_B T\}
\label{fri2}
\end{equation}
is a steady state solution (cf. (\ref{2.18'})). Under the sufficient 
(but not necessary) condition that $\eta(x,t)>0$ for all $x$ and $t$ 
(and that $T>0$) this long time asymptotics can be proven to be 
furthermore unique 
\cite{lan54,ber55,leb57,sch80,kam92}.
The remarkable feature of the steady state distribution (\ref{fri2}) is that
the friction coefficient $\eta(x,t)$ does not appear at all.
For $\eta = const.$ we are dealing with an equilibrium system and
the second law of thermodynamics implies the result
\begin{equation}
\langle \dot x \rangle =0
\label{fri3}
\end{equation}
for similar reasons as in \sect \ref{sec2.1}.
Considering that (\ref{fri2})
does not depend on the friction coefficient, it is quite plausible
that the result (\ref{fri3}) carries over
to arbitrary $\eta(x,t)$.
The same conclusion is corroborated \cite{jay96,jay96a,mah96,luc00}
by a more detailed calculation similarly as in \sects \ref{sec2.1.3'}
and \ref{sec2.1.4}.

The basic physical reason behind the result (\ref{fri3}) is
that the model (\ref{2.1}), (\ref{2.3}) describes an equilibrium 
system for arbitrary $\eta(x,t)$:
In fact, we have noticed below \eq (\ref{2.3}) that the
friction coefficient can also be considered as the coupling strength
between the system and its thermal environment.
In the absence of other perturbations, the model
(\ref{2.1}) thus continues to describe an equilibrium system even for a 
non-constant coupling $\eta(x,t)$.
Since an equilibrium system reaches an equilibrium state
in the long time limit, the second law of thermodynamics can be invoked and
(\ref{fri3}) follows.
(Only the transient dynamics depends on the details of $\eta(x,t)$.)
Thus, there is no contradiction to Curie's principle:
The current-prohibiting symmetry, which
is easily overlooked at first glance,
is in fact once again the detailed balance symmetry.

Returning finally to the overdamped limit $m\to 0$,
we only state here the outcome of a more rigorous analysis
\cite{ryt81,san82,kam88,jay96a,mah96}, 
namely that this limit cannot
be consistently carried out in the stochastic dynamics (\ref{2.1})
itself but only on the level of the Fokker-Planck equation (\ref{fri1}),
with the result of a probability current in (\ref{2.10a}) of the
form like in (\ref{2.10d}) with $\eta(x,t)$ in place of
$\eta$.
The conclusion (\ref{fri3}) then follows
along the same line of reasoning as in \sect \ref{sec2.1.4}.

\subsection{Inhomogeneous and pulsating friction}\label{sec4.8.4b}
The microscopic origin of a time-independent inhomogeneous friction
mechanism $\eta(x)$ has been discussed in \sect \ref{sec3.2.1}, namely
a broken translation invariance of the thermal environment
with respect to the relevant (slow) state variable(s) of interest.
Physical examples are the Brownian motion near geometrical 
confinements of the fluid due to deviations from Stokes friction 
\cite{bre86,luc94,luc00,lin00a,lan01},
phase dependent dissipation in Josephson junctions 
due to the interference of pair and quasiparticle tunneling currents 
\cite{fal76},
generic chemical reactions \cite{kri92a,kri92b}
(cf. \sect \ref{sec3.2.1}), and
protein friction in molecular motors, see \sect \ref{sec5.3}.
In the following, we restrict ourselves to the most important case that
$\eta (x)$ is strictly positive and exhibits the same periodicity $L$ 
as the potential $V(x)$.

As mentioned in the preceding subsection,
the overdamped limit in the presence of an inhomogeneous friction
amounts \cite{ryt81,san82,kam88,jay96a,mah96}
to replacing $\eta$ by $\eta(x)$ in (\ref{2.10a}), (\ref{2.10d}).
By means of the transformation
\begin{eqnarray}
& & \bar x(x) := \int_0^x dx'\, \sqrt{\eta(x')/\bar \eta}
\label{q101} \\
& & \bar\eta := \left[\int_0^L \frac{dx'}{L}\, \sqrt{\eta(x')}\right]^2
\label{q102}
\end{eqnarray}
the Fokker-Planck equation for $P(\bar x,t)$ takes exactly the
constant friction form (\ref{2.10}) if one replaces $x$ by $\bar x$
and $\eta $ by $\bar\eta$.
Including the ``perturbations'' $f(t)$, $y(t)$, and $F$ (cf. (\ref{4a})),
the transformed equivalent Langevin equation takes the form
\begin{equation}
\bar \eta\,\dot{\bar x} (t) = 
-\bar V '(\bar x(t),f(t))
+\sqrt{\bar\eta/\eta (\bar x)}\,[y(t)+F]
+\xi(t)
\label{q103}
\end{equation}
where $\eta(\bar x):=\eta (x(\bar x))$ and
\begin{equation}
\bar V(\bar x,f(t)) := V(x(\bar x),f(t)) + (k_BT/2)\, 
\ln (\eta (\bar x)/\bar \eta) \ .
\label{q104}
\end{equation}
With (\ref{q101}), (\ref{q102}) one sees that $\bar V(\bar x)$ and $\eta (\bar x)$ 
exhibit again the same periodicity $L$ as $V(x)$ and $\eta(x)$.

In other words, {\em we have mapped the original overdamped
ratchet dynamics with inhomogeneous friction to our standard working
model} (\ref{4a}) with the only exception that the homogeneous
external perturbation $[y(t)+F]$ acquires a spatially periodic 
multiplicative factor.
Namely, an originally pure tilting ratchet now picks up
some pulsating potential admixture, while a static force $F$
is now accompanied by a modification of the static part of the 
periodic potential profile.
As a consequence, the basic qualitative features of such 
inhomogeneous friction ratchet models
can be readily understood on the basis of our previously
discussed results.
For instance, a ratchet effect may now arise even if both
$V(x,f(t))$ and $\eta (x)$ are symmetric according
to (\ref{s1}) but each with a different 
$\Delta x$-value, i.e. they are out of phase,
since this gives rise to a genuine effective ratchet potential
with broken symmetry in (\ref{q104}).
Regarding various interesting quantitative results for several
specific models we refer to 
\cite{mil95,jay96,jay96a,mah96,bao98b,bao99c,dan00,dan01a,dan01b}.

An additional time-dependence of the friction $\eta (x)$
may arise under certain temporal
variations of the system-plus-environment which are sufficiently slow
in comparison with the characteristic relaxation time of the environment in order
to always maintain an (approximate) accompanying equilibrium state of the bath. 
In such a case, the time dependence of $\eta (x,t)$ may be absorbed into
the potential and the forces appearing on the right hand side of 
the properly rewritten original stochastic dynamics (\ref{4a})
similarly as in \sect \ref{sec4.8.3}. 
Afterwards, the remaining $x$-dependence can again
be transformed away as in (\ref{q103}).

The special instance of a pulsating potential
$V(x,f(t))$ in combination with a {\em pulsating friction 
coefficient} $\eta (x,f(t))$, both with the same periodicity in $x$,
has been studied in the case of a dichotomous driving $f(t)$ in
\cite{luc00}. Since detailed balance symmetry gets lost
in this way and in the absence of special symmetries, a ratchet 
effect is recovered \cite{luc00} for such a 
{\em pulsating friction ratchet}. 
Especially, both
$V(x,f(t))$ and $\eta (x,f(t))$ may be symmetric according
to (\ref{s1}) but each with a different $\Delta x$-value, i.e.
they are out of phase. 
As a further generalization, the transition probabilities
between the two states of $f(t)$ may also periodically vary with $x$.
Unless both of them are in phase with $\eta (x,f(t))$, a ratchet
effect is then generically observed even in the absence of
the potential\footnote{A trivial example is:
if $f(t)$ is in state $1$ then for $x\in[0,L]$ the friction $\eta(x)$
is non zero only within $[0,3L/4]$ and the transition probability into state $2$
only within $[L/2,3L/4]$; if $f(t)$ is in state $2$ then everything is 
shifted by $L/2$.} $V(x,f(t))$.

\subsection{Memory friction and correlated thermal noise}\label{sec4.8.4c}
Instead of forcing an unbiased ($F=0$) system of the general form (\ref{4a})
by means of the perturbations $f(t)$ or $y(t)$ away from thermal
equilibrium, one may as well consider a modification of the friction
term $\eta\dot x(t)$ (while $f(t)=y(t)=0$).
Much like in the previous subsection, the overdamped limit becomes then
rather subtle and one better keeps a finite mass $m$ in the original 
description (\ref{2.1}).
The simplest such generalization 
\cite{nyq28,joh28,cal51,mag59,zwa73,gra80,gra82a,cal83,for88,han90,wei99a}
includes a so-called linear memory friction of the form
\begin{equation}
m\, \ddot x(t) + V'(x(t)) = 
- \int_{-\infty}^t \hat \eta (t-t')\, \dot x(t')\, dt' + \xi(t) \ .
\label{fri4}
\end{equation}
see also \sects \ref{sec3.2.1} and \ref{sec6.5.1}
(the lower integration limit $0$ in (\ref{qm1.7}) is 
recovered from (\ref{fri4}) by observing that $\dot x(t)\equiv 0$ 
for times smaller than the initial time $t=0$). 
The proper generalization (cf. (\ref{2.3}), (\ref{4b})) of the
{\em fluctuation-dissipation relation} then reads
\cite{nyq28,joh28,cal51,mag59,zwa73,gra80,gra82a,cal83,for88,han90,wei99a}
\begin{eqnarray}
\langle\xi(t)\xi(s)\rangle & = & 
\hat \eta(t-s) \, k_B T \ ,
\label{fri5}
\end{eqnarray}
see also (\ref{q11}), (\ref{q12}).
Unless $\xi(t)$ is a stationary {\em Gaussian} process with zero mean and correlation
(\ref{fri5}), the environment responsible for the dissipation and fluctuations
in (\ref{fri4}) cannot be a thermal equilibrium bath \cite{rei01a} 
and therefore a ratchet
effect is expected generically (and indeed observed), 
as exemplified in \cite{iba97}.
Especially, the fact that some noise (Gaussian or not)
is uncorrelated (white) does {\em not} necessarily imply that
its origin is a thermal equilibrium environment 
nor does a correlated noise exclude thermal equilibrium.

\section{Ratchet models with an internal degree of freedom}\label{sec5.7}
In this section we briefly review Brownian motors which
posses -- in addition to the mechanical coordinate $x$ -- an 
``internal degree of freedom'' analogous to the chemical state 
variable of molecular motors (cf. \sect \ref{sec4.5} and
\ch \ref{cha5}),
but without the main intention of representing a faithful 
modeling of such intacellular transport processes.
Another closely related model class are the two-dimensional tilting
ratchet systems from \sect \ref{sec6.2.4}.

So-called {\em active Brownian particles} \cite{ste94,schw97,sch98}
with an ``energy depot'' as additional internal variable have been
considered in \cite{til99,sch00b} under the influence of a static ratchet potential.
The internal energy depot models the capability to take up
energy from the environment, store it, and (partially) convert it into directed 
motion.
While the original, phenomenological model dynamics from \cite{til99,sch00b} does not
fit into the generalized pulsating ratchet scheme from (\ref{4.2.1}),
it is possible to transform it into an equivalent form
closely related to (\ref{4.2.1}),
namely a combined fluctuating potential and temperature ratchet with
a back-coupling mechanism.
Upon variation of the noise strength or of the energy supply,
a remarkably rich behavior of the particle current $\langle\dot x\rangle$,
both in magnitude and sign, is recovered \cite{til99,sch00b}.

A different type of ``active Brownian particles'', namely a 
{\em reaction-diffusion system} 
with one species of particles possessing a fluctuating potential ratchet
type internal degree of freedom (chemical reaction cycle), 
has been demonstrated in \cite{sch00} to induce a pattern forming process.
Note also the connection of this setup with the collective ratchet 
models from \ch \ref{cha7}.

A precursor of a two-headed motor enzyme model is 
the {\em ``elastic dumb-bell''} 
from \cite{ajd94}, consisting of two point-like Brownian particles
which are linked by a (passive) elastic spring, and which move in 
either the same or two different (shifted) on-off ratchet potentials,
see also \cite{klu01} and \cite{li00a}
for the cases of fluctuating and traveling potential ratchet schemes,
respectively.
The corresponding ``rocking-ratchet''
situation, i.e. a static ratchet potential but a periodically
varying external driving force, has been studied in \cite{cil98}
and extended in \cite{cil98a} and \cite{cil01} to the cases when each
of the two partcles moves in a two-dimensional ratchet potential and
in a one-dimensional symmetric periodic potential, respectively.
The analogous ``fluctuating force ratchet'' in the limit of a 
``rigid dumb-bell'' has been considered in \cite{dia96,dia97,klu01}.
Note that there exists a close connection to the models for
single molecular motors in \sect \ref{sec5.6}, especially those
in \cite{der96,mog98,els00a,els00b}.

A first {\em experimental} realization of such
a two-head-like system with an active, 
spring-like element was reported as early as 1992 \cite{osa92}: A curved strip
of gel with periodically varying curvature
(by externally applied electric fields) moves in a worm-like fashion with
its two ends (``heads'') along a ratchet-shaped substrate.
A second experimental ratchet system with an
``internal degree of freedom'' was presented
in \cite{san99}:
A water droplet in oil is positioned on a ratchet shaped surface and its shape
(internal degree of freedom) is periodically changed by means of externally applied
electric fields. With the shape also the contact angles between the droplet
and the surface change, with the result of a systematic directed motion.
Since the droplet covers several periods of the ratchet, the
rough picture is a somewhat similar worm-like motion as before,
though the actual systems and their possible applications are of
course completely different.

A Brownian particle in a periodic electric potential with
an autonomously rotating {\em ``internal electric dipole''}
has been theoretically analyzed in \cite{li98}.
Since the direction of this rotation breaks the spatial symmetry, the
periodic potential may be chosen symmetric in this model.
While there exists a close formal analogy with 
the traveling potential ratchet scheme from \sect \ref{sec4.3.1},
the physical picture is different \cite{li00}.

\section{Drift ratchet}\label{sec6.4}
In this section we discuss in some detail
the so-called drift ratchet scheme \cite{ket00} which resembles
a rocking ratchet but at the same time goes substantially beyond
our original tilting ratchet model from (\ref{6.1}).
We will outline the theoretical framework of a particle separation device 
based on this drift ratchet scheme, presently under construction \cite{mul00}
in the laboratories of the Max-Planck-Institut in Halle (Germany).

The system basically consists of
a piece of silicon -- a so-called silicon wafer -- 
pierced by a huge number of identical pores with a ratchet-shaped
(periodic but asymmetric) variation of the diameter along the pore-axis \cite{mul00},
see \fig \ref{figket2}.
The pores are filled with a liquid (e.g. water) 
which is periodically pumped back and forth in an unbiased
fashion, i.e. such that no net motion of the liquid 
is produced on the average.
Suspended into the liquid are particles of micrometer size and the  
objective is to separate them according to their size. 


\figketzwei

For a theoretical description of the particle
motion 
we consider
a single, infinitely long pore under the idealizing assumptions
that the particles have spherical shape,
that the suspension is sufficiently diluted such that particle
interaction effects are negligible, and that the interaction with the
pore walls can be captured by perfectly reflecting boundary conditions.
For the typical 
parameter values of the real experiment, buoyancy 
effects due to the influence of gravitation as well as inertia
effects of the particle are negligibly small, i.e., the particle dynamics 
in the viscous liquid is strongly overdamped.
Assuming that the three-dimensional time-dependent 
velocity field $\vec v (\vec x,t)$ of the liquid is known, the
particle $\vec x(t)$ is governed by
the deterministic dynamics 
$\dot{\vec{x}}  \left (t \right) = \vec v  \left (\vec{x}(t),t \right )$.
Here, $\vec v (\vec x,t)$ is, stictly speaking, not the velocity field of
the fluid alone but rather the speed with which a spherical particle
with center at $\vec x(t)$ and a small but finite radius is carried
along by the surrounding liquid. 
This deterministic dynamics induced by the streaming liquid has to
be complemented by 
the diffusion of the micrometer sized particle due to random 
thermal fluctuations $\vec \xi (t)$, which are caused 
by the impacts of the surrounding
liquid molecules, and which we model in the usual way as 
Gaussian white noise. We thus end up with the following  
stochastic dynamics for the trajectory  
$\vec{x}(t)$ of a microsphere inside a single pore: 
\begin{equation} 
\dot{\vec{x}}  \left (t \right) = \vec v  \left (\vec{x}(t),t \right) 
+ \vec{\xi}  \left (t \right ) \ . 
\label{ket1} 
\end{equation} 
The vector components $\xi_i(t),i=1,2,3$, of the noise $\vec\xi(t)$ 
are unbiased Gaussian processes with correlation
\begin{equation}
\langle \xi_i(t) \xi_j(s) \rangle   =  \frac{2 \, k_B T}{\eta}\, 
\delta_{ij}\, \delta(t-s) \ .
\label{ket2}
\end{equation} 
The friction coefficient $\eta$ is in very good approximation
given by Stokes law
$6 \pi  R \, \nu$, where $R$ is the particle radius and
$\nu$ the viscosity of the liquid. 

In view of the external, time-periodic pumping of the
liquid through the pores, the above so-called drift-ratchet 
scheme has a certain similarity to a rocking ratchet system.
On the other hand, it also remids one of the
hydrodynamic ratcheting mechanism based on the the so-called 
Stokes drift \cite{bor98,jan98,vdb99,li00,ben00} as discussed in the
context of traveling potential ratchets in \sects
\ref{sec4.3} and \ref{sec4.4}.
However, in contrast to both, the rocking as well as the traveling 
potential ratchet paradigms, in the present case (\ref{ket1})
no  ``ratchet-{\em potential}'' is involved\footnote{Under the
assumption of an incompressible fluid, i.e. $\vec\nabla \cdot \vec{v} = 0$,
the velocity field $\vec{v}$
appearing in (\ref{ket1}) can be written as the curl of some
vector potential, but never as the gradient of a scalar potential.}.
Furthermore, the dynamics within a single pore is still a 
complicated three-dimensional 
problem that cannot be reduced in a straightforward manner
to an effective one-dimensional model. 

After one period of driving, the liquid in the
pore returns to the same position from where it started out.
Why should we not expect the same null-effect for the 
suspended particles?
The basic reason is as usual the far from
equilibrium situation, created in the present case by the
periodic pumping, in combination with Curie's principle, which
predicts the generic appearance of a preferential direction
of the stochastic particle dynamics with broken
spatial symmetry (\ref{ket1}).
The physical mechanism for the emergence of such a 
non-vanishing net particle current are the thermal diffusion between 
``liquid layers'' of different speed  
and the collisions with the pore walls:
Through the asymmetry of the pore-profile, an asymmetry between  
pumping forth and back arises for both the thermal inter-layer 
diffusion and the collisions with the pore-walls, 
resulting in a non-vanishing particle displacement on average after one 
driving period. 
The fact that the excursions of the particles during one driving period 
are typically much larger than the net displacement after one 
period (see 
\fig \ref{figket3})
motivates the name ``drift ratchet''. 

\figketdrei

The calculation of the velocity field $\vec{v} $ in (\ref{ket1})
is a rather involved hydrodynamic problem in itself.
For details of the necessary approximations (and their justification)
in order to make the
problem tractable at least by numerical methods we refer to
\cite{ket00}.
Once such an approximation for $\vec{v} $ is available, the numerical
simulation of the stochastic dynamics is straightforward.
Typical results for realistic parameter values
are depicted in 
\fig \ref{figket3}, demonstrating that
the direction of the particle current depends very sensitively 
on the size of the particles. 

\figketvier
\figketfunf

While, according to \sect \ref{sec3.5},
such current inversions are a rather common phenomenon,
the distinguishing feature of our present device is its
{\em highly parallel architecture}\footnote{We remark that also the
experimental systems from \cite{rou94,fau95a,gor98} 
(discussed in \sect \ref{sec4.1.1} and at the end of 
\sect \ref{sec4.3.1}) include a parallelization in two dimensions, 
while in the present case three dimensions are exploited.}: 
a typical silicon waver
contains about one million pores per square centimeter.
On the other hand, the pores in a real silicon wafer are
not of infinite length -- as so far assumed -- but rather the wafer 
is connected at both ends to basins of the liquid-plus-particle 
suspension and the actual pumping device.
For practical applications, not the steady state 
current in an infinite pore is of main importance, 
but rather the time needed to achieve reasonably large 
concentration differences between the two basins
(see also the discussion below \eq (\ref{4.3})).
We now focus on the case of two identical basins,
each of an extension $\Delta z$ along the $z$ axis
and of the same cross section as the wafer 
(perpendicular to the $z$ axis).
The typical time evolution of the particle density for such a
setup is depicted in \figs \ref{figket4} and \ref{figket5}.
These calculations 
predict a remarkable theoretical separating power of the
device.
Its experimental realization -- presently under 
construction \cite{mul00} -- thus appears to be a promising new particle 
separation device, possibly superior to existing 
methods for particles-sizes on the micrometer scale.

\section{Spatially discrete models and Parrondo's game}\label{sec3.9}
The spatially discretized counterpart of our working model (\ref{4a})
arises when the state variable
$x(t)$ is restricted to a set of discrete values $x_i$.
In the simplest case, the time evolution is given by a so-called 
{\em Markov-chain dynamics}, i.e. transitions are only possible between
neighboring states $x_i$ and $x_{i\pm 1}$, and they are governed by
transition rates $k_{i\to i\pm 1}(t)$, which in general
may still depend on time.
As a consequence, the probability distribution
$P_i:=\langle \delta( x_i-x(t)) \rangle$ evolves in time
according to a master equation of the form
\begin{equation}
\dot P_i(t) = -[k_{i\to i+1}(t)+k_{i\to i-1}(t)]\, P_i(t)
+ k_{i+1 \to i}(t)\, P_{i+1}(t)
+ k_{i-1 \to i}(t)\, P_{i-1}(t) \ .
\label{dis1}
\end{equation}
The spatial periodicity of the system implies that there is an integer
$l$ with the properties that 
\begin{eqnarray}
x_{i+l} &  = & x_i+L\label{dis1a}\\
k_{i+l\to j+l}(t) & = & k_{i\to j}(t)
\label{dis2}
\end{eqnarray}
for all $i$ and $j$.

A periodic Markov-chain model (\ref{dis1})-(\ref{dis2}) may arise
in several different contexts.
The most prominent is the {\em activated barrier crossing limit} as discussed in
\sect \ref{sec3.6}, i.e., the spatially continuous dynamics (\ref{4a})
is characterized by rare transition events between metastable states $x_i$.
In the simplest (and most common) case $l=1$,
i.e. there is only one metastable 
state $x_i$ per spatial period $L$ and the rates are -- possibly after
temporal coarse graining (see \sect \ref{sec3.6}) --
independent of time.
While the actual calculation of those rates $k_\pm := k_{i\to i\pm 1}$
is in general highly non-trivial, once they are given, the determination
of the current and the diffusion coefficient is straightforward, 
see (\ref{4i}), (\ref{4j}) and the footnote on page \pageref{fot51}.

The problem of calculating the rates $k_{i\to i\pm 1}(t)$
simplifies a lot if the characteristic
time scale of the driving $f(t)$ and/or $y(t)$ in (\ref{4a}) is
much larger than the intrawell
relaxation time within any metastable state
(but not necessarily larger than the characteristic 
interwell transition 
times $1/k_{i\to i\pm1}(t)$ themselves).
Under these circumstances, an {\em adiabatic approximation} like in
\sect \ref{sec2.4.1} can be adopted, with the result that at 
any given time $t$, the rates $k_{i\to i\pm 1}(t)$ are given by
a Kramers-Smoluchowski type expression analogous to (\ref{2.23t}).
Comparing (\ref{4a}) with (\ref{2.21}), we see that in those rate 
expressions not only the instantaneous effective potential
$\veff (x,t)=V(x,f(t))-x y(t) -xF$ (cf. (\ref{2.21'})) depends
on $f(t)$ and/or $y(t)$, but also the locations $\xmin =x_i$
of the metastable states (local minima) and of the activated
states (local maxima) $\xmax$.
Besides the slow variations of $f(t)$ and/or $y(t)$,
the implicit assumptions of this approximation are that
the number of metastable states within one spatial period $L$ is
the same for all times $t$, that their position changes in the course of
time continuously or with not too big jumps,
and that the potential barriers between any two of them is much
larger than the thermal energy $k_B T$.
Within these restrictions,
{\em any spatially continuous class of ratchets from \sect \ref{sec3.1.3}
immediately entails a spatially discretized counterpart}.
Especially, we note that the characteristic features of the diffusion 
ratchet scheme will be a time-dependent temperature $T(t)$
in the Kramers-Smoluchowski rates (\ref{2.23t}), while
for a Seebeck ratchet (\sect \ref{sec4.8.1}), the effective barriers
$\Delta\veff$ and pre-exponential factors in (\ref{2.23t})
have to be calculated along the lines of \sect \ref{sec4.2.2}.

Along this general ideology,
the spatially discretized on-off ratchet scheme (see \sect \ref{sec4.1})
has been worked out in \cite{fre99,amb99}, while a modified
on-off description of a Feynman ratchet (see \sect \ref{sec4.8.2})
is due to \cite{jar99}.
As another variation, an {\em asynchronously pulsating} on-off
model (cf. \sect \ref{sec3.2.2}) has been put forward in \cite{schi97}.
In such a model, (\ref{dis2}) is no longer satisfied
and instead within each spatial period $L$ the potential 
switches independently between its on- and an off-state.
If these switching events within neighboring periods are correlated
or anticorrelated, the current is enhanced as compared to the completely
uncorrelated case \cite{schi97}.
Related, spatially continuous, asynchronously pulsating ratchet
models have been studied in \cite{par98a,par98b,hoh99,par00,hoh01}.

Spatially discretized pulsating ratchet models have been addressed in
\cite{sok98,sok99},
temperature ratchets in \cite{sok97,sok98},
traveling potential ratchets in \cite{kol98},
and rocking ratchets in \cite{sok98,sok99,fre99},
see also \cite{der82,der83,keh97,der98z,keh00} for the case of extremely
slow rocking.

For biological intracellular transport processes
(cf. sect \ref{sec4.5} and \ch \ref{cha5}), spatially
discretized descriptions arise naturally and
have been analyzed in detail e.g. in 
\cite{tso86,wes86,lei91,lei93,duk96,ast96a,ast96b,fis99,fis99a,kel00,kol00a,kol00b,how01,fis01}.

In all those works, the above mentioned approximation of the
rates $k_{i\to i\pm 1}(t)$ by instantaneous Kramers-Smoluchowski type
expressions (\ref{2.23t}) have been exploited.
The advantage of such an approach is that closed analytical
solutions can often be obtained, especially if the driving
$f(t)$, $y(t)$, and/or $T(t)$ jumps (either periodically
or randomly) between only a few different values.
Since the main qualitative findings are very similar as for the
spatially continuous case (see \chs \ref{cha4}-\ref{cha6})
we do not discuss these features
in any further detail at this place.
We only remark that if the spatially continuous model leads to a 
vanishing current in the adiabatically slow driving case
(e.g. for fluctuating potential and temperature ratchets), then
at least two metastable states $x_i$ per period $L$ (i.e. $l\geq 2$)
are required
for a ratchet effect in the spatially discrete counterpart \cite{amb99}.
In any other case, one metastable state $x_i$ per period $L$ 
(i.e. $l=1$)
is sufficient. Such spatially discretized, adiabatically driven
models with a minimal number $l$ of states
per period are sometimes called {\em minimal ratchets} in
view of their mathematical and conceptual simplicity.

We emphasize again that while discrete models are usually easier to
analyze than their spatially continuous counterparts, the actual 
hard problem has now been shifted to justifying such a 
discretized modeling and to determine the rates
(``phenomenological model parameters'')
either from a more detailed (usually continuous) description
(cf. \sect \ref{sec3.6}) or from experimental observations.

A second context in which a spatially discretized dynamics (\ref{dis1})
arises is the {\em numerical method} for solving the originally continuous 
problem (\ref{4a}) which has been introduced in \cite{els96}
and applied to various specific models in \cite{rei96ra,ari96,ari98,san98,ari99}.
Chosing the rates $k_{i\to i\pm 1}(t)$ according to the
recipe from \cite{els96},
this numerical scheme approximates the solution of the
continuous system
better and better as the number of states $l$ per period increases.
Conversely, the often analytically solvable
models with only very few states $x_i$
per period $L$ may be still considered as a first rough
approximation of the spatially continuous problem.

Another cute application of the discretized on-off ratchet scheme
has been invented by Parrondo 
\cite{parrondo,vdb99a,har99a,har99b,har00x1,har00x2,har00x3}.
Namely, the spatially discretized random dynamics for
both the on- and the off-configurations of the potential are 
re-interpreted as games, and by construction each of these two 
games in itself is fair (unbiased).
The astonishing phenomenon of the ratchet effect then translates
into the surprising observation
that {\em by randomly switching between two fair games one ends up with
a game which is no longer fair}.
This so-called {\em Parrondo paradox} is thus in some sense the
game theoretic transfiguration of Brillouin's paradox
from \sect \ref{sec2.2.3c}.
Generalizations are obvious:
For instance, by switching between two games, each (weakly)
biased into the same direction,
the resulting game may be biased just in the
opposite direction. Another option is to take as starting 
point for the translation into a game
a ratchet model different from the on-off scheme \cite{par00a,tor01}, and so on.

\section{Influence of disorder}\label{sec3.7}
In this section we briefly review some basic effects which arise
if the periodicity of the potential (\ref{4e}) is modified by a certain amount
of quenched spatial disorder.
Our starting point is an ``unperturbed'', strictly periodic system in the activated 
barrier crossing limit as discussed in \sect \ref{sec3.6}, i.e. transitions 
between neighboring spatial ``cells'' of length $L$ can be described by 
``hopping''-rates $k_+$ and $k_-$ to the right and left, respectively.
Without loss of generality we furthermore assume that the unperturbed
current $\langle\dot x\rangle$ in (\ref{4i}) is positive, i.e.
\begin{equation}
k_+ > k_- \ .
\label{diso1}
\end{equation}

In the simplest case we may now introduce a quenched randomness as follows:
For each pair of neighboring
cells we interchange with a certain probability $p$
the original transition rates $k_+$ and $k_-$ to the right and left.
For instance, in a piecewise linear ``sawtooth-potential'' as depicted in 
\fig \ref{fig4.1}, such an interchange of the transition rates can be realized by
randomly inverting the orientation of each single saw-tooth with probability $p$
{\em independently} of each other.
Without loss of generality we can restrict ourselves to
probabilities
\begin{equation}
0\leq p \leq 1/2 \ .
\label{diso2}
\end{equation}

The following basic effects have been unraveled by Derrida and Pomeau 
\cite{der82,der83}: Upon increasing $p$ the particle current
$\langle\dot x\rangle$ monotonically decreases from its initial
value (\ref{4i}) and vanishes for $p\geq p_1$, where
\begin{equation}
p_1:=\frac{k_-}{k_+ + k_-} \ .
\label{dis3}
\end{equation}
More precisely, for $p\geq p_1$ the mean displacement $\langle x(t)\rangle$ 
grows asymptotically slower than linearly with $t$.
The effective diffusion coefficient (\ref{4c2}) increases monotonically from
its unperturbed value (\ref{4j}) and diverges at $p=p_2$, where
\begin{equation}
p_2:=\frac{k^2_-}{k^2_+ + k^2_-} \ .
\label{dis4}
\end{equation}
For $p_2\leq p\leq p_3$, where
\begin{equation}
p_3:=\frac{\sqrt{k_-}}{\sqrt{k_+} + \sqrt{k_-}} \ ,
\label{dis5}
\end{equation}
a superdiffusive behavior arises ($D_{\rm{eff}}=\infty$),
i.e. the dispersion $\langle[x(t)-\langle x(t)\rangle]^2\rangle$ grows 
asymptotically faster than linearly with $t$, switching 
over \cite{bou90} to a subdiffusive behavior
(slower than linear growth of the dispersion,  i.e.
$D_{\rm{eff}}=0$) for $p_3<p\leq 1/2$.
(Note that $0<p_2<p_1<p_3<1/2$).
At least in the regimes where they are finite, the
quantities $\langle\dot x\rangle$ and $D_{\rm{eff}}$ are self-averaging,
i.e. the same (finite) value is observed with probability $1$ for any given 
realization of the quenched disorder.
A simple intuitive explanation of these results does not
seem possible, which may not be so surprising in view of the
above mentioned self-averaging issue and other subtle problems of commuting
limits in this context, see \cite{bou90,koz99,den00} and references therein.

More general, but still uncorrelated randomizations of the transition rates 
between pairs of neighboring ``cells'' of length $L$ are given already
by Derrida in \cite{der83}. A variety of cases with {\em correlated} 
randomization has been
discussed in \cite{mar97} (see also the review \cite{bou90})
together with several interesting physical applications.

A bold but apparently quite satisfactory approximative extension 
{\em beyond the activated barrier crossing limit} 
has been proposed in \cite{har97}:
The basic idea is to evaluate, either analytically or numerically, for the
unperturbed (strictly periodic) ratchet dynamics
both the current $\langle\dot x\rangle$ and the diffusion coefficient
$D_{\rm{eff}}$. Introducing these results for 
$\langle\dot x\rangle$ and  $D_{\rm{eff}}$ into (\ref{4i}) and (\ref{4j})
yields formal expressions for the rates $k_\pm$ even though
these rates no longer adequately describe the actual transitions between 
neighboring ``cells''. Assuming that a randomization of the ratchet potential can still be
captured by a corresponding randomization of the formal forward and backward
rates $k_\pm$, one thus can continue to use
Derrida and Pomeau's formulas \cite{der82,der83} for an approximative
description of such a randomized ratchet dynamics.
For the example of an on-off ratchet scheme, a fair
agreement of this approximative approach with accurate numerical
simulations has been reported in \cite{har97}.

Another, more systematic first step beyond the activated
barrier crossing limit is due to \cite{ali99}, considering
a fluctuating force ratchet with a very general
disordered potential $V(x)$ that is (additively)
driven by asymptotically weak symmetric Poissonian shot
noise (cf. \sect \ref{sec6.1}).

A deterministic ($T=0$) rocking ratchet model 
with quenched spatial disorder has been addressed in \cite{pop00}.
Similarly as before, the current decreases and the (deterministic)
diffusion accelerates with increasing disorder, but apparently
these quantities no longer exhibit the experimentally important
self-averaging property.
Results more in accordance with the above described standard scenario of 
Derrida and Pomeau are recovered upon including inertia 
effects \cite{ari01}.
A similar overdamped case but with finite $T$ and adiabatically 
slow rocking has been addressed in \cite{jia01}.

\section{Efficiency}\label{sec3.8}
The issue of efficiency of Brownian motors has recently
developed into an entire subfield of its own right.
Here, we resrict ourselves to a very short overview.

The most widely accepted {\em definition of the efficiency} for a ratchet
dynamics of the general form (\ref{4a}) is given by the ratio 
$R$ of the average mechanical
work per time unit $\langle\dot x\rangle\, F$ produced by the the 
``energy transducer'' $x(t)$ and
the average net power input $\langle\pin\rangle$ stemming from the external
driving $f(t)$ and/or $y(t)$, i.e.
\begin{equation}
R:=\frac{\langle\dot x\rangle\, F}{\langle\pin\rangle} \ .
\label{e1}
\end{equation}
Both averages in this equation are meant with respect to all random processes
and time-periodicities involved in (\ref{4a}) and
transients are assumed to have died out.
For ergodicity reasons, both averages can then also be rewritten as long time
averages for a single realizations of the stochastic dynamics (\ref{4a}),
cf. (\ref{4c1}).
In order to quantitatively calculate the efficiency (\ref{e1}) for the different
classes of ratchet models (\ref{4a}), a very general and elegant framework has
been developed by Sekimoto \cite{sek97a,sek97b,sek98,sas98,hon00},
unifying and putting on firm grounds the various previously proposed,
model-specific expressions for $\langle\pin\rangle$ in (\ref{e1}).

As pointed out in \sect \ref{sec3.2.2}, the origin of a {\em random}
external driving $f(t)$ and/or $y(t)$ may be conceived as a thermal heat bath,
very weakly coupled to the system variable $x(t)$ in order that back-coupling 
(friction-type) effects are negligible\footnote{We remark that Sekimoto's framework
\cite{sek97a,sek97b,sek98,sas98,hon00} for evaluating the 
efficiency in (\ref{e1}) remains
valid even when such back-coupling effects are included, as it is the
case e.g. for the Feynman ratchet model in \sect \ref{sec4.8.2}
or the molecular motors studied in \ch \ref{cha5}.},
but at a temperature much higher than the temperature $T$ of the thermal 
noise $\xi(t)$.
From the viewpoint of a Carnot machine, the temperature $T$ is thus 
to be associated with the cooler
heat bath and 
{\em the maximally achievable Carnot efficiency is practically 100\%.}
If the external driving is not random but periodic, it is quite suggestive
that the same conclusion with respect to the achievable efficiency still 
carries over.

While it is not yet clear whether efficiency issues are of major relevance in
practical applications of the ratchet effect or for intracellular transport
processes, their principle interest has stimulated a considerable amount of
theoretical studies.
Apparently the first such discussion goes once again back to
Feynman's lectures \cite{fey63}, though the conclusion that under certain conditions
the maximal Carnot efficiency is reached, cannot be upheld\footnote{The same
misguided method of calculating efficiencies has been adopted in \cite{val90},
see also \cite{sak98,hon98}.}
against more detailed studies of microscopic Feynman ratchet models
\cite{par96,sek97a,mag98,hon98,vel01}, see also \sect \ref{sec4.8.2}.
For a representative example, a maximal efficiency of the order
of $10^{-4}$ has been obtained in \cite{sek97a}.

Efficiency aspects of photovoltaic and photorefractive effects in non-centrosymmetric
materials (see \sect \ref{sec6.2.1}) are surveyed in  \cite{bel80}.
Typical values of the efficiency in real materials are found to be $10^{-3}-10^{-2}$,
while theoretical situations with up to 100\% efficiency are conceivable \cite{bel80}.
The existence of so-called {\em reversible ratchet} models which achieve 
in certain 
limits the maximal possible efficiency of 100\% has also been demonstrated 
for rocking ratchets\footnote{However, under typical conditions only a few
percent are actually reached \cite{bao98a,dan01b}.}
in \cite{kam98,sok01}, for generalized pulsating ratchets
(i.e. neither of the purely fluctuating nor traveling potential type) in
\cite{par98a,par98b,par00}, and for an even more general class of models
in \cite{par99}.
A first condition for reaching the maximal
Carnot efficiency of 100\% is that the system is at every given time instant 
in a quasi equilibrium state \cite{sok00}. 
Especially, all temporal variations due to the external driving
$f(t)$ and/or $y(t)$ must be adiabatically slow. 
A second condition is \cite{par98a,par98b,par00} that the current
$\langle\dot x\rangle$ approaches zero not faster than proportional to
the inverse characteristic time scales of $f(t)$ and $y(t)$ in the adiabatic limit.
E.g. for fluctuating potential ratchets and the closely related
temperature ratchets (cf. \sect \ref{sec4.8.3}), the maximal Carnot
efficiency can not be reached since the latter condition is 
not met (see \sect \ref{sec4.1a}).

The efficiency of a Seebeck ratchet model (see \sect \ref{sec4.8.1}) with a
temperature $T(x)$ which varies periodically in space between two different
values $T_1$ and $T_2>T_1$, has been addressed in \cite{der99b,mat00a,hon00,sek00}.
According to \cite{der99b,mat00a}, 
for a suitable choice of the model parameters, an efficiency
arbitrarily close to the maximal Carnot efficiency $(T_2-T_1)/T_2$ can be 
reached. The above mentioned requirement that the system is
at any time in a quasi equilibrium state may then be 
granted by the overdamped limit $m\to 0$, entailing a vanishingly small
relaxation time of the particle $x(t)$ towards local thermal equilibrium.
On the other hand, in \cite{hon00,sek00} it is argued that
Carnot efficiency is unattainable in such systems.
A related model of a non-isothermal electrical circuit with a diode
(thermogenerator, see \sect \ref{sec4.8.1})
has been analyzed from the viewpoint of efficiency in
\cite{sok98a}.
While this one-diode setup, much like its mechanical Feynman ratchet counterpart,
always leads to an efficiency smaller than the Carnot value,
an extended model with two diodes may approach this theoretical upper limit
for the efficiency arbitrarily close \cite{sok99b},
demonstrating that even a system
which is simultaneously in contact with more than one heat bath
may still operate reversibly, see also \cite{sok01}.

Universal, i.e. largely model-independent features of efficiencies for
ratchet models close to thermal equilibrium (Onsager regime of
linear response) have been worked out in \cite{sek97b,jul97a,par99,jar99}.
Remarkably, by moving out of the linear response regime into the
far from equilibrium realm the efficiency may not necessarily decrease
\cite{par99}.
Similarly, for some ratchet models, the efficiency may even increase upon 
increasing the temperature $T$ of the thermal heat bath both, for systems near
\cite{par99} and far from equilibrium \cite{sok98,ari99,tak99}, in
contrast to what one would expect from a Carnot efficiency
point of view.

As already mentioned, fluctuating potential ratchets  
and temperature ratchets cannot reach the maximal Carnot efficiency.
Specifically, the on-off scenario leads under typical conditions to
efficiencies of a few percent \cite{par98a,par99,san98,ari99,amb99}.
However, in the case that many on-off ratchet are coupled together
(see \sect \ref{sec5.4.4}) the efficiency may again reach values
of 50\% and beyond \cite{jul95,jul97a}.
Efficiencies of at most a few percent have also been reported for
fluctuating potential ratchets (see \sect \ref{sec4.2}) \cite{sok98},
temperature ratchets (see \sect \ref{sec4.8.3}) \cite{sok97,sok98,bao00},
and coupled rocking ratchets \cite{buc00} (see \eq (\ref{7.31})).

Based on experimental measurements of intracellular transport
processes, the possibility that the molecular motor kinesin 
(cf. \ch \ref{cha5}) may
reach an efficiency as high as 50\% or even
80-95\% is discussed in \cite{how01} and \cite{fis99,fis99a}, respectively,
see also \cite{lip00a,li00}.

Other definitions of efficiencies than in (\ref{e1}) have been introduced 
and discussed in \cite{mag94,bie96b,tar98,der99a,par99,li00,bie01,her01,vel01,hum01b}.
Related quantities like entropy production, Kolmogorov
information entropy, and algorithmic complexity have been explored in 
\cite{mil95,san98,ari99,jar99}.
Evidently, with respect to the definitions of such alternative 
efficiency-type quantities it does not make sense to ask 
whether they are ``right'' or ``wrong'' (apart from the trivial
requirement that they are ``well-defined'' in the mathematical sense).
Rather, the crucial question regards their usefulness \cite{par99}.
For instance, it may be possible to agree on one such quantity as 
being a particularly 
appropriate quality measure in a certain context \cite{mag94}.
In many cases this will indeed be the standard 
``efficiency for generating force'' (\ref{e1}).
However, in other cases, it may be important to accomplish a certain
task not only by means of a minimal amount of input energy --
as in (\ref{e1}) -- but in addition within a prescribed, 
finite amount of time. 
This constrained optimization task is the basis of the
alternative ``efficiency of transprotation'' 
concept from \cite{der99a,li00,bie01},
which has been to some extent anticipated in \cite{bie96b},
and which is also closely related to the issue of
finite-time thermodynamics \cite{cur75,and83,ber99}.
For further details, we refer to the above cited original works,
see also at the end of \sect \ref{sec6.5.5}.

\chapter{Molecular motors}\label{cha5}


In this chapter we exemplify in detail the typical stochastic modeling
procedure by elaborating the general scheme
from \sect (\ref{sec4.5})
for a particularly important special case of intracellular transport,
namely so-called motor enzymes or molecular motors which are able to
travel along polymer filaments inside a cell.
Specifically, we shall focus on molecular motors 
from a subfamily of the the so-called kinesin superfamily,
which are capable of operating individually.
For the two other main superfamilies of motor enzymes 
(myosin and dynein) many of the basic 
qualitative modeling ingredients remain the same as for kinesin,
while the details are different \cite{how97a,meh99};
we will briefly address the case of molecular motors which
only can operate collectively, e.g. the 
so-called myosin II subfamily, later in this chapter.
More involved variants of intracellular transport 
like rotary mechanochemical energy transducers are
treated e.g. in \cite{mei89,doe95a,els97,els98,ber98}.
Also not covered by the present chaper are ``Brownian ratchets'' --
a notion which has been coined in a rather differen context, 
namely as a possible operating 
principle for the translocation of proteins accross 
membranes \cite{sim92,pes93,kuo00,els00c,lie01}.
A collection of computer animations which visualize
several of these intracellular transport processes
is available on the internet under \cite{bormovie}.

\section{Biological setup}\label{sec5.1}
The most primitive living cells are the so-called prokariotes, {\em i.e.}
cells without a nucleus (mostly bacteria) \cite{alb94}. 
Their interior is basically one 
large soup without any internal partition.
Since prokaryotic cells are at the same time very small,
the intracellular transport of various substances can be accomplished
passively, namely through thermal diffusion.
In contrast, eucaryotic cells (the constituents of any multicellular organism)
are not only higher organized 
but also considerably larger so that passive
diffusive transport becomes too inefficient \cite{how97b}.
Their distinguishing features are the existence of a cell nucleus
(responsible for the storage and transcription of the genetic material),
many other internal compartments, called organelles, and a 
network of polymer filaments
-- the ``cytoskeleton'' -- 
which organizes and interconnects them.
These filaments radiate from a structure near the nucleus 
called the centrosome to the periphery of the cell and so support
the shape of the cell.
Besides several other intracellular functions, which go
beyond our present scope,
they act as a circulatory system, connecting and feeding
distinct regions of the cell. They are paths along which nutrients, wastes,
proteins, etc. are transported in packages, called vesicles,
by specific motor proteins (mechanoenzymes).

One major type of such polymer filaments are fibers of proteins
called microtubuli, with the constituent protein ``tubulin''
-- a dimer of two very similar globular 
proteins ($\alpha$-tubulin and $\beta$-tubulin)
about $4\, nm$ in diameter and $ 8\, nm$ long \cite{alb94}.
The microtubulus is composed of typically 13 protofilaments 
(rows of tubulin-dimers) that run parallel to the axis of the fiber.
The emerging shape of the 
microtubulus resembles that of a hollow, moderately flexible tube
with an outer diameter of about $25\, nm$, and inner diameter
of about $17\, nm$, and an overall length of up to a few $\mu m$.
Due to the asymmetry of the tubulin-dimers,
the tube has a polarity, one end exposes only $\alpha$-tubulin, 
and the other only $\beta$-tubulin.
On top of that, the tube exhibits a definite chirality
or helicity  since the dimer-rows of neighboring parallel
protofilaments are shifted against each other.

One specific motor enzyme which can travel on a microtubulus
and pull along various objects like chromosomes, viruses, or vesicles
with chemicals in it, is the protein ``kinesin'' \cite{alb94,how97a}.
The necessary energy to move against the viscous drag is supplied by 
the so-called ATPase, i.e. the exothermic chemical hydrolysis of
ATP (adenosine triphosphate) into ADP (adenosine diphosphate)
and P$_i$ (inorganic phosphate).
The shape of a single kinesin molecule is rather elongated, about $110\, nm$
in length and about $10\, nm$ in the other two spatial directions.
One of its ends consists of a bifurcated ``tail'', capable of grasping
the cargo to be carried, then follows a very long rod-shaped middle segment, the
$\alpha$-helical coiled-coil stalk, while the other end bifurcates into
two identical globular ``heads'' or ``motor domains'' \cite{how96,blo98}.
In spite of the nomenclature, the functioning of the heads is actually quite 
similar to that of human legs, proceeding along the microtubulus in a ``step-by-step''
or ``hand-over-hand'' fashion \cite{svo93}.
We emphasize that the comparison with the walking of a human is common
but should not be overstretched:
There is evidence that the bound head
in fact produces a rotation that ``swings'' the second head 
towards its next binding site \cite{how96,how97a}.
The reason is that the kinesin as a hole seems to
possess an (approximate) axis of
rotational symmetry by 180$^o$, implying that we should think of the
two heads not as ``right and left feet'' but rather as ``two left feet'' 
\cite{how96}. Each single foot, on the other hand, 
does not share this (spatial inversion-) symmetry, it has well 
distinguishable ``heel'' and ``fingers''.

Each head comprises in particular a microtubulus-binding site as well as an 
ATP-binding site, called the ATP-binding pocket.
Accordingly, each head can bind and hydrolyze ATP on its own.
The underlying chemical reaction cycle consists of the following
four\label{fot5.2}\footnote{Additional intermediate steps can be identified \cite{gil95}
but are usually neglected due to their short lifetimes.}
basic steps (and corresponding states)
with the result of about $20\, k_B T$ energy gain per cycle \cite{lei93}:
State 1: The motor domain is interacting with the
environment and attached to the microtubulus, but without anything
else  bound to it. Transition into state 2: The head binds one ATP 
molecule out of the environment in its ATP binding pocket.
Transition into state 3: 
The ATP is broken up into ADP and P$_i$ -- the so-called power-stroke --
with the above mentioned energy gain of about $20\, k_B T$.
Transition into state 4:
The P$_i$ is released from the ATP binding pocket
and simultaneously the affinity to the microtubulus decreases dramatically,
so that the head typically detaches.
Transition into state 1:
The ADP is released,
the affinity to the binding sites ($\beta$-tubulin) of the
microtubulus becomes again large, with the result
that the head will, after some random diffusion, attach to one of them,
and we are back in state 1.

The ``energy factories'' of the cell are constantly supplying fresh ATP
and removing the used ADP and P$_i$, thereby keeping
the concentration of ATP inside the cell
about 6 decades above its thermal equilibrium (detailed balance)
value, so that the probability of an inverse (endothermic)
chemical cycle, transforming 
ADP and P$_i$ back into ATP is completely negligible.

It is noteworthy that the heads do not hydrolyze ATP at any appreciable rate unless
they interact with the microtubulus, indicating that at least part of the chemical
cycle is intimately coupled to the binding to a microtubulus \cite{svo93}.
The hydrolyzing step takes place while the head
is attached to the microtubulus; the subsequent release of P$_i$ enables the
head to release its hold so that it can take another step 
on its journey along the
microtubulus. The key to the energy transduction is thus the large change in affinity
between the heads of the motor protein and the protein filament
on which it walks. A particularly strong affinity develops between the
microtubulus-binding site of a head and the $\beta$-tubulin monomers.
As a consequence, each tubulin dimer can bind at most one head and thus a 
single head has to cover the length of two dimers (about $16\, nm$) during each 
step of the motor enzyme along the microtubulus. To complete the picture, 
it should be mentioned that the motor enzyme proceeds along the microtubulus in 
a straight way, it does not ``spiral'' around the hollow tube during its 
journey \cite{svo93,gel88,how96,cil98a}.
Rather it follows with high fidelity a path parallel to the protofilaments 
so that the
helicity of the microtubulus most likely plays no essential  role;
the main origin of the spatial asymmetry as far as the kinesin walk is concerned
is that of the constituent dimers of the microtubulus together with that of
the binding sites of the single heads.
Remarkably, each given species of the kinesin superfamily can travel only
in one preferential direction along the microtubulus,
but different species may move in opposite directions
though they may be of striking structural similarity 
\cite{sab96,fin94,how96,cro97,hen97,blo98}.

Kinesin is a so-called processive motor enzyme, that is,
it can operate individually.
A single kinesin molecule can cover a distance of the
order of $1\mu m$ before it may loose contact with the microtubulus
and diffuses away \cite{how89,how97a,blo98,man99}, and this possibly 
even against an opposing force of up to $5$ piconewtons \cite{svo94,hun94}.
The reason seems to be on the one hand that the time-interval
during which a single head is detached from the microtubulus while
``stepping forward'' is relatively short (one speaks of a high ``duty ratio'')
and on the other, that the two heads coordinate their
actions so that at least one head is always attached \cite{gil95}.
A striking manifestation of this coordination is the fact 
\cite{how96,how97a} that apparently it is
the energy gain out of the power stroke of the ``front'' head which
triggers the ``rear'' head to make a step forward.

For a more detailed exposition of the biophysical basics and
experimental findings we refer to the excellent recent 
monograph by Howard \cite{how01}.

\section{Basic modeling-steps}\label{sec5.2}
\subsection{Biochemical framework}\label{sec5.2.1}
Our first step in modeling a motor enzyme consists in 
recalling the description
of a general biochemical reaction 
\cite{fuk81,mil83,kop84,atk86,tru92,dau92,mic92,schu99}:
In principle, 
the starting point should be a quantum chemical {\em ab initio} treatment 
of all the electrons
and atomic nuclei of the molecules involved in the reaction.
Due to the clear cut separation of electron and nuclei masses,
the electron dynamics can be adiabatically eliminated 
for each fixed geometrical configuration of the nuclei
(Born-Oppenheimer approximation \cite{kop84,tru92})
with the result of an effective potential energy 
landscape for the nuclei's motion alone.
In principle, there are many quantum mechanical energy eigenstates of the 
electrons for any fixed configuration of the nuclei, giving rise to
a multitude of possible ``potential energy surfaces'' in the configuration space of 
the nuclei \cite{kop84,tru92,dau92,mic92}. 
We assume that only one of them (the ground state
energy of the accompanying electrons) is relevant in our case and 
especially is always well separated
from all the other potential energy surfaces. In other words, the effective potential
landscape governing the dynamics of the nuclei is single valued
 and no excitations of the electronic states are involved in the reaction cycle.
Since the nuclei are already fairly massive objects,
quantum mechanical effects will often play only a minor role for their dynamics,
and we can focus on an approximate classical treatment.
Indeed, while for very simple chemical reactions, a semiclassical or fully
quantum chemical treatment may be necessary and still feasible,
classical molecular dynamics is the only practically realistic approach in the 
case of a complex biomolecular system with hundreds or thousands of atoms, as
we consider it here.
In other words, all the relevant quantum mechanics of the system is 
assumed to be already encapsulated in the effective potential in which 
the nuclei move\footnote{Several of the above assumptions are in fact 
not necessary for the validity of our final reduced description
(see below), i.e. after the elimination of the (fast) bath 
degrees of freedom and the discretization of the chemical 
state variables.}.

So far our description still comprises both the molecular 
motor\label{fot5.3}\footnote{More precisely: 
the compound motor-filament system, see below.}
and its environment, typically some aqueous solution 
containing in particular
ATP, ADP, and P$_i$ molecules in certain concentrations.
The role of the environment is twofold:
On the one hand, it acts as a heat bath, giving rise to randomly 
fluctuating forces and to the associated damping (energy dissipation)
mechanism in the molecular motor's dynamics.
On the other hand, it represents a source and sink of the
reactants (ATP molecules) and products (ADP and P$_i$)
of the chemical reaction cycle.

The influence of that part of the environment which acts as thermal
heat bath can be taken into account along the same line of reasoning
as in \sects \ref{sec2.1.2} and \ref{sec3.2.1}.
The result is a classical stochastic dynamics for the motor enzyme
with a certain type of random noise and dissipation term, possibly
supplemented by a renormalization of the effective
potential landscape and the nuclear masses \cite{gra80,gra82a}.
Under the assumption that the typical potential barriers are
large compared to the thermal energy $k_B T$, the configurations of the
motor enzyme (defined by the coordinates of the nuclei) will be
restricted for most of the time to the local minima 
(metasable states) of the potential 
landscape and small fluctuations 
there about\label{fot5.4}\footnote{Trivial neutral translational and rotational 
degrees of freedom
are assumed to have been eliminated already.},
while transitions between different local minima are rare events.

In the case of simple chemical reactions, these transitions furthermore
occur practically always along the same ``most probable escape path'',
called also ``chemical pathway'', ``reaction path'', 
or ``intrinsic reaction coordinate''
in this context \cite{fuk81,dau92,mic92}. 
One thus can describe all the essential configurations of
the reaction in terms of this single intrinsic reaction coordinate
and small (thermal) fluctuations there about. The latter can again 
be taken into account by means of dissipation and fluctuation terms
in complete analogy to the above mentioned modeling
of the thermal heat bath \cite{gra80,gra82a,mil83}. 
As a result, a renormalization of the potential, the noise,
the dissipation mechanism etc. in the stochastic dynamics
of the ``intrinsic reaction coordinate'' will arise, but the main
point is that finally an effective  description of the entire reaction
in terms of a single generalized coordinate
(also called collective coordinate, state variable or 
reaction coordinate) can been achieved, see also \sect \ref{sec3.2.1}.

In the case of complex biomolecules such as a motor enzyme, different 
possible paths between the various metastable states may be realized with
non-negligible probability \cite{gil95,schu99}. 
In such a case, more than one collective coordinate (state variable) 
has to be kept in order to admit a faithful representation of all 
the possible pathways in the reduced description.
Moreover, only some of those state variables can be identified
with chemical reaction coordinates, while others are of
a more mechanical or geometrical nature (see below).
Finally, these concepts can also be generalized to cases without
a clear cut distinction of metastable states and rare transition
events, i.e. some of the (non-chemical)
state variables may be governed by
a predominantly relaxational or diffusive dynamics.

Often, an equivalent way to discriminate relevant
(generalized) coordinates which should be explicitly kept
from ``irrelevant noise'' which can be savely eliminated is according
to their characteristic time scale \cite{gre52,gra82a,kel00}
(see also \sect \ref{sec3.2.1}):
On the smallest time scales (femtoseconds) the motion of the molecule 
consists of fast but small fluctuations,
while significant conformational changes will develop only on a
much slower time scale of milliseconds.

\subsection{Mechanical and chemical state variables}\label{sec5.2.2}
For realistic systems, the above 
program -- starting with the full quantum mechanical
problem and ending with a simple approximate dynamics in terms of a few 
relevant classical stochastic variables -- cannot be practically carried 
out. Therefore, a phenomenological modeling, roughly based on the
above intuitive
picture and supplemented by experimental evidence is necessary,
see also \sect \ref{sec3.2.1}.
In the case of a motor enzyme like kinesin, the picture one has in mind is 
the following:
The actual chemical conversion of ATP into ADP and P$_i$ takes place in 
relatively well defined and small regions of the enzyme -- 
the ATP binding pockets of the two heads. 
This chemical cycle is captured by a set of chemical reaction
coordinates or state variables $y$. 
On the other hand, the much larger conformational changes of 
the enzyme as a whole are represented by a different set of 
``mechanical''\label{fot5.5}\footnote{``Mechanical'' may refer here either to the fact
that $x$ represents the global geometrical shape of the molecule,
or to the fact that some mechanical ``strains'' in the molecule,
which have been created by the chemical transitions, may be released 
through a relaxational dynamics of $x$.}
collective coordinates or state variables $x$.
Note that both $y$ and $x$ are ultimately describing nothing else than
the geometrical configuration of the nuclei, but the distinction
between chemical and mechanical coordinates are both conventional 
and suggestive\footnote{Note that the same applies for the (already
eliminated) ``irrelevant'' degrees of freedom both of the environment
and of the molecular motor itself: The dissipation and fluctuation
effects, to which they give rise, may be either due to ``mechanical''
processes (vibrations, elasic and/or inselastic collisions etc.) or due
to chemical processes (making and braking of chemical bonds etc.).}.

Particularly difficult to explicitly derive from first principles is
the central feature of the enzymatic chemical reaction cycle, namely 
that reactant and product molecules can be exchanged with the environment.
Typically, such events are possible (with non-negligible probability)
only in certain specific configurations of the enzyme
and it is assumed that the collective coordinates $(x,y)$ are capable
to faithfully monitor such events and, in particular, of whether 
some reactant/product molecule is presently attached to one of the 
heads or 
not\label{fot5.6}\footnote{It is indeed plausible that the set of possible geometrical shapes
of the enzyme while a reactant molecule is bound will be satisfactorily
disjoint from the corresponding set in the absence of the reactant,
and similarly for the products, provided the coordinates $(x,y)$ have
been suitably chosen.}.
The binding probabilities for both reactants and products depend on their
concentrations in the environment of the enzyme. The fact that 
these dependences
should be simply proportional to the respective concentrations
is very suggestive and we will take it for granted in the following
without any further derivation from a more fundamental description.

It is quite plausible that whether or not one or both heads
of the kinesin are attached to the microtubulus will have a significant
influence on both, the chemical reaction process and
the mechanical behavior \cite{svo94,how01}.
A priori, we should therefore not speak of an isolated kinesin but rather
of the compound kinesin-microtubulus system.
However, similar as for the previously discussed attachment and detachment of reactants
and products, the attachment and detachment of the
heads as well as the influence of the microtubulus in the attached state can
be represented by the relevant collective coordinates $(x,y)$
of the motor enzyme alone, if they have been appropriately chosen.

\subsection{Discrete chemical states}\label{sec5.2.4}
We recall that the ``mechanical coordinates'' $x$ describe
configurational changes of the enzyme as a whole, while the 
actual chemical ATPase is monitored by the ``chemical coordinates''
$y$ and takes place in the rather restricted spatial regions of the
ATP binding pockets.
One therefore expects that transitions between different ``chemical
states'' $y$ are accomplished during rather short time intervals
in comparison with the typical time scales on which the global geometrical
configuration $x$ notably changes \cite{how01}.
As a consequence, one can neglect the details of the transitions between
chemical states itself and focus on a discrete
number of states, $m=1,2,...\mtot$, with certain ``instantaneous'' 
transition rates $k_{m\to m'}(x)$ between 
them, which in general still depend on the
configuration $x$. Similarly, the potential landscape, which $x$ 
experiences, still depends on the ``chemical state'' $m$. Formally, 
the transition rates $k_{m\to m'}(x)$ are those between the local 
minima with respect to the $y$-coordinates at fixed $x$, which,
however, need not necessarily be local minima in the full $x-y$-space.

In doing so, it is taken for granted that
a well-defined, relatively samll number $\mtot$ of discrete
``chemical states'' exists and that all
transitions between them can be described in terms of 
rates $k_{m\to m'}(x)$.
Though such an approach is known to be problematic
in other types of proteins due to their general ``glass-like'' properties
and especially for
the binding and unbinding processes of reactants and products
in ``pockets'' of the proteins \cite{fra85,fra91},
in the context of motor enzymes like kinesin it has to
our knowledge not been
theoretically or experimentally challenged 
so far and we
will therefore follow the general belief in its adequacy.

At this point it should be emphasized again that
one motor enzyme incorporates 
two ``heads'', each endowed with an ATP-binding pocket and able to loop 
through its own chemical reaction cycle.
Thus the set of ``chemical coordinates'' (vectors)
$y$ is in fact composed of two subsets (scalars), $y=(y_1,y_2)$, one for each head,
and similarly the discretized states are of the form 
\begin{equation}
m=(m_1,m_2)\ \ ,\ \  m_{1,2}\in\{ 1,2,...,M\}\ .
\label{4.2.-1}
\end{equation}
For instance, for the standard model for the ATP reaction cycle
consisting of $M=4$ distinct states (cf. \sect \ref{sec5.2.1}), 
the compound set of states $m$ will comprise $\mtot=M^2=16$ elements. 
Note that exactly simultaneous reaction steps in both heads have negligible
probability, i.e., only indices $m=(m_1,m_2)$
with $m'=(m_1',m_2)$ or $m'=(m_1,m_2')$ are possible in $k_{m\to m'}(x)$.
In order to further reduce the 
number of nontrivial transition rates $k_{m\to m'}(x)$,
one common and suggestive assumption  \cite{how01}
is that only transitions between ``neighboring'' states within either of
the two chemical cycles occur with non-negligible probability, 
i.e., 
\begin{eqnarray}
k_{m\to m'}(x) = 0 \ \ \mbox{if} \ \ m'\not \in\{ (m_1\pm 1,m_2),\ 
(m_1,m_2\pm 1)\}\ , \label{4.2.0}
\end{eqnarray}
where states $m_1$ which differ by a multiple of $M$ are identified, 
and similarly for $m_2$.
In other words, each of the two chemical reaction cycles loops through 
a definite sequence of states, bifurcations into
different chemical pathways are ruled out.

Note that the cooperativity between the two heads,
mentioned at the end of \sect \ref{sec5.1},
is mediated by the geometrical configuration $x$
and will manifest itself
in the $x$-dependence of the rates, possibly 
reducing the number of non-trivial transition rates (\ref{4.2.0})
once again.

\section{Simplified stochastic model}\label{sec5.3}
While the so far reasoning and approximations have been
relatively systematic and microscopically 
well founded, further possible simplifications 
are necessarily of a more drastic and phenomenological nature.

In concrete models, the geometrical configuration of the motor is usually 
assumed to be characterized by a {\em single}\footnote{In other words,
each ``head'' has its own (discrete) chemical state variable (cf. \eq
(\ref{4.2.-1})), but the geometrical shape of the entire motor
(including the two ``heads'') is described by a
single (continuous) state variable $x$.} relevant state variable $x$.
One convenient choice for $x$ turns out to be the position of the molecular 
motor along the microtubulus.
To be precise, $x$ may for instance be chosen to represent the position of
the common center of mass of the two heads.
Indeed, knowing that the motor enzyme walks in a step-by-step fashion 
straight along 
the $\beta$-tubulin sites of one and the same 
protofilament, it is suggestive that the geometrical configuration 
of each of the two heads can be reconstructed quite faithfully
from the knowledge of the position $x$ (the rest of the motor molecule (``tail''
and ``middle segment'', cf. \sect \ref{sec5.1}) does not seem to play a 
significant role for the actual ``motor function'').

Once the relevant collective chemical and mechanical state variables have
been identified along the above line of reasoning, their ``thermal environment''
consits of two parts (cf. \sect \ref{sec5.2.1}): Namely, on the one
hand there are the huge number of ``irrelevant'' degrees of freedom of the
liquid which surrounds the protein, and on the other hand there are those
of the protein itself and of the microtubulus with which it interacts. 
Upon eliminating them along the lines of \sect \ref{sec3.2.1},
their effects on the discretized
chemical state variables are captured by the phenomenological rates (\ref{4.2.0}).
However, their effects on the mechanical state variable $x$ are more involved due to the
fact that $x$ does not simply represent the cartesian coordinate of a point 
particle but rather the complicated geometrical configurations of the entire
motor protein and in this sense is a generalized coordinate.
As a consequence, the so-called {\em solvent friction},
caused by the eliminated
degrees of freedom of the surrounding fluid, comprises not only 
a Stokes-type viscous
friction against straigh translational motion but also a damping force
against configurational changes of the geometrical enzyme structure.
Similarly, the so-called {\em protein friction} \cite{taw91,lei93,how01}, 
caused by the eliminated degrees of freedom of the
enzyme and the microtubulus, is composed of two analogous partial effects: 
on the one hand, a viscous drag against straight translational motion due to 
the continuous making and breaking of bonds between the motor and the 
microtubulus; on the other hand an effective ``internal''
frictional force against changes of the geometrical configuration.
All these friction mechanisms are in general not invariant 
under arbitrary translations of $x$ and are therefore explicitly 
$x$-dependent\footnote{Regarding the Stokes-type viscous friction,
we recall that $x$ represents not only the position but also the
changing geometrical shape of the motor molecule.
As mentioned in \sect \ref{sec4.8.4b} we furthermore expect
corrections of Stokes friction due to the nearby microtubulus,
which are again in general $x$-dependent.}.
The same carries over to the thermal fluctuations which they
bring along (``solvent noise'' and ``protein noise''), 
see also \sect \ref{sec3.2.1}.
Since quantitatively the effects of protein friction are typically
comparable or even more important than those of solvent 
friction \cite{lei93,how01}, a quite significant spatial 
inhomogeneity of the friction and the thermal noise is expected \cite{bie96b}.
We recall that the microscopic origin of both solvent and protein friction
is partly of a mechanical (geometrical) nature
(mainly collisions and vibrations, respectively)
and partly due to the making and braking of numerous weak chemical bonds,
as detailed in \sect \ref{sec5.2.1}.
On an even more basic level, all these distinctions become again blurred 
since the ultimate origin of friction is always the ``roughness'' of some 
effective potential energy landscape.

\subsection{Stochastic ratchet dynamics}\label{sec5.3.1}
On the basis of the above considerations,
the simplest working model for the stochastic dynamics
governing the mechanical coordinate (position) $x(t)$ is of the form
\begin{equation}
\eta\,\dot x(t) = - V'_m(x(t)) + F + \xi (t)\ ,
\label{4.2.1}
\end{equation}
where $m=m(t)$ is understood as a stochastic process, with states (\ref{4.2.-1})
and transition rates $k_{m\to m'}(x)$.
The assumption of a first order (overdamped) dynamics in time is justified
as usual by the fact that on these small scales inertia effects can be 
safely neglected \cite{kel00}.
The damping coefficient $\eta$ and the random noise $\xi(t)$ model the effects
of the environment and of the eliminated
fast degrees of freedom of the molecular motor itself (possibly also of
the microtubulus) and both these contributions are treated as a single 
thermal bath.
Under the assumption that the origin of $\xi(t)$ is a very large number of 
very fast processes (on the time scale of $x$) we can model those fluctuations
as a Gaussian noise of zero mean and negligible correlation time
\begin{equation}
\langle\xi (t)\,\xi(s)\rangle = 2 \, \eta\, k_B T\, \delta(t-s)  \ .
\label{4.2.2}
\end{equation}
In fact, already the very form of the dissipation assumed on the left hand side
of (\ref{4.2.1}) leaves no other choice for the noise $\xi(t)$
at equilibrium, see \sect \ref{sec3.2.1}.
A further assumption implicit in (\ref{4.2.1}) is the independence
of the coupling to the heat bath $\eta$ (see below (\ref{2.3}))
from the chemical state $m$ and the geometrical configuration $x$.
The former simplification is plausible in view of the fact that the chemical
processes only involve a very restricted region of the entire motor enzyme.
On the other hand, the $x$-independence of $\eta$
is not obvious in view of our above considerations about
solvent and protein friction, but can be justified as follows:
First, inhomogeneous friction, and in particular protein friction, can
be modeled quite well by means of potentials $V_m(x)$ in (\ref{4.2.1}) with
a suitably chosen ``roughness'' on a very ``fine'' spatial scale.
After a spatial coarse graining, only the broader structures of the potential
survive while the initially homogeneous ``bare'' friction is dressed by an
inhomogeneous renormalization contribution.
A second possibility consists in a change of 
variable\footnote{In this case, the transformed potentials in (\ref{4.2.1})
remain periodic but in general pick up an $F$-dependence, which we 
neglect for the sake of simplicity (see below).}
as detailed in \sect \ref{sec4.8.4b}. 


Note that $\eta$ accounts for the coupling of the thermal environment
(fast degree of freedom)
of the molecular motor only. The additional slow variable representing
the cargo of the motor can be accounted for \cite{svo94}
via a contribution of the 
form $-\langle\dot x\rangle \,\eta_{cargo}$ to the force $F$ in
(\ref{4.2.1}) under the tacit but apparently realistic
assumption that its connection to the motor 
(via ``tail'' and middle segment'', cf. \sect \ref{sec5.1})
is sufficiently elastic \cite{mei89,els00a,els00b}.
Although the cargo is typically much bigger than the motor 
itself, this viscous drag force seems
negligibly small \cite{svo94} in comparison with the intrinsic 
friction of the motor, modeled by $\eta\, \dot x(t)$ in (\ref{4.2.1}).

The deterministic mechanical forces in (\ref{4.2.1}) on the one hand 
derive from an effective, free-energy like potential $V_m(x)$ and on 
the other hand leave room
for the possibility of an externally applied extra force $F$.
Originating from the potential energy landscape in which the nuclei of the
motor and its environment move, the effective
(renormalized) potential $V_m(x)$ in addition
accounts for some of the effects of the eliminated fast degrees 
of freedom.
The approximate independence of this effective potential $V_m(x)$
from the external load $F$ is assumed here for the sake of 
simplicity\footnote{As far as $x$ describes the center of mass of the
molecular motor, the simple $F$-dependence on the right hand side
of (\ref{4.2.1}) is fully justified. 
However, in so far as $x$ at the same time 
accounts for the geometrical shape of the motor molecule, the
relation between position and shape and hence the effective potentials
$V_m(x)$ are expected to change upon application of a force $F$.}
\cite{lip00a,lip00b,par01,lat01}.
On the other hand, 
the dependence of the potential on the chemical state $m$ is crucial.
The latter in conjunction with the $x$-dependence of the chemical reaction
rates $k_{m\to m'}(x)$ is called the
mechanochemical coupling mechanism of the model motor enzyme,
decisive for the chemical to mechanical energy 
transduction\footnote{The $F$-independence of the rates 
$k_{m\to m'}(x)$ (and {\em a forteriori} of the number $\mtot$
of chemical states) is plausible on the basis of the physical
picture from \sect \ref{sec5.2.4} (the chemical processes are
spatially localized and thus involve negligibly small changes of the
geometrical configuration of the motor molecule).}.
The underlying picture is that certain chemical reaction
steps take place preferably or even exclusively while the molecular motor
has a specific geometrical shape $x$. In turn, certain mechanical relaxations
of strains or thermally activated configurational transitions may be
triggered or made possible only after a certain chemical reaction step 
has been accomplished.

Clearly, the dynamical behavior of the motor enzyme is invariant
after a step of one head has been completed if at the same time the chemical
states $m_1 ,\, m_2$ of the two heads are exchanged \cite{how96,jul99}.
This invariance under a displacement $x\mapsto x+L$ and simultaneously
$(m_1,m_2)\mapsto (m_2,m_1)$ has to be respected by the potentials
$V_m(x)$ and the rates $k_{m\to m'}(x)$,
\begin{eqnarray}
& & V_{m}(x+L) = V_{\overline{m}}(x)\label{4.2.2a}\\
& & k_{m \to m'}(x+L) = 
k_{(\overline{m} \to \overline{m'})}(x)\ , \label{4.2.2b}
\end{eqnarray}
where the bar denotes the exchange of the vector-components:
\begin{equation}
\overline{(m_1,m_2)} := (m_2,m_1) \ ,
\label{4.2.2c}
\end{equation}
and where the spatial period $L$ is given by
the length of one tubulin dimer (about $8\, nm$).
Consequently, the functions $V_m(x)$ and $k_{m\to m'}(x)$
are invariant under $x\mapsto x+2L$ without any change of the chemical states.

The polarity of the microtubulus, on which the motor walks, reflects itself 
in a generic {\em spatial asymmetry} 
of the potential $V_m(x)$ as well as of the rates
$k_{m\to m'}(x)$.
Note that on top of that, there is also an intrinsic asymmetry of the 
motor domains (but not of the entire enzyme, see \sect \ref{sec5.1}):
If one detaches a motor domain
from the microtubulus, turns it around by $180^o$, and
puts it back on the microtubulus, no invariance arises \cite{fin94,how96,cro97}, 
that is, reflection symmetry is broken. 
In other words, the asymmetry of the microtubulus is necessary to make manifest 
the asymmetry of the motor, while the asymmetry of the compound system
is caused and maybe even mutually enhanced by both \cite{pes94,kik95}.

The stochastic dynamics (\ref{4.2.1}) as it stands is a convenient 
starting point for numerical simulations (cf. \sect \ref{sec2.1.3}) 
but not for quantitative analytical calculations.
Exactly like for the fluctuating potential ratchet model in \eqs (\ref{4.4}), (\ref{4.5}),
one obtains the following master equation 
(reaction-diffusion equation) equivalent to (\ref{4.2.1}):
\begin{eqnarray}
\frac{\partial}{\partial t} P_m(x,t) & = & 
\frac{\partial}{\partial x}\frac{[V_m'(x)-F]\, P_m(x,t)}{\eta}
+ \frac{k_B T}{\eta}\, \frac{\partial^2}{\partial x^2} P_m(x,t)\nonumber\\
&-& P_m(x,t)\sum_{m'} k_{m\to m'}(x) +  \sum_{m'} P_{m'} (x,t) k_{m'\to m}(x)\ ,
\label{4.2.6}
\end{eqnarray}
where $P_m(x,t)$ is the joint probability density that at time $t$ the 
chemical state is $m$ and the motor enzyme is at the position $x$, with
normalization $\sum_m\int dx\, P_m(x,t) = 1$.
In order to technically simplify matters one defines similarly as in (\ref{2.12})
reduced densities 
\begin{equation}
\hat P_m(x,t) := \frac{1}{2} \sum_{n=-\infty}^\infty\left\{
P_m(x+ 2\, n\, L,t) + P_{\overline{m}}(x + 2\, (n+1)\, L,t) \right\} \ .
\label{4.2.6b}
\end{equation}
The reduced densities satisfy the same master equation (\ref{4.2.6})
but are periodic in $x$ with period $2\, L$ and normalization
\begin{equation}
\sum_m\int_0^{2L} dx\, \hat P_m(x,t) = 1 \ .
\label{4.2.6c}
\end{equation}
The symmetries (\ref{4.2.2a}), (\ref{4.2.2b}) furthermore imply that
\begin{equation}
\hat P_m(x,t)= P_{\overline{m}}(x+L,t)= \hat P_m(x+2\,L,t)\ .
\label{4.2.6a}
\end{equation}

Once $\hat P_m(x,t)$ is determined, the average speed of the motor enzyme
follows along the same line of reasoning as in \sect \ref{sec2.1.3'} 
as\footnote{We recall that the argument $t$ in $\langle\dot x\rangle$ is omitted 
(cf. (\ref{4c'})) since in most cases one is interested in the steady state
behavior with $\hat P_m(x,t)=\hat P^{st}_m(x)$. The same applies for the 
rate $r_{ATP} = r_{ATP}(t)$ in (\ref{4.2.7a}).}
\begin{equation}
\langle\dot x\rangle = 
\frac{1}{\eta}\,\left[ F - 
 \sum_m \int_0^{2 L} dx\,  V'_m(x) \, \hat P_m(x,t)\right]\ .
\label{4.2.7}
\end{equation}
A further interesting quantity is the rate $r_{ATP}= r_{ATP}(t)$ 
of ATP-consumption per time unit, given by
\begin{equation}
r_{ATP} = \sum_{m,m'} \chi_{m,m'}^{ATP}\, 
\int_0^{2L} dx\, \left\{ \hat P_m(x,t)\, k_{m\to m'}(x) -
\hat P_{m'}(x,t)\, k_{m' \to m}(x) \right\} \ ,
\label{4.2.7a}
\end{equation}
where $\chi_{m,m'}^{ATP}$ is the indicator function for ATP-binding transitions
$m\to m'$. For example, using the labeling of the chemical states from
\sect \ref{sec5.2.1} for the standard ATP-hydrolysis cycle with $M=4$ states, we have
\begin{equation}\chi_{m,m'}^{ATP} = 
\left \{ \begin{array}{clcrcl} 
& 1  & \mbox{if} & m=(1,m_2) & \mbox{and}\ \ m'=(2,m_2)\\ 
& 1  & \mbox{if} & m=(m_1,1) & \mbox{and}\ \ m'=(m_1,2)\\ 
& 0  & \mbox{otherwise}  & &   
\end{array} \right . 
\label{4.2.7b} 
\end{equation}

A comparison of the above model setup
with the working model from \sect \ref{sec3.1.1}
very obviously establishes a close connection between 
our present section
about molecular motors and the general framework for our studies of ratchet 
models,
especially the class of {\em pulsating ratchets}\footnote{The
driving $f(t)$ of the pulsating potential
$V(x,f(t))$ is denoted here by $m(t)$ and the pulsating potential itself
by $V_m(x)=V_{m(t)}(x)$ (cf. (\ref{4a}) and below (\ref{4.2.1}), respectively).
Moreover, $m(t)$ is here a discrete and -- in general -- two-dimensional
state variable (cf. (\ref{4.2.-1})), though in most concrete models (see \sects
\ref{sec5.4}-\ref{sec5.6}) again a simplified, effectively one-dimensional
description will be adopted.} according to the classification scheme 
from\footnote{It may be worth to recall that for traveling potential ratchets 
and their descendants (\sects \ref{sec4.3} and \ref{sec4.4}) a broken symmetry of
the potential is not necessary for directed transport, though for real 
molecular motors this symmetry will be typically broken.} 
\sect \ref{sec3.1.3}.
However, there is also one important point in which the present
model goes beyond the latter general framework.
Namely, there is a {\em back-coupling} of the state-variable
$x(t)$ to the ``potential fluctuations'' $m(t)$ through the $x$-dependence
of the transition rates $k_{m\to m'}(x)$. Especially, the statistical
properties of the potential fluctuations $m(t)$ can no longer be assumed
{\em a priori} as stationary. We will show later in \sect \ref{sec5.4.2}
that far away from equilibrium an effective $x$-independence of the
potential fluctuations $m(t)$ may arise nevertheless,
entailing stationarity of their statistical properties in the
long time limit, i.e. a veritable pulsating ratchet scheme is recovered.

\subsection{Nonequilibrium chemical reaction}\label{sec5.3.2}
At thermal equilibrium, the concentrations of ATP, ADP, and P$_i$
are not independent, their ratio $C^0_{ATP}/C^0_{ADP} C^0_{P_i}$
satisfies the so-called mass action law.
Especially, the numerical value of this ratio must be independent
of whether any motor enzymes (acting as catalyst) are present or not.
Since this represents a single constraint for three variables, there still remains
a freedom in the choice of two out of the three equilibrium concentrations
$C^0_{ATP}$, $C^0_{ADP}$, and $ C^0_{P_i}$. We consider an arbitrary but fixed
such choice from now on.
Since the system is an equilibrium system,
the stochastic dynamics has furthermore to respect the so-called condition
of detailed balance 
\cite{ons31,gre52,kam57,gra71a,gra71b,han82b,gar83,ris84,kam92}.
For our specific model (\ref{4.2.1}) this condition can be readily shown to imply
the following 
relation between the transition rates $k_{m\to m'}(x)$ and the 
corresponding potentials $V_m(x)$ and $V_{m'}(x)$ for any pair of chemical
states $m$ and $m'$:
\begin{equation}
\frac{k_{m\to m'}(x)}{k_{m'\to m}(x)} = 
\exp\left\{\frac{V_{m}(x)-V_{m'}(x)}{k_BT}\right\} \ .
\label{4.2.3}
\end{equation}
Thus, one of the two rates in (\ref{4.2.3})
can be considered as a free, phenomenological function 
of the model, while the other rate is then fixed.
Note that the appearance of negligibly small rates in (\ref{4.2.0})
as well as the symmetry relations (\ref{4.2.2a}), (\ref{4.2.2b}) are
still compatible with (\ref{4.2.3}).

The salient point is now to clarify what is meant by saying that
one goes ``away from equilibrium'' in our present context.
Meant is, that as far as the heat bath properties of the
environment (random fluctuations and
energy dissipation mechanism) are concerned, 
nothing is changed as compared to the thermal equilibrium case.
The only things which change are the concentrations of 
reactants and/or products \cite{fox98}.

For instance, if the ATP concentration $C_{ATP}$ 
is changed away from its equilibrium
value $C^0_{ATP}$, then all the rates $k_{m\to m'}(x)$
remain unchanged except those which describe the binding of ATP
to one of the two heads of the molecular motor.
As discussed in \sect \ref{sec5.2.2}
these rates simply acquire an extra multiplicative factor of the form 
$C_{ATP}/C^0_{ATP}$, i.e., (\ref{4.2.3}) is 
generalized\footnote{Without discretizing the chemical state variable(s)
(or equivalently, assuming that a separation of time-scales
exists such that a rate description is justified) the proper 
reformulation of a relation like in (\ref{4.2.4})
does not seem possible, see also \sect \ref{sec4.5} and \cite{kel00}.}
to
\begin{equation}
\frac{k_{m\to m'}(x)}{k_{m' \to m}(x)} = 
\left[ 1 + \left(\frac{C_{ATP}}{C^0_{ATP}} - 1 \right)\,\chi_{m,m'}^{ATP}\right] 
\exp\left\{\frac{V_{m}(x)-V_{m'}(x)}{k_BT}\right\} \ ,
\label{4.2.4}
\end{equation}
where $\chi_{m,m'}^{ATP}$ is the ATP-binding indicator function from (\ref{4.2.7b}).
Similar modifications arise if the concentrations of ADP and P$_i$ are changed.
However, in order to describe the real situation one may
without loss of generality assume that these concentrations
have already their correct value due to our choice of $C^0_{ADP}$ and $C^0_{P_i}$.
In doing so, it follows from the quantitative
biological findings mentioned in \sect \ref{sec5.1} 
that $C_{ATP}$ has to be chosen about 6 decades beyond its equilibrium 
value $C^0_{ATP}$:
\begin{equation}
\frac{C_{ATP}}{C_{ATP}^0} \simeq 10^6 \ .
\label{4.2.5}
\end{equation}

From the conceptual viewpoint we are thus facing the following interesting
setup of a far from equilibrium system:
On one hand, the system is in contact with a thermal equilibrium heat reservoir 
as far as dissipation and fluctuations are concerned.
On the other hand, it is in contact with several reservoirs of 
reactant and products with concentrations which are externally
kept far away from equilibrium.
All these various reservoirs are physically localized at 
the same place but the effects due to their direct interaction
with each other is practically negligible.
Only the indirect interaction by way of the motor 
molecules (catalysts) is relevant.

\section{Collective one-head models}\label{sec5.4}
At this stage, the number of free, phenomenological functions 
in (\ref{4.2.6}) is still very large. 
There is little chance to make a convincing guess for each of them on the basis
of our present knowledge about the structure and functioning
of the real motor enzyme, while for fitting the dynamical behavior
of the model to experimental curves, the available variety and accuracy
of measurements is not sufficient.
Our next goal must therefore be to reduce the effective number $\mtot$ of relevant
chemical states.

\subsection{A.F. Huxley's model}\label{sec5.4.1}
The most prominent such simplification goes back to A.F. Huxley's 1957 paper
\cite{hux57} and consists in the assumption of one instead
of two heads per motor enzyme.
In our model (\ref{4.2.1}) this means that $m$ is no longer composed of two
``substates'', see \eq  (\ref{4.2.-1}),
but rather is a scalar state variable with $\mtot = M$ values 
\begin{equation}
m \in\{1,2,...,M\} \ .
\label{4.2.7c}
\end{equation}
Likewise, $x$ now represents the center of mass of a single head.
As a consequence, 
the symmetries (\ref{4.2.2a}), (\ref{4.2.2b}) become
\begin{equation}
V_m(x+L)=V_m(x)\ ,\ \ k_{m\to m'}(x+L)= k_{m\to m'}(x)
\label{4.2.8a}
\end{equation}
and \eqs  (\ref{4.2.6b})-(\ref{4.2.6a}) are replaced by
\begin{eqnarray}
& & \hat P_m(x,t) := \sum_{n=-\infty}^\infty P_m(x+n\, L ,t)
\label{4.2.8b}\\
& & \hat P_m(x+L,t)  = \hat P_m(x,t)
\label{4.2.8b1}\\
& & \sum_{m=1}^M \int_0^L dx\, \hat P_m(x,t) = 1 \label{4.2.8b2}\ .
\end{eqnarray}
Furthermore, \eqs  (\ref{4.2.7}), (\ref{4.2.7a}) assume the form
\begin{eqnarray}
& & 
\langle\dot x\rangle = 
\frac{1}{\eta}\,\left[ F - 
 \sum_{m=1}^M \int_0^{L} dx\,  V'_m(x) \, \hat P_m(x,t)\right] 
\label{4.2.8c}\\
& & 
r_{ATP} = 
\int_0^{L} dx\, \left\{ \hat P_1(x,t)\, k_{1\to 2}(x) -
\hat P_{2}(x,t)\, k_{2 \to 1}(x) \right\} \ .
\label{4.2.8d}
\end{eqnarray}
Finally, all rates $k_{m\to m'}(x)$ with $m'\not = m\pm1$ are zero
according to (\ref{4.2.0}), and for $m'=m\pm 1$ \eq  (\ref{4.2.4}) 
takes the form
\begin{equation}
\frac{k_{m\to m'}(x)}{k_{m' \to m}(x)} = 
\left[ 1 + \left(\frac{C_{ATP}}{C^0_{ATP}} - 1 \right)
\, \delta_{m,1}\delta_{m',2}\right]\, 
\exp\left\{\frac{V_{m}(x)-V_{m'}(x)}{k_BT}\right\}  \ ,
\label{4.2.8}
\end{equation}
where the ATP binding transition is assumed to be $m= 1 \to m'=2$
and where states which differ by a multiple of $M$ are identified.

A second ingredient of Huxley's model is a ``backbone'' to which a 
{\em number $N$ of such single headed motors} is permanently attached.
The emerging intuitive picture is a centipede, walking along the polymer filament.
The interaction of the single-headed motors  is mediated by the common backbone,
assumed rigid and moving with constant speed $\langle\dot x\rangle$,
but otherwise they are considered as operating independently of each other.
We may then concentrate on any of the single heads and
without loss of generality denote the site where this specific head 
is rooted in the backbone by $\langle\dot x \rangle\, t$.
In physical terms, we are dealing with a 
{\em mean field model ($N\to\infty$)}, described
by an arbitrary but fixed reference head according to (\ref{4.2.1}),
where the potentials $V_m(x)$ and the rates $k_{m\to m'}(x)$ may, in general,
acquire an additional dependence on the backbone site 
$\langle\dot x \rangle\, t$.
The possible difference between the center of mass of the head $x$ and
the point $\langle\dot x \rangle\, t$ where it is attached to the backbone may,
for instance, reflect a variable angle between the head's length axis and the
polymer filament, similarly to a human leg while walking.
Thus, we may also look upon $\langle\dot x \rangle\, t$ as an additional
relevant (slow) mechanical state variable of the motor.
However, no extra equation of motion for this coordinate is needed since it
already follows in the spirit of a mean field approach from the behavior
of the other relevant mechanical state variable $x$.
For instance, a term of the form $\kappa\, (x-\langle\dot x \rangle\, t)^2$
in the potentials $V_m(x)$ models a harmonic coupling of the head to the
uniformly advancing backbone, with spring constant $\kappa$.
As in any mean field model, the characteristic feature is the appearance
of  an a priori unknown {\em ``order parameter'',
$\langle\dot x \rangle$ in our case, which has to be determined 
self-consistently}
in the course of the solution of the model (\ref{4.2.1}), (\ref{4.2.7})
(for an explicit example see (\ref{j1a}), (\ref{j2}) below).
We emphasize that for a rigid backbone, in the limit $N\to\infty$ Huxley's mean
field approach is not an approximation but rather an exact description because
the interaction between the single motors is of infinite range.

We remark that such 
a model of $N$ single headed motors with a mean field coupling through a rigid
``backbone'' may even be acceptable as a rough approximation in the case of 
a single kinesin molecule. Admittedly, 
the number $N=2$ of involved heads makes a mean 
field approximation somewhat questionable. On the other hand,
taking into account that a rather ``heavy'' load is attached to the
motor, may render the assumption of an uniformly moving backbone 
not so bad \cite{svo94}.
On top of that, the cooperativity of the two heads in the real kinesin
is at least roughly incorporated  into
the model through their interaction via the backbone and through the
implicit assumption that the motor will not diffuse away from the
microtubulus even if both heads happen to take a step at the
same time.

More suggestive is the case when an appreciable number $N$ of single motor
molecules truly cooperate.
This may be a couple of kinesins which drag a common ``big'' cargo.
More importantly, there exist motor enzymes different from kinesin
which indeed are interconnected by a backbone-like structure by nature.
Examples are so-called myosin enzymes, walking on polymer filament tracks
called actin, thereby not carrying loads but rather
playing a central role in muscular contraction \cite{alb94,how97a,how97b}.
While the quantitative and structural details are different
from the kinesin-microtubulus system, the main
qualitative features of the myosin-actin system
are sufficiently similar \cite{fin94,spu94,kit99}
such that the same general framework 
(\ref{4.2.0})-(\ref{4.2.7}) is equally appropriate in both 
cases\footnote{Intriguingly enough, certain species of the myosin superfamily
(e.g. the so-called myosin V subfamily)
show again a behavior similar to kinesin \cite{how97a,meh99a}. In the
following we always have in mind collectively operating myosin
species (the myosin II subfamily).}.
Though a single myosin enzyme again consists of two individual
motor domains, their cooperativity seems not so highly developed as for kinesin 
\cite{fin94,sch95} and therefore the above mentioned 
mean field approximation for a large number $N$ of interacting single heads
appears indeed quite convincing.

In his landmark paper \cite{hux57}, Huxley proposed
a model of this type without any knowledge about the structural features
of an individual motor enzyme!
It is not difficult to map the slightly different language
used in his model to our present framework, but since the details
of his setting cannot be upheld in view of later
experimental findings, we desist from explicitly carrying
out this mapping here.
While the model is apparently in satisfactory agreement with the 
main experimental
facts available at that time, Huxley himself points out \cite{hux57} that
``there is little doubt  that equally good agreement could be reached
on very different sets of assumptions, all equally consistent with 
the structural, physical and chemical data to which this set has been fitted.
The agreement does however show that this type of mechanism deserves to be
seriously considered and that it is worth looking for direct evidence of the
side pieces.''

\subsection{Free choice of chemical reaction rates}\label{sec5.4.2}
One specific point of Huxley's model is worth a more
detailed discussion since it illustrates a much more general
line of reasoning in the construction of such models.
In doing so, we first recall that we are dealing with a single head motor
model described by $M=4$ (scalar) chemical states $m$:
1. the head without anything bound to it; 
2. the head with an ATP bound;
3. the head with an ADP and a P$_i$ bound;
4. the head with an ADP bound.
The chemical state variable travels back and forth between neighboring states of this 
cycle according to the transition rates $k_{m\to m'}(x)$,
respecting (\ref{4.2.8}) if $m'=m \pm 1$ and being zero otherwise.
For the case of kinesin, we have discussed at the beginning
of \sect \ref{sec5.1} 
in addition the ``affinity'' between head and filament in each state,
which essentially tells us whether the head is attached to the filament
or not in the respective state, and which has to be taken into account in the
concrete choice of the respective model potentials $V_m(x)$.
We remark that this correspondence between states and affinity is somewhat
different for myosin \cite{lei93} and again different in Huxley's model, but
will not play any role in the following, since it only regards
quantitative, but not qualitative properties of the potential $V_m(x)$.

As a first simplification, Huxley postulates a $3$-state model, in which
$m=2$ and $m=3$ in our above scheme are treated as a single state, 
and the question
arises of whether and how this can be justified, at least in principle.
One possible line of reasoning goes as follows:
Aiming at a unification of 
$m=2$ and $m=3$ means in particular that we should choose
$V_2(x)=V_3(x)$ and thus $k_{2\to 3}(x)= k_{3 \to 2}(x)$ 
according to (\ref{4.2.8}).
Since there are no further {\em a priori}
restrictions on the choice of these rates,
we may take them as independent of $x$ and very large\label{fot5.8}\footnote{Such
a choice is obviously admissible within our general modeling framework;
how to justify it against experimental findings is a different matter
\cite{sch97,gil95,svo94b,sch95b}.}.
Thus, as soon as the system reaches either state $2$ or $3$ it will be 
practically instantaneously distributed among both states with equal probability.
One readily sees that the two states can now be treated as a single
``superstate'' if the two transition rates out of this state are defined as half the
corresponding original values $k_{2\to 1}(x)$ and $k_{3 \to 4}(x)$.
At first glance, it may seem that in this reduced $3$-state model, the
condition (\ref{4.2.8}), in the case that $m$ represents the new ``superstate'',
has to be modified by a factor $1/2$. However, since in the stochastic dynamics
(\ref{4.2.1}) only the derivative of the potential $V_m(x)$ appears, this factor
$1/2$ can be readily absorbed into an additive constant of that potential.

Given the reduced model with $M=3$ states, Huxley furthermore
assumes that the $3$ ``forward'' rates $k_{m\to m+1}(x)$ can be freely
chosen, while the $3$ ``backward'' rates $k_{m+1 \to m}(x)$ are negligibly small.
On the other hand, \eq (\ref{4.2.8}) tells us that
the $3$ forward rates can indeed be chosen freely,
but once they are fixed, the $3$ backwards rates are also fixed.
At this point, one may exploit once again the observation that only
the derivatives $V'_m(x)$ enter the dynamics (\ref{4.2.1}) and
therefore we still can add an arbitrary constant to any of the three model
potentials $V_m(x)$. Under the additional assumption that
$\exp\{[V_m(x)-V_{m+1}(x)]/k_B T\}$ varies over one spatial period $L$
at most by a factor significantly smaller than 
$(C_{ATP}/C^0_{ATP})^{1/3}\simeq 10^2$ (see \ref{4.2.5}),
one readily sees that by adding appropriate constants to the $3$ potentials 
$V_m(x)$ one can make the ratios $k_{m\to m+1}(x)/ k_{m+1\to m}(x)$ rather 
small for all $3$ values of $m$ according to (\ref{4.2.8}).
Pictorially speaking, by adding proper constants to the potentials
$V_m(x)$ one can split and re-distribute the factor $C_{ATP}/C_{ATP}^0$
from (\ref{4.2.5}) along the entire chemical reaction cycle.
In this way, all $3$ backward rates
$k_{m+1\to m}(x)$, though not exactly zero,
can indeed be practically neglected.
Generalizations to more than $3$ states and to the neglection of
only some, but not all, backward rates are obvious.

A final important observation concerns the case $M=3$ with all the forward rates
still at our disposition and all the backward rates approximately neglected.
Specifically, one may assume that $V'_2(x) = V'_3(x)$ and that the
corresponding forward rate $k_{2\to 3}$ is $x$-independent and very large.
The two states $m=2$ and $m=3$ can then again be lumped into
a single superstate. The result is \cite{ast96a} a model (\ref{4.2.1})
with $M=2$ effective chemical states but with {\em both},
the forward and backward rates between these two states,
still free to choose.

We have thus achieved by way of various simplifying assumptions our goal
to substantially reduce the number of free, phenomenological functions 
in the model (\ref{4.2.1}).
Still, even for the minimal number $M=2$ of chemical states 
the shape of the two
potentials and especially the choice of the
two rates \cite{lei93,svo94,duk96,jul99} are very difficult
to satisfactory justify on the basis of experimental findings.
Accordingly, the existing literature does not seem to indicate
that a common denominator of how these functions
should be realistically chosen is within hands reach.

\subsection{Generalizations}\label{sec5.4.3}
Huxley's choice of model parameters and functions 
(\ref{4.2.7c}), (\ref{4.2.8a}) in the general setup (\ref{4.2.1})
has been subsequently modified and
extended in various ways in order to maintain agreement with
new experimental findings. Most of the following
works include verifications of the theoretical models against 
measurements, though we will not repeat this fact each time.
Moreover, a detailed discussion of the specific choices and 
justifications of the free, phenomenological
parameters and functions in the general model (\ref{4.2.1}) 
in those various studies goes beyond the scope of our review.
Our main focus in this section
will be on the character of the mechanochemical coupling
(cf. \sect \ref{sec2.2.3a})
and the relevance of the thermal noise for the dynamics of
the mechanical state variable
in (\ref{4.2.1}), see also
\sect \ref{sec5.8} for a more systematic discussion of these points.

With more structural data of the actin-myosin system
on the molecular level becoming available, A.F. Huxley and Simmons
\cite{hux71} already in 1971 came up with a more realistic modification of 
the original model, featuring a ``fast'' (chemical) variable
with a small number of discrete states, 
tightly coupled to a ``slow'' (mechanical) continuous 
coordinate. For more recent studies along these lines see also 
\cite{smi87,pat89,pat91,pat93,sek95,tho98,bar99} and references therein.
A very recent, analytically solvable model, closely resembling 
A. F. Huxleys original setup and in quantitative
agreement with a large body of experimental data,
is due to \cite{how01}.

The issue of the chemical to mechanical coupling
has been for the first time addressed in detail by
Mitsui and Oshima \cite{mit88}, pointing out that deviations from a 
simple and rigid one-to-one coupling may play an important role.

A connection between a model of the Huxley type
with Feynman's ratchet-and-pawl 
gadget has apparently been realized and
worked out for the first time by 
Braxton and Yount \cite{bra88,bra89},
though their model was later proven unrealistic
by more detailed quantitative considerations \cite{hun94,how96}.
A similar Feynman-type approach has been
independently elaborated by Vale and Oosawa \cite{val90}.
More importantly, they seem to have been the first to bring into play
the crucial question of the relative importance of the thermal
fluctuations appearing in the dynamics of the mechanical
coordinate (\ref{4.2.1}) as compared to conformational (relaxational)
changes powered by the chemical cycle (that is, ultimately by the
power stroke).

One extreme possibility is characterized by barriers of the potentials $V_m(x)$
which can be crossed only with the help of the thermal noise $\xi (t)$ in
(\ref{4.2.1}), independently of how the chemical state $m$ evolves in the
course of time.
An example is the fluctuating potential ratchet (\ref{4.1}) with $f(t)$ restricted to
a discrete number ($M$)
of possible values, all smaller than unity in modulus.
In such a case, the role of the chemical cycle is merely the breaking of the
detailed balance, necessary for a manifestation of the ratchet effect in the
$x$-dynamics. Moreover, the mechanochemical coupling 
is typically (i.e. unless the rates $k_{m\to m'}(x)$ exhibit a very 
special, strong $x$-dependence, see below) loose, the number
of chemical cycles per mechanical cycle randomly varies over a wide range.

The opposite possibility is represented by the traveling potential ratchet mechanism, 
see \sect \ref{sec4.3}. Each chemical transition $m\to m'= m+1$ induces a strain in the
mechanical coordinate via $V'_{m'}(x)$ in (\ref{4.2.1}) which then is 
released while $x$ relaxes towards the closest local minimum of $V'_{m'}(x)$. As
$m$ proceed through the chemical loop, also the local minima of $V_m(x)$
are shifting forward in sufficiently small steps such that $x$ typically 
advances by one period $L$ after one chemical cycle.
In this case, the thermal noise has only an indirect effect through the
chemical rates $k_{m\to m'}(x)$, but as far as the mechanical dynamics
(\ref{4.2.1}) is concerned, almost nothing changes in comparison
with a purely deterministic ($\xi(t)\equiv 0$) behavior.
In other words, the mechanochemical coupling is very rigid,
the mechanical coordinate $x$ is
almost exclusively powered by the chemical reaction
and its behavior is basically
``slaved'' by the chemical transitions.
The mechanical coordinate $x$ may at most play a role
in that the practically deterministic relaxation of $x$ after a chemical transition
$m\to m'$ may delay the occurrence of the next transition $m'\to m''$ until the new
local minimum of $V_{m'}(x)$ has been reached. Essentially, the system
can thus be described by the chemical reaction cycle alone,
possibly augmented by appropriate deterministic refractory periods
(waiting times) after each reaction step \cite{fis99,fis99a,kol00b}.
It then does not seem any more appropriate to speak of a noise induced transport 
in the closer sense and the ratchet effect only enters the picture via the
somewhat trivial traveling potential ratchet mechanism from 
\sect \ref{sec4.3}, see also \sect \ref{sec2.2.3a}.

A situation intermediate between these two extreme cases 
arises if the potential $V_m(x)$ exhibits (approximately)
flat segments, requiring diffusion but no activated barrier crossing for being
transversed.
An example is the on-off ratchet scheme from \sect \ref{sec4.1}.

Another compromise between the two extremes 
consists in the following scenario:
Thermally activated barrier crossing is unavoidable for the advancement
of $x$. Yet, due to the choice of the rates $k_{m\to m'}(x)$,
the next chemical step becomes only possible after the
respective barrier crossing has been accomplished.
In other words, though thermal noise effects are an indispensable ingredient
for the working of the
motor enzyme model, the stepping of $x$ and $m$ is tightly coupled.
Since the thermal activation processes can be considered as rate processes,
such a model can be mapped in very good approximation
to an augmented reaction cycle, with some mechanical states
added to the chemical ones.
The proper notion for such a situation seem to be 
``mechanochemical reaction cycle''.
For a more systematic treatment of such issues see \sect \ref{sec5.8}.

The conclusion of Vale and Oosawa \cite{val90} is that, within
their Huxley-type model (\ref{4.2.1}), thermal 
noise in the mechanical coordinate $x$ plays an important
role; specifically, a mechanism similar to a fluctuating potential ratchet
(\sect \ref{sec4.2}), a temperature ratchet model (\sects \ref{sec2.2.2} and
\ref{sec4.8.3}), or
a combination of both is postulated (later criticized 
as being unrealistic in \cite{lei93,hun94,how96}).

While the latter conclusion is mainly of a qualitative nature,
a more quantitative investigation of the same question is due to 
Cordova, Ermentrout, and Oster \cite{cor92}, with the result that for
cooperating motor enzymes like myosin, thermal activation processes are 
-- within their choice of model parameters and functions in 
(\ref{4.2.1}) -- crucial in the dynamics of the mechanical variable $x$,
while for kinesin such processes may be of somewhat less importance.
In deriving the latter conclusion, these authors go one step
beyond Huxley's framework in that they also analyze
the motion of a single head (no backbone),
and especially of two heads without invoking a mean field approximation 
for the motion of the backbone, which, in this context
should then rather be viewed as a ``hinge'' connecting the two heads.

A further refined variation of Huxley's model has been worked out by 
Leibler and Huse \cite{lei91,lei93}, together with a few-head model
(beyond mean field) in a general spirit similar to that of 
Cordova, Ermentrout, and Oster \cite{cor92}.
In this model, however, a tight mechanochemical coupling is built-in
from the beginning, namely the choice of the parameters and functions
in (\ref{4.2.1}) is such that thermal noise effects
on the mechanical coordinate $x$ play a minor role by construction.
Furthermore, all the transition rates $k_{m\to m'}$ are assumed to be
independent of $x$. Within such a model, it is shown that at
least $M=4$ chemical states are required to avoid incompatibilities with known
experimental findings.
The main achievement of these studies \cite{lei91,lei93}
is a unified description of ``porter'' motor proteins, e.g. kinesin,
operating individually and spending a relatively short
time detached from the polymer filament (moderate-to-large duty ratio),
and of ``rowers'', e.g. myosin, which operate collectively and are
characterized by a small duty ratio.
Thus, ``porters'' are essentially processive, and ``rowers'' non-processive
motor enzymes. 
A refined model similar in spirit has been put forward in \cite{duk96}.

\subsection{J\"ulicher-Prost model}\label{sec5.4.4}
One of the most striking statistical mechanical features of interacting
many body systems, both at and far away from thermal equilibrium,
is the possibility of spontaneous
ergodicity-breaking, entailing phase transitions, the coexistence
of different (meta-) stable phases, and a hysteretic behavior in 
response to the variation of appropriate parameters.
There is no reason why such genuine collective effects
should not be expected also in Huxley-type mean field models, 
but it was not before 1995 that J\"ulicher and 
Prost \cite{jul95}
explicitly demonstrated the occurrence of those phenomena in such a model,
see also \cite{jul97a,jul98,jul99,mar99,jul99a}.
Specifically, they focused on the dependence of the average velocity 
$\langle\dot x\rangle$ upon (parametric) variations of the external force $F$ in
(\ref{4.2.1}), henceforth 
called $\langle\dot x\rangle$-versus-$F$ characteristics.
As already mentioned, formally the crucial point in such a mean field approach is 
the appearance of a self-consistency equation for the ``order parameter''
$\langle\dot x\rangle$. Typically,
this equation is 
nonlinear\footnote{For an example, see \eqs  (\ref{j1a}), (\ref{j2})
below.} and the existence of multiple (stable)
solutions signals the breaking of ergodicity.

After having observed such a situation in their model,
J\"ulicher and Prost pointed out in a subsequent work \cite{jul97b} the following
remarkable consequence of the hysteretic 
$\langle\dot x\rangle$-versus-$F$ characteristics:
If the rigid backbone is coupled to a spring, then an effective
external force $F$ depending on the position of the backbone arises.
If the spring is sufficiently soft, then the changes of $F$ are 
sufficiently slow such that the parametric 
$\langle\dot x\rangle$-versus-$F$ characteristics can be used.
If this relation furthermore exhibits a hysteresis loop with the two 
$\langle\dot x\rangle$-versus-$F$ branches confined to either side of 
$\langle\dot x\rangle=0$,
then a permanent periodic back-and-forth motion of the backbone is the result.
Remarkably, strong indications for both,
spontaneous ergodicity breaking (dynamical phase transition in
the velocity-froce-relationship)
as well as spontaneous oscillations can indeed be
observed in motility assays \cite{riv98} and
in muscle cells under suitable conditions 
\cite{yas96,jul97a,jul97b,jul98,fuj98,jul99,mar99,jul99a,cam99}, respectively.

We recall that spontaneous breaking of ergodicity with its above
mentioned consequences is a common phenomenon already at equilibrium.
In contrast, a finite current $\langle\dot x\rangle\not = 0$ at $F=0$ as well as
spontaneous oscillations \cite{win80,nic81,kur84,vid88,gla88}
represent genuine collective non-equilibrium effects
which are excluded at thermal equilibrium by the second law of thermodynamics.

Both, from the conceptual viewpoint and with regard to the mechanochemical
coupling issue, the J\"ulicher-Prost model exhibits a couple of noteworthy features.
A first crucial assumption of the model is
that not only the backbone itself but also
the positions of the $N$ individual motors
with respect to the backbone are perfectly rigid.
Since the backbone moves with a speed $\langle\dot x\rangle$ 
it follows that for any single motor
\begin{equation}
\dot x = \langle\dot x\rangle\ .
\label{j0}
\end{equation}
Much like any intensive state variable in equilibrium thermodynamics,
the ``order parameter'' $\langle \dot x\rangle$ within such a mean
field approach is a macroscopic state variable and is {\em not}
any more subject to any kind of random fluctuations in the thermodynamic
limit $N\to \infty$.
In other words, the {\em stochastic} equation for the single (uncoupled) motors
(\ref{4.2.1}) simplifies to an equivalent 
{\em deterministic} (noise-free) dynamics
(\ref{j0}) for every single motor in the presence of a mean field 
(perfectly rigid all-to-all) coupling.
Here $\langle\dot x\rangle$ plays the role of a formal (not yet explicitly known)
deterministic force and -- as already pointed out in section \ref{sec5.4.1} --
our next goal must now be to derive a self consistency equation for
this order parameter $\langle\dot x\rangle$ if we wish to determine its
explicit value.
To this end we first notice that working with (\ref{j0}) instead of
(\ref{4.2.1}) is tantamount to setting 
$T=0$ and $-V'_m(x)+F = \eta\, \langle\dot x\rangle$ in (\ref{4.2.1}).
Accordingly, the first two terms on the right hand side of the master equation
(\ref{4.2.6}) may be replaced by the equivalent simplified expression
$-\langle\dot x\rangle\,\partial\, P_m(x,t)/\partial x$.

A second essential assumption is that the $N$ individual motor
enzymes are rooted in the backbone either at random positions or 
-- biologically more realistic --
with a constant spacing which is
incommensurate with the period $L$ of the polymer filament.
As a consequence,
the reduced spatial distribution of particles $\sum_m\hat P_m(x,t)$ approaches
an $x$- and $t$-independent constant value for $N\to\infty$.

As a final assumption,
a one-head description of the individual motor enzymes with $M=2$
chemical states is adopted\label{fot5.9}\footnote{According 
to \sect  \ref{sec5.4.1}, this model may equally
well be viewed as a $M=2$ state model of enzymes with two highly coordinated heads.
See also \sect  \ref{sec5.4.2} 
for references and more details regarding such a model.}.
Exploiting the above mentioned fact
that $\hat P_1(x,t)+\hat P_2(x,t)$ is a constant and normalized on
$[0,L]$ according to (\ref{4.2.8b2}), one can eliminate $\hat P_2(x,t)$ from
the master equation (\ref{4.2.6}), yielding in the steady 
state\footnote{The convergence towards a steady state in the long time limit 
is tacitly taken for granted. A partial justification of this ansatz 
can be given {\em a posteriori} by showing that such a solution indeed
exists and moreover satisfies certain stability conditions against 
perturbations.
Especially, the task to prove that no additional (non-stationary) long
time solutions co-exist is a delicate issue. 
In practice, the only viable way consists in a direct numerical
simulation of a large number $N$ of coupled stochastic equations.}
(superscript $st$)
the ordinary first order equation \cite{jul95}
\begin{equation}
\langle\dot x\rangle\,\frac{d}{d x}\hat P^{st}_1(x) = 
-\hat P^{st}_1(x)\, k_{1\to 2}(x) + [1/L - \hat P^{st}_1(x) ] k_{2\to 1}(x)\ ,
\label{j1}
\end{equation}
supplemented by the periodic boundary 
condition\footnote{There is no normalization condition for $\hat P^{st}_1(x)$ alone.}
$\hat P^{st}_1(x+L) = \hat P^{st}_1(x)$.
The unique solution is
\begin{equation}
\hat P^{st}_1(x) =
\frac{
\int_x^{x+L} dy\,  k_{2\to 1}(y)\, 
\exp\left\{\int_x^y dz\, 
\frac{k_{1\to 2}(z)+k_{2\to 1}(z)}{\langle\dot x\rangle}\right\}
}
{
L\, \langle\dot x\rangle\left[
\exp\left\{\int_0^L dz \, 
\frac{k_{1\to 2}(z)+k_{2\to 1}(z)}{\langle\dot x\rangle}\right\}
 - 1 \right] 
} \ .
\label{j1a}
\end{equation}
Note that the non-negativity of $\hat P^{st}_1(x)$ is guaranteed 
if $k_{1\to 2}(x)\geq 0$ and $k_{2\to 1}(x)\geq 0$ for all $x$.
Finally, by eliminating in the same way $\hat P_2(x,t)$
in the self-consistency equation (\ref{4.2.8c}) for $\langle\dot x\rangle$
one finds that
\begin{equation}
\langle\dot x\rangle = 
\frac{1}{\eta}\,\left[ F - 
\int_0^{L} dx\,  \left( V'_1(x) -  V'_2(x) \right) \, \hat P_1^{st}(x)\right]\ . 
\label{j2}
\end{equation}
By introducing (\ref{j1a}) into (\ref{j2}) a closed (transcendental)
self consistency equation for the order parameter $\langle \dot x\rangle$
is obtained.
Much like in the elementary mean field theory (Weiss theory) for 
a ferromagnet, the occurrence of multiple solutions will signal
the breaking of ergodicity and thus a phase transition.
Apart from the need of solving a transcendental equation at the very end,
the above model is one of the very rare special cases (cf. section
\ref{sec4.2.1}) of an analytically exactly tractable fluctuating potential ratchet.
We finally recall that by interpreting the $M=2$ state model
as a reduced $M=4$ state description, both rates $k_{1\to 2}(x)$ and $k_{2\to 1}(x)$ 
are still at our disposition (see \sect \ref{sec5.4.2}).

Besides the tremendous technical simplification of the problem,
the most remarkable feature of the J\"ulicher-Prost model (\ref{j1a}), (\ref{j2})
is that only the difference $V_1(x)-V_2(x)$ of the two potentials counts (one may
thus choose one of them identically zero without loss of generality).
It follows that the emerging qualitative results for a generic ($L$-periodic and
asymmetric) choice of $V_1(x)-V_2(x)$ will be valid independently of whether
the mechanochemical coupling is loose
(e.g. a dichotomously
fluctuating potential  ratchet with $V_2(x)\propto V_1(x)$, 
see \sect \ref{sec4.2})
or tight (e.g. a traveling two-state ratchet with $V_2(x)=V_1(x+L/2)$, see 
\sect \ref{sec4.3.2}).
Whether this feature should be considered as a virtue (robustness)
or shortcoming (oversimplification) of the model is not clear.

Modified Huxley-J\"ulicher-Prost type models
have been explored by Vilfan, Frey, and Schwabl \cite{vil98,vil99}.
Their basis is a $M=2$ state description of the single motors with a built-in 
tight mechanochemical coupling through the choice of the rates and potentials,
but, at variance with J\"ulicher and Prost, without a completely rigid
shape of the motors with respect to the backbone:
unlike in (\ref{j0}), the center of mass of an individual motor may differ
from the position where it is rooted in the backbone, say $\langle\dot x\rangle\, t$.
Similarly as in Huxley's original work, the possibility of ``strain''
dependent (i.e. $x-\langle\dot x\rangle\, t$ dependent) rates $k_{m\to m'}(x)$
plays an important role.
With a rigid backbone, a mean field approach is still exact for $N\to\infty$ but 
technically more involved than in the J\"ulicher-Prost model, 
while resulting in qualitative similar collective phenomena \cite{vil99}.
In contrast, by admitting an elastic instead of a rigid 
backbone\footnote{A computer animation (Java applet) which graphically
visualizes the effect is available on the internet under
\cite{vilmovie}.}, 
the interaction between the motors is no longer of infinite range and corrections
to a mean field approximation may become relevant under certain experimental 
conditions
\cite{vil98}.

Another refined version of the Huxley-J\"ulicher-Prost setup, taking 
into account an extended number of biological findings, is due to
Derenyi and Vicsek \cite{der98}.
While $M=4$ chemical states are included, only two different potential shapes
$V_m(x)$ are proposed, one of them being identically zero, and
a tight mechanochemical coupling is built in through the choice of the rates
$k_{m\to m'}(x)$.
While a very good agreement with different experimentally measured curves
is obtained, the issue of genuine collective phenomena is not 
specifically addressed.

Further studies of collective effects in coupled Brownian motors
will be discussed in \ch \ref{cha7}.

\section{Coordinated two-head model}\label{sec5.5}
In this subsection we return to the description of a single
motor enzyme with two heads within the general modeling framework 
(\ref{4.2.-1})-(\ref{4.2.5}).
Especially, we recall that this model respects an invariance under a spatial 
displacement by one period if simultaneously the chemical
states of the two heads are exchanged, see (\ref{4.2.2a}), (\ref{4.2.2b}).
We furthermore recall that for a  processive motor enzyme,
i.e., one which can operate individually (for instance kinesin),
the two heads need to coordinate their actions in order that at least
one of them is always attached to the polymer filament.

Our goal is to approximately
boil down the two-dimensional chemical state vector
$m=(m_1,m_2)$ into an effective one-dimensional (scalar) description.
To this end, we make the assumption that the two
heads are so strongly coordinated that between subsequent steps there
exists a time instant at which not only both heads are attached to the
filament, but  on top of that, the heads are in the same chemical
state, $m_1=m_2$.
Taking such a configuration as reference state, one of the two
heads will be the first to make a chemical transition into another
state. This may be, with certain, generically unequal probabilities,
either the front or the rear head, and the chemical state may, 
again with typically unequal probabilities
(cf. (\ref{4.2.4}), (\ref{4.2.5})), either go one step forward
or backward in its reaction cycle as time goes 
on\footnote{There seems to be no general agreement upon whether such
inverse processes are possible with finite (however small)
probability \cite{pes95} or not \cite{how96}.}.
Our central assumption is now, that once one of the heads has left 
the reference state $m_1=m_2$, the other head will not change its chemical state until
the first one has returned into the reference state.

We are not aware of experimental observations which indicate that 
such a property
is strictly fulfilled, but it appears to be an acceptable 
approximation, especially
in view of the great simplification of the model it entails.
Moreover, if one starts with a reduced description of the
chemical cycle in each head in terms of only two effective 
states, based on a similar
line of reasoning as in the preceding \sect \ref{sec5.4.2}, then
necessarily one of these two states must correspond to the head being
attached to the polymer filament and the other to the detached situation.
Since both heads cannot be detached simultaneously, our assumption
is thus automatically fulfilled in such a two-state description
for each head.

If one makes the additional simplifying assumption that, starting from the
reference state with both heads attached, 
only the rear head is allowed to detach,
then an effective one-dimensional description of the chemical states of the
two heads is straightforward:
After the rear head has returned into the reference state, it either
will have attached at the same binding site ($\beta$-tubulin)
from which it started out or
it will have advanced to the next free binding site at a distance $2L$.
In the former case, it is again the same head which will make the next chemical
reaction out of the reference state, while the other head continues to be stuck.
In the latter case, the rear head has completed a step\label{fot5.10}\footnote{We
recall that the stepping head itself advances by $2L$, but the center of mass $x$
of the two heads only by $L$.}, $x\mapsto x+L$,
and is now the new front head.
If we additionally exchange the chemical labels $m_1$ and $m_2$ of the two heads 
then we are back in the original situation due to the
symmetry of the system (\ref{4.2.2a}), (\ref{4.2.2b}).
In other words, we have obtained an effective description in which only one of the
chemical state variables, say $m_1$, can change, while $m_2$ is stuck all the time.
Dropping the index of $m_1$, one readily sees that the effective
potentials $V_m(x)$ and rates $k_{m\to m'}(x)$ in this new description
are now indeed 
$L$-periodic in $x$ 
and satisfy (\ref{4.2.8}).

The more general case that out of the reference state $m_1=m_2$ both,
the front and the rear head may detach from the filament with certain probabilities,
can only be captured approximately by means of an effective one-dimensional
chemical state variable\footnote{The problem is that now the information about which 
of the two heads is chemically active (detached) must be uniquely encapsulated in 
$x$ in addition to the position of the center of mass.}:
Namely, one has to assume that if the rear head detaches, then $x$ can only take 
values larger than the initial reference position $x_{{\rm ref}}$
(but smaller than $x_{{\rm ref}}+L$). Likewise, if the front head detaches, $x$ is
restricted to $[x_{{\rm ref}}-L , x_{{\rm ref}}]$.
These two possibilities can be imitated by ``splitting probabilities''
with which $x(t)$ in (\ref{4.2.1}) will evolve into the positive
or negative direction after detachment by way of an appropriate choice of the
potentials $V_m(x)$. Especially, these potentials have to be
chosen such that a recrossing of $x_{{\rm ref}}$ after detachment is practically 
impossible\label{fot5.11}\footnote{To be specific, we may model the chemical reference 
vector-state $m_{{\rm ref}}$ with both heads attached by a potential
$V_{m_{{\rm ref}}}(x)$ with a very deep and narrow minimum at $x_{{\rm ref}}$.
If $m_{{\rm ref}}$ goes over into one of the ``neighboring'' states, say $m'$,
then $V_{m'}(x)$ should have pronounced maxima on either side of
$x_{{\rm ref}}$ such that $x(t)$ will proceed rather quickly and irrevocably 
away from $x_{{\rm ref}}$, either 
to the right or left. The actual direction into which $x(t)$ disappears
decides a posteriori whether it was the front or the rear head
which has detached.}. For the rest, the mapping to an effective one-dimensional
chemical state variable $m$ with $L$-periodic potentials $V_m(x)$ and
rates $k_{m\to m'}(x)$ satisfying (\ref{4.2.8}) can be accomplished
exactly like before.

Two noteworthy features which can be described within such a general modeling
framework are thus (i) the possibility not only of ``forward'' but
also of ``backward'' steps and (ii) the possibility that a head re-attaches 
to the same binding site from where it started out.
Both these possibilities may be realized only with a small 
probability\footnote{We remark
that there are also models which rule out
such backward steps {\em a priori} \cite{how96}.}
under ``normal'' conditions \cite{svo94,sch97,hua97} 
but could become increasingly
important \cite{svo94,pes95,der96,cop97,ric99}
as the load force $F$ in (\ref{4.2.1}) approaches
the ``stopping force'' or ``stall load'', characterized by zero net motion
$\langle\dot x\rangle = 0$ 
(cf. \sect \ref{sec2.2.2.2}).

\section{Further models for a single motor enzyme}\label{sec5.6}
The above interpretation of (\ref{4.2.6}), (\ref{4.2.7c})-(\ref{4.2.8}) 
as a model
for a single motor enzyme with {\em two} highly cooperative heads
has, to our knowledge, not been pointed out and derived in detail 
before\footnote{Somewhat similar ideas can be found in 
\cite{pes95,par99,jul99,lip00a}.}.
However,
practically the same model dynamics (\ref{4.2.1}) has been
used to describe the somewhat artificial scenario\label{fot5.13}\footnote{Whether
or not manipulated, single-headed kinesin can travel over appreciable
distances on a microtubulus seems to be still controversial
\cite{gell95,pes95,ber95,val96,how96,oka99,meh99,ric99}. Remarkably, the experimental data from 
\cite{oka99} could be fitted very well by an on-off ratchet model 
(A video illustrating motility data can be viewed on the internet under 
\cite{okamovie}).
There seems to be evidence \cite{blo98} that single-headed motion is fundamentally 
different from two-headed motion.}
of a {\em single} head moving along a polymer filament 
\cite{cor92,lei93}.
By changing the interpretation, such results can immediately be 
translated into our two-head setting.

As mentioned in \sect \ref{sec5.4.3}, 
models with two completely independently operating heads, 
except that they  are connected  
by a ``hinge'', have been briefly addressed numerically in
\cite{cor92}.
A refined model of this type has later been put forward and analyzed by 
Vilfan, Frey, and Schwabl \cite{vil99} exhibiting good agreement with 
a variety of experimental curves and structural results.

Valuable contribution to the general conceptual 
framework \cite{mag93} of single motor modeling and
especially of the mechanochemical coupling \cite{mag94} are due to Magnasco,
see also \cite{kel00}.
At variance with our present setup, the chemical processes within
the entire motor enzyme are described from the beginning by
a single, continuous chemical state variable \cite{mag94} 
(see also \sects \ref{sec4.5} and \ref{sec5.3.2}).
Published practically at the same time, models similar in spirit, 
but with only two
discrete chemical states have been proposed by Astumian and Bier \cite{ast94},
by Prost, Chauwin, Peliti, and Ajdari \cite{pro94},
and by Peskin, Ermentrout, and Oster \cite{pes94}.
The underlying picture is that, essentially, the motor enzyme as a whole
is either ``attached'' to or ``detached'' from the protein filament.
The emphasis in all these works \cite{mag93,mag94,ast94,pro94,pes94} 
(see also \cite{zho96}) is put on the 
fundamental aspects and generic properties of motion generation in such systems;
apart from the general features of spatial periodicity and broken symmetry,
no contact with any further biological ``details'' is established.
Yet, by using reasonable parameter values in a fluctuating potential
sawtooth ratchet model, measured data for the average speed $\langle\dot x\rangle$
and the rate of
ATP-consumption (cf. (\ref{4.2.8c}) and (\ref{4.2.8d}))
could be reproduced within an order of magnitude \cite{ast94,ast96a}.
On the other hand, it was demonstrated in \cite{fis99,fis99a} that even within the
simplest two-state models for a single motor ($M=2$ in (\ref{4.2.7c})), 
a large variety of even
qualitatively contrasting results can be produced upon varying the
model parameters. Not only a realistic choice of the model parameters
but also of the details of the model itself is therefore indispensable.

A biologically well founded description of a motor enzyme with
two cooperating heads, similar to our present setup with 
$M=2$ chemical states,
has been introduced by Peskin and Oster \cite{pes95}.
A central point in this study is once more 
the relative roles of the thermal fluctuations
and the relaxational processes due to release of chemically generated strain in 
the dynamics of the spatial coordinate $x$ in (\ref{4.2.1}).
Another important feature of the model is that, besides regular forward steps
of the heads, also backward steps after detachment of the front head are 
admitted with a certain probability.
The result, after fitting the model parameters to the experiment,
is that -- within this specific model -- thermal
fluctuations play a minor role.
Furthermore, it is found that backward steps are about 20 times less probable
than forward steps\footnote{See also the discussion at the end of the 
previous subsection.}.

The model by Derenyi and Vicsek \cite{der96} is to some extent
similar in spirit to the one by Peskin and Oster.
Especially, backward steps are admitted and the two heads act highly cooperatively.
The built-in mechanochemical coupling is a compromise in that thermal activation 
is indispensable but the rates $k_{m\to m'}(x)$ are tailored such that the next 
chemical step can only occur after $x$ has crossed the
respective barrier and is basically undergoing a purely
mechanical relaxation.
The model can be mapped almost exactly to a $M=2$ state model from 
\sect \ref{sec5.2.4},
though the original formulation \cite{der96}
in terms of two rigid heads, coupled by a hinge and an ``active'' spring
with variable rest length 
is admittedly more natural in this specific instance.
In either case, the model can be described in very
good approximation by an augmented reaction cycle with mechanical states 
properly added to the chemical states.
The distinguishing feature of the model, the experimental justification of which
remains unclear \cite{blo95,how97a,vis99,meh99}
is that the two heads cannot pass each other:
The distance between the front and the rear head 
(in other words, of the spring) can change but never become zero
so that the heads never exchange their roles.
The virtue of the model is its ability to fit very well
various measured curves.
The limiting case of a very strong ``active'' spring, such 
that thermal activation is no longer 
important, has been explored in \cite{str99}.
A somewhat related model with two effectively asymmetric heads
is due to \cite{mog98}, see also \sect \ref{sec5.7}.

The conceptual framework \cite{ast94}
of Astumian and Bier for modeling single molecular motors 
has been further developed and refined in a remarkable
series of works 
\cite{bie96b,ast96a,ast96b,ast97,bie97,ast98,ast99a,ast00a}.
Various aspects and results of their central study \cite{ast96a} have
been repeatedly referred to already in the outline of our 
general modeling framework.
The chief points in \cite{ast96a} are a comprehensive discussion of the
mechanochemical coupling problem and the conclusion that many experimental
indications and theoretical arguments seem to be compatible with
a rather loose coupling, especially when a suitably augmented 
cooperative two-head model is invoked \cite{ast99a,ast00a}. 
A complementary discussion along a closely 
related spirit is given in \cite{zho96}.
Especially worth mentioning is
that the fluctuational analysis of measured single motor protein trajectories
in \cite{svo94b} is incompatible with a certain class of very simple
(fluctuating potential) 
ratchet models but have been demonstrated in \cite{ast99a,ast00a} to be
perfectly reproducible by means of a more elaborated and 
refined description.

Non-cooperative discrete-state models with a built-in tight 
mechanochemical coupling in the spirit of \cite{lei91,lei93,duk96,kol98} 
(see at the end of \sect \ref{sec5.4.3})
have been addressed in \cite{fis99,fis99a,kel00,kol00a,kol00b,how01,fis01}, 
especially with respect to their behavior under the influence of an
external load $F$.
Notwithstanding the conclusion in \cite{lei91,lei93}
that at least four states are necessary for 
a realistic model, the agreement of the two- and three-state 
models proposed in \cite{fis99,fis99a,fis01} and \cite{how01}, respectively, 
with experimental observations is quite good.
Various generalizations of these
``mechanochemical reaction cycle models''
(cf. \sects \ref{sec5.4.3} and \ref{sec5.8}) 
are due to \cite{kol00a}, while the extension to general waiting
time distributions has been addressed in \cite{fis99,fis99a,kol00b},
admitting in addition to thermally activated mechanochemical rate 
processes e.g. also the description of mechanical relaxation of strain.
Their drawback is a large number of additional phenomenological model 
parameters.

The viewpoint \cite{pro94} of Prost and collaborators with respect to modeling 
single motor enzymes has been further elaborated in \cite{jul97a,jul98}
and especially in \cite{par99,jul99,par01,lat01}.
While the general framework has much in common with that of Astumian and
Bier, these workers put special emphasize on the possible relevance
of ``active sites'', i.e. a pronounced dependence of the transition rates
$k_{m\to m'}(x)$
on the mechanical state $x$, such that transitions are practically
excluded outside of certain small $x$-regions.
They furthermore leave room to the possibility that a 
traveling potential ratchet
mechanism may dominate over a possibly coexisting fluctuating potential 
ratchet mechanism, in which case 
the mechanochemical coupling might be rather tight.
An explicit modeling of cooperative
two-headed motor enzymes along somewhat similar lines as in 
\sect \ref{sec5.5} is briefly mentioned in \cite{par99,jul99}. The resulting
description with $M=2$ effective chemical states associates each state with one 
of the heads being bound to the filament and the other detached.
One thus recovers the traveling potential ratchet model from \cite{cha94},
advancing in discrete steps of $L/2$ as detailed in \sect \ref{sec4.3.2}.
The influence of an external load $F$ on velocity and processivity
(detachment rate of the molecular motor from the microtubulus) has been
addressed in \cite{par01}, see also \cite{kol00a}. 
A related study due to \cite{lat01} suggests
a loose mechanochemical coupling at least under heavy load.
The case that the load is not an externally imposed constant force but
rather is due to the ``cargo'', modeled as additonal relevant dynamical
variable that interacts with the motor via an elastic coupling,
has been addressed in \cite{els00a,els00b}, see also below
\eq (\ref{4.2.2}).

A detailed analysis of a somewhat extended model class with 
pronounced ``active sites'' and a strong
traveling potential component has been carried out 
in \cite{lip00a,lip00b,lip00c}.
In agreement with experiments, these models reproduce a ``saturation''
of the current $\langle\dot x\rangle$ as a function of the 
ATP-concentration \cite{pro94,jul97a,jul98},
captured by a Michaelis-Menten relation for a large
class of moderately
and strongly cooperative models under zero lead $F$
\cite{lip00a,lip00b,lip00c}, while for finite load a somewhat
modified quantitative behavior is expected 
\cite{vis99,lip00a,lip00b,lat01}.
We remark that while for cooperative two-head models with only $M=2$
chemical (or ``internal'') states per head,
the assumption of ``active sites''
is indeed indispensable for such a saturation
of the current, the same is no longer true as soon as 
$M>2$.

\section{Summary and discussion}\label{sec5.8}

We close with some general remarks regarding the modeling
of molecular motors as reviewed in this chapter.
Previously introduced notions and facts are freely used without
explaining them or citing the original literature again.
To some extent, this discussion continues and makes more precise
those from \sects \ref{sec4.5} and \ref{sec5.4.3}.

The general importance of asymmetry induced rectification,
thermal fluctuations, and the coupling of non-equilibrium enzymatic 
reactions to mechanical currents according to Curie's principle
for intracellular transport processes is long known \cite{bio83,fri86}.
The present framework has the virtue that it
is based on a quantitative microscopic modeling and as such is not
restricted to the linear response regime close to thermal equilibrium.
Within this general framework,
roughly speaking two approaches of modeling molecular motors 
may be distinguished:
The first, ``traditional'' one is a bottom-up-type strategy,
starting with a certain set of biological facts (measurements
and more or less ``basic'' conclusions therefrom) and then
constructing an ``ad hoc'' model on this basis.
The second is the top-down-type approach, followed 
to some extent in more recent works based on the 
``ratchet paradigm'' and elaborated in full detail 
in our present chapter.

Our first main conclusion is that
all models known to this author
are compatible (possibly after some
mapping or transformation of state variables)
with the basic framework from \sect \ref{sec5.2}, 
and most of them also with
the simplified description in terms of a single mechanical 
state variable $x$ and the corresponding model dynamics
(\ref{4.2.1}), identified below (\ref{4.2.7b}) as a
{\em (generalized) pulsating ratchet scheme}.
In other words, such an approach is 
{\em not in contradiction with ``traditional'' biological models},
but may well offer a fresh and more systematic 
(top-down) view of things \cite{mag94,how97b,lib98,ast99a,kel00,ast00a}.

Within this still very general class of models (\ref{4.2.1})
the most realistic choice of model parameters and model
functions is still under debate and certainly also depends
on the specific type of molecular motor under consideration
(especially whether it is of processive (individually
acting) or non-processive (collectively acting) nature
or even consists of a single motor domain (head) only).
Conversely, it is remarkable that all these different
species can be treated within one general framework.

Three basic questions in this respect, which are not always
sufficiently clearly separated from each other, regard:
{\em (i) The possibility of an (approximate) description
in terms of a single (effective) chemical state variable.
(ii) The relative importance of the thermal
fluctuations appearing in the dynamics of the mechanical
coordinate (\ref{4.2.1}) as compared to conformational 
changes powered by the chemical cycle.
(iii) The character (loose or tight) of the mechanochemical 
coupling.}

The answer to these questions may not only depend on the
type of molecular motor under consideration (see above)
but also on whether an external
load $F$ is acting and possibly on still other external conditions.
For example, it may well be that, 
as the load $F$ increases, the relative importance
in (\ref{4.2.1}) of thermal activated barrier crossing and
deterministic relaxation processes 
(i.e. the answer to question (ii) above)
considerably changes.
The reason is that, as the force $F$ increases, 
existing effective potential barriers in the stochastic 
dynamics (\ref{4.2.1}) may disappear and new ones 
appear\label{fot5.12}\footnote{The ``total'' or ``effective''
potential in (\ref{4.2.1}) is given by $V_m(x)-x\,F$.}.
Similarly, the external load $F$ may also significantly change
the character of the mechanochemical 
coupling\footnote{Especially, $F$ may change the shape of the
potentials $V_m(x)$, as discussed below \eq (\ref{4.2.2}).}  
\cite{svo94,mey95,ast96a,der96,cop97,lip00a,lip00b,how01,hou01,par01,lat01}
(i.e. the answer to question (iii) above).

The answer to the first of the above questions depends on
the cooperativity of the two heads: If they act completely
independently of each other, they can obviously be described
individually, and a single (scalar)
chemical state variable $m$ for each head is then sufficient.
That the same may be possible for a very strong coordination
of the heads has been demonstrated in \sect \ref{sec5.5}.
On the other hand, 
if the cooperativity is loose but non-negligible, then
a reduction of the two-dimensional chemical state space 
(cf. (\ref{4.2.-1})) is impossible.

The second question is sometimes also discussed under the label of
power-stroke versus motor-diffusion modeling strategy \cite{how96,how01,hou01}.
In the first case, the chemical cycle ``slaves'' the mechanical
cycle by creating a sequence of strong mechanical strains
(power strokes) that are released by concomitant, basically deterministic
changes of the mechanical state (geometrical shape).
Typically, there is little back reaction of the mechanical to
the chemical coordinate, and we are thus essentially dealing with
a genuine traveling potential ratchet scheme.
In the second case, thermal fluctuations play a major role in the
dynamics of the mechanical state variable (\ref{4.2.1}).
The first model of this type goes once again back to Huxley \cite{hux57}
and the apparent lack of strong experimental support for the power 
stroke concept \cite{hou01} has served as a motivation for various
other such models ever since.
Especially, this controversy has a long history already within
the realm of ``traditional'' biological modeling
and the gain of new insight in this repect 
from an approach based on the ``ratchet paradigm'' may be 
limited.
Also, we may emphasize once more that in either 
case thermal noise
plays a crucial role with respect to the chemical 
reaction cycle -- in this sense {\em any} model of a
molecular motor (not only those of the motor-diffusion type)
``rectifies'' thermal fluctuations. We further remark that
also within a motor-diffusion modeling, the mechanochemical
coupling may still be either tight (e.g. Huxleys model) 
or loose (e.g. the on-off ratchet).
On the other hand, a power-stroke model 
always implies a tight mechanochemical coupling.

Another related question within a motor-diffusion modeling
is whether the thermal fluctuations acting on the
mechanical state variable can be treated within the
activated barrier crossing limit (see \sect \ref{sec3.6})
or whether free diffusion-like behavior plays a significant
role. Only in the former case, a description of the mechanical
state variable in terms of discrete states and transition rates
between them is admissible, see \sect \ref{sec3.9} and
\cite{fis99,fis99a,kol00b,fis01}. Note that both options are still
compatible with either a tight or a loose mechanochemical 
coupling. In the case of a tight coupling in combination with
an activated barrier crossing description, a so-called ``mechanochemical
reaction cycle'' arises (cf. \sect \ref{sec5.4.3}).
Since from a fundamental viewpoint, the distinction between
chemical and mechanical state variables is somewhat 
arbitrary anyway (see \sect \ref{sec5.2.2}), we are 
then basically recovering an effective power-stroke model.

We finally come to the question (iii) of the mechanochemical coupling
(see also \sect \ref{sec2.2.3a}).
We first remark that a tight coupling not necessarily means
that the chemical state variable always ``slaves'' the mechanical one
(genuine power-stroke model) but that one variable ``slaves'' the other at
each stage of the mechanochemical reaction cycle 
(the chemical reaction may be blocked -- 
due to active sites, i.e. stronly $x$-dependent rates $k_{m\to m'}(x)$ --
until some mechanical
transition between different geometrical shapes of the motor
has been accomplished, e.g. in the above effective power-stoke model).
Restricting ourselves to 
the simplest case of a one-dimensional chemical state variable,
{\em a tight mechanochemical coupling means that a description 
in terms of a single effective state variable is possible,
and a loose coupling means that such a description is impossible}.
In other words, the state space is either essentially one- or two-dimensional.
In the first case, there is a unique ``pathway'' in the $x$-$m$-space,
in the second case bifurcations exist. 
Examples are genuine traveling potential ratchets
and fluctuating potential ratchets, respectively.
We, however, emphasize that the conclusion suggested by the latter
example, namely that a loose mechanochemical coupling implies that 
thermal fluctuations play an essential
role in the dynamics of the mechanical coordinate (\ref{4.2.1}),
can be easily demonstrated as incorrect by counterexamples. 
In other words, the thermally
induced randomness of the chemical reactions suffices to
produce bifurcations in the ``pathway'' through the full 
$x$-$m$-space.

The possibility of a loose mechanochemical coupling is widely
considered as one of the
main conceptually new aspect of the ``ratchet paradigm'' as compared to
``traditional'' biological models\footnote{Sometimes, also the
possibility of a motor-diffusion modeling approach is considered
as such.}.
However, in its simplest and most pronounced form, namely
the fluctuating potential ratchet scheme from \sect \ref{sec4.2}
(i.e. with $x$-independent rates $k_{m\to m'}$)
it is apparently incompatible with the fluctuational analysis
of single (two-headed) motor protein trajectories \cite{svo94b}.
On the other hand, the experimental data for single-headed kinesin
from \cite{oka99} could be fitted very well to an on-off ratchet model.
The currently prevailing opinion seems to be that
a loose coupling is unlikely for processive motors like
two-headed kinesin but a realistic option in the case
of non-processive (cooperative or single-headed) motors
\cite{svo94b,oos95,sch97,how97a,hua97,kit99,oka99,meh99,meh99a,vis99,bar99,oos00,yan00,how01,hou01}. 
However, 
room for the possibility of a loose coupling even 
in the case of kinesin is still left e.g. in 
\cite{tay93,svo94,ast96a,cop97}.
If one considers the concept of a loose mechanochemical coupling
as the only substantial new contribution of the ``ratchet paradigm''
to the modeling of molecular motors, then -- in the so far absence
of striking experimental indications of such a 
coupling --
the merits of this paradigm may still be considered as
questionable.
However, such a viewpoint may not do due justice to
other noteworthy achievements like the prediction of
new collective effects from \sect \ref{sec5.4.4} or the 
unified new view and working model.

\chapter{Quantum ratchets}\label{sec6.5}
For many of the so far discussed ratchet systems, especially those
for which thermal fluctuations play any significant role,
the characteristic length-, energy- etc. scales are very
small and it is thus just one more natural step forward to
also take into account quantum mechanical effects.

Before we enter the actual discussion of such
effects, two remarks are in place:
First, we have encountered in \sects \ref{sec6.2.3.1} and \ref{sec6.2.4a}
theoretical models and experimental realizations of Josephson and SQUID
ratchet systems. Since the basic state variables in such devices are phases
of macroscopic quantum mechanical wave functions, it is tempting
to classify them as quantum ratchet systems.
Our present viewpoint, however, it that the decisive criterion
should be the classical or quantum mechanical character of the
effective dynamics governing the relevant state variables of a system,
independently of whether the microscopic basis of this effective dynamics
is of classical or quantum mechanical nature, see also \cite{cal83}.
For instance, the existence of stable atoms, molecules, and solids
is clearly a genuine quantum mechanical phenomenon,
yet a classical theory of gases, liquids, and solids can be established.
From this viewpoint, the Josephson and SQUID ratchet systems as discussed in 
\sects \ref{sec6.2.3.1} and \ref{sec6.2.4a} are thus classical ratchets.
The realization of a full-fledged quantum mechanical ratchet according
to our present understanding in SQUID systems
will be discussed later in \sect \ref{sec6.5.5}.
As a second remark we mention that the proper quantum mechanical
treatment, e.g. of the Seebeck effect
(\sect \ref{sec4.8.1}) or the photovoltaic effects 
(\sect \ref{sec6.2.1}),
may arguably be considered as very early quantum ratchet studies
of considerable practical relevance.
However, in the present section we put our main emphasis not on
a faithful quantum mechanical modeling of such specific systems but
rather on the exploration of the basic features
of much simpler models.
Namely, {\em our main focus will be on the interplay between 
tunneling and the
effects induced by the thermal environment (i.e. dissipation and
thermal noise) in the quantum mechanical counterparts of the
classical tilting ratchet dynamics (\ref{6.1})}.

\section{Model}\label{sec6.5.1}
In the case of classical Brownian motion, we have introduced
in \sect \ref{sec2.1.2} a model which takes into account the
influence of the thermal environment 
along a rather heuristic line of reasoning, 
see also \sects \ref{sec2.2.3c} and \ref{sec3.2.1}.
In contrast, on a quantum mechanical level, such a heuristic
modeling of dissipation and thermal noise, e.g. on the level
of the Schr\"odinger equation, is much more problematic and
liable to {\em subtle inconsistencies} for instance
with the second law of thermodynamics 
or some basic principles of quantum mechanics, see
\cite{dum79,tal86,amb91,kam92,pec94,for96,gnu96,cap99} and further references therein.
To avoid such problems, we follow here the common route
\cite{mag59,fey63b,cal83,for88,gra88a,han90,ben94,ing98,gri98,wei99a}
to describe both the system and its thermal environment within a
common Hamiltonian framework, with the heat bath being
modeled by an infinite set of harmonic oscillators.
Especially, within a quantum mechanical approach, keeping a finite
mass of the system is unavoidable, i.e. {\em a quantum ratchet
is by nature endowed with finite inertia}.
If one insists in considering the overdamped limit $m\to 0$
then this limit usually has to be postponed to the very end 
of the calculations.

Similarly as in \sect \ref{sec3.2.1}, our 
starting point is a one-dimensional
quantum particle with mass $m$ in
an asymmetric, periodic ratchet-potential
$V(x)$ of period $L$ in the presence
of a tilting force field $y(t)$ that is unbiased on average.
This bare system is furthermore coupled via coupling strengths $c_j$
to a {\em model heat bath of infinitely many harmonic oscillators} with
masses $m_j$ and frequencies $\omega_j$ 
($\omega_j >0$ without loss of generality)
yielding the compound (system-plus-environment) Hamiltonian
\begin{eqnarray}
{\bf H} (t) & = & \frac{{\bf p} ^2 }{2m}+V({\bf x}) - {\bf x}\, y(t)+ {\bf
H}_B
\label{qm1.1}\\
{\bf H}_B & := & \sum_{j=1}^\infty
\frac{{\bf p}_j^2}{2\, m_j}+
\frac{1}{2}m_j\omega_j^2\left({\bf x}_j-\frac{c_j{\bf x}}{m_j\omega_j^2}
\right)^2\ .
\label{qm1.2}
\end{eqnarray}
Here, ${\bf x}$ and ${\bf p}$ are the one-dimensional
coordinate and momentum operators of the
quantum Brownian particle of interest, while  ${\bf x}_j$ and ${\bf p}_j$
are those of the bath oscillators.
As initial condition at time $t=0$ we assume that the bath is at
thermal equilibrium and is decoupled from the 
system.
The infinite number of oscillators guarantees an infinite heat capacity and 
thus a reasonable model of a heat bath that keeps its initial temperatures 
for all later 
times\footnote{Further shortcomings of a heat bath with a finite number of
oscillators are: 
(i) Both the memory kernel (\ref{qm1.8}) and the
noise-correlation (\ref{qm1.10}) do not decay to zero for large times,
rather they are quasi-periodic.
(ii) The future behavior of the ``noise'' (\ref{qm1.9}) becomes
perdictable from its past, at least in the classical limit, see 
section 11-5 in \cite{pap65}.}
$t>0$.
For the rest, it turns out 
\cite{for88,gra88a,han90,ben94,ing98,gri98,wei99a}
that the effect of the environment on the system
is completely fixed by the frequencies $\omega_j$ and the ratios $c_j^2/m_j$, or
equivalently, by the so called spectral density
\begin{equation}
J(\omega) := \frac{\pi}{2}
\sum_{j=1}^\infty \frac{c_j^2}{m_j\omega_j}\,\delta(\omega-\omega_j) \ .
\label{qm1.3}
\end{equation}

By way of integrating out the bath degrees of freedom in (\ref{qm1.1})
one obtains \cite{for88,gra88a,han90,ben94,ing98,gri98,wei99a} 
the following one-dimensional Heisenberg
equation for the position operator ${\bf x} (t)$:
\begin{equation}
m\, \ddot{\bf x}(t) + V^\prime ({\bf x}(t)) - y(t)
= - \int_{0}^t
\hat\eta (t-t^\prime )\, \dot{\bf x}(t^{\prime})\, dt^{\prime} + \bldx (t) \ .
\label{qm1.7}
\end{equation}
Like in (\ref{2.1}),
the left hand side can be associated to the bare system dynamics, while
the right hand side accounts for the influence of the environment through
the damping kernel
\begin{equation}
\hat\eta (t) := \frac{2}{\pi}\,
\int_0^\infty d\omega \,\omega^{-1}\, J(\omega)\, \cos (\omega t).
\label{qm1.8}
\end{equation}
and the operator valued quantum noise
\begin{equation}
\bldx(t)=
\sum_{j=1}^\infty  c_j\, \left(
\frac{{\bf p}_j(0)}{m_j \omega_j} \sin (\omega_j t)
+ \left( {\bf x}_j(0)-\frac{c_j{\bf x}(0)}{m_j\omega_j^2} \right)
\cos (\omega_j t) \right) \ ,
\label{qm1.9}
\end{equation}
containing the initial conditions of the bath and of the
particle's position. Exploiting the assumed thermal
distribution of the bath ${\bf H}_B$ at $t=0$ one finds 
\cite{for88,gra88a,han90,ben94,ing98,gri98,wei99a}
that $\bldx (t)$ becomes
a stationary Gaussian noise with mean value zero.
Moreover, one recovers the usual connection
(via $J(\omega )$)
between the random and the friction effects of the bath
on the right hand side of (\ref{qm1.7})
in the form of the fluctuation-dissipation relation
\begin{equation}
\langle \bldx(t+\tau) \bldx (t) \rangle = \frac{\hbar}{\pi}
\int_0^\infty d\omega J(\omega ) \left [
\mbox{coth} \left ( {\hbar \omega \over 2k_BT} \right ) \cos (\omega
\tau) - i\sin( \omega \tau)\right ] \ ,
\label{qm1.10}
\end{equation}
where $\langle\cdot\rangle$ indicates the thermal average (quantum statistical
mechanical expectation value), $i:=\sqrt{-1}$, and $\tau\geq 0$.

In the following, we will focus on a so called Ohmic bath,
characterized by a linear initial growth of the spectral sensity $J(\omega)$,
a ``cutoff'' frequency $\omega_c$, and 
a ``coupling parameter'' $\eta$:
\begin{equation}
J(\omega ) =\eta\, \omega \,\exp\{-\omega/\omega_c\} \ .
\label{qm1.5}
\end{equation}
The cutoff $\omega_c$ is introduced in order to avoid unphysical 
ultraviolet divergences but will always be chosen much larger than 
any other relevant characteristic frequency of the model.
The special role of such an Ohmic heat bath becomes 
apparent by observing that the corresponding
damping kernel (\ref{qm1.8}) approaches
\begin{equation}
\hat\eta (t)=2\, \eta\,\delta (t) 
\label{qm1.10a}
\end{equation}
when the cutoff $\omega_c$ goes to infinity.
The integral in (\ref{qm1.7})
thus boils down to the memory-less viscous friction
$-\eta\,\dot{\bf x}(t)$. In other words, 
$\eta$ in (\ref{qm1.5}) has the meaning
of a damping coefficient due to viscous friction.

In the {\em classical limit}, {\em i.e.}, for 
$\hbar/k_BT$ much smaller than any other
characteristic time scale of the noiseless system (\ref{qm1.7}),
the correlation (\ref{qm1.10}) with (\ref{qm1.8}) correctly 
approaches the classical fluctuation-dissipation relation
from (\ref{fri5}). Furthermore, in this limit
all quantum fluctuations vanish, so that q-numbers go over into c-numbers and
(\ref{qm1.7}) reproduces (for $y(t)\equiv 0$) the
classical model (\ref{fri4}) of a real valued stochastic process $x(t)$ in the
presence of Gaussian noise $\xi(t)$ and (\ref{2.1})
in the special case of a memoryless damping (\ref{qm1.10a}),
see also \sect \ref{sec3.2.1}.

For later purposes, it is useful to distinguish
between two different variants of the classical limit:
The first one, which we call {\em formal classical limit},
consists in letting $\hbar\to 0$, i.e. {\em quantum effects are simply ignored}
within such a description, independent of how relevant they are in the true 
system under study.
This limit is formal in so far as in reality $\hbar$ is a natural constant.
A second possibility,
which we call {\em physical classical limit},
consists in focusing on large temperatures $T$ such that $\hbar/k_BT$ is
sufficiently small and thus {\em quantum effects become indeed negligible}
in the real system.

As suggested by the above mentioned findings in the classical limit, 
the harmonic oscillator model for the thermal environment
(\ref{qm1.2}), (\ref{qm1.3}), (\ref{qm1.5}),
provides a {\em rather satisfactory description
in a large variety of real situations} 
\cite{cal83,for88,han90,sch90,ben94,con96,ing98,wei99a}, even though
for many complex systems, one does not have a very clear understanding 
of the actual microscopic origin of the damping and fluctuation effects.
In fact, it seems to be widely believed that once the dissipation mechanism
is known to be of the general form
appearing on the right hand side of \eq (\ref{qm1.7}), i.e. to
be a linear functional of the system velocity,
then for a heat bath {\em at thermal equilibrium} all the statistical properties
of the quantum noise $\bldx (t)$ in (\ref{qm1.7})
are uniquely fixed, i.e. 
{\em independent of any further microscopic
details of the thermal heat bath}.
Arguments in favor of this conjecture have been given e.g. in 
\cite{cal51,sen60,fey63b,cal83,for88}, but a veritable proof does not 
seem to exist yet,
see also \sects \ref{sec2.1.2}, \ref{sec3.2.1} 
and \cite{nyq28,gra80,gra82a,rei01a} for the classical limit.
Under the assumption 
that the conjecture holds, it can be inferred \cite{for88} that
{\em any} dissipative dynamics of the form (\ref{qm1.7}) which is in
contact to an {\em equilibrium} heat bath can be represented by a harmonic
oscillator model (\ref{qm1.1}), (\ref{qm1.2}).
This does not mean that in every such physical system the actual bath is a
harmonic oscillator bath, but only 
that one cannot tell the difference as far as the
behavior of the system ${\bf x}(t)$ is concerned \cite{for88}.
We finally remark that the damping kernel (\ref{qm1.8})
does not change in the classical limit, it is the same
for both a quantum mechanical or classical treatment of the system
dynamics.
In other words,
the knowledge of the dissipation term in the classical limit
appears to be sufficient to completely
fix the quantum mechanical stochastic dynamics.

{\em Historically}, the harmonic oscillator model has apparently
been invoked for the first time by Einstein and Hopf \cite{ein10}
for the description of an oscillating electrical dipole under blackbody
irradiation and subjected to radiation 
damping\footnote{A preliminary toy-model, somewhat related to
the problem considered by Einstein and Hopf is due to Lamb \cite{lam00}.
It can be mapped onto a harmonic oscillator model \cite{for88}
but does not involve fluctuations of any kind.
The same proviso applies for further related early works,
like e.g. \cite{kam51,sch64}.}.
A classical model with a harmonic oscillator potential $V(x)$,
but otherwise exactly like in (\ref{qm1.1})-(\ref{qm1.3})
has been put forward by Bogolyubov
\cite{bog45}, however, without explicitly working out the 
statistical properties of the fluctuations $\xi(t)$, 
especially their Gaussian character and the 
classical counterpart (\ref{fri4}) of the
fluctuation-dissipation relation (\ref{qm1.8}), (\ref{qm1.10}).
The latter issues, together with a quantum mechanical transcription of 
the model, has been accomplished by Magalinskii \cite{mag59}.
Subsequent re-inventions, refinements, and generalizations of the model
have been worked out e.g. in \cite{rub60,fey63b,ull66,zwa73,cal83,for88,gra88a}.

\section{Adiabatically tilting quantum ratchet}\label{sec6.5.2}
For general driving $y(t)$, \eq (\ref{qm1.7}) gives rise to a very complicated
non-equilibrium quantum dynamics. To simplify matters \cite{rei97,rei97a,rei98}, 
we restrict
ourselves to very slowly varying tilting forces $y(t)$ such that the system can 
always adiabatically adjust to the instantaneous thermal equilibrium 
state (accompanying equilibrium). We furthermore assume that $y(t)$ is 
basically restricted to the values $\pm F$, i.e., the transitions
between $\pm F$  occur on a time scale of negligible duration
in comparison with the time the particle in (\ref{qm1.7}) is exposed 
to either of the ``tilted washboard'' potentials
\begin{equation}
U^{\pm}(x) := V(x)\mp F\, x \ , 
\label{qm3}
\end{equation}
cf. \fig \ref{figqm1}.
As a final assumption we require a positive but not too large $F$, such that
$U^\pm(x)$ still display a local maximum and minimum within each
period $L$.
Apart from this, the 
tilting force $y(t)$ may still 
be either of stochastic or of deterministic nature.

\figqmeins

Within the so defined model, we are essentially left with six model 
parameters\footnote{Throughout this section the cutoff $\omega_c$ in
(\ref{qm1.5}) is chosen much larger than any other characteristic frequency 
of the system and therefore does not appear any more in the following.}, 
namely
the particle mass $m$, the ``potential parameters'' $V_0$, $L$, and $F$ (see
\fig \ref{figqm1}), 
and the ``thermal environment parameters'' $\eta$ and $T$.
We now make the assumption that these parameters
are chosen such that a classical particle which starts at rest close 
to any local maximum of
$U^{\pm}(x)$ will {\em deterministically} slide down the corresponding slope
but will not be able to subsequently surmount any further potential barrier
and so is bound to end in the next local minimum.
Differently speaking, a moderate-to-strong friction
dynamics is considered and
deterministically  ``running solutions'' are excluded.

We further assume weak thermal noise, that is,
any potential barrier appearing in  (\ref{qm3})
is much larger than the thermal energy, i.e.
\begin{equation}
\Delta U^{{\rm min}}\gg k_B T \ ,
\label{qm3a}
\end{equation}
where $\Delta U^{{\rm min}}$ denotes the smallest of those potential barriers.
As a consequence, we are dealing with an barrier crossing problem
(see \sect \ref{sec3.6})
and thus the average particle current in either of the two potentials $U^\pm(x)$
can be expressed in terms of two rates according to (\ref{4i}), 
see also (\ref{6.5}).
Moreover, the assumption of rare jumps of $y(t)$ between the two values $\pm F$
makes it possible to express the net current by way of an adiabatic limit
argument analogous to (\ref{6.2}), (\ref{6.8}) in terms of these two
partial currents.
In this way, one finally arrives at the following expression for the
averaged net particle current in terms of two rates:
\begin{equation}
\langle\dot x\rangle = \frac{L}{2}\, (1-e^{-FL/k_BT})\, (k_r^+-k_l^-)\ .
\label{qm4}
\end{equation}
Here, $k_r^+$ indicates the escape rate from one local minimum of $U^+(x)$ 
to its neighboring local minimum to the right, and similarly $k_l^-$ denotes
the rate to the left in the potential $U^-(x)$. 
We also recall that the average
on the left hand side of (\ref{qm4}) indicates
a thermal averaging (quantum statistical mechanical expectation value)
together with an averaging over the driving $y(t)$.

Within a purely classical treatment of the problem,
i.e. within the formal classical limit $\hbar\to 0$, 
any of the two rates $k$
in (\ref{qm4}) describe thermally activated transitions ``over'' a
certain potential barrier $\Delta U$
between neighboring local minima of
the corresponding potential. 
Due to the weak noise condition (\ref{qm3a}), 
such a rate $k$ is given in very good approximation by the well 
known Kramers-rate expression \cite{han90}
\begin{eqnarray}
& & k=\frac{\mu \sqrt{U''(x_0)}}{2\pi\sqrt{|U''(x_b)|}}\,e^{-\Delta U/k_BT}
\label{qm4a}\\
& & \mu:=\frac{\sqrt{\eta^2+4m|U''(x_b)|}-\eta}{2\, m} \ ,
\label{qm4a'}
\end{eqnarray}
where $x_b$ and $x_0$ denote the above mentioned local potential-maximum 
and -minimum, respectively, and
where indices $r$, $l$, and $\pm$ have been dropped. 
Note that in the overdamped limit $m\to 0$, the Kramers-Smoluchowski 
rate-expression from (\ref{2.23t}) is recovered.

Turning to a quantum mechanical treatment of the problem, the rates in (\ref{qm4})
in addition have to account for quantum tunneling ``through'' the potential
barriers. Especially, due to our assumption that
moderate-to-strong friction is acting, the tunneling dynamics is incoherent
and a quantum rate description of the current (\ref{qm4}) is valid.
To evaluate these rates, a sophisticated line of reasoning has been
elaborated \cite{han90,ben94,wei99a}. 
Starting with the Hamiltonian 
system-plus-reservoir model (\ref{qm1.1}) and adopting the so-called
``imaginary free energy method'' \cite{gra87,han90} 
or, equivalently, the
``multidimensional quantum transition state theory'' \cite{fri86a,hon88,han90},
it is possible to express the escape rate $k$
in terms of functional path integrals. After integration over
the bath modes and a steepest descent approximation,  
one obtains the semiclassical form
\begin{equation}
k=A\,e^{-S/\hbar}\ .
\label{qm7}
\end{equation}
Here, the exponentially dominating contribution $S$ is defined 
via the nonlocal action
\begin{equation}
S_b[q] := \int\limits_0^{\hbar/k_BT}\! \!d \tau \left[ \frac{m\dot q^2(\tau)}{2}+U(q(\tau))
+\frac{\eta}{4\pi}\!\int\limits_{-\infty}^{\infty}
\!\!d \tau'\left(\frac{q(\tau)-q(\tau')}{\tau-\tau'}\right)^2\right]\ .
\label{qm8}
\end{equation}
This action has to be extremized for paths $q(\tau)$ under the constraints that 
$q(\tau+\hbar/k_BT)=q(\tau)$
for all $\tau$, and that there exists a $\tau$-value such that $q(\tau)=x_b$.
A trivial such extremizing $q(\tau)$ is always $q(\tau)\equiv x_b$. Among this and
the possibly existing further extrema  one selects the one 
that minimizes $S_b[q]$, say $q_b(\tau)$, to obtain 
\begin{equation}
S := S_b[q_b]-\hbar\beta U(x_0) \ .
\label{qm8'}
\end{equation}
The pre-exponential factor $A$ in (\ref{qm7})
accounts for fluctuations about the semiclassically dominating
path $q_b(\tau)$.

For a numerical exemplification \cite{rei97,rei97a,rei98} we 
use $T$ as control parameter and fix the 
five remaining model parameters $m$, $\eta$, $V_0$, $F$, and $l:=L/2\pi$.
Without specifying a particular unit system this can be
achieved by prescribing the following five dimensionless numbers: 
First we fix $V_0$, $F$, $l$ and thus $U^\pm(x)$ through 
$Fl/V_0=0.2$, $\Delta U^{{\rm min}}/V_0=1.423$, and $|{U^+}''(x_b)|\, l^2/V_0=1.330$
corresponding to the situation depicted in \fig \ref{figqm1}. 
Next we choose $\eta/m\Omega_0=1$ with
$\Omega_0:=[V_0/l^2 m]^{1/2}$, meaning a moderate damping as 
compared to inertia effects. To see this we notice that $\Omega_0$  
approximates rather well the true ground state
frequency $\omega_0^+:=[{U^+}''(x_0)/m]^{1/2}$ in the potential
$U^+(x)$, $\omega_0^+=1.153\,\Omega_0$, and similarly for $U^-(x)$.
In particular, $\eta/m\Omega_0=1$ strongly forbids deterministically
running solutions. 
In order to specify our last dimensionless number we remark that 
within the weak noise assumption (\ref{qm3a})
it can be shown \cite{han90} that in the potential $U^+(x)$
genuine quantum tunneling events ``through'' the potential barrier
are rare above a so-called crossover temperature
\begin{equation}
T^+_c=\frac{\hbar\,\mu^+}{2\pi k_B}\ ,\qquad 
\label{qm8a}
\end{equation}
while for $T<T_c^+$ tunneling yields the dominant contribution to
the transition rates.
An analogous crossover temperature $T_c^-$ arises for the potential
$U^-(x)$ which is typically not identical but rather close to $T_c^+$.
With the definitions
\begin{equation}
T_c^{{\rm max}}=\max\{T_c^+,\, T_c^-\}\ ,\qquad T_c^{{\rm min}}=\min\{T_c^+,\, T_c^-\}
\label{qm8b}
\end{equation}
we now fix our last dimensionless quantity through
$\Delta U^{{\rm min}}/k_BT_c^{{\rm max}}=10$. In this way, the weak noise condition 
(\ref{qm3a}) is safely fulfilled for $T\leq 2 T_c^{{\rm max}}$, i.e., 
up to temperatures
well above both $T_c^+$ and $T_c^-$. At the same time, the so-called 
semiclassical condition \cite{han90} 
\begin{equation}
\Delta U^{{\rm min}}\gg k_B T_c^{{\rm max}} \ ,
\label{qm8c}
\end{equation}
can be taken for granted when evaluating the
quantum mechanical transition rates (\ref{qm7})
for all $T\leq 2 T_c^{{\rm max}}$. Specifically, the prefactor $A$ appearing in
(\ref{qm7}) can be evaluated within a saddle
point approximation scheme \cite{han90}
if the semiclassical condition (\ref{qm8c}) holds.
Moreover, the implicit assumption in (\ref{qm4}) that not only
thermally activated barrier crossings are rare (see (\ref{qm3a})) but 
also tunneling probabilities are small, is self-consistently
fulfilled if (\ref{qm8c}) holds.
For more details regarding the actual numerical calculation of those rates 
we refer to \cite{rei98}.

\figqmzwei

Representative results \cite{rei97,rei97a}
for the above specified quantum ratchet model
are depicted in \fig \ref{figqm2}. 
Shown are the current $\langle\dot x\rangle_{{\rm qm}}$ 
following from (\ref{qm4}) within the above sketched
quantum mechanical treatment of the rates according to (\ref{qm7})
together with the result $\langle\dot x\rangle_{{\rm cl}}$  
that one would obtain by means of a purely classical calculation
(formal classical limit $\hbar\to 0$)
according to (\ref{qm4a}).
The small dashed part in
$\langle\dot x\rangle_{{\rm qm}}$  
in a close vicinity of the crossover temperatures $T_c^{{\rm max}}$ and
$T_c^{{\rm min}}$ from (\ref{qm8b}) signifies an increased uncertainty of the 
semiclassical rate theory in this temperature domain.

Our first observation is that even above $T_c^{{\rm max}}$, quantum effects
may enhance the classical transport by more than a decade.
They become negligible, that is, the physical classical limit is approached,
only beyond several $T_c^{{\rm max}}$.
In other words, significant quantum corrections of the classically 
predicted particle current set in already well above the cross-over 
temperature $T_c^{{\rm max}}$, where tunneling processes are still rare.
(They can be associated to quantum effects other than genuine tunneling
``through'' a potential barrier.)
With decreasing temperature, $T<T_c^{{\rm min}}$,
quantum transport is even much more enhanced in comparison with
the classical results.
The most remarkable feature caused by the intriguing interplay
between thermal noise and quantum tunneling is the inversion of 
the quantum current direction at very low temperatures
\cite{rei97,rei97a,jor97,rei98,han99,bro00,lin00,wir00}.
Working within a formal a classical limit ($\hbar\to 0$),
such a reversal for adiabatically slow driving 
is ruled out. Finally, $\langle\dot x\rangle_{{\rm qm}}$ approaches
a finite (negative) limit when $T\to 0$, implying a finite
(positive) stopping force\footnote{Recalling the
definition from  \sect \ref{sec2.2.2.2}, the stopping force is that
external force $F$ in (\ref{4a}) which leads to a cancellation of
the ratchet effect, i.e. $\langle\dot x\rangle =0$.} also at $T=0$.
In contrast, the classical prediction
$\langle\dot x\rangle_{{\rm cl}}$  remains positive but becomes arbitrarily 
small with decreasing $T$. 
A curious detail in \fig \ref{figqm2} is the non-monotonicity of 
$\langle\dot x\rangle_{{\rm qm}}$ around
$T_c^{{\rm max}}/T\simeq 2.5$, caused via (\ref{qm4})
by a similar resonance-like
$T$-dependence in the prefactor $A$ of one of the underlying quantum 
mechanical transition rates (\ref{qm7}).

\subsection{Tunneling induced current inversion}\label{sec6.5.3}
The most remarkable result of the preceding subsection
(see also \fig \ref{figqm2}) is the inversion of the
current upon decreasing the temperature.
On the other hand, within the formal classical limit ($\hbar\to 0$)
the current never changes its direction.
Since at high temperatures the physical classical limit is approached,
i.e. the formal classical limit provides a more and more accurate
approximation for the true physical system,
{\em the temperature controlled current inversion represents a
new signature of genuine quantum mechanical effects}.
In the following we provide a simple heuristic explanation 
of this finding \cite{rei97,rei98,bro00}.

As a first simplification, we exclusively focus in the exponentially
leading contribution in the semiclassical rate expression (\ref{qm7}),
i.e. the sign of the current in (\ref{qm4})
is given by that of $S_{l}^- - S_{r}^+$.

For sufficiently large temperatures, quantum mechanical effects become
negligible and the exponentially leading part in (\ref{qm7}) goes over 
into that of (\ref{qm4a}).
Indeed, one can show \cite{han90,wei99a} 
that for $T>T_c^{{\rm max}}$ only the
trivial extremizing paths $q(\tau)\equiv x_b$ in (\ref{qm8}) exist
for both potentials $U^\pm(x)$, and thus we recover with (\ref{qm8'}) 
that
$S_{r}^+/\hbar=\Delta U_r^+/k_BT$ and 
$S_{l}^-/\hbar=\Delta U_l^-/k_BT$.
In other words, the lower of the two barriers 
$\Delta U_r^+$ and $\Delta U_l^-$
determines the direction of the current.

A second case for which the extremization of the action (\ref{qm8})
can be readily carried out is the combined limit $T\to 0$ and
$\eta\to 0$ (no heat bath), resulting in the familiar Gamow formula
for the exponentially leading tunneling contribution in (\ref{qm8'}),
namely
\begin{equation}
S=2 \sqrt{2\, m} \, \left|\int_{x_0}^{x_1}dq \, 
[ U(q)-U(x_0) ]^{1/2}\right|\ .
\label{gamo}
\end{equation}
As before, $x_0$ denotes a local minimum of $U(x)$ and $x_1$ is the first
point beyond the considered potential barrier
with the property that $U(x_1)=U(x_0)$.
The absolute value in (\ref{gamo})
is needed since $x_1<x_0$ for the escapes
to the left, i.e., across
$\Delta U^-_l$.
Thus, the smaller of the two Gamow-factors
$S_{r}^+$ and $S_{l}^-$ determines the direction of the current.
Strictly speaking, by letting $\eta\to 0$ we of course
violate our previously made assumption that deterministically 
running solutions should be ruled out. 
However, it is plausible that small but finite $\eta$ will exist
for which our qualitative arguments can be adapted self-consistently.

From \fig \ref{figqm1} one can see by naked eye that
the activation energy barrier $\Delta U_r^+$ to proceed
in the potential $U^+(x)$ from one local minimum to the 
neighboring local minimum to the right
is smaller than the corresponding
barrier $\Delta U_l^-$. Hence the current is positive for
sufficiently large temperatures.
In contrast, the fact that $S_{r}^+$ is larger than $S_{l}^-$
cannot definitely be read off by eye directly from \fig \ref{figqm1} 
since the two quantities are rather similar, but it can be readily
verified numerically. In other words, for very small $T$ indeed a 
negative current is predicted.
A change of sign of the current at some intermediate temperature 
is thus a necessary consequence.

\figqmdrei

Things become even more obvious by considering instead of the
smooth potential from \fig \ref{figqm1} a stilized sawtooth
profile\footnote{For such a singular potential shape the
$\mu$-factor in the crossover temperature (\ref{qm8a})
is no longer given by (\ref{qm4a'}). Instead of changing
the definition of $\mu$, one may also slightly smoothen out
the singularities of the potential.}
as sketched in \fig \ref{figqm3}.
Focusing on the local minimum $x_0 =0$, the fact that 
$\Delta U_r^+<\Delta U_l^-$ is 
read off immediately from
\fig \ref{figqm3}.
Denoting by $\lambda:=|x_1-x_0|$ the ``tunneling-length'',
the Gamow-factor (\ref{gamo}) takes the simple form
\begin{equation}
S =\frac{4}{3}\, \sqrt{2\, m\, \Delta U} \, \lambda\ ,
\label{gamo'}
\end{equation}
where indices $r$, $l$, and $\pm$ have been omitted as usual.
From \fig \ref{figqm3} one reads off that $\lambda_+=L$,
$\lambda_- = L/(1+LF/V_0)$, $\Delta U^-_l=V_0$, and
$\Delta U^+_r=V_0\, (1 - LF/V_0)$, 
where we have assumed without loss of generality
that $0\leq F\leq V_0/L$. It readily follows that
for small-to-moderate tilting forces 
$F\in [0,\, 0.618\, V_0/L]$
we have that $S_{r}^+>S_{l}^-$
and the current is therefore negative.

In conclusion, the basic physical mechanism
behind the opposite sign of the current at high and
low temperatures is apparently rather simple and robust,
suggesting that this feature should be very common in tilting
quantum ratchet systems.
Since the decrease of temperature is accompanied by a
transition from thermally activated to tunneling dominated transport, 
the concomitant change of the transport direction may be considered
as tunneling induced current inversion, see \sect \ref{sec6.5.5}.

\section{Beyond the adiabatic limit}\label{sec6.5.4}
For an non-adiabatic tilting force $y(t)$ in (\ref{qm1.1})
the determination of the average particle current is in 
general very difficult.
An approximative analytical approach becomes possible within
a so-called {\em tight-binding model} description.
The starting point consists in the observation that the first two terms on the
right hand side of (\ref{qm1.1}) define a time-independent
particle dynamics in a periodic potential and can thus be treated within the
standard Bloch-theory
for independent (quasi-)particles in a one-dimensional lattice \cite{ash76}.
Under the assumptions that both the external tilting force $y(t)$ and the
thermal fluctuations of the environment, entering through the last two terms
in (\ref{qm1.1}), are sufficiently weak, one can focus 
on a {\em single-band truncation}
of the problem, i.e., the Hilbert-space accessible to the particle is 
spanned solely by the Bloch-states of the lowest energy-band.
Especially, both the thermal energy $k_BT$ and the energy $\hbar\omega_c$
associated to the cutoff in (\ref{qm1.5}) have to be restricted to
values much smaller than the excitation energy into the second band
(or the continuum).
Upon going over from these Bloch-states of the lowest band
to a new basis $\{ | n \rangle \}_{n=-\infty}^\infty$
of so-called localized- or Wannier-states \cite{ash76},
the truncated model Hamiltonian (\ref{qm1.1}) takes the standard single-band
tight-binding form \cite{hol96}
\begin{equation}
{\bf H} (t) = - \frac{\hbar\Delta}{2}\sum_{n=-\infty}^\infty
\left( |n\rangle \langle n+1| + |n+1\rangle \langle n| \right)  - {\bf x}\, y(t)+ {\bf H}_B \ ,
\label{tb1}
\end{equation}
where both in (\ref{tb1}) and (\ref{qm1.2}) the operator ${\bf x}$
is defined as
\begin{equation}
{\bf x} := L\, \sum_{n=-\infty}^\infty n\, |n\rangle \langle n| \ .
\label{tb2}
\end{equation}
The quantity $\hbar\Delta$ in (\ref{tb1}) is the so-called tunneling coupling 
energy between neighboring potential minima. In principle, its explicit
value can be determined from the Bloch-states and the potential $V(x)$ 
\cite{ash76}.
Alternatively, the tunneling coupling energy
may be considered as an adjustable model parameter. An additional 
approximation implicit in (\ref{tb1}) is the assumption that
only tunneling between neighboring potential minima of $V(x)$
plays an appreciable role. In other words, so-called
coherent tunneling (co-tunneling) is neglected.

By construction, the single-band tight-binding model (\ref{tb1}) cannot
capture thermally activated transport across the energy barriers
between neighboring minima of $V(x)$; its validity
is restricted to quantum mechanical tunneling processes at low
energies and temperatures.
Furthermore, the model does not exhibit any traces of a possible
asymmetry in the periodic potential $V(x)$.
One is therefore restricted to effectively symmetric
potentials $V(x)$ and a ratchet effect may only be studied
within an asymmetrically tilting ratchet scheme (see \sect \ref{sec6.3}).
Besides these restrictions, the tight-binding model also goes beyond
the approach from \sects \ref{sec6.5.2}, \ref{sec6.5.3} in that
the semiclassical condition (\ref{qm8a}) is not required
and the tilting force $y(t)$ needs not be adiabatically slow (see below).
In this sense, the approaches from 
\sects \ref{sec6.5.2}, \ref{sec6.5.3} 
and of the present \sect are complementary.

A non-adiabatically tilting quantum ratchet within the above single-band
tight-binding approximation has been considered in \cite{goy98} for
a rather general class of unbiased, asymmetric {\em random} drivings
$y(t)$, including asymmetric dichotomous noise as a special case.
In the absence of the heat bath ${\bf H}_B$ in (\ref{tb1}), the
average particle current is found to vanish in all cases
(for the same model, but with a periodic driving $y(t)$, see also
\cite{goy00,goy01}).
In the presence of the heat bath, the occurrence of a finite current
is generically observed. Current inversions upon variation of
different model parameters are also reported. Especially, such
an inversion may occur when the temperature is changed, which,
for reasons detailed above, cannot be explained by the
heuristic argument from \sect \ref{sec6.5.4} and thus
represent a genuine feature of the non-adiabatic driving.
Regarding a more detailed discussion of the effective
diffusion coefficient (\ref{4c2})
within this model we refer to the original paper \cite{goy98}.

The same model, but with an asymmetric {\em periodic} driving
$y(t)$ of the harmonic mixing form (\ref{asy3}) has been
addressed in \cite{goy98b}. The emerging quantum current
exhibits multiple reversals, characteristic for the non-adiabatic
nature of the driving, and a stochastic resonance-like,
bell-shaped behavior upon variation of the temperature.
Via {\em control of the phase} and the amplitudes of the driving
signal (\ref{asy3}) it is furthermore possible to
selectively control the magnitude of both the quantum current
and diffusion, as well as the current direction.
For further theoretical and experimental works along
related lines see \cite{shm85,ent89,ata96,hac97,ale99,goy00,goy01}
and references therein\footnote{Closer inspection indicates 
\cite{goy00,goyp} that the conclusions from \cite{ale99} 
in the case of a dissipationless (collisionless) 
single-band model 
are at most valid for very special (non-generic) initial 
conditions.}.

While the rich behavior of the single-band tight-binding
ratchet model can be obtained by means of sophisticated 
analytical approximations \cite{goy98,goy98b} which go
beyond our present scope, simple intuitive explanations
can usually not be given.

{\em Generalizations} of the single-band tight-binding model (\ref{tb1})
have been addressed in \cite{yuk97,ron98,yuk00}.
The main new ingredient is an extra ``potential''-term ${\bf H}_V$
of the form
\begin{equation}
{\bf H}_{V} = \sum_{n=-\infty}^\infty 
|n\rangle \langle n|\, V_{n\,\mbox{mod}\, N}[1+f(t)]
\label{tb3}
\end{equation}
on the right hand side of (\ref{tb1}), reminiscent of a
spatially discretized, asymmetric ratchet potential with period 
$N\geq 3$. 
The case with $y(t)\equiv 0$ in (\ref{tb1}), corresponding to a
fluctuating potential ratchet, has been treated in \cite{yuk97,yuk00}.
The opposite case with $f(t)\equiv 0$ but again with an adiabatically
slow, symmetric rocking force $y(t)$ has been addressed 
in \cite{ron98}.
At first glance, 
such an extra term (\ref{tb3}) in order to upgrade the single-band
tight-binding model (\ref{tb1}) into a veritable
ratchet system with a broken spatial symmetry is indeed suggestive.
However, the present author was only able to figure out {\em very artificial}
actual physical situations which may be captured by such a
model with a non-trivial term (\ref{tb3}).
In particular, 
due to the nearest neighbor hopping term in (\ref{tb1}),
such a description clearly cannot properly account for more than
one single band.

Further quantum ratchet works, some of which go
beyond the adiabatic limit, and which are not 
based on the single-band tight-binding approximation (\ref{tb1})
of the original dynamics (\ref{qm1.1}), are \cite{tat98,bao98a,bao98b,yuk00}.
However, the present author finds these studies
questionable with respect to the conceptual basis
and/or the technical methodology.
A quantum Smoluchowski-Feynman-type model (equilibrium system) has been
investigated on the basis of a widely used standard approximation 
in \cite{cap99} and further references therein.
The observed appearance of a ratchet effect in such an equilibrium model
underlines once more the warning at the beginning of
\sect \ref{sec6.5.1} that
even well-established {\em ad hoc} approximations for a quantum thermal environment
may easily lead to inconsistencies with fundamental principles
of statistical mechanics.

For a periodic driving $y(t)$ and
in the absence of the heat bath ${\bf H}_B$ in (\ref{qm1.1})
the quantum mechanical counterpart of the Hamiltonian rocking
ratchet model from \eq (\ref{ham1}) in
\sect \ref{sec6.2.3.2} is recovered.
Within a single-band tight-binding approximation (\ref{tb1})
this type of model has been solved in closed analytical form
in \cite{goy98,goy00,goy01}.
Though the chaotic features of the classical counterpart
cannot be captured in this way, 
a dependence of the current on the initial conditions is found 
\cite{goy00,goy01}
which is quite similar to the classical results from \cite{yev00},
while strongly non-classical features \cite{goy98} arise in the presence
of a finite static tilt $F$ (i.e. $y(t)$ in (\ref{tb1}) is replaced by $y(t)+F$).
A first step into the direction of a chaotic (Hamiltonian)
quantum ratchet system has been taken in \cite{dit00,sch01} and \cite{lin01a}
with the main focus on the semiclassical regime and on mesoscopic electron
billiard devices, respectively, see also \sect \ref{sec6.2.3.2} for the 
classical limit.

\section{Experimental quantum ratchet systems}\label{sec6.5.5}
As a first candidate for an experimental realization of a quantum
ratchet we consider the {\em SQUID rocking ratchet} model 
\cite{zap96} from \eq (\ref{j2'}).
As argued at the beginning of this chapter,
this stochastic dynamics (\ref{j2'}) as it stands represents a classical
ratchet system \cite{cal83}.
The question of how to properly ``quantize'' such a ``classical''
dynamics, which itself arises as an effective description of
characteristic quantum effects, has been discussed extensively in the
literature, see \cite{cal83,sch90} and references therein.
Leaving aside devices which contain ultra small tunnel junctions,
ample theoretical \cite{amb82,lar83,eck84,sch90} as well as experimental
(see references in \cite{han90}) justification has been given
that, after proper renaming of symbols, \eqs (\ref{qm1.1}), (\ref{qm1.5})
provide the basis for an adequate quantum mechanical extension of the 
classical model (\ref{j2'}) when the temperature is decreased below a 
few times $T_c^{{\rm max}}$ from (\ref{qm8b}).
Conceptionally, it is interesting to note \cite{cal83}
that we are dealing here with quantum effects which manifest 
themselves via the {\em macroscopic} phase-variable $\varphi$.
In other words, the observation of
transport properties characteristic for a quantum ratchet is
not necessarily restricted to the realm of microscopic systems.
So far, an actual experimental realization of
a SQUID ratchet system \cite{wae67,wae69,wei99,wei00}
is only available for the two-dimensional modification 
(\ref{j7'}), (\ref{j8'}) of the archetypal rocking ratchet 
setup (\ref{j2'}).
While the experiment from \cite{wei99,wei00} works with high-$T_c$ SQUIDS 
at temperatures too large to see any traces of quantum mechanical tunneling
of the phase $\varphi$, an analogous experiment with conventional 
superconductors is presently under construction, 
with the intention to reveal such quantum mechanical effects.

A second potential realization of a tilting quantum ratchet system is based
on the motion of 
{\em ultracold atoms in the presence of standing electromagnetic waves}, 
creating a ratchet potential through the radiation-pressure forces
of the counterpropagating light beams \cite{let77,hem93,gry93,pre93,wil96}.
For sufficiently weak potentials and low temperatures, quantum effects will 
clearly play a dominant role in the atomic motion and may be roughly captured by
a model like in (\ref{qm1.1}). Especially, the tilting force may be created
by exploiting the mapping of the model onto an improper traveling potential
ratchet from \sect \ref{sec4.3.2}. The corresponding accelerating optical
potentials have been experimentally realized e.g. in \cite{wil96}.
As detailed in \sect \ref{sec4.4.2}, a somewhat
related system has indeed been experimentally studied in \cite{men99}.
Due to the remaining considerable differences between this
system and the theoretical model (\ref{qm1.1}), a direct comparison
is, however, not possible.

A third promising class of experimental tilting quantum ratchet 
devices are {\em semiconductor heterostructures}.
The lacking periodicity of a single {\em diode} (n-p juction)
can be readily remedied, in the simplest case by 
connecting identical diodes by normal conducting wires.
Similarly as in the above discussed case of SQUID ratchets,
such a simple array of diodes realizes a 
classical ratchet system in so far as (at the usual
working temperatures) the essential
transport processes across the junctions are governed
by classical thermal diffusion rather than quantum 
mechanical tunneling (see also \sects \ref{sec2.2.3c},
\ref{sec6.2.1}, and \ref{sec4.8.1}).
Closely related devices are spatially
periodic {\em semiconductor superlattices}.
Examples with broken spatial symmetry 
(so-called {\em sawtooth superlattices})
have been experimentally realized since long \cite{all80,car83}
but have never been studied so far from the viewpoint
of the ratchet effect.
Heterostructures consisting of alternating layers of
GaAs and AlGaAs in quantum mechanically dominated temperature regimes
have been experimentally explored e.g. in \cite{ign94,kea95a,kea95b}.
The motion of a (quasi-) particle (dressed electron) in such a superlattice
may be roughly described by an effective, one-dimensional model
of the form (\ref{qm1.1}), where the heat bath takes into account
the effects of the crystal phonons 
\cite{kea95a,kea95b,goy98,goy98b}.
In the simplest case of a semiconductor superlattice
with only two different alternating layers, a {\em symmetric} periodic potential
$V(x)$ in (\ref{qm1.1}) arises, thus the asymmetrically tilting quantum ratchet
scheme from \sect \ref{sec6.5.4} has to be employed.
The quantitative estimates from \cite{goy98b} furthermore show that the one-band
tight-binding model (\ref{tb1}) may be a valid approximation for a 
typical experimental setting \cite{ign94,kea95a,kea95b}.
Moreover, if the driving $y(t)$ is provided by the usual 
electromagnetic waves in the THz-regime,
one is indeed dealing with the non-adiabatic 
regime from \sect \ref{sec6.5.4}.
On the other hand, the semiclassical theory from \sect \ref{sec6.5.2} cannot
be applied to such an experimental situation not only because
the driving is not adiabatically slow, but also since the semiclassical 
condition (\ref{qm8a}) is typically not satisfied.

Important progress towards an adiabatically rocking quantum 
ratchet in one-dimensional {\em Josephson junction arrays}, 
consisting of three ``cells'' (effective periods) with 
broken spatial symmetry, 
has been achieved very recently in \cite{maj01}: somewhat similar as
in the systems from \sect \ref{sec6.2.3.1}, but operating in the
quantum mechanical regime, the voltage due to the dynamical
response of the vortices (directed transport of quasi particles)
against an applied bias current
exhibits an asymmetry when the sign of this bias is inverted,
cf. \fig \ref{fig4}.
In particular, the theoretically expected asymptotic 
temperature independence of the effect in the deep quantum 
cold is experimentally recovered.

A molecular rectifier for electrons, combining the quantum ratchet
with the {\em Coulomb blockade effect}, has been proposed in \cite{sto01}.
For additional experiments which may be considered to some extent
as quantum ratchet systems we also refer to the applications
of the genuine traveling potential ratchet scheme discussed in
\sect \ref{sec4.3.1}. 

We close this section with the experimental
realization of an adiabatically rocking quantum ratchet 
by Linke and colleagues on the basis of a 
{\em quantum dot array with broken spatial symmetry}.
We skip the preliminary experiment on ac-driven electron transport 
through a single triangular shaped quantum dot 
\cite{lin98,lin99b,lin99c,lin99d,han99,rau99}
(see also \cite{ven96})
and immediately turn to the exploration of an entire array
of such triangular dots \cite{lin99a,hum99,bro00,lin00,wir00,lin01b}.

\figqmvier

The basic setup \cite{lin99a} is depicted in \fig \ref{figqm4}:
A two-dimensional conducting electron gas is constricted by two 
insulating boundary-regions (dark areas in \fig \ref{figqm4}).
In other words, the ``conducting channel'' along the $x$-axis
is laterally confined to a width of about $1\mu m$.
Roughly speaking, the corresponding lateral
confinement energy creates an effective
ratchet-shaped potential $V(x)$ for the particle dynamics along the $x$-axis
of a qualitatively similar character as in \fig \ref{figqm1}.
The two ``side gates'' in \fig \ref{figqm4} allow one to externally modify
this effective potential by putting them on different electrical potentials.
The actual rocking force $y(t)$ is created by applying an ac-voltage along the
$x$-axis, periodically switching between
the two values $\pm F$, with a typical voltage $F$ of about $1 mV$. 
The driving frequency of 191Hz used in the experiment is definitely deep
within the adiabatic regime.
The ``bottlenecks'' which connect neighboring triangles in \fig \ref{figqm4}
are chosen such that quantum tunneling dominates at low 
temperatures, while for higher temperatures the conduction
electrons can also substantially proceed by way of thermal activation
across the corresponding effective potential barriers.

\figqmfunf

The measured \cite{lin99a}
current through the quantum dot array as a function of
temperature is exemplified in \fig \ref{figqm5}.
The two main features theoretically predicted in \sect \ref{sec6.5.2}
are clearly reproduced, namely a current inversion upon decreasing
the temperature and a saturation of the current as temperature approaches
absolute zero\footnote{Note that \fig \ref{figqm2} is an Arrhenius plot
($\log \langle\dot x\rangle$ versus $1/T$), while \fig \ref{figqm5} depicts
the bare quantities ``electrical current'' versus ``temperature''.}.
Though a model along the lines of (\ref{qm1.1}) is obviously a very crude
description of the experimental situation, the basic features and thus
the heuristic explanation of the current inversion from \sect \ref{sec6.5.3},
are apparently still qualitatively correct.
Note that also the direction of the current is in agreement with
\fig \ref{figqm2}  by taking into account that the relevant effective
potential for the experiment is of the same qualitative shape as in
\fig \ref{figqm1} and that the electrical current is opposite to the
particle current for the negatively charged electrons.
For a somewhat more realistic theoretical model, which also reproduces 
the main qualitative features of the experiment, we refer to \cite{lin99a,hum99}.

Based on the observation from \sect \ref{sec6.5.3}, namely that ``cold''
and ``hot'' particles move in opposite directions (as long as their 
individual ``temperatures'' (kinetic energy)
change sufficiently little), an interesting 
idea is \cite{hum01a,hum01b} to apply the above quantum rocking
ratchet setup in the absence of a net particle transport, 
i.e. operating at the current inversion point,
for ``cooling'' 
purposes\footnote{Of foremost interest in 
this context are ``single-period setups''
(e.g. a single triangular quantum dot) 
in contact with two electron-reservoirs at 
either the same or different temperatures.}.
However, the quantitative analysis of the 
experiment in \cite{lin99a} shows that the 
heating due to the external rocking force exceeds the
cooling effect due to the above separation of 
particles with different temperatures \cite{hum01a}.
On the other hand, assuming the existence of ``ideal filters'',
which let pass in both directions only particles with one specific 
energy, it is possible to modify the original setup 
such that it can act either as refrigerator or as heat engine
arbitrarily close to the maximal Carnot efficiency \cite{hum01b}.
In contrast to the standard framework for 
considerations on the efficiency of particle
transport from \sect \ref{sec3.8}, here a zero particle current situation 
is addressed, and the relevant mechanical work is now associated with
the external driving force. In other words, the particle motion
is now considered as an internal part of the engine under 
consideration, and no longer as the resulting effect of the 
engine.

\chapter{Collective effects}\label{cha7}
At the focus of this chapter are collective effects that arise when several 
copies of ``single'' classical ratchet systems, 
as considered {\em in extenso} in the
previous chapters \ref{cha2}-\ref{cha06}, start to interact with each other.

Accordingly, the general working model (\ref{4a}) goes over into
$N$ coupled stochastic differential equations of the form
\begin{equation}
\eta\,\dot x_i(t) = -V'(x_i(t),f_i(t)) + y_i(t) + F + \xi_i(t) 
- \frac{\partial \psi (x_1(t),...,x_N(t))}{\partial x_i} 
\label{7.1}
\end{equation}
with $i=1,...,N$.
The last term accounts for the interaction through an interaction
potential $\psi$ which is
assumed to be spatially homogeneous and inversion symmetric,
i.e. no preferential direction is 
introduced through the interaction.
The assumption of a thermal equilibrium environment implies that the
thermal noises $\xi_i(t)$ are mutually independent 
Gaussian white noises with correlation
\begin{equation}
\langle \xi_i(t)\xi_j(s)\rangle 
= 2\, \eta\, k_B T \, \delta_{ij}\, \delta (t-s) \ .
\label{7.1a}
\end{equation}
The drivings $f_i(t)$ and/or $y_i(t)$ are usually assumed
to be either mutually independent random processes or equal to
the same periodic function for all $i$.
In any case, these drivings as well as the interaction
$\psi$ in (\ref{7.1}) have to respect the equivalence of all the ``single
particles'' $i$.
Therefore, the average particle current $\langle\dot x\rangle$ from
(\ref{4c1}) will be the same for each particle $i$ and consequently
independent of whether or not an additional average over $i$ 
is performed.
The case of foremost interest is usually the zero 
load ($F=0$) situation, but also
the response when a finite $F$ is acting will lead to
quite remarkable observations in \sect \ref{sec7.2}.

We recall that various examples with $N=2$ interacting systems (\ref{7.1})
have been discussed already in 
\sects \ref{sec6.2.4}, \ref{sec5.7}, and \ref{sec5.6}.
In the present chapter, our main interest will concern collective
effects in the case of a large number $N\to\infty$ of interacting
systems\footnote{A deterministic
collective model which does not fit into this general
framework is due to \cite{por00a,por01}: It works for $N\geq 3$ particles
with finite mass $m$ in a static, not necessarily asymmetric potential 
$V(x)$. A worm-like deterministic motion is generated by active changes of
the interaction in a wave-like manner along the chain of particles
$x_i(t)$.
A similar model with non-Newtonian interaction forces (actio $\not =$ reactio)
is due to \cite{zhe01}.
Moreover, reaction-diffusion model for interacting Brownian motors 
has been discussed in \sect \ref{sec5.7}.} (thermodynamic limit).
In doing so, two basic types of questions can be addressed.
First, one may consider cases for which already in the absence
of the interaction in (\ref{7.1}) each single system exhibits a 
ratchet effect, i.e. both, thermal equilibrium and spatial symmetry 
are broken.
In such a case, one may study the modification of 
the current $\langle\dot x\rangle$
in magnitude and possibly even in sign when the interaction is included.
A survey of such explorations will be presented in \sect \ref{sec7.1}.

A second type of questions regards {\em genuine collective effects}, namely
spontaneous ergodicity breaking, entailing phase transitions,
the coexistence of different (meta-) stable
phases, and hysteretic behavior in response
to the variation of certain parameters.
While all these collective phenomena are well known also in
equilibrium systems, the second law of
thermodynamics precludes a finite particle current
in such systems even if their spatial symmetry is broken.
We thus focus on interacting systems (\ref{7.1})
out of equilibrium, which for the usual interactions is the case
if and only if already the uncoupled
systems (\ref{7.1}) are out of equilibrium.
Such genuine non-equilibrium collective effects have already been encountered
in the context of the {\em Huxley-J\"ulicher-Prost model}
for cooperating molecular motors in \sect \ref{sec5.4.4}.
There, the main emphasize was put on systems with a built-in
spatial asymmetry already of the single (uncoupled) systems in (\ref{7.1}),
which is then inherited by the coupled model.
In contrast, in \sect \ref{sec7.2} we will
address coupled non-equilibrium systems (\ref{7.1})
which are fully symmetric under spatial inversion.
The essential idea is then that instead of a built-in asymmetry,
a perfectly symmetric system may create the asymmetry,
which is necessary for the manifestation of a ratchet effect,
by itself, namely through spontaneous symmetry breaking.
While the occurrence of such a ``spontaneous current'' has been
pointed out already for a spatially symmetric special case of the
J\"ulicher-Prost model in \cite{jul95},
we will focus in \sect \ref{sec7.2}
on a simpler model which admits a partial analytical treatment and
exhibits additional, quite remarkable collective non-equilibrium 
features.

We close with two remarks:
First, the 
subject under study in this chapter 
is intimately related with many 
other topics, like for instance
non-equilibrium phase transitions, 
reaction-diffusion systems,
pattern formation,
driven diffusive systems,
Frenkel-Kontorova models,
Josephson junction arrays,
sine-Gordon equations,
and coupled phase oscillators.
A detailed discussion of any of these adjacent
topics goes, however, beyond the scope of
our present review.
Second, while we feel that very interesting and unexpected
theoretical discoveries 
are still to come, 
on the experimental side the field is even more so 
at a very underdeveloped stage.

\section{Coupled ratchets}\label{sec7.1}
In this section we review investigations of $N\to\infty$ coupled
ratchet systems in the case that each single particle 
$i$ exhibits a ratchet effects already in the absence of the interaction
in (\ref{7.1}).
For related discussions of models for cooperative molecular motors we also
refer to \sect \ref{sec5.4}.
Throughout this section, we restrict ourselves to the case
$F=0$ in (\ref{7.1}) and to potentials $V(x)$ with a broken symmetry,
i.e. ratchet potentials, as exemplified by \figs \ref{fig2} and 
\ref{fig4.1}.

The case of {\em interacting rocking ratchets}
\begin{equation}
\eta\,\dot x_i(t) = -V'(x_i(t)) + y (t) + \xi_i(t) 
+ I_b (x_{i+1}(t),x_{i}(t),x_{i-1}(t))
\label{7.2}
\end{equation}
with a hard core repulsive interaction $I_b$ such as to guarantee 
$x_{i+1}(t)>x_i(t)+b$ for all $i$ and $t$ has been explored in \cite{der95}.
Pictorially speaking, all particles are thus moving in
the same one-dimensional, periodically rocked ratchet potential $V(x)-xy(t)$
and they have a finite extension $b$ which sets a lower limit for
their mutual distance.
The central (numerical) finding in \cite{der95} is a current
inversion upon variation of the average density of particles along the
$x$-axis. This inversion is robust against various modifications,
especially of the driving $y(t)$ in (\ref{7.2})
[e.g. stochastic instead of periodic, or with small, $i$-dependent variations
of the driving-period $\ttt$] and implies according to \sect \ref{sec3.5}
analogous inversions upon variation of practically any other parameter of 
the model (\ref{7.2}).
For adiabatically slow driving $y(t)$ and simultaneously almost densely
packed particles, an analytical treatment is possible,
revealing an extremely complex dependence of the current upon the
particle extension $b$.
Somewhat similarly as in the J\"ulcher-Prost model \cite{jul95},
also in the present case the magnitude of the current $\langle\dot x \rangle$
depends sensitively on whether the spatial period $L$ is
commensurate or not with the average interparticle distance
$\langle x_{i+1}(t)-x_i(t)\rangle$.
Such effects may become practically relevant for separating
particles at high densities e.g. according to the
drift ratchet scheme from \sect \ref{sec6.4}.
A related, spatially discrete model with an adiabatically slow driving
has been considered in \cite{keh00},
thus establishing contact with the methods and concepts of so-called
driven diffusive systems \cite{sch95}.

A second basic model consists of a chain of 
{\em linearly coupled fluctuating force ratchets}
\begin{equation}
\eta\,\dot x_i (t) = -V'(x_i(t)) + y_i(t) + \xi_i(t) 
+ \kappa\, [x_{i+1}(t) - 2 x_{i}(t) + x_{i-1}(t)] \ ,
\label{7.2a}
\end{equation}
where $\kappa$ is the spring constant (interaction strength) and 
$y_i(t)$ are independent Ornstein-Uhlenbeck noise sources 
(cf. \eqs (\ref{6l}), (\ref{6m})).
In the continuum limit, one obtains a sine-Gordon type model,
which has been analyzed by means of the sophisticated analytical machinery
in this field in \cite{mar96}.
The main result is the appearance of a
ratchet effect in the form of a stationary directed transport
of kinks and antikinks in opposite directions.
As a rule, the kink and hence the entire particle chain move into the
same direction as in the uncoupled, Ornstein-Uhlenbeck noise driven
fluctuating force ratchet (cf. \sect  \ref{sec6.1}),
however with a highly non-trivial modification of the
quantitative behavior of the current.
Similar results for models of the type (\ref{7.2a}) have been reached also
in \cite{csa97,sav97a,sav97b,zol99} and for an analogous coupled
temperature ratchet model in \cite{sak98}.
Possible applications include the dynamics
of dislocations in solids,
solitonic fluxes in long Josephson junction arrays
and magnetically ordered crystals, and
models for friction and stick-slip motion such as the
Frenkel-Kontorova model.
For related studies in the context of
coupled Josephson junction arrays see also \sect \ref{sec6.2.3.1}.

Next we turn to the {\em interacting on-off ratchet} counterpart of
(\ref{7.2}), i.e. 
\begin{equation}
\eta\,\dot x_i(t) = -V'(x_i(t))\, [1+f_i(t)]  + \xi_i(t) 
+ I_b (x_{i+1}(t),x_{i}(t),x_{i-1}(t))
\label{7.3}
\end{equation}
with $f_i(t)\in\{\pm 1\}$.
In this case \cite{der96z}, the direction of the particle current
$\langle\dot x\rangle$ may even change many times as the density
of particles is varied. For high particle densities and slow on-off 
cycles, an extremely complex dependence of $\langle\dot x\rangle$
on the particle size $b$ similarly as for the model (\ref{7.2})
is recovered.
Such effects clearly become relevant for the various experiments
from \sect \ref{sec4.1.1} at high particle densities.
An {\em experiment} which may be considered to some extent as related
to the theoretical model (\ref{7.3}) has been realized in \cite{der98c,far99}.
In this work, the horizontal transport of granular particles in a vertically
vibrated system, whose base has a ratchet-shaped profile,
has been measured\footnote{A computer animation (Java applet) which 
graphically visualizes a somewhat related effect is available 
on the internet under \cite{rapmovie}.}.
The resulting material flow exhibits current
inversions and other complex collective behavior as a function
of the particle density and the driving frequency,
displaying a rough qualitative similarity with 
the theoretical model (\ref{7.3}).

A coupled rocking ratchet model, but in contrast to (\ref{7.2}) 
{\em with a global, Kuramoto-type interaction} \cite{kur75,kur84,str94}
with the same period $L$ as the ratchet potential $V(x)$, i.e.
\begin{equation}
\eta\,\dot x_i(t) = -V'(x_i(t)) + y (t) + \xi_i(t) 
+ \frac{K}{N}\sum\limits_{j=1}^N \sin\left(\frac{2\pi}{L}\,
[x_{j}(t) - x_{i}(t)]\right)
\label{7.4}
\end{equation}
has been addressed in \cite{hau97}.
Upon increasing the coupling strength $K$, the current may change 
direction and moreover the effect of the noise becomes weaker and weaker:
For $K\to\infty$ all particles in (\ref{7.4}) are lumped (modulo $L$)
into one single effective ``superparticle'' subjected to an effectively
{\em deterministic} single-particle rocking ratchet dynamics like in
\sect \ref{sec6.2}.
The existence of current inversions upon variation of other model parameters
than the coupling strength immediately
follows from \sect \ref{sec3.5}.
Considering that a single particle ($N=1$) rocking ratchet can be realized 
by means of three Josephson junctions (see \eq (\ref{j2'})),
the coupled model (\ref{7.4})
may well be of relevance for
Josephson junction arrays\footnote{A nearest neighbor instead of the 
global coupling in (\ref{7.4}) may then be a more realistic choice.
Such a modification is, 
however, not expected to change the basic 
qualitative features of the model (at least in $d\geq 2$ dimensions),
see also \figs \ref{fig7.2} and \ref{fig7.3} below.} 
\cite{mar87}.

Universal properties of particle density fluctuations at long
wavelengths and times for a large class of {\em short-range interaction}
ratchet models like for instance in (\ref{7.2}), (\ref{7.3})
have been revealed in \cite{agh99}.
More precisely, the steady state density-density correlation function
exhibits {\em dynamical scaling}
according to the Kadar-Parisi-Zhang universality class \cite{agh99}.

\section{Genuine collective effects}\label{sec7.2}
For non-interacting periodic systems, the basic result of the
previous chapters \ref{cha2}-\ref{cha06}
is that necessary, and generically also sufficient
conditions for the occurrence of directed transport are that
the system is out of thermal equilibrium and that its spatial
symmetry is broken.
The essential idea of this section is to abandon the latter condition
of a built-in asymmetry. Instead, 
{\em the system may create an
asymmetry by itself as a collective effect, namely
by way of spontaneous symmetry breaking}.
As a consequence, according to Curie's principle, a collective
ratchet effect in the form of a {\em ``spontaneous current''} 
is then expected\footnote{In contrast to ``permanent currents'', 
appearing for instance in mesoscopic rings {\em at thermal equilibrium},
the ``spontaneous currents'' which we have here in mind can
be exploited to do useful work and are moreover a purely classical 
phenomenon.}.
It turns out that this idea can indeed be realized, and 
in fact even in several different ways 
\cite{jul95,jul97a,jul97b,rei99a,vdb99a,rei99b,buc00,vdb00,man01,cle01}.
Here, we will focus on a particularly simple example of
globally coupled fluctuating potential ratchets 
\cite{rei99a,vdb99a,man01}.
We finally remark that the appearance of a ``spontaneous current'' 
has also been predicted in a rather different theoretical mean field model
for {\em driven semiconductor superlattices} in \cite{ale98,can00}.

\subsection{Model}\label{sec7.2.1}
As a combination of the fluctuating potential ratchet scheme from (\ref{4.1})
and of our general working model for interacting systems
(\ref{7.1}) we take as
starting point the following set of $i=1,...,N$ coupled stochastic equations
\begin{equation}
\dot x_i(t) = - V'(x_i(t))\, [1+f_i(t)]+\xi_i(t)+\frac{K}{N}
\sum\limits_{j=1}^N \sin(x_j(t)-x_i(t)) \ .
\label{7.10}
\end{equation}
For the sake of simplicity only, 
we consider a Kuramoto-type, sinusoidal {\em global coupling} 
\cite{kur75,kur84,str94},
and we will restrict ourselves to attractive interactions $K>0$.
Furthermore, we have adopted dimensionless units
(see \sect \ref{sec2.1.2.4} in Appendix A) with 
\begin{equation}
\eta=k_B=1,\ L=2\pi\ .
\label{7.10a1}
\end{equation}
[For esthetical reasons we will often continue to use
the symbol $L$.]
In particular, the potential $V(x)$ and the interaction respect
the same periodicity\footnote{Mathematically, we avoid in this way additional
complications due to incommensurability effects.
Physically, this assumption is especially natural if the state variables $x_i$
are originally of a phase-like nature, see \sect \ref{sec3.2.2}.
Some generalizations will be addresses in \sect \ref{sec7.2.5} 
below.} 
$L=2\pi$.
However, in contrast to ``conventional'' fluctuating potential 
ratchets without interaction (see \sect \ref{sec4.2}),
we {\em exclude any built-in spatial asymmetry} of the
system (\ref{7.10}), which can be achieved if the potential 
$V(x)$ respects the symmetry condition
\begin{equation}
V(-x)= V(x) \ ,
\label{7.10a}
\end{equation}
independently of any further properties of $f_i(t)$,
see the discussion below (\ref{s4}).
Finally, in view of the analytic tractability in the absence of 
interaction (see \sect \ref{sec4.2.2}) we specialize to
potential fluctuations $f_i(t)$ which are given by independent
Ornstein-Uhlenbeck processes (\ref{6l}), (\ref{6m}) of strength
\begin{equation}
\int_{-\infty}^\infty dt\, \langle f_i(t) f_j(s)\rangle 
= 2\, Q\, \delta_{ij} 
\label{7.10b}
\end{equation}
(cf. \eq (\ref{4.7a}))
and a negligibly small correlation time $\tau$ in 
comparison to all the other relevant time scales of 
the system.

\subsection{Spontaneous symmetry breaking}\label{sec7.2.2}
In this section, we first present a somewhat formal analytical
demonstration of the existence of spontaneous symmetry
breaking for a system (\ref{7.10}) in the thermodynamic limit
$N\to\infty$, followed quantitative numerical illustrations
and an intuitive explanation of the
basic physical mechanism at work.

The main collective features of (\ref{7.10}) are captured 
by the {\em particle density} 
\begin{equation}
P (x,t):= \frac{1}{N}\, \sum\limits_{i=1}^N \delta(x - x_i(t)) \ .
\label{7.11}
\end{equation}
In contrast to the definition for non-interacting systems in (\ref{2.8a}),
the average over the noise is omitted in (\ref{7.11}) and instead
an average over the particles $i$ is included.
Being an intensive quantity, $P(x,t)$ becomes 
independent of the
specific realization of the 
noises\footnote{To be precise, this means that a convolution
(average) $\int P(x,t)\, h(x)\, dx$ of the particle density
with an arbitrary but fixed, smooth test function $h(x)$
that vanishes as $x\to\pm\infty$, gives the same 
result with probability $1$ for
$N\to\infty$, independent of the realization of the noises 
$\xi_i(t)$ and $f_i(t)$.}  $\xi_i(t)$ and $f_i(t)$
when $N\to\infty$ (self-averaging), as demonstrated in 
detail in \cite{des78,daw83,bon87,str91}. 
In other words, it does actually not matter whether we consider an
average over the noise as included or not in the definition of
$P(x,t)$ in (\ref{7.11}). Finally, we go over to
the reduced density $\hat P(x,t)$ as usual, cf. \sect \ref{sec2.1.4}.

By rewriting the interaction term in (\ref{7.10}) as
$K\,[S\,\cos(x_i(t))- C\,\sin(x_i(t))]$,
where\footnote{For later convenience, the argument $t$ is
suppressed in $S$ and $C$.}
\begin{equation}
S:= \perint dx \,  \hat P(x,t) \sin x\, \ \ 
C:= \perint dx \,  \hat P(x,t) \cos x \ ,
\label{7.12}
\end{equation}
the dynamics of each particle (\ref{7.10}) is
exactly of the type which we have considered in \sect \ref{sec4.2.2}. 
By summing the corresponding single particle
Fokker-Planck equations (\ref{2.10a}), (\ref{4.8}) 
according to (\ref{7.11}) one recovers
\begin{eqnarray}
& &\frac{\partial}{\partial t}\,  \hat P(x,t)  =   
\frac{\partial}{\partial x} \left\{\hat V'(x)
+ g(x) \frac{\partial}{\partial x} g(x)\right\} \hat P(x,t)
\label{7.13}\\
& &\hat V(x)  :=  V(x) + K\, (S\, \sin x + C\, \cos x)
\label{7.14}\\
& & g(x)  :=  [ T + Q\, V'(x)^2]^{1/2}
\label{7.15}
\end{eqnarray}

Note that (\ref{7.13}) represents a {\em non-linear} Fokker-Planck equation
due to the implicit $\hat P(x,t)$-dependence of $\hat V(x)$
via (\ref{7.12}) and (\ref{7.14}).
Especially, the linear superposition principle is not respected.
This feature reflects the fact that while $P(x,t)$ in (\ref{7.11})
is self-averaging with respect to the noises $f_i(t)$ and
$y_i(t)$, it describes the particle density for a system with an arbitrary but 
{\em fixed initial distribution of particles } $P(x,t_0)$.
A statistical ensemble average over different initial particle
distributions is no longer captured by (\ref{7.13}), in
clear contrast to single-particle systems described e.g. by a {\em linear}
master equation of the form (\ref{2.10a}), or more general, finite-$N$ 
particle systems.
As usual in the context of phase transitions, the basic reason for this structural
difference is the thermodynamic limit $N\to\infty$ in concert with
the mean field coupling in (\ref{7.10}),
entailing the exact self-averaging property of the particle distribution 
(\ref{7.11}) in this limit $N\to\infty$.
The non-linear character of the Fokker-Planck equation opens the
possibility that different initial conditions $P(x,t_0)$ display a different
long time behavior, again in contrast to the typical asymptotic uniqueness
(ergodicity) of linear Fokker-Planck equations 
\cite{lan54,ber55,leb57,sch80,kam92}.
The reason for this possibility of ergodicity breaking with all its
consequences (spontaneous symmetry breaking,
phase transitions, etc.) is that the thermodynamic
limit $N\to\infty$ does not commute with the ``ergodicity limit''
$t\to\infty$.
In conclusion, \eqs (\ref{7.11})-(\ref{7.15}) display the typical structure
of a {\em mean field theory}, with $S$ and $C$ in (\ref{7.12}) playing the role
of order parameters which have to be determined self-consistently with 
the mean field \eq (\ref{7.13}) for the particle density.

Next, we discuss the behavior of $\hat P(x,t)$ in (\ref{7.13}) for 
asymptotically {\em large coupling strengths} $K$ in (\ref{7.10}).
To keep things simple,
we further assume that multiples of $L$ are the only
minima of $V(x)$.
As a consequence, all particles in (\ref{7.10}) are forced to 
occupy practically the same position $\mu(t)$ modulo $L$ and hence
$\hat P(x,t)$ takes the form
\begin{equation}
\hat P(x,t) = \sum\limits_{n=-\infty}^\infty 
\delta(x-\mu(t) + nL) \ .
\label{7.18a}
\end{equation}
Introducing (\ref{7.18a}) into (\ref{7.13}) and operating on both
sides with $\int_{\mu(t) - L/2}^{\mu(t) + L/2} dx\, x\dots$
the equation of motion for $\mu(t)$ takes the form of a simple 
relaxation dynamics
\begin{eqnarray}
& & \dot \mu(t)=-\bar U'(\mu(t))
\label{7.20}\\
& & \bar U(x)  :=  V(x)-Q\, V'(x)^2/2 \ .
\label{7.17}
\end{eqnarray}

For small $Q$, the extrema of $\bar U(x)$ in (\ref{7.17}) are
identical to those of $V(x)$. So, for any initial
condition $\mu(t_0)\in(-L/2,L/2)$, 
the center of mass $\mu(t)$ in (\ref{7.20})
moves  for $t\to\infty$ towards the minimum $x=0$ of $V(x)$,
and $\hat P(x,t)$ approaches a {\em stationary},
{\em symmetric} limit $\hat P^{st}(x)=P^{st}(-x)$.
However, this stationary solution $\mu(t)\equiv 0$ of (\ref{7.20}) 
looses stability and two new stable fixed points appear
when $Q$ in (\ref{7.17}) exceeds the critical value
\begin{equation}
Q_c:=1/V''(0) \ .
\label{7.21}
\end{equation}
One thus recovers a so-called
{\em noise induced nonequilibrium phase transition}
\cite{gar92,vdb94,bek94,ram95,gri96,vdb97,kim97,mul97,man97,zai98,gar99}
with a concomitant
{\em spontaneous symmetry breaking} of $P^{st}(x)$.

If the coupling strength $K$ is no longer assumed to be very
large, one has to solve the non-linear Fokker-Planck 
equation (\ref{7.17}) numerically until transients have died out 
and for a representative sample of different initial conditions.
In this way, a {\em stationary} and -- apart from the obvious
degeneracy when the symmetry is spontaneously broken --
{\em unique} long time limit $\hat P^{st}(x)$ is obtained.
In the symmetric phase ($P^{st}(-x)= P^{st}(x)$), the order 
parameter $S$ from (\ref{7.11}) vanishes,
while a spontaneously broken symmetry is generically
monitored by a non-zero $S$-value, see \figs \ref{fig7.1}.
Moreover, for large $K$, the above analytical prediction 
is confirmed by the numerics, for moderate $K$, one
recovers a {\em re-entrant} behavior as a function of
the potential fluctuation strength $Q$, and for small
$K$, a phase with broken symmetry ceases to exist 
\cite{vdb94,vdb97}.

\figsiebeneins

For an intuitive understanding of why the system-intrinsic
symmetry can be spontaneously broken, we return to a one-particle 
dynamics of the form (\ref{4.7d}).
By averaging over the noise, this equation takes the form
$\langle\dot x\rangle = - \langle V'(x(t))/\eta\rangle 
+ \langle g(x(t))\xi (t)\rangle $. On the other hand, 
evaluating the particle current by means of the probability
current (\ref{4.8}) according to (\ref{2.10c}),
one obtains $\langle\dot x\rangle = - \langle V'(x(t))/\eta\rangle 
+ \langle g'(x(t))g(x(t))\rangle /2$. Upon comparison of these
two expressions one recovers that
\begin{equation}
\langle g(x(t))\xi (t)\rangle =  
\langle 
\mbox{\small $\frac{1}{4}$ \normalsize}
\!
\mbox{\small $\frac{d}{dx}$ \normalsize}
\!
[g(x(t))]^2 \rangle \ .
\label{7.22}
\end{equation}
In other words, the white noise $\xi(t)$ induces a systematic
drift into the direction of increasing 
effective local temperature $T_{{\rm eff}}(x):=g^2(x)$ (see (\ref{7.15})).
To get a rough heuristic picture of how this so-called
{\em Stratonovich drift} term \cite{ris84} comes about, we imagine a
force-free,
overdamped Brownian particle starting at $x(0)=0$ in the
presence of a high temperature in the region $x>0$ and a 
low temperature for $x<0$. Though the particle spends on the
average the same amount of time on either side of $x=0$,
the thermal random motion within $x>0$ is enhanced, leading to
a net bias of the average particle position 
$\langle x(t)\rangle$ towards the right\footnote{Strictly speaking,
the issue is rather subtle with respect to the correct order
of the overdamped limit $m\to 0$ in (\ref{2.1}), 
the white noise limit $\tau\to 0$ in (\ref{4.7b}), and the
limit of a discontinuous temperature at $x=0$. Only if the limits are
taken in the latter order ($m\to 0$ first,
discontinuous temperature last), this explanation of the
Stratonovich drift can be applied, see also 
(\ref{2.4c}) in Appendix A 
and the corresponding discussion in \sect \ref{sec4.2.2}.}.

One can readily see by comparison with (\ref{7.15}) 
that this {\em noise induced drift} term (\ref{7.22}) is indeed 
the origin of the second term on the right hand side
of the effective potential (\ref{7.17}), which governs
the relaxation dynamics of the particle peak $\mu(t)$
in (\ref{7.20}). Since the intensity
of the multiplicative noise $f_i(t)$ in (\ref{7.10}) has a minimum
at the origin (modulo $L$), 
the noise induced drift pushes the particles away
from this point $x=0$ and may lead, if the noise is strong 
enough and the
particles cluster together sufficiently strongly, to
a spontaneous dislocation of the peak of particles $\mu(t)$
towards one or the other side of the origin.
If, on the other hand, the interaction is too weak in comparison
to either the thermal or the potential fluctuations, then
the random motions of the single particles are not sufficiently
coordinated and a collective spontaneous symmetry breaking
is therefore not expected.
These heuristic arguments are confirmed by,
and essentially explain the numerical phase diagram 
in \fig \ref{fig7.1}.

\subsection{Spontaneous ratchet effect}\label{sec7.2.3}
We start by rewriting (\ref{7.10}) in the form
\begin{equation}
\dot{x} (t) = - \veff'(x(t),f (t)) + \xi (t)  \ ,
\label{7.23}
\end{equation}
where we dropped the subscript $i$ and where
\begin{equation}
\veff(x,f(t)) := V(x)\, [1+f(t)] 
+ K\, (C \cos x + S \sin x) \ .
\label{7.24}
\end{equation}
If there is no spontaneous symmetry breaking 
($\hat P^{st}(-x) = \hat P^{st}(x)$),
then (\ref{7.12}) implies $S=0$ and hence the pulsating potential 
(\ref{7.24})
respects the symmetry condition (\ref{s1}) with $\Delta x=0$
due to (\ref{7.10a}).
If, on the other hand, the symmetry of the system is spontaneously
broken, then -- in the generic case -- we have that
$ S \not = 0 $. Hence the symmetry condition (\ref{s1}) is generically
violated
and the occurrence of a ratchet effect with 
$\langle\dot x\rangle\not = 0$ is expected according to Curie's principle.
There is, however, one prominent exception, namely a 
{\em supersymmetric potential (\ref{7.24}) excludes a current 
even if the symmetry of the system is spontaneously broken.}
For our present purposes it is sufficient to
focus on the supersymmetry condition (\ref{ss7}).
Since the white noise $f(t)$ is time-inversion invariant,
we see that for instance a pure cosine-potential $V(x)$ indeed
leads to a supersymmetric effective potential in (\ref{7.24}),
whatever the values of $S$ and $C$ are.
In order to break this supersymmetry, we can either modify the interaction
in (\ref{7.10}), or consider a colored noise $f_i(t)$, or, as we
will do in the following, choose an augmented cosine potential of the form
\begin{equation}
V(x) = - \cos x - A\,\cos (2x)
\label{7.25}
\end{equation}
with $A\not = 0$.

Given that the potential (\ref{7.24}) respects neither symmetry nor supersymmetry,
each particle (\ref{7.23}) is expected to exhibit a ratchet effect 
$\langle\dot x\rangle\not =0$ in the
generic case \cite{rei99a,vdb99a,man01}, as confirmed by the numerical 
result in\footnote{A computer animation of this collective phenomenon
is available on the
internet under \cite{ryovideo}.
It is based on simulations of (\ref{7.10})-(\ref{7.10b}), (\ref{7.25}) 
with $N=1000$, $T=2$, $Q=4$, $K=10$, $A=0.15$.
Out of the $1000$ particles, $100$ are shown as green dots and
one ``tracer-particle'' as a red dot. The position $x=-\pi$ is 
identified with $x=\pi$ (periodic boundary conditions). 
The initial particle distribution is symmetric about $x=0$. After a 
spontaneous breaking of the symmetry ``to the right'' ($S>0$)
an average particle current ``to the left'' ($\langle\dot x\rangle <0$)
can be observed.}  
\fig \ref{fig7.1}.
The underlying mechanism is clearly of the general {\em pulsating ratchet} type, 
and according to (\ref{7.24}) similar but not exactly identical to a 
{\em fluctuating potential ratchet} scheme from \sect \ref{sec4.2}.

With the notation from (\ref{7.14}), (\ref{7.15}) we can rewrite (\ref{7.23})
in yet another from, namely
\begin{eqnarray}
& & \dot x(t) = -\hat V'(x(t))+\hat\xi (t)
\label{7.25a}\\
& & \langle\hat \xi (t)\hat\xi(s)\rangle = 
2\, g(x(t))\, \delta(t-s) \ . 
\label{7.25b}
\end{eqnarray}
While $g(x)$ from (\ref{7.15}) has its minima at the integer multiples of
$L/2$, the potential $\hat V(x)$ from (\ref{7.14}) exhibits for $S\not = 0 $
not only an asymmetric, ratchet-shaped profile, but also its extrema are 
generically shifted with respect to those of $g(x)$.
From this viewpoint, the ratchet mechanism to which every single
particle is subjected in the symmetry broken phase is thus of the
{\em Seebeck ratchet} type from \sect \ref{sec4.8.1}.

Quantitatively, once the values of the order parameters
$S$ and $C$ in the long time limit
are known, the current follows readily along the lines of 
\sect \ref{sec4.2.2} with the result
\begin{eqnarray}
& & \hat P^{st}(x) = {\cal N}
\frac{e^{-\phi(x)}}{g(x)}
\int_x^{x+L} dy\,
\frac{e^{\phi(y)}}{g(y)}
\label{7.26}\\
& & \langle\dot x\rangle = L\, {\cal N}\, [ 1 - e^{\phi(L)}]
\label{7.26a}\\
& & \phi(x):=\int_0^x d\bar x\, \hat V '(\bar x)/g^2(\bar x) \ ,
\label{7.27}
\end{eqnarray}
where the normalization ${\cal N}$ is fixed through (\ref{2.14}).
Thus, the current is finite unless $\phi (L)=0$, and its
sign is given by  that of $-\phi (L)$. 

Specializing once again to large coupling strengths,
we can exploit (\ref{7.18a}) to recast (\ref{7.27}) into the
simplified form
\begin{eqnarray}
& & \phi(L)= - \hat K_{1}\sin\mu
\label{7.28}\\
& & \hat K_{n}:=\perint
dx\, [K\, \cos x]^n /g(x)^2\ ,
\label{7.28'}
\end{eqnarray}
where $\mu:=\mu(t\to\infty)$ follows from (\ref{7.20}).
Here, a remarkable feature arises, entailing even more
striking consequences later on.
Namely, if $Q>Q_c$, and $\hat K_{1} <0$, which is the case
whenever $A>0$ in (\ref{7.25}),
then the sign of $\mu (t)$ from (\ref{7.20}) will,
in the long time limit,
be opposite to that of $\langle\dot x\rangle$.
In other words, for a symmetry broken $P(x)$ with a peak
to one side of $x=0$, the flux of particles
will move just in the {\em opposite} direction!
On average the particles surprisingly prefer to travel from 
their typical position, say $\mu(t\to\infty)<0$ down to the potential 
minimum of $V(x)$ at $x=0$ and
then over the full barrier to their right rather than to directly surmount
the partial remaining barrier that they typically see to their left.

\subsection{Negative mobility and anomalous hysteresis}\label{sec7.2.4}
We now come to the response of the steady state current 
$\langle\dot x\rangle$ 
when an additional external force $F$ is added on the right hand side of
(\ref{7.10}).
After making the replacement
\begin{equation}
V(x) \, \mapsto\,  V(x) - x F
\label{7.29a}
\end{equation}
the entire analysis from \sects {\ref{sec7.2.1}-\ref{sec7.2.3}
can be repeated basically unchanged.
For small $F$ in combination with $Q<Q_c$ and large $K$ we can then
infer from (\ref{7.20}), (\ref{7.21}), (\ref{7.26a}), (\ref{7.28}) 
after some calculations that
\begin{equation}
\langle\dot x\rangle =
F\,L\, {\cal N}\, \left( \hat K_{0} +
\frac{\hat K_{1}}{[V''(0)]^{2}\, [Q_c-Q]}\right) + O(F^3).
\label{7.29}
\end{equation}
Thus, for sufficiently large, negative $\hat K_1$,
a {\em negative zero-bias mobility} (also called absolute negative mobility)
is predicted\footnote{Note that such a current 
$\langle \dot x\rangle$ opposite to the applied force $F$ is not in 
contradiction with any kind of ``stability criteria'', cf. 
the discussion below (\ref{2.23p}).}
\cite{rei99a,vdb99a}. 
A numerical example
for this remarkable behavior is shown in \fig \ref{fig7.2} (solid line).
Apparently, the effect of pulling the particles to one side is
analogous to that of a spontaneous symmetry breaking: it 
generates an effective, coupling-induced ratchet dynamics (\ref{7.23})
in which the nonequilibrium
fluctuations promote a current opposite to $F$.
Upon approaching the phase boundary,
the linear response of $P(x)$ to variations of $F$ diverges,
hence the denominator $Q_c -Q$ in (\ref{7.29})
and the very steep response curve in \fig \ref{fig7.2}.

\figsiebenzwei

We remark that for networks with dead-ends (see \cite{bal95}
and further references therein) 
and in the ratchet works \cite{cec96,ven96,sla97,sto01}, a negative 
{\em differential} mobility (far away from $F=0$) 
has been reported, but {\em not a current
opposite to the applied force} as in \fig \ref{fig7.2}.
Further, as illustrated by \fig \ref{fig5}, in the current-versus-force
characteristics for ``standard'', non-interacting ratchet models,
a current opposite to the applied force is possible as well.
However, as discussed in \sect \ref{sec2.2.2.2}, the ratchet effect is 
characterized by a current $\langle\dot x\rangle$ which is 
non-zero for $F=0$ and does not change its direction within 
an entire neighborhood of $F=0$. Accordingly, it
inevitably involves some kind of symmetry breaking (for $F=0$),
cf. \sect \ref{sec3.1.2}. 
In contrast, 
according to the characteristics of
negative zero-bias mobility exemplified in \fig \ref{fig7.2}.
the current $\langle\dot x\rangle$  is always opposite to the 
(not too large) force $F$, independently of whether 
$F$ is positive or negative.
Furthermore, the symmetry of the system (for $F=0$) is
neither externally, nor intrinsically, nor spontaneously broken.
In other words, the negative zero-bias mobility and the
ratchet effect exhibit some striking similarities but also
some fundamental differences.
We also mention that so-called 
{\em absolute negative conductance} has been
theoretically and experimentally studied in detail in the
context of 
semiconductor devices
\cite{ban71,pav76,poh81,sol90,kea95b,ign95,dag95,agu97,har97a,goy98c,can00},
photovoltaic effects in ruby crystals 
\cite{kro59,mat59,lia80,stu92},
tunnel junctions between superconductors with unequal energy gaps
\cite{aro75,ger86,ger88},
and has been theoretically predicted for certain
ionized gas mixtures \cite{dya87,roz88,gol89}.
While these effects are in fact basically identical to 
negative zero-bias mobility, their origin is of a 
genuine {\em quantum mechanical} character
which does not leave room for any kind of classical 
counterpart\footnote{Moreover, in the last three examples spatial 
periodicity is either not crucial or absent and in the case of 
tunnel junctions the spatial symmetry is intrinsically broken}.

For more general $F$- and $Q$-values but still large $K$, 
the qualitative dependence of $\langle\dot x\rangle$
on $F$ follows from (\ref{7.28}) by observing how $\mu$
moves in the adiabatically
changing potential $\bar U(x)$ from (\ref{7.17}), (\ref{7.20}). 
In this way, not only the
continuation of the zero-bias negative
conductance beyond $F\simeq 0$ in \fig \ref{fig7.2}
can be readily understood, but
also its even more spectacular counterpart when
$Q > Q_c$, namely an
{\em anomalous hysteresis-loop} \cite{rei99a,vdb99a},
see \fig \ref{fig7.3}.
Its striking difference in comparison with a ``normal'' 
hysteresis-cycle,
as observed e.g. in a ferromagnet or in the J\"ulicher-Prost model
\cite{jul95}, is as follows:
Given a spontaneous current in one or the other direction,
we can apply a small additional force $F$ in the {\em same} direction,
with the expected result of an increased current in that direction. But upon
further increasing $F$, the current will, all of a sudden, switch its
direction and run opposite to the applied force! 

\figsiebendrei

In short, the anomalous
response curves in \figs \ref{fig7.2} and \ref{fig7.3}
are basically the result of a
competition between the effect of the bias $F$,
favoring a current in that direction,
and the ratchet-effect, 
which arises as a collective property and pumps
particles in the opposite direction for $\hat K_1<0$.
The coexistence of two solutions $\hat P^{st}(x)$ over a
certain $F$-interval when $Q>Q_c$ gives rise
to the hysteresis,
and the destabilization of one of them
to the jumps of $\langle\dot x\rangle$ in \fig \ref{fig7.3}.

\subsection{Perspectives}\label{sec7.2.5}
In this section we briefly discuss some generalizations and potential
applications of our above considerations.

A first natural modification of the model (\ref{7.10}) 
consist in replacing the global coupling in by a 
{\em nearest neighbor} coupling in $d$ dimensions, i.e.
\begin{equation}
\frac{K}{N} \sum\limits_{j=1}^N \sin(x_j(t)-x_i(t))
\ \mapsto \ 
\frac{K}{2d} \sum\limits_{<ij>} \sin(x_j(t)-x_i(t)) \ ,
\label{7.30}
\end{equation}
by associating the indices $i$  with the vertices of some 
$d$-dimensional lattice with periodic 
boundary conditions. 
As \figs \ref{fig7.2}, \ref{fig7.3} demonstrate, e.g. for
a square lattice ($d=2$), the same 
qualitative phenomena as for global coupling are recovered, though
the quantitative details are of course different.

{\em Further generalizations} \cite{rei99a}
are: (i) The bare potential, represented by the
``$1$'' in the first term  on the right hand side of
(\ref{7.10}) plays a very minor role; even without this
term all results remain qualitatively unchanged.
Similarly, the thermal noise strength $T$ is arbitrary, except
that it must not vanish in the present model, but may
even vanish in a somewhat modified setup \cite{rei99b}.
(ii) A strictly periodic interaction $K(x)$ is not necessary.
For instance,
one may add on top of the periodic a (not too strong)
attractive interaction such as to keep the ``cloud'' of
particles $x_i$ in (\ref{7.10}) always well clustered.

Closely related studies on nonequilibrium phase transitions
\cite{vdb98,ben99} suggest that also {\em periodic instead of
stochastic drivings} $f_i(t)$ in (\ref{7.10}) will lead
to qualitatively similar results, see also \cite{ala00}.
Furthermore, a 
fluctuating force or rocking ratchet scheme instead of
a fluctuating potential model, amounting in (\ref{7.10})
to a substitution
\begin{equation}
- V'(x_i(t))\, [1+f_i(t)] 
\ \mapsto \ 
- V'(x_i(t)) + y_i(t) \ , 
\label{7.31}
\end{equation}
can apparently be employed as well \cite{buc00}.
Especially, the characteristic time scale of the driving $f_i(t)$ in
(\ref{7.10}) \cite{vdb00} and of $y_i(t)$ in (\ref{7.31}) \cite{buc00}
may become asymptotically large.
 
It might appear \cite{doe87,vdb94,vdb97}
that taking the overdamped limit $m\to 0$ in
(\ref{2.1}) before the white noise limit $\tau\to 0$ of the 
Ornstein-Uhlenbeck noise $f_i(t)$ in (\ref{7.10}), (\ref{7.10b})
(see also (\ref{4.7b}) and (\ref{4.7c})) is an
indispensable prerequisite for spontaneous symmetry 
breaking and spontaneous current,
since only in this way \cite{gra82,ris84} a white noise $f_i(t)$
in the sense of Stratonovich and a concomitant noise induced drift term can arise.
Our detailed analysis, however, reveals that the same phenomena can in
fact still be encountered even if the white noise limit $\tau\to 0$
is performed prior to $m\to 0$, see also \cite{iba01}.
In other words, {\em finite inertia terms} are also admissible 
on the left hand side of (\ref{7.10}).

While a spontaneous ergodicity-breaking with all its above discussed
consequences is clearly possible only in the thermodynamic 
limit $N\to\infty$, the same {\em a priori}
restriction does not hold for 
the phenomenon of negative zero-bias mobility. 
Indeed, a stylized, spatially discretized descendant of
the above discussed working model (\ref{7.10}) with 
negative mobility for $N\geq 4$ has been
presented in \cite{cle01}.
A different, experimentally realistic single
particle system ($N=1$) in two dimensions with 
negative mobility has been introduced in \cite{eic01},
while a game theoretic counterpart of the effect 
(cf. \sect \ref{sec3.9}) is due to \cite{vdb01}.

In conclusion, the above revealed {\em main phenomena
seem to be rather robust} against modifications and extensions
of the considered model (\ref{7.10}).
Much like in equilibrium phase transitions, such an 
extremely simple
model is thus expected to be of interest for a variety of
different systems, corresponding to a
``normal form'' description that subsists after the 
irrelevant terms have been eliminated. 
Models of this type may be of relevance
not only in the context of molecular motors 
(see \sect \ref{sec5.4}), but also for
coupled phase oscillators \cite{kur75,kur84,str94},
active rotator systems \cite{shi86}, 
charge density waves \cite{str89},
and many other physical, chemical, and biological 
systems \cite{win80,som91,swi92,han93}. 
For instance, one may also look at (\ref{7.10}) as a planar
XY-spin-model \cite{are94} exposed to a strong (but incoherent)
electromagnetic irradiation \cite{bel80,but87,pro94,jan94,pal98,kou99,kos00a,kre00},
with the various effects of the photon-impacts
(scattering, excitations of the host-crystal ions, etc.) 
roughly described by the non-equilibrium fluctuations $f_i(t)$.
An experimental realization in a granular gas system is presently 
under construction.

\chapter{Conclusions}\label{cha9}
The central theme of our review are transport phenomena
in spatially periodic ``Brownian motors'' or ``ratchet systems'' 
induced by unbiased perturbations of the thermal equilibrium. 
Letting aside variations and extensions
like diffusive transport, quenched spatial disorder, or questions
of efficiency, our extensive discussions may be summarized
under three main categories:
i) Understanding and predicting the ``ratchet effect'' {\em per se}, 
i.e. the occurrence (or not) of a directed average long-time
current $\langle \dot x \rangle$.
ii) Exploring qualitative features of the current 
as a function of various parameters,
for example 
the sign of the current and the possible
appearance of current inversions,
monotonic versus non-monotonic 
``resonance-like'' behavior with some type of ``optimum'', 
or the asymptotic behavior for 
fast, slow, and weak perturbations etc.
iii) On the one hand, 
identifying particularly simple or counterintuitive 
``minimal models'' and very general ``normal forms''
exhibiting a ratchet effect and/or current 
inversions. On the other hand, 
elaborating realistic models and their quantitative
features with some specific experimental situation 
in mind.

For several of these questions, symmetry considerations
play an important role.
This is so basically due to Curies principle, stating that 
in the absence of prohibiting ``systematic'' symmetries, the
appearance of a certain phenomenon (here: the ratchet effect) 
will be the rule, while its absence will be the exception.
In our case, there are three such ``systematic'' symmetry conditions,
each of which is sufficient to rule out the appearance of a ratchet effect:
1. Detailed balance symmetry, implying that we are dealing with an equilibrium system
and that a thermal equilibrium state will thus be approached in the long time limit.
2. (Spatial) symmetry as detailed in \sect \ref{sec3.1.2}.
3. Supersymmetry as detailed in \sect \ref{sec3.4} in the overdamped limit
and its counterpart (\ref{ham2}) in the underdamped (deterministic Hamiltonian) limit.
Closely related to these symmetry conditions, there are in addition a couple of
``systematic'' no go theorems for certain classes of ratchet systems, 
see at the beginning of \sect \ref{sec4.2}, at the end of 
\sect \ref{sec4.3.2}, and in \sect \ref{sec4.8.4a}.

If all three above systematic symmetry conditions are violated,
then a vanishing current is the exception, which may be termed an
``accidental symmetry'', and which is usually connected with a current 
inversion.
A very general method of tailoring such current inversions has been
elaborated in \sect \ref{sec3.5} together with a very simple
and in fact obvious necessary and sufficient condition for 
their existence.
Our ratchet classification scheme from \sect \ref{sec3.1.3} is mainly based
on the specific manner in which the second of the above systematic symmetries 
is broken.
Depending on whether current inversions exist or not, we may speak of a
``non trivial'' and an ``obvious'' ratchet effect,
exemplified by fluctuating potential and tilting ratchets and 
by (proper) traveling potential ratchets, respectively.
In the first case, the direction of the current is obvious in some simple cases,
but not at all in general, while in the second it is always rather clear.

We remark that for both, ``systematic'' and ``accidental'' symmetries,
the result $\langle\dot x \rangle = 0$ is 
unstable against completely
general, generic variations of the model, while the property
$\langle\dot x \rangle \not = 0$ is 
robust against such variations, i.e.
``a finite current is the rule''.
The only difference is that for ``systematic'' symmetries, the hyperspace 
of parameters with $\langle\dot x \rangle = 0$ 
(and thus the definition of the symmetry itself) 
can be easily expressed in
terms of ``natural'' model parameters, while for ``accidental'' symmetries
such a hyperspace exists as well but is very difficult to characterize.
In this sense, there are actually no ``accidental'' symmetries, they are
only very difficult to define and therefore ``overlooked'' within any
``natural'' invariance-considerations of the problem.

We note that the above symmetries 1.-3. refer strictly speaking
to the (asymptotic) state and not to the system itself.
Since the thermodynamic limit of infinitely many interacting 
subsystems may not commute with the long-time limit, and so
an asymmetry of the initial condition may never disappear,
some symmetry property of the system dynamics alone does not yet imply 
the corresponding asymptotic symmetry of the state (solution)
in the case of extended systems.
While this implication is still correct (leaving asside glass-like systems)
for the first of the above mentioned symmetries
(an equilibrium system implies an asymptotic equilibrium state),
it may be incorrect in the second case of (spatial) symmetry:
Even in a perfectly symmetric system, a spontaneous symmetry
breaking of the asymptotic state may occur, leading to a spontaneous
ratchet effect, see \ch \ref{cha7}.

Regarding future perspectives of the field, the fact
that many of the above symmetry considerations became clear
only very recently suggest that further
new theoretical results on a very basic conceptual 
level may still be to come.
If a specific direction has to be named then
the still rather fresh topic of coupled Brownian motors
appears to be a particularly promising candidate, 
both theoretically (\ch \ref{cha7})
and with respect to biological applications (\sect \ref{sec5.4}).
Further, there is a remarkable large and rapidly increasing number
of exciting experimental studies, some of them
with promising perspectives regarding technological applications.
Whether Brownian motors offer just a new view or an entirely new
paradigm with respect to the modeling of molecular motors (\ch \ref{cha5})
remains to be seen as well.

\chapter*{Acknowledgment}
The present review would not have been possible without the
continuous support by 
P.~H\"anggi 
and his group and the very fruitful collaborations with 
C.~Van den Broeck, 
R.~Kawai,
R.~Bartussek, 
R.~H\"aussler,
M.~Schreier,
E.~Pollak, 
C.~Kettner,
F.~M\"uller,
T.C.~Elston,
B.~Lindner,
L.~Schimansky-Geier,
I.~Bena,
M.~Nagaoka,
M.~Grifoni,
G.J.~Schmid,
and J.~Lehmann.
In preparing this work I have furthermore
profited a lot from the scientific interaction with
I.~Goychuk,
J.~Luczka,  
P.~Talkner, 
R.D.~Astumian, 
M.~Bier,
J.~Prost,
F.~J\"ulicher,
C.R.~Doering, 
P.~Jung, 
G.-L.~Ingold,
H.~Linke,
D.~Koelle,
S.~Weiss,
M.I.~Sokolov,
T.~Dittrich,
R.~Ketzmerick,
H.~Schanz,
S.~Flach,
O.~Yevtushenko,
A.~Lorke,
C.~Mennerat-Robilliard,
C.M.~Arizmendi,
F.~Sols,
I.~Zapata, 
T.~T\'el,
M.~Rubi,
A. Perez-Madrid, 
M.~Thorwart,
R.~Eichhorn,
and J.M.R.~Parrondo.
Special thanks is due to
P. H\"anggi and I. Goychuk,
for providing Refs. \cite{smo12} and \cite{bog45}, 
respectively, to
J.M.R.~Parrondo and
H.~Linke 
for providing figures 
\ref{fig1}, \ref{figfey1} and figure \ref{figqm4}, 
respectively,
and to N.~G. van Kampen for providing a copy of his thesis
\cite{kam51}.
Financial support by the
DFG-Sachbeihilfe HA1517/13-2,13-4 and
the Graduiertenkolleg GRK283
is gratefully acknowledged.

\appendix
\chapter{Supplementary material regarding section \ref{sec2.1.1}}
We have modeled the two effects of the environment
on the right hand side of \eq (\ref{2.1})
phenomenologically, and we will discuss in the next three
subsections the rather far reaching
implications of this specific phenomenological {\em ansatz}.
Especially, we will argue that the assumptions of the environment being
at thermal equilibrium and of a dissipation mechanism of the form $-\eta\dot x(t)$
competely fix the statistical properties of the 
additive fluctuations $\xi (t)$ in (\ref{2.1}).
While our line of reasoning will be conducted on a heuristic physical level, 
it still captures the essential ideas of mathematically more sophisticated
and rigorous approaches 
\cite{ein05,ein10,joh28,nyq28,cal51,ber55,leb57,mag59,rub60,leb63,res64,ull66,zwa73,hyn75,spo78,gra80,gra82a,cal83,for88,han90,wei99a,rei01a},
see also \sects \ref{sec3.2.1}, \ref{sec4.8.4c}, and \ref{sec6.5.1}.

\section{Gaussian white noise}\label{sec2.1.2.1}
The fact that the friction force on the right hand side of (\ref{2.1}) 
is linear in $\dot x(t)$, i.e. no spatial direction is preferred,
suggests that -- due to their common origin -- also the thermal fluctuations
are unbiased, that is (cf. (\ref{2.2}))
\begin{eqnarray}
\langle\xi(t)\rangle & = & 0 \ , 
\label{2.2'}
\end{eqnarray}
for all times $t$, where $\langle\cdot\rangle$ indicates the average over
independent realizations of the random process $\xi (t)$.
Similarly, the fact that the friction force only depends on the
present state of the system and not on what happened in the past
has its counterpart in the assumption that the random fluctuations are 
uncorrelated in time, i.e.
\begin{equation}
\langle\xi (t) \xi (s)\rangle = 0 \ \ \mbox{if}\ t\not = s \ .
\label{2.4a}
\end{equation}
Furthermore, the fact that the friction involves no explicit time dependence
has its correspondence in the time-translation invariance of all statistical 
properties of the fluctuations, i.e. the noise $\xi (t)$ is a stationary
random process.
Finally, the fact that the friction force acts permanently in time indicates 
that the same will be the case for the fluctuations.
In other words, a noise $\xi (t)$ exhibiting rare but relatively 
strong ``kicks'', caused e.g. by 
impacts of single molecules in a diluted gas, is excluded.
Technically speaking, one says that $\xi (t)$ cannot contain a shot 
noise component \cite{ric54,han78,han80,vdb83,vdb84}.
During a small time interval, the effect of the
environment thus consists of a large number of small and, according
to (\ref{2.4a}) practically independent, contributions.
Due to the central limit theorem\footnote{In its
simplest version -- sufficient for our present purposes --
the central limit theorem  \cite{kam92}
states that if $r_1,...,r_N$ are
independent, identically distributed random variables with zero mean and unit variance
then the sum
$N^{-1/2}[r_1+...+r_N]$ converges for $N\to\infty$
towards a Gaussian random
variable of zero mean and unit variance.}
the net effect of all these contributions on the particle
$x(t)$ will thus be Gaussian distributed.
Such a Gaussian random process which is unbiased (\ref{2.2'}) and
uncorrelated in time (\ref{2.4a}) is called {\em Gaussian white noise}.

\section{Fluctuation-dissipation relation}\label{sec2.1.2.2}
A crucial implicit assumption in (\ref{2.1}) is the independence
of the friction force, and hence also of the fluctuation force,
from the system
$x(t)$, i.e.\footnote{We
remark that $m>0$ (cf. (\ref{2.1})) is understood in (\ref{2.4c}).
The properties (\ref{2.4a}) and (\ref{2.4c})
lead for $m\to 0$ to a Gaussian 
white noise $\xi(t)$ in the so-called Ito-sense 
\cite{han82b,ris84}.}
\begin{equation}
\langle\xi (t) x(s)\rangle = 0
\label{2.4c}
\end{equation}
for all 
times\footnote{The case $t<s$ is somewhat subtle and not needed in the following.} 
$t\geq s$.
It reflects the assumption that the environment is given by a ``huge''
heat bath so that its properties are practically not influenced by the behavior
of the ``small'' system $x(t)$. Especially,
the statistical properties of the
fluctuations will not depend on the choice of the potential $V(x)$ and
we may set $V'(x) \equiv 0$ in the 
following.
One readily verifies that in this case the equation of motion (\ref{2.1})
is solved by
\begin{equation}
\dot x(t) = \dot x(t_0)\, e^{-\frac{\eta}{m}(t-t_0)} + \frac{1}{m}
\int_{t_0}^t dt'\, e^{-\frac{\eta}{m}(t-t')}\, \xi(t')\ .
\label{2.4d}
\end{equation}
Choosing as initial time $t_0 = -\infty$ it follows that
\begin{equation}
\langle \dot x^2(t) \rangle =  \frac{1}{m^2}
\int_{-\infty}^t dt'\, \int_{-\infty}^t dt''\, 
e^{-\frac{\eta}{m}(2t-t'-t'')}\, \langle \xi(t') \xi(t'')\rangle \ .
\label{2.4e}
\end{equation}
In view of (\ref{2.4a}), the integrand only contributes 
if $t'=t''$ and the upper limit $t$ in the second integral 
can be furthermore extended to $+\infty$, i.e.
\begin{equation}
\langle \dot x^2(t) \rangle =  \frac{1}{m^2}
\int_{-\infty}^t dt'\, e^{-\frac{2\eta}{m}(t-t')}
\int_{-\infty}^\infty dt''\, \langle \xi(t') \xi(t'')\rangle \ .
\label{2.4f}
\end{equation}
Since the statistical properties of the 
fluctuations $\xi(t)$ are time-translation invariant, the second integral 
has the same value for all times $t'$ and we can conclude that
\begin{equation}
\int_{-\infty}^{\infty} ds\, \langle\xi(t)\xi(s)\rangle 
= 2\, \eta\,  m\, \langle\dot x^2(t)\rangle
\label{2.4g}
\end{equation}
for all times $t$. The left hand side of this equation
is called the {\em intensity} of the noise $\xi(t)$ or
the {\em noise strength}.

At this point, we make use of the fact that the environment
is a heat bath at thermal equilibrium with temperature $T$.
Since we have chosen as initial time $t_0=-\infty$, all transients have
died out and the particle is in thermal equilibrium with the bath,
satisfying the equipartition principle 
(for a one-dimensional dynamics)
\begin{equation}
\frac{m}{2}\,\langle\dot x^2(t)\rangle = \frac{1}{2}\, k_BT \ ,
\label{2.4h}
\end{equation}
where $k_B$ is Boltzmann's constant.
Collecting (\ref{2.4a}), (\ref{2.4g}), (\ref{2.4h})
we obtain the so-called {\em fluctuation-dissipation relation}
\cite{nyq28,joh28,cal51} (cf. (\ref{2.3}))
\begin{eqnarray}
\langle\xi(t)\xi(s)\rangle & = & 
2\,\eta\, k_B T \, \delta(t-s) \ ,
\label{2.3'}
\end{eqnarray}
where $\delta (t)$ is Dirac's delta function.
In other words, $\xi(t)$ is a Gaussian white noise
of intensity $2\eta k_B T$.
Note that since $\xi(t)$ is a Gaussian random process,
all its statistical properties are completely determined 
\cite{han82b,ris84,kam92} 
already by the mean value (\ref{2.2'}) and the correlation (\ref{2.3'}).

\section{Einstein relation}\label{sec2.1.2.3}
In the absence of the potential $V(x)$ in (\ref{2.1}),
we know that the particle exhibits a {\em free thermal diffusion}
in one dimension with a diffusion constant $D$,
i.e. for asymptotically large times $t$ we have that\footnote{Corrections
of order $o(t)$ are omitted in (\ref{2.5a})
and we will tacitly assume that their time derivative 
approaches zero for $t\to\infty$ \cite{han95}. 
Furthermore, we note that this asymptotic result (\ref{2.5a}) 
is independent of the initial condition $x(0)$.}
\begin{equation}
\langle x^2(t)\rangle = 2\, D\, t\ .
\label{2.5a}
\end{equation}
On the other hand, upon multiplying \eq (\ref{2.1}) by $x(t)$,
averaging, and exploiting (\ref{2.4c}), we obtain
\begin{equation}
m\,\langle\ddot x(t) x(t)\rangle 
= -\eta\,\langle \dot x(t) x(t)\rangle\ .
\label{2.5b}
\end{equation}
The left hand side of this equation can be rewritten as
\begin{equation}
m\,\langle\ddot x(t) x(t)\rangle 
= m\,\frac{d}{dt}\langle \dot x(t) x(t)\rangle - m \,
\langle\dot x^2 (t)\rangle \ .
\label{2.5b'}
\end{equation}
By differentiating (\ref{2.5a})
we have (for large $t$) that $\langle\dot x(t) x(t)\rangle = D$ and
hence $d \langle\dot x(t) x(t)\rangle / dt = 0$.
Observing (\ref{2.4h}) we finally obtain from (\ref{2.5b}),
(\ref{2.5b'}) the so-called 
{\em Einstein relation} \cite{ein05} (cf. (\ref{2.11}))
\begin{equation}
D = k_B T / \eta \ .
\label{2.4'}
\end{equation}
Its most remarkable feature is that the diffusion in
(\ref{2.5a}) does not depend on the mass $m$ of the particle $x(t)$
for asymptotically large times $t$.

\section{Dimensionless units and overdamped dynamics}\label{sec2.1.2.4}
The objective of this section is to recast the stochastic dynamics (\ref{2.1}), 
(\ref{2.3'}) into a dimensionless form, useful for qualitative theoretical
considerations and indispensable for a numerical implementation.

We start with defining the barrier height
\begin{equation}
\Delta V := \max_{x}\left\{ V(x)\right\} - 
\min_{x}\left\{ V(x)\right\} 
\label{2.6a}
\end{equation}
between adjacent local minima of the periodic potential $V(x)$.
Next, we introduce for the three dimensionful quantities
$\eta$, $L$, and $\Delta V$ dimensionless counterparts
$\hat \eta$, $\hat L$, and $\Delta \hat V$, which for the moment
can still be freely chosen.
With the definitions of the dimensionless quantities
\begin{eqnarray}
\hat t & := & \alpha\, t \ \ , \ \ 
\alpha:=\frac{\Delta V}{\eta L^2}\,
\frac{\hat\eta\hat L ^2}{\Delta\hat V}\label{2.6b}\\
\hat x (\hat t) & := & \frac{\hat L}{L}\, x(\hat t/\alpha)\label{2.6c}\\
\hat V(\hat x) & := & 
\frac{\Delta\hat V}{\Delta V}\, V(\hat x\, L/\hat L)
\label{2.6d}
\end{eqnarray}
we can rewrite (\ref{2.1}) in the dimensionless form
\begin{equation}
\hat m\,\frac{d^2\,\hat x(\hat t)}{d\hat t ^2} 
+ \hat\eta\, \frac{d\,\hat x(\hat t)}{d\hat t} 
= - \frac{d\,  \hat V(\hat x(\hat t))}{d\hat x}
+\hat\xi (\hat t)\ ,
\label{2.6e}
\end{equation}
where $\hat\xi(\hat t)$ is a dimensionless Gaussian white noise
with correlation
\begin{eqnarray}
\langle\hat\xi(\hat t)\hat \xi(\hat s)\rangle & = & 
2\,\hat\eta\, \hat k_B \hat T\,  \delta(\hat t-\hat s) \ .
\label{2.6f}
\end{eqnarray}
Furthermore, the dimensionless mass in (\ref{2.6e}) is defined as
\begin{equation}
\hat m := m\,\frac{\Delta V}{\eta^2L^2}\,
\frac{\hat L^2\hat\eta^2}{\Delta \hat V}
\label{2.6g}
\end{equation}
and the dimensionless temperature in (\ref{2.6f}) as
\begin{equation}
\hat T := \frac{k_B T}{\Delta V}\,\frac{\Delta\hat V}{\hat k_B}\ ,
\label{2.6h}
\end{equation}
where $\hat k_B$ may be chosen arbitrarily, e.g. $\hat k_B =1$.

Next we choose $\hat\eta$, $\hat L$, $\Delta\hat V$, and $\hat k_B$ 
all equal to unity. For the typically very small systems one has in mind, 
and for which thermal fluctuations play any notable role at all,
the rescaled mass (\ref{2.6g}) then often turns out to be smaller than 
unity by many orders of magnitude, see e.g. in \cite{kul98a}, while
the dimensionless temperature (\ref{2.6h}) is of order unity or
smaller\footnote{In the opposite case, i.e. if $\Delta V/k_B T$ is 
a small quantity (especially if $V(x)=\mbox{const.}$) one has to replace 
$\Delta V$ by $k_BT$ in the definition (\ref{2.6b}) of $\alpha$,
and similarly in (\ref{2.6d}), (\ref{2.6g}).
On condition that $m\,k_BT/\eta^2L^2$ is small, one can
then drop the inertia term.
The condition for arbitrary $\Delta V/ k_BT$ is thus that
the dimensionless quantity $m\,\eta^{-2}L^{-2}\max\{\Delta V,\, k_BT\}$ 
has to be a small quantity.}.
On the other hand, the period $\hat L$ and the barrier height 
$\Delta\hat V$ of the 
potential $\hat V(\hat x)$ are both unity, so the
derivative of this potential is typically of order unity as well.
It is therefore quite plausible that in (\ref{2.6e}) the inertia term
$\hat m\, d^2\hat x(\hat t)/ d\hat t^2$ can be dropped in very
good approximation.
Admittedly, from a mathematical viewpoint, dropping the highest order
derivative in a differential equation, especially in the presence
of such an elusive object as the Gaussian white noise $\hat \xi(\hat t)$,
may rise some concerns.
A more careful treatment of this problem has been worked out e.g. in
\cite{ryt81,gra82,san82,kam88,jay96a,mah96,sek99}
with the same conclusion as along our simple heuristic argument.
We finally note that letting $m\to 0$ affects neither the 
fluctuation-dissipation relation (\ref{2.3'}) nor Einstein's relation 
(\ref{2.4'}).

Finally, we turn to the the typical case that
$\hat m$ is known to be a small quantity and we thus can set
formally $m=0$ in (\ref{2.1}).
We thus recover the ``minimal'' {\em Smoluchowski-Feynman ratchet}
model from (\ref{2.5}).
Introducing dimensionless units like before,
one arrives again at (\ref{2.6e}) but now
with $\hat m=0$ right from the beginning.
In principle, $\hat\eta$, $\hat L$, and $\Delta\hat V$
may still be chosen arbitrarily. 
However, in most
concrete cases it is convenient to assume that $\hat L$ and
$\Delta \hat V$ are of order unity, but not necessary equal to $1$
(e.g. $\hat L=2\pi$ or $\hat V_0=1$ in (\ref{2.1''}) may 
sometimes be a more convenient choice), while $\hat\eta$ may still
be a variable ``control parameter'' of the model.
The implication of a dimensionless solution $\hat x(\hat t)$ for the
original, dimensionful system $x(t)$ is obvious. Especially,
varying one parameter (e.g. $\hat\eta$ or $\hat T$)
and keeping the others fixed, corresponds to exactly the same 
parameter-variation in the dimensionful system. 

We finally remark that 
in the end one usually drops again 
the ``hat''-symbols of the dimensionless quantities.
Depending on the context,
\eq (\ref{2.5}) may thus represent either the dimensionful or the 
dimensionless version of the model.

\chapter{Alternative derivation of the Fokker-Planck equation}
In this appendix we give a derivation of the Fokker-Planck equation
(\ref{2.10}) by considering the corresponding
overdamped stochastic dynamics
(\ref{2.5}) as limiting case of the discretized dynamics (\ref{2.7}) when
$\Delta t\to 0$.

To simplify notation, we use dimensionless units (see below \eq (\ref{2.5}))
with $k_B=\eta=1$.
Next we recall that $\xi_n$ in (\ref{2.7}) are independent, Gaussian
distributed random variables with $\langle\xi_n\rangle =0 $ and
$\langle\xi_n^2\rangle =2T/\Delta t$ (see (\ref{2.7'})).
It follows that for a particle (\ref{2.7}),
the conditional probability $P(x|y)$
to start out at time $t=t_n=n\,\Delta t$ from the point $x_n=y$ and
to arrive one time step $\Delta t$ later at the point $x_{n+1}=x$ 
is Gaussian distributed about $x=y-\Delta t\, V'(y)$ with 
variance $\langle(\Delta t\, \xi_n)^2\rangle =2T \Delta t$,
i.e.
\begin{equation}
P(x|y) = (4\pi T\Delta t)^{-1/2}\ 
\exp\left\{-\frac{[x-y+\Delta t\, V'(y)]^2}{4\, T\, \Delta t}\right\} \ .
\label{A1}
\end{equation}
Furthermore, the probability distribution $P(x,t+\Delta t)$ at
time $t+\Delta t$ is obviously related to that at time $t$ 
through the so-called Chapman-Kolmogorov equation \cite{kam92}
\begin{equation}
P(x,t+\Delta t) = \int_{-\infty}^\infty dy\, P(x|y)\, P(y,t) \ .
\label{A2}
\end{equation}
After a change of the integration variable according to
$z=(x-y)/\sqrt{\Delta t}$ we obtain
\begin{eqnarray}
\!\!\!\!\!\!\!\!& & \!\!\!\!\!\!\!\!
P(x,t+\Delta t) = \nonumber\\
\!\!\!\!\!\!\!\!& & \!\!\!\!\!\!\!\!
\int\limits_{-\infty}^\infty \frac{dz}{(4\pi T)^{1/2}}
\exp\left\{-\frac{[z+\sqrt{\Delta t}\,\, 
V'(x- \sqrt{\Delta t}\,z)]^2}{4\, T}\right\} 
\ P(x- \sqrt{\Delta t}\, z,t) \ .
\label{A3}
\end{eqnarray}
Under the assumption that $P(x,t)$ behaves sufficiently well as $\Delta t\to 0$,
we can expand the right hand side of (\ref{A3}) in powers of
$\sqrt{\Delta t}$ and perform the remaining Gaussian integrals,
with the result
\begin{eqnarray}
\!\!\!\!\!\!\!\!& & P(x,t+\Delta t) = \nonumber\\
\!\!\!\!\!\!\!\!& & P(x,t ) 
+ \Delta t\,\frac{\partial}{\partial x}\left\{V'(x)\, P(x,t)\right\}
+ \Delta t\, T\, \frac{\partial^2}{\partial x^2} P(x,t) 
+ o(\Delta t) \ .
\label{A4}
\end{eqnarray}
In particular, there is no contribution proportional to 
$\sqrt{\Delta t}$.
In the limit $\Delta t \to 0$, 
the Fokker-Planck equation (\ref{2.10}) now readily follows.

\chapter{Perturbation analysis}
In this appendix we solve the Fokker-Planck equation
(\ref{2.24'}) perturbatively for small time-periods $\ttt$
in (\ref{2.23''}) and zero load $F=0$.
We recall that for evaluating the particle current (\ref{2.24''})
we can focus on probability densities 
$\hat P(x,t)$ which are $L$-periodic in space and $\ttt$-periodic
in time and that the function $\hat T(h)$ from (\ref{2.3.1'})
is assumed to be $\ttt$-independent.
The latter assumption suggests to introduce
\begin{equation}
W_\ttt(x,h) := \hat P(x,h\ttt) \ ,
\label{b1}
\end{equation}
so that the Fokker-Planck equation (\ref{2.24'}) takes the form
\begin{equation}
\frac{\partial}{\partial h} W_\ttt(x,h) = \ttt\left\{
\frac{\partial}{\partial x}\left[\frac{V'(x)}{\eta} 
\, W_\ttt(x,h)\right]
+ \frac{k_B\, \hat T(h)}{\eta} \, 
\frac{\partial^2}{\partial x^2} W_\ttt(x,h)\right\} \ .
\label{b2}
\end{equation}
The small quantity $\ttt$ on the right hand side of this
equation furthermore suggest a power series
{\em ansatz}
\begin{equation}
W_\ttt(x,h) = \sum_{n=0}^\infty \ttt^n\, W_n(x,h)
\label{b3}
\end{equation}
with $\ttt$-{\em independent functions} $W_n(x,h)$.
From the periodicity and normalization of $\hat P(x,t)$
one readily finds that
\begin{eqnarray}
& & W_n(x+L,h)=W_n(x,h+1)=W_n(x,h)\label{b4}\\
& & \int_0^L dx\,  W_n(x,h) = \delta_{n,0}  \label{b5}
\end{eqnarray}
for $n\geq 0$, where $\delta_{i,j}$ is the Kronecker delta.

Next the usual perturbation analysis argument is invoked:
Introducing the {\em ansatz} ({\ref{b3}) into the
Fokker-Planck equation (\ref{b2}) and observing that this
equation is supposed to hold for arbitrary $\ttt$
it follows that the coefficients of each power of $\ttt$
must be equal to zero separately.
In the lowest order $\ttt^0$ it follows that
\begin{equation}
\frac{\partial}{\partial h} W_0(x,h) = 0 \ ,
\label{b6}
\end{equation}
i.e. $W_0(x,h)$ is equal to a $h$-independent
but otherwise still unknown function $W_0(x)$.
By introducing this function into (\ref{b3}),
equating order $\ttt^1$-terms, and averaging over one time
period $\ttt$ one obtains
\begin{equation}
0 = \frac{\partial}{\partial x}\left[\frac{V'(x)}{\eta} 
\, W_0(x)\right]
+ \frac{k_B\, \overline{T}}{\eta} \, 
\frac{\partial^2}{\partial x^2} W_0(x)\ ,
\label{b7}
\end{equation}
where the time-averaged temperature
$\overline{T}$ is defined in (\ref{2.3.1}).
This ordinary second order equation for $W_0(x)$ can 
now be readily solved, with the two emerging
integration constants being determined 
by the periodicity and normalization conditions
(\ref{b4}), (\ref{b5}). The result is
\begin{eqnarray}
& & W_0(x) = Z^{-1}\, e^{-V(x)/k_B \overline{T}}\label{b8}\\
& & Z := \int_0^L dx\,  e^{-V(x)/k_B \overline{T}} 
\label{b9}
\end{eqnarray}
and the corresponding contribution of
order $\ttt^0$ to the particle current (\ref{2.24''})
is found to vanish.
In other words, we have recovered in the limit $\ttt\to 0$
the same results as for a constant, time-averaged
temperature $\overline{T}$ in \sect \ref{sec2.1.4}, 
in accordance with what one may have expected.

Proceeding in exactly the same way up to the next order
$\ttt^1$ still gives a zero contribution to the
particle current.
It is only in the second order $\ttt^2$ that the first
nontrivial contribution (\ref{2.3.2}) is encountered.


\begin{thebibliography}{100}
\expandafter\ifx\csname url\endcsname\relax
  \def\url#1{\texttt{#1}}\fi
\expandafter\ifx\csname urlprefix\endcsname\relax\def\urlprefix{URL }\fi

\bibitem{smo12}
M.~v.~Smoluchowski, Experimentell nachweisbare, der \"ublichen {T}hermodynamik
  widersprechende {M}olekularph\"anomene, Physik. Zeitschr. 13 (1912) 1069.

\bibitem{fey63}
R.~P. Feynman, R.~B. Leighton, M.~Sands, The {F}eynman Lectures on Physics,
  Vol. 1, chapter 46, Addison Wesley, Reading MA, 1963.

\bibitem{bri50}
L.~Brillouin, Can the rectifier become a thermodynamical demon?, Phys. Rev. 78
  (1950) 627.

\bibitem{hux57}
A.~F. Huxley, Muscle structure and theories of contraction, Prog. Biophys. 7
  (1957) 255.

\bibitem{bra88}
S.~M. Braxton, Synthesis and use of a novel class of {ATP} carbamates and a
  ratchet diffusion model for directed motion in muscle, PhD thesis, Washington
  State University, Pullman, WA, 1988.

\bibitem{bra89}
S.~Braxton, R.~G. Yount, A ratchet diffusion model for directed motion in
  muscle, Biophys. J. 55 (1989) 12a (abstr.).

\bibitem{val90}
R.~D. Vale, F.~Oosawa, Protein motors and {M}axwell's demons: Does
  mechanochemical transduction involve a thermal ratchet?, Adv. Biophys. 26
  (1990) 97.

\bibitem{lei91}
S.~Leibler, D.~A. Huse, A physical model for motor proteins, C.~R. Acad. Sci.
  Paris t. 313, Serie III (1991) 27.

\bibitem{lei93}
S.~Leibler, D.~A. Huse, Porters versus rowers: A unified stochastic model of
  motor proteins, J. Cell Biol. 121 (1993) 1357.

\bibitem{cor92}
N.~J. Cordova, B.~Ermentrout, G.~F. Oster, Dynamics of single-motor molecules:
  The thermal ratchet model, Proc. Natl. Acad. Sci. USA 89 (1992) 339.

\bibitem{mag93}
M.~O. Magnasco, Forced thermal ratchets, Phys. Rev. Lett. 71 (1993) 1477.

\bibitem{mag94}
M.~O. Magnasco, Molecular combustion motors, Phys. Rev. Lett. 72 (1994) 2656.

\bibitem{pro94}
J.~Prost, J.-F. Chauwin, L.~Peliti, A.~Ajdari, Asymmetric pumping of particles,
  Phys. Rev. Lett. 72 (1994) 2652.

\bibitem{jul97a}
F.~J\"ulicher, A.~Ajdari, J.~Prost, Modeling molecular motors, Rev. Mod. Phys.
  69 (1997) 1269.

\bibitem{ast94}
R.~D. Astumian, M.~Bier, Fluctuation driven ratchets: molecular motors, Phys.
  Rev. Lett. 72 (1994) 1766.

\bibitem{ast96a}
R.~D. Astumian, M.~Bier, Mechanochemical coupling of the motion of molecular
  motors to {ATP} hydrolysis, Biophys. J. 70 (1996) 637.

\bibitem{pes94}
C.~S. Peskin, G.~B. Ermentrout, G.~F. Oster, The correlation ratchet: a novel
  mechanism for generating directed motion by {ATP} hydrolysis, in: V.~C. Mov,
  F.~Guilak, R.~{Tran-Son-Tay}, R.~M. Hochmuth (Eds.), Cell Mechanics and
  Cellular Engineering, Springer, New York, 1994.

\bibitem{pes95}
C.~S. Peskin, G.~Oster, Coordinated hydrolysis explains the mechanical behavior
  of kinesin, Biophys. J. 68 (1995) 202s.

\bibitem{ser83}
E.~H. Serpersu, T.~Y. Tsong, Stimulation a {O}ubain-sensitive {R}b$^+$ uptake
  in human erthrocytes with an external electric field, J. Membr. Biol. 74
  (1983) 191.

\bibitem{ser84}
E.~H. Serpersu, T.~Y. Tsong, Activation of electrogenic {R}b$^+$ transport of
  {(Na,K)-ATP}ase by an electric field, J. Biol. Chem. 259 (1984) 7155.

\bibitem{tso86}
T.~Y. Tsong, R.~D. Astumian, Absorption and conversion of electric field energy
  by membrane bound {ATP}ase, Bioelectrochem. Bioenerg. 15 (1986) 457.

\bibitem{wes86}
H.~V. Westerhoff, T.~Y. Tsong, P.~B. Chock, Y.~Chen, R.~D. Astumian, How
  enzymes can capture and transmit free energy from an oscillating electric
  field, Proc. Natl. Acad. Sci. USA 83 (1986) 4734.

\bibitem{bio83}
W.~Hoppe, W.~Lohmann, H.~Markl, H.~{Ziegler (Editors)}, Biophysics, Springer,
  Berlin, 1983.

\bibitem{fri86}
M.~H. Friedman, Principles and models of biological Transport, Springer,
  Berlin, 1986.

\bibitem{wae67}
A.~de~Waele, W.~H. Kraan, R.~de~Bruin~Ouboter, K.~W. Taconis, On the dc voltage
  across a double point contact between two superconductors at zero applied dc
  current in situations in which the junction is in the resistive region due to
  the circulating current of flux quantization, Physica (Utrecht) 37 (1967)
  114.

\bibitem{wae69}
A.~de~Waele, R.~de~Bruin~Ouboter, Quantum-interference phenomena in point
  contacts between two superconductors, Physica (Utrecht) 41 (1969) 225.

\bibitem{gla74}
A.~M. Glas, D.~van~der Linde, T.~J. Negran, High-voltage bulk photovoltaic
  effect and the photorefractive process in {LiNbO}$_3$, Appl. Phys. Lett. 25
  (1974) 233.

\bibitem{bel80}
V.~I. Belinicher, B.~I. Sturman, The photogalvanic effect in media lacking a
  center of symmetry, Sov. Phys. Usp. 23 (1980) 199, [Usp. Fiz. Nauk. 130
  (1980) 415].

\bibitem{stu92}
B.~I. Sturman, V.~M. Fridkin, The photovoltaic and photorefractive effects in
  noncentrosymmetric materials, Gordon and Breach, Philadelphia, 1992.

\bibitem{see78}
K.~Seeger, W.~Maurer, Nonlinear electronic transport in {TTF}-{TCNQ} observed
  by microwave harmonic mixing, Solid State Commun. 27 (1978) 603.

\bibitem{won79}
W.~Wonneberger, Stochastic theory of harmonic microwave mixing in periodic
  potentials, Solid State Commun. 30 (1979) 511.

\bibitem{bug87}
A.~L.~R. Bug, B.~J. Berne, Shaking-induced transition to a nonequilibrium
  state, Phys. Rev. Lett. 59 (1987) 948.

\bibitem{but87}
M.~B\"uttiker, Transport as a consequence of state-dependent diffusion, Z.
  Phys. B 68 (1987) 161.

\bibitem{ajd92}
A.~Ajdari, J.~Prost, Mouvement induit par un potentiel periodique de basse
  symmetrie: dielectrophorese pulsee, C. R. Acad. Sci. Paris t. 315, S\'erie II
  (1992) 1635.

\bibitem{doe94}
C.~R. Doering, W.~Horsthemke, J.~Riordan, Nonequilibrium fluctuation-induced
  transport, Phys. Rev. Lett. 72 (1994) 2984.

\bibitem{hon94}
T.~Hondou, Symmerty breaking by correlated noise in a multistable system, J.
  Phys. Soc. Jpn. 63 (1994) 2014.

\bibitem{mil94}
M.~M. Millonas, M.~I. Dykman, Transport and current reversal in stochastically
  driven ratchets, Phys. Lett. A 185 (1994) 65.

\bibitem{rou94}
J.~Rousselet, L.~Salome, A.~Ajdari, J.~Prost, Directional motion of {B}rownian
  particles induced by a periodic asymmetric potential, Nature 370 (1994) 446.

\bibitem{ajd94a}
A.~Ajdari, D.~Mukamel, L.~Peliti, J.~Prost, Rectified motion induced by ac
  forces in periodic structures, J. Phys. I France 4 (1994) 1551.

\bibitem{cha94}
J.-F. Chauwin, A.~Ajdari, J.~Prost, Force-free motion in asymmetric structures:
  a mechanism without diffusive steps, Europhys. Lett. 27 (1994) 421.

\bibitem{ajd94}
A.~Ajdari, Force-free motion in an asymmetric environment: a simple model for
  structured objects, J. Phys. I France 4 (1994) 1577.

\bibitem{bar94}
R.~Bartussek, P.~H\"anggi, J.~G. Kissner, Periodically rocked thermal ratchets,
  Europhys. Lett. 28 (1994) 459.

\bibitem{mad1}
J.~Maddox, Making models of muscle contraction, Nature 365 (1993) 203.

\bibitem{mad2}
J.~Maddox, More models of muscle contraction, Nature 368 (1994) 287.

\bibitem{mad3}
J.~Maddox, Directed motion from random noise, Nature 369 (1994) 181.

\bibitem{lei94}
S.~Leibler, Moving forward noisily, Nature 370 (1994) 412.

\bibitem{pop94}
C.~P\"oppe, Die ordnende {K}raft der {A}symmetrie, Spektrum der Wissenschaft,
  November issue (1994) 38.

\bibitem{bar95a}
R.~Bartussek, P.~H\"anggi, {B}rownsche {M}otoren, Phys. Bl. 51 (1995) 506.

\bibitem{doe95}
C.~R. Doering, Randomly rattled ratchets, Il Nuovo Cimento D 17 (1995) 685.

\bibitem{faz95}
C.~Ettl, Perpetuum mobile zweiter {A}rt, Frankfurter Allgemeine Zeitun{g,} 5.
  April (1995) 3.

\bibitem{han96}
P.~H\"anggi, R.~Bartussek, {B}rownian rectifiers: how to convert {B}rownian
  motion into directed transport, in: J.~Parisi, S.~C. M\"uller, W.~Zimmermann
  (Eds.), Lecture notes in Physics Vol. 476: Nonlinear Physics of Complex
  Systems, Springer, Berlin, 1996.

\bibitem{kos96}
K.~Kostur, J.~Luczka, Transport in ratchet-type systems, Acta Phys. Polon. B 27
  (1996) 663.

\bibitem{luc96}
J.~Luczka, Ratchets, molecular motors, and noise-induced transport, Cell. Mol.
  Biol. Lett. 1 (1996) 311.

\bibitem{ast97}
R.~D. Astumian, Thermodynamics and kinetics of a {B}rownian motor, Science 276
  (1997) 917.

\bibitem{bie97}
M.~Bier, {B}rownian ratchets in physics and biology, Contemp. Phys. 38 (1997)
  371.

\bibitem{bie97z}
M.~Bier, A motor protein model and how it relates to stochastic resonance,
  {F}eynman's ratchet, and {M}axwell's demon, in: L.~Schimansky-Geier,
  T.~P\"oschel (Eds.), Lecture Notes in Physics, Vol. 484, Springer, Berlin,
  1997.

\bibitem{cha98}
R.~D. {Astumian and F. Moss (editors)}, Focus issue: The constructive role of
  noise in fluctuation driven transport and stochastic resonance, Chaos 8
  (1998) 533--664.

\bibitem{luc99}
J.~Luczka, Application of statistical mechanics to stochastic transport,
  Physica A 274 (1999) 200.

\bibitem{ast00}
R.~D. Astumian, Ratchets, rectifiers, and demons: the constructive role of
  noise in free energy and signal transduction, in: J.~Walleczek (Ed.),
  Self-organized biological dynamics and nonlinear control, Cambridge
  University Press, Cambridge, 2000.

\bibitem{nzz01}
C.~Speicher, Die {K}analisierung des {Z}ufalls, Neue Z\"urcher Zeitun{g,} 9.
  Mai (2001) 49.

\bibitem{ast01a}
R.~D. Astumian, Making molecules into motors, Scientific American, July issue
  (2001) 56.

\bibitem{gam98}
L.~Gammaitoni, P.~H\"anggi, P.~Jung, F.~Marchesoni, Stochastic resonance, Rev.
  Mod. Phys. 70 (1998) 223.

\bibitem{hor84}
W.~Horsthemke, R.~Lefever, Noise-induced transitions, Springer, Berlin, 1984.

\bibitem{vdb97}
C.~{Van den Broeck}, J.~M.~R. Parrondo, R.~Toral, R.~Kawai, Nonequilibrium
  phase transitions induced by multiplicative noise, Phys. Rev. E 55 (1997)
  4084.

\bibitem{gar99}
J.~Garcia-Ojalvo, J.~M. Sancho, Noise in spatially extended systems, Springer,
  New York, 1999.

\bibitem{han90}
P.~H\"anggi, P.~Talkner, M.~Borkovec, Reaction rate theory: fifty years after
  {K}ramers, Rev. Mod. Phys. 62 (1990) 251.

\bibitem{han95}
P.~H\"anggi, P.~Jung, Colored noise in dynamical systems, Adv. Chem. Phys. 89
  (1995) 239.

\bibitem{rei97ra}
P.~Reimann, P.~H\"anggi, Surmounting fluctuating barriers: basic concepts and
  results, in: L.~Schimansky-Geier, T.~P\"oschel (Eds.), Lecture Notes in
  Physics, Vol. 484, Springer, Berlin, 1997.

\bibitem{sch95c}
B.~Schmittmann, R.~K.~P. Zia, Statistical mechanics of driven diffusive
  systems, in: C.~Domb, J.~L. Lebowitz (Eds.), Phase Transitions and Critical
  Phenomena, Vol. 17, Academic Press, London, 1995.

\bibitem{sch00c}
G.~M. Sch\"utz, Exactly solvable models for many-body systems far from
  equilibrium, in: C.~Domb, J.~L. Lebowitz (Eds.), Phase Transitions and
  Critical Phenomena, Vol. 19, Academic Press, London, 2000.

\bibitem{max71}
J.~C. Maxwell, Theory of Heat, Longmans, Green and Co., London, 1872.

\bibitem{lef90}
H.~S. Leff, A.~F. Rex, {M}axwell's demon, entropy, information, computing, Adam
  Hilger, Bristol, 1990.

\bibitem{kel97}
T.~R. Kelly, I.~Tellitu, J.~P. Sestelo, In search of molecular ratchets, Angew.
  Chem. Int. Ed. Engl. 36 (1997) 1866.

\bibitem{kel98}
T.~R. Kelly, J.~P. Sestelo, I.~Tellitu, New molecular devices: In search of a
  molecular ratchet, J. Org. Chem. 63 (1998) 3655.

\bibitem{dav98}
A.~P. Davis, Tilting at windmills? {T}he second law survives, Angew. Chem. Int.
  Ed. Engl. 37 (1998) 909.

\bibitem{seb00}
K.~L. Sebastian, Molecular ratchets: verification of the principle of detailed
  balance and the second law of dynamics, Phys. Rev. E 61 (2000) 937.

\bibitem{ein05}
A.~Einstein, \"{U}ber die von der molekularkinetischen {T}heorie der {W}\"arme
  geforderte {B}ewegung von in ruhenden {F}\"ussigkeiten suspendierten
  {T}eilchen, Ann. Phys. (Leipzig) 17 (1905) 549.

\bibitem{ein10}
A.~Einstein, L.~Hopf, Statistische {U}ntersuchung der {B}ewegung eines
  {R}esonators in einem {S}trahlungsfeld, Ann. Phys. (Leipzig) 33 (1910) 1105.

\bibitem{joh28}
J.~B. Johnson, Thermal agitation of electricity in conductors, Phys. Rev. 32
  (1928) 97.

\bibitem{nyq28}
H.~Nyquist, Thermal agitation of electric charge in conductors, Phys. Rev. 32
  (1928) 110.

\bibitem{cal51}
H.~B. Callen, T.~A. Welton, Irreversibility and generalized noise, Phys. Rev.
  83 (1951) 34.

\bibitem{ber55}
P.~G. Bergmann, J.~L. Lebowitz, New approach to nonequilibrium processes, Phys.
  Rev. 99 (1955) 578.

\bibitem{leb57}
J.~L. Lebowitz, P.~G. Bergmann, Irreversible {G}ibbsian ensembles, Ann. Phys.
  (New York) 1 (1957) 1.

\bibitem{mag59}
V.~B. Magalinskii, Dynamical model in the theory of the {B}rownian motion, Sov.
  Phys. JETP 9 (1959) 1381, [JETP 36 (1959) 1942].

\bibitem{rub60}
R.~J. Rubin, Statistical dynamics of simple cubic lattices. {M}odel for the
  study of {B}rownian motion, J. Math. Phys. 1 (1960) 309.

\bibitem{leb63}
J.~L. Lebowitz, E.~Rubin, Dynamical study of {B}rownian motion, Phys. Rev. 131
  (1963) 2381.

\bibitem{res64}
P.~Resibois, H.~T. Davis, Transport equation of a {B}rownian particle in an
  external field, Physica (Utrecht) 30 (1964) 1077.

\bibitem{ull66}
P.~Ullersma, An exactly solvable model for {B}rownian motion, Physica (Utrecht)
  32 (1966) 27, 56, 74, and 90.

\bibitem{zwa73}
R.~Zwanzig, Nonlinear generalized {L}angevin equations, J. Stat. Phys. 9 (1973)
  215.

\bibitem{hyn75}
J.~T. Hynes, J.~M. Deutch, Nonequilibrium problems -- projection operator
  techniques, in: D.~Henerson (Ed.), Physical Chemistry, and advanced treatise,
  Academic Press, New York, 1975.

\bibitem{spo78}
H.~Spohn, J.~L. Lebowitz, Irreversible thermodynamics for quantum systems
  weakly coupled to thermal reservoirs, Adv. Chem. Phys. 38 (1978) 109.

\bibitem{gra80}
H.~Grabert, P.~H\"anggi, P.~Talkner, Microdynamics and nonlinear stochastic
  processes of gross variables, J. Stat. Phys. 22 (1980) 537.

\bibitem{gra82a}
H.~Grabert, Projection operator techniques in nonequilibrium statistical
  mechanics, Springer, Berlin, 1982.

\bibitem{cal83}
A.~O. Caldeira, A.~J. Leggett, Quantum tunneling in dissipative systems, Ann.
  Phys. (New York) 149 (1983) 374, erratum: Ann. Phys. (New York) 153 (1984)
  445.

\bibitem{for88}
G.~W. Ford, J.~T. Lewis, R.~F. O'Connell, Quantum {L}angevin equation, Phys.
  Rev. A 37 (1988) 4419.

\bibitem{wei99a}
U.~Weiss, Quantum dissipative systems, second enlarged Edition, World
  Scientific, Singapore, 1999.

\bibitem{rei01a}
P.~Reimann, A uniqueness-theorem for ``linear'' thermal baths, Chem. Phys. 268
  (2001) 337.

\bibitem{gar83}
C.~W. Gardiner, Handbook of stochastic methods for physics, chemistry, and the
  natural sciences, Springer, Berlin, 1983.

\bibitem{ris84}
H.~Risken, The {F}okker-{P}lanck Equation, Springer, Berlin, 1984.

\bibitem{kam92}
N.~G. van Kampen, Stochastic processes in physics and chemistry, revised and
  enlarged Edition, North-Holland, Amsterdam, 1992.

\bibitem{han82b}
P.~H\"anggi, H.~Thomas, Stochastic processes: time evolution, symmetries, and
  linear response, Phys. Rep. 88 (1982) 207.

\bibitem{kra40}
H.~A. Kramers, {B}rownian motion in a field of force and the diffusion model of
  chemical reactions, Physica (Utrecht) 8 (1940) 284.

\bibitem{han84}
P.~H\"anggi, H.~Grabert, P.~Talkner, H.~Thomas, Bistable systems: master
  equation versus {F}okker-{P}lanck modeling, Phys. Rev. A 29 (1984) 371.

\bibitem{zwa90}
R.~Zwanzig, Rate processes with dynamical disorder, Acc. Chem. Res. 23 (1990)
  148.

\bibitem{rys97}
G.~Ryskin, Simple procedure for correcting equations of evolution: application
  to {M}arkov processes, Phys. Rev. E 56 (1997) 5123.

\bibitem{kam97}
N.~G. van Kampen, Die {F}okker-{P}lanck-{G}leichung, Phys. Bl. 53 (1997) 1012.

\bibitem{cha43}
S.~Chandrasekhar, Stochastic problems in physics and astronomy, Rev. Mod. Phys.
  15 (1943) 1.

\bibitem{lan54}
P.~T. Landsberg, Method of transition probabilities in quantum mechanics and
  quantum statistics, Phys. Rev. 96 (1954) 1420.

\bibitem{sch80}
F.~Schl\"ogl, Stochastic measures in nonequilibrium thermodynamics, Phys. Rep.
  62 (1980) 267.

\bibitem{par96}
J.~M.~R. Parrondo, P.~Espanol, Criticism of {F}eynman's analysis of the ratchet
  as an engine, Am. J. Phys. 64 (1996) 1125.

\bibitem{mag98}
M.~O. Magnasco, G.~Stolovitzky, Feynman's ratchet and pawl, J. Stat. Phys. 93
  (1998) 615.

\bibitem{str58}
R.~L. Stratonovich, Oscillator synchronization in the presence of noise,
  Radiotekhnika i elektronika 3 (1958) 497, english translation in {\em
  Non-linear transformations of stochastic processes}, edited by P.~I.
  Kuznetsov, R.~L Stratonovich, V.~I. Tikhonov, Pergamon press, Oxford 1965.

\bibitem{iva69}
Y.~M. Ivanchenko, L.~A. Zil'berman, The {J}osephson effect for small tunnel
  contacts, Sov. Phys. JETP 28 (1969) 1272, [Zh. Eksp. Teor. Fiz 55 (1968)
  2395].

\bibitem{amb69}
V.~Ambegaokar, B.~I. Halperin, Voltage due to thermal noise in the dc
  {J}osephson effect, Phys. Rev. Lett. 22 (1969) 1364.

\bibitem{str69}
R.~L. Stratonovich, Theory of Random Noise, Gordon and Breach, 1969.

\bibitem{cec96}
G.~Cecchi, M.~O. Magnasco, Negative resistance and rectification in {B}rownian
  transport, Phys. Rev. Lett. 76 (1996) 1968.

\bibitem{rei01}
P.~Reimann, C.~{Van den Broeck}, H.~Linke, P.~H\"anggi, J.~M. Rubi,
  A.~P\'erez-Madrid, Giant acceleration of free diffusion by use of tilted
  periodic potentials, Phys. Rev. Lett. 87 (2001) 010602.

\bibitem{rei96}
P.~Reimann, R.~Bartussek, R.~H\"aussler, P.~H\"anggi, {B}rownian motors driven
  by temperature oscillations, Phys. Lett. A 215 (1996) 26.

\bibitem{iba97}
L.~Ibarra-Bracamontes, V.~Romero-Rochin, Stochastic ratchets with colored
  noise, Phys. Rev. E 56 (1997) 4048.

\bibitem{doe98a}
C.~R. Doering, Stochastic ratchets, Physica A 254 (1998) 1.

\bibitem{sim92}
S.~M. Simon, C.~S. Peskin, G.~F. Oster, What drives the translocation of
  proteins, Proc. Natl. Acad. Sci. USA 89 (1992) 3770.

\bibitem{pes93}
C.~S. Peskin, G.~M. Odell, G.~F. Oster, Cellular motions and thermal
  fluctuations: the {B}rownian ratchet, Biophys. J. 65 (1993) 316.

\bibitem{kuo00}
S.~C. Kuo, J.~L. McGrath, Steps and fluctuations of {L}isteria monocytogenes
  during actin-based motility, Nature 407 (2000) 1026.

\bibitem{els00c}
T.~C. Elston, Models of post-translational protein translocation, Biophys. J.
  79 (2000) 2235.

\bibitem{lie01}
W.~Liebermeister, T.~A. Rapoport, R.~Heinrich, Ratcheting in post-translational
  protein translocation : a mathematical model, J. Mol. Biol. 305 (2001) 643.

\bibitem{luc97}
J.~Luczka, T.~Czernik, P.~H\"anggi, Symmetric white noise can induce directed
  current in ratchets, Phys. Rev. E 56 (1997) 3968.

\bibitem{li97}
Y.-X. Li, Transport generated by fluctuating temperature, Physica A 238 (1997)
  245.

\bibitem{sok97}
M.~I. Sokolov, A.~Blumen, Non-equilibrium directed diffusion and inherently
  irreversible heat engines, J. Phys. A 30 (1997) 3021.

\bibitem{sok98}
I.~M. Sokolov, A.~Blumen, Thermodynamical and mechanical efficiency of a
  ratchet pump, Chem. Phys. 235 (1998) 39.

\bibitem{bao00}
J.-D. Bao, Efficiency of energy transformation in an underdamped diffusion
  ratchet, Phys. Lett. A 267 (2000) 122.

\bibitem{bao99a}
J.~D. Bao, S.~J. Liu, Broad-band colored noise: digital simulation and
  dynamical effect, Phys. Rev. E 60 (1999) 7572.

\bibitem{ral84}
K.~S. Ralls, W.~J. Skocpol, L.~D. Jackel, R.~E. Howard, L.~A. Fetter, R.~W.
  Epworth, D.~M. Tennant, Discrete resistance switching in submicrometer
  silicon inversion layers: individual interface traps and low-frequency (1/f?)
  noise, Phys. Rev. Lett. 52 (1984) 228.

\bibitem{mul92}
C.~J. M\"uller, J.~M. van Ruitenbeek, L.~J. de~Jongh, Conductance and
  supercurrent discontinuities in atomic-scale metallic constrictions of
  variable width, Phys. Rev. Lett. 69 (1992) 140.

\bibitem{gol92}
B.~Golding, N.~M. Zimmerman, S.~N. Coppersmith, Dissipative quantum tunneling
  of a single microscopic defect in a mesoscopic metal, Phys. Rev. Lett. 69
  (1992) 998.

\bibitem{ral92}
D.~C. Ralph, R.~A. Buhrman, Observation of {K}ondo-scattering without magnetic
  impurities: a point contact study of two-level tunneling systems in metals,
  Phys. Rev. Lett. 69 (1992) 2118.

\bibitem{kei96}
R.~J. Keijsers, O.~I. Shklyarevskii, H.~van Kempen, Point contact study of fast
  and slow two-level fluctuators in metallic glasses, Phys. Rev. Lett. 77
  (1996) 3411.

\bibitem{kog96}
S.~Kogan, Electronic noise and fluctuations in solids, Cambridge University
  Press, Cambridge, 1996.

\bibitem{smi96}
J.~C. Smith, C.~Berven, S.~M. Goodnick, M.~N. Wybourne, Nonequilibrium ransom
  telegraph switching in quantum point contacts, Physica B 227 (1996) 197.

\bibitem{bri97}
J.~Brini, P.~Chenevier, P.~d'Onofrino, P.~Hruska, Higher order statistics of
  the thermal noise of ultrasmall {MOSFET}'s, in: C.~Claeys, E.~Simeon (Eds.),
  Noise in Physical Systems and 1/f fluctuations, World Scientific, Singapore,
  1997.

\bibitem{mul83}
A.~W. M\"uller, Thermoelectric energy conversion could be an energy source of
  living organisms, Phys. Lett. A 96 (1983) 319.

\bibitem{mul95}
A.~W. M\"uller, Were the first organisms heat engines? {A} new model for
  biogenesis and the early evolution of biological energy conversion, Prog.
  Biophys. Molec. Biol. 63 (1995) 193.

\bibitem{hun94}
A.~J. Hunt, F.~Gittes, J.~Howard, The force exerted by a single kinesin
  molecule against a viscous load, Biophys. J. 67 (1994) 766.

\bibitem{how96}
J.~Howard, The movement of kinesin along microtubules, Annu. Rev. Physiol. 58
  (1996) 703.

\bibitem{mit88}
T.~Mitsui, H.~Oshima, A self-induced translational model of myosin head motion
  in contracting muscle. {I}. {F}orce-velocity relation and energy liberation,
  J. Musc. Res. Cell Motil. 9 (1988) 248.

\bibitem{bie00a}
M.~Bier, M.~Kostur, Nonlinearly coupled chemical reactions, in: J.~A. Freund,
  T.~P\"oschel (Eds.), Lecture Notes in Physics, Vol. 557, Springer, Berlin,
  2000.

\bibitem{bie00}
M.~Bier, M.~Kostur, I.~Der\'enyi, R.~D. Astumian, Nonlinearly coupled flows,
  Phys. Rev. E 61 (2000) 7184.

\bibitem{curie}
P.~Curie, Sur la sym\'etrie dans les ph\'enomenes physiques, sym\'etrie d'un
  champ \'electrique et d'un champ magn\'etique, J. Phys. (Paris) 3. S\'erie
  (th\'eorique et appliqu\'e) t. III (1894) 393.

\bibitem{gra71a}
R.~Graham, H.~Haken, Generalized thermodynamic potential for {M}arkoff systems
  in detailed balance and far from thermal equilibrium, Z. Phys. 243 (1971)
  289.

\bibitem{ons31}
L.~Onsager, Reciprocal relations in irreversible processes {I}, Phys. Rev. 37
  (1931) 405.

\bibitem{gre52}
M.~S. Green, {M}arkoff random processes and the statistical mechanics of
  time-dependent phenomena, J. Chem. Phys. 20 (1952) 1281.

\bibitem{kam57}
N.~G. van Kampen, Derivation of the phenomenological equations from the master
  equation, Physica (Utrecht) 23 (1957) 707 and 816.

\bibitem{gra71b}
R.~Graham, H.~Haken, Fluctuations and stability of stationary non-equilibrium
  systems in detailed balance, Z. Phys. 245 (1971) 141.

\bibitem{kam65}
N.~G. van Kampen, Fluctuations in nonlinear systems, in: R.~E. Burgess (Ed.),
  Fluctuation phenomena in solids, Academic Press, New York, 1965.

\bibitem{mcf71}
R.~McFee, Self-rectification in diodes and the second law of thermodynamics,
  Am. J. Phys. 39 (1971) 814.

\bibitem{str92}
R.~L. Stratonovich, Nonlinear nonequilibrium thermodynamics {I}, Springer,
  Berlin, 1992.

\bibitem{lan98}
P.~S. Landa, Noise-induced transport of {B}rownian particles with consideration
  for their mass, Phys. Rev. E 58 (1998) 1325.

\bibitem{sok98a}
I.~M. Sokolov, On the energetics of a nonlinear system rectifying thermal
  fluctuations, Europhys. Lett. 44 (1998) 278.

\bibitem{ast87}
R.~D. Astumian, P.~B. Chock, T.~Y. Tsong, Y.~Chen, H.~V. Westerhoff, Can free
  energy be transduced from electric noise?, Proc. Natl. Acad. Sci. USA 84
  (1987) 434.

\bibitem{koc75}
W.~T.~H. Koch, R.~Munser, W.~Ruppel, P.~W\"urfel, Bulk photovoltaic effect in
  {BaTiO}$_3$, Solid State Commun. 17 (1975) 847.

\bibitem{asn79}
V.~M. Asnin, A.~A. Bakun, A.~M. Danishevskii, E.~L. Ivchenko, G.~E. Pikus,
  A.~A. Rogachev, ``{C}ircular'' photogalvanic effect in optically active
  crystals, Solid State Commun. 30 (1979) 565.

\bibitem{rei98}
P.~Reimann, P.~H\"anggi, Quantum features of {B}rownian motors and stochastic
  resonance, Chaos 8 (1998) 629.

\bibitem{ari96}
C.~M. Arizmendi, F.~Family, Approach to steady state current in ratchets,
  Physica A 232 (1996) 119.

\bibitem{han97a}
K.~Handrich, F.-P. Ludwig, Friction coefficients and directed motion of
  asymmetric test particles, J. Stat. Phys. 86 (1997) 1067.

\bibitem{kol98}
A.~Kolomeisky, B.~Widom, A simplified ``ratchet'' model of molecular motors, J.
  Stat. Phys. 93 (1998) 633.

\bibitem{yev00}
O.~Yevtushenkov, S.~Flach, K.~Richter, ac-driven phase-dependent directed
  current, Phys. Rev. E 61 (2000) 7215.

\bibitem{goy00}
I.~Goychuk, P.~H\"anggi, Directed current without dissipation: re-incarnation
  of a {M}axwell-{L}oschmidt-demon, in: J.~A. Freund, T.~P\"oschel (Eds.),
  Lecture Notes in Physics, Vol. 557, Springer, Berlin, 2000.

\bibitem{goy01}
I.~Goychuck, P.~H\"anggi, Minimal quantum {B}rownian rectifiers, J. Phys. Chem.
  105 (2001) 6642.

\bibitem{cox67}
R.~D. Cox, Renewal Theory, Methuen and Co., London, 1967.

\bibitem{vdb89}
C.~{Van den Broeck}, A glimpse into the world of random walks, in: J.~L.
  Munoz-Cobo, F.~C. Difilippo (Eds.), Noise and nonlinear phenomena in nuclear
  systems, Plenum Publishing Corporation, 1989.

\bibitem{jun96}
P.~Jung, J.~G. Kissner, P.~H\"anggi, Regular and chaotic transport in
  asymmetric periodic potentials: inertia ratchets, Phys. Rev. Lett. 76 (1996)
  3436.

\bibitem{har97}
T.~Harms, R.~Lipowsky, Driven ratchets with disordered tracks, Phys. Rev. Lett.
  79 (1997) 2895.

\bibitem{fre99}
J.~A. Freund, L.~Schimansky-Geier, Diffusion in discrete ratchets, Phys. Rev. E
  60 (1999) 1304.

\bibitem{con99}
G.~Constantini, F.~Marchesoni, Threshold diffusion in a tilted washboard
  potential, Europhys. Lett. 48 (1999) 491.

\bibitem{lin01}
B.~Lindner, M.~Kostur, L.~Schimansky-Geier, Optimal diffusive transport in a
  tilted periodic potential, Fluct. Noise Lett. 1 (2001) R25.

\bibitem{ket00}
C.~Kettner, P.~Reimann, P.~H\"anggi, F.~M\"uller, Drift ratchet, Phys. Rev. E
  61 (2000) 312.

\bibitem{kly77}
V.~I. Klyatskin, Dynamic systems with parameter fluctuations of the
  telegraphic-process type, Radiophys. Quantum Electron. 20 (1978) 382,
  [Radiofizika 20 (1977) 562].

\bibitem{han83}
P.~H\"anggi, P.~Riseborough, Activation rates in bistable systems in the
  presence of correlated noise, Phys. Rev. A 27 (1983) 3379.

\bibitem{vdb84}
C.~{Van den Broeck}, P.~H\"anggi, Activation rates for nonlinear stochastic
  flows driven by non-{G}aussian noise, Phys. Rev. A 30 (1984) 2730.

\bibitem{han96a}
P.~H\"anggi, R.~Bartussek, P.~Talkner, J.~Luczka, Noise-induced transport in
  symmetric periodic potentials: white shot noise versus deterministic noise,
  Europhys. Lett. 35 (1996) 315.

\bibitem{chi97}
D.~R. Chialvo, M.~I. Dykman, M.~M. Millonas, Fluctuation-induced transport in a
  periodic potential: noise versus chaos, Phys. Rev. Lett. 78 (1997) 1605.

\bibitem{neu01}
E.~Neumann, A.~Pikovsky, Quasiperiodically driven {J}osephson junctions:
  strange nonchaotic attractors, symmetries, and transport, {S}ubmitted for
  publication.

\bibitem{wei99}
S.~Weiss, D.~Koelle, J.~M\"uller, K.~Barthel, R.~Gross, Ratchet effect in dc
  {SQUIDs}, Europhys. Lett. 51 (2000) 499.

\bibitem{wei00}
S.~Weiss, Ratscheneffekt in supraleitenden {Q}uanteninterferenzdetektoren, PhD
  thesis (in German), Shaker Verlag, Aachen, 2000.

\bibitem{cil01}
S.~Cilla, L.~M. Floria, Mirror symmetry breaking through an internal degree of
  freedom leading to directional motion, Phys. Rev. E 63 (2001) 031110.

\bibitem{mil83}
W.~H. Miller, Reaction-path dynamics for polyatomic systems, J. Chem. Phys. 87
  (1983) 3811.

\bibitem{kel00}
D.~Keller, C.~Bustamante, The mechanochemistry of molecular motors, Biophys. J.
  78 (2000) 541.

\bibitem{ast96b}
R.~D. Astumian, Adiabatic theory for fluctuation-induced transport on a
  periodic potential, J. Phys. Chem. 100 (1996) 19075.

\bibitem{liu90}
D.~S. Liu, R.~D. Astumian, T.~Y. Tsong, Activation of the {N}a$^+$ and
  {R}b$^+$-pumping modes of {(Na,K)-ATP}ase by an oscillating electric field,
  J. Biol. Chem. 265 (1990) 7260.

\bibitem{fey63b}
R.~P. Feynman, F.~L. Vernon, The theory of a general quantum systems
  interacting with a linear dissipative system, Ann. Phys. (New York) 24 (1963)
  118.

\bibitem{mil95}
M.~M. Millonas, Self-consistent microscopic theory of fluctuation-induced
  transport, Phys. Rev. Lett. 74 (1995) 10, erratum: Phys. Rev. Lett. 75 (1995)
  3027.

\bibitem{jay96}
A.~M. Jayannavar, Simple model for {M}axwell's-demon-type information engine,
  Phys. Rev. E 53 (1996) 2957.

\bibitem{han97}
P.~H\"anggi, Generalized {L}angevin equations: a useful tool for the perplexed
  modeler of nonequilibrium fluctuations?, in: L.~Schimansky-Geier,
  T.~P\"oschel (Eds.), Lecture Notes in Physics, Vol. 484, Springer, Berlin,
  1997.

\bibitem{zap98}
I.~Zapata, J.~Luczka, F.~Sols, P.~H\"anggi, Tunneling center as a source of
  voltage rectification in {J}osephson junctions, Phys. Rev. Lett. 80 (1998)
  829.

\bibitem{pos96}
D.~E. Postnov, A.~P. Nikitin, V.~S. Anishchenko, Control of the probability
  flux in a system of phase-controlled frequency self-tuning, Tech. Phys. Lett.
  22 (1996) 352.

\bibitem{nik98}
A.~P. Nikitin, D.~E. Postnov, Effect of particle mass on the behavior of
  stochastic ratchets, Tech. Phys. Lett. 24 (1998) 61.

\bibitem{arr00}
M.~Arrayas, R.~Mannella, P.~V.~E. McClintock, A.~J. McKane, N.~D. Stein,
  Ratchet driven by quasimonochromatic noise, Phys. Rev. E 61 (2000) 139.

\bibitem{schi97}
L.~Schimansky-Geier, M.~Kschischo, T.~Fricke, Flux of particles in sawtooth
  media, Phys. Rev. Lett. 79 (1997) 3335.

\bibitem{par98a}
J.~M.~R. Parrondo, Reversible ratchets as {B}rownian particles in an
  adiabatically changing periodic potential, Phys. Rev. E 57 (1998) 7297.

\bibitem{par98b}
J.~M.~R. Parrondo, J.~M. Blanco, F.~J. Cao, R.~Brito, Efficiency of {B}rownian
  motors, Europhys. Lett. 43 (1998) 248.

\bibitem{hoh99}
E.~M. H\"ohberger, Magnetotransport in lateralen {H}albleiter\"ubergittern
  unter {E}influss von {S}ymmetriebrechung, Diploma thesis (in German),
  Ludwig-Maximilian-Universit\"at M\"unchen (Germany), unpublished, 1999.

\bibitem{par00}
J.~M.~R. Parrondo, B.~{Jimenez de Cisneros}, R.~Brito, Thermodynamics of
  isothermal {B}rownian motors, in: J.~A. Freund, T.~P\"oschel (Eds.), Lecture
  Notes in Physics, Vol. 557, Springer, Berlin, 2000.

\bibitem{hoh01}
E.~M. H\"ohberger, A.~Lorke, W.~Wegscheider, M.~Bichler, Adiabatic pumping of
  two-dimensional electrons in a ratchet-type lateral superlattice, Appl. Phys.
  Lett. 78 (2001) 2905.

\bibitem{wit81}
E.~Witten, Dynamical breaking of supersymmetry, Nucl. Phys. B188 (1981) 513.

\bibitem{dut88}
R.~Dutt, A.~Khare, U.~P. Sukhatme, Supersymmetry, shape invariance, and exactly
  solvable potentials, Am. J. Phys. 56 (1988) 163.

\bibitem{ben83}
C.~M. Bender, F.~Cooper, B.~Freedman, A new strong-coupling expansion for
  quantum field theory based on the {L}angevin equation, Nucl. Phys. B219
  (1983) 61.

\bibitem{ber84}
M.~Bernstein, L.~S. {B}rown, Supersymmetry and the bistable {F}okker-{P}lanck
  equation, Phys. Rev. Lett. 52 (1984) 1933.

\bibitem{mar88}
F.~Marchesoni, P.~Sodano, M.~Zanetti, Supersymmetry and bistable soft
  potentials, Phys. Rev. Lett. 61 (1988) 1143.

\bibitem{jun96a}
G.~Junker, Supersymmetric methods in quantum and statistical physics, Springer,
  Berlin, 1996.

\bibitem{fav67}
L.~F. Favella, Brownian motions and quantum mechanics, Ann. Inst. Henri
  Poincar\'e 7 (1967) 77.

\bibitem{tom76}
H.~Tomita, A.~Ito, H.~Kidachi, Eigenvalue problem of metastability in
  macrosystems, Prog. Theor. Phys. 56 (1976) 786.

\bibitem{jun91a}
P.~Jung, P.~H\"anggi, Amplification of small signals via stochastic resonance,
  Phys. Rev. A 44 (1991) 8032.

\bibitem{lei87}
T.~Leibler, F.~Marchesoni, H.~Risken, Colored noise and bistable
  {F}okker-{P}lanck equations, Phys. Rev. Lett. 59 (1987) 1381, erratum: Phys.
  Rev. Lett. 60 (1988) 659.

\bibitem{lei88}
T.~Leibler, F.~Marchesoni, H.~Risken, Numerical analysis of stochastic
  relaxation in bistable systems driven by colored noise, Phys. Rev. A 38
  (1988) 983.

\bibitem{kan99}
R.~Kanada, K.~Sasaki, Thermal ratchets with symmetric potentials, J. Phys. Soc.
  Jpn. 68 (1999) 3759.

\bibitem{yev01}
O.~Yevtushenko, S.~Flach, Y.~Zolotaryuk, A.~A. Ovchinikov, Rectification of
  current in ac-driven nonlinear systems and symmetry properties of the
  {B}oltzmann equation, Europhys. Lett. 54 (2001) 141.

\bibitem{yan01}
B.~Yan, R.~M. Miura, Y.-D. Chen, Direction reversal of fluctuation-induced
  biased {B}rownian motion in distorted ratchets, J. Theor. Biol. 210 (2001)
  141.

\bibitem{rei01c}
P.~Reimann, Supersymmetric ratchets, Phys. Rev. Lett. 86 (2001) 4992.

\bibitem{bre82}
H.-J. Breymayer, H.~Risken, H.~D. Vollmer, W.~Wonneberger, Harmonic mixing in a
  cosine potential for large damping and arbitrary field strengths, Appl. Phys.
  B 28 (1982) 335.

\bibitem{won84}
W.~Wonneberger, H.-J. Breymayer, Broadband current noise and ac induced current
  steps by a moving charge density wave domain, Z. Phys. B 56 (1984) 241.

\bibitem{bre84}
H.-J. Breymayer, Harmonic mixing in a cosine potential for arbitrary damping,
  Appl. Phys. A 33 (1984) 1.

\bibitem{fla00}
S.~Flach, O.~Yevtushenko, Y.~Zolotaryuk, Directed current due to broken
  time-space symmetry, Phys. Rev. Lett. 84 (2000) 2358.

\bibitem{lin97}
B.~Lindner, L.~Schimansky-Geier, P.~Reimann, P.~H\"anggi, Mass separation by
  ratchets, in: J.~B. Kadtke, A.~Bulsara (Eds.), Applied Nonlinear Dynamics and
  Stochastic Systems near the Millennium, AIP Proc. 411, 1997.

\bibitem{lin99}
B.~Lindner, L.~Schimansky-Geier, P.~Reimann, P.~H\"anggi, M.~Nagaoka, Inertia
  ratchets: a numerical study versus theory, Phys. Rev. E 59 (1999) 1417.

\bibitem{bie96c}
M.~Bier, Reversal of noise induced flow, Phys. Lett. A 211 (1996) 12.

\bibitem{der96z}
I.~Der\'enyi, A.~Ajdari, Collective transport of particles in a ``flashing''
  periodic potential, Phys. Rev. E 54 (1996) R5.

\bibitem{ber97}
C.~Berghaus, U.~Kahlert, J.~Schnakenberg, Current reversal induced by a cyclic
  stochastic process, Phys. Lett. A 224 (1997) 243.

\bibitem{bar97}
R.~Bartussek, P.~H\"anggi, B.~Lindner, L.~Schimansky-Geier, Ratchets driven by
  harmonic and white noise, Physica D 109 (1997) 17.

\bibitem{sch98}
M.~Schreier, P.~Reimann, P.~H\"anggi, E.~Pollak, Giant enhancement of diffusion
  and particle separation in rocked periodic potentials, Europhys. Lett. 44
  (1998) 416.

\bibitem{aba98}
E.~Abad, A.~Mielke, {B}rownian motion in fluctuating periodic potentials, Ann.
  Phys. (Leipzig) 7 (1998) 9.

\bibitem{mat00}
J.~L. Mateos, Chaotic transport and current reversal in deterministic ratchets,
  Phys. Rev. Lett. 84 (2000) 258.

\bibitem{mat01}
J.~L. Mateos, Current reversals in chaotic ratchets, Acta Physica Polonica B 32
  (2001) 307.

\bibitem{kos01}
M.~Kostur, J.~Luczka, Multiple current reversals in {B}rownian ratchets, Phys.
  Rev. E 63 (2001) 021101.

\bibitem{der82}
B.~Derrida, Y.~Pomeau, Classical diffusion on a random chain, Phys. Rev. Lett.
  48 (1982) 627.

\bibitem{der83}
B.~Derrida, Velocity and diffusion constants of a periodic one-dimensional
  hopping model, J. Stat. Phys. 31 (1983) 433.

\bibitem{koz99}
Z.~Koza, General technique of calculating the drift velocity and diffusion
  coefficient in arbitrary periodic systems, J. Phys. A 32 (1999) 7637.

\bibitem{keh97}
K.~W. Kehr, K.~Mussawisade, T.~Wichmann, W.~Dieterich, Rectification by hopping
  motion through nonsymmetric potentials with strong bias, Phys. Rev. E 56
  (1997) R2351.

\bibitem{der98z}
I.~Der\'enyi, C.~Lee, A.-L. Barabasi, Ratchet effect in surface
  electromigration: smoothing surfaces by an ac field, Phys. Rev. Lett. 80
  (1998) 1473.

\bibitem{doe92}
C.~R. Doering, J.~C. Gadoua, Resonant activation over a fluctuating barrier,
  Phys. Rev. Lett. 69 (1992) 2318.

\bibitem{han80ra}
P.~H\"anggi, Dynamics of nonlinear oscillators with fluctuating parameters,
  Phys. Lett. A 78 (1980) 304.

\bibitem{ste90}
D.~L. Stein, C.~R. Doering, R.~G. Palmer, J.~L. van Hemmen, R.~M. McLaughlin,
  Escape over fluctuating barrier: the white noise limit, J. Phys. A 23 (1990)
  L203.

\bibitem{zur93}
U.~Z\"urcher, C.~R. Doering, Thermally activated escape over fluctuating
  barriers, Phys. Rev. E 47 (1993) 3862.

\bibitem{bie93}
M.~Bier, R.~D. Astumian, Matching a diffusive and a kinetic approach for escape
  over a fluctuating barrier, Phys. Rev. Lett. 71 (1993) 1649.

\bibitem{pec94}
P.~Pechukas, P.~H\"anggi, Rates of activated processes with fluctuating
  barriers, Phys. Rev. Lett. 73 (1994) 2772.

\bibitem{han94ra}
P.~H\"anggi, Escape over fluctuating barriers driven by colored noise, Chem.
  Phys. 180 (1994) 157.

\bibitem{rei94ra}
P.~Reimann, Surmounting fluctuating barriers: A simple model in discrete time,
  Phys. Rev. E 49 (1994) 4938.

\bibitem{rei95ra1}
P.~Reimann, Thermally driven escape with fluctuating potentials: A new type of
  resonant activation, Phys. Rev. Lett. 74 (1995) 4576.

\bibitem{rei95ra2}
P.~Reimann, Thermally activated escape with potential fluctuations driven by an
  {O}rnstein-{U}hlenbeck process, Phys. Rev. E 52 (1995) 1579.

\bibitem{mad95}
A.~J.~R. Madureira, P.~H\"anggi, V.~Buonamano, W.~A. Rodriguez, Escape from a
  fluctuating double well, Phys. Rev. E 51 (1995) 3849.

\bibitem{bar95}
R.~Bartussek, A.~J.~R. Madureira, P.~H\"anggi, Surmounting a fluctuating double
  well: a numerical study, Phys. Rev. E 52 (1995) R2149.

\bibitem{rei96ra}
P.~Reimann, T.~C. Elston, {K}ramers rate for thermal plus dichotomous noise
  applied to ratchets, Phys. Rev. Lett. 77 (1996) 5328.

\bibitem{iwa96}
J.~Iwaniszewski, Escape over a fluctuating barrier: limits of small and large
  correlation times, Phys. Rev. E 54 (1996) 3173.

\bibitem{rei98ra}
P.~Reimann, R.~Bartussek, P.~H\"anggi, Reaction rates when barriers fluctuate:
  a singular perturbation approach, Chem. Phys. 235 (1998) 11.

\bibitem{rei99ra}
P.~Reimann, G.~J. Schmid, P.~H\"anggi, Universal equivalence of mean-first
  passage time and {K}ramers rate, Phys. Rev. E 60 (1999) R1.

\bibitem{ank99}
J.~Ankerhold, P.~Pechukas, Mathematical aspects of the fluctuating barrier
  problem. {E}xplicit equilibrium and relaxation solutions, Physica A 261
  (1999) 458.

\bibitem{che87}
Y.~Chen, Asymmetry and external noise-induced free energy transduction, Proc.
  Natl. Acad. Sci. USA 84 (1987) 729.

\bibitem{xie94}
T.~D. Xie, P.~Marszalek, Y.~Chen, T.~Y. Tsong, Recognition and processing of
  randomly fluctuating electric signals by {Na},{K}-{ATP}ase, Biophys. J. 67
  (1994) 1247.

\bibitem{xie97}
T.~D. Xie, Y.~Chen, P.~Marszalek, T.~Y. Tsong, Fluctuation-driven directional
  flow in biochemical cycles: further study of electric activation of {Na},{K}
  pumps, Biophys. J. 72 (1997) 2496.

\bibitem{jun93}
P.~Jung, Periodically driven stochastic systems, Phys. Rep. 234 (1993) 175.

\bibitem{dyk97}
M.~I. Dykman, H.~Rabitz, V.~N. Smelyanskiy, B.~E. Vugmeister, Resonant directed
  diffusion in nonadiabatically driven systems, Phys. Rev. Lett. 79 (1997)
  1178.

\bibitem{sme99}
V.~N. Smelyanskiy, M.~I. Dykman, B.~Golding, Time oscillations of escape rates
  in periodically driven systems, Phys. Rev. Lett. 82 (1999) 3193.

\bibitem{tal99}
P.~Talkner, Stochastic resonance in the semiadiabatic limit, New J. Phys. 1
  (1999) 4.

\bibitem{gra84}
R.~Graham, T.~T\'el, On the weak-noise limit of {F}okker-{P}lanck models, J.
  Stat. Phys. 35 (1984) 729.

\bibitem{leh00}
J.~Lehmann, P.~Reimann, P.~H\"anggi, Surmounting oscillating barriers, Phys.
  Rev. Lett. 84 (2000) 1639.

\bibitem{leh00a}
J.~Lehmann, P.~Reimann, P.~H\"anggi, Surmounting oscillating barriers:
  Path-integral approach for weak noise, Phys. Rev. E 62 (2000) 6282.

\bibitem{bar96}
R.~Bartussek, P.~Reimann, P.~H\"anggi, Precise numerics versus theory for
  correlation ratchets, Phys. Rev. Lett. 76 (1996) 1166.

\bibitem{mie95b}
A.~Mielke, Transport in a fluctuating potential, Ann. Phys. (Leipzig) 4 (1995)
  721.

\bibitem{bao00a}
J.~D. Bao, Y.~Abe, Y.~Z. Zhuo, Competition and cooperation between thermal
  noise and external driving force, Physica A 277 (2000) 127.

\bibitem{pla98}
J.~Plata, Rocked thermal ratchets: the high frequency limit, Phys. Rev. E 57
  (1998) 5154.

\bibitem{mil99}
G.~N. Milstein, M.~V. Tretyakov, Mean velocity of noise-induced transport in
  the limit of weak periodic forcing, J. Phys. A 32 (1999) 5795.

\bibitem{elmvideo}
{\tt http://monet.physik.unibas.ch/\verb+~+elmer/bm}.

\bibitem{cha95}
J.-F. Chauwin, A.~Ajdari, J.~Prost, Current reversal in asymmetric pumping,
  Europhys. Lett. 32 (1995) 373, erratum: Europhys. Lett. 32 (1995) 699.

\bibitem{che99}
Y.~Chen, B.~Yan, R.~Miura, Asymmetry and direction reversal in
  fluctuation-induced biased {B}rownian motion, Phys. Rev. E 60 (1999) 3771.

\bibitem{fau95a}
L.~P. Faucheux, A.~Libchaber, Selection of {B}rownian particles, J. Chem. Soc.
  Faraday Trans. 91 (1995) 3163.

\bibitem{sch00d}
T.~Schnelle, T.~M\"uller, G.~Gradl, S.~G. Shirley, G.~Fuhr, Dielectrophoretic
  manipulation of suspended submicron particles, Electrophoresis 21 (2000) 66.

\bibitem{fau95b}
L.~P. Faucheux, L.~S. Bourdieu, P.~D. Kaplan, A.~Libchaber, Optical thermal
  ratchet, Phys. Rev. Lett. 74 (1995) 1504.

\bibitem{gor97}
L.~Gorre-Talini, S.~Jeanjean, P.~Silberzan, Sorting of {B}rownian particles by
  pulsed application of an asymmetric potential, Phys. Rev. E 56 (1997) 2025.

\bibitem{gor98}
L.~Gorre-Talini, J.~P. Spatz, P.~Silberzan, Dielectrophoretic ratchets, Chaos 8
  (1998) 650.

\bibitem{bad99}
J.~S. Bader, R.~W. Hammond, S.~A. Henck, M.~W. Deem, G.~A. McDermott, J.~M.
  Bustillo, J.~W. Simpson, G.~T. Mulhern, J.~M. Rothberg, {DNA} transport by a
  micromachined {B}rownian ratchet device, Proc. Natl. Acad. Sci. USA 96 (1999)
  13165.

\bibitem{ham00}
R.~W. Hammond, J.~S. Bader, S.~A. Henck, M.~W. Deem, G.~A. McDermott, J.~M.
  Bustillo, J.~M. Rothberg, Differential transport of {DNA} by a rectified
  {B}rownian motion device, Electrophoresis 21 (2000) 74.

\bibitem{row97}
L.~Rowen, G.~Mahairas, L.~Hood, Sequencing the human genome, Science 278 (1997)
  605.

\bibitem{3inchaos}
E.~Lai, B.~W. Birren (Eds.), Electrophoresis of large {DNA} molecules, Cold
  Spring Harbor Laboratory Press, Cold Spring Harbor, NY, 1990.

\bibitem{4inchaos}
G.~F\"uhr, U.~Zimmermann, S.~Shirley, Cell motion in time varying fields:
  principles and potential, in: U.~Zimmermann, S.~Neil (Eds.),
  Electromanipulation of Cells, CRC Press, Boca Raton, 1996, p. 259.

\bibitem{ert98}
D.~Ertas, Lateral separation of macromolecules and polyelectrolytes in
  microlithographic arrays, Phys. Rev. Lett. 80 (1998) 1548.

\bibitem{tar98}
M.~B. Tarlie, R.~D. Astumian, Optimal modulation of a {B}rowinan ratchet and
  enhanced sensitivity to a weak external force, Proc. Natl. Acad. Sci. USA 95
  (1998) 2039.

\bibitem{ajd00}
A.~Ajdari, Pumping liquids using asymmetric electrode arrays, Phys. Rev. E 61
  (2000) R45.

\bibitem{jan94}
I.~Janossy, Molecular interpretation of the absorbtion-induced optical
  reorientation of nematic liquid crystals, Phys. Rev. E 49 (1994) 2957.

\bibitem{pal98}
P.~{Palffy-Muhoray}, E.~Weinan, Orientational ratchets and angular momentum
  balance in the {J}anossy effect, Mol. Cryst. Liq. Cryst. 320 (1998) 193.

\bibitem{kos00a}
T.~Kosa, E.~Weinan, P.~{Palffy-Muhoray}, Brownian motors in the photoalignment
  of liquid crystals, Int. J. Eng. Sci. 38 (2000) 1077.

\bibitem{kre00}
M.~Kreuzer, L.~Marrucci, D.~Paparo, Light-induced modification of kinetic
  molecular properties: enhancement of optical {K}err effect in absorbing
  liquids, photoinduced torque and molecular motors in dye-doped nematics, J.
  Nonlin. Opt. Phys. Mat. 9 (2000) 157.

\bibitem{gol01}
E.~Goldobin, A.~Sterck, D.~Koelle, Josephson vortex in a ratchet potential:
  {T}heory, Phys. Rev. E 63 (2001) 031111.

\bibitem{kul98a}
J.~Kula, M.~Kostur, J.~Luczka, {B}rownian motion controlled by dichotomic and
  thermal fluctuations, Chem. Phys. 235 (1998) 27.

\bibitem{bie96a}
M.~Bier, R.~D. Astumian, Biasing {B}rownian motion in different directions in a
  3-state fluctuating potential and an application for the separation of small
  particles, Phys. Rev. Lett. 76 (1996) 4277.

\bibitem{rei97c}
P.~Reimann, Current reversal in a white noise driven flashing ratchet, Phys.
  Rep. 290 (1997) 149.

\bibitem{gra82}
R.~Graham, A.~Schenzle, Stabilization by multiplicative noise, Phys. Rev. A 26
  (1982) 1676.

\bibitem{bao96a}
J.-D. Bao, Y.-Z. Zhuo, X.-Z. Wu, Diffusion current for a system in a periodic
  potential driven by additive colored noise, Phys. Lett. A 215 (1996) 154.

\bibitem{bao96b}
J.-D. Bao, Y.-Z. Zhuo, X.-Z. Wu, Effect of multiplicative noise on
  fluctuation-induced transport, Phys. Lett. A 217 (1996) 241.

\bibitem{lee99a}
K.~Lee, W.~Sung, Effects of nonequilibrium fluctuations on ionic transport
  through biomembranes, Phys. Rev. E 60 (1999) 4681.

\bibitem{arc250}
{Archimedes of Syracuse}, Ca. 250 b.~c., unpublished.

\bibitem{bor98}
M.~Borromeo, F.~Marchesoni, {B}rownian surfers, Phys. Lett. A 249 (1998) 8457.

\bibitem{jan98}
K.~M. Jansons, G.~D. Lythe, Stochastic {S}tokes drift, Phys. Rev. Lett. 81
  (1998) 3136.

\bibitem{vdb99}
C.~{Van den Broeck}, Stokes' drift: an exact result, Europhys. Lett. 46 (1999)
  1.

\bibitem{bor99}
M.~Borromeo, F.~Marchesoni, Thermal conveyers, Appl. Phys. Lett. 75 (1999)
  1024.

\bibitem{li00}
Y.-X. Li, X.-Z. Wu, Y.-Z. Zhuo, Brownian motors: solitary waves and efficiency,
  Physica A 286 (2000) 147.

\bibitem{ben00}
I.~Bena, M.~Copelli, C.~{Van den Broeck}, Stokes' drift: a rocking ratchet, J.
  Stat. Phys. 101 (2000) 415.

\bibitem{sto47}
G.~G. Stokes, On the theory of oscillatory waves, Trans. Camb. Philos. Soc. 8
  (1847) 441.

\bibitem{mes92}
O.~N. Mesquita, S.~Kane, J.~P. Gollub, Transport by capillary waves:
  fluctuating {S}tokes drift, Phys. Rev. A 45 (1992) 3700.

\bibitem{tho83}
D.~J. Thouless, Quantization of transport, Phys. Rev. B 27 (1983) 6083.

\bibitem{swi99}
M.~Switkes, C.~M. Marcus, K.~Campman, A.~C. Gossard, An adiabatic electron
  pump, Science 283 (1999) 1905.

\bibitem{wag99}
M.~Wagner, F.~Sols, Subsea electron transport: pumping deep within the {F}ermi
  sea, Phys. Rev. Lett. 83 (1999) 4377.

\bibitem{sol00}
F.~Sols, M.~Wagner, Pipeline model of a {F}ermi-sea electron pump, Ann. Phys.
  (Leipzig) 9 (2000) 776.

\bibitem{ast01}
R.~D. Astumian, I.~Der\'enyi, Towards a chemically driven molecular electron
  pump, Phys. Rev. Lett. 86 (2001) 3859.

\bibitem{kou91a}
L.~P. Kouwenhoven, A.~T. Johnson, N.~C. van~der Vaart, C.~P.~M. Harmans,
  Quantized current in a quantum-dot turnstile using oscillating tunnel
  barriers, Phys. Rev. Lett. 67 (1991) 1626.

\bibitem{kou91b}
L.~P. Kouwenhoven, A.~T. Johnson, N.~C. van~der Vaart, A.~van~der Enden,
  C.~P.~M. Harmans, C.~T. Foxton, Quantized current in a quantum dot turnstile,
  Z. Phys. B 85 (1991) 381.

\bibitem{pot92}
H.~Pothier, P.~Lafarge, C.~Urbina, D.~Esteve, M.~H. Devoret, Single-electron
  pump based on charging effects, Europhys. Lett. 17 (1992) 249.

\bibitem{kel96}
M.~W. Keller, J.~M. Martinis, N.~M. Zimmerman, A.~H. Steinbach, Accuracy of
  electron counting using a 7-junction electron pump, Appl. Phys. Lett. 69
  (1996) 1804.

\bibitem{wei95}
J.~Weis, R.~J. Haug, K.~von Klitzing, K.~Poog, Single-electron tunneling
  transistor as a current rectifier with potential-controlled current polarity,
  Semicond. Sci. Technol. 10 (1995) 877.

\bibitem{xia99}
X.~Wang, T.~Junno, S.-B. Carlsson, C.~Thelander, L.~Samuelson, Coulomb blockade
  ratchet, cond-mat/9910444.

\bibitem{lan85}
R.~Landauer, M.~B\"uttiker, Drift and diffusion in reversible computation,
  Physica Scripta T9 (1985) 155.

\bibitem{fau95}
L.~P. Faucheux, G.~Stolovitzky, A.~Libchaber, Periodic forcing of a {B}rownian
  particle, Phys. Rev. E 51 (1995) 5239.

\bibitem{tal97}
V.~I. Talyanskii, J.~M. Shilton, M.~Pepper, C.~G. Smith, C.~J.~B. Ford, E.~H.
  Linfield, D.~A. Ritchie, G.~A.~C. Jones, Single electron transport in a
  one-dimensional channel by high frequency surface acoustic waves, Phys. Rev.
  B 56 (1997) 15180.

\bibitem{roc97}
C.~Rocke, S.~Zimmermann, A.~Wixforth, J.~P. Kotthaus, G.~B\"ohm, G.~Weinmann,
  Acoustically driven storage of light in a quantum well, Phys. Rev. Lett. 78
  (1997) 4099.

\bibitem{pos98}
D.~E. Postnov, A.~P. Nikitin, V.~S. Anishchenko, Synchronization of the mean
  velocity of a particle in stochastic ratchets with a running wave, Phys. Rev.
  E 58 (1998) 1662.

\bibitem{mal98}
A.~N. Malakhov, A new model of {B}rownian transport, Izv. VUZ ``AND'' 6 (1998)
  105.

\bibitem{sas98}
S.~Sasa, T.~Shibata, Brownian motors driven by particle exchange, J. Phys. Soc.
  Jpn. 67 (1998) 1918.

\bibitem{fuk98}
K.~Fukui, J.~H. Frederick, J.~I. Cline, Chiral dissociation dynamics of
  molecular ratchets: {P}referential sense of rotatory motion in microscopic
  systems, Phys. Rev. E 58 (1998) 929.

\bibitem{ric54}
S.~O. Rice, in: N.~Wax (Ed.), Selected papers on noise and stochastic
  processes, Dover, New York, 1954.

\bibitem{han78}
P.~H\"anggi, Correlation functions and master equations of generalized
  (non-{M}arkovian) {L}angevin equations, Z. Phys. B 31 (1978) 407.

\bibitem{han80}
P.~H\"anggi, Langevin description of {M}arkovian integro-differential master
  equations, Z. Phys. B 36 (1980) 271.

\bibitem{vdb83}
C.~{Van den Broeck}, On the relation between white shot noise, {G}aussian white
  noise and the dichotomic {M}arkov process, J. Stat. Phys. 31 (1983) 467.

\bibitem{luc95}
J.~Luczka, R.~Bartussek, P.~H\"anggi, White-noise-induced transport in periodic
  structures, Europhys. Lett. 31 (1995) 431.

\bibitem{cze97}
T.~Czernik, J.~Kula, J.~Luczka, P.~H\"anggi, Thermal ratchets driven by
  {P}oissonian white shot noise, Phys. Rev. E 55 (1997) 4057.

\bibitem{cze00}
T.~Czernik, J.~Luczka, Rectified steady flow induced by white shot noise:
  diffusive and non-diffusive regimes, Ann. Phys. (Leipzig) 9 (2000) 721.

\bibitem{cze01}
T.~Czernik, M.~Niemiec, J.~Luczka, Brownian motors driven by {P}oissonian
  fluctuations, Acta Physica Polonica B 32 (2001) 321.

\bibitem{li00b}
Y.-X. Li, Y.-Z. Zhuo, Directed motion induced by shifting ratchet, Int. J. Mod.
  Phys. B 14 (2000) 2609.

\bibitem{che97}
Y.~Chen, Asymmetric cycling and biased movement of {B}rownian particles in
  fluctuating symmetric potentials, Phys. Rev. Lett. 79 (1997) 3117.

\bibitem{li97a}
Y.-X. Li, Directed motion induced by a cyclic stochastic process, Mod. Phys.
  Lett. B 11 (1997) 713.

\bibitem{gor97a}
L.~Gorre-Talini, P.~Silberzan, Force-free motion of a mercury drop
  alternatively submitted to shifted asymmetric potentials, J. Phys. I France 7
  (1997) 1475.

\bibitem{por00}
M.~Porto, M.~Urbakh, J.~Klafter, Molecular motor that never steps backwards,
  Phys. Rev. Lett. 85 (2000) 491.

\bibitem{men99}
C.~Mennerat-Robilliard, D.~Lucas, S.~Guibal, J.~Tabosa, C.~Jurczak, J.-Y.
  Courtois, G.~Grynberg, Ratchet for cold {R}ubidium atoms: The asymmetric
  optical lattice, Phys. Rev. Lett. 82 (1999) 851.

\bibitem{kel99}
T.~R. Kelly, H.~{De Silva}, R.~A. Silva, Unidirectional rotary motion in a
  molecular system, Nature 401 (1999) 150.

\bibitem{dav99}
A.~P. Davis, Synthetic molecular motors, Nature 401 (1999) 120.

\bibitem{kel01}
T.~R. Kelly, Progress towards rationally designed molecular motors, Acc. Chem.
  Res. 34 (2001) 514.

\bibitem{kou99}
N.~Koumura, R.~W.~J. Zijistra, R.~A. van Delden, N.~Harada, B.~L. Feringa,
  Light-driven monodirectional molecular motor, Nature 401 (1999) 152.

\bibitem{gim98}
J.~K. Gimzewski, C.~Joachim, R.~R. Schlittler, V.~Langlais, H.~Tang,
  I.~Johannsen, Rotation of a single molecule within a supramolecular bearing,
  Science 281 (1998) 531.

\bibitem{gim99}
J.~K. Gimzewski, C.~Joachim, Nanoscale science of single molecules using local
  probes, Science 283 (1999) 1683.

\bibitem{alb94}
B.~Alberts, D.~Bray, J.~Lewis, M.~Raff, K.~Roberts, J.~D. Watson, The molecular
  biology of the cell, Garland, New York, 1994.

\bibitem{ast89a}
R.~D. Astumian, P.~B. Chock, T.~Tsong, H.~V. Westerhoff, Effects of
  oscillations and energy-driven fluctuations on the dynamics of enzyme
  catalysis and free-energy transduction, Phys. Rev. A 39 (1989) 6416.

\bibitem{ast89b}
R.~D. Astumian, B.~Robertson, Nonlinear effect of an oscillating electric field
  on membrane proteins, J. Chem. Phys. 91 (1989) 4891.

\bibitem{ful94}
A.~Fulinski, Noise-stimulated active transport in biological cell membranes,
  Phys. Lett. A 193 (1994) 267.

\bibitem{ful97}
A.~Fulinski, Active transport in biological membranes and stochastic resonance,
  Phys. Rev. Lett. 79 (1997) 4926.

\bibitem{ful98}
A.~Fulinski, Barrier fluctuations and stochastic resonance in membrane
  transport, Chaos 8 (1998) 549.

\bibitem{ast98}
R.~D. Astumian, I.~Der\'enyi, Fluctuation driven transport and models of
  molecular motors and pumps, Eur. Biophys. J. 27 (1998) 474.

\bibitem{tso00}
T.~Y. Tsong, Cellular transduction of periodic and stochastic signals by
  electroconformational coupling, in: J.~Walleczek (Ed.), Self-organized
  biological dynamics and nonlinear control, Cambridge University Press,
  Cambridge, 2000.

\bibitem{rob90}
B.~Robertson, R.~D. Astumian, Michaelis-{M}enten equation for an enzyme in an
  oscillating electric field, Biophys. J. 58 (1990) 969.

\bibitem{mie95a}
A.~Mielke, Noise induced transport, Ann. Phys. (Leipzig) 4 (1995) 476.

\bibitem{chi95}
D.~R. Chialvo, M.~M. Millonas, Asymmetric unbiased fluctuations are sufficient
  for the operation of a correlation ratchet, Phys. Lett. A 209 (1995) 26.

\bibitem{zap96}
I.~Zapata, R.~Bartussek, F.~Sols, P.~H\"anggi, Voltage rectification by a
  {SQUID} ratchet, Phys. Rev. Lett. 77 (1996) 2292.

\bibitem{mil96}
M.~M. Millonas, D.~R. Chialvo, Nonequilibrium fluctuation-induced phase
  transport in {J}osephson junctions, Phys. Rev. E 53 (1996) 2239.

\bibitem{sar99}
A.~Sarmiento, H.~Larralde, Deterministic transport in ratchets, Phys. Rev. E 59
  (1999) 4878.

\bibitem{lan99}
P.~S. Landa, P.~V.~E. McClintock, Changes in the dynamical behavior of
  nonlinear systems induced by noise, Phys. Rep. 323 (1999) 1.

\bibitem{els96}
T.~C. Elston, C.~R. Doering, Numerical and analytical studys of nonequilibrium
  fluctuation-induced transport processes, J. Stat. Phys. 83 (1996) 359.

\bibitem{for73}
W.~Forst, Theory of unimolecular reactions, Academic Press, New York, 1973.

\bibitem{doe98}
C.~R. Doering, L.~A. Dontcheva, M.~M. Klosek, Constructive role of noise: fast
  fluctuation asymptotics of transport in stochastic ratchets, Chaos 8 (1998)
  643.

\bibitem{koh98}
H.~Kohler, A.~Mielke, Noise-induced transport at zero temperature, J. Phys. A
  31 (1998) 1929.

\bibitem{man00}
R.~Mankin, A.~Ainsaar, Current reversals in ratchets driven by trichotomous
  noise, Phys. Rev. E 61 (2000) 6359.

\bibitem{klo99}
M.~M. Klosek, R.~W. Cox, Steady-state currents in sharp stochastic ratchets,
  Phys. Rev E 60 (1999) 3727.

\bibitem{kul96}
J.~Kula, T.~Czernik, J.~Luczka, Transport generated by dichotomic fluctuations,
  Phys. Lett. A 214 (1996) 14.

\bibitem{kul98b}
J.~Kula, T.~Czernik, J.~Luczka, {B}rownian ratchets: transport controlled by
  thermal noise, Phys. Rev. Lett. 80 (1998) 1377.

\bibitem{ari98}
C.~M. Arizmendi, F.~Family, Memory correlation effect on thermal ratchets,
  Physica A 251 (1998) 368.

\bibitem{bar98}
R.~Bartussek, Stochastische {R}atschen, PhD thesis (in German), Logos Verlag,
  Berlin, 1998.

\bibitem{dia97}
T.~E. Dialynas, K.~Lindenberg, G.~P. Tsironis, Ratchet motion induced by
  deterministic and correlated stochastic forces, Phys. Rev. E 56 (2000) 3976.

\bibitem{bao99b}
J.~D. Bao, Rectification of different colored noise, Phys. Lett. A 256 (1999)
  356.

\bibitem{cor00}
E.~Cortes, Ratchet motion induced by a correlated stochastic force, Physica A
  275 (2000) 78.

\bibitem{bar97b}
R.~Bartussek, Ratchets driven by colored {G}aussian noise, in:
  L.~Schimansky-Geier, T.~P\"oschel (Eds.), Lecture Notes in Physics, Vol. 484,
  Springer, Berlin, 1997.

\bibitem{mar98}
F.~Marchesoni, Conceptional design of a molecular shuttle, Phys. Lett. A 237
  (1998) 126.

\bibitem{lan01}
P.~Lancon, G.~Batrouni, L.~Lobry, N.~Ostrowsky, Drift without flux: {B}rownian
  walker with a space-dependent diffusion coefficient, Europhys. Lett. 54
  (2001) 28.

\bibitem{bal81}
R.~von Baltz, W.~Krauth, Theory of the bulk photovoltaic effect in pure
  crystals, Phys. Rev. B 23 (1981) 5590.

\bibitem{mag01}
L.~I. Magarill, Photogalvanic effect in asymmetric lateral superlattice,
  Physica E 9 (2001) 625.

\bibitem{fal89}
V.~I. Fal'ko, D.~E. Khmel'nitskii, Mesoscopic photovoltaic effect in
  microjunctions, Sov. Phys. JETP 68 (1989) 186, [Zh. Eksp. Teor. Fiz. 95
  (1989) 328].

\bibitem{liu92}
J.~Liu, M.~A. Pennington, N.~Giordano, Mesoscopic photovoltaic effect, Phys.
  Rev. B 45 (1992) 1267.

\bibitem{dal95}
G.~Dalba, Y.~Soldo, F.~Rocca, V.~M. Fridkin, P.~Sainctavit, Giant bulk
  photovoltaic effect under linearly polarized x-ray synchrotron radiation,
  Phys. Rev. Lett. 74 (1995) 988.

\bibitem{kra93}
V.~E. Kravtsov, V.~I. Yudson, Directed current in mesoscopic rings induced by
  high-frequency electromagnetic field, Phys. Rev. Lett. 70 (1993) 210.

\bibitem{aro93}
A.~G. Aronov, V.~E. Kravtsov, Nonlinear properties of disordered normal-metal
  rings with magnetic flux, Phys. Rev. B 47 (1993) 13409.

\bibitem{shm85}
G.~M. Shmelev, N.~H. Song, G.~I. Tsurkan, Photostimulated even acoustoelectric
  effect, Sov. Phys. J. (USA) 28 (1985) 161.

\bibitem{ent89}
M.~V. Entin, Theory of the coherent photogalvanic effect, Sov. Phys. Semicond.
  23 (1989) 664.

\bibitem{ata96}
R.~Atanasov, A.~Hach\'e, J.~L.~P. Hughes, H.~M. van Driel, J.~E. Sipe, Coherent
  control of photocurrent generation in bulk semiconductors, Phys. Rev. Lett.
  76 (1996) 1703.

\bibitem{hac97}
A.~Hach\'e, Y.~Kostoulas, R.~Atanasov, J.~L.~P. Hughes, J.~E. Sipe, H.~M. van
  Driel, Observation of controlled photocurrent in unbiased bulk {GaAs}, Phys.
  Rev. Lett. 78 (1997) 306.

\bibitem{ale99}
K.~N. Alekseev, M.~V. Erementchouk, F.~V. Kusmartsev, Direct-current generation
  due to wave mixing in semiconductors, Europhys. Lett. 47 (1999) 595.

\bibitem{rei00a}
P.~Reimann, Rocking ratchets at high frequencies, in: J.~A. Freund,
  T.~P\"oschel (Eds.), Lecture Notes in Physics, Vol. 557, Springer, Berlin,
  2000.

\bibitem{sha63}
S.~{S}hapiro, {J}osephson currents in superconducting tunneling: the effect of
  microwaves and other observations, Phys. Rev. Lett. 11 (1963) 80.

\bibitem{jun91}
P.~Jung, P.~H\"anggi, Effect of periodic driving on the escape in periodic
  potentials, Ber. Bunsenges. Phys. Chem. 95 (1991) 311.

\bibitem{gor96}
L.~Gorre, E.~Ioannidis, P.~Silberzan, Rectified motion of a mercury drop in an
  asymmetric structure, Europhys. Lett. 33 (1996) 267.

\bibitem{fal99}
F.~Falo, P.~J. Martinez, J.~J. Mazo, S.~Cilla, Ratchet potential for fluxons in
  {J}osephson-junction arrays, Europhys. Lett. 45 (1999) 700.

\bibitem{tri99}
E.~Trias, J.~J. Mazo, F.~Falo, T.~P. Orlando, Depinning of kinks in a
  {J}osephson-junction ratchet array, Phys. Rev. E 61 (2000) 2257.

\bibitem{car01}
G.~Carapella, Relativistic flux quantum in a field-induced deterministic
  ratchet, Phys. Rev. B 63 (2001) 054515.

\bibitem{lee99}
C.-S. Lee, B.~Janko, I.~Der\'enyi, A.-L. Barabasi, Reducing vortex density in
  superconductors using the ``ratchet effect'', Nature 400 (1999) 337.

\bibitem{wam99}
J.~F. Wambaugh, C.~Reichhardt, C.~J. Olson, F.~Marchesoni, F.~Nori,
  Superconducting fluxon pumps and lenses, Phys. Rev. Lett. 83 (1999) 5106.

\bibitem{pab00}
P.~J. {de Pablo}, J.~Colchero, J.~{Gomez-Herrero}, A.~Asenjo, M.~Luna, P.~A.
  Serena, A.~M. Baro, Ratchet effect in surface electromigration detected with
  scanning force microscopy in gold micro-stripes, Surf. Sci. 464 (2000) 123.

\bibitem{bar00}
M.~Barbi, M.~Salerno, Phase locking effect and current reversals in
  deterministic underdamped ratchets, Phys. Rev. E 62 (2000) 1988.

\bibitem{bar01}
M.~Barbi, M.~Salerno, Stabilization of ratchet dynamics by weak periodic
  signals, Phys. Rev. E 63 (2001) 066212.

\bibitem{ari01}
C.~M. Arizmendi, F.~Family, A.~L. Salas-Brito, Quenched disorder effects on
  deterministic inertia ratchets, Phys. Rev. E 63 (2001) 061104.

\bibitem{fuj82}
H.~Fujisaka, S.~Grossmann, Chaos-induced diffusion in nonlinear discrete
  dynamics, Z. Phys. B 48 (1982) 261.

\bibitem{gei82}
T.~Geisel, J.~Nierwetberg, Onset of diffusion and universal scaling in chaotic
  systems, Phys. Rev. Lett. 48 (1982) 7.

\bibitem{sch82}
M.~Schell, S.~Fraser, R.~Kapral, Diffusive dynamics in systems with
  translational symmetry: a one-dimensional-map model, Phys. Rev. A 26 (1982)
  504.

\bibitem{gei84}
T.~Geisel, J.~Nierwetberg, Statistical properties of intermittent diffusion in
  chaotic systems, Z. Phys. B 56 (1984) 59.

\bibitem{rei94a}
P.~Reimann, Suppression of deterministic diffusion by noise, Phys. Rev. E 50
  (1994) 727.

\bibitem{rei94b}
P.~Reimann, C.~{Van den Broeck}, Intermittent diffusion in the presence of
  noise, Physica D 75 (1994) 509.

\bibitem{kla95}
R.~Klages, J.~R. Dorfman, Simple maps with fractal diffusion coefficient, Phys.
  Rev. Lett. 74 (1995) 387.

\bibitem{far98}
O.~Farago, Y.~Kantor, Directed chaotic motion in a periodic potential, Physica
  A 249 (1998) 151.

\bibitem{mei92}
J.~D. Meiss, Symplectic maps, variational principles, and transport, Rev. Mod.
  Phys. 64 (1992) 795.

\bibitem{zas99}
G.~M. Zaslavsky, Chaotic dynamics and the origin of statistical laws, Phys.
  Today, August issue (1999) 39.

\bibitem{kov00}
S.~Kovalyov, Phase space structure and anomalous diffusion in a rotational
  fluid experiment, Chaos 10 (2000) 153.

\bibitem{dit00}
T.~Dittrich, R.~Kretzmerick, M.-F. Otto, H.~Schanz, Classical and quantum
  transport in deterministic {H}amiltonian ratchets, Ann. Phys. (Leipzig) 9
  (2000) 755.

\bibitem{sch01}
H.~Schanz, M.-F. Otto, R.~Ketzmerick, T.~Dittrich, Classical and quantum
  {H}amiltonian ratchets, Phys. Rev. Lett. 87 (2001) 070601.

\bibitem{den01}
S.~Denisov, S.~Flach, Dynamical mechanism of dc current generation in driven
  {H}amiltonian systems, {S}ubmitted for publication.

\bibitem{bao98c}
J.-D. Bao, Y.-Z. Zhuo, Langevin simulation approach to a two-dimensional
  coupled flashing ratchet, Phys. Lett. A 239 (1998) 228.

\bibitem{gho00}
A.~W. Ghosh, S.~V. Khare, Rotation in an asymmetric multidiomensional periodic
  potential due to colored noise, Phys. Rev. Lett. 84 (2000) 5243.

\bibitem{qia98}
H.~Qian, Vector field formalism and analysis for a class of thermal ratchets,
  Phys. Rev. Lett. 81 (1998) 3063.

\bibitem{kos00}
M.~Kostur, L.~Schimansky-Geier, Numerical study of diffusion induced transport
  in 2d systems, Phys. Lett. A 265 (2000) 337.

\bibitem{han99}
P.~H\"anggi, P.~Reimann, Quantum ratchet reroute electrons, Physics World 12
  (1999) 21.

\bibitem{bro00}
M.~Brooks, Quantum clockwork, New Scientist 2222 (2000) 29.

\bibitem{bal95}
V.~Balakrishnan, C.~{Van den Broeck}, Transport properties on a random comb,
  Physica A 217 (1995) 1.

\bibitem{sla97}
G.~W. Slater, H.~L. Guo, G.~I. Nixon, Bidirectional transport of
  polyelectrolytes using self-modulating entropic ratchets, Phys. Rev. Lett. 78
  (1997) 1170.

\bibitem{tur90}
C.~Turmel, E.~Brassard, R.~Forsyth, K.~Hood, G.~W. Slater, J.~Noorlandi, High
  resolution zero intergated field electrophoresis of {DNA}, in: E.~Lai, B.~W.
  Birren (Eds.), Electrophoresis of large {DNA} molecules, Cold Spring Harbor
  Laboratory Press, Cold Spring Harbor, NY, 1990.

\bibitem{des98}
C.~Desruisseaux, G.~W. Slater, T.~B. Kist, Trapping electrophoresis and
  ratchets: a theoretical study for {DNA}-protein complexes, Biophys. J. 75
  (1998) 1228.

\bibitem{sla00}
G.~W. Slater, C.~Desruisseaux, S.~J. Hubert, J.~F. Mercier, J.~Labrie,
  J.~Boileau, F.~Tessier, M.~P. Pepin, Theory of {DNA} electrophoresis: a look
  at some current challenges, Electrophoresis 21 (2000) 3873.

\bibitem{gri01}
G.~A. Griess, E.~Rogers, P.~Serwer, Application of the concept of an
  electrophoretic ratchet, Electrophoresis 22 (2001) 981.

\bibitem{sto01}
M.~Stopa, Charging ratchets, {S}ubmitted for publication.

\bibitem{ven96}
M.~{Di Ventra}, G.~Papp, C.~Coluzza, A.~Baldereschi, P.~A. Schulz, Indented
  barrier resonant tunneling rectifiers, J. Appl. Phys. 80 (1996) 4174.

\bibitem{duk98}
T.~A.~J. Duke, R.~H. Austin, Microfabricated sieve for the continuous sorting
  of macromolecules, Phys. Rev. Lett. 80 (1998) 1552.

\bibitem{duk98a}
T.~Duke, Separation techniques, Curr. Opin. Chem. Biol. 2 (1998) 592.

\bibitem{der98b}
I.~Der\'enyi, R.~D. Astumian, ac-separation of particles by biased {B}rownian
  motion in a two-dimensional sieve, Phys. Rev. E 58 (1998) 7781.

\bibitem{vol92}
W.~D. Volkmuth, R.~H. Austin, {DNA} electrophoresis in microlithographic
  arrays, Nature 358 (1992) 600.

\bibitem{oud99}
A.~van Oudenaarden, S.~G. Boxer, {B}rownian ratchet: molecular separation in
  lipid bilayers supported on patterned arrays, Science 285 (1999) 1046.

\bibitem{lor98}
A.~Lorke, S.~Wimmer, B.~Jager, J.~P. Kotthaus, W.~Wegschneider, M.~Bichler,
  Far-infrared and transport properties of antidot arrays with broken symmetry,
  Physica B 249 (1998) 312.

\bibitem{ear95}
E.~A. Early, A.~F. Clark, C.~J. Lobb, Physical basis for half-integral
  {S}hapiro steps in a dc {SQUID}, Physica C 245 (1995) 308.

\bibitem{lif62}
S.~Lifson, J.~L. Jackson, On the self-diffusion of ions in polyelectrolytic
  solution, J. Chem. Phys. 36 (1962) 2410.

\bibitem{ajd91}
A.~Ajdari, J.~Prost, Free-flow electrophoresis with trapping by a transverse
  inhomogeneous field, Proc. Natl. Acad. Sci. USA 88 (1991) 4468.

\bibitem{gho94}
A.~Ghosh, Diffusion rate for a {B}rownian particle in a cosine potential in the
  presence of colored noise, Phys. Lett. A 187 (1994) 54.

\bibitem{mal98b}
A.~N. Malakhov, Acceleration of {B}rownian particle diffusion parallel to a
  fast random field with a short spatial period, Tech. Phys. Lett. 24 (1998)
  833.

\bibitem{cla91}
I.~Claes, C.~{Van den Broeck}, Stochastic resonance for dispersion in
  oscillatory flows, Phys. Rev. A 44 (1991) 4970.

\bibitem{cla93}
I.~Claes, C.~{Van den Broeck}, Dispersion of particles in periodic media, J.
  Stat. Phys. 70 (1993) 1215.

\bibitem{kim98}
Y.~W. Kim, W.~Sung, Does stochastic resonance occur in periodic potentials?,
  Phys. Rev. E 57 (1998) R6237.

\bibitem{gan96}
H.~Gang, A.~Daffertshofer, H.~Haken, Diffusion of periodically forced
  {B}rownian particles moving in space-periodic potentials, Phys. Rev. Lett. 76
  (1996) 4874.

\bibitem{mah95}
M.~C. Mahato, A.~M. Jayannavar, Synchronized first-passages in a double-well
  system driven by an asymmetric periodic field, Phys. Lett. A 209 (1995) 21.

\bibitem{mah97}
M.~C. Mahato, A.~M. Jayannavar, Asymmetric motion in a double well under the
  action of zero-mean {G}aussian white noise and periodic forcing, Phys.~Rev. E
  55 (1997) 3716.

\bibitem{vid85}
A.~K. Vidybida, A.~A. Serikov, Electrophoresis by alternating fields in a
  non-{N}ewtonian fluid, Phys. Lett. A 108 (1985) 170.

\bibitem{ser98}
P.~Serwer, G.~A. Griess, Adaptation of pulsed-field gel electrophoresis for the
  improved fractionation of spheres, Analytica Chimica Acta 372 (1998) 299.

\bibitem{ser99}
P.~Serwer, G.~A. Griess, Advances in the separation of bacteriophages and
  related particles, J. Chromatography B 722 (1999) 179.

\bibitem{cha97}
M.~J. Chacron, G.~W. Slater, Particle trapping and self-focusing in temporally
  asymmetric ratchets with strong field gradients, Phys. Rev. E 56 (1997) 3446.

\bibitem{mog98}
A.~Mogliner, M.~Mangel, R.~J. Baskin, Motion of molecular motor ratcheted by
  internal fluctuations and protein friction, Phys. Lett. A 237 (1998) 297.

\bibitem{zol00}
A.~V. Zolotaryuk, P.~L. Christiansen, B.~Norden, A.~V. Savin, Y.~Zolotaryuk,
  Pendulum as a model system for driven rotation in moleculear nanoscale
  machines, Phys. Rev. E 61 (2000) 3256.

\bibitem{luc00a}
D.~G. Luchinsky, M.~J. Greenall, P.~McClintock, Resonant rectification of
  fluctuations in a {B}rownian ratchet, Phys. Lett. A 273 (2000) 316.

\bibitem{hon96}
T.~Hondou, Y.~Sawada, Comment on ``{W}hite-noise-induced transport in periodic
  structures'' by {J}. {L}uczka et al., Europhys. Lett. 35 (1996) 313.

\bibitem{wei93}
G.~H. Weiss, M.~Gitterman, Motion in a periodic potential driven by rectangular
  pulses, J. Stat. Phys. 70 (1993) 93.

\bibitem{ber97a}
V.~Berdichevsky, M.~Gitterman, {J}osephson junction with noise, Phys. Rev. E 56
  (1997) 6340.

\bibitem{li98b}
J.~Li, Z.~Huang, Transport of particles caused by correlation between additive
  and multiplicative noise, Phys. Rev. E 57 (1998) 3917.

\bibitem{li98a}
J.~Li, Z.~Huang, Net voltage caused by correlated symmetric noises, Phys. Rev.
  E 58 (1998) 139.

\bibitem{li98c}
J.~Li, Z.~Huang, Flux in the case of {G}aussian white noises, Comm. Theor.
  Phys. 30 (1998) 527.

\bibitem{cao00}
L.~Cao, D.~Wu, Fluctuation induced transport in a spatially symmetric periodic
  potential, Phys. Rev. E 62 (2000) 7478.

\bibitem{jia00}
Y.~A. Jia, J.~R. Li, Effects of correlated noises on current, Int. J. Mod.
  Phys. B 14 (2000) 507.

\bibitem{arg87}
F.~Argoul, A.~Arneodo, P.~Collet, A.~Lesne, Transition to chaos in presence of
  an external periodic field: cross-over effects in the measure of critical
  exponents, Europhys. Lett. 3 (1987) 643.

\bibitem{col89}
P.~Collet, A.~Lesne, Renormalization group analysis of some dynamical systems
  with noise, J. Stat. Phys. 57 (1989) 967.

\bibitem{bec90}
C.~Beck, Brownian motion from deterministic dynamics, Physica A 169 (1990) 324.

\bibitem{hon95}
T.~Hondou, S.~Sawada, Dynamical behavior of a dissipative particle in a
  periodic potential subjected to chaotic noise: Retrieval of chaotic
  determinism with broken parity, Phys. Rev. Lett. 75 (1995) 3269.

\bibitem{gre84}
C.~Grebogi, E.~Ott, S.~Pelikan, J.~Yorke, Strange attractors that are not
  chaotic, Physica D 13 (1984) 261.

\bibitem{pol92}
D.~D. Pollock, Thermoelectricity, in: Encyclopedia of Physical Science and
  Technology, Vol.~16, Academic Press, San Diego, 1992.

\bibitem{ash76}
N.~W. Ashcroft, N.~D. Mermin, Solid State Physics, Saunders College,
  Philadelphia, 1976.

\bibitem{kam88}
N.~G. van Kampen, Relative stability in nonuniform temperature, IBM J. Res.
  Develop. 32 (1988) 107.

\bibitem{lan88}
R.~Landauer, Motion out of noisy states, J. Stat. Phys. 53 (1988) 233.

\bibitem{jay96a}
A.~M. Jayannavar, M.~C. Mahato, Macroscopic equation of motion in inhomogeneous
  media: a microscopic treatment, Pramana-J. Phys. 45 (1996) 369.

\bibitem{mah96}
M.~C. Mahato, T.~P. Pareek, A.~M. Jayannavar, Enslaving random fluctuations in
  nonequilibrium systems, Int. J. Mod. Phys. B 10 (1996) 3857.

\bibitem{bao99c}
J.~D. Bao, Y.~Abe, Y.~Z. Zhuo, Inhomogeneous friction leading to current in
  periodic systems, Physica A 265 (1999) 111.

\bibitem{luc00}
R.~H. Luchsinger, Transport in nonequilibrium systems with position-dependent
  mobility, Phys. Rev. E 62 (2000) 272.

\bibitem{sek99}
K.~Sekimoto, Temporal coarse graining for systems of {B}rownian particles with
  non-constant temperature, J. Phys. Soc. Jpn. 68 (1999) 1448.

\bibitem{mat00a}
M.~Matsuo, S.~Sasa, Stochastic energetics of non-uniform temperature systems,
  Physica A 276 (2000) 188.

\bibitem{kam87}
N.~G. van Kampen, Diffusion in inhomogeneous media, Z. Phys. B 68 (1987) 135.

\bibitem{bla98}
Y.~M. Blanter, M.~B\"uttiker, Rectification of fluctuations in an underdamped
  ratchet, Phys. Rev. Lett. 81 (1998) 4040.

\bibitem{ris79}
H.~Risken, Vollmer, {B}rownian motion in periodic potentials in the
  low-friction-limit; linear response to an external force, Z. Phys. B 35
  (1979) 177.

\bibitem{sek97a}
K.~Sekimoto, Kinetic characterization of heat bath and the energetics of
  thermal ratchet models, J. Phys. Soc. Jpn. 66 (1997) 1234.

\bibitem{hon98}
T.~Hondou, F.~Takaga, Irreversible operation in a stalled state of {F}eynman's
  ratchet, J. Phys. Soc. Jpn. 67 (1998) 2974.

\bibitem{sak98}
H.~Sakaguchi, {L}angevin simulation for the {F}eynman ratchet model, J. Phys.
  Soc. Jpn. 67 (1998) 709.

\bibitem{sak00}
H.~Sakaguchi, Fluctuation theorem for a {L}angevin model of the {F}eynman
  ratchet, J. Phys. Soc. Jpn. 69 (2000) 104.

\bibitem{jar99}
C.~Jarzynski, O.~Mazonka, Feynman's ratchet and pawl: An exactly solvable case,
  Phys. Rev. E 59 (1999) 6448.

\bibitem{bao99d}
Y.-D. Bao, Directed current of {B}rownian ratchet randomly circulating between
  two thermal sources, Physica A 273 (1999) 286.

\bibitem{bao00b}
Y.-D. Bao, Transport induced by dichotomic temperature fluctuations, Comm.
  Theor. Phys. 34 (2000) 441.

\bibitem{han81}
P.~H\"anggi, Nonlinear effects of colored nonstationary noise: exact results,
  Phys. Lett. A 83 (1981) 196.

\bibitem{ryt81}
D.~Ryter, {B}rownian motion in inhomogeneous media and with interacting
  particles, Z. Phys. B 41 (1981) 39.

\bibitem{san82}
J.~M. Sancho, M.~{San Miguel}, D.~D\"urr, Adiabatic elimination for systems of
  {B}rownian particles with nonconstant damping coefficients, J. Stat. Phys. 28
  (1982) 291.

\bibitem{bre86}
J.~Happel, H.~Brenner, Low {R}eynolds number hydrodynamics, Martinuis {N}ijhoff
  {P}ublishers, Dordrecht, 1986.

\bibitem{luc94}
L.~P. Faucheux, A.~J. Libchaber, Confined {B}rownian motion, Phys. Rev. E 49
  (1994) 5158.

\bibitem{lin00a}
B.~Lin, J.~Yu, S.~A. Rice, Diffusion of an isolated colloidal sphere confined
  between flat plates, Colloids and Surfaces A 174 (2000) 121.

\bibitem{fal76}
C.~M. Falco, Phase-space of a driven, damped pendulum {(Josephson weak link)},
  Am. J. Phys. 44 (1976) 733.

\bibitem{kri92a}
R.~Krishnan, S.~Singh, G.~W. Robinson, Space-dependent friction in the theory
  of activated rate processes: The {H}amiltonian approach, J. Chem. Phys. 97
  (1992) 5516.

\bibitem{kri92b}
R.~Krishnan, S.~Singh, G.~W. Robinson, Space-dependent friction in the theory
  of activated rate processes, Phys. Rev. A 45 (1992) 5408.

\bibitem{bao98b}
J.-D. Bao, Y.~Abe, Y.~Z. Zhuo, Rocked quantum periodic systems in the presence
  of coordinate-dependent friction, Phys. Rev. E 58 (1998) 2931.

\bibitem{dan00}
D.~Dan, A.~M. Jayannavar, M.~C. Mahato, Efficiency and current reversals in
  spatially inhomogeneous ratchets, Int. J. Mod. Phys. 14 (2000) 1585.

\bibitem{dan01a}
D.~Dan, M.~C. Mahato, A.~M. Jayannavar, Multiple current reversals in forced
  inhomogeneous ratchets, Phys. Rev. E 63 (2001) 056307.

\bibitem{dan01b}
D.~Dan, M.~C. Mahato, A.~M. Jayannavar, Motion in a rocked ratchet with
  spatially periodic friction, Physica A 296 (2001) 375.

\bibitem{ste94}
O.~Steuernagel, W.~Ebeling, V.~Calenbuhr, An elementary model for directed
  active motion, Chaos, Solitons, and Fractals 4 (1994) 1917.

\bibitem{schw97}
F.~Schweitzer, Active {B}rownian particles: Artificial agents in physics, in:
  L.~Schimansky-Geier, T.~P\"oschel (Eds.), Lecture Notes in Physics, Vol. 484,
  Springer, Berlin, 1997.

\bibitem{til99}
B.~Tilch, F.~Schweitzer, W.~Ebeling, Directed motion of {B}rownian particles
  with internal energy depot, Physica A 273 (1999) 294.

\bibitem{sch00b}
F.~Schweitzer, B.~Tilch, W.~Ebeling, Uphill motion of active {B}rownian
  particles in piecewise linear potentials, Eur. Phys. B 14 (2000) 157.

\bibitem{sch00}
L.~Schimansky-Geier, S.~Seefeld, V.~Buchholtz, Making spatial structures by
  ratchets, Ann. Phys. (Leipzig) 9 (2000) 705.

\bibitem{klu01}
S.~Klump, A.~Mielke, C.~Wald, Noise-induced transport of two coupled particles,
  Phys. Rev. E 63 (2001) 031914.

\bibitem{li00a}
Y.~Li, X.~Wu, Y.~Zhou, Directed motion of two-headed {B}rownian motors, Mod.
  Phys. Lett. B 14 (2000) 479.

\bibitem{cil98}
S.~Cilla, L.~M. Floria, Internal degrees of freedom in a thermodynamical model
  for intercellular transport, Physica D 113 (1998) 157.

\bibitem{cil98a}
S.~Cilla, L.~M. Floria, A two-dimensional model for kinesin and dynein stepping
  along microtubules, Il Nuovo Cimento D 20 (1998) 1761.

\bibitem{dia96}
T.~E. Dialynas, G.~Tsironis, Vectorial stochastic motion driven by dichotomous
  noise, Phys. Lett. A 218 (1996) 292.

\bibitem{der96}
I.~Der\'enyi, T.~Vicsek, The kinesin walk: a dynamic model with elastically
  coupled heads, Proc. Natl. Acad. Sci. USA 93 (1996) 6775.

\bibitem{els00a}
T.~C. Elston, C.~S. Peskin, The role of flexibility in molecular motor
  function: coupled diffusion in a tilted periodic potential, SIAM J. Appl.
  Math. 60 (2000) 842.

\bibitem{els00b}
T.~C. Elston, D.~You, C.~S. Peskin, Protein flexibility and the correlation
  ratchet, SIAM J. Appl. Math. 61 (2000) 776.

\bibitem{osa92}
Y.~Osada, H.~Okuzaki, H.~Hori, A polymer gel with electrically driven motility,
  Nature 355 (1992) 242.

\bibitem{san99}
O.~Sandre, L.~Gorre-Talini, A.~Ajdari, J.~Prost, P.~Silberzan, Moving droplets
  on asymmetrically structured surfaces, Phys. Rev. E 60 (1999) 2964.

\bibitem{li98}
Y.-X. Li, {B}rownian motors possessing internal degree of freedom, Physica A
  251 (1998) 382.

\bibitem{mul00}
F.~M\"uller, A.~Birner, J.~Schilling, U.~G\"osele, C.~Kettner, P.~H\"anggi,
  Membranes for micropumps from macroporous silicon, Phys. Stat. Sol. A 182
  (2000) 585.

\bibitem{amb99}
H.~Ambaye, K.~W. Kehr, Toy model for molecular motors, Physica A 267 (1999)
  111.

\bibitem{sok99}
I.~Sokolov, A perturbation approach to transport in discrete ratchet systems,
  J. Phys. A 32 (1999) 2541.

\bibitem{keh00}
K.~W. Kehr, Z.~Koza, Hopping motion of lattice gases through nonsymmetric
  potentials under strong bias conditions, Phys. Rev. E 61 (2000) 2319.

\bibitem{duk96}
T.~Duke, S.~Leibler, Motor protein mechanics: a stochastic model with minimal
  mechanochemical coupling, Biophys. J. 71 (1996) 1235.

\bibitem{fis99}
M.~E. Fisher, A.~B. Kolomeisky, The force exerted by a molecular motor, Proc.
  Natl. Acad. Sci. USA 96 (1999) 6597.

\bibitem{fis99a}
M.~E. Fisher, A.~B. Kolomeisky, Molecular motors and the forces they exert,
  Physica A 274 (1999) 241.

\bibitem{kol00a}
A.~B. Kolomeisky, M.~E. Fisher, Periodic sequential kinetic models with
  jumping, branching and deaths, Physica A 279 (2000) 1.

\bibitem{kol00b}
A.~B. Kolomeisky, M.~E. Fisher, Extended kinetic models with waiting-time
  distributions: exact results, J. Chem. Phys. 113 (2000) 10867.

\bibitem{how01}
J.~Howard, Mechanics of motor proteins and the cytoskeleton, Sinauer
  Associates, Sunderland, 2001.

\bibitem{fis01}
M.~E. Fisher, A.~B. Kolomeisky, Simple mechanochemistry describes the dynamics
  of kinesin molecules, Proc. Natl. Acad. Sci. USA 98 (2001) 7748.

\bibitem{san98}
J.~R. Sanchez, F.~Family, C.~M. Arizmendi, Algorithmic complexity of thermal
  ratchet motion, Phys. Lett. A 249 (1998) 281.

\bibitem{ari99}
C.~M. Arizmendi, F.~Family, Algorithmic complexity and efficiency of a ratchet,
  Physica A 269 (1999) 285.

\bibitem{parrondo}
{\tt http://seneca.fis.ucm.es/parr/}.

\bibitem{vdb99a}
C.~{Van den Broeck}, P.~Reimann, R.~Kawai, P.~H\"anggi, Coupled {B}rownian
  motors, in: D.~Reguera, J.~M. Rubi, J.~M.~G. Vilar (Eds.), Lecture Notes in
  Physics, Vol. 527, `Statistical Mechanics and Biocomplexity', Springer,
  Berlin, 1999.

\bibitem{har99a}
G.~P. Harmer, D.~Abbott, Losing strategies can win by {P}arrondo's paradox,
  Nature 402 (1999) 864.

\bibitem{har99b}
G.~P. Harmer, D.~Abbott, {P}arrondo's paradox, Stat. Sci. 14 (1999) 206.

\bibitem{har00x1}
G.~P. Harmer, D.~Abbott, P.~G. Taylor, J.~M.~R. Parrondo, {P}arrondo's
  paradoxical games and the discrete {B}rownian ratchet, in: D.~Abbott, L.~Kish
  (Eds.), Proc. 2nd Int. Conf. on Unsolved Problems of Noise, AIP Proc. 511,
  2000.

\bibitem{har00x2}
G.~P. Harmer, D.~Abbott, P.~G. Taylor, C.~E.~M. Pearce, J.~M.~R. Parrondo,
  Information entropy and {P}arrondo's discrete-time ratchet, in: D.~S.
  Broomhead, E.~A. Luchinskaya, P.~V.~E. McClintock (Eds.), Proc. Stochaos, AIP
  Proc. 502, 2000.

\bibitem{har00x3}
G.~P. Harmer, D.~Abbott, P.~G. Taylor, The paradox of {P}arrondo's games, Proc.
  R. Soc. Lond. A 456 (2000) 1.

\bibitem{par00a}
J.~M.~R. Parrondo, G.~P. Harmer, D.~Abbott, New paradoxical games based on
  {B}rownian ratchets, Phys. Rev. Lett. 85 (2000) 5226.

\bibitem{tor01}
R.~Toral, Cooperative {P}arrondo's games, Fluct. Noise Lett. 1 (2001) L7.

\bibitem{bou90}
J.-P. Bouchaud, A.~Georges, Anomalous diffusion in disordered media:
  statistical mechanisms, models and physical applications, Phys. Rep. 195
  (1990) 127.

\bibitem{den00}
S.~I. Denisov, W.~Horsthemke, Mean first-passage time for an overdamped
  particle in a disordered force field, Phys. Rev. E 62 (2000) 3311.

\bibitem{mar97}
F.~Marchesoni, Transport properties in disordered ratchet potentials, Phys.
  Rev. E 56 (1997) 2492.

\bibitem{ali99}
R.~Alicki, Disordered {M}arkovian {B}rownian ratchets, Phys. Rev. E 60 (1999)
  2559.

\bibitem{pop00}
M.~N. Popescu, C.~M. Arizmendi, A.~L. Salas-Brito, F.~Family, Disorder induced
  diffusive transport in ratchets, Phys. Rev. Lett 85 (2000) 3321.

\bibitem{jia01}
Y.~Jia, S.~N. Yu, J.~R. Li, Effects of random potential on transport, Phys.
  Rev. E 63 (2001) 052101.

\bibitem{sek97b}
K.~Sekimoto, S.~Sasa, Complementary relations for irreversible process derived
  from stochastic energetics, J. Phys. Soc. Jpn. 66 (1997) 3326.

\bibitem{sek98}
K.~Sekimoto, {L}angevin equation and thermodynamics, Prog. Theor. Phys.
  Supplement 130 (1998) 17.

\bibitem{hon00}
T.~Hondou, K.~Sekimoto, Unattainability of {C}arnot efficiency in the
  {B}rownian heat engine, Phys. Rev. E 62 (2000) 6021.

\bibitem{vel01}
S.~Velasco, J.~M.~M. Roco, A.~Medina, A.~C. Hernandez, Feynman's ratchet
  optimization: maximum power and maximum efficiency regimes, J. Phys. D 34
  (2001) 1000.

\bibitem{bao98a}
J.-D. Bao, Variational path-integral approach to current and efficiency with
  quantum correction, Phys. Lett. A 247 (1998) 380.

\bibitem{kam98}
H.~Kamegawa, T.~Hondou, F.~Takagi, Energetics of forced thermal ratchets, Phys.
  Rev. Lett. 80 (1998) 5251.

\bibitem{sok01}
I.~Sokolov, Irreversible and reversible modes of operation of deterministic
  ratchets, Phys. Rev. E 63 (2001) 021107.

\bibitem{par99}
A.~Parmeggiani, F.~J\"ulicher, A.~Ajdari, J.~Prost, Energy transduction of
  isothermal ratchets: generic aspects and specific examples close to and far
  from equilibrium, Phys. Rev. E 60 (1999) 2127.

\bibitem{sok00}
I.~M. Sokolov, Ideally efficient irreversible molecular gears,
  cond-mat/0002251.

\bibitem{der99b}
I.~Der\'enyi, R.~D. Astumian, Efficiency of {B}rownian heat engines, Phys. Rev.
  E 59 (1999) R6219.

\bibitem{sek00}
K.~Sekimoto, F.~Takagi, T.~Hondou, Carnot's cycle for small systems:
  irreversibility and cost of operations, Phys. Rev. E 62 (2000) 7759.

\bibitem{sok99b}
I.~M. Sokolov, Reversible fluctuation rectifier, Phys. Rev. E 60 (1999) 4946.

\bibitem{tak99}
F.~Takagi, T.~Hondou, Thermal noise can facilitate energy conversion by a
  ratchet system, Phys. Rev. E 60 (1999) 4954.

\bibitem{jul95}
F.~J\"ulicher, J.~Prost, Cooperative molecular motors, Phys. Rev. Lett. 75
  (1995) 2618.

\bibitem{buc00}
J.~Buceta, J.~M. Parrondo, C.~{Van den Broeck}, F.~J. de~la Rubia, Negative
  resistance and anomalous hysteresis in a collective molecular motor, Phys.
  Rev. E 61 (2000) 6287.

\bibitem{lip00a}
R.~Lipowsky, T.~Harms, Molecular motors and nonuniform ratchets, Eur. Biophys.
  J. 29 (2000) 542.

\bibitem{bie96b}
M.~Bier, R.~D. Astumian, Biased {B}rownian motors as the operating principle
  for microscopic engines, Bioelectrochem. Bioenerg. 39 (1996) 67.

\bibitem{der99a}
I.~Der\'enyi, M.~Bier, R.~D. Astumian, Generalized efficiency and its
  application to microscopic engines, Phys. Rev. Lett. 83 (1999) 903.

\bibitem{bie01}
M.~Bier, Motor proteins: mechanochemical energy transduction on the microscopic
  scale, Acta Physica Polonica B 32 (2001) 287.

\bibitem{her01}
A.~C. Hernandez, A.~Medina, J.~M.~M. Roco, J.~A. White, S.~Velasco, Unified
  optimization criterion for energy converters, Phys. Rev. E 63 (2001) 037102.

\bibitem{hum01b}
T.~Humphrey, R.~Newbury, R.~Taylor, H.~Linke, Reversible quantum heat engines,
  {S}ubmitted for publication.

\bibitem{cur75}
F.~L. Curzon, B.~Ahlborn, Efficiency of a {C}arnot engine at maximum power
  output, Am. J. Phys. 43 (1975) 22.

\bibitem{and83}
B.~Andresen, Finite-Time Thermodynamics, University of Copenhagen Press,
  Copenhagen, 1983.

\bibitem{ber99}
R.~S. Berry, V.~A. Kazakov, S.~Sieniutycz, Z.~Szwast, A.~M. Tsirlin,
  Thermodynamics Optimization of Finite-Time Processes, Wiley, Chichester,
  1999.

\bibitem{how97a}
J.~Howard, Molecular motors: structural adaptation to cellular functions,
  Nature 389 (1997) 561.

\bibitem{meh99}
A.~D. Mehta, M.~Rief, J.~A. Spudich, D.~A. Smith, R.~M. Simmons,
  Single-molecule biomechanics with optical methods, Science 283 (1999) 1689.

\bibitem{mei89}
M.~Meister, S.~R. Caplan, H.~C. Berg, Dynamics of a tightly coupled mechanism
  for flagellar rotation, Biophys. J. 55 (1989) 905.

\bibitem{doe95a}
C.~Doering, B.~Ermentrout, G.~Oster, Rotary {DNA} motors, Biophys. J. 69 (1995)
  2256.

\bibitem{els97}
T.~C. Elston, G.~Oster, Protein turbines. {I}: {T}he bacterial flagellar motor,
  Biophys. J. 73 (1997) 703.

\bibitem{els98}
T.~Elston, H.~Wang, G.~Oster, Energy transduction in {ATP} synthase, Nature 391
  (1998) 510.

\bibitem{ber98}
H.~C. Berg, Keeping up with the {F}$_1$-{ATP}ase, Nature 394 (1998) 324.

\bibitem{bormovie}
{\tt http://www.borisylab.nwu.edu/pages/supplemental/mtfr.html}.

\bibitem{how97b}
J.~Howard, F.~Gittes, Motor proteins, in: H.~Flyvbjerg, J.~Hertz, M.~J. Jensen,
  O.~G. Mouritsen, K.~Sneppen (Eds.), Physics of Biological Systems; from
  Molecules to Species. Lecture Notes in Physics, Vol. 366, Springer, Berlin,
  1997, p. 155.

\bibitem{blo98}
S.~M. Block, Leading the procession: new insights into kinesin motors, J. Cell.
  Biol. 140 (1998) 1281.

\bibitem{svo93}
K.~Svoboda, C.~F. Schmidt, B.~J. Schnapp, S.~M. Block, Direct observation of
  kinesin stepping by optical trapping interferometry, Nature 365 (1993) 721.

\bibitem{gil95}
S.~P. Gilbert, M.~R. Webb, M.~Brune, K.~A. Johnson, Pathway of processive {ATP}
  hydrolysis by kinesin, Nature 373 (1995) 671.

\bibitem{gel88}
J.~Gelles, B.~J. Schnapp, M.~P. Sheetz, Tracking kinesin-driven movements with
  nanometer-scale precision, Nature 331 (1988) 450.

\bibitem{sab96}
E.~P. Sablin, F.~J. Kull, R.~Cooke, R.~D. Vale, R.~J. Fletterick, Crystal
  structure of the motor domain of the kinesin-related motor ncd, Nature 380
  (1996) 555.

\bibitem{fin94}
J.~T. Finer, R.~S. Simmons, J.~A. Spudich, Single myosin molecule mechanics:
  Piconewton forces and nanometer steps, Nature 368 (1994) 113.

\bibitem{cro97}
R.~A. Cross, Reversing the kinesin ratchet -- a diverting tail, Nature 389
  (1997) 15.

\bibitem{hen97}
U.~Henningsen, M.~Schliwa, Reversal in the direction of movement of a molecular
  motor, Nature 389 (1997) 93.

\bibitem{how89}
J.~Howard, A.~J. Hudspeth, R.~D. Vale, Movement of microtubules by single
  kinesin molecules, Nature 342 (1989) 154.

\bibitem{man99}
E.~Mandelkow, A.~Hoenger, Structure of kinesin and kinesin-microtubule
  interactions, Current Opinion in Cell Biology 11 (1999) 34.

\bibitem{svo94}
K.~Svoboda, S.~M. Block, Force and velocity measured for single kinesin
  molecules, Cell 77 (1994) 773.

\bibitem{fuk81}
K.~Fukui, The path of chemical reactions -- the {IRC} approach, Acc. Chem. Res.
  14 (1981) 363.

\bibitem{kop84}
H.~K\"oppel, W.~Domcke, L.~S. Cederbaum, Multimode molecular dynamics beyond
  the {B}orn-{O}ppenheimer approximation, Adv. Chem. Phys. 57 (1984) 59.

\bibitem{atk86}
P.~W. Atkins, Physical Chemistry, 3rd Edition, Oxford University Press, Oxford,
  1986.

\bibitem{tru92}
D.~G. Truhlar, Potential energy surfaces, in: Encyclopedia of Physical Science
  and Technology, Vol.~13, Academic Press, San Diego, 1992.

\bibitem{dau92}
R.~Daudel, Quantum chemistry, in: Encyclopedia of Physical Science and
  Technology, Vol.~13, Academic Press, San Diego, 1992.

\bibitem{mic92}
J.~Michl, Organic chemical systems, theory, in: Encyclopedia of Physical
  Science and Technology, Vol.~12, Academic Press, San Diego, 1992.

\bibitem{schu99}
C.~Sch\"utte, Conformational dynamics: modeling, theory, algorithm, and
  applications to biomolecules, Habilitation thesis, Konrad-Zuse-Zentrum f\"ur
  Informationstechnik Berlin (Germany), unpublished, 1999.

\bibitem{fra85}
H.~Frauenfelder, P.~G. Wolynes, Rate theories and puzzles of hemeprotein
  kinetics, Science 229 (1985) 337.

\bibitem{fra91}
H.~Frauenfelder, S.~G. Sligar, P.~G. Wolynes, The energy landscapes and motions
  of proteins, Science 254 (1991) 1598.

\bibitem{taw91}
K.~Tawada, K.~Sekimoto, Protein friction exerted by motor enzymes through a
  weak-binding interaction, J. Theor. Biol. 150 (1991) 193.

\bibitem{lip00b}
R.~Lipowsky, Molecular motors and stochastic models, in: J.~A. Freund,
  T.~P\"oschel (Eds.), Lecture Notes in Physics, Vol. 557, Springer, Berlin,
  2000.

\bibitem{par01}
A.~Parmeggiani, F.~J\"ulicher, L.~Peliti, J.~Prost, Detachment of molecular
  motors under tangential loading, {S}ubmitted for publication.

\bibitem{lat01}
G.~Lattanzi, A.~Maritan, Force dependence of the {M}ichaelis constant in a
  two-state ratchet model for molecular motors, Phys. Rev. Lett 86 (2001) 1134.

\bibitem{jul99}
F.~J\"ulicher, Force and motion generation of molecular motors: a generic
  description, in: S.~C. M\"uller, J.~Parisi, W.~Zimmermann (Eds.), Lecture
  Notes in Physics: Transport and structure: Their competitive roles in
  Biophysics and Chemistry, Springer, Berlin, 1999.

\bibitem{kik95}
M.~Kikkawa, T.~Ishikawa, T.~Wakabayashi, N.~Hirokawa, Three-dimensional
  structure of the kinesin head-microtubule complex, Nature 376 (1995) 274.

\bibitem{fox98}
R.~F. Fox, Rectified {B}rownian movement in molecular and cell biology, Phys.
  Rev. E 57 (1998) 2177.

\bibitem{spu94}
J.~A. Spudich, How molecular motors work, Nature 372 (1994) 515.

\bibitem{kit99}
K.~Kitamura, M.~Tokunaga, A.~H. Iwane, T.~Yanagida, A single myosin head moves
  along an actin filament with regular steps of 5.3 nanometers, Nature 397
  (1999) 129.

\bibitem{meh99a}
A.~D. Mehta, R.~S. Rock, M.~Rief, J.~A. Spudich, M.~S. Mooseker, R.~E. Cheney,
  Myosin-{V} is a processive actin-based motor, Nature 400 (1999) 590.

\bibitem{sch95}
B.~J. Schnapp, Two heads are better than one, Nature 373 (1995) 655.

\bibitem{sch97}
M.~J. Schnitzer, S.~M. Block, Kinesin hydrolyses one {ATP} per 8-$nm$ step,
  Nature 388 (1997) 386.

\bibitem{svo94b}
K.~Svoboda, P.~P. Mitra, S.~M. Block, Fluctuation analysis of motor protein
  movement and single enzyme kinetics, Proc. Natl. Acad. Sci. USA 91 (1994)
  11782.

\bibitem{sch95b}
M.~J. Schnitzer, S.~M. Block, Statistical kinetics of processive enzymes, Cold
  Spring Harb. Symp. Quant. Biol. 60 (1995) 793.

\bibitem{hux71}
A.~F. Huxley, R.~M. Simmons, Proposed mechanism of force generation in striated
  muscle, Nature 233 (1971) 533.

\bibitem{smi87}
D.~A. Smith, S.~Sicilia, The theory of sliding filament models for muscle
  contraction. {I}. {T}he two-state model, J. Theor. Biol 127 (1987) 1.

\bibitem{pat89}
E.~Pate, R.~Cooke, A model of crossbridge action: the effecs of {ATP}, {ADP}
  and {P}$_i$, J. Muscle Res. Cell Motil. 10 (1989) 181.

\bibitem{pat91}
E.~Pate, R.~Cooke, Simulation of stochastic processes in motile crossbridge
  systems, J. Muscle Res. Cell Motil. 12 (1991) 376.

\bibitem{pat93}
E.~Pate, R.~Cooke, H.~White, Determination of the myosin step size from
  mechanical and kinetic data, Proc. Natl. Acad. Sci. USA 90 (1993) 2451.

\bibitem{sek95}
K.~Sekimoto, K.~Tawada, Extended time correlation of in vitro motility by motor
  protein, Phys. Rev. Lett. 75 (1995) 180.

\bibitem{tho98}
N.~Thomas, R.~A. Thornhill, The physic of biological molecular motors, J. Phys.
  D 31 (1998) 253.

\bibitem{bar99}
C.~J. Barclay, A weakly coupled version of the {H}uxley crossbridge model can
  simulate energetics of amphibian and mammalian skeletal muscle, J. Muscle.
  Res. Cell M. 20 (1999) 163.

\bibitem{jul98}
F.~J\"ulicher, J.~Prost, Molecular motors: from individual to collective
  behavior, Prog. Theor. Phys. Supplement 130 (1998) 9.

\bibitem{mar99}
J.~L. Marin, M.~Huerta, J.~Muniz, X.~Trujillo, Comment on ``{C}oopertaive
  molecular motors'', Phys. Rev. Lett. 83 (1999) 5403.

\bibitem{jul99a}
F.~J\"ulicher, J.~Prost, J\"ulicher and {P}rost reply, Phys. Rev. Lett. 83
  (1999) 5404.

\bibitem{jul97b}
F.~J\"ulicher, J.~Prost, Spontaneous oscillations in collective molecular
  motors, Phys. Rev. Lett. 78 (1997) 4510.

\bibitem{riv98}
D.~Riveline, A.~Ott, F.~J\"ulicher, D.~Winkelmann, O.~Cardoso, J.~Lacapere,
  S.~Magnusdottir, J.~Viovy, L.~Gorre-Tallini, J.~Prost, Acting on actin: the
  electric motility assay, Eur. Biophys. J. 27 (1998) 403.

\bibitem{yas96}
K.~Yasuda, Y.~Shindo, S.~Ishiwata, Synchronous behavior of spontaneous
  oscillations of sacromeres in skeletal myofibrils under isotonic conditions,
  Biophys. J. 70 (1996) 1823.

\bibitem{fuj98}
H.~Fujita, S.~Ishiwata, Spontaneous oscillatory contraction without regulatory
  proteins in actin filament-reconstituted fibres, Biophys. J. 75 (1998) 1439.

\bibitem{cam99}
S.~Camalet, F.~J\"ulicher, J.~Prost, Self-organized beating and swimming of
  internally driven filaments, Phys. Rev. Lett. 82 (1999) 1590.

\bibitem{win80}
A.~T. Winfree, The Geometry of Biological Time, Springer, Berlin, 1980.

\bibitem{nic81}
G.~Nicolis, I.~Prigogine, Self Organization in Nonequilibrium Systems,
  Springer, Berlin, 1981.

\bibitem{kur84}
Y.~Kuramoto, Chemical Oscillations, Waves, and Turbulence, Springer, Berlin,
  1984.

\bibitem{vid88}
C.~Vidal, A.~Pacault, Nonequilibrium Dynamics in Chemical Systems, Springer,
  Berlin, 1988.

\bibitem{gla88}
L.~Glass, M.~C. Mackey, From Clocks to Chaos, Princeton University Press,
  Princeton NJ, 1988.

\bibitem{vil98}
A.~Vilfan, E.~Frey, F.~Schwabl, Elastically coupled molecular motors, Eur.
  Phys. J. B 3 (1998) 535.

\bibitem{vil99}
A.~Vilfan, E.~Frey, F.~Schwabl, Force-velocity relation for a two-state
  crossbridge model for molecular motors, Europhys. Lett. 45 (1999) 283.

\bibitem{vilmovie}
{\tt http://www.physik.tu-muenchen.de/\verb+~+avilfan/ecmm/}.

\bibitem{der98}
I.~Der\'enyi, T.~Vicsek, Realistic models of biological motion, Physica A 249
  (1998) 397.

\bibitem{hua97}
W.~Hua, E.~C. Young, M.~L. Fleming, J.~Gelles, Coupling of kinesin steps to
  {ATP} hydrolysis, Nature 388 (1997) 390.

\bibitem{cop97}
C.~M. Coppin, D.~W. Pierce, L.~Hsu, R.~D. Vale, The load dependence of
  kinesin's mechanical cycle, Proc. Natl. Acad. Sci. USA 94 (1997) 8539.

\bibitem{ric99}
S.~Rice, A.~W. Lin, D.~Safer, C.~L. Hart, N.~Naber, B.~O. Carragher, S.~M.
  Cain, E.~Pechatnikova, E.~M. Wilson-Kubalek, M.~Wittaker, E.~Pate, R.~Cooke,
  E.~W. Taylor, R.~A. Milligan, R.~D. Vale, A structural change in the kinesin
  motor protein that drives motility, Nature 402 (1999) 778.

\bibitem{gell95}
J.~Gelles, E.~Berliner, E.~C. Young, H.~K. Mahtani, B.~Perez-Ramirez,
  K.~Anderson, Structural and functional features of one- and two-headed
  biotinated kinesin derivatives, Biophys. J. 68 (1995) 276s.

\bibitem{ber95}
E.~Berliner, E.~C. Young, K.~Anderson, H.~K. Mahtani, J.~Gelles, Failure of a
  single-headed kinesin to track parallel to microtubule protofilaments, Nature
  373 (1995) 718.

\bibitem{val96}
R.~D. Vale, T.~Funatsu, D.~W. Pierce, L.~Romberg, Y.~Harada, T.~Yanagida,
  Direct observation of single kinesin molecules moving along microtubules,
  Nature 380 (1996) 451.

\bibitem{oka99}
Y.~Okada, N.~Hirokawa, A processive single-headed motor: kinesin superfamily
  protein {KIF1A}, Science 283 (1999) 1152.

\bibitem{okamovie}
{\tt http://www.sciencemag.org/feature/data/985876.shl}.

\bibitem{zho96}
H.-X. Zhou, Y.~Chen, Chemically driven motility of {B}rownian particles, Phys.
  Rev. Lett. 77 (1996) 194.

\bibitem{blo95}
S.~M. Block, K.~Svoboda, Analysis of high resolution recordings of motor
  movement, Biophys. J. 68 (1995) 230s.

\bibitem{vis99}
K.~Visscher, M.~J. Schnitzer, S.~M. Block, Single kinesin molecules studied
  with a molecular force clamp, Nature 400 (1999) 184.

\bibitem{str99}
G.~N. Stratopoulos, T.~E. Dialynas, G.~Tsironis, Directional {N}ewtonian motion
  and reversal of molecular motors, Phys. Lett. A 252 (1999) 151.

\bibitem{ast99a}
R.~D. Astumian, I.~Der\'enyi, A chemically reversible {B}rownian motor:
  application to kinesin and ncd, Biophys. J. 77 (1999) 993.

\bibitem{ast00a}
R.~D. Astumian, The role of thermal activation in motion and force generation
  by molecular motors, Phil. Trans. R. Soc. London B 355 (2000) 511.

\bibitem{lip00c}
R.~Lipowsky, Universal aspects of the chemomechanical coupling for molecular
  motors, Phys. Rev. Lett. 85 (2000) 4401.

\bibitem{lib98}
A.~Libchaber, Genome stability, cell motility, and force generation, Prog.
  Theor. Phys. Supplement 130 (1998) 1.

\bibitem{mey95}
E.~Meyerh\"ofer, J.~Howard, The force generated by a single kinesin molecule
  against an elastic load, Proc. Natl. Acad. Sci. USA 92 (1995) 574.

\bibitem{hou01}
A.~Houdusse, H.~L. Sweeney, Myosin motors: missing structures and hidden
  springs, Curr. Opin. Struct. Biol. 11 (2001) 182.

\bibitem{oos95}
F.~Oosawa, Sliding and {ATP}ase, J. Biochem. 118 (1995) 863.

\bibitem{oos00}
F.~Oosawa, The loose coupling mechanism in molecular machines of living cells,
  Genes to Cells 5 (2000) 9.

\bibitem{yan00}
T.~Yanagida, K.~Kitamura, H.~Tanaka, A.~H. Iwane, S.~Esaki, Single molecule
  analysis of the actomyosin motor, Curr. Opin. Cell Biol. 12 (2000) 20.

\bibitem{tay93}
E.~W. Taylor, Variations on the theme of movement, Nature 361 (1993) 115.

\bibitem{dum79}
R.~D\"umcke, H.~Spohn, The proper form of the generator in the weak coupling
  limit, Z. Phys. B 34 (1979) 419.

\bibitem{tal86}
P.~Talkner, The failure of the quantum regression hypothesis, Ann. Phys. (New
  York) 167 (1986) 390.

\bibitem{amb91}
V.~Ambegaokar, Quantum {B}rownian motion and its classical limit, Ber.
  Bunsenges. Phys. Chem. 95 (1991) 400.

\bibitem{for96}
G.~W. Ford, R.~F. O'Connell, There is no quantum regression theorem, Phys. Rev.
  Lett. 77 (1996) 798.

\bibitem{gnu96}
S.~Gnutzmann, F.~Haake, Positivity violation and initial slips in open systems,
  Z. Phys. B 101 (1996) 263.

\bibitem{cap99}
V.~Capek, T.~Mancal, Isothermal {M}axwell daemon as a molecular rectifier,
  Europhys. Lett. 48 (1999) 365.

\bibitem{gra88a}
H.~Grabert, P.~Schramm, G.-L. Ingold, Quantum {B}rownian motion: The functional
  integral approach, Phys. Rep. 168 (1988) 115.

\bibitem{ben94}
V.~A. Benderskii, D.~E. Makarov, C.~A. Wight, Chemical dynamics at low
  temperatures, Adv. Chem. Phys. 88 (1994) 1.

\bibitem{ing98}
T.~Dittrich, P.~H\"anggi, G.-L. Ingold, B.~Kramer, G.~Sch\"on, W.~Zwerger,
  Quantum transport and dissipation, Wiley-VCH Verlag GmBH, Weinheim, 1998,
  Ch.~4.

\bibitem{gri98}
M.~Grifoni, P.~H\"anggi, Driven quantum tunneling, Phys. Rep. 304 (1998) 229.

\bibitem{pap65}
A.~Papoulis, Probability, random variables, and stochastic processes, McGraw
  Hill, New York, 1965.

\bibitem{sch90}
G.~Sch\"on, A.~D. Zaikin, Quantum coherent effects, phase transitions and the
  dissipative dynamics of ultra small tunnel junctions, Phys. Rep. 198 (1990)
  237.

\bibitem{con96}
R.~F. O'Connell, Dissipative and fluctuation phenomena in quantum mechanics
  with applications, Int. J. Quant. Chem. 58 (1996) 569.

\bibitem{sen60}
I.~R. Senitzky, Dissipation in quantum mechanics. {T}he harmonic oscillator,
  Phys. Rev. 119 (1960) 670.

\bibitem{lam00}
H.~Lamb, On a peculiarity of the wave-system due to the free vibrations of a
  nucleus in an extended medium, Proc. London. Math. Soc. 32 (1900) 208.

\bibitem{kam51}
N.~G. van Kampen, Contribution to the quantum theory of light scattering, Dan.
  Mat. Fys. Medd. 26 (no. 15) (1951) 1.

\bibitem{sch64}
F.~Schwabl, W.~Thirring, Quantum theory of laser radiation, Ergeb. exakt.
  {N}aturwiss. 36 (1964) 219.

\bibitem{bog45}
N.~N. Bogolyubov, Elementary example for establishing statistical equilibrium
  in a system coupled to a thermostat, in: On some statistical methods in
  mathematical physics, Publ. Acad. Sci. Ukr. SSR, 1945, pp. 115--137, [In
  Russian].

\bibitem{rei97}
P.~Reimann, M.~Grifoni, P.~H\"anggi, Quantum ratchets, Phys. Rev. Lett. 79
  (1997) 10.

\bibitem{rei97a}
P.~Reimann, M.~Grifoni, P.~H\"anggi, Adiabatically rocked quantum ratchets, in:
  J.~B. Kadtke, A.~Bulsara (Eds.), Applied Nonlinear Dynamics and Stochastic
  Systems near the Millennium, AIP Proc. 411, 1997.

\bibitem{gra87}
H.~Grabert, P.~Olschowski, U.~Weiss, Quantum rates for dissipative systems at
  finite temperatures, Phys. Rev. B 36 (1987) 1931.

\bibitem{fri86a}
E.~Freidkin, P.~S. Riseborough, P.~H\"anggi, Quantum tunneling at low
  temperatures: results for weak damping, Z. Phys. B 64 (1986) 237, erratum: Z.
  Phys. B 67 (1987) 271.

\bibitem{hon88}
P.~H\"anggi, W.~Hontscha, Unified approach to the quantum-{K}ramers reaction
  rate, J. Chem. Phys 88 (1988) 4094.

\bibitem{jor97}
S.~Jorda, Quanten auf der {K}ippratsche, Phys. Bl. 53 (1997) 975.

\bibitem{lin00}
H.~Linke, Von {D}\"amonen und {E}lektronen, Phys. Bl. 56 (2000) 45.

\bibitem{wir00}
A farewell to wire?, Wire Industry, March issue (2000) 137--140.

\bibitem{hol96}
M.~Holthaus, D.~W. Hone, Localization effects in ac-driven tight-binding
  lattices, Philos. Mag. B 74 (1996) 105.

\bibitem{goy98}
I.~Goychuk, M.~Grifoni, P.~H\"anggi, Nonadiabatic quantum {B}rownian
  rectifiers, Phys. Rev. Lett. 81 (1998) 649, erratum: Phys. Rev. Lett. 81
  (1998) 2837.

\bibitem{goy98b}
I.~Goychuk, P.~H\"anggi, Quantum rectifiers from harmonic mixing, Europhys.
  Lett. 43 (1998) 503.

\bibitem{goyp}
I.~Goychuck, private communication.

\bibitem{yuk97}
S.~Yukawa, M.~Kikuchi, G.~Tatara, H.~Matsukawa, Quantum ratchets, J. Phys. Soc.
  Jpn. 66 (1997) 2953.

\bibitem{ron98}
R.~Roncaglia, G.~Tsironis, Discrete quantum motors, Phys. Rev. Lett. 81 (1998)
  10.

\bibitem{yuk00}
S.~Yukawa, G.~Tatara, M.~Kikuchi, H.~Matsukawa, Quantum ratchet, Physica B
  284-288 (2000) 1896.

\bibitem{tat98}
G.~Tatara, M.~Kikuchi, S.~Yukawa, H.~Matsukawa, Dissipation enhanced asymmetric
  transport in quantum ratchets, J. Phys. Soc. Jpn. 67 (1998) 1090.

\bibitem{lin01a}
H.~Linke, T.~E. Humphrey, R.~P. Taylor, A.~P. Micolich, R.~Newbury, Chaos in
  quantum ratchets, Physica Scripta T90 (2001) 54.

\bibitem{amb82}
V.~Ambegaokar, U.~Eckern, G.~Sch\"on, Quantum dynamics of tunneling between
  superconductors, Phys. Rev. Lett. 48 (1982) 1745.

\bibitem{lar83}
A.~I. Larkin, Y.~N. Ovchinikov, Decay of supercurrent in tunnel junctions,
  Phys. Rev. B 28 (1983) 6281.

\bibitem{eck84}
U.~Eckern, G.~Sch\"on, V.~Ambegaokar, Quantum dynamics of a superconducting
  tunnel junction, Phys. Rev. B 30 (1984) 6419.

\bibitem{let77}
V.~S. Letokhov, V.~G. Minogin, B.~D. Pavlik, Cooling and capture of atoms and
  molecules by a resonant light field, Sov. Phys. JETP 45 (1977) 698.

\bibitem{hem93}
A.~Hemmerich, T.~W. H\"ansch, Two-dimensional atomic crystals bound by light,
  Phys. Rev. Lett. 70 (1993) 410.

\bibitem{gry93}
G.~Grynberg, B.~Lounis, P.~Verkerk, J.-Y. Courtois, C.~Salomon, Quantized
  motion of cold {C}esium atoms in two- and three-dimensional optical
  potentials, Phys. Rev. Lett. 70 (1993) 2249.

\bibitem{pre93}
M.~G. Prentiss, Bound by light, Science 260 (1993) 1078.

\bibitem{wil96}
S.~R. Wilkinson, C.~F. Bharucha, K.~W. Madison, Q.~Niu, M.~G. Raizen,
  Observation of atomic {W}annier-{S}tark ladders in an accelerating optical
  potential, Phys. Rev. Lett. 76 (1996) 4512.

\bibitem{all80}
C.~L. Allyn, A.~C. Gossard, W.~Wiegmann, A new rectifying semiconductor
  structure by molecular epitaxy, Appl. Phys. Lett. 36 (1980) 373.

\bibitem{car83}
F.~Capasso, S.~Luryi, W.~T. Tsang, C.~G. Bethea, B.~F. Levine, New transient
  electrical polarization phenomenon in sawtooth superlattices, Phys. Rev.
  Lett. 51 (1980) 2318.

\bibitem{ign94}
A.~A. Ignatov, E.~Schomburg, K.~F. Renk, W.~Schatz, J.~F. Palmier, F.~Mollot,
  Response of a {B}loch oscillator to a {THz}-field, Ann. Phys. (Leipzig) 3
  (1994) 137.

\bibitem{kea95a}
B.~J. Keay, S.~J. {Allen Jr.}, J.~Galan, J.~P. Kaminski, K.~L. Campman, A.~C.
  Gossard, U.~Bhattacharya, M.~J.~W. Rodwell, Photon-assisted electric field
  domains and multiphoton-assisted tunneling in semiconductor superlattices,
  Phys. Rev. Lett. 75 (1995) 4098.

\bibitem{kea95b}
B.~J. Keay, S.~Zeuner, S.~J. {Allen Jr.}, K.~D. Maranowski, A.~C. Grossard,
  U.~Bhattacharya, M.~J.~W. Rodwell, Dynamic localization, absolute negative
  conductance and stimulated multiphoton emission in sequential resonant
  tunneling semiconductor superlattices, Phys. Rev. Lett. 75 (1995) 4102.

\bibitem{maj01}
J.~B. Majer, M.~Grifoni, M.~Tusveld, J.~E. Mooij, Quantum ratchet effect for
  vortices, {S}ubmitted for publication.

\bibitem{lin98}
H.~Linke, W.~Sheng, A.~L\"ofgren, H.~Xu, P.~Omling, P.~E. Lindelof, A quantum
  dot ratchet: experiment and theory, Europhys. Lett. 44 (1998) 341, erratum:
  Europhys. Lett. 45 (1999) 406.

\bibitem{lin99b}
H.~Linke, Experimental quantum ratchets based on solid state nanostructures,
  Aust. J. Phys. 52 (1999) 895.

\bibitem{lin99c}
H.~Linke, H.~Xu, A.~L\"ofgren, W.~Sheng, A.~Svensson, P.~Omling, P.~E.
  Lindelof, R.~Newbury, R.~P. Taylor, Voltage and temperature limits for the
  operation of a quantum dot ratchet, Physica B 272 (1999) 61.

\bibitem{lin99d}
H.~Linke, W.~Sheng, A.~L\"ofgren, A.~Svensson, H.~Xu, P.~Omling, P.~E.
  Lindelof, Electron quantum dot ratchets, Microelectr. Eng. 47 (1999) 265.

\bibitem{rau99}
M.~Rauner, Einbahnstrasse {Q}uantenpunkt, Phys. Bl. 55 (1999) 16.

\bibitem{lin99a}
H.~Linke, T.~E. Humphrey, A.~L\"ofgren, A.~O. Sushkov, R.~Newbury, R.~P.
  Taylor, P.~Omling, Experimental tunneling ratchets, Science 286 (1999) 2314.

\bibitem{hum99}
T.~Humphrey, A numerical simulation of a quantum ratchet, Master's thesis,
  University of New South Wales (Sydney, Australia), unpublished, 1999.

\bibitem{lin01b}
H.~Linke, P.~Omling, From linear to non-linear transport in asymmetric
  mesoscopic devices, Acta Physica Polonica B 32 (2001) 267.

\bibitem{hum01a}
T.~Humphrey, H.~Linke, R.~Newbury, Pumping heat with quantum ratchets,
  {S}ubmitted for publication.

\bibitem{por00a}
M.~Porto, M.~Urbakh, J.~Klafter, Atomic scale engines: cars and wheels, Phys.
  Rev. Lett. 84 (2000) 6058.

\bibitem{por01}
M.~Porto, Atomic scale engines: taking a turn, Acta Physica Polonica B 32
  (2001) 295.

\bibitem{zhe01}
Z.~Zheng, G.~Hu, B.~Hu, Collective directional transport in coupled nonlinear
  oscillators without external bias, Phys. Rev. Lett. 86 (2001) 2273.

\bibitem{der95}
I.~Der\'enyi, T.~Vicsek, Cooperative transport of {B}rownian particles, Phys.
  Rev. Lett. 75 (1995) 374.

\bibitem{mar96}
F.~Marchesoni, Thermal ratchets in 1$+$1 dimensions, Phys. Rev. Lett. 77 (1996)
  2364.

\bibitem{csa97}
Z.~Csahok, F.~Family, T.~Vicsek, Transport of elastically coupled particles in
  an asymmetric periodic potential, Phys. Rev. E 55 (1997) 5179.

\bibitem{sav97a}
A.~V. Savin, G.~Tsironis, A.~Zolotaryuk, Ratchet and switching effects in
  stochastic kink dynamics, Phys. Lett. A 229 (1997) 279.

\bibitem{sav97b}
A.~V. Savin, G.~Tsironis, A.~Zolotaryuk, Reversal effects in stochastic kink
  dynamics, Phys. Rev. E 56 (1997) 2457.

\bibitem{zol99}
A.~V. Zolotaryuk, P.~L. Christiansen, B.~Norden, A.~V. Savin, Soliton and
  ratchet motions in helices, Cond. Mat. Phys. 2 (1999) 293.

\bibitem{der98c}
I.~Der\'enyi, P.~Tegzes, T.~Vicsek, Collective transport in locally asymmetric
  periodic structures, Chaos 8 (1998) 657.

\bibitem{far99}
Z.~Farkas, P.~Tegzes, A.~Vukics, T.~Vicsek, Transitions in the horizontal
  transport of vertically vibrated granular layers, Phys. Rev. E 60 (1999)
  7022.

\bibitem{rapmovie}
{\tt http://www.ph.biu.ac.il/\verb+~+rapaport/java-apps/vibseg.html}.

\bibitem{kur75}
Y.~Kuramoto, Self-entrainment of a population of coupled non-linear
  oscillators, in: Lecture Notes in Physics, Vol. 39, Springer, Berlin, 1975,
  p. 420.

\bibitem{str94}
S.~H. Strogatz, Norbert {W}iener's brain waves, in: S.~Levine (Ed.), Frontiers
  in Mathematical Biology, Springer, Berlin, 1994.

\bibitem{hau97}
R.~H\"aussler, R.~Bartussek, P.~H\"anggi, Coupled {B}rownian rectifiers, in:
  J.~B. Kadtke, A.~Bulsara (Eds.), Applied Nonlinear Dynamics and Stochastic
  Systems near the Millennium, AIP Proc. 411, 1997.

\bibitem{mar87}
P.~Martinoli, P.~Lerch, C.~Leemann, H.~Beck, Arrays of {J}osephson junctions:
  Model systems for two-dimensional physics, Jpn. J. Appl. Phys. supplement
  26-3 (1987) 1999.

\bibitem{agh99}
Y.~Aghababaie, G.~Menon, M.~Plischke, Universal properties of {B}rownian
  motors, Phys. Rev. E 59 (1999) 2578.

\bibitem{rei99a}
P.~Reimann, R.~Kawai, C.~{Van den Broeck}, P.~H\"anggi, Coupled {B}rownian
  motors: Anomalous hysteresis and zero-bias negative conductance, Europhys.
  Lett. 45 (1999) 545.

\bibitem{rei99b}
P.~Reimann, C.~{Van den Broeck}, R.~Kawai, Nonequilibrium noise in coupled
  phase oscillators, Phys. Rev. E 60 (1999) 6402.

\bibitem{vdb00}
C.~{Van den Broeck}, I.~Bena, P.~Reimann, J.~Lehmann, Ann. Phys. (Leipzig) 9
  (2000) 713.

\bibitem{man01}
S.~E. Mangioni, R.~R. Deza, H.~S. Wio, Transition from anomalous to normal
  hysteresis in a system of coupled {B}rownian motors: a mean-field approach,
  Phys. Rev. E 63 (2001) 041115.

\bibitem{cle01}
B.~Cleuren, C.~{Van den Broeck}, Ising model for a {B}rownian donkey, Europhys.
  Lett. 54 (2001) 1.

\bibitem{ale98}
K.~Alekseev, E.~Cannon, J.~McKinney, F.~Kusmartsev, D.~Campbell, Spontaneous dc
  current generation in a resistively shunted semiconductor superlattice driven
  by a terahertz field, Phys. Rev. Lett. 80 (1998) 2669.

\bibitem{can00}
E.~H. Cannon, F.~V. Kusmartsev, K.~N. Alekseev, D.~K. Cambell, Absolute
  negative conductivity and spontaneous current generation in semiconductor
  superlattices with hot electrons, Phys. Rev. Lett. 85 (2000) 1302.

\bibitem{des78}
R.~C. Desai, R.~Zwanzig, Statistical mechanics of a nonlinear stochastic model,
  J. Stat. Phys. 19 (1978) 1.

\bibitem{daw83}
D.~A. Dawson, Critical dynamics and fluctuations for a mean-field model of
  cooperative behavior, J. Stat. Phys. 31 (1983) 29.

\bibitem{bon87}
L.~L. Bonilla, Stable nonequilibrium probability densities and phase
  transitions for mean-field models in the thermodynamic limit, J. Stat. Phys.
  46 (1987) 659.

\bibitem{str91}
S.~Strogatz, R.~Mirollo, Stability of incoherence in a population of coupled
  oscillators, J. Stat. Phys. 63 (1991) 613.

\bibitem{gar92}
J.~Garcia-Ojalvo, J.~M. Sancho, L.~Ramirez-Piscina, A nonequilibrium phase
  transition with colored noise, Phys. Lett. A 168 (1992) 35.

\bibitem{vdb94}
C.~{Van den Broeck}, J.~M.~R. Parrondo, R.~Toral, Noise induced nonequilibrium
  phase transitions, Phys. Rev. Lett. 73 (1994) 3395.

\bibitem{bek94}
A.~Becker, L.~Kramer, Linear stability analysis for bifurcations in spatially
  extended systems with fluctuating control parameter, Phys. Rev. Lett. 73
  (1994) 955.

\bibitem{ram95}
S.~Ramaswami, Comment on ``{N}oise-induced nonequilibrium phase transitions'',
  Phys. Rev. Lett. 75 (1995) 4786.

\bibitem{gri96}
G.~Grinstein, M.~A. Munoz, Y.~Tu, Phase structure of systems with
  multiplicative noise, Phys. Rev. Lett. 76 (1996) 4376.

\bibitem{kim97}
S.~Kim, S.~H. Park, C.~S. Ryu, Noise-enhanced multistability in coupled
  oscillator systems, Phys. Rev. Lett. 78 (1997) 1616.

\bibitem{mul97}
R.~M\"uller, K.~Lippert, A.~K\"uhnel, U.~Behn, First-order nonequilibrium phase
  transition in a spatially extended system, Phys. Rev. E 56 (1997) 2658.

\bibitem{man97}
S.~Mangioni, R.~Deza, H.~S. Wio, R.~Toral, Disordering effects of colored noise
  in nonequilibrium phase transitions induced by multiplicative noise, Phys.
  Rev. Lett. 79 (1997) 2389.

\bibitem{zai98}
A.~A. Zaikin, L.~Schimansky-Geier, Spatial patters induced by additive noise,
  Phys. Rev. E 58 (1998) 4355.

\bibitem{ryovideo}
{\tt http://www.kawai.phy.uab.edu/research/motor}.

\bibitem{ban71}
T.~J. Banys, I.~V. Parshelyunas, Y.~K. Pozhela, Absolute negative resistance of
  {G}allium-{A}rsenide in a strong microwave field, Sov. Phys. Semicond. 5
  (1972) 1727, [Fiz. Tekh. Poluprovodn. 5 (1971) 1990].

\bibitem{pav76}
V.~V. Pavlovich, E.~M. Epstein, Conductivity of a superlattice semiconductor in
  strong electric fields, Sov. Phys. Semicond. 10 (1976) 1196.

\bibitem{poh81}
J.~Pozhela, Plasma and Current Instabilities in Semiconductors, Pergamon Press,
  Oxford, 1981.

\bibitem{sol90}
T.~C. Sollner, E.~R. {B}rown, W.~D. Goodhue, H.~Q. Le, Microwave and
  millimeter-wave resonant-tunneling devices, in: F.~Carpasso (Ed.), Springer
  Series in Electronics and Photonics, Vol. 28: Physics of Quantum electron
  devices, Springer, Berlin, 1990.

\bibitem{ign95}
A.~A. Ignatov, E.~Schomburg, J.~Grenzer, K.~F. Renk, E.~P. Dodin, {THz}-field
  induced nonlinear transport and dc voltage generation in a semiconductor
  superlattice due to {B}loch oscillations, Z. Phys. B 98 (1995) 187.

\bibitem{dag95}
Y.~Dakhnovskii, H.~Metiu, Absolute negative resistance in double-barrier
  heterostructures in a strong laser field, Phys. Rev. B 51 (1995) 4193.

\bibitem{agu97}
R.~Aguado, G.~Platero, Dynamical localization and absolute negative conductance
  in an ac-driven double quantum well, Phys. Rev. B 55 (1997) 12860.

\bibitem{har97a}
L.~Hartmann, M.~Grifoni, P.~H\"anggi, Dissipative transport in dc-ac-driven
  tight-binding lattices, Europhys. Lett. 38 (1997) 497.

\bibitem{goy98c}
I.~Goychuk, E.~Petrov, V.~May, Noise-induced current reversal in a
  stochastically driven dissipative tight-binding model, Phys. Lett. A 238
  (1998) 59.

\bibitem{kro59}
H.~Kr\"omer, Proposed negative-mass microwave amplifier, Phys. Rev. 109 (1959)
  1856.

\bibitem{mat59}
D.~C. Mattis, M.~J. Stevenson, Theory of negative-mass cyclotron resonance,
  Phys. Rev. Lett. 3 (1959) 18.

\bibitem{lia80}
P.~F. Liao, A.~M. Glass, L.~M. Humphrey, Optically generated pseudo-{S}tark
  effect in ruby, Phys. Rev. B 22 (1980) 2276.

\bibitem{aro75}
A.~G. Aronov, B.~Z. Spivak, Photoeffect in a {J}osephson junction, JETP Lett.
  22 (1975) 101.

\bibitem{ger86}
M.~E. Gershenzon, M.~I. Falei, Absolute negative resistance of a tunnel contact
  between superconductors with a nonequilibrium quasiparticle distribution
  function, JETP Lett. 44 (1986) 682.

\bibitem{ger88}
M.~E. Gershenzon, M.~I. Falei, Absolute negative resistance in tunnel junctions
  of nonequilibrium superconductors, Sov. Phys. JETP 67 (1988) 389.

\bibitem{dya87}
N.~A. Dyatko, I.~V. Kochetov, A.~P. Napartovich, Absolute negative conductivity
  of a low-temperature plasma, Sov. Tech. Phys. Lett. 13 (1987) 610.

\bibitem{roz88}
Z.~Rozenberg, M.~Lando, M.~Rokni, On the possibility of steady state negative
  mobility in externally ionized gas mixtures, J. Phys. D 21 (1988) 1593.

\bibitem{gol89}
P.~M. Golovinskii, A.~I. Shchedrin, Weak-field absolute negative conductivity
  in the mixture {Xe:F}$_2$ ionized by a beam of fast electrons, Sov. Phys.
  Tech. Phys. 34 (1989) 159.

\bibitem{vdb98}
C.~{Van den Broeck}, R.~Kawai, Absorption-desorption phase transition induced
  by parametric modulation, Phys. Rev. E 57 (1998) 3866.

\bibitem{ben99}
I.~Bena, C.~{Van den Broeck}, Coupled parametric oscillators, Europhys. Lett.
  48 (1999) 498.

\bibitem{ala00}
T.~Alarcon, A.~Perez-Madrid, J.~M. Rubi, Energy transduction in periodically
  driven non-{H}ermitian systems, Phys. Rev. Lett. 85 (2000) 3995.

\bibitem{doe87}
C.~R. Doering, A stochastic partial differential equation with multiplicative
  noise, Phys. Lett. A 122 (1987) 133.

\bibitem{iba01}
M.~Ibanes, J.~Garcia-Ojalvo, R.~Toral, J.~M. Sancho, Noise-induced scenario for
  inverted phase diagrams, Phys. Rev. Lett. 87 (2001) 020601.

\bibitem{eic01}
R.~Eichhorn, P.~Reimann, P.~H\"anggi, Brownian motion with absolute negative
  mobility, {S}ubmitted for publication.

\bibitem{vdb01}
C.~{Van den Broeck}, unpublished.

\bibitem{shi86}
S.~Shinomoto, Y.~Kuramoto, Phase transitions in active rotator systems, Prog.
  Theor. Phys. 75 (1986) 1105.

\bibitem{str89}
S.~H. Strogatz, C.~M. Marcus, R.~M. Westervelt, R.~E. Mirollo, Collective
  dynamics of coupled oscillators with random pinning, Physica D 36 (1989) 23.

\bibitem{som91}
H.~Sompolinsky, D.~Golomb, D.~Kleinfeld, Cooperative dynamics in visual
  processing, Phys. Rev. A 43 (1991) 6990.

\bibitem{swi92}
J.~W. Swift, S.~Strogatz, K.~Wiesenfeld, Averaging of globally coupled
  oscillators, Physica D 55 (1992) 239.

\bibitem{han93}
D.~H.~G. Mato, C.~Meunier, Clustering and slow switching in globally coupled
  phase oscillators, Phys. Rev E 48 (1993) 3470.

\bibitem{are94}
A.~Arenas, C.~J. P\'erez-Vicente, Exact long-time behavior of a network of
  phase oscillators under random fields, Phys. Rev. E 50 (1994) 949.

\end{thebibliography}
\end{document}